\documentclass[11pt,a4paper,oneside]{book}
\usepackage{epsfig}
\usepackage{amssymb}
\usepackage{amsmath}
\usepackage{amsfonts}
\usepackage{dutchcal}
\usepackage{subfigure}
\usepackage{color}
\usepackage{multirow}
\usepackage{hyperref}
\usepackage{amssymb}
\usepackage{amsfonts}
\usepackage{epsf}
\usepackage{rotating}
\usepackage{graphicx}
\usepackage{setspace}
\usepackage{amsmath}
\usepackage{fancyhdr}
\usepackage{lineno}
\usepackage[english]{babel}
\usepackage{graphics}
\usepackage{hhline}
\usepackage[top=3cm,bottom=2.8cm,left=2.8cm,right=2.6cm]{geometry}
\usepackage{color}
\usepackage{multirow}
\def\beq{\begin{equation}}
\def\eeq{\end{equation}}
\newcommand{\be}{\begin{equation}}
\newcommand{\ee}{\end{equation}}
\def\bea{\begin{eqnarray}}
\def\eea{\end{eqnarray}}
\def\ksl{\hbox{\hbox{${k}$}}\kern-1.9mm{\hbox{${/}$}}}
\def\a{\alpha}
\def\b{\beta}

\def\g{\gamma}

\def\etmiss{\not\!\!{E_T}}
\def\ptmiss{\not\!\!{p_T}}

\def\th{\theta}
\def\slashed{\ds}
\def\ds#1{#1\kern-1ex\hbox{/}}
\newcommand{\eps}{\varepsilon}
\newcommand{\beqa}{\begin{eqnarray}}
\newcommand{\eeqa}{\end{eqnarray}}

\newcommand{\pd}{\partial}
\newcommand{\nn}{\nonumber}
\def\lsim{\raise0.3ex\hbox{$\;<$\kern-0.75em\raise-1.1ex\hbox{$\sim\;$}}} 

\def\gsim{\raise0.3ex\hbox{$\;>$\kern-0.75em\raise-1.1ex\hbox{$\sim\;$}}}

\newcommand{\HRule}{\rule{\linewidth}{1mm}}

\begin{document}
\pagestyle{myheadings}
\thispagestyle{empty}

\begin{titlepage}
\begin{center}
\includegraphics[scale=.12]{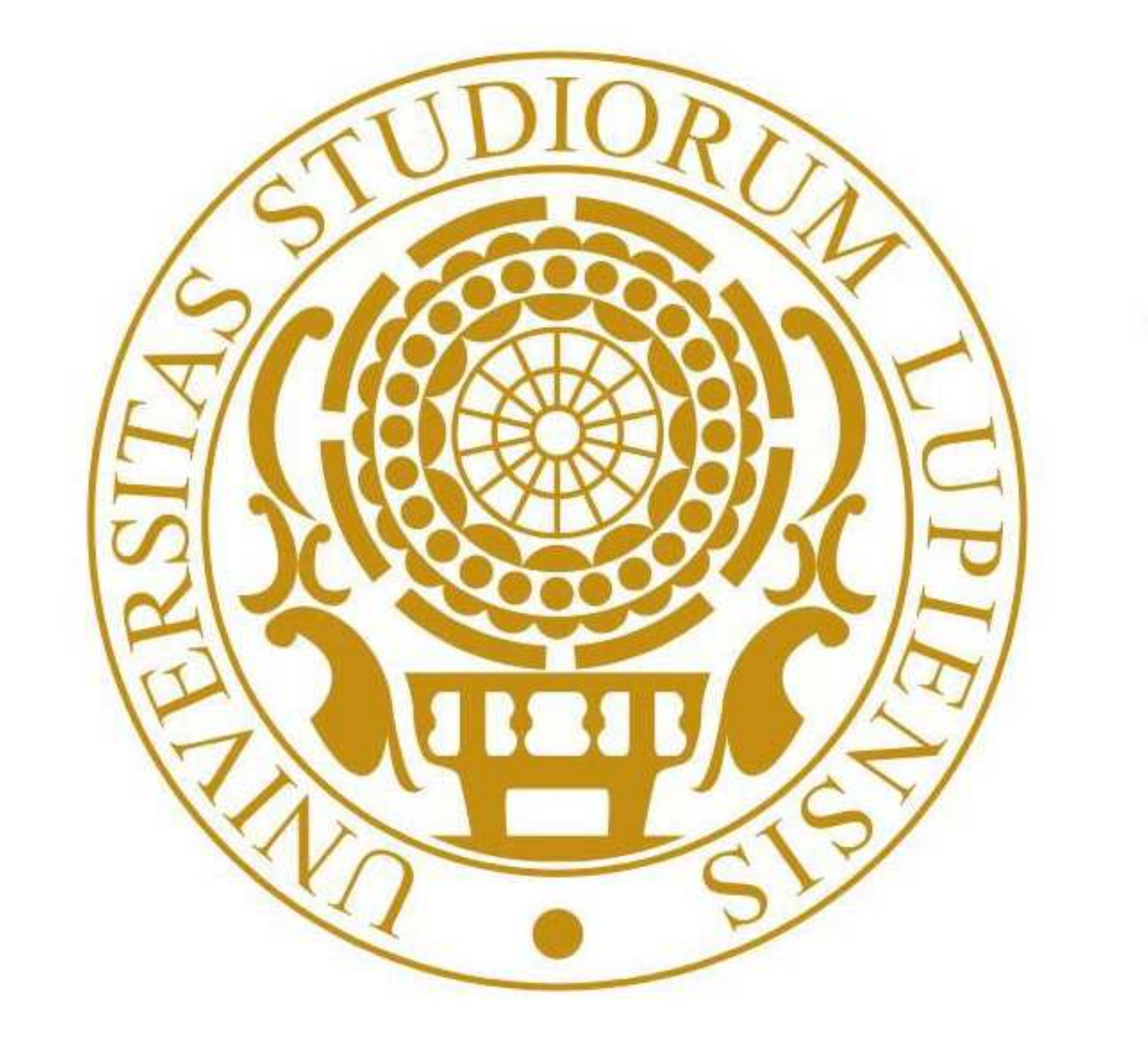}\\
\HRule\\[0.3cm]
\textrm{\LARGE \bf  Universit\`a del Salento}\\[0.3cm]
\textrm{\Large Facolt\`a di Scienze Matematiche, Fisiche e Naturali}\\[0.3cm]
\textrm{\Large Dipartimento di Matematica e Fisica ``Ennio De Giorgi''}\\[0.1cm]

\HRule \\[3.5cm]

\begin{spacing}{2}
{\Large { \bf{Studies on Conformal and Superconformal Extensions of the Standard Model with an Application to Gravity}}}\\[6cm]
\end{spacing}

% Author and supervisor
\begin{minipage}{0.4\textwidth}
\begin{flushleft}\large
\textit{Supervisor}\\
Prof.~Claudio Corian\`o
\end{flushleft}
\end{minipage}
\begin{minipage}{0.4\textwidth}
\begin{flushright} \large
\textit{Candidate}\\
Antonio Costantini
\end{flushright}
\end{minipage}\\[0.5cm]
\begin{minipage}{0.4\textwidth}
\begin{flushright} \large
\textit{}\\
\end{flushright}
\end{minipage}\\[4cm]
Tesi di Dottorato di Ricerca in Fisica -- XXVIII ciclo
\vfill

\end{center}
\end{titlepage}
\tableofcontents

\chapter*{List of Publications}
\centerline{ \bf Published and submitted articles}

\vspace{0.5cm}
\begin{description}
%\cite{Bandyopadhyay:2016fad}
\item
  P.~Bandyopadhyay, C.~Corian\`o, A.~Costantini and L.~Delle Rose\\
  \textit{Bounds on the Conformal Scale of a Minimally Coupled Dilaton and Multi-Leptonic Signatures at the LHC}\\
  arXiv:1607.01933 [hep-ph]
  %%CITATION = ARXIV:1607.01933;%%

%\cite{Bandyopadhyay:2016xgn}

%\cite{Bandyopadhyay:2015ifm}
\item
  P.~Bandyopadhyay, C.~Corian\`o and A.~Costantini\\
  \textit{A General Analysis of the Higgs Sector of the $Y=0$ Triplet-Singlet Extension of the MSSM at the LHC}\\
  arXiv:1512.08651 [hep-ph]. Accepted for publication in Phys. Rev. D (2016).
  %%CITATION = ARXIV:1512.08651;%%
  %2 citations counted in INSPIRE as of 08 Aug 2016  
  
%\cite{Bandyopadhyay:2015tva}
\item
  P.~Bandyopadhyay, C.~Corian\`o and A.~Costantini\\
  \textit{Probing the hidden Higgs bosons of the $Y = 0$ triplet- and singlet-extended Supersymmetric Standard Model at the LHC}\\
  JHEP {\bf 1512} (2015) 127,
  doi:10.1007/JHEP12(2015)127\\
  arXiv:1510.06309 [hep-ph]
  %%CITATION = doi:10.1007/JHEP12(2015)127;%%
  %3 citations counted in INSPIRE as of 08 Aug 2016
  
%\cite{Bandyopadhyay:2015oga}
\item
  P.~Bandyopadhyay, C.~Corian\`o and A.~Costantini\\
  \textit{Perspectives on a supersymmetric extension of the standard model with a Y = 0 Higgs triplet and a singlet at the LHC}\\
  JHEP {\bf 1509} (2015) 045,
  doi:10.1007/JHEP09(2015)045\\
  arXiv:1506.03634 [hep-ph]
  %%CITATION = doi:10.1007/JHEP09(2015)045;%%
  %4 citations counted in INSPIRE as of 08 Aug 2016  
  
 %\cite{Coriano:2015mva}
\item
  C.~Corian\`o, A.~Costantini, M.~Dell'Atti and L.~Delle Rose\\
  \textit{Neutrino and Photon Lensing by Black Holes: Radiative Lens Equations and Post-Newtonian Contributions}\\
  JHEP {\bf 1507} (2015) 160,
  doi:10.1007/JHEP07(2015)160\\
  arXiv:1504.01322 [hep-ph]
  %%CITATION = doi:10.1007/JHEP07(2015)160;%% 
  
%\cite{Costantini:2014gga}

%\cite{Coriano:2014gja}
\item
  C.~Corian\`o, A.~Costantini, L.~Delle Rose and M.~Serino\\
  \textit{Superconformal sum rules and the spectral density flow of the composite dilaton (ADD) multiplet in $\mathcal{N}=1$ theories}\\
  JHEP {\bf 1406} (2014) 136,
  doi:10.1007/JHEP06(2014)136\\
  arXiv:1402.6369 [hep-th]
  %%CITATION = doi:10.1007/JHEP06(2014)136;%%
  %4 citations counted in INSPIRE as of 08 Aug 2016  
  
\end{description}

\begin{description}
\centerline{\bf Conference Proceedings}
\item
  A.~Costantini, L.~Delle Rose and M.~Serino\\
  \textit{Sum rules and spectral density flow in QCD and in superconformal theories},
  EPJ Web Conf.\  {\bf 80} (2014) 00017,
  doi:10.1051/epjconf/20148000017\\
  arXiv:1409.5075 [hep-th]
  %%CITATION = doi:10.1051/epjconf/20148000017;%%  
 \item
  P.~Bandyopadhyay, C.~Corian\`o and A.~Costantini\\
  \textit{Higgs bosons: discovered and hidden, in extended Supersymmetric Standard Models at the LHC}\\
  arXiv:1604.00228 [hep-ph]
  %%CITATION = ARXIV:1604.00228;%%

\end{description}

\chapter*{Introduction}

Our description of the laws of Nature, from the point of view of fundamental physics, is tightly 
 connected with the concept of symmetry. In fact, symmetries play a crucial role in our understanding of the fundamental interactions, and the Standard Model of the elementary particles (SM), which provides the best phenomenological description of the subatomic world, is entirely built 
 around the two fundamental principles of Lorentz and of gauge symmetries, formulated according to a well established quantum field theory (QFT) language.\\
The gauge structure of the SM has been tested, by now, for almost half a century with an incredible success and it is expected that additional gauge simmetries will likely emerge, as the experimental search moves towards even higher energy. This search may not be quick and simple, 
rather it will require an extraordinary effort and several years to be formulated in a consistent framework, as shown by the enormous scientific effort which has already gone into the construction of the LHC and of the experimental detectors. \\ 
Given the daisy chain of successes built around the gauge principle, it is therefore obvious that, at theoretical level, the study of possible symmetric extensions of this model has to engage with the study of realistic patterns of their breakings and of their restorations. \\
The recent validation of the Higgs mechanism  for the generation of the mass of the particles of the SM, indicates that we should anchor our experimental efforts to the banks of this principle, keeping however an open view about possible unexpected events which may appear along the way. More exotic 
scenarios, with the appearence of new symmetries, unrelated to the gauge structure, such as the conformal symmetry, remain an open possibility and will be central to this work.

Theorists are still allowed wide options for molding new scenarios, since there are several and important features of the current phenomenology which find no answer within the SM. 
Such are the gauge hierarchy problem or the unsolved question of the origin of the masses of the light neutrinos, not to mention the issue of triviality for fundamental scalars at high energy.
At this time, we still do not have definitive results about the structure of the scalar field which appears in the Higgs mechanism, for instance regarding its fundamental or composite nature. Likewise we also cannot exclude that a new symmetry may be playing an important role in it. These considerations provide the leitmotif for the first part of this thesis work, which collects five studies of specific extensions of the SM dealing with direct analysis of the scalar sector of gauge theories. This part is organised as follows.\\
The first two chapters, one of theoretical and the the other of phenomenological nature, address the role of a dilaton and of conformal symmetries together with their breaking, at phenomenological level. A dilaton is a state 
that appears as a signature of the breaking of a conformal symmetry, and the approximate scale invariance of the SM as we run towards the UV, where one could reach a scale of symmetry 
restoration, gives realistic motivations for such analysis. \\
In particular, the issue whether a dilaton can be fundamental or composite remains quite open. These two chapters both address this specific topic though in two different contexts, 
the first one supersymmetric, where fundamental open questions about the nature of this state are addressed, and the other non supersymmetric and phenomenological. \\
A dilaton is the Nambu-Goldstone mode which results from the breaking of conformal symmetry and couples to the trace anomaly of the SM.
We just recall that the anomalous breaking of the conformal symmetry, as for a chiral symmetry, is characterised by the appearance of an intermediate state which interpolates between the anomalous current and the other (vector) currents of a trilinear correlator.  
As we show in chapter 2, the interaction of a dilaton with the anomaly provides a significant enhancement of the production rates of massless vector states in a typical trilinear vertex where this particle is involved. \\
Similarly, in the chiral case it is well known that such an anomalous interaction is responsible for the enhancement of the decay of the pion into two photons, and one wonders whether this  enhancement could be true on a more general basis, and valid for all the anomalous vertices. 
Chapter 1 gives an affirmative answer to this question, showing the presence of a theoretical link between chiral and conformal anomaly interactions on a very rigorous basis and by a direct perturbative analysis. 
This test, as we are going to show, can be directly performed in the case of a superconformal theory, which also provides the natural framework for addressing this point, thanks to the appearance of a superconformal anomaly multiplet. Here trace and chiral anomalies appear in a similar way. \\
We show that there is a complete analogy between the chiral and the conformal cases. In particular we prove that the origin of either a chiral or a conformal anomaly is in the emergence of effective scalar or pseudoscalar degrees of freedom in the corresponding effective action. This interpretation is supported by the identification of a sum rule - which is completely fixed by the anomaly - for all the components of a supersymmetric multiplet and by the appearance of what we call a "spectral density flow" as we reach the conformal limit. We show, using a mass deformation of the superconformal theory, that the form factors responsible for the generation of the anomalies exhibit the same spectral density, with a branch cut that turns into a pole as the mass deformation parameter is removed. In other words, the flow generates a sequence of spectral functions which become a delta function in the conformal limit, signalling the exchange of a massless state in the intermediate channel. The intermediate exchange of scalar or pseudoscalar states, identified as an axion, a dilaton and a dilatino, as mediator of either chiral or conformal anomalies, is therefore firmly proved. \\
The study of dilaton interactions is furtherly discussed in Chapter 2, where we investigate the current limits  on the possible discovery of a conformal scale at the LHC. An in depth phenomenological analysis of the possible channels mediated by an intermediate dilatons is presented, showing that such a scale is currently constrained 
by a lower bound of 5 TeV, which is not stringent enough to exclude it all together from the future experimental analysis at ATLAS and CMS. \\
In Chapters 3 and 4 we move to an analysis of a supersymmetric model with an extended Higgs sector, the TNMSSM, which is a scale invariant scalar theory with a Higgs superfield in a triplet representation of weak $SU(2)$. The model represents a significant departure respect to the NMSSM, manifesting a wider 
scalar spectrum which we have investigated in great detail. The inclusion of Higgs scalars belonging to higher representation of the gauge structure remains an open possibility of considerable theoretical and experimental interests, which the current and future analysis at the LHC have to confront.
The presence of a light pseudoscalar in its spectrum  is for sure one of its most significant features, which has been addressed in Chapter 4. 
The identification of the relevant region of parameter space of this model which is currently allowed at the LHC has been investigated by comparing the signal with direct simulations of the background. \\
Part 2 of this thesis can be read independently and presents an application of correlators which are affected by a conformal anomaly in a gravitational context. The same $TVV$ correlator (with T-denoting the stress-energy tensor of a gauge theory and V the vector current) which is studied in chapters 1 and 2, is used in the analysis of semiclassical lensing in gravity. As typical in the course of a theoretical analysis, fundamental results in a certain area may be quickly applied to other areas which seem to be apparently unrelated. Conformal anomalies are indeed fundamental, and the propagation of a photon in a curved gravitational background is affected by the same anomaly which shows up in the interaction of a dilaton at the LHC via its coupling to two photons. 
We just recall that the coupling of the SM to gravity occurs via the energy momentum tensor of the theory and that the dilatation current $J_D$ of a given theory has a divergence which is given by the trace of the same tensor. It is therefore not a big surprise that the TVV correlator can be used to describe the propagation of photons and neutrinos in gravitational backgrounds. This analysis hinges on a previous study in which it has been shown that the anomaly form factor of the $TVV$ vertex in the SM, where V is the electromagnetic current, induces a small change in Einstein's formula for the deflection, which has been quantified.  In this final chapter 
we develop a complete formulation of these radiative effects in the interaction of photons and neutrinos in a gravitational background extending the formalism of gravitational lensing to the semiclassical case. 

\part{ Theoretical and Phenomenological Aspects of a Superconformal theory}\label{pI}

\chapter{The Superconformal anomaly multiplet in an $\mathcal{N}=1$ theory}

\section{Synopsis} 
In this chapter we investigate a supersymmetric Yang-Mills theory in its superconformal phase and its corresponding anomalies. We show that there is a unifying feature of the chiral and conformal anomalies appearing in correlators involving the the Ferrara-Zumino supercurrent and two vectors supercurrent. These are characterised by the presence of massless anomaly poles in their related anomaly form factors. The states associated to these massless poles are interpreted as a dynamical realization of a scalar (dilaton), a pseudoscalar (an axion) and a fermion (dilatino) interpolating between such a current and the two vector currents of the anomaly vertex. 
The appearance of  a dilaton in a scale invariant theory is connected to the breaking of a conformal symmetry, a fact that can be simply illustrated with a realistic example. \\
A dilaton may appear in the spectrum of different extensions of the Standard Model not 
only as a result of the compactification of extra spacetime dimensions, but also as an effective state, related to the breaking of a 
dilatation symmetry.  The Standard Model is not a scale-invariant theory, but can be such in its defining classical Lagrangian if we slightly modify the scalar potential with the 
introduction of a dynamical field $\Sigma$. This extension allows to restore this symmetry, which must be broken at a certain scale, where $\Sigma$ acquires a vacuum expectation value. This task is accomplished by the replacement of every 
dimensionfull parameter $m$ of the defining Lagrangian according to the prescription $m \rightarrow m \frac{\Sigma}{\Lambda}$, where $\Lambda$ is the classical conformal 
breaking scale. Establishing the size of this scale is a fundamental issue which may require considerable effort at phenomenological level. \\
In the case of the SM, classical scale invariance can be easily restored by the simple change in the scalar potential briefly described above. This is defined modulo a constant, therefore we may consider, for instance, two equivalent choices 
\beqa
V_1(H, H^\dagger)&=& - \mu^2 H^\dagger H +\lambda(H^\dagger H)^2 =
\lambda \left( H^\dagger H - \frac{\mu^2}{2\lambda}\right)^2 - \frac{\mu^4}{4 \lambda}\nonumber \\
V_2(H,H^\dagger)&=&\lambda \left( H^\dagger H - \frac{\mu^2}{2\lambda}\right)^2
\eeqa
which generate two different scale-invariant extensions 
\beqa
V_1(H,H^\dagger, \Sigma)&=&- \frac{\mu^2\Sigma^2}{\Lambda^2} H^\dagger H +\lambda(H^\dagger H)^2 \nonumber \\
V_2(H,H^\dagger, \Sigma)&=& \lambda \left( H^\dagger H - \frac{\mu^2\Sigma^2}{2\lambda \Lambda^2}\right)^2 \,,
\eeqa 
where $H$ is the Higgs doublet, $\lambda$ is its dimensionless coupling constant, while $\mu$ has the dimension of a mass and, 
therefore, is the only term involved in the scale invariant extension.%
The invariance of the potential under the addition of constant terms, typical of any Lagrangian, is lifted once we 
require the presence of a dilatation symmetry. One can immediately check that only the second choice $(V_2)$ guarantees the existence of a stable ground state with a spontaneously 
broken phase. In $V_2$ we parameterize the Higgs, as usual, around the electroweak vev $v$ and indicate with $\Lambda$ the vev of the dilaton field $\Sigma = \Lambda + \rho$, 
setting all the Goldstone modes generated in the breaking of the gauge symmetry to zero, as 
customary in the unitary gauge. \\
The potential $V_2$ has a massless mode due to the existence of a flat direction. 
Performing a diagonalization of the mass matrix we define the two mass eigenstates $\rho_0$ and $h_0$, which are given by 
\beq
 \left( \begin{array}{c}
 {\rho_0}\\
  h_0 \\
  \end{array} \right)
 =\left( \begin{array}{cc}
\cos\alpha & \sin\alpha \\
-\sin\alpha & \cos\alpha  \\
 \end{array} \right)
 \left( \begin{array}{c}
  \rho\\
 {h} \\
  \end{array} \right)
\eeq
with 
\beq
\cos\alpha=\frac{1}{\sqrt{1 + v^2/\Lambda^2}}\qquad \qquad  \sin\alpha=\frac{1}{\sqrt{1 + \Lambda^2/v^2}}.
\eeq
We denote with ${\rho_0}$ the massless dilaton generated by this potential, while 
$h_0$ will describe a massive scalar, interpreted as a new Higgs field, whose mass is given by  
\beq 
m_{h_0}^2= 2\lambda v^2 \left( 1 +\frac{v^2}{\Lambda^2}\right) \qquad \textrm{with} \qquad v^2=\frac{\mu^2}{\lambda},
\eeq
and with $m_h^2=2 \lambda v^2$ being the mass of the Standard Model Higgs.
The Higgs mass, in  this case, is corrected by the new scale of the spontaneous breaking of the dilatation symmetry ($\Lambda$), 
which remains a free parameter. 
 
The vacuum degeneracy of the scale-invariant model  can be lifted by the introduction of 
extra (explicit breaking) terms which give a small mass to the dilaton field.
To remove such degeneracy, one can introduce, for instance, the term
\beq
\mathcal{L}_{break} 
= \frac{1}{2} m_{\rho}^2 {\rho}^2 + \frac{1}{3!}\, {m_{\rho}^2} \frac{{\rho}^3}{\Lambda} + \dots \, ,
\eeq
where $m_{\rho}$ represents the dilaton mass.

It is clear that in this approach the coupling of the dilaton to the anomaly has to be added by hand.
The obvious question to address, at this point, is if one can identify in the effective action of the Standard Model 
an effective state which may interpolate between the dilatation current of the same model and the final state with two
neutral currents, for example with two photons. Such a state can be identified in ordinary perturbation theory in the form of an anomaly pole.
We are entitled to interpret this scalar exchange as a composite state whose interactions with the rest of the Standard Model are 
defined by the conditions of scale and gauge invariance.
\subsection{Dilaton coupling to the anomaly}

We will show rigorously, in the supersymmetric case, that this state couples to the conformal anomaly by a direct analysis of the $J_DVV$ correlator, 
in the form of an anomaly pole, with $J_D$ and $V$ being the dilatation and the vector currents respectively. The correlator is extracted from the more general supersymmetric 3-point function involving the FZ current and two vector supercurrents, as already mentioned in the introduction. 

Poles  in a correlation function are usually there to indicate that a specific state can be created by a field operator in the Lagrangian of the theory, or, alternatively, as a composite particle of the same elementary fields. Obviously, a perturbative hint of the existence of such intermediate state does not correspond to a complete 
description of the state, in the same way as the discovery of an anomaly pole in the $AVV$ correlator of QCD (with $A$ being the 
axial current) is not equivalent to a proof of the existence of the pion. Nevertheless, massless poles extracted from the perturbative effective action do not appear for no reasons, and this should be sufficient to justify a more complete analysis of the 1-loop effective action of classical conformal theories. \\
Originally, the appearance of classical scalar degrees of freedom in the context of gravitational interactions has been pointed out starting from several analysis of the $TVV$ vertex, where $T$ stands for the energy momentum tensor of a given theory and $V$ denotes the vector current \cite{Giannotti:2008cv, Armillis:2009pq}. Subsequently it has been shown that the same pole is inherited by the dilatation current, in the $J_D VV$ 
vertex, being the two vertices very closely related. We recall that the dilatation current can be defined as 
\beq
J_D^\mu(z)= z_\nu T^{\mu \nu}(z) \qquad \textrm{with}  \qquad \partial\cdot J_D = {T^\mu}_\mu. 
\label{def}
\eeq
The $T^{\mu\nu}$ has to be symmetric and on-shell traceless for a classical scale-invariant theory, and includes, at
quantum level, the contribution from the trace anomaly together with the additional terms describing the explicit breaking of the 
dilatation symmetry. The contribution of the conformal anomaly, in flat space, is summarised by the equation 
\beq
\label{anomz}
T^\mu_\mu= \beta F_{\alpha\beta} F^{\alpha\beta}
\eeq
which holds for a classical scale invariant theory (i.e. with $T_\mu^\mu=0$), with the right hand side of 
(\ref{anomz}) related to the $\beta$ function $(\beta)$ of the gauge theory and to $F_{\mu\nu}$, the field strength of the vector particle (V). A similar equation holds in the case of chiral anomaly
\beq
\partial_\rho j_5^\rho= a_n F\tilde F  
\eeq
for the chiral anomaly, with $j_5^\rho$ denoting the axial vector current, and with $\tilde{F}=1/2 \epsilon^{\mu\nu\alpha\beta} F_{\alpha\beta}$ being the dual of the field strength of the gauge field.
We recall that the $U(1)_A$ current is characterized by an anomaly pole which describes the interaction between the 
Nambu-Goldstone mode, generated by the breaking of the chiral symmetry, and the gauge currents. 
In momentum space this corresponds to the nonlocal vertex 
\beq
\label{AVVpole}
V_{\textrm{anom}}^{\lambda \mu\nu}(k,p,q)=  \frac{k^\lambda}{k^2}\epsilon^{\mu \nu \alpha \beta}p_\alpha q_\beta +...
\eeq
with $k$ being the momentum of the axial-vector current and $p$ and $q$ the momenta of the two photons.
In the equation above, the ellipsis refer to terms which are suppressed at large energy. 
In this regime, this allows to distinguish the operator accounting for the chiral anomaly (i.e. $\square^{-1}$ in coordinate space)
from the contributions due to mass corrections. 
Polology arguments can be used to relate the appearance of such a pole to the pion state 
around the scale of chiral symmetry breaking.  We refer to \cite{Giannotti:2008cv} 
and \cite{Armillis:2009pq, Armillis:2010qk} for more details concerning the analysis of ths correlator in the QED and QCD cases, while the discussion of the $J_D VV$ vertex can be found in \cite{CDS}.
Using the relation between $J_D^\mu$ and the EMT $T^{\mu\nu}$ we introduce the $J_DVV$ correlator 
\beqa
\Gamma_D^{\mu\alpha\beta}(k,p)
&\equiv& 
\int d^4 z\, d^4 x\, e^{-i k \cdot z + i p \cdot x}\,
\left\langle J^\mu_D(z) V^\alpha(x)V^\beta(0)\right\rangle 
\label{gammagg}
\eeqa
which can be related to the $TVV$ correlator 
\beqa
\Gamma^{\mu\nu\alpha\beta}(k,p)&\equiv& \int d^4 z\, d^4 x\, e^{-i k \cdot z + i p \cdot x}\, 
\left\langle T^{\mu \nu}(z) V^\alpha(x) V^\beta(0)\right\rangle 
\eeqa
according to
\beqa
\Gamma_D^{\mu\alpha\beta}(k,p)&=& 
i \frac{\partial}{\partial k^\nu}\Gamma^{\mu\nu\alpha\beta}(k,p) \,.
\eeqa
As we have already mentioned, this equation allows us to identify a pole term in the $J_DVV$ diagram from the corresponding pole 
structure in the $TVV$ vertex. The analysis presented below shows that supersymmetry provides the natural framework where this pole-like behaviour is reproduced both in the chiral and the conformal anomaly parts of the superconformal anomaly vertex.

\section{Spectral analysis of supersymmetric correlators}
Dilaton fields are expected to play a very important role in the dynamics of the early universe and are present in almost any model which attempts to unify gravity with the ordinary gauge interactions (see for instance \cite{Gasperini:2007ar}). Important examples of these constructions are effective field theories derived from strings, describing their massless spectra, but also theories of gravity compactified on extra dimensional spaces, where the dilaton (graviscalar) emerges in 4 spacetime dimensions from the extra components of the higher dimensional metric (see for instance \cite{LopesCardoso:1991zt,LopesCardoso:1992yd,Derendinger:1991kr,Derendinger:1991hq,Derendinger:1985cv}). In these formulations, due to the geometrical origin of these fields, the dilaton is, in general, a fundamental (i.e. not a composite) field. 
Other extensions, also of significant interest,
in which a fundamental dilaton induces a gauge connection for the abelian Weyl symmetry in a curved spacetime, have been considered (see the discussion in \cite{Codello:2012sn,Buchmuller:1988cj, Coriano:2013nja}). However, also in this case, the link of this fundamental particle to gravity renders it a crucial player in the physics of the early universe, and not a particle to be searched for at colliders. In fact, its interaction with ordinary matter should be suppressed by the Planck scale, except if one entails scenarios with large extra dimensions.

More recently, following an independent route, several extensions of the Standard Model with an {\em effective} dilaton have been considered. They conjecture the existence of a scale-invariant extension of the Higgs sector \cite{Goldberger:2007zk, CDS,Coriano:2012dg}. In this case the breaking of the underlying conformal dynamics, in combination with the spontaneous breaking of the electroweak symmetry \cite{Coriano:2013nja}, suggests, in fact, that the dilaton could emerge as a composite field, appearing as a Nambu-Goldstone mode of the broken conformal symmetry. A massless dilaton of this type could acquire a mass by some explicit potential and could mix with the Higgs of the Standard Model.\\
By reasoning in terms of the conformal symmetry of the Standard Model, which should play a role at high energy, the dilaton would be the physical manifestation of the trace anomaly in the Standard Model, in analogy to the pion, which is interpolated by the $U(1)_A$ chiral current and the corresponding $\langle AVV \rangle$ (axial-vector/vector/vector) interaction in QCD. 
As in the $\langle AVV \rangle$ case,  this composite state should be identified with the anomaly 
pole of the related anomaly correlator (the $\langle TVV \rangle$ diagram, with $T$ the energy momentum tensor (EMT)), at least at the level of the 1-particle irreducible (1PI) anomaly effective action \cite{Coriano:2012nm}. Considerations of this nature brings us to the conclusion that the effective massless Nambu Goldstone (NG) modes which should appear as a result of the existence of global anomalies, should be looked for in specific perturbative form factors under special kinematical limits. For this reason they are easier to investigate in the on-shell anomaly effective action, with a single mass parameter which drives the conformal/superconformal deformation. This action has the advantage of being gauge invariant and easier to compute than its off-shell relative.

To exploit at a full scale the analogy between chiral and conformal anomalies, one should turn to supersymmetry, where the correlation between poles and anomalies should be more direct. In fact, in an ordinary quantum field theory, the $\langle TVV \rangle$ diagram (and the corresponding anomaly action) is characterized, as we are going to show, by pole structures both in those form factors which multiply tensors that contribute to the trace anomaly  and in those which don't. For this reason we turn our attention to the effective action of the superconformal (the Ferrara-Zumino, FZ) multiplet, where chiral and conformal anomalies share similar signatures, being part of the same multiplet. Therefore one would expect that supersymmetry should help in clarifying the significance of these singularites in the effective action. \\
We are going to prove rigorously in perturbation theory that the anomaly of the FZ multiplet is associated with the exchange of three composite states in the 1PI superconformal anomaly action. These have been discussed in the past, in the context of the spontaneous breaking of the superconformal symmetry \cite{Dudas:1993mm}. They are identified with 
the anomaly poles present in the effective action, extracted from a supersymmetric correlator containing the superconformal hypercurrent and two vector currents, and correspond to the dilaton, the dilatino and the axion.  
This exchange is identified by a direct analysis of the anomalous correlators in perturbation theory or by the study of the flow of their spectral densities under massive deformations.
The flow describes a 1-parameter family of spectral densities - one family for each component of the correlator - which satisfy mass independent sum rules, and are, therefore, independent of the superpotential. This behaviour turns a dispersive cut of the spectral density $\rho(s,m^2)$ into a pole (i.e. a $\delta(s)$ contribution) as the deformation parameter $m$ goes to zero. 
Moreover, denoting with $k^2$ the momentum square of the anomaly vertex, each of the spectral densities induces on the corresponding form factor a $1/k^2$ behaviour also at large $k^2$, as a consequence of the sum rule. \\
We also recall that the partnership between dilatons and axions is not new in the context of anomalies, and it has been studied in the past - for abelian gauge anomalies - in the case of the supersymmetric St\"uckelberg multiplet \cite{Kors:2004ri, Coriano:2008xa, Coriano:2008aw, Coriano:2010ws}. \\
The three states associated to the three anomaly poles mentioned above, are described - in the perturbative picture - by the exchange of two collinear particles. These are a fermion/antifermion pair in the axion case, a fermion/antifermion pair and a pair of scalar particles in the dilaton case, and a collinear scalar/fermion pair for the dilatino. 
The Konishi current will be shown to follow an identical pattern and allows the identification of extra states, 
one for each fermion flavour present in the theory. \\ 
This pattern appears to be general in the context of anomalies, and unique in the case of supersymmetry. In fact, we are going to show that in a supersymmetric theory anomaly correlators have a single pole in each component of the anomaly multiplet, a single spectral flow and a single sum rule, proving the existence of a one-to-one correspondence between anomalies and poles in these correlators.

\section{Theoretical framework}
In this section we review the definition and some basic properties of the Ferrara-Zumino supercurrent multiplet, which from now on we will denote also as the \emph{hypercurrent}, in order to distinguish it from its fermionic component, usually called the {\em supercurrent}.  \\
We consider a $\mathcal N=1$ supersymmetric Yang-Mills theory with a chiral supermultiplet in the matter sector. In the superfield formalism the action is given by
\bea
\label{SUSYactionSF}
S = \left( \frac{1}{16 g^2 T(R)} \int d^4 x \, d^2 \theta \, \textrm{Tr} W^2 + h.c.  \right) 
+ \int d^4 x \, d^4 \theta \, \bar \Phi e^V \Phi 
+ \left( \int d^4 x \, d^2 \theta \, \mathcal W(\Phi) +h.c. \right)
\eea
where the supersymmetric field strength $W_A$ and gauge vector field $V$ are contracted with the hermitian generators $T^a$ of the gauge group to which the chiral superfield $\Phi$ belongs. 
In particular
\bea
V =2 g  V^a T^a \,, \qquad \mbox{and} \qquad W_A = 2 g  W^a_A T^a = -\frac{1}{4} \bar D^2 e^{- V} D_A \, e^V \,, 
\eea
with $\textrm{Tr} \, T^a T^b = T(R) \delta^{ab}$. \\
In order to clarify our conventions we give the component expansion of the chiral superfield $\Phi$
\bea
\label{PHIexpansion}
\Phi_i = \phi_i + \sqrt{2} \theta \chi_i + \theta^2 F_i \,,
\eea
and of the superfields $W_A^a$ and $V^a$ in the Wess-Zumino gauge
\bea
\label{GAUGEexpansion}
W^a_A &=&  \lambda^a_A +\theta_A \, D^a - (\sigma^{\mu \nu} \theta)_A F^a_{\mu\nu} + i \theta^2 \, \sigma^{\mu}_{ A \dot B} \mathcal D_\mu \bar \lambda^{a \, \dot B} \,, \\
V^a &=&  \theta \sigma^\mu \bar \theta A^a_\mu + \theta^2 \bar \theta \bar \lambda^a + \bar \theta^2 \theta \lambda^a + \frac{1}{2} \theta^2 \bar \theta^2 \left( D^a + i \partial_\mu A^{a \, \mu} \right) \,,
\eea
where $\phi_i$ is a complex scalar and $\chi_i$ its superpartner, a left-handed Weyl fermion, $A^a_\mu$ and $\lambda^a$ are the gauge vector field and the gaugino respectively, $F^a_{\mu\nu}$ is the gauge field strength while $F_i$ and $D^a$ correspond to the $F$- and $D$-terms. Moreover, we have defined $\sigma^{\mu\nu}=(i/4)(\sigma^\mu \bar \sigma^\nu -\sigma^\nu \bar \sigma^\mu )$. \\
Using the component expansions introduced in Eq.(\ref{PHIexpansion}) and (\ref{GAUGEexpansion}) we obtain the supersymmetric lagrangian in the component formalism, which we report for convenience
\bea
\label{SUSYlagrangianCF}
\mathcal L &=& - \frac{1}{4} F^a_{\mu\nu} F^{a \, \mu\nu} + i \lambda^a \sigma^\mu \mathcal D_\mu^{ab} \bar \lambda^b
+ ( \mathcal D_{ij}^\mu \phi_j )^\dag (\mathcal D_{ik \, \mu} \phi_k) + i \chi_j \sigma_\mu \mathcal D_{ij}^{\mu \, \dag} \bar \chi_i \nn \\
&& - \sqrt{2} g \left( \bar \lambda^a \bar \chi_i T^a_{i j} \phi_j  + \phi_i^\dag T^a_{ij} \lambda^a \chi_j \right) - V(\phi, \phi^\dag) - \frac{1}{2} \left( \chi_i \chi_j \mathcal W_{ij}(\phi) + h.c.  \right) \,,
\eea
where the gauge covariant derivatives on the matter fields and on the gaugino are defined respectively as
\bea
\mathcal D^\mu_{ij} = \delta_{ij} \partial^\mu + i g A^{a \, \mu} T^a_{ij} \,, \qquad
\mathcal D_\mu^{ac} = \delta^{ac} \partial^\mu -g \, t^{abc} A^b_\mu \,,
\eea
with $t^{abc}$ the structure constants of the adjoint representation, and the scalar potential is given by
\bea
V(\phi, \phi^\dag) = \mathcal W^\dag_i(\phi^\dag) \mathcal W_i(\phi) + \frac{1}{2} g^2 \left( \phi_i^\dag T^a_{ij} \phi_j \right)^2 \,.
\eea
For the derivatives of the superpotential we have been used the following definitions
\bea
\mathcal W_i(\phi) = \frac{\partial \mathcal W(\Phi)}{\partial \Phi_i} \bigg|  \,, \qquad \mathcal W_{ij}(\phi) = \frac{\partial^2 \mathcal W(\Phi)}{\partial \Phi_i \partial \Phi_j} \bigg| \,,
\eea
where the symbol $|$ on the right indicates that the quantity is evaluated at $\theta = \bar \theta = 0$. \\
Notice that in the above equations the $F$- and $D$-terms have been removed exploiting their equations of motion.
Having defined the model, we can introduce the Ferrara-Zumino hypercurrent
\bea
\label{Hypercurrent}
\mathcal J_{A \dot A} = \textrm{Tr}\left[ \bar W_{\dot A} e^V W_A e^{- V}\right]
- \frac{1}{3} \bar \Phi \left[  \stackrel{\leftarrow}{\bar \nabla}_{\dot A}  e^V \nabla_A - e^V \bar D_{\dot A} \nabla_A +  \stackrel{\leftarrow}{\bar \nabla}_{\dot A} \stackrel{\leftarrow}{D_A} e^V \right] \Phi \,,
\eea
where $\nabla_A$ is the gauge-covariant derivative in the superfield formalism whose action on chiral superfields is given by
\bea
\nabla_A \Phi = e^{-V} D_A \left( e^V \Phi \right)\,, \qquad  \bar \nabla_{\dot A} \bar \Phi = e^{V} \bar D_{\dot A} \left( e^{-V} \bar \Phi \right)\,.
\eea
The conservation equation for the hypercurrent $\mathcal J_{A \dot A}$ is
\bea
\label{HyperAnomaly}
\bar D^{\dot A} \mathcal J_{A \dot A} = \frac{2}{3} D_A \left[ - \frac{g^2}{16 \pi^2} \left( 3 T(A) - T(R)\right) \textrm{Tr}W^2 - \frac{1}{8} \gamma \, \bar D^2 (\bar \Phi e^V \Phi)+ \left( 3 \mathcal W(\Phi) - \Phi \frac{\partial \mathcal W(\Phi)}{\partial \Phi} \right) \right] \,,
\eea
where $\gamma$ is the anomalous dimension of the chiral superfield. \\
The first two terms in Eq. (\ref{HyperAnomaly}) describe the quantum anomaly of the hypercurrent, while the last is of classical origin and it is entirely given by the superpotential. In particular, for a classical scale invariant theory, in which $\mathcal W$ is cubic in the superfields or identically zero, this term identically vanishes. If, on the other hand, the superpotential is quadratic the conservation equation of the hypercurrent acquires a non-zero contribution even at classical level. This describes the explicit breaking of the conformal symmetry.

We can now project the hypercurrent $\mathcal J_{A \dot A}$ defined in Eq.(\ref{Hypercurrent}) onto its components. The lowest component is given by the $R^\mu$ current, the $\theta$ term is associated with the supercurrent $S^\mu_A$, while the $\theta \bar \theta$ component contains the  energy-momentum tensor $T^{\mu\nu}$. In the $\mathcal N=1$ super Yang-Mills theory described by the Lagrangian in Eq. (\ref{SUSYlagrangianCF}), these three currents are defined as
\bea
\label{Rcurrent}
R^\mu &=& \bar \lambda^a \bar \sigma^\mu \lambda^a 
+ \frac{1}{3} \left( - \bar \chi_i \bar \sigma^\mu \chi_i + 2 i \phi_i^\dag \mathcal D^\mu_{ij} \phi_j - 2 i (\mathcal D^\mu_{ij} \phi_j)^\dag \phi_i \right) \,, \\
\label{Scurrent}
S^\mu_A &=& i (\sigma^{\nu \rho} \sigma^\mu \bar \lambda^a)_A F^a_{\nu\rho}
 - \sqrt{2} ( \sigma_\nu \bar \sigma^\mu \chi_i)_A (\mathcal D^{\nu}_{ij} \phi_j)^\dag - i \sqrt{2} (\sigma^\mu \bar \chi_i) \mathcal W_i^\dag(\phi^\dag) \nn \\
&-&  i g (\phi^\dag_i T^a_{ij} \phi_j) (\sigma^\mu \bar \lambda^a)_A + S^\mu_{I \, A}\,, \\
\label{EMT}
T^{\mu\nu} &=&  - F^{a \, \mu \rho} {F^{a \, \nu}}_\rho 
+ \frac{i}{4} \left[ \bar \lambda^a \bar \sigma^\mu (\delta^{ac} \stackrel{\rightarrow}{\partial^\nu} - g \, t^{abc} A^{b \, \nu} ) \lambda^c + 
\bar \lambda^a \bar \sigma^\mu (- \delta^{ac} \stackrel{\leftarrow}{\partial^\nu} - g \, t^{abc} A^{b \, \nu} ) \lambda^c + (\mu \leftrightarrow \nu) \right] \nn \\
&+&  ( \mathcal D_{ij}^\mu \phi_j )^\dag (\mathcal D_{ik}^\nu \phi_k)  +   ( \mathcal D_{ij}^\nu \phi_j )^\dag (\mathcal D_{ik}^\mu \phi_k) +
\frac{i}{4} \left[ \bar \chi_i \bar \sigma^\mu ( \delta_{ij} \stackrel{\rightarrow}{\partial^\nu} + i g T^a_{ij} A^{a \, \nu} ) \chi_j \right. \nn \\
&+& \left.  \bar \chi_i \bar \sigma^\mu ( - \delta_{ij} \stackrel{\leftarrow}{\partial^\nu} + i g T^a_{ij} A^{a \, \nu} ) \chi_j + (\mu \leftrightarrow \nu) \right]  - \eta^{\mu\nu} \mathcal L + T^{\mu\nu}_I \,, 
\eea
where $\mathcal L$ is given in Eq.(\ref{SUSYlagrangianCF}) and $S^\mu_I$ and $T^{\mu\nu}_I$ are the terms of improvement in $d=4$ of the supercurrent and of the EMT respectively. As in the non-supersymmetric case, these terms are necessary only for a scalar field and, therefore, receive contributions only from the chiral multiplet. They are explicitly given by
\bea
S^\mu_{I \, A} &=& \frac{4 \sqrt{2}}{3} i \left[ \sigma^{\mu\nu} \partial_\nu (\chi_i \phi_i^\dag) \right]_A \,, \\
T^{\mu\nu}_I &=& \frac{1}{3} \left( \eta^{\mu \nu} \partial^2 - \partial^\mu \partial^\nu \right) \phi^\dag_i \phi_i \,.
\eea
The terms of improvement are automatically conserved and guarantee, for $\mathcal W(\Phi) = 0$, upon using the equations of motion, the vanishing of the classical trace of $T^{\mu\nu}$ and of the classical gamma-trace of the supercurrent $S^\mu_A$. %

The anomaly equations in the component formalism, which can be projected out from Eq. (\ref{HyperAnomaly}), are
\bea
\label{AnomalyR}
\partial_\mu R^\mu &=& \frac{g^2}{16 \pi^2} \left( T(A) - \frac{1}{3} T(R) \right) F^{a \, \mu\nu} \tilde F^a_{\mu\nu} \,, \\
\label{AnomalyS}
\bar \sigma_\mu S^\mu_A &=&  - i \frac{3 \, g^2}{8 \pi^2} \left( T(A) -\frac{1}{3} T(R) \right) \left( \bar \lambda^a \bar \sigma^{\mu\nu} \right)_A F^a_{\mu\nu }\,, \\
\label{AnomalyT}
\eta_{\mu\nu} T^{\mu\nu} &=& -  \frac{3 \, g^2}{32 \pi^2} \left(T(A) - \frac{1}{3} T(R) \right) F^{a \, \mu\nu}  F^a_{\mu\nu} \,.
\eea
The first and the last equations are respectively extracted from the imaginary and the real part of the $\theta$ component of Eq.(\ref{HyperAnomaly}), while the gamma-trace of the supercurrent comes from the lowest component.

\section{The perturbative expansion in the component formalism}
In this section we will present the one-loop perturbative analysis of the one-particle irreducible correlators, built with a single current insertion contributing - at leading order in the gauge coupling constant - to the anomaly equations previously discussed. \\
We define the three correlation functions, $\Gamma_{(R)}$, $\Gamma_{(S)}$ and $\Gamma_{(T)}$ as
\bea
\label{RSTCorrelators}
\delta^{ab} \, \Gamma_{(R)}^{\mu\alpha\beta}(p,q) &\equiv& \langle R^{\mu}(k)\, A^{a \, \alpha}(p) \, A^{b \, \beta}(q) \rangle \qquad \langle RVV \rangle \,, \nn \\
\delta^{ab} \, \Gamma_{(S) \, A\dot B}^{\mu\alpha}(p,q) &\equiv& \langle S^{\mu}_A (k) \, A^{a \, \alpha}(p) \, \bar \lambda^b_{\dot B}(q) \rangle \qquad \langle SVF \rangle \,, \nn \\
\delta^{ab} \, \Gamma_{(T)}^{\mu\nu\alpha\beta}(p,q) &\equiv& \langle T^{\mu\nu}(k) \, A^{a \, \alpha}(p) \, A^{b \, \beta}(q)\rangle \qquad \langle TVV \rangle  \,,
\eea
with $k = p+q$ and where we have factorized, for the sake of simplicity, the Kronecker delta on the adjoint indices. These correlation functions have been computed at one-loop order in the dimensional reduction scheme (DRed). The Feynman rules used for the computation are given in \cite{CCDS}. We recall that in this scheme the tensor and scalar loop integrals are computed in the analytically continued spacetime while the sigma algebra is restricted to four dimensions. \\
We will present the results for the matter chiral and gauge vector multiplets separately, for on-shell gauge external lines.

The one-particle irreducible correlation functions of the Ferrara-Zumino multiplet are ultraviolet (UV) divergent, as one can see from a direct computation, and we need a suitable renormalization procedure in order to get finite results. In particular we have explicitly checked that, at one-loop order, among the three correlators defined in Eq. (\ref{RSTCorrelators}), only those with $S^\mu_A$ and $T^{\mu\nu}$ require a UV counterterm. 
The renormalization of the correlation functions is guaranteed by replacing the bare operators in Eq. (\ref{Scurrent}) and Eq. (\ref{EMT}) with their renormalized counterparts. This introduces the renormalized parameters and wave-function renormalization constants which are fixed by some conditions that specify the renormalization scheme. In particular, for the correlation functions we are interested in, the bare $S^\mu_A$ and $T^{\mu\nu}$ current become
\bea
\label{RenormalizedST}
S^\mu_A &=&  i Z_\lambda^{1/2} Z_A^{1/2}(\sigma^{\nu \rho} \sigma^\mu \bar \lambda^a_R)_A F^a_{R \, \nu\rho} + \ldots \,, \nn \\
T^{\mu\nu} &=&   Z_A \left( - F_R^{a \, \mu \rho} {F^{a \, \nu}}_{R \, \rho} + \frac{1}{4} \eta^{\mu\nu}  F_R^{a \, \rho \sigma} F^a_{R \, \rho \sigma}  \right)  + \ldots \,, 
\eea 
where the suffix $R$ denotes renormalized quantities. $Z_A$ and $Z_\lambda$ are the wave-function renormalization constants of the gauge and gaugino field respectively, while the ellipses stand for all the remaining operators. In the previous equations we have explicitly shown only the contributions from which, at one-loop order, we can extract the counterterms needed to renormalize our correlation functions. All the other terms, not shown, play a role at higher perturbative orders.\\ Expanding the wave-function renormalization constants at one-loop as $Z = 1 + \delta Z$ we obtain the vertices of the counterterms
\bea
\label{counterterms}
\delta[S^{\mu}_A(k) A^{a \, \alpha}(p) \bar \lambda^b_{\dot B}(q)] &=&  \left( \delta Z_A + \delta Z_\lambda \right) \,  p_\rho \, \left( \sigma^{\alpha \rho} \sigma^{\mu} \right)_{A \dot B} \,, \nn \\
\delta[T^{\mu\nu}(k)A^{a \, \alpha}(p) A^{b \, \beta}(q)] &=& \delta Z_A \, \delta^{ab} \left\{ p \cdot q \, C^{\mu\nu\alpha\beta} + D^{\mu\nu\alpha\beta}(p,q) \right\} \,,
\eea
with $p$ and $q$ outgoing momenta. The $\delta Z$ counterterms can be defined, for instance, by requiring a unit residue of the full two-point functions on the physical particle poles. This implies that 
\bea
\delta Z_A = - \frac{\partial}{\partial p^2} \Sigma^{(AA)}(p^2) \bigg|_{p^2 = 0}  \qquad \mbox{and} \qquad \delta Z_\lambda = - \Sigma^{(\lambda \bar \lambda)}(0) \,, 
\eea
where the one-loop corrections to the gauge and gaugino two-point functions are defined as
\bea
\Gamma^{(AA)}_{\mu\nu}(p) &=& - i \delta^{ab} \left( \eta_{\mu\nu} - \frac{p_\mu p_\nu}{p^2} \right) \Sigma^{(AA)}(p^2) \,, \\
\Gamma^{(\lambda \bar \lambda)}_{A \dot B}(p) &=& i \delta^{ab} \, p_{\mu} \sigma^{\mu}_{A \dot B} \, \Sigma^{(\lambda \bar \lambda)}(p^2) \,,
\eea 
with
\bea
\Sigma^{(AA)}(p^2) &=& \frac{g^2}{16 \pi^2} p^2 \left\{ T(R) \, \mathcal B_0(p^2,m^2)  - T(A) \, \mathcal B_0(p^2,0) \right\}   \,, \\
\Sigma^{(\lambda \bar \lambda)}(p^2) &=&  \frac{g^2}{16 \pi^2} \left\{ T(R) \, \mathcal B_0(p^2,m^2)  + T(A) \, \mathcal B_0(p^2,0)\right\} \,.
\eea
Using the previous expressions we can easily compute the wave-function renormalization constants
\bea
\delta Z_A &=& -  \frac{g^2}{16 \pi^2}  \left\{ T(R) \, \mathcal B_0(0,m^2)  - T(A) \, \mathcal B_0(0,0) \right\}   \,, \nn \\
\delta Z_\lambda &=& - \frac{g^2}{16 \pi^2}  \left\{ T(R) \, \mathcal B_0(0,m^2)  + T(A) \, \mathcal B_0(0,0)\right\} \,,
\eea
and therefore obtain the one-loop counterterms needed to renormalize our correlators. In the following we will always present results for the renormalized correlation functions. \\ It is interesting to observe that, accordingly to Eq. (\ref{counterterms}), the one-loop counterterm to the supercurrent correlation function is identically zero for the vector gauge multiplet, due to a cancellation between $\delta Z_A$ and $\delta Z_\lambda$. Therefore we expect a finite result for the vector supermultiplet contribution to the $\Gamma^{\mu\alpha}_{(S)}$. Indeed this is the case as we will show below.\\
The correctness of our computations is secured by the check of some Ward identities. These arise from gauge invariance, from the conservation of the energy-momentum tensor and of the supercurrent. In particular, for the three point correlators defined above, we have
\bea
\label{VectorWI}
&& p_\alpha \, \Gamma_{(R)}^{\mu\alpha\beta}(p, q) = 0 \,, \qquad \qquad q_\beta \, \Gamma_{(R)}^{\mu\alpha\beta}(p, q) = 0 \,, \nn \\
&& p_\alpha \, \Gamma_{(S)}^{\mu\alpha}(p, q) = 0 \,, \nn \\
&& p_\alpha \, \Gamma_{(T)}^{\mu\nu\alpha\beta}(p, q) = 0 \,, \qquad \qquad q_\beta \, \Gamma_{(T)}^{\mu\nu\alpha\beta}(p, q) = 0 \,.
\eea
from the conservation of the vector current, and 
\bea
\label{TensorWI}
i \, k_\mu \, \Gamma_{(S)}^{\mu\alpha}(p, q) &=& - 2 p_\mu \, \sigma^{\mu \alpha} \hat \Gamma_{(\lambda \bar \lambda)}(q) - i \sigma_\mu \hat \Gamma^{\mu \alpha}_{(AA)}(p)\,, \nn \\
i \, k_\mu \, \Gamma_{(T)}^{\mu\nu\alpha\beta}(p, q) &=& q_\mu \hat \Gamma^{\alpha \mu}_{(AA)}(p) \eta^{\beta \nu} + p_\mu \hat \Gamma^{\beta \mu}_{(AA)}(q) \eta^{\alpha \nu} - q^\nu \hat \Gamma^{\alpha \beta}_{(AA)}(p) - p^\nu \hat \Gamma^{\alpha \beta}_{(AA)}(q) \,, 
\eea
for the conservation of the supercurrent and of the EMT, where $\hat \Gamma_{(AA)}$ and $\hat \Gamma_{(\lambda \bar \lambda)}$ are the renormalized self-energies. Their derivation follows closely the analysis presented in \cite{Armillis:2010qk}.
Notice that, for on-shell gauge and gaugino external lines, the two identities in Eq. (\ref{TensorWI}) simplify considerably because their right-hand sides vanish identically. 

\section{The supercorrelator in the on-shell and massless case} 
In this section we discuss the explicit results of the computation of supercorrelator when the components of the external vector supercurrents are on-shell and the superpotential of the chiral multiplet is absent. We will consider first the contributions due to the exchange of the chiral multiplet, followed by a subsection in which we address the exchange of a virtual vector multiplet. 

\subsection{The chiral multiplet contribution}
We start from the chiral multiplet, presenting the result of the computation for massless fields and on shell gauge and gaugino external lines. 
\begin{figure}[t]
\centering
\subfigure[]{\includegraphics[scale=0.5]{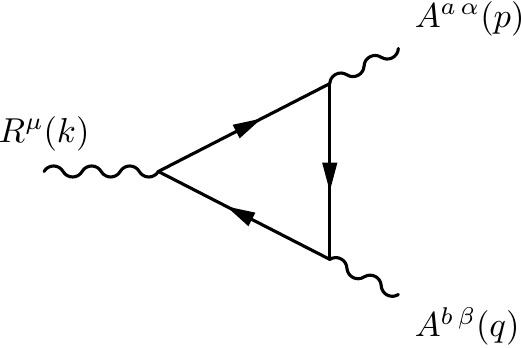}} \hspace{.5cm}
\subfigure[]{\includegraphics[scale=0.5]{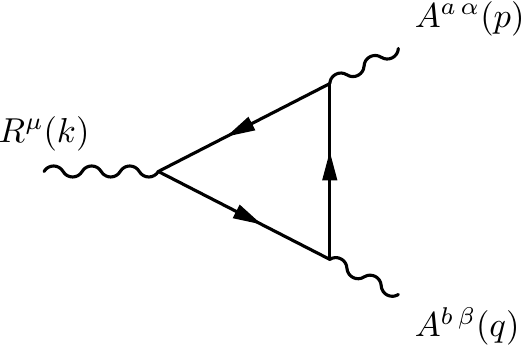}} \hspace{.5cm}
\subfigure[]{\includegraphics[scale=0.5]{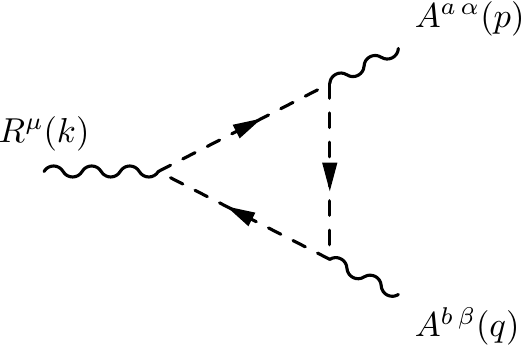}} \hspace{.5cm}
\subfigure[]{\includegraphics[scale=0.5]{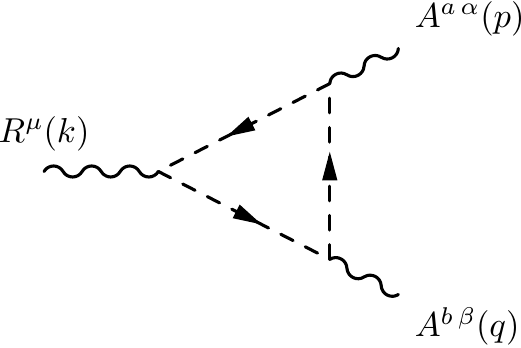}} \\
\subfigure[]{\includegraphics[scale=0.5]{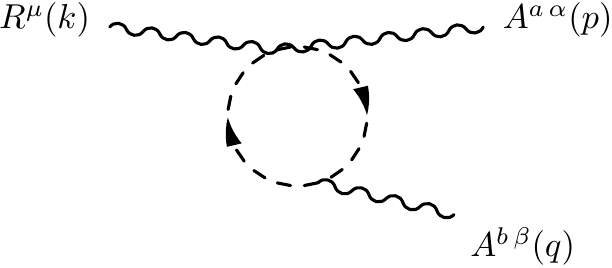}} \hspace{.5cm}
\subfigure[]{\includegraphics[scale=0.5]{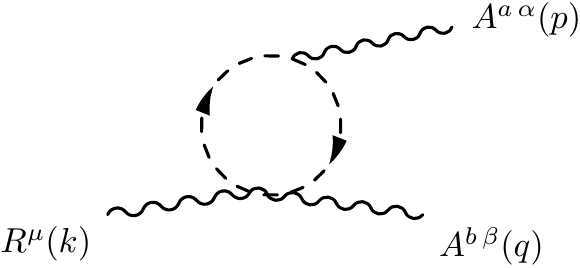}} \hspace{.5cm}
\subfigure[]{\includegraphics[scale=0.5]{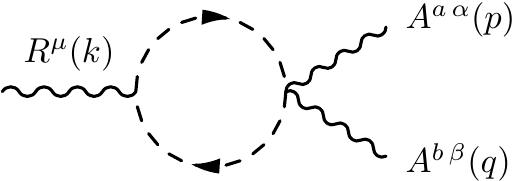}} 
\caption{The one-loop perturbative expansion of the $\langle RVV \rangle$ correlator with a massless chiral multiplet running in the loops. \label{Fig.Rchiral}}
\end{figure}
\begin{figure}[t]
\centering
\subfigure[]{\includegraphics[scale=0.5]{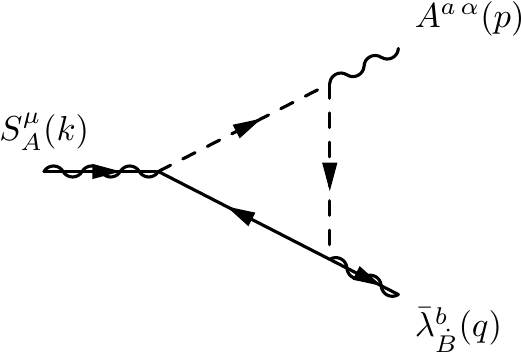}} \hspace{.5cm}
\subfigure[]{\includegraphics[scale=0.5]{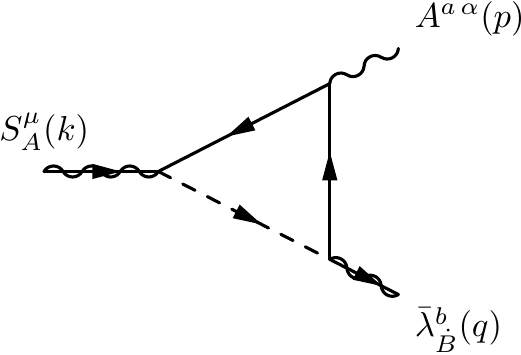}} \hspace{.5cm}
\subfigure[]{\includegraphics[scale=0.5]{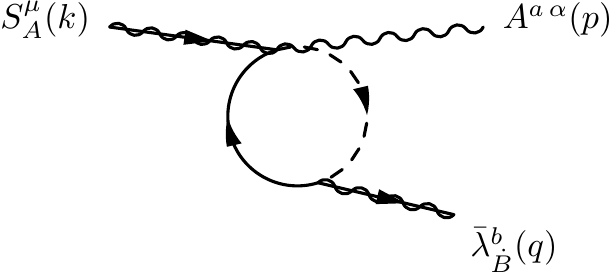}} \hspace{.5cm}
\subfigure[]{\includegraphics[scale=0.5]{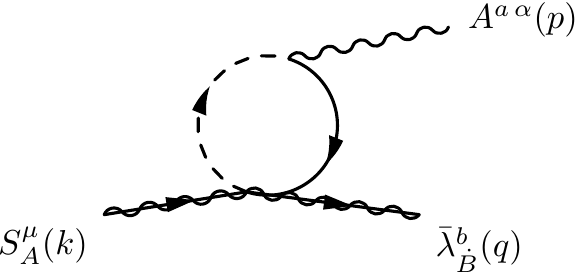}}
\caption{The one-loop perturbative expansion of the $\langle SVF \rangle$ correlator with a massless chiral multiplet running in the loops. \label{Fig.Schiral}}
\end{figure}
\begin{figure}[t]
\centering
\subfigure[]{\includegraphics[scale=0.5]{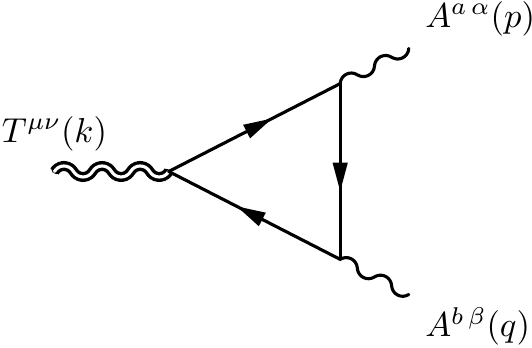}} \hspace{.5cm}
\subfigure[]{\includegraphics[scale=0.5]{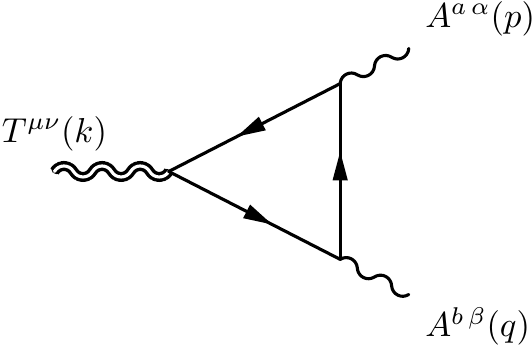}} \hspace{.5cm}
\subfigure[]{\includegraphics[scale=0.5]{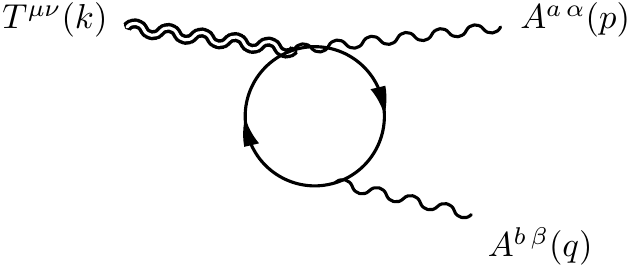}} \hspace{.5cm}
\subfigure[]{\includegraphics[scale=0.5]{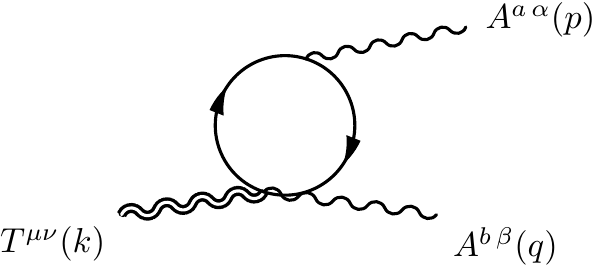}} \\
\subfigure[]{\includegraphics[scale=0.5]{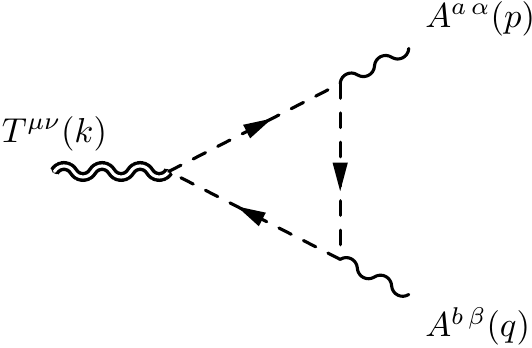}} \hspace{.5cm}
\subfigure[]{\includegraphics[scale=0.5]{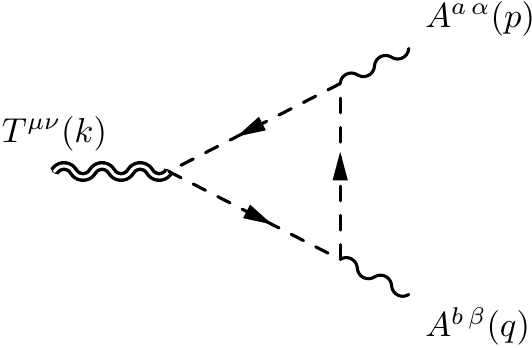}} \hspace{.5cm}
\subfigure[]{\includegraphics[scale=0.5]{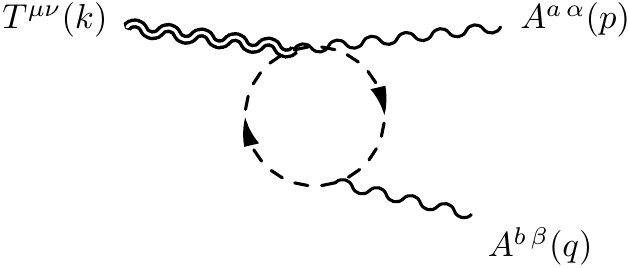}} \hspace{.5cm}
\subfigure[]{\includegraphics[scale=0.5]{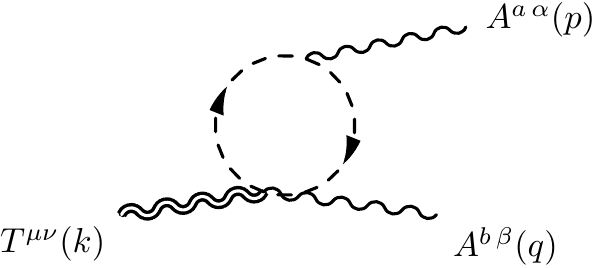}} \\
\subfigure[]{\includegraphics[scale=0.5]{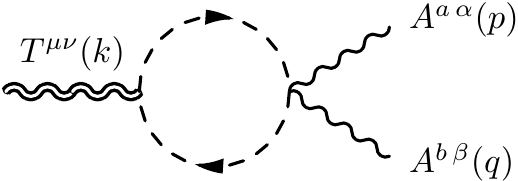}} \hspace{.5cm}
\subfigure[]{\includegraphics[scale=0.5]{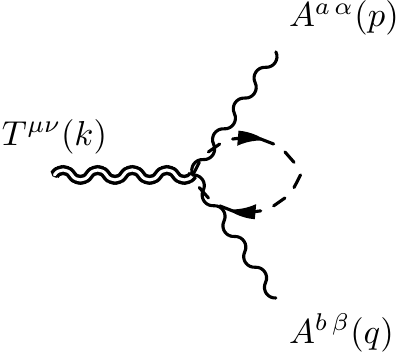}} \hspace{.5cm}
\caption{The one-loop perturbative expansion of the $\langle TVV \rangle$ correlator with a massless chiral multiplet running in the loops. The last diagram, being a massless tadpole, is identically zero in dimensional regularization. \label{Fig.Tchiral}}
\end{figure}

The diagrams defining the one-loop expansion of the $\Gamma_{(R)}$ correlator are shown in Fig.~(\ref{Fig.Rchiral}). They consist of triangle and bubble topologies with fermions, since the scalars do not contribute. The explicit result for a massless chiral multiplet with on-shell external gauge bosons is given by
\bea
\label{RChiralOSMassless}
\Gamma_{(R)}^{\mu\alpha\beta}(p,q) = - i \frac{g^2 \, T(R)}{12 \pi^2} \frac{k^\mu}{k^2} \eps[p, q, \alpha ,\beta] \,,
\eea
The correlator in Eq.(\ref{RChiralOSMassless}) satisfies the vector current conservation constraints given in Eq.(\ref{VectorWI}) and the anomalous equation of Eq.(\ref{AnomalyR}) 
\bea
\label{AnomalyRmom}
i k_\mu \, \Gamma_{(R)}^{\mu\alpha\beta}(p,q) = \frac{g^2 \, T(R)}{12 \pi^2} \, \eps[p, q, \alpha ,\beta] \,.
\eea
There is no much surprise, obviously, for the anomalous structure of Eq. (\ref{RChiralOSMassless}) which is characterized by a pole $1/k^2$ term, since in the on-shell case and for massless fermions (which are the only fields contributing to the $\langle RVV \rangle$ at this perturbative order), we recover the usual structure of the $\langle AVV \rangle$ diagram.

The perturbative expansion of the $\Gamma^{\mu\alpha}_{(S) \, A \dot B}$ correlation function is depicted in Fig.~(\ref{Fig.Schiral}). For simplicity we will remove, from now on, the spinorial indices from the corresponding expressions. The explicit result for a massless chiral supermultiplet with on-shell external gauge and gaugino lines is then given by
\bea
\label{SChiralOSMassless}
\Gamma^{\mu\alpha}_{(S)}(p,q) = - i \frac{g^2 T(R)}{6 \pi^2 \, k^2} s_1^{\mu\alpha}
+ i \frac{g^2 T(R)}{64 \pi^2} \Phi_2(k^2,0) \, s_2^{\mu\alpha} \,,
\eea
where the form factor $\Phi_2(k^2,0)$ is defined as
\bea
\label{Phi2massless}
\Phi_2(k^2,0) = 1 - \mathcal B_0(0,0) + \mathcal B_0(k^2,0) \,,
\eea
and the two tensor structures are 
\bea
s_1^{\mu\alpha} &=&   \sigma^{\mu \nu} k_\nu \, \sigma^\rho k_\rho \,  \bar \sigma^{\alpha \beta} p_\beta \,,\nn \\ 
s_2^{\mu\alpha} &=&  2 p_\beta \, \sigma^{\alpha \beta} \sigma^\mu \,.
\eea 
The $\mathcal B_0$ function appearing in Eq.(\ref{Phi2massless}) is a two-point scalar integral defined in Appendix \ref{AppScalarIntegrals}. Notice that the form factor multiplying the second tensor structure $s_2$ is ultraviolet finite, due to the renormalization procedure, but has an infrared singularity inherited by the counterterms in Eq.~(\ref{counterterms}).  \\
It is important to observe that the only pole contribution comes from the anomalous structure $s_1^{\mu\alpha}$, which shows that 
the origin of the anomaly has to be attributed to a unique fermionic pole ($\sigma^\rho k_\rho / k^2$) in the correlator, in the form factor multiplying $s_1^{\mu\alpha}$. 
It is easy to show that Eq.~(\ref{SChiralOSMassless}) satisfies the vector current and EMT conservation equations. Moreover, the anomalous equation reads as
\bea
\label{AnomalySmom}
\bar \sigma_{\mu} \, \Gamma^{\mu\alpha}_{(S)}(p,q) =  \frac{g^2 T(R)}{ 4 \pi^2}  \bar \sigma^{\alpha \beta} p_\beta \,,
\eea
where only the first tensor structure contributes to the $\sigma$-trace of the correlator. This result is clearly in agreement with Eq.(\ref{AnomalyS}), after Fourier transform $(\mathcal{F.T.})$ owing to
\bea
\mathcal{F.T.} \left\{ \frac{i}{2} \frac{\delta^2  F_{\mu\nu} \bar \sigma^{\mu\nu} \bar \lambda }{\delta A_\alpha(x) \delta \bar \lambda (y)} \right\} =  \bar \sigma^{\alpha \beta}  p_\beta \,.
\eea
Notice also that
\bea
\mathcal{F.T.} \left\{ \frac{\delta^2 S^\mu}{\delta A_\alpha(x) \delta \bar \lambda (y)} \right\} = s_{2}^{\mu\alpha} \,.
\eea

The diagrams appearing in the perturbative expansions of the $\Gamma_{(T)}$ are depicted in Fig.(\ref{Fig.Tchiral}). They consist of triangle and bubble topologies. There is also a tadpole-like contribution, Fig.(\ref{Fig.Tchiral}j), which is non-zero only in the massive case. \\
The explicit expression of the  $\Gamma_{(T)}$ correlator for a massless chiral supermultiplet and on-shell gauge lines is given by
\bea
\label{TChiralOSMassless}
\Gamma_{(T)}^{\mu\nu\alpha\beta}(p,q) = - \frac{g^2 \, T(R)}{24 \pi^2 \, k^2} t_{1S}^{\mu\nu\alpha\beta}(p,q)  + \frac{g^2 \, T(R)}{16 \pi^2} \Phi_2(k^2,0) \, t_{2S}^{\mu\nu\alpha\beta}(p,q) \,,
\eea
where the $\Phi_2$ is defined in Eq.(\ref{Phi2massless}) and
\bea
t_{1S}^{\mu\nu\alpha\beta}(p,q) &\equiv& \phi_1^{\mu\nu\alpha\beta}(p,q) = (\eta^{\mu\nu} k^2 - k^\mu k^\nu) u^{\alpha\beta}(p,q)\,, \\
t_{2S}^{\mu\nu\alpha\beta}(p,q) &\equiv& \phi_3^{\mu\nu\alpha\beta}(p,q) = (p^\mu q^\nu + p^\nu q^\mu) \eta^{\alpha\beta} + p \cdot q (\eta^{\alpha\nu} \eta^{\beta\mu} + \eta^{\alpha\mu} \eta^{\beta\nu}) - \eta^{\mu\nu} u^{\alpha \beta}(p,q) \nn \\
&-&  (\eta^{\beta\nu}p^\mu + \eta^{\beta\mu}p^\nu)q^\alpha - (\eta^{\alpha\nu}q^\mu + \eta^{\alpha\mu}q^\nu)p^\beta\,,
\eea
where $\phi_1^{\mu\nu\alpha\beta}, \phi_3^{\mu\nu\alpha\beta}$ and $u^{\alpha\beta}$ are given in Eqs. (\ref{phitensors}) and (\ref{utensor}).
As in the previous cases we have explicitly checked all the Ward identities originating from gauge invariance and conservation of the energy-momentum tensor. As one can easily verify by inspection, only the first one of the two tensor structures is traceful and contributes to the anomaly equation of the $\Gamma_{(T)}$ correlator
\bea
\label{AnomalyTmom}
\eta_{\mu\nu} \, \Gamma_{(T)}^{\mu\nu\alpha\beta}(p,q) = - \frac{g^2 \, T(R)}{8 \pi^2} u^{\alpha\beta}(p,q) \,.
\eea 
The comparison of Eq.(\ref{AnomalyTmom}) to Eq.(\ref{AnomalyT}) is evident if one recognizes that
\bea
\mathcal{F.T.} \left\{ - \frac{1}{4} \frac{\delta^2 F_{\mu\nu}F^{\mu\nu}}{\delta A_\alpha(x) \delta A_\beta(y)} \right\} = u^{\alpha\beta}(p,q) \,.
\eea
For completeness we give also the inverse Fourier transform of $t_{2S}^{\mu\nu\alpha\beta}(p,q)$ which is obtained from
\bea
\mathcal{F.T.} \left\{ \frac{\delta^2 T^{\mu\nu}_{gauge}}{\delta A_\alpha(x) \delta A_\beta(y)} \right\} = t_{2S}^{\mu\nu\alpha\beta}(p,q) \,,
\eea 
where $T^{\mu\nu}_{gauge}$ is the pure gauge part of the energy-momentum tensor. Notice that $t_{2S}$ is nothing else than the tree-level vertex with two onshell gauge fields on the external lines. 

As in the previous subsection, concerning the supersymmetric current $S^\mu_A$, also in the case of this correlator there is only one structure containing a pole term, which appears in the only form factor (which multiplies $t_{1S}$) with a nonvanishing trace. Differently from the non supersymmetric case, such as in QED and QCD, with fermions or scalars running in the loops, as shown in Eqs.~(\ref{RVectorOS}), (\ref{SVectorOS}), and (\ref{TVectorOS}), there are {\em no extra poles} in the traceless structures of the decomposition of the correlators proving that in a  supersymmetric theory the signature of all the anomalies in the 
$\langle \mathcal{J} \mathcal{V} \mathcal{V} \rangle$ correlator are only due to anomaly poles in each channel.
 
\subsection{The vector multiplet contribution}
Finally, we come to a discussion of the perturbative results for the vector (gauge) multiplet to the three anomalous correlation functions presented in the previous sections. Notice that due to the quantization of the gauge field, gauge fixing and ghost terms must be taken into account, increasing the complexity of the computation. This technical problem is completely circumvented with on-shell gauge boson and gaugino, which is the case analyzed in this work. \\
Concerning the diagrammatic expansion, the topologies of the various contributions defining the three correlators is analogous to those illustrated in massless chiral case. The explicit results are given by
\bea
\label{RVectorOS}
\Gamma_{(R)}^{\mu\alpha\beta}(p,q) &=&  i \frac{g^2 \, T(A)}{4 \pi^2} \frac{k^\mu}{k^2} \eps[p, q, \alpha ,\beta] \,,  \\
\label{SVectorOS}
\Gamma_{(S)}^{\mu\alpha}(p,q) &=&   i \frac{g^2 T(A)}{2 \pi^2 \, k^2} s_1^{\mu\alpha} + i \frac{g^2 T(A)}{64 \pi^2} V(k^2) \, s_2^{\mu\alpha} \,, \\
\label{TVectorOS}
\Gamma_{(T)}^{\mu\nu\alpha\beta}(p,q) &=&  \frac{g^2 \, T(A)}{8 \pi^2 \, k^2} t_1^{\mu\nu\alpha\beta}(p,q)  + \frac{g^2 \, T(A)}{16 \pi^2} V(k^2) \, t_{2}^{\mu\nu\alpha\beta}(p,q) \,,
\eea
where
\bea
V(k^2) = -3 + 3 \, \mathcal B_0(0,0) - 3 \, \mathcal B_0(k^2,0) - 2 k^2 \, \mathcal C_0(k^2,0) \,.
\eea
The tensor expansion of the correlators is the same as in the previous cases. The only differences are in the form factors. In particular, the first in each of them is the only one responsible for the anomaly and is multiplied, respect to the chiral case, by a factor $-3$ and by a different group factor. The result reproduces exactly the anomaly Eqs (\ref{AnomalyR},\ref{AnomalyS},\ref{AnomalyT}). Concerning the ultraviolet divergences of these correlators, the explicit computation shows that the vector multiplet contribution to $\Gamma_{(S)}^{\mu\nu}$ is indeed finite at one-loop order before any renormalization. This confirms a result obtained in the analysis of the renormalization properties of these correlators presented in a previous section, where it was shown the vanishing of the counterterm of $\Gamma^{\mu\alpha}_{(S)}$ for the vector multiplet.

Also for the vector multiplet, the result is similar, since the only anomaly poles present in the three correlators 
(\ref{RVectorOS}), (\ref{SVectorOS}) and (\ref{TVectorOS}) are those belonging to anomalous structures. We conclude that in all the cases discussed so far, the signature of an anomaly, in a superconformal theory, are anomaly poles.

\section{ The supercorrelator in the on-shell and massive case}
We now extend our previous analysis to the case of a massive chiral multiplet. This will turn out to be extremely useful in order to discuss the general behaviour of the spectral densities away from the conformal point.

\begin{figure}[t]
\centering
\subfigure[]{\includegraphics[scale=0.6]{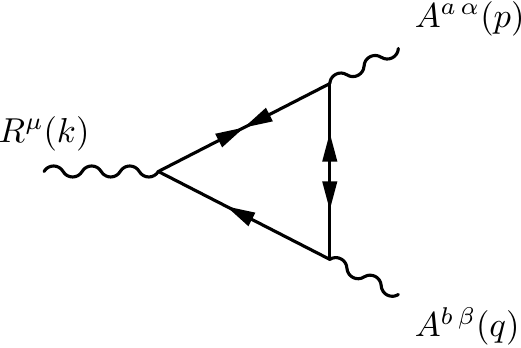}} \hspace{.5cm}
\subfigure[]{\includegraphics[scale=0.6]{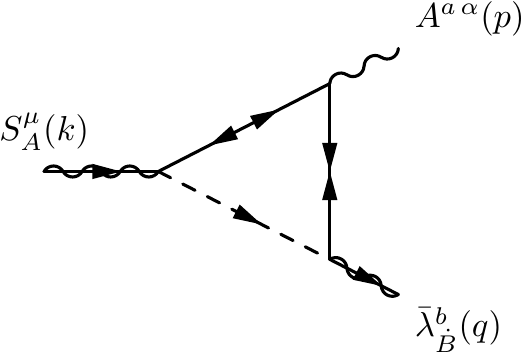}} \hspace{.5cm}
\subfigure[]{\includegraphics[scale=0.6]{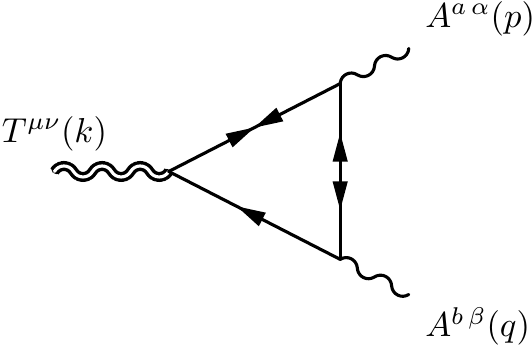}} 
\caption{A sample of diagrams, for a massive chiral multiplet, mass insertions in the fermion propagators. \label{Fig.Massivechiral}}
\end{figure}
The diagrammatic expansion of the three correlators for a massive chiral multiplet in the loops get enlarged by a bigger set of contributions characterized by mass insertions on the $S^\mu_A$ and $T^{\mu\nu}$ vertices and on the propagators of the Weyl fermions. A sample of them are shown in Fig. (\ref{Fig.Massivechiral}). An explicit computation, in this case, gives
\bea
\label{RChiralOSMassive}
\Gamma_{(R)}^{\mu\alpha\beta}(p,q) &=&  i \frac{g^2 \, T(R)}{12 \pi^2} \, \Phi_1(k^2,m^2) \, \frac{k^\mu}{k^2} \eps[p, q, \alpha ,\beta]  \,, \\
\label{SChiralOSMassive}
\Gamma^{\mu\alpha}_{(S)}(p,q) &=&  i \frac{g^2 T(R)}{6 \pi^2 \, k^2} \, \Phi_1(k^2,m^2) \, s_1^{\mu\alpha}
+ i \frac{g^2 T(R)}{64 \pi^2}  \, \Phi_2(k^2,m^2) \, s_2^{\mu\alpha} \,, \\
\label{TChiralOSMassive}
\Gamma_{(T)}^{\mu\nu\alpha\beta}(p,q) &=& \frac{g^2 \, T(R)}{24 \pi^2 \, k^2} \, \Phi_1(k^2,m^2) \, t_{1S}^{\mu\nu\alpha\beta}(p,q) + \frac{g^2 \, T(R)}{16 \pi^2} \, \Phi_2(k^2,m^2) \, t_{2S}^{\mu\nu\alpha\beta}(p,q) \,, 
\eea
with
\bea
\Phi_1(k^2,m^2) &=& - 1 - 2\, m^2 \, \mathcal C_0(k^2,m^2) \,, \nn \\
\Phi_2(k^2,m^2) &=& 1 - \mathcal B_0(0,m^2) + \mathcal B_0(k^2,m^2) + 2 m^2  \mathcal C_0(k^2,m^2) \,.
\label{exp1}
\eea
The expressions above show that the only modification introduced by the mass corrections is in the form factors, while the tensor structure remains unchanged. \\
As we have previously discussed, if the superpotential is quadratic in the chiral superfield, the hypercurrent conservation equation develops a classical (non-anomalous) contribution describing the explicit breaking of the conformal symmetry. Therefore, in this case, the anomaly equations (\ref{AnomalyRmom}),(\ref{AnomalySmom}), and (\ref{AnomalyTmom}) must be modified in order to account for the mass dependence. The new conservation equations for a massive chiral supermultiplet become
\bea
i k_\mu \, \Gamma^{\mu\alpha\beta}_{(R)}(p,q) &=& -\frac{g^2 T(R)}{12\pi^2} \Phi_1(k^2,m^2) \eps[p,q,\alpha,\beta] \,, \\
\bar \sigma_{\mu} \, \Gamma^{\mu\alpha}_{(S)}(p,q) &=& - \frac{g^2 T(R)}{ 4 \pi^2}  \Phi_1(k^2,m^2) \bar \sigma^{\alpha \beta} p_\beta \,, \\
\eta_{\mu\nu} \, \Gamma^{\mu\nu\alpha\beta}_{(T)}(p,q) &=&  \frac{g^2 T(R)}{8\pi^2} \Phi_1(k^2,m^2) u^{\alpha\beta}(p,q) \,.
\eea
It is interesting to observe that supersymmetry prevents the appearance of new structures in the conservation equations, at least for these correlation functions, being the explicit classical breaking terms just a correction to the anomaly coefficient. This is not the case for non-supersymmetric theories  \cite{Giannotti:2008cv, Armillis:2009pq}.

\section{Comparing supersymmetric and non supersymmetric cases: sum rules and extra poles in the Standard Model}
In this section and in the following one, we compare the structure of the spectral densities between supersymmetric and non supersymmetric theories in the presence of mass terms, looking for the additional sum rules not directly related to the anomalies, which may be present in the $\langle TVV \rangle$ and $\langle AVV \rangle$ correlators. We anticipate that these are found in the $\langle TVV \rangle$ in the non suspersymmetric case in all the gauge invariant sectors of the Standard Model. We start our analysis with the conformal anomaly action of QCD, described by the EMT-gluon-gluon vertex and then move to the EMT-$\gamma\gamma$  vertex in the complete electroweak theory. Obviously, the spectral densitites develope anomaly poles in the limit in which all the second scales of the vertices turn to zero. By this we mean fermion masses, the $W$ mass and external virtualities. Moreover, we identify the explicit form of the sum rules satisfied in perturbation theory. 
\subsection{The extra pole of QCD}
For definiteness we focus our attention on a specific gauge theory, QCD.   
We write the whole amplitude $\Gamma^{\mu\nu\a\b}(p,q)$ of the $\langle TVV \rangle$ diagram in QCD as
\bea
\Gamma^{\mu\nu\a\b}(p,q) = \Gamma_q^{\mu\nu\a\b}(p,q) + \Gamma_g^{\mu\nu\a\b}(p,q),
\eea
having separated the quark $(\Gamma_q)$ and the gluons/ghosts $(\Gamma_g)$ contributions. We have omitted the colour indices for simplicity, being the correlator diagonal in colour space.
As described before in Section \ref{TVVsection} for the massless case, also in the massive case the amplitude $\Gamma$ is expressed in terms of 3 tensor structures. In the $\overline{MS}$ scheme these are given by \cite{Armillis:2010qk}
\beq
\Gamma^{\mu\nu\alpha\beta}_{q/g}(p,q) =  \, \sum_{i=1}^{3} \Phi_{i\,q/g} (k^2,m^2)\, \phi_i^{\mu\nu\alpha\beta}(p,q)\,.
\label{Gamt}
\eeq
For on-shell and transverse gluons, only 3 invariant amplitudes contribute, which for the quark loop case are given by
\bea
\Phi_{1\, q} (k^2,m^2) &=&
\frac{g^2}{6 \pi^2  k^2} \bigg\{  - \frac{1}{6}   +  \frac{ m^2}{k^2}  - m^2 \mathcal C_0(k^2,m^2)
\bigg[\frac{1}{2 \,  }-\frac{2 m^2}{ k^2}\bigg] \bigg\} \,,  \\
\Phi_{2\, q} (k^2,m^2)  &=&
- \frac{g^2}{4 \pi^2 k^2} \bigg\{   \frac{1}{72} + \frac{m^2}{6  k^2}
+  \frac{ m^2}{2  k^2} \mathcal D (k^2,m^2) +  \frac{ m^2}{3 } \mathcal C_0(k^2,m^2 )\, \left[ \frac{1}{2} + \frac{m^2}{k^2}\right] \bigg\} \,,  \\
\Phi_{3\,q} (k^2,m^2) &=& 
\frac{g^2}{4 \pi^2} \bigg\{ \frac{11}{72}  +   \frac{ m^2}{2 k^2}
 +  m^2  \mathcal C_0(k^2,m^2) \,\left[ \frac{1}{2 } + \frac{m^2}{k^2}\right]  +  \frac{5  \, m^2}{6 k^2}  \mathcal D (k^2,m^2) + \frac{1}{6} \mathcal B_0^{\overline{MS}}(k^2, m^2) \bigg\},\nn\\
\label{masslesslimit}
\eea
where the on-shell scalar integrals $\mathcal D (k^2,m^2)$, $\mathcal C_0(k^2, m^2)$ and $\mathcal B_0^{\overline{MS}}(k^2, m^2)$ are given in Appendix \ref{AppScalarIntegrals}. \\
Here we concentrate on the two form factors which are unaffected by renormalization, namely $\Phi_{1,2 q}$. Both admit convergent dispersive integrals of the form 
\bea
\Phi_{1,2 q}(k^2,m^2) &=& \frac{1}{\pi} \int_0^{\infty} ds \frac{\rho_{1,2 q}(s,m^2)}{s-k^2} \,,
\eea
in terms of spectral densities ${\rho_{1,2 q}(s,m^2)}$. From the explicit expressions of these two form factors, the corresponding spectral densities are obtained using the relations 
\bea
&& \textrm{Disc}\left( \frac{1}{s^2}\right) = 2i\pi \delta'(s),\nn\\
&& \textrm{Disc}\left( \frac{\mathcal C_0(s,m^2)}{s^2} \right)= - \frac{2i \pi}{s^3} \log \frac{1+\sqrt{\tau(s,m^2)}}{1-\sqrt{\tau(s,m^2)}}\theta(s-4 m^2) +   i\pi \delta'(s) A(s), \label{disc1}
\eea
where $A(s)$ is defined in Eq.(\ref{As}) and we have used the general relation 
\beq
\left( \frac{1}{x +i \epsilon}\right)^n - \left( \frac{1}{x -i \epsilon}\right)^n=(-1)^n \frac{2 \pi i}{(n-1)!}\delta^{(n-1)}(x) \,,
\eeq
with $\delta^{(n)}(x)$ the $n$-th derivative of the delta function. 
The contribution proportional to $\delta'(s)$ in Eq.(\ref{disc1}) can be rewritten in the form
\bea
\delta'(s) A(s) = -\delta(s) A'(0)+ \delta'(s) A(0), \qquad  \mbox{with} \quad A(0)=-\frac{1}{m^2} \,, \quad A'(0)=-\frac{1}{12 m^4} \,,
\eea
giving for the spectral densities
\bea
\label{rhoq}
\rho_{1q}(s,m^2) &=& \frac{g^2}{12 \pi} \frac{m^2}{s^2} \tau(s,m^2) \log \frac{1+\sqrt{\tau(s,m^2)}}{1-\sqrt{\tau(s,m^2)}} \theta(s-4m^2) \,, \nn \\
\rho_{2q}(s,m^2) &=& \frac{-g^2}{12 \pi} \left[ \frac{3 m^2}{2 s^2} \sqrt{\tau(s,m^2)} - \frac{m^2}{s} \left( \frac{1}{2 s} + \frac{m^2}{s^2}   \right) \log \frac{1+\sqrt{\tau(s,m^2)}}{1-\sqrt{\tau(s,m^2)}} \right] \theta(s-4m^2)
\eea
Both functions are characterized by a two particle cut starting at $4m^2$, with $m$ the quark mass.
Notice also that in this case there is a cancellation of the localized contributions related to the $\delta(s)$, showing that for nonzero mass there are no pole terms in the dispersive integral. The crucial difference, respect to the supersymmetric case discussed above, is that now we have two independent sum rules   
\bea
 \frac{1}{\pi} \int_{0}^{\infty} ds \, \rho_{1 q}(s,m^2) = \frac{g^2}{36 \pi^2} \,, \qquad \qquad
 \frac{1}{\pi} \int_0^{\infty} ds \, \rho_{2q}(s,m^2) = \frac{g^2}{288 \pi^2} \,,
\eea
one for each form factor, as it can be verified by a direct integration. We can normalize both densities as
\beq
\bar{\rho}_{1 q}(s,m^2)\equiv  \frac{36 \pi^2}{g^2}\rho_{1 q}(s,m^2) \qquad \bar{\rho}_{2 q}(s,m^2) \equiv  \frac{288 \pi^2}{g^2}\rho_{2 q}(s,m^2)
\eeq
in order to describe the two respective flows, which are homogeneuos, since both densities carry the same physical dimension and both converge to a $\delta(s)$ as the quark mass $m$ is sent to zero 
\beq
\lim_{m\to 0}\bar{ \rho}_{1 q}=\lim_{m\to 0}\bar{ \rho}_{2 q}=\delta(s). 
\eeq
Indeed at $m=0$, $\Phi_{1,2 q}$ are just given by pole terms, while $\Phi_{3q}$ is logarithmic in momentum 
\bea
\Phi_{1\,q} (k^2,0) &=& - \frac{g^2}{36 \pi^2  k^2}, \qquad
\Phi_{2\,q} (k^2,0) = - \frac{g^2}{288 \pi^2 \, k^2}, \\
\Phi_{3\,q} (k^2,0) &=& - \frac{g^2}{288 \pi^2} \, \left( 12 \log \left( -\frac{k^2}{\mu^2} \right ) - 35\right), \qquad \mbox{for} \quad k^2<0.
\eea
It is then clear, from this comparative analysis, that the supersymmetric and the non supersymmetric anomaly correlators can be easily differentiated in regards to their spectral behaviour. In the non supersymmetric case 
the spectral analysis of the $\langle TVV \rangle$ correlator shows the appearance of two flows, one of them anomalous, the other not. 
A similar pattern is found in the gluon sector, which obviously is not affected by the mass term. 
In this case the on-shell and transverse condition on the external gluons brings to three very simple form factors  
whose expressions are
\bea
\Phi_{1\,g}(k^2) &=& \frac{11 \, g^2}{72 \pi^2 \, k^2} \, C_A \,, \qquad 
\Phi_{2\,g}(k^2) = \frac{g^2}{288 \pi^2 \, k^2} \, C_A \,, \\
\Phi_{3\,g}(k^2) &=& - \frac{g^2}{8 \pi^2}  C_A \bigg[ \frac{65}{36} + \frac{11}{6} \mathcal{ B}_0^{\overline{MS}}(k^2,0) - \mathcal {B}_0^{\overline{MS}}(0,0) +  k^2  \,\mathcal C_0(k^2,0) \bigg].
\label{gl2}
\eea
The $\overline{MS}$ renormalized scalar integrals can be found in Appendix \ref{AppScalarIntegrals}.
Also in this case, it is clear that the simple poles in $\Phi_{1\,g}$ and $\Phi_{2\,g}$, the two form factors which are not affected by the renormalization, are accounted for by two spectral densities which are proportional to $\delta(s)$. The anomaly pole in $\Phi_{1\,g}$ is accompanied by a second pole in the non anomalous form factor $\Phi_{2\,g}$. Notice that $\Phi_{3 g}$ is affected by renormalization, and as such it is not considered relevant in the spectral analysis. 

\subsection{$\langle TVV \rangle$ and the two spectral flows of the electroweak theory} 
\begin{figure}[t]
\centering
\includegraphics[scale=0.8]{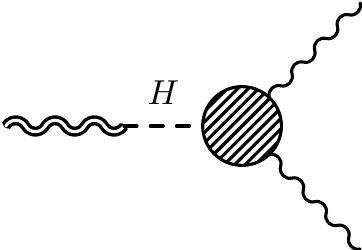}
\caption{Amplitude with the graviton - Higgs mixing vertex generated by the term of improvement. The blob represents the SM Higgs -VV' vertex at one-loop.}
 \label{HVVImpr}
\end{figure}

The point illustrated above can be extended to the entire electroweak theory by looking at some typical diagrams which manifest a trace anomaly. The simplest case is the $\langle TVV \rangle$ in the full electroweak theory, where $V$, in this case, denotes on-shell photons. At one loop level it is given by the vertex $\Gamma^{\mu\nu\alpha\beta}$ and expanded onto two terms
 \beq
\label{TVVew}
 \Gamma^{\mu\nu\alpha\beta}(p,q)=\Sigma^{\mu\nu\alpha\beta}(p,q) +\Delta^{\mu\nu\alpha\beta}(p,q)\,,
 \eeq
where $\Sigma^{\mu\nu\alpha\beta}(p,q)$ is a full irreducible contribution, corresponding to topologies of triangles, bubbles and tadpoles. %
In this case $\Sigma^{\mu\nu\alpha\beta}(p,q)$ is given by the expression%
 \cite{Coriano:2011ti,Coriano:2011zk,CDS}
\bea
\Sigma ^{\mu\nu\a\b}(p,q) = \Sigma_{F}^{\mu\nu\a\b}(p,q) + \Sigma_{B}^{\mu\nu\a\b}(p,q) + \Sigma_{I}^{\mu\nu\a\b}(p,q),
\eea
corresponding to the exchange of fermions ($\Sigma_{F}$), gauge bosons ($\Sigma_{B}$) and to a term of improvement $(\Sigma_{I})$. The latter is generated by an EMT of the form 
\bea
T^I_{\mu\nu} = - \frac{1}{3} \bigg[ \partial_{\mu} \partial_{\nu} - \eta_{\mu\nu} \, \Box \bigg] \mathcal H^\dag \mathcal H = - \frac{1}{3} \bigg[ \partial_{\mu} \partial_{\nu} - \eta_{\mu\nu} \, \Box \bigg] \bigg( \frac{H^2}{2} + \frac{\phi^2}{2} + \phi^{+}\phi^{-} + v \, H \bigg).
\eea
and is responsible for a bilinear mixing between the EMT and the Higgs field. \\
The term $\Delta^{\mu\nu\alpha\beta}(p,q)$ in Eq.(\ref{TVVew}) comes from the insertion of the EMT of improvement given above with the Standard Model $H\gamma\gamma$ vertex. The relevant diagram is reported in Fig. (\ref{HVVImpr}). The inclusion of this term 
is necessary, from a careful analysis of the Ward identities, as shown in \cite{Coriano:2011zk}. \\
They full irreducible contributions are expanded as 
\bea
\Sigma ^{\mu\nu\alpha\beta}_{F}(p,q) &=&  \, \sum_{i=1}^{3} \Phi_{i\,F} (s,0, 0,m_f^2) \, \phi_i^{\mu\nu\alpha\beta}(p,q)\,, \\
\Sigma ^{\mu\nu\alpha\beta}_{B}(p,q) &=&  \, \sum_{i=1}^{3} \Phi_{i\,B} (s,0, 0,M_W^2) \, \phi_i^{\mu\nu\alpha\beta}(p,q)\,, \\
\Sigma ^{\mu\nu\alpha\beta}_{I}(p,q) &=&  \Phi_{1\,I} (s,0, 0,M_W^2) \, \phi_1^{\mu\nu\alpha\beta}(p,q) + \Phi_{4\,I} (s,0, 0,M_W^2) \, \phi_4^{\mu\nu\alpha\beta}(p,q) \,.
\eea
with $s=k^2=(p+q)^2$, $\phi_i^{\mu\nu\alpha\beta}(p,q)$ given in Eq. (\ref{phitensors}) and 
\beq 
\phi_4^{\mu\nu\alpha\beta}(p,q) = (s \, \eta^{\mu\nu} - k^{\mu}k^{\nu}) \, \eta^{\alpha\beta},
\eeq
while the $\Delta$ term reads as
\bea
\Delta^{\mu\nu\alpha\beta}(p,q) &=& \Delta^{\mu\nu\alpha\beta}_I (p,q) \nn \\
&=&  \Psi_{1\, I} (s,0, 0,m_f^2,M_W^2,M_H^2) \, \phi_1^{\mu\nu\alpha\beta}(p,q) + \Psi_{4 \, I} (s,0, 0,M_W^2)  \, \phi_4^{\mu\nu\alpha\beta}(p,q)\, .
\label{DAA}
\eea
This is built by combining the tree level vertex for EMT/Higgs mixing, coming from the improved EMT, and the Standard Model $H\gamma\gamma$ correlator at one-loop.\\
The spectral densities of the fermion contributions, related to $\Sigma_F$  have structure similar to those computed above in Eq. (\ref{rhoq}), with $\rho_{\Phi_{1F}} \sim \rho_{1 q}(s)$ and $\rho_{\Phi_{2F}} \sim \rho_{2 q} (s)$. Therefore we have two sum rules and two spectral flows also in this case, 
following the pattern discussed before for the spectral densities in Eq. (\ref{rhoq}).\\
A similar analysis on the two form factors $\Phi_B$ in the gauge boson sector gives 
\beq
\rho_{\phi_{1 B}}(s)=\frac{ 2 M_W^2}{s^3} (2 M_W^2 - s) \alpha \log\left(\frac{1+ \sqrt{\tau(s,M_W^2}) }{1-\sqrt{\tau(s,M_W^2} )}\right)\theta(s- 4 M_W^2)
\eeq
while $\rho_{\phi_{2 B}}$ has the same functional form of $\rho_{\phi_{2 F}}$, modulo an overall factor, with $m$, the fermion mass, replaced by the $W$ mass $M_W$. Notice that both $\rho_{\phi_{1 B}}$ and $\rho_{\phi_{2 B}}$, as well as $\rho_{\phi_{1 F}}$ and $\rho_{\phi_{2 F}}$ are deprived of resonant contributions, being the diagrams massive. \\
Coming to the form factors in $\Sigma_I$, one realizes that the spectral density of $\Phi_{1I}$ 
shares the same functional form of $\rho_\chi$, extracted from Eq. (\ref{spectralrho}), and there is clearly a sum rule associated to it. Also in this case, this result is accompanied by the $1/k^2$ behaviour of the corresponding form factor, due to the anomaly. \\
Finally, for the case of $\psi_{1I}$, one can also show that the spectral density finds support only above the two particle cuts. The cuts are linked to $2 m$ and $2 M_W$. In this case there is no sum rule and the contribution is not affected by an anomaly pole, as expected, being the virtual loop connected with the $H\gamma\gamma$ vertex (see Fig. \ref{HVVImpr}).

\subsection{The non-transverse $\langle AVV \rangle$ correlator } 
Before closing the analysis on the spectral densitites of non supersymmetric theories, we pause for few comments 
on the structure of the $\langle AVV \rangle$ diagram, which, as we are going to show, is affected by a single flow even if we do not impose the transversality condition on the two photons. We consider once more the anomaly vertex as parameterized in 
Eq. (\ref{anom1}), and consider the second form factor $A_{4+6}\equiv A_4 + A_6$, which contributes to the anomaly loop for non transverse (but on-shell) photons. The expression of $A_6$, the anomalous form factor, has been given in 
Eq. (\ref{a6}), while $A_4$ is given by 
\beq
A_4(k^2,m^2)=-\frac{1}{2 \pi^2 k^2}\left[ 2 - \sqrt{\tau(k^2,m^2)}\log \frac{\sqrt{\tau(k^2,m^2)}+1}{\sqrt{\tau(k^2,m^2)}-1} \right], \qquad k^2<0
\label{a4}
\eeq
and $A_{4+6}$ takes the form 
\beq
A_{4+6}(k^2,m^2)=\frac{1}{2\pi^2 k^2}\left[ -1 + \sqrt{\tau(k^2,m^2)} \log\frac{\sqrt{\tau(k^2,m^2)} + 1}{\sqrt{\tau(k^2,m^2)} - 1}  + \frac{m^2}{k^2} \log^2 \frac{\sqrt{\tau(k^2,m^2)} + 1}{\sqrt{\tau(k^2,m^2)} - 1} \right].
\eeq
Its discontinuity is given by 
\beq
\textrm{Disc}\,A_{4+6}(k^2,m^2)= - 2 i \pi\left[ \frac{\sqrt{\tau(k^2,m^2)}}{k^2} + \frac{2 m^2}{(k^2)^2}\log \frac{\sqrt{\tau(k^2,m^2)}+1}{\sqrt{\tau(k^2,m^2)}-1} \right] \theta(k^2 - 4m^2).
\eeq
Notice that in this case there is no sum rule satisfied by this spectral density, being non-integrable along the cut.  
Coming to the spectral density for the anomaly coefficient $A_6$, this is proportional to the density of 
$\chi(s,m^2)$ given in Eq. (\ref{spectralrho})
and shares the same behaviour found for $\rho_{\chi}(s,m^2)$, as expected. This analysis shows that in the $\langle AVV \rangle$ case one encounters a single sum rule and a single massive flow which degenerates into a $\delta(s)$ behaviour, as in the supersymmetric case. This condition remains valid also for non-transverse vector currents. It is then clear that the crucial difference between the non supersymmetric and the supersymmetric case is carried by the $\langle TVV \rangle$ diagram, due to the extra sum rule discussed above.

\subsection{Cancellations in the supersymmetric case}
In order to further clarify how the cancellation of the extra poles occurs for the supersymmetric $\langle TVV \rangle$, we consider the non-anomalous form factor $f_2$ in a general theory (given in Eqs. (\ref{FFfermions},\ref{FFscalars},\ref{FFgauge})), with $N_f$ Weyl fermions, $N_s$ complex scalars and $N_A$ gauge fields. We work, for simplicity, in the massless limit. In this case the non anomalous form factor $f_2$, which is affected by pole terms, after combining scalar, fermions and gauge contributions can be written in the form
\bea
\label{F2total}
f_2(k^2) &=& \frac{N_f}{2} f_2^{(f)}(k^2) + N_s \, f_2^{(s)}(k^2) + N_A \,  f_2^{(A)}(k^2) \nn \\
&=& \frac{g^2}{144 \pi^2 \, k^2} \left[ -  \frac{N_f}{2} T(R_f) + N_s \, \frac{T(R_s)}{2} + N_A \, \frac{T(A)}{2} \right] \,, 
\eea
where the fermions give a negative contribution with respect to scalar and gauge fields. 
If we turn to a $\mathcal N=1$ Yang-Mills gauge theory, which is the theory that we are addressing, we need to consider in the anomaly diagrams the virtual exchanges both of a chiral and of a vector supermultiplet. In the first case the multiplet is built out of one Weyl fermion and one complex scalar, therefore in Eq.(\ref{F2total}) we have $N_f = 1, N_s = 1, N_A = 0$ with $T(R_f) = T(R_s)$. With this matter content, the form factor is set to vanish. \\
For a vector multiplet, on the othe other end, we have one vector field and one Weyl fermion, all belonging to the adjoint representation and then we obtain $N_f=1, N_s=0, N_A=1$ with $T(R_f) = T(A)$. Even in this case all the contributions in the $f_2$ form factor sum up to zero. It is then clear that the cancellation of the extra poles in the $\langle TVV \rangle$ is a specific tract of supersymmetric Yang Mills theories, due to their matter content, not shared by an ordinary gauge theory. A corollary of this is that in a supersymmetric theory we have just one spectral flow driven by the deformation parameter $m$, accompanied by one sum rule for the entire deformation.

\section{The anomaly effective action and the pole cancellations for $\mathcal{N}=4$ }
\begin{figure}[t]
\centering
\subfigure{\includegraphics[scale=0.6]{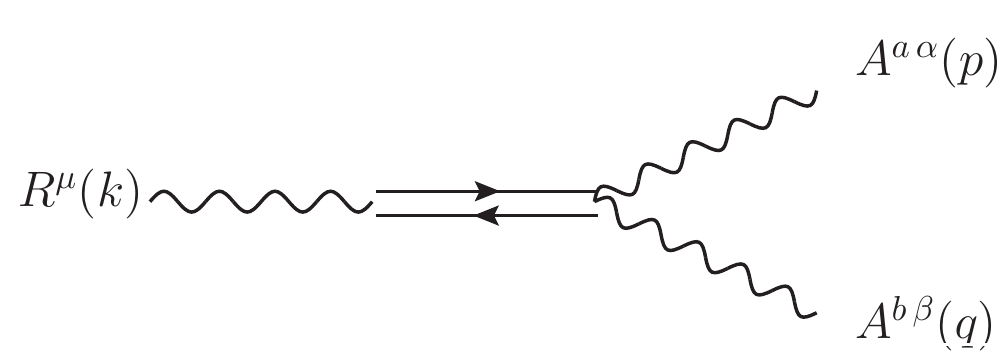}} \hspace{.5cm}
\subfigure{\includegraphics[scale=0.6]{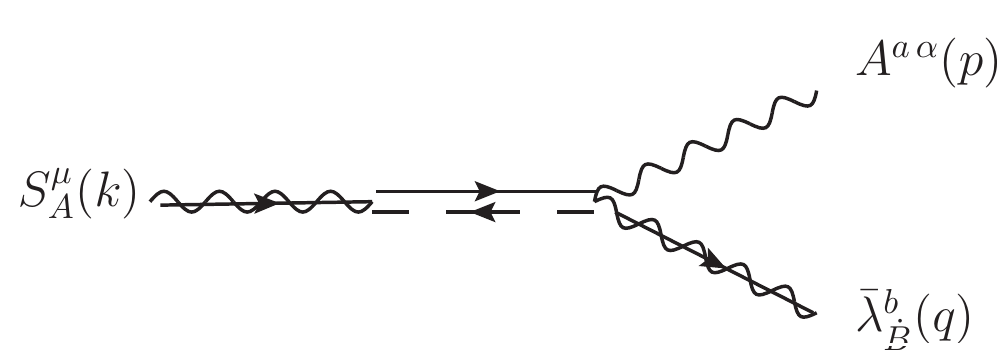}} 
\subfigure{\includegraphics[scale=0.6]{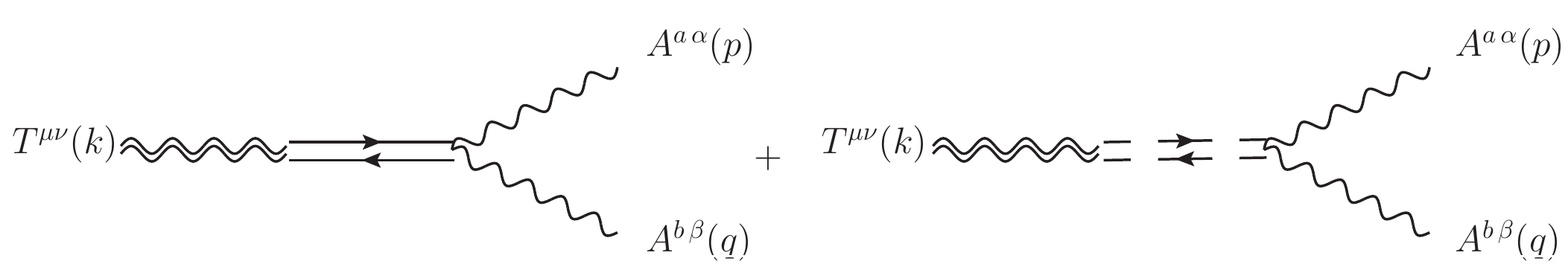}} 
\caption{The collinear diagrams corresponding to the exchange of a composite axion (top right), a dilatino (top left) and the two sectors of an intermediate dilaton (bottom). Dashed lines denote intermediate scalars.}
\label{RST}
\end{figure}

The presence of poles in the effective action is associated either with fundamental fields in the defining Lagrangian or with the exchange of intermediate bound states. Here we present the quantum effective action obtained from the three-point correlation functions discussed previously. We consider the massless case for the chiral supermultiplet and on-shell external gauge bosons and gauginos. 
The anomalous part is given by the three terms 
\beq
S_{\textrm{anom}}= S_{\textrm{axion}} +  S_{\textrm{dilatino}} + S_{\textrm{dilaton}}
\eeq
which are
\bea
S_{\textrm{axion}}&=& - \frac{g^2}{4 \pi^2} \left(  T(A) - \frac{T(R)}{3} \right)  \int d^4 z \, d^4 x \,  \partial^\mu B_\mu(z) \, \frac{1}{\Box_{zx}} \, \frac{1}{4} F_{\alpha\beta}(x)\tilde F^{\alpha\beta}(x) \\
S_{\textrm{dilatino}} &=&  \frac{g^2}{2 \pi^2} \left(T(A) - \frac{T(R)}{3} \right)  \int d^4 z \, d^4 x \bigg[  \partial_\nu \Psi_\mu(z) \sigma^{\mu\nu} \sigma^\rho \frac{\stackrel{\leftarrow}{\partial_\rho}}{\Box_{zx}} \,  \bar \sigma^{\alpha\beta} \bar \lambda(x) \frac{1}{2} F_{\alpha\beta}(x) + h.c. \bigg]  \\
S_{\textrm{dilaton}}&=&- \frac{g^2}{8 \pi^2} \left(T(A)  - \frac{T(R)}{3} \right) \int d^4 z \, d^4 x \, \left(\Box h(z) - \partial^\mu \partial^\nu h_{\mu\nu}(z) \right) \,  \frac{1}{\Box_{zx}} \,\frac{1}{4} F_{\alpha\beta}(x) F^{\alpha\beta}(x) 
\eea
We show in Figs.\ref{RST} the three types of intermediate states which interpolate between the Ferrara-Zumino hypercurrent and the gauge $(A)$ and the gaugino ($\lambda$) of the final state. The axion is identified by the collinear exchange of a bound fermion/antifermion pair in a pseudoscalar state, generated in the $\langle RVV \rangle$ correlator. In the case of the $\langle SVF \rangle$ correlator, the intermediate state is a collinear scalar/fermion pair, interpreted as a dilatino. In the $\langle TVV \rangle$ case, the collinear exchange is a linear combination of a fermion/antifermion and scalar/scalar pairs.  

The non-anomalous contribution is associated with the extra term $S_0$ which is given by
\bea
S_0 &=&  \frac{g^2}{16 \pi^2}  \int d^4 z \, d^4 x \,  h_{\mu\nu}(z) \left( T(R) \, \tilde \Phi_2(z-x) + T(A) \, \tilde V(z-x) \right) T^{\mu\nu}_{gauge}(x) \nn \\
&+&   \frac{g^2}{64 \pi^2}  \int d^4 z \, d^4 x \bigg[ i \, \Psi_\mu(z) \left( T(R) \, \tilde \Phi_2(z-x) + T(A) \, \tilde V(z-x) \right) S^{\mu}_{gauge}(x)  + h.c. \bigg] \,,
\eea
where $\tilde \Phi_2(z-x)$ and $\tilde V(z-x)$ are the Fourier transforms of $\Phi_2(k^2,0)$ and $V(k^2)$ respectively. 

Their contributions in position space correspond to nonlocal logarithmic terms.

The relation between anomaly poles, spectral density flows and sum rules appear to be a significant feature of supersymmetric theories affected by anomalies. It is then clear that supersymmetric anomaly-free theories should be free of such contributions in the anomaly effective action. In this respect, it natural to turn to the $\mathcal N = 4$ theory, which is free of anomalies, in order to verify and validate this reasoning. Indeed the $\beta$ function of the gauge coupling constant in this theory has been shown to vanish up to three loops, and there are several arguments about its vanishing to all perturbative orders. As a consequence, the anomaly coefficient in the trace of the energy-momentum tensor, being proportional to the $\beta$ function, must vanish identically and the same occurs for the other anomalous component, related to the $R$ and to the $S$ currents in the Ferrara-Zumino supermultiplet. \\
We recall that in the $\mathcal{N}=4$ theory the spectrum contains a gauge field $A^\mu$, four complex fermions $\lambda^i$ ($i=1,2,3,4$) and six real scalars $\phi_{ij} = - \phi_{ji}$ $(i,j=1,2,3,4)$. All fields are in the adjoint representation of the gauge group. 

From the point of view of the $\mathcal N=1$ SYM, this theory can be interpreted as describing a vector and three massless chiral supermultiplets, all in the adjoint representation. Therefore the $\langle TVV \rangle$ correlator in $\mathcal N=4$ can be easily computed from the general expressions in Eqs (\ref{TChiralOSMassless}) and (\ref{TVectorOS}) which give
\bea
\label{residualV}
\Gamma^{\mu\nu\alpha\beta}_{(T)}(p,q) = \frac{g^2 \, T(A)}{16 \pi^2} \left[ V(k^2) + 3 \Phi_2(k^2,0)\right] t_{2S}^{\mu\nu\alpha\beta}(p,q) 
= - \frac{g^2 \, T(A)}{8 \pi^2} k^2 \, \mathcal C_0(k^2,0) \, t_{2S}^{\mu\nu\alpha\beta}(p,q) \,.
\label{gamma}
\eea
One can immediately observe from the expression above the vanishing of the anomalous form factor proportional to the tracefull tensor structure 
$t_{1S}^{\mu\nu\alpha\beta}$. The partial contributions to the same form factor, which can be computed using Eqs. (\ref{TChiralOSMassless}) and (\ref{TVectorOS}) for the various components, are all affected by pole terms, but they add up to give a form factor whose residue at the pole is proportional to the $\beta$ function of the $\mathcal N=4$ theory. It is then clear that the vanishing of the conformal anomaly, via a vanishing $\beta$ function, is equivalent to the cancellation of the anomaly pole for the entire multiplet. \\
Notice also that the only surviving contribution in Eq. (\ref{gamma}), proportional to the traceless tensor structure 
$t_{2S}^{\mu\nu\alpha\beta}$, is finite. This is due to the various cancellations between the UV singular terms from $V(k^2)$ and $\Phi_2(k^2,0)$ which give a finite correlator without the necessity of any regularization. 

We recall that the cancellation of infinities and the renormalization procedure, as we have already seen in the $\mathcal N=1$ case, involves only the form factor of tensor $t_{2S}^{\mu\nu\alpha\beta}$, which gets renormalized with a counterterm 
 proportional to that of the two-point function $\langle AA \rangle$, and hence to the gauge coupling. 
 For this reason the finiteness of the second form factor and then of the entire $\langle TVV \rangle$ in $\mathcal N=4$ is directly connected to the vanishing of the anomalous term, because its non-renormalization naturally requires that the $\beta$ function has to vanish.

\chapter{ Dilaton Phenomenology at the LHC with the $TVV$ vertex}

%%%%%%%%%% dilaton &&&&&&&&&&&&&&&&&&
\section{Synopsis} 
In this chapter we explore the potential for the discovery of a dilaton $O(200-500)$ GeV
in a classical scale/conformal invariant extension of the Standard Model by investigating
the size of the corresponding breaking scale $\Lambda$ at the LHC.
In particular, we address the recent bounds on $\Lambda$ derived from Higgs boson searches. We
investigate if such a dilaton can be produced via gluon-gluon fusion, presenting rates for its
decay either into a pair of Higgs bosons or into two heavy gauge bosons, which can give rise
to multi-leptonic final states. We include a detailed analysis via PYTHIA-FastJet 
of the dominant Standard Model backgrounds, at a centre of mass energy of 14 TeV. We
show that early data of $\sim 20$ fb$^{-1}$
can certainly probe the region of parameter space where
such a dilaton is allowed. A conformal scale of 5 TeV is allowed by the current data, for
almost all values of the dilaton mass investigated.

\section{Introduction}
 An important feature of the electroweak sector of the Standard Model (SM) is its approximate scale invariance which holds if the quadratic terms of the Higgs potential are absent. 
These terms are obviously necessary in order for the theory to be in a spontaneously broken phase with a vacuum expectation value (vev) $v$ which is fixed by the experiments. \\
The issue of incorporating a mechanism of spontaneous symmetry breaking of a gauge symmetry while preserving the scale invariance of the Lagrangian is a subtle one, 
which naturally brings to the conclusion that the breaking of this symmetry has to be dynamical, with the inclusion of a dilaton field. In this case the mass of the dilaton should be attributed to a specific symmetry-breaking potential, probably of non-perturbative origin. A dilaton, in this case, is likely to be a composite \cite{CDS} state, with a conjectured behaviour which can be partly discussed using the conformal anomaly action.  

The absence of any dimensionful constant 
in a tree level Lagrangian is, in fact, a necessary condition in order to guarantee the scale invariance of the theory.  This is also the framework that we will consider, which is based on the requirement of {\em classical} scale invariance. A stricter condition, for instance, lays in the (stronger) requirement of quantum scale invariance, with correlators which, in some cases, are completely fixed by the symmetry and incorporate the anomaly \cite{OP, EO, BMS1,BMS2,BMS3}.  
In the class of theories that we consider, the invariance of the Lagrangian under special conformal transformations are automatically fulfilled by the condition of scale invariance. For this reason we will 
refer to the breaking of such symmetry as to a conformal breaking.\\
 Approaching a scale invariant theory from a non scale-invariant one requires all the dimensionful couplings of the model to be turned into dynamical fields, with a compensator ($\Sigma(x)$) which is rendered dynamical by the addition of a scalar kinetic term. It is then natural to couple such a field both to the anomaly and to the explicit (mass-dependent) extra terms which appear in the classical trace of the stress-energy tensor. \\
 The inclusion of an extra $\Sigma$-dependent potential in the scalar sector of the new theory is needed in order to break the conformal symmetry at the TeV scale, with a dilaton mass which remains, essentially, a free parameter.   We just mention that for a classically scale invariant extension of the SM Lagrangian, the choice of the scalar potential has to be appropriate, in order to support a spontaneously broken phase of the theory, such as the electroweak phase \cite{CDS}. For such a reason, the two mechanisms of electroweak and scale breaking have to be directly related, with the electroweak scale $v$ and the conformal breaking scale $\Lambda$ linked by a simple expression. At the same time, the invariance of the action under a change induced by a constant shift of the potential, which remains unobservable in a non scale-invariant theory, 
becomes observable and affects the vacuum energy of the model and its stability.  \\
The goal of our work is to elaborate on a former theoretical analysis \cite{CDS} of dilaton interactions, by discussing the signatures and the phenomenological bounds on a possible state of this type at the LHC, using the current experimental constraints. Some of the studies carried so far address a state of geometrical origin ({\em the radion}) \cite{GRW}, which shares several of the properties of a (pseudo) Nambu-Goldstone mode of a broken conformal symmetry, except, obviously, its geometric origin and its possible compositeness. Other applications are in inflaton physics (see for instance \cite{AR}). \\ 
The production and decay mechanisms of a dilaton, either as a fundamental or a composite state, are quite similar to those of the Higgs field, except for the presence of a suppression related to a conformal scale ($\Lambda$) and of a direct contribution derived from the conformal anomaly. As we are going to show, the latter causes an enhancement of the dilaton decay modes into massless states, which is maximized if its coupling $\xi$ is conformal.

In the phenomenological study that we present below we do not consider possible modifications of the production and decay rates of this particle typical of the dynamics of a bound state, if a dilaton is such. 
This point would require a separate study that will be addressed elsewhere. We just mention that there are significant indications from the study of conformal anomaly actions \cite{CDS,CCDS} both in ordinary and in supersymmetric theories, that the conformal anomaly manifests with the appearance of anomaly poles in specific channels. These interpolate with the dilatation current \cite{CDS}, similarly to the behaviour manifested by an axial-vector current in $AVV$ diagrams.
The exchange of these massless poles are therefore the natural signature of anomalies in general, being them either chiral or conformal  \cite{ACD0}. 
Concerning the conformal ones, these analyses have been fully worked out in perturbation theory in a certain class of correlators ($TVV$ diagrams) \cite{Giannotti:2008cv,Armillis:2009pq}, starting from QED. We have included one  
section (section \ref{non0xi}) where we briefly address these points, in view of some recent developements and prospects for future studies. In this respect, the analysis that we present should be amended with the inclusion of corrections coming from a possible wave function of the dilaton in the production/decay processes involving such a state. These possible developments require  specific assumptions which we are not going  to discuss in great detail in the current study but on which we will briefly comment prior to our conclusions.

\section{Classical scale invariant extensions of the Standard Model and dilaton interactions}
\label{revv}
A scale invariant extension of the SM, at tree level, can be trivially obtained by promoting all the 
dimensionful couplings in the scalar potential, which now includes quartic and quadratic Higgs terms,  to dynamical fields. The new field ($\Sigma(x)=\Lambda e^{\rho(x)/\Lambda}$) is accompanied by a conformal scale ($\Lambda$) and introduces a dilaton field $\rho(x)$, as a fluctuation around the vev of $\Sigma(x)$
 \beq
 \Sigma(x)= \Lambda + \rho(x) + O(\rho^2), \qquad \qquad \langle \Sigma(x) \rangle=\Lambda, \qquad \qquad \langle \rho(x) \rangle =0.
 \eeq

 The leading interactions of the dilaton with the SM fields are obtained through the divergence of the dilatation current. This corresponds to the trace of the energy-momentum tensor $T^\mu_{\mu \, SM}$ computed on the SM fields
 
\bea\label{tmunu}
\mathcal L_{int} = -\frac{1}{\Lambda}\rho T^\mu_{\mu\,SM}.
\eea

\begin{figure}[t]
\begin{center}
\hspace*{-2cm}
\mbox{\subfigure[]{
\includegraphics[width=0.5\linewidth]{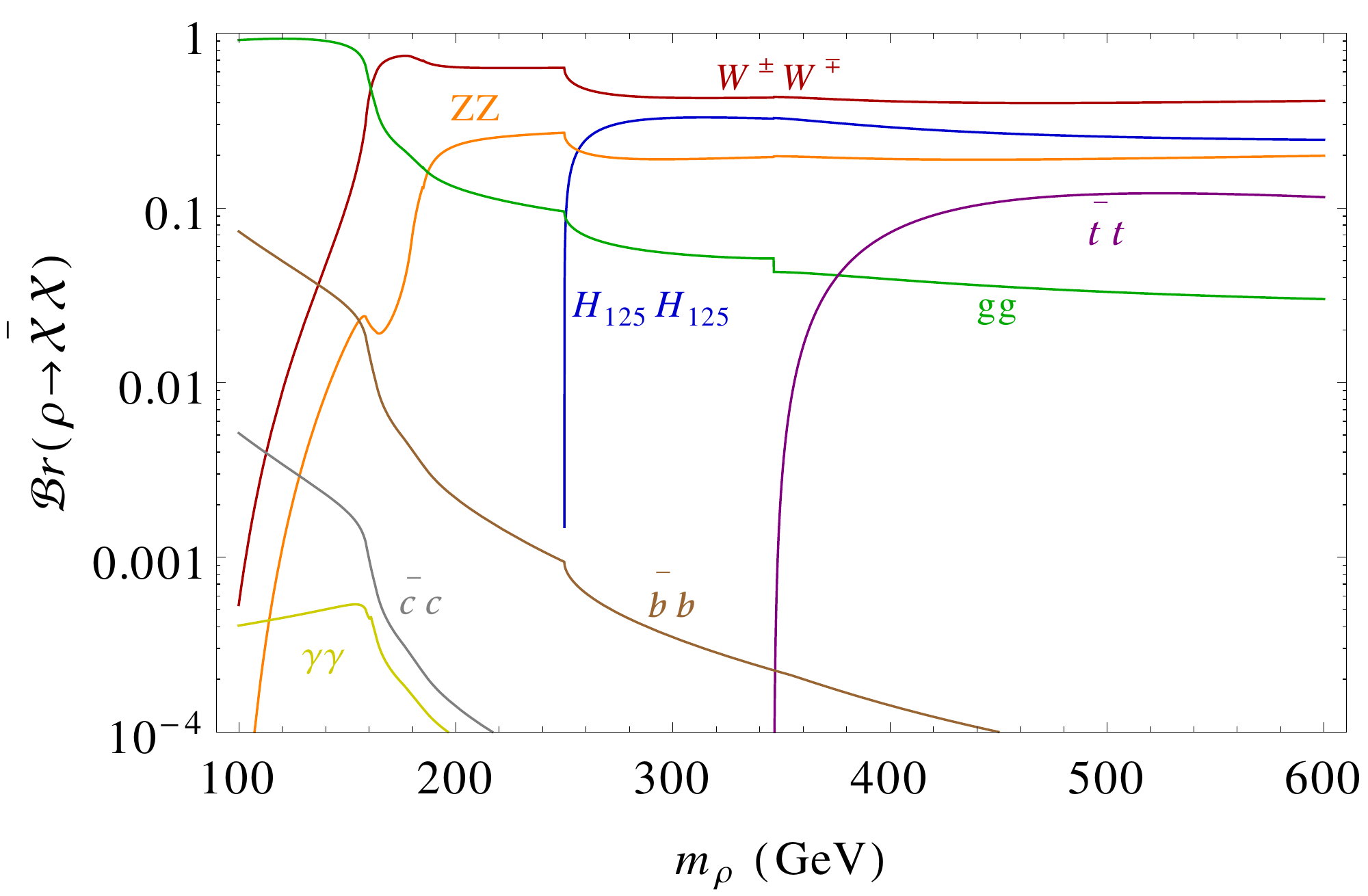}}\hskip 15pt
\subfigure[]{\includegraphics[width=0.5\linewidth]{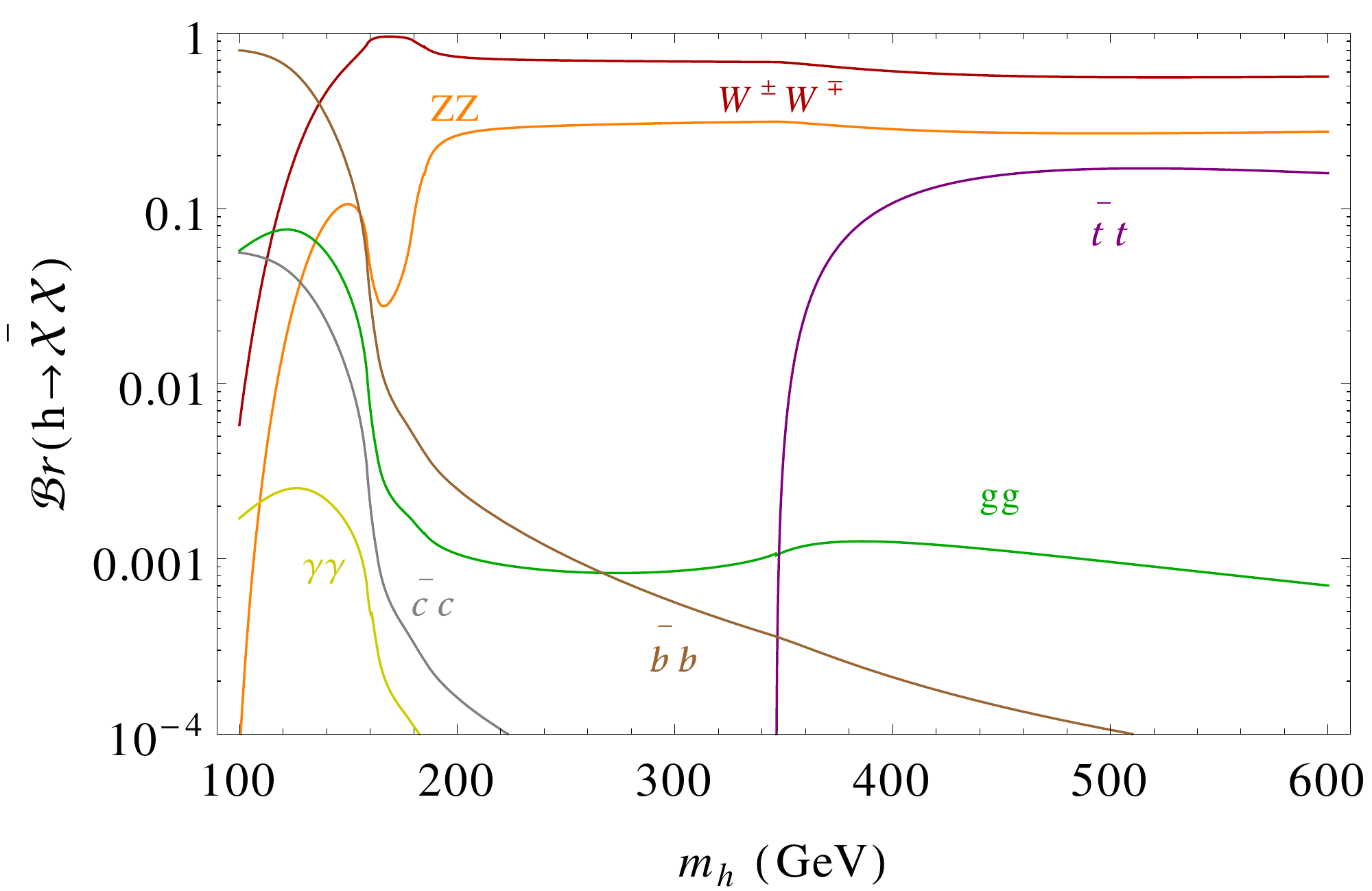}}}
\caption{The mass dependence of the branching ratios of the dilaton (a) and of the Higgs boson (b).}\label{brrhoh}
\end{center}
\end{figure}

% \section{Decays of the dilaton}\label{decays} 
The interactions of the dilaton to the massive states are very similar to those of the Higgs, except that $v$ is replaced by $\Lambda$. The distinctive feature between the dilaton and the SM Higgs emerges in the coupling with photons and gluons.
One-loop expressions for the decays into all the neutral currents sector has been given in \cite{CDS}, while leading order decay widths of $\rho$ in some relevant channels (fermions, vector and Higgs pairs) are easily written in the form (for a minimally coupled dilaton, with $\xi=0$)
\begin{align}
&\Gamma_{\rho\to \bar ff}=N_f^c\frac{m_\rho}{8\pi}\frac{m_f^2}{\Lambda^2}\left(1-4\frac{m_f^2}{m_\rho^2}\right)^{3/2},\label{ffWidth}\\
&\Gamma_{\rho\to VV}=\delta_V\frac{1}{32\pi}\frac{m_\rho^3}{\Lambda^2}\left(1-4\frac{m_V^2}{m_\rho^2}+12\frac{m_V^4}{m_\rho^4}\right)\sqrt{1-4\frac{m_V^2}{m_\rho^2}},\label{vvWidth}\\
&\Gamma_{\rho\to HH}=\frac{1}{32\pi}\frac{m_\rho^3}{\Lambda^2}\left(1+2\frac{m_H^2}{m_\rho^2}\right)^2\sqrt{1-4\frac{m_H^2}{m_\rho^2}}\label{hhWidth}.
\end{align}
The one-loop expression for decays into $\gamma\gamma$ is
\beqa
\Gamma(\rho \rightarrow \gamma\gamma) 
&=&
\frac{\alpha^2\,m_{\rho}^3}{256\,\Lambda^2\,\pi^3} \, \bigg| \beta_{2} + \beta_{Y} 
 -\left[ 2 + 3\, x_W  +3\,x_W\,(2-x_W)\,f(x_W) \right] \nn \\
&& + \frac{8}{3} \, x_t\left[1 + (1-x_t)\,f(x_t) \right] \bigg|^2. \,
\label{PhiGammaGamma}\nn\\ 
\eeqa
Here, the contributions to the decay, beside the anomaly term, come from the $W$ and the fermion (top) loops. $\beta_2 (= 19/6)$ and $\beta_Y (= -41/6)$ are the $SU(2)_L$ and $U(1)_Y$ $\beta$ functions, while the $x_i$'s  are proportional to the ratios between the mass of each  particle in the loops $m_i$ and the $\rho$ mass. In general, we have defined the variable 
\beq \label{x}
x_i = \frac{4\, m_i^2}{m^2_\rho} \, ,
\eeq
with the index "$i$" labelling the corresponding massive virtual particles. The leading fermionic contribution in the loop comes from the top quark via $f(x_t)$, while $f(x_W)$ denotes the contribution of the $W$-loop. The function $f(x)$ is given by
\beqa
\label{fx}
f(x) = 
\begin{cases}
\arcsin^2(\frac{1}{\sqrt{x}})\, , \quad \mbox{if} \quad \,  x \geq 1 \\ 
-\frac{1}{4}\,\left[ \ln\frac{1+\sqrt{1-x}}{1-\sqrt{1-x}} - i\,\pi \right]^2\, , \quad \mbox{if} \quad \, x < 1.
\end{cases}
\eeqa
related to the scalar three-point master integral through the relation 
\beq \label{C03m}
C_0(s,m^2) = - \frac{2}{s} \, f(\frac{4\,m^2}{s}) \, .
\eeq
The decay rate of a dilaton into two gluons is given by
\beqa
\Gamma(\rho \rightarrow gg) 
&=&
\frac{\alpha_s^2\,m_\rho^3}{32\,\pi^3 \Lambda^2} \, \bigg| \beta_{QCD} +  x_t\left[1 + (1-x_t)\,f(x_t) \right] \bigg|^2 \,,
\label{ggWidth}
\eeqa
where $\beta_{QCD}$ is the QCD $\beta$ function and we have taken the top quark as the only massive fermion, with $x_i$ and $f(x_i)$ defined in Eq. (\ref{x}) and Eq. (\ref{fx}) respectively. \\
Differently from the cross section case,
 the dependence of the decay amplitudes Eq.~(\ref{ffWidth}) - Eq.~(\ref{hhWidth}) on the conformal scale $\Lambda$, which amounts to an overall factor,  
the branching ratios
\bea
\mathit{Br}(\rho\to\bar X X)=\frac{\Gamma_{\rho\to\bar X X}}{\sum_X\Gamma_{\rho\to \bar X X}},
\eea
are $\Lambda$-independent. \\
We show in Fig.~\ref{brrhoh}(a) the decay branching ratios of the dilation as a function of its mass, while in Fig.~\ref{brrhoh}(b) we plot the  corresponding decay branching ratios for a SM-like heavy Higgs boson, here assumed to be of a variable mass. For a light dilaton with $m_\rho < 200$ GeV the dominant decay mode is into two gluons ($gg$), while for a dilaton of larger mass ($m_\rho > 200$ GeV) the same channels which are available for the SM-like Higgs ($ZZ, WW, \bar{t} t$) are now accompanied by a significant $gg$ mode. From the two figures it is easily observed that the 2 gluon rate in the Higgs case is at the level of few per mille, while in the dilaton case is just slightly below 10$\%$.

%%%%%%%%%%%%%%%%%%%%%%%%%
%%%%%%%%%%
\section{Production of the dilaton}\label{prod}
The main production process of the dilaton at the LHC is through gluon fusion, as for the Higgs boson, with a suppression induced by the conformal breaking scale $\Lambda$, which lowers the production rates. Even in this less favourable situation, if confronted with the Higgs production rates of the SM, the dilaton phenomenology can still be studied al the LHC. \\
We calculate the dilaton production cross-section via gluon fusion by weighting the Higgs boson to gluon-gluon decay widths with the corresponding dilaton decay width. The dilaton production cross-section with the incoming gluons thus can be written as
\bea\label{ggrho}
\sigma_{gg\to\rho}=\sigma_{gg\to H}\,\,\frac{\Gamma_{\rho\to gg}}{\Gamma_{H\to gg}} ,
\eea
where we use the same factorization scale in the DGLAP evolution of the parton distribution functions (PDF) of \cite{HCWG}.
The width of $\rho\to gg$ is given in Eq.~(\ref{ggWidth}) and we can use the same expression to calculate the width of $H\to gg$, replacing the breaking scale $\Lambda$ with $v$ and setting $\beta_{QCD}\equiv 0$. The ratio of the two widths appearing in Eq.~(\ref{ggrho}) is then given by
\bea\label{Wrhoh}
\frac{\Gamma_{\rho\to gg}}{\Gamma_{H\to gg}}=\frac{v^2}{\Lambda^2}\frac{m^3_\rho}{m^3_H}\frac{\left|\beta_{QCD}+ x_t\left[1 + (1-x_t)\,f(x_t)\right] \right|^2}{\left| x_t\left[1 + (1-x_t)\,f(x_t)\right] \right|^2} .
\eea

\begin{figure}[thb]
\begin{center}
\hspace*{-1cm}
\mbox{\subfigure[]{
\includegraphics[width=0.5\linewidth]{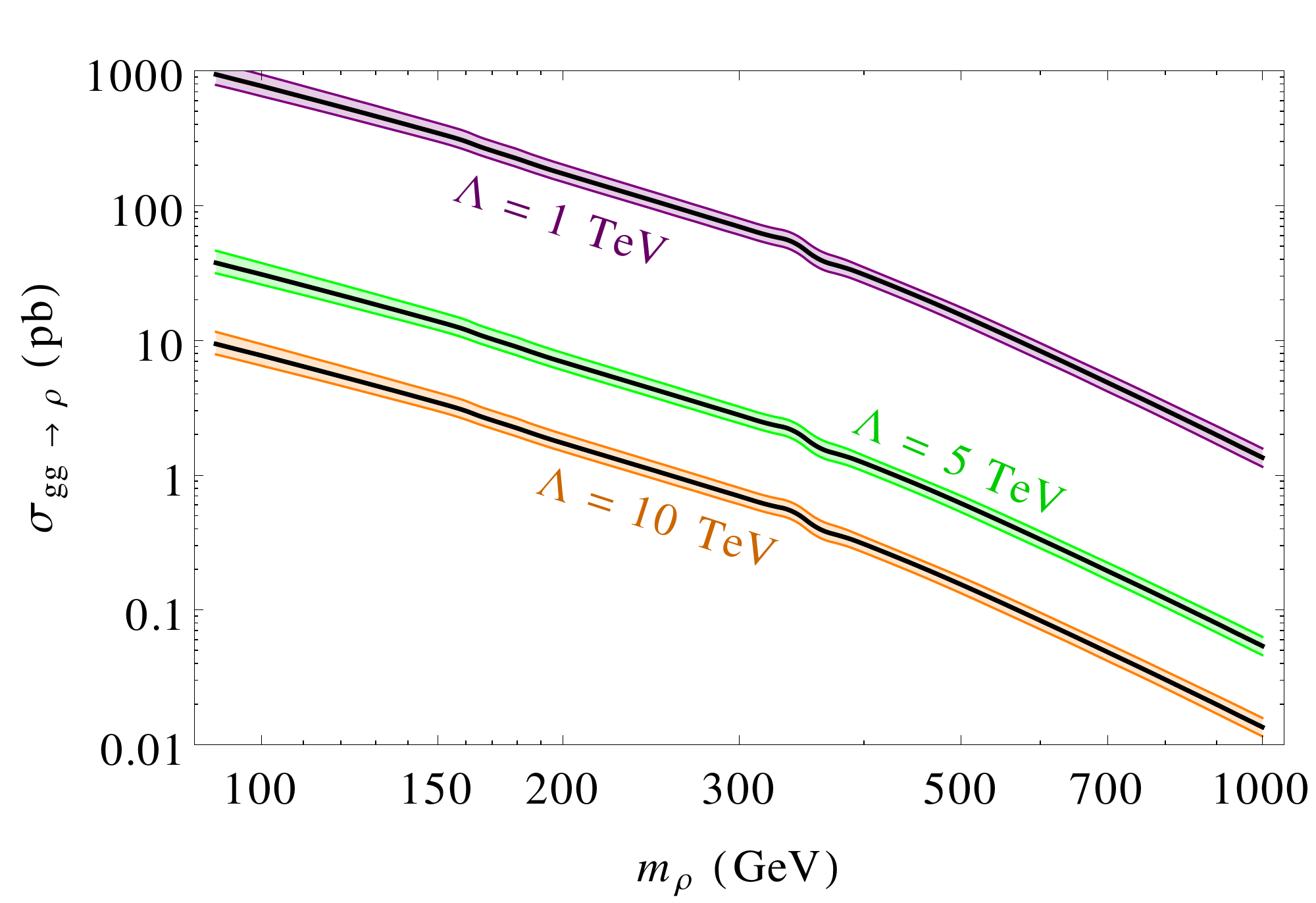}}\hskip 15pt
\subfigure[]{\includegraphics[width=0.5\linewidth]{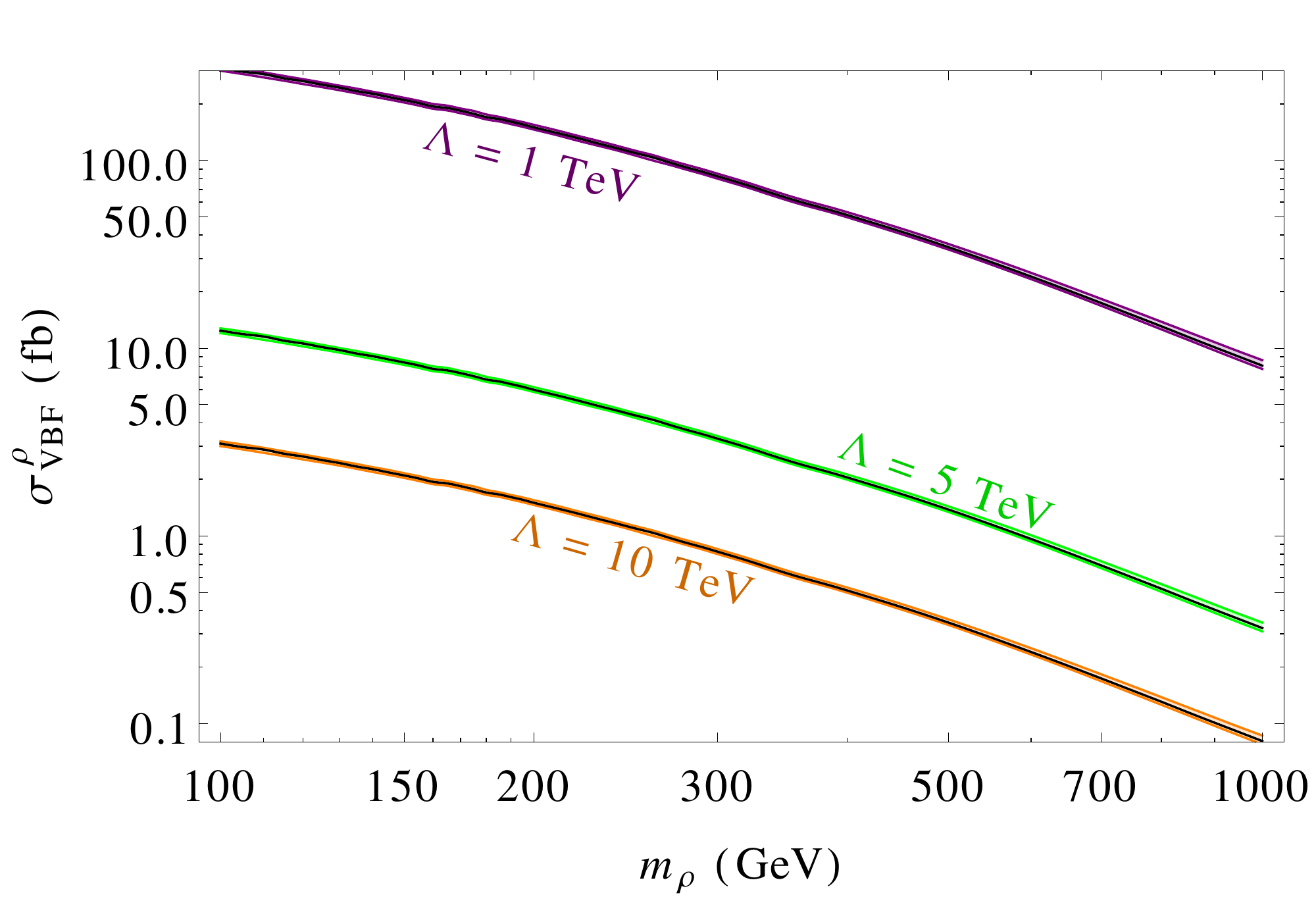}}}
\caption{The mass dependence of the dilaton cross-section via gluon fusion (a) and vector boson fusion (b) for three different choices of the conformal scale, $\Lambda=1, 5, 10$ TeV respectively.}\label{ggvbf}
\end{center}
\end{figure}

In Fig.~\ref{ggvbf} we present the production cross-section of the dilaton at the LHC at 14 TeV centre of mass energy mediated by (a) gluon fusion and (b) vector boson fusion, versus $m_\rho$. Shown are the variations of the same observables for three conformal breaking scales with $\Lambda=1, 5, 10$ TeV. Notice that the contribution from the gluon fusion is about a factor $10^4$ larger than the vector boson fusion.
%%%%%%%%%%%%%%%%%%%%%%%%%
%%%%%%%%%%
\subsection{Bounds on the dilaton from heavy Higgs searches at the LHC}
%%%%%%%%%%%%%%%%%%%%%%%%%
\begin{figure}%[thb]
\begin{center}
\hspace*{-1cm}
\mbox{\subfigure[]{
\includegraphics[width=0.5\linewidth]{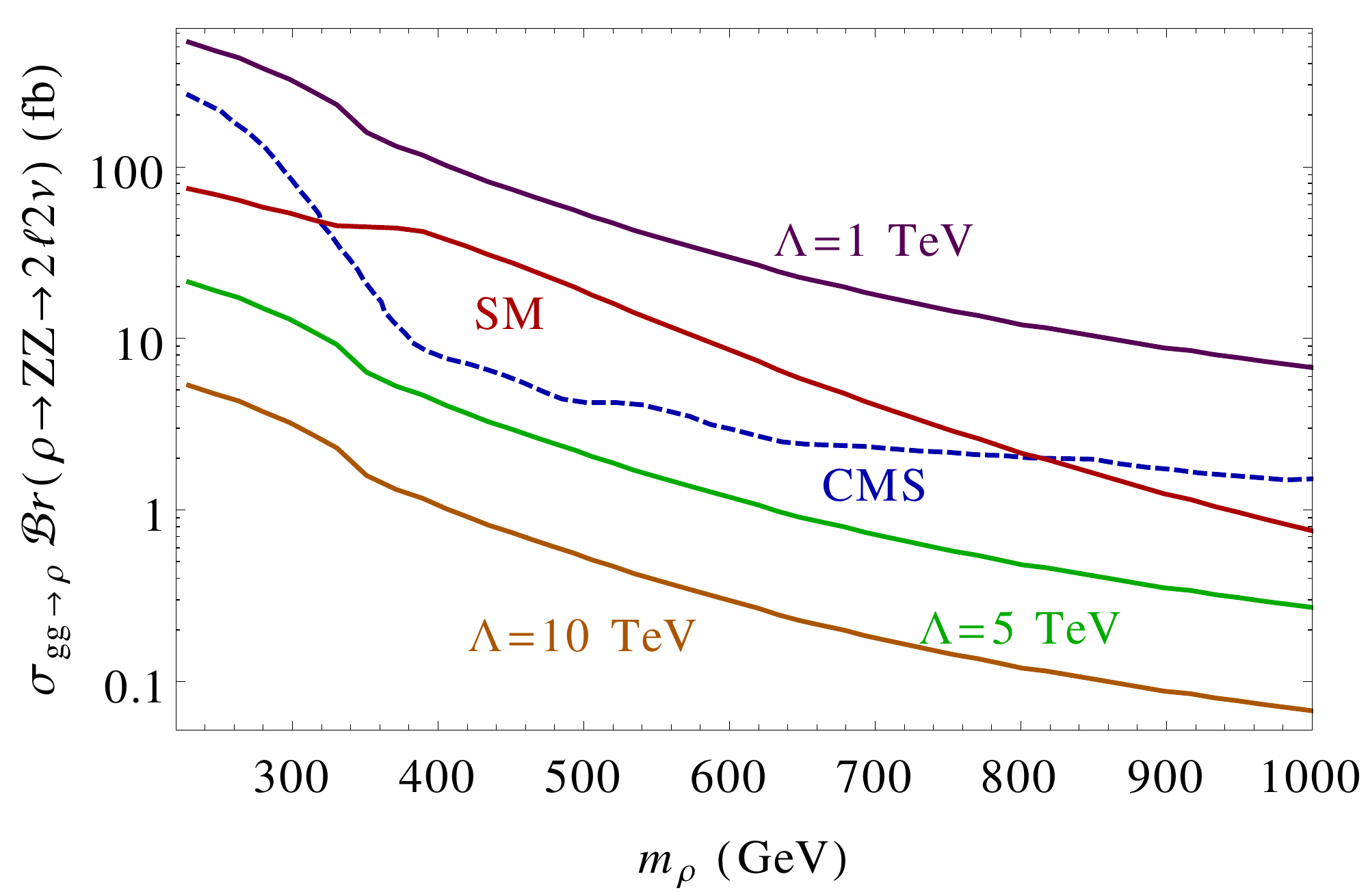}}\hskip 15pt
\subfigure[]{\includegraphics[width=0.5\linewidth]{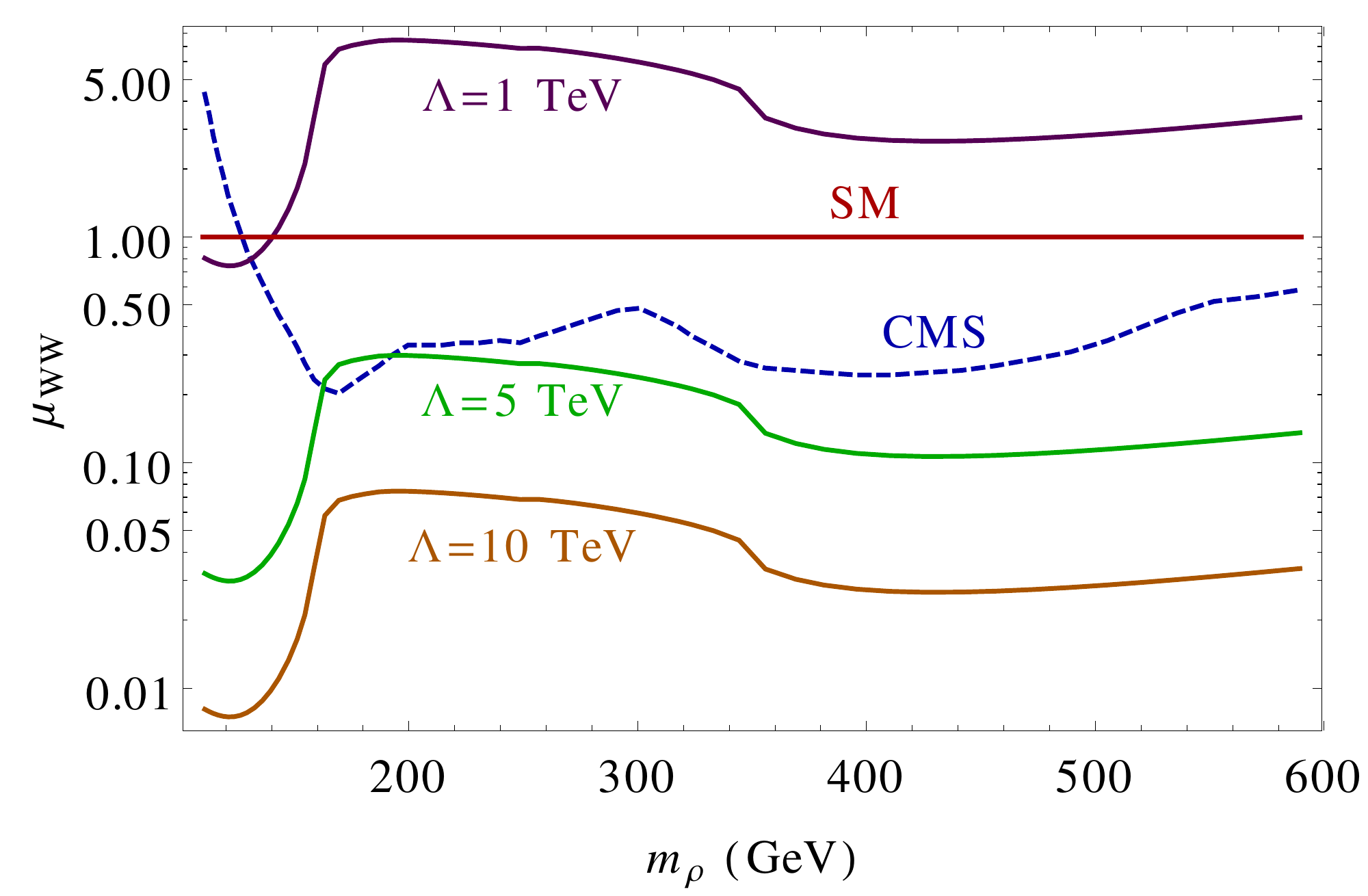}}}
\hspace*{-1cm}
\mbox{\subfigure[]{\includegraphics[width=0.5\linewidth]{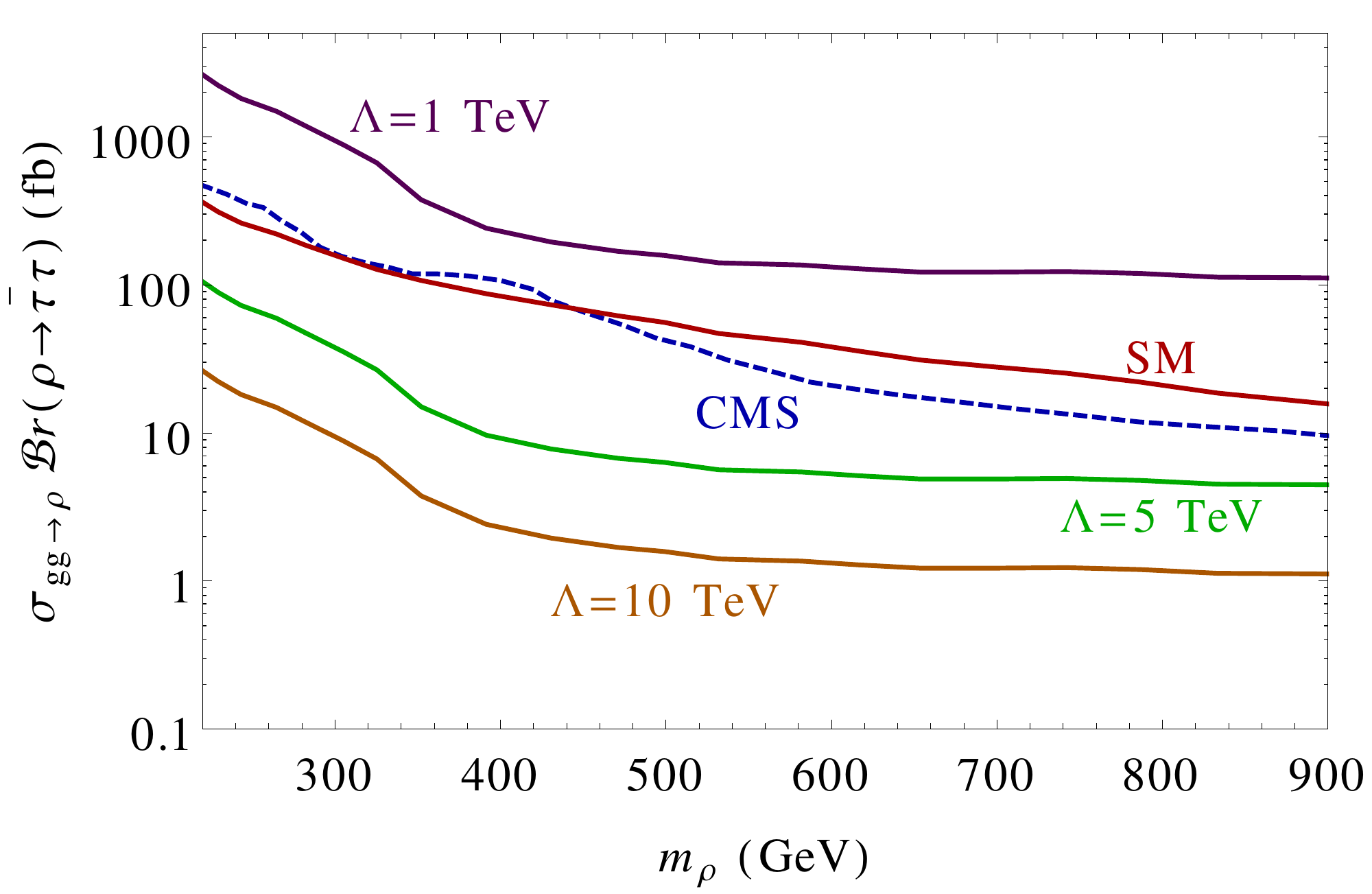}}\hskip 15pt
\subfigure[]{\includegraphics[width=0.5\linewidth]{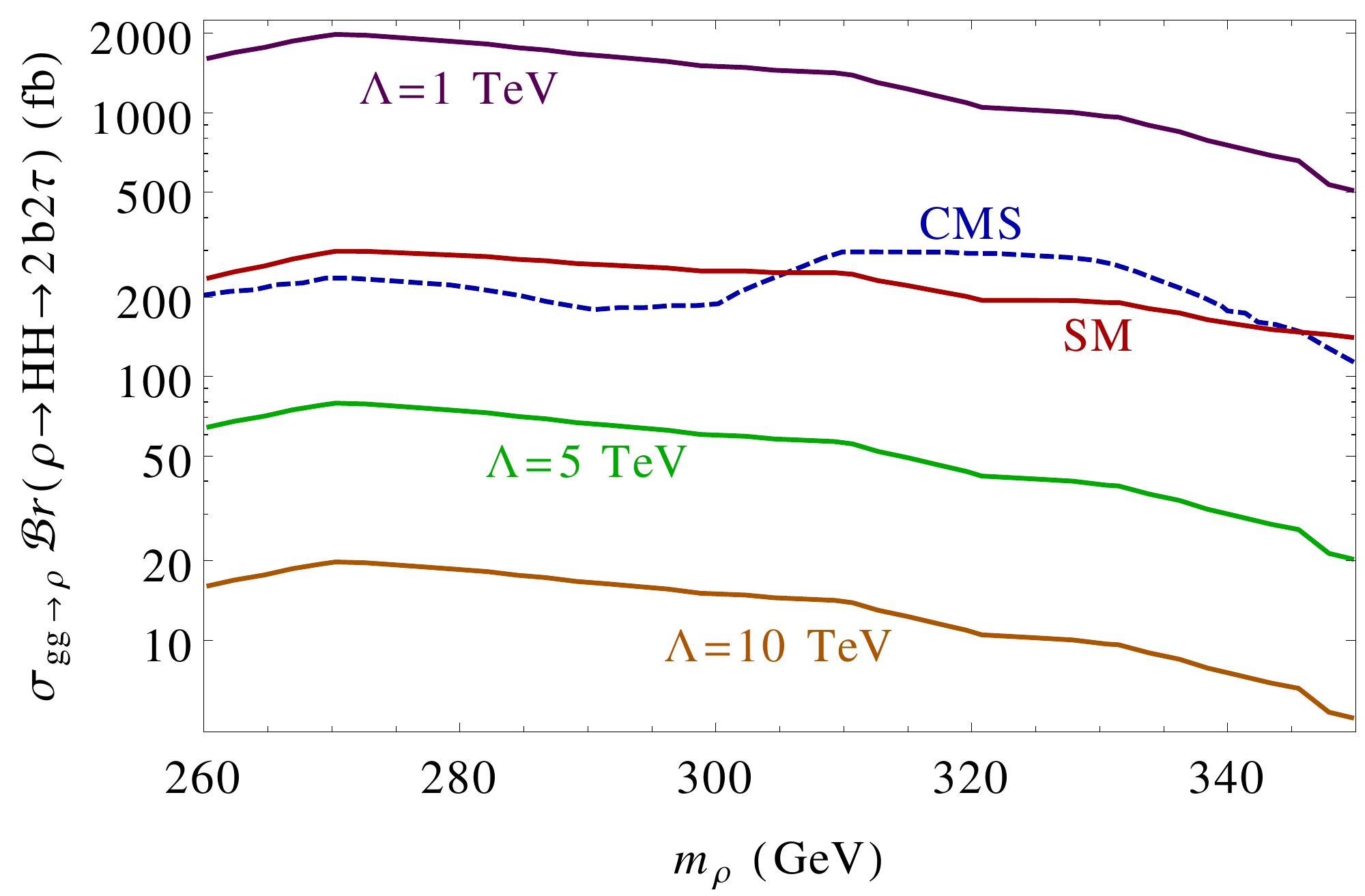}}}
\caption{The mass bounds on the dilaton from heavy scalar decays to (a) $ZZ$ \cite{CMSzz}, (b) $W^\pm W^\mp$ \cite{CMSww}, (c) $\bar\tau\tau$ \cite{CMStautau} and (d) to $H\,H$ \cite {CMShh} for three different choices of conformal scale, $\Lambda=1, 5, 10$ TeV respectively.}\label{bzzww}
\end{center}
\end{figure}
%%%%%%%%%%%%%%%%%%%%%%
Since the mass of the dilaton is a free parameter, and given the similarities with the main  production and decay channels of this particle with the Higgs boson, several features of the production and decay channels in the Higgs sector, with the due modifications, are shared also by the dilaton case.  

As we have already mentioned, the production cross-section depends sensitively on $\Lambda$, as shown in Eqs.~(\ref{ggrho}) and (\ref{Wrhoh}). Bounds on this breaking scale has been imposed by the experimental searches for a heavy, SM-like Higgs boson at the LHC, heavier than the $125$ GeV Higgs, $H_{125}$.  \\
We have investigated the bounds on $\Lambda$ coming from the following datasets
\begin{itemize}
\item{}
the 4.9 $\rm{fb}^{-1}$ (at 7 TeV) and 19.7 $\rm{fb}^{-1}$ (at 8 TeV) datasets  for a heavy Higgs decaying into $Z\,Z$ \cite{CMSzz}, $W^\pm W^\mp$ \cite{CMSww}, $\bar\tau\tau$ \cite{CMStautau} and 
\item{}
the 19.7 fb$^{-1}$ datasets (at 8 TeV) for the decay in $H\,H$ \cite{CMShh} from CMS  
\item{}
the 20.3 fb$^{-1}$ at 8 TeV data from ATLAS for the decay of the heavy Higgs into $Z\,Z$ \cite{ATLASzz}
and $W^\pm W^\mp$ \cite{ATLASww}. 
\end{itemize}

\begin{figure}%[thb]
\begin{center}
\hspace*{-2cm}
\mbox{\subfigure[]{
\includegraphics[width=0.55\linewidth]{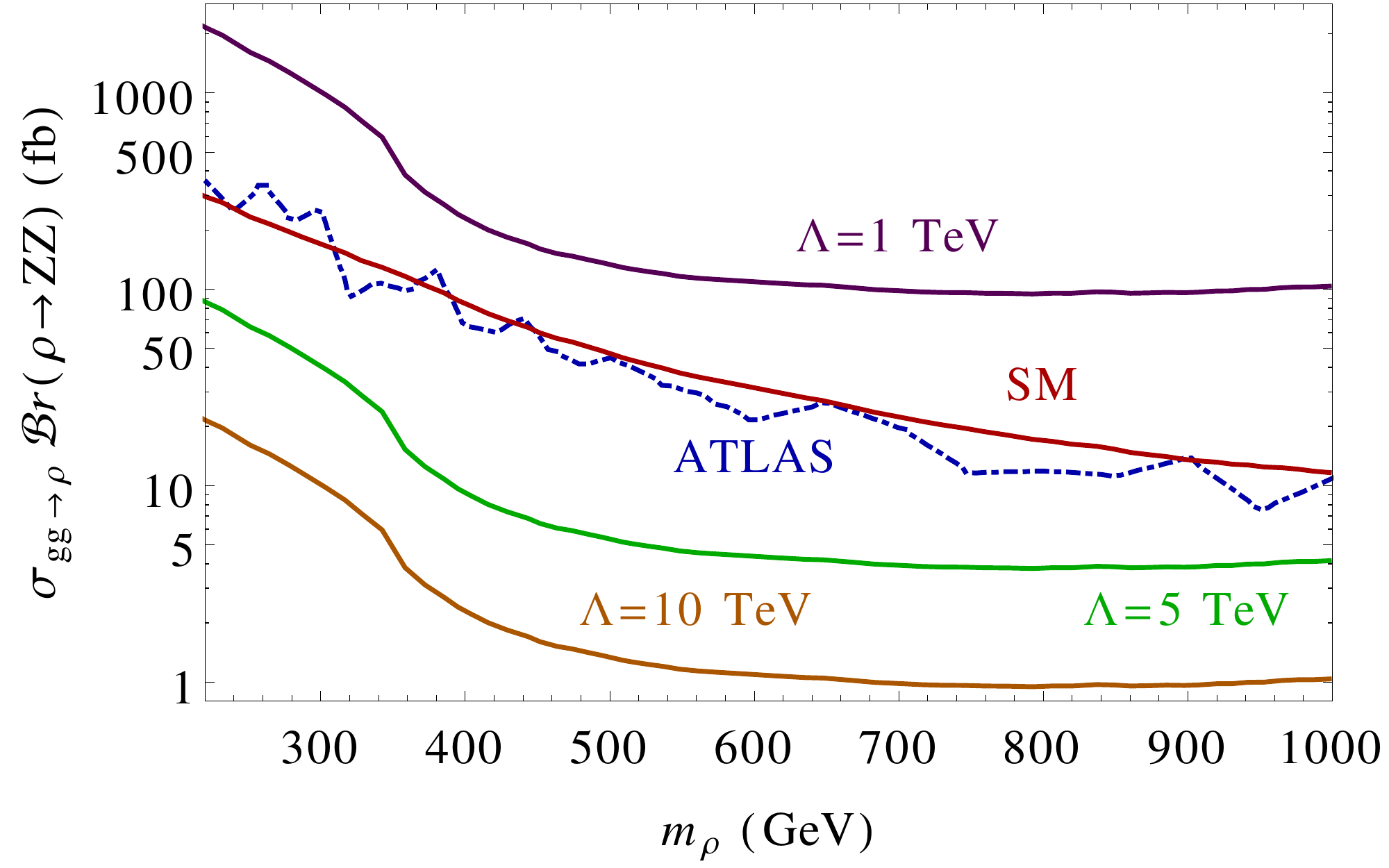}}\hskip 15pt
\subfigure[]{\includegraphics[width=0.55\linewidth]{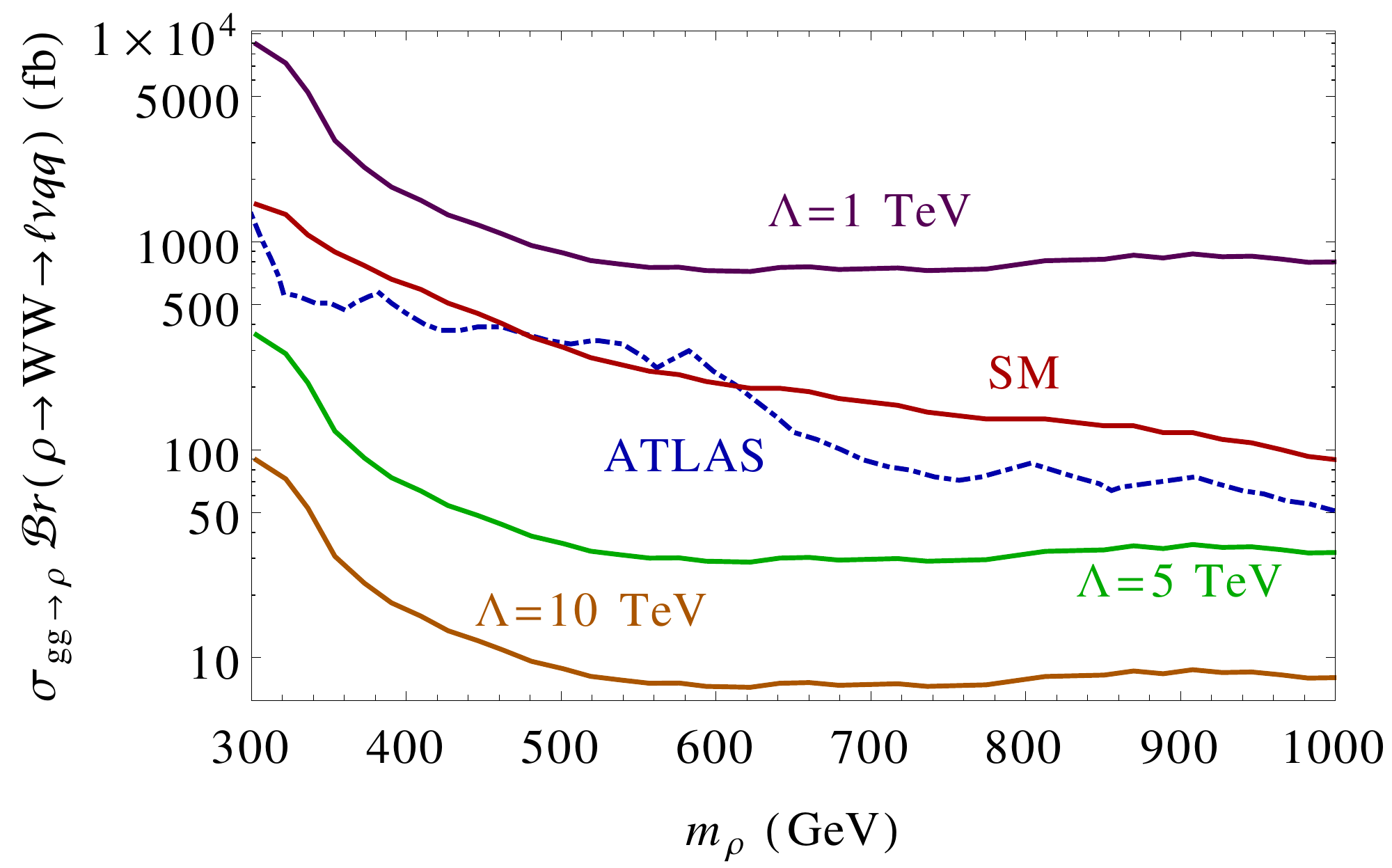}}}
\caption{The mass bounds on the dilaton from heavy scalar decays to (a) $ZZ$ \cite{ATLASzz} and (b) $W^\pm W^\mp$ \cite{ATLASww} for three different choices of conformal scale, $\Lambda=1, 5$ and $10$ TeV respectively.}\label{bzzwwA}
\end{center}
\end{figure}

The dotted line in each plot presents the upper bound on the cross-section, i.e. the $\mu$ parameter in each given modes defined as
\bea\label{mu}
\mu_{XY}=\frac{\sigma_{gg\to H}{{Br}(H\to XY)}}{{{\sigma_{gg \to H}}_{SM}}{{Br}(H \to XY)_{SM}}}.
\eea
In Fig. \ref{bzzww} we show the dependence of the 4-lepton ($2 l\, 2\nu$) channel on the mass of the $\rho$ at its peak, assuming $Z\,Z$, $W^\pm W^\mp$, $\bar\tau\tau$ and $H\,H$ intermediate states. 
The three continuous lines in violet, green and brown correspond to 3 diffferent values of the conformal scale, equal to 1, 5 and 10 TeV respectively. The SM predictions are shown in red. The dashed blue line 
separates the excluded and the admissible regions, above and below the blue curve respectively, which sets an upper bound of exclusion obtained from a CMS analysis.
A similar study is shown in  Fig. \ref{bzzwwA}, limited to the $Z\,Z$ and $W^\pm W^\mp$ channels, where we report the corresponding bound presented, in this case, by the ATLAS collaboration. Both the ATLAS and CMS data completely exclude the $\Lambda=1$ TeV case whereas the $\Lambda = 5$ TeV case has only a small tension with the CMS analysis of the $W^\pm W^\mp$ channel if $m_\rho\sim 160$ GeV.  Any value of $\Lambda \geq 5$ TeV is not  ruled out by the current data.\\
In Table \ref{cross} we report the values of the gluon fusion cross-section for three benchmark points 
(BP) that we have used in our phenomenological analysis. We have chosen $\Lambda = 5$ TeV, and the factorization in the evolution of the parton densities has been performed in concordance with those of the Higgs working group \cite{HCWG}. In the following subsection we briefly 
discuss some specific features of the dilaton phenomenology at the LHC, which will be confronted with a PYTHIA based simulation of the SM background.

%%%%%%%%%%%%%%%%%%%%%%%%%
%%%%%%%%%%%%%%%%%%%%%%

%%%%%%%%%% gg-> rho %%%%%%%%%%%%%%%%    
 \begin{table}[t]
\begin{center}
\hspace*{-1.0cm}
\renewcommand{\arraystretch}{1.2}
\begin{tabular}{|c||c|c|c||}
\hline\hline
Benchmark& $m_\rho$& $gg\to \rho$\\
Points &GeV& in fb \\
\hline
BP1 &200&6906.62\\
\hline
BP2 &260&3847.45\\
\hline
BP3 &400&1229.25\\
\hline
\hline
%&&&&&&&&&\\
\end{tabular}
\caption{Dilation production cross-section via gluon fusion at the LHC at 14 TeV, for the 3 selected benchmark points, with $\Lambda=5$ TeV.}\label{cross}
\end{center}
\end{table}
%%%%%%%%%%%%%%%%%%%%%%%%%%%%%%%%%%%%%%%%%%%%%%%%%%%%%%%    

\subsection{Dilaton phenomenology at the LHC}
\begin{figure}[thb]
\begin{center}
%\hspace*{-2cm}
\mbox{\subfigure[]{
\includegraphics[width=0.4\linewidth]{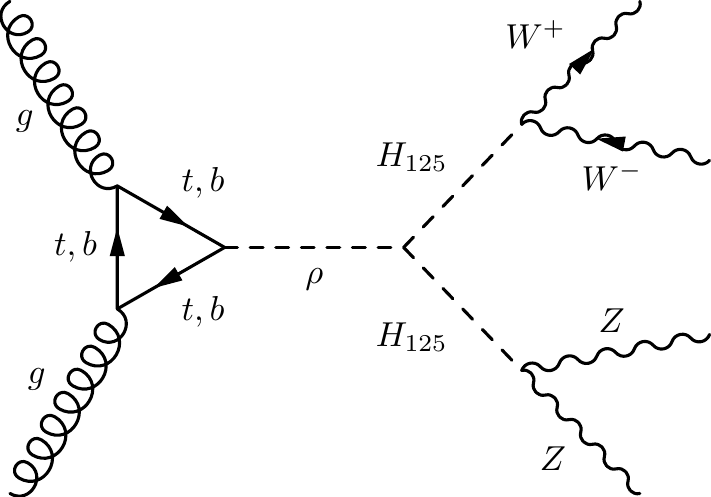}}
\hspace*{1cm}
\subfigure[]{\includegraphics[width=0.35\linewidth]{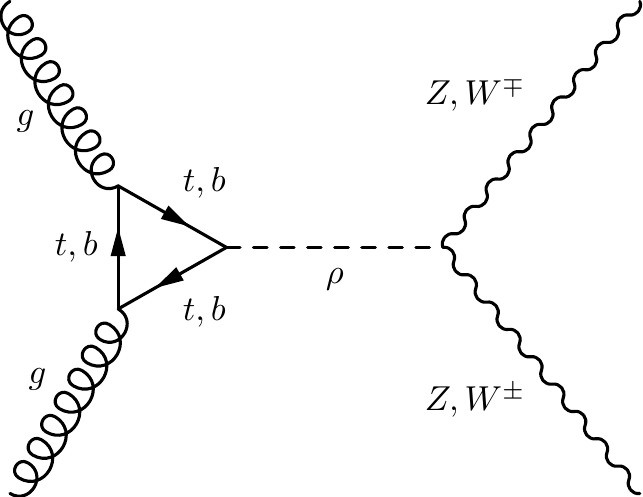}}}
\caption{The Feynman diagrams showing the dilaton production via gluon-gluon fusion and its decay to (a) pair of Higgs boson which further decays into gauge boson pairs and (b) a pair of gauge bosons.}\label{ggd}
\end{center}
\end{figure}

Fig.~\ref{ggd} shows the production and decay amplitudes mediated by an intermediate dilaton at the LHC.
We can see from Fig.~\ref{brrhoh}(a) that some of the main interesting decays of the dilaton are into two on-shell SM Higgs bosons $H\,H$, or into a real/virtual pair $H\,H^*$ and gauge boson pairs. The corresponding SM Higgs boson then further decays into $WW^*$ and/or $ZZ^*$.  Certainly these gauge bosons and their leptonic decays will give rise to multi-leptonic final states with missing transverse energy ($\etmiss$) via the chain 
\bea \label{dcy}
pp &\to& \rho \to  H\,H^* \nn\\
  &\to & WW^*, WW^* \nn \\
  & \to & 4\ell +\etmiss, \, 3\ell + 2j +\etmiss.
\eea
%
%%%%%%%%%%%%%%%%%%%%%%%%%
%%%%%%%%%%%%%%%%%%%%%%
As shown above, there are distinct intermediate states mediating the decay of the dilaton into four $W^\pm$ bosons on/off-shell which give rise to $3\ell +\etmiss$ and $4\ell +2j +\etmiss$ final states. When we demand that one of the SM Higgs bosons $h$ decays to $ZZ^*$ and the other to $WW^*$, we gain a factor of two in multiplicity and generate a final state of the form $ 6\ell +\etmiss$, $ 4\ell + \geq 2j +\etmiss$ and $3\ell + 4j +\etmiss$ (i.e. 4 leptons, plus at least 2 jets accompanied by missing $E_T$) as in 
  \bea \label{dcy1}
pp &\to& \rho \to  H H^* \nn\\
  &\to & WW^*, ZZ^* \nn \\
  & \to & 6\ell +\etmiss, \, 4\ell + \geq 2j +\etmiss,\, 3\ell + 4j +\etmiss. 
\eea
Though the SM Higgs boson decay branching ratios to
$ZZ^*$ are relatively small $\sim 3\%$, when the dilaton decays via an intermediate $ZZ^*$, final states 
with several leptons are expected as in
\bea \label{dcy2}
pp &\to& \rho \to  H H^* \nn\\
  &\to & ZZ^*, ZZ^* \nn \\
  & \to & 8\ell,\, 6\ell + 2j,\, 4\ell + 4j. 
\eea
From the last decay channel, final states with multiple charged leptons and zero missing energy are now allowed, a case which we will explore next. \\
 The SM gauge boson branching ratios to charged leptons are very small, specially for channels mediated by a $Z$, due to the small rates. Therefore leptonic final states of higher multiplicities will be suppressed compared to those of a low number.  For this reason we will restrict the choice of the leptonic final states in our  simulation to $\geq 3\ell +X$ and $\geq 4\ell +X$. The requirement of 
$\geq 3\ell$ and $\geq 4\ell$  already allow to reduce most of the SM backgrounds,  although not completely, due to some some irreducible components, as we are going to discuss next.

 \section{Collider simulation}\label{colsim}
 We analyse dilaton production by gluon-gluon fusion, followed by its decay either to a pair of SM-like Higgs bosons ($\rho \to H_{125} H_{125}$) or to 
 a pair of gauge bosons ($WW$, $ZZ$). The $H_{125}$ thus produced will further decay into gauge boson pairs, i.e. $W^\pm W^\mp$ and $ZZ$, giving rise to mostly leptonic final states, as discussed above.  When the intermediate decays into one or more gauge bosons in the hadronic modes are considered, then we get leptons associated with extra jets in the final states. For $m_{\rho} < 2m_{H_{125}}$ the dilaton decays to two on-shell $H_{125}$ states are not kinematically allowed. In that case we consider its direct decay into gauge boson pairs, $W^\pm W^\mp, ZZ$. In the following subsections we consider the two case separately, where we analyze final states at the LHC at 14 TeV and simulate the contributions coming from the SM backgrounds. \\ 
For this goal we have implemented the model in SARAH \cite{sarah}, generated the model files for CalcHEP \cite{calchep}, later used to produce the decay file SLHA containing the decay rates and the corresponding mass spectra. The generated events have then been simulated with {\tt PYTHIA} \cite{pythia} via the the SLHA interface \cite{slha}. The simulation at hadronic level has been performed using the {\tt Fastjet-3.0.3} \cite{fastjet} with the {\tt CAMBRIDGE AACHEN} algorithm with a jet size $R=0.5$ for the jet formation, chosen according to the following criteria:
\begin{itemize}
  \item the calorimeter coverage is $\rm |\eta| < 4.5$

  \item minimum transverse momenta of the jets $ p_{T,min}^{jet} = 20$ GeV and the jets are ordered in $p_{T}$
  \item leptons ($\rm \ell=e,~\mu$) are selected with
        $p_T \ge 20$ GeV and $\rm |\eta| \le 2.5$
  \item no jet should be accompanied by a hard lepton in the event
   \item $\Delta R_{lj}\geq 0.4$ and $\Delta R_{ll}\geq 0.2$
  \item Since an efficient identification of the leptons is crucial for our study, we additionally require  
a hadronic activity within a cone of $\Delta R = 0.3$ between two isolated leptons. This is defined by the condition on the transverse momentum $\leq 0.15\, p^{\ell}_T$ GeV in the specified cone.

\end{itemize}

\subsection{Benchmark points}
\begin{figure}[bht]
\begin{center}
\hspace*{-2cm}
\mbox{\subfigure[]{
\includegraphics[width=0.55\linewidth]{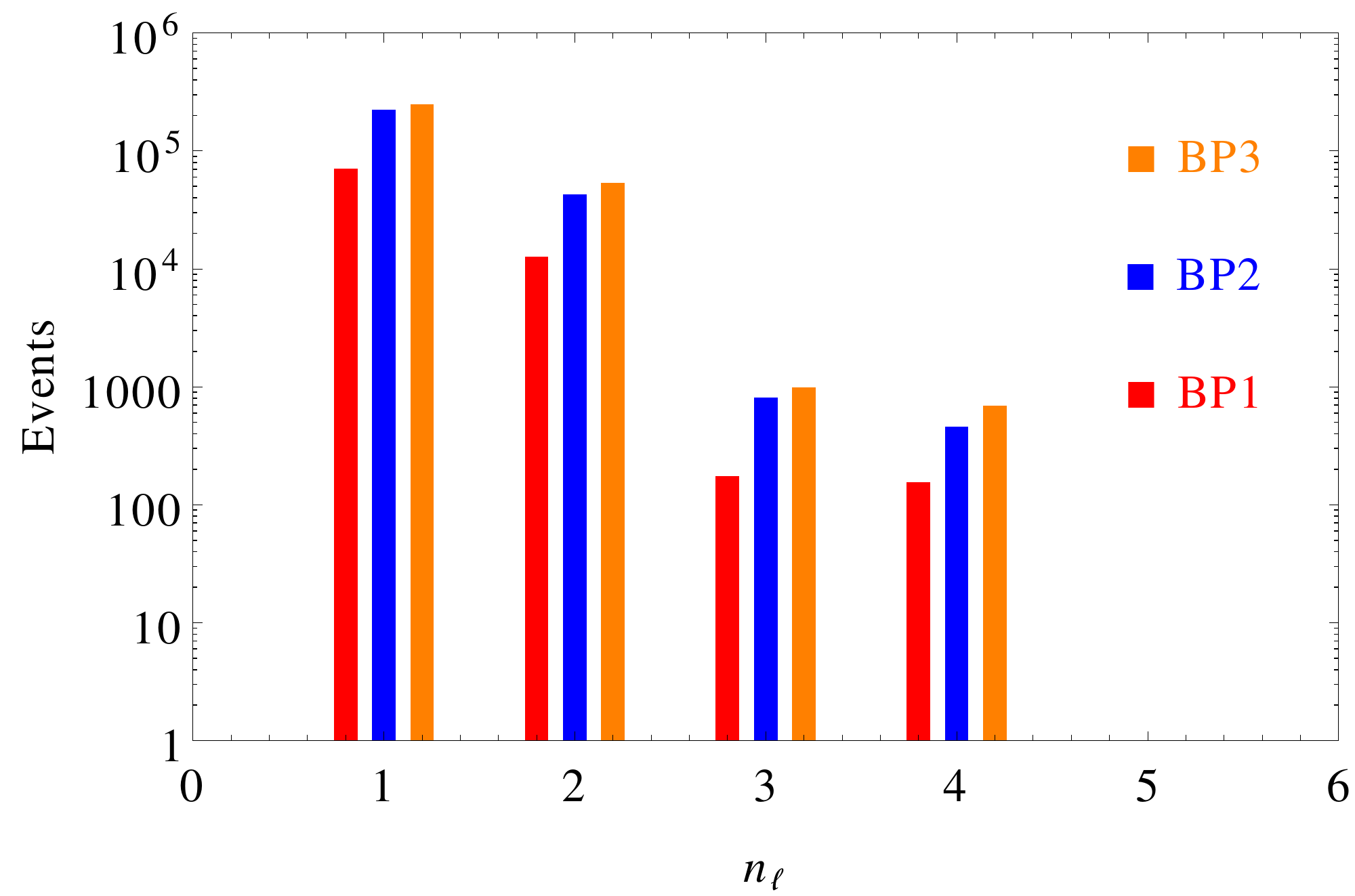}}
\hspace*{.5cm}
\subfigure[]{\includegraphics[width=0.55\linewidth]{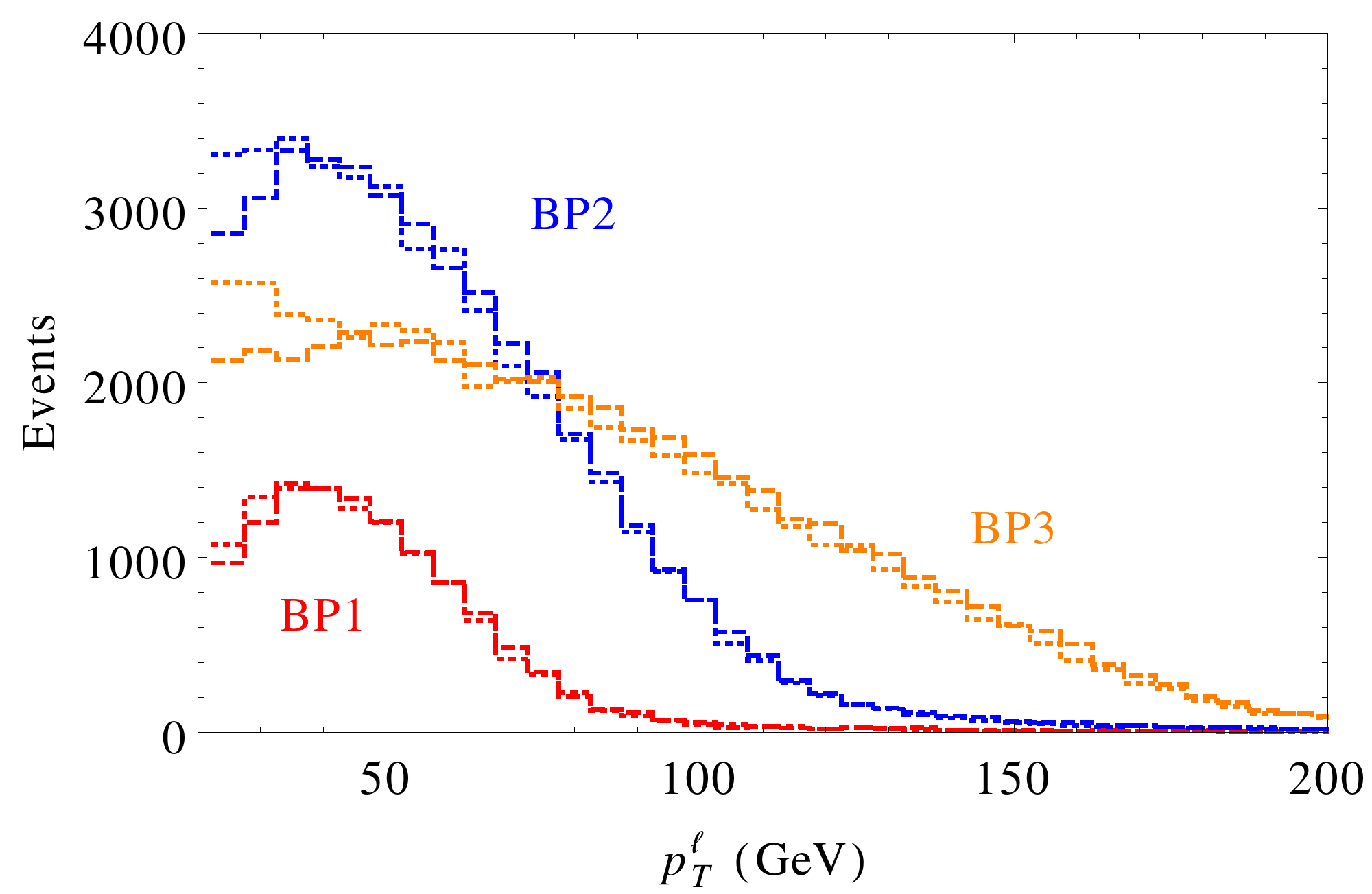}}}
\caption{The (a) lepton multiplicity and (b) lepton $p_T$ distribution for the benchmark points.}\label{lpf}
\end{center}
\end{figure}

\begin{figure}[thb]
\begin{center}
\hspace*{-2cm}

\includegraphics[width=0.6\linewidth]{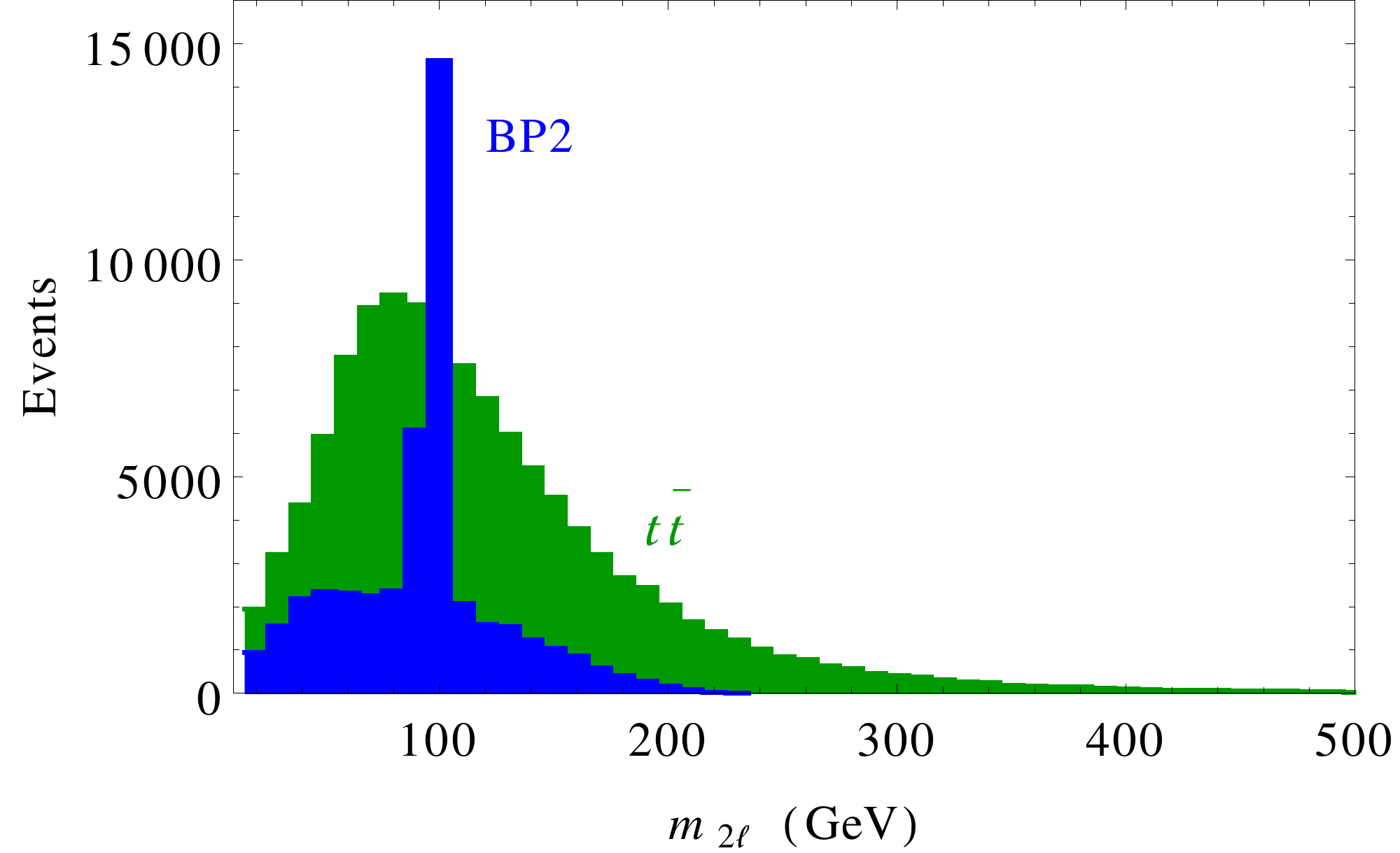}
\caption{The di-lepton invariant mass distribution for the signal BP2 and the background $t\bar{t}$.}\label{mll12}
\end{center}
\end{figure}

We have carried out a detailed analysis of the signal and of the background in a possible search for a light dilaton. For this purpose we have selected three benchmark points as given in Table~\ref{diltnbr}. The decay branching ratios given 
in Table~\ref{diltnbr} are independent of the conformal scale. For the benchmark point 1 (BP1), the dilaton is assumed to be of light mass of $200$ GeV, and its decay to the $H_{125}$ pair is not 
kinematically allowed. For this reason, as already mentioned, we look for slightly different final states in the analysis of such points. It appears evident that the dilaton may decay into gauge boson pairs when they are kinematically allowed. Such decays still remain dominant even after that the $t\bar{t}$ mode is open. This prompts us to study dilaton decays into $ZZ$, $WW$ via $3\ell$ and $4\ell$ final states. In the alternative case in which the dilaton also decays into a SM Higgs pair ($H_{125}$) along with gauge boson pairs, we have additional jets or leptons in the final states. This is due to the fact that 
the $H_{125}$ Higgs decays to the $WW$ and $ZZ$ pairs with one of the two gauge bosons off-shell (see Table~\ref{hbr}). We select two of such points when this occurs, denoted as BP2 and BP3, which are shown in Table~\ref{diltnbr}. Below we are going to present a separate analysis for each of the two cases. \\
%%%%%%%%%% BPs %%%%%%%%%%%%%%%%    
 \begin{table}[t]
\begin{center}
\hspace*{-1.0cm}
\renewcommand{\arraystretch}{1.2}
\begin{tabular}{|c||c|c|c||}
\hline\hline
Decay&BP1 & BP2&BP3 \\
Modes &$m_\rho$ = 200 GeV&$m_\rho$ = 260 GeV&$m_\rho$ = 400 GeV\\
\hline
HH&-&0.245&0.290\\
\hline
$W^\pm W^\mp$&0.639&0.478&0.408\\
\hline
ZZ&0.227&0.205&0.191\\
\hline
$\tau\tau$&$2.54\times10^{-4}$&$7.8\times10^{-5}$&$2.05\times10^{-5}$\\
\hline
$\gamma \gamma$&$9.28\times10^{-5}$&$2.88\times10^{-5}$&$4.33\times10^{-6}$\\
\hline
$gg$&$0.131$&$0.0691$&$0.0390$\\
\hline
\hline
%&&&&&&&&&\\
\end{tabular}
\caption{The benchmark points for a light dilaton with their mass-dependent decay branching ratios.}\label{diltnbr}
\end{center}
\end{table}
%%%%%%%%%%%%%%%%%%%%%%%%%%%%%%%%%%%%%%%%%%%%%%%%%%%%%%%    
%%%%%%%%%% Higgs branching %%%%%%%%%%%%%%%%    
 \begin{table}%[t]
\begin{center}
\renewcommand{\arraystretch}{1.2}
\begin{tabular}{||c||c|c|c|c|c|c||}
\hline\hline
Decay Modes &$W^\pm W^\mp$&$Z\,Z$&$\bar b b$&$\bar \tau \tau$&$gg$&$\gamma\,\gamma$\\
\hline
$H_{125}$&0.208&0.0259&0.597&0.0630&0.0776&$2.30\times10^{-3}$\\
\hline
\hline
%&&&&&&&&&\\
\end{tabular}
\caption{The corresponding branching ratios of the SM Higgs boson with a mass of 125 GeV.}\label{hbr}
\end{center}
\end{table}
%%%%%%%%%%%%%%
The leptons in the final state are produced from the decays of the gauge bosons, which can come, in turn, either from the decay of the dilaton or from that of the $H_{125}$.  In such cases, for a dilaton sufficiently heavy, the four lepton signature ($4\ell$) of the final state is quite natural and their momentum configuration will be boosted. In Fig.~\ref{lpf}(a) we show the multiplicity distribution of the leptons and  in Fig.~\ref{lpf}(b) their $p_T$ distribution for the chosen benchmark points. Here the lepton multiplicity has been subjected to some basic cuts on their transverse momenta ($p_T\geq 20$) GeV and isolation criteria given earlier in this section. Thus soft and non-isolated leptons are automatically cut out from the distribution. From Fig.~\ref{lpf}(b) it is clear that the leptons in BP3 can have a very hard transverse momentum ($p_T \sim 200$ GeV), as the corresponding dilaton is of $400$ GeV. Notice that the di-lepton invariant mass distribution in Fig.~\ref{mll12} presents a mass peak around $m_Z$ for the signal (BP2) but not for the dominant SM top/antitop ($t\bar{t}$) background. This will be used later as a potential selection cut in order to reduce some of the SM backgrounds.

%%%%%%%%%%Lepton multiplicity and pt distribution%%%%%%%%%%%%%%%
%%%%%%%%%%%%%%%%%%%%%%

%%%%%%%%%%di-Lepton invariant mass distribution %%%%%%%%%%%%%%%
%%%%%%%%%%%%%%%%%%%%%%

\subsection{Light dilaton: $m_{\rho} < 2m_{H_{125}}$}
 In this subsection we analyse final states with at least three  ($\geq 3\ell +X +\etmiss$) and 4 ($\geq 4\ell +X +\etmiss$) leptons (inclusive) and missing transverse energy that can result from the decays of the dilaton into $ZZ$, where we consider the potential  SM backgrounds. The reason for considering the $3\ell$ final states is because one of the four leptons ($4\ell$) could be missed. This is in general possible due to the presence of additional kinematical cuts introduced when hadronic final states are accompanied by leptons. We present a list of the number of events 
 for the $3\ell$ and $4\ell$ final states in Table~\ref{BP1n} for BP1, and the dominant SM backgrounds at integrated luminosity of 100 fb$^{-1}$ at the LHC. The potential SM backgrounds come from the $t\bar{t}Z$ and  $tZW$ sectors, from intermediate gauge boson pairs ($VV$) and from the triple gauge boson vertices $VVV$ ($V: W^\pm, Z$). Due to the large $t\bar{t}$ cross-section, with the third and fourth lepton - which can originate from the corresponding $b$ decays -  this background appears to be an irreducible one. For this reason we are going to apply successive cuts for its further reduction, as described in Table~\ref{BP1n}.
   %%%%%%%%%% 3l final states %%%%%%%%%%%%%%%%    
 \begin{table}[t]
\begin{center}
\hspace*{-1.0cm}
\renewcommand{\arraystretch}{1.3}
\begin{tabular}{|c||c||c|c|c|c|c||}
\hline\hline
Final states&\multicolumn{1}{|c||}{Benchmark}&\multicolumn{5}{|c||}{Backgrounds }
\\
\hline
&BP1  & $t\bar{t}$& $t\bar{t}Z$ &$tZW$&$VV$& $VVV$\\
\hline
\hline
$\geq 3\ell \,+\, \ptmiss \leq 30\, \rm{GeV}$&494.97&275.52&65.17&22.29&6879.42&765.11\\
$\,+\,|m_{ll}-m_Z|<5\,\rm{GeV}$&384.47&68.88&62.68&20.93&2514.92&16.16\\
$\,+\,n_{\rm{b_{jet}}}=0$&377.56&9.84&17.64&10.08&2479.66&15.13\\
\hline
Significance&7.00&\multicolumn{5}{|c||}{}\\
\hline
$\mathcal L_5$&51 fb$^{-1}$&\multicolumn{5}{|c||}{}\\
\hline
\hline
$\geq 4\ell \,+\, \ptmiss \leq 30\, \rm{GeV}$&273.96&0.00&3.32&1.36&1655.99&34.18\\
$\,+\,|m_{ll}-m_Z|<5\,\rm{GeV}$&218.71&0.00&3.11&1.16&627.38&4.44\\
\hline
Significance&7.48&\multicolumn{5}{|c||}{}\\
\hline
$\mathcal L_5$&45 fb$^{-1}$&\multicolumn{5}{|c||}{}\\
\hline
\hline
\end{tabular}
\caption{Numbers of events for the $3\ell+\ptmiss$  and $4\ell$ final states for the BP1 and 
the dominant SM backgrounds, at an integrated luminosity of $100$ fb $^{-1}$.}\label{BP1n}
\end{center}
\end{table}
%%%%%%%%%%%%%%%%%%%%%%%%%%%%%%%%%%%%%%%%%%%%%%%%%%%%%%%%

The primary signal that is considered is characterised  by the kinematical cut $3\ell +\ptmiss \leq 30$ GeV. The choice of a very low missing $p_T$ is justified because when both $Z$'s decay to charged lepton pairs they give rise to $\geq 3\ell$ and $\geq 4\ell$ final states which are neutrinoless. The theoretical prediction of no missing energy, however, cannot be fully satisfied as the missing transverse momentum $\ptmiss$ is calculated by estimating the total visible $p_T$ of the jets and of the leptons after the threshold cuts. Next we demand the di-lepton be characterised by an invariant mass around $Z$ mass i.e., $\,|m_{ll}-m_Z|<5\,\rm{GeV}$, which reduces the $t\bar{t}$, 
$VV$ and $VVV$ backgrounds quite significantly. A further requirement of no $b$-jet ( i.e., $n_b=0$) reduces the $t\bar{t}, $$t\bar{t}Z$ and $tZW$ backgrounds. By looking at the signal, we observe that these cuts do not affect the signal number for BP1. 
After imposing all the cuts, we find that an integrated luminosity of $\mathcal{O}(51)$ fb$^{-1}$ is required for 
a $5\sigma$ reach in this final state.  The demand of $4\ell$ of course reduces the background but also reduces the signal event numbers. In this case  $\mathcal{O}(45)$ fb$^{-1}$ of integrated luminosity 
is required for a $5\sigma$ discovery.

 \subsection{Heavy dilaton: $m_{\rho} > 2m_{H_{125}}$}
  In this case we consider points where $m_{\rho} > 2m_{H_{125}}$, allowing decays of the dilaton to $H_{125}$ pairs. For this purpose we have chosen two benchmark points, one with $m_{\rho}=260$ GeV - where the channel $\rho \to\, H_{125}H_{125}$ is just open - and another one with  $m_{\rho}=400$ GeV, where even the $\rho \to t\bar{t}$ channel is open.  The decay mode via a $H_{125}$ pair, in turn decaying into gauge boson pairs, gives additional jets which accompany the $3\ell$ and $4\ell$ final states and help in a further reduction of the SM backgrounds.
  
Table~\ref{4l}  presents the number of expected events generated at the BP2  and BP3 benchmark points for the signal and for the dominant SM backgrounds. Here we have considered $\geq 3\ell$ GeV and $\geq 4\ell$ final states respectively, at an integrated luminosity of $1000$ fb$^{-1}$. The dominant backgrounds are as before, and listed in Table~\ref{4l}. Notice that if we demand the tagging of at least two additional jets and the $b$-jet veto, we can reduce the backgrounds even further. The result shows that in the case of BP2 and BP3 a dilaton signal could be discovered at an integrated luminosity of $\mathcal{O}(130)$ and $\mathcal{O}(570)$ fb$^{-1}$ respectively for the $\geq 3\ell$ final state. For the  $\geq 4\ell$  f a $5\sigma$ discovery reach can be achieved even with 114 fb$^{-1}$ and 374 fb$^{-1}$ of integrated luminosity for BP2 and BP3 respectively.

%%%%%%%%%% 4l final states %%%%%%%%%%%%%%%%    
 \begin{table}[t]
\begin{center}
\hspace*{-1.0cm}
\renewcommand{\arraystretch}{1.3}
\begin{tabular}{|c||c|c||c|c|c|c|c||}
\hline\hline
Final states&\multicolumn{2}{|c||}{Benchmark}&\multicolumn{5}{|c||}{Backgrounds }
\\
\hline
& BP2&BP3 & $t\bar{t}$& $t\bar{t}Z$  &$tZW$&$VV$& $VVV$\\
\hline
\hline
$\geq 3\ell$&3882.08&1642.28&10725.9&4790.19&1364.73&177140&53660.2\\
$\,+\, n_{\rm{b_{jet}}}=0$&3812.82&1627.53&5510.54&1550.38&664.92&176167&53604.8\\
$\,+\,n_{\rm{jet}}\geq2$&2677.82&1255.06&2952.08&1469.43&579.62&29165.5&324.28\\
\hline
Significance&13.89&6.64&\multicolumn{5}{|c||}{}\\
\hline
$\mathcal L_5$&130 fb$^{-1}$&568 fb$^{-1}$&\multicolumn{5}{|c||}{}\\
\hline
\hline
$\geq 4\ell$&1400.47&678.55&0.00&502.26&149.27&17338.1&2379.06\\
$\,+\,n_{\rm{jet}}\geq2\,+\, n_{\rm{b_{jet}}}=0$&865.68&448.68&0.00&147.36&48.46&2334.44&36.13\\
\hline
Significance&14.78&8.17&\multicolumn{5}{|c||}{}\\
\hline
$\mathcal L_5$&114 fb$^{-1}$&374 fb$^{-1}$&\multicolumn{5}{|c||}{}\\
\hline
\hline
%&&&&&&&&&\\
\end{tabular}
\caption{We present the final state numbers for $4\ell+\ptmiss$ final states for the benchmark points and 
the dominant SM backgrounds at an integrated luminosity of $1000$ fb $^{-1}$.}\label{4l}
\end{center}
\end{table}
%%%%%%%%%%%%%%%%%%%%%%%%%%%%%%%%%%%%%%%%%%%%%%%%%%%%%%%%%%%%

%%%%%%%%%%%%%
%%%%%%%%%%%%%%%%%%%%%%%%%
\begin{figure}[thb]
\begin{center}
\hspace*{-2cm}
\mbox{\subfigure[]{
\includegraphics[width=0.54\linewidth]{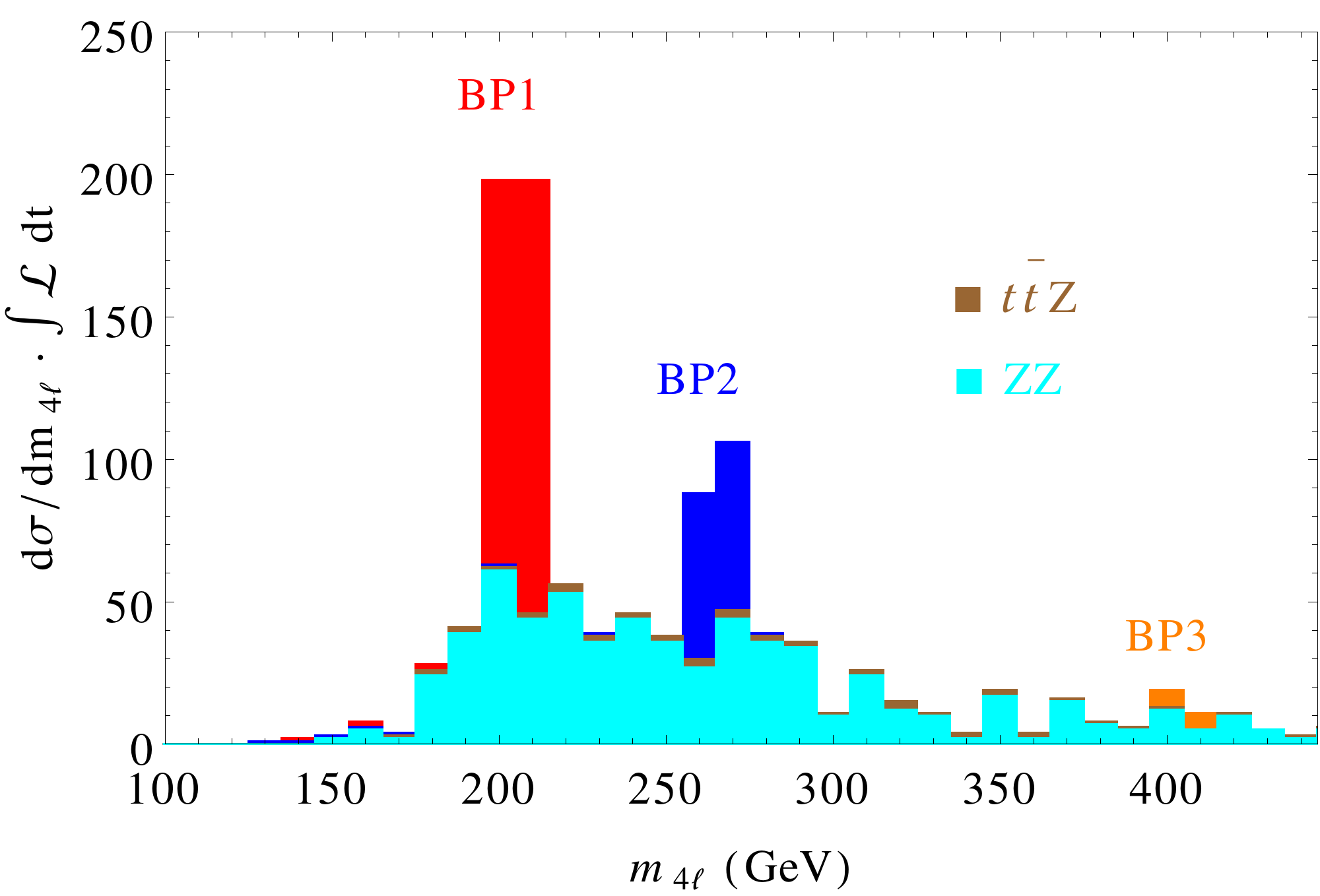}}
\hspace*{.5cm}
\subfigure[]{\includegraphics[width=0.57\linewidth]{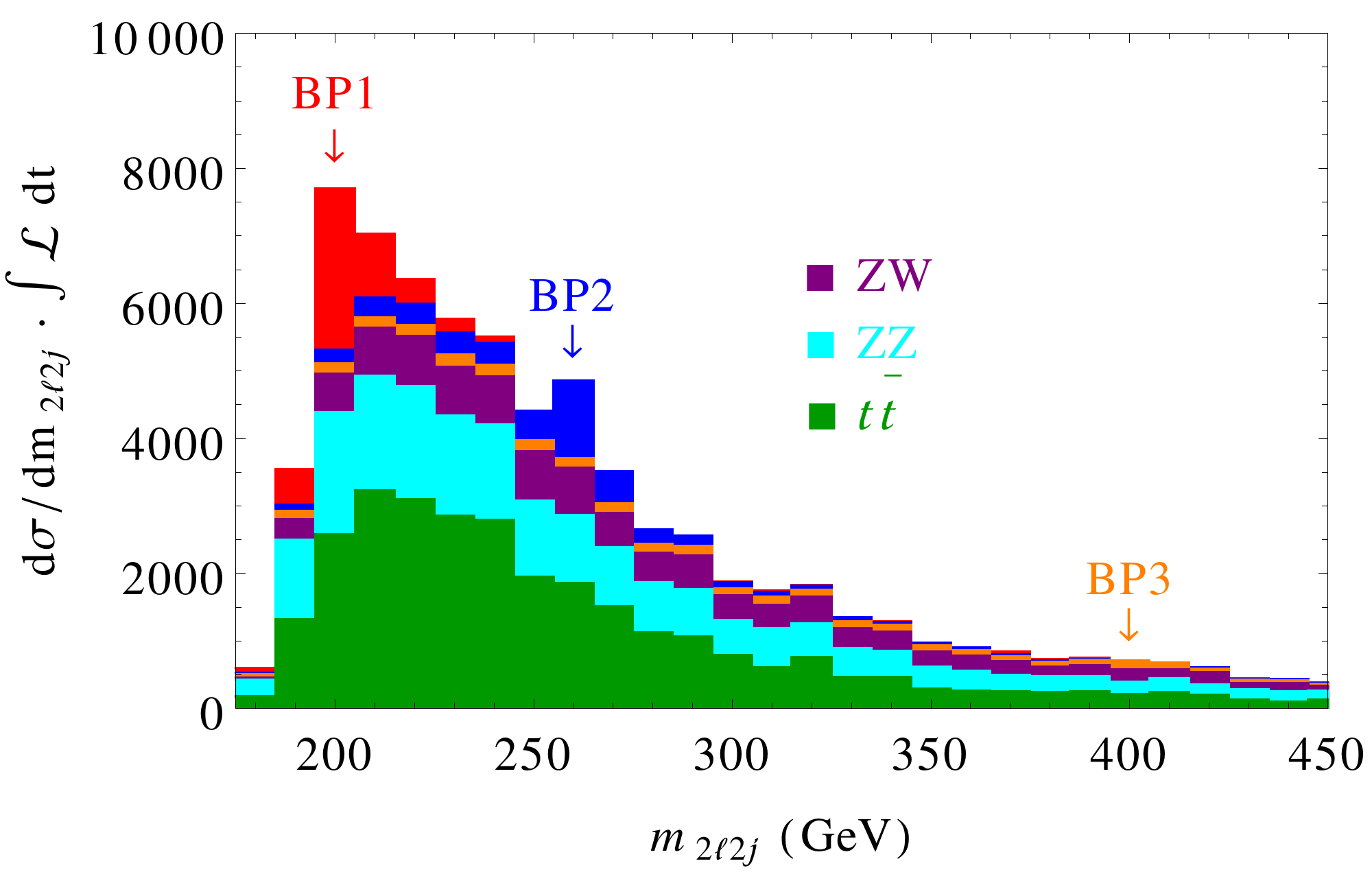}}}
\caption{The invariant mass distribution for the benchmark points and the dominant SM backgrounds 
for $4\ell$ and $2\ell2j$ final state respectively at an integrated luminosity of 100 fb$^{-1}$.}\label{invmass}
\end{center}
\end{figure}
%%%%%%%%%%%%%%%%%%%%%%
Next we try to reconstruct the dilaton mass peak from the $\geq 4\ell$ and $2\ell\, 2j$ channels. 
In the first case we consider the isolated $4\ell$'s after enforcing the basic cuts, and then demand that the di-leptons are coming from the $Z$ boson mass peak. This guarantees that we are reconstructing either the $\rho\to ZZ$ or the $\rho \to H_{125}H_{125}\to ZZ+X$ incoming channel. Fig.~\ref{invmass}(a) shows the plot of the invariant 
mass distributions $m_{4\ell}$ for all three benchmark points, along with the dominant backgrounds. The presence of a clear mass peak certainly allows the reconstruction of the dilaton mass. We have selected the number of events around the mass peaks, i.e., $|m_{4\ell} -m_{\rho}|\leq 10$ GeV for the benchmark points, which are shown in Table~\ref{4lpeak} at an integrated luminosity of 100 fb$^{-1}$. It is clear that for the BP1 and BP2 benchmark points the mass peak can be resolved with very early data at the LHC, with a 14 TeV run.\\
Fig.~\ref{invmass}(b) shows the invariant mass distribution, where we consider a pair of charged leptons around the $Z$ mass peak, i.e., $|m_{\ell\ell}-m_Z|<5\,\rm{GeV}$ as well as a pair of jets, i.e., $|m_{jj}-m_Z|< 10\,\rm{GeV}$. Such di-jet pairs and di-lepton pairs are then taken in all possible combinatorics to evaluate the $m_{\ell\ell jj}$ mass distribution, as shown in Fig.~\ref{invmass}(b). Clearly the $Y$ axis of the figure shows such possible pairings and the $X$ axis indicates the mass scale. We see the right combinations peak, which sits around the benchmark points. We have also taken the dominant backgrounds with their combinatorics to reproduce the invariant mass $m_{\ell\ell jj}$. In Table~\ref{2l2jlpeak}
we list the results around the mass peak, i.e. for $|m_{2\ell2j} -m_{\rho}|\leq 10$ GeV. It is easily observed that such constraint can be a very handy guide to identify the resonance mass peak using very early data at the LHC with 14 TeV.

%%%%%%%%%% 4l final states %%%%%%%%%%%%%%%%    
 \begin{table}[t]
\begin{center}
\hspace*{-1.0cm}
\renewcommand{\arraystretch}{1.5}
\begin{tabular}{c||c|c|c||}
\hhline{~===}
&\multicolumn{3}{c||}{Number of events in}\\
&\multicolumn{3}{c||}{$|m_{4\ell} -m_{\rho}|\leq 10$ GeV}\\
\hhline{~===}
& BP1&BP2&BP3\\
\hline
\hline
\multicolumn{1}{||c||}{Signal}&396&194&30\\
\hline
\multicolumn{1}{||c||}{Background}&108&77&18\\
\hline
\multicolumn{1}{||c||}{Significance}&17.64&11.78&4.33\\
\hline
\hline
\end{tabular}
\caption{We present the events number for $\geq 4\ell$ final state around the dilaton mass peak, i.e. $|m_{4\ell} -m_{\rho}|\leq 10$ GeV, for the benchmark points and the backgrounds at an integrated luminosity of 100 fb$^{-1}$.}\label{4lpeak}
\end{center}
\end{table}
%%%%%%%%%%%%%

%%%%%%%%%% 4l final states %%%%%%%%%%%%%%%%    
%%%%%%%%%% 4l final states %%%%%%%%%%%%%%%%    
 \begin{table}[t]
\begin{center}
\hspace*{-1.0cm}
\renewcommand{\arraystretch}{1.5}
\begin{tabular}{c||c|c|c||}
\hhline{~===}
&\multicolumn{3}{c||}{Number of events in}\\
&\multicolumn{3}{c||}{$|m_{\ell\ell jj} -m_{\rho}|\leq 10$ GeV}\\
\hhline{~===}
& BP1&BP2&BP3\\
\hline
\hline
\multicolumn{1}{||c||}{Signal}&14727&8371&1390\\
\hline
\multicolumn{1}{||c||}{Background}&10887&6706&1234\\
\hline
\multicolumn{1}{||c||}{Significance}&92.02&68.17&27.13\\
\hline
\hline
\end{tabular}
\caption{We present the events number for $\geq 2\ell$ final state around the dilaton mass peak, i.e. $|m_{2\ell2j} -m_{\rho}|\leq 10$ GeV, for BP1, BP2, BP3 and the backgrounds  at an integrated luminosity of 100 fb$^{-1}$.}\label{2l2jlpeak}
\end{center}
\end{table}
%%%%%%%%%%%%%

\section{Perspectives on compositeness and $\xi$ dependence }
\label{non0xi}
\begin{figure}[thb]
\begin{center}
\hspace*{-1.5cm}
\mbox{\subfigure[]{
\includegraphics[width=0.55\linewidth]{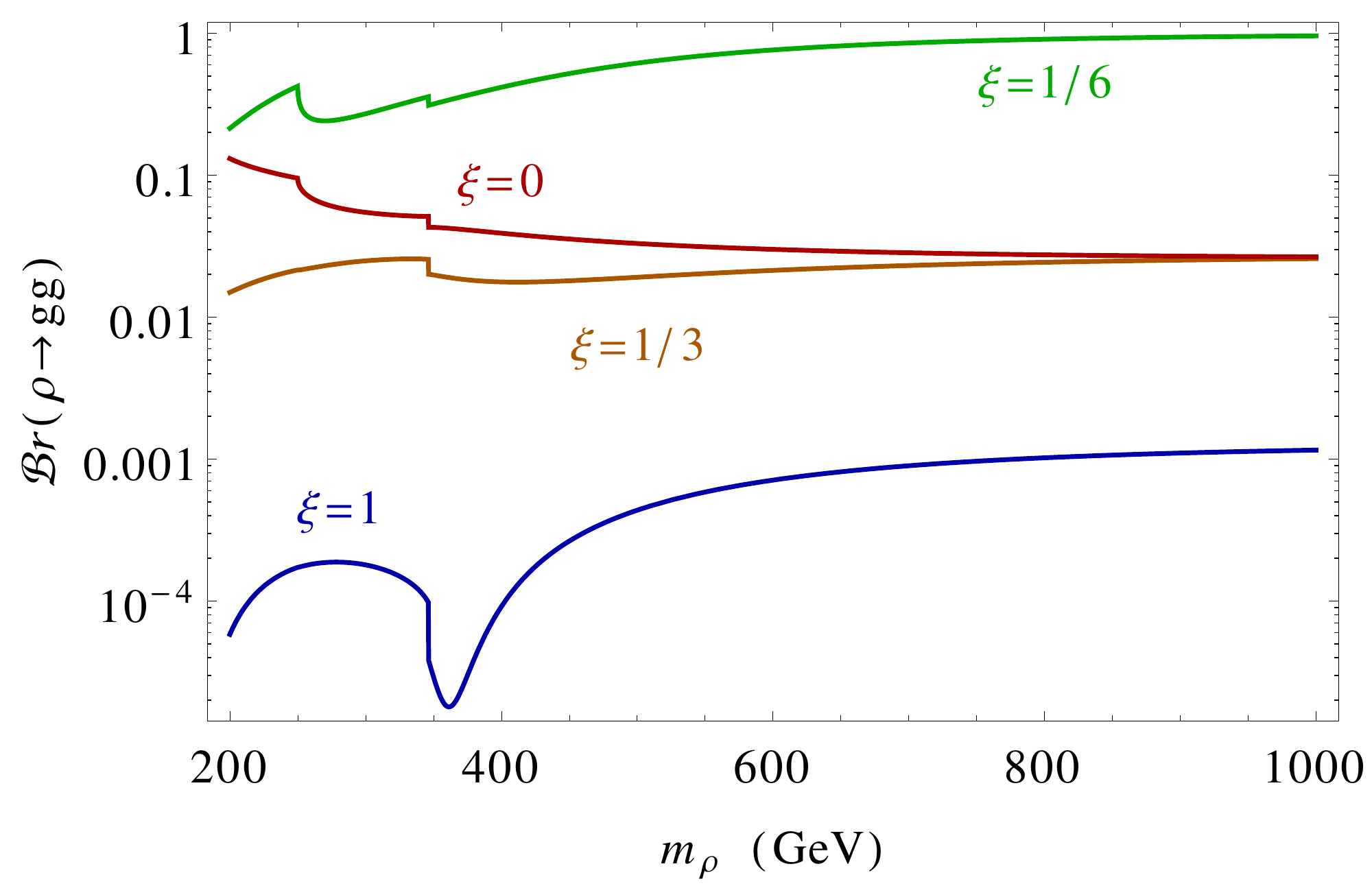}}
\hspace*{0.3cm}
\subfigure[]{\includegraphics[width=0.55\linewidth]{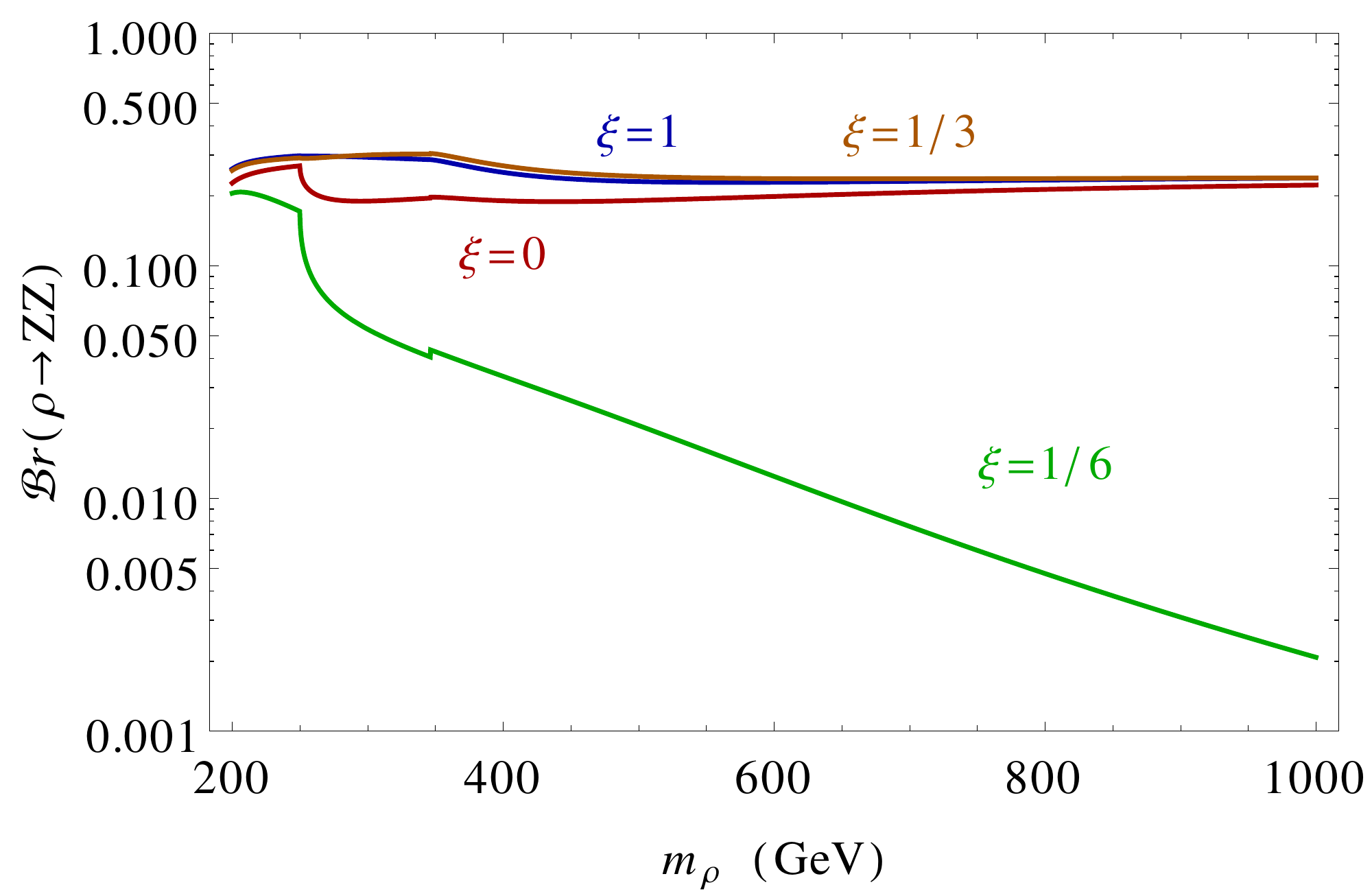}}}
\hspace*{-1.5cm}
\mbox{\subfigure[]{
\includegraphics[width=0.55\linewidth]{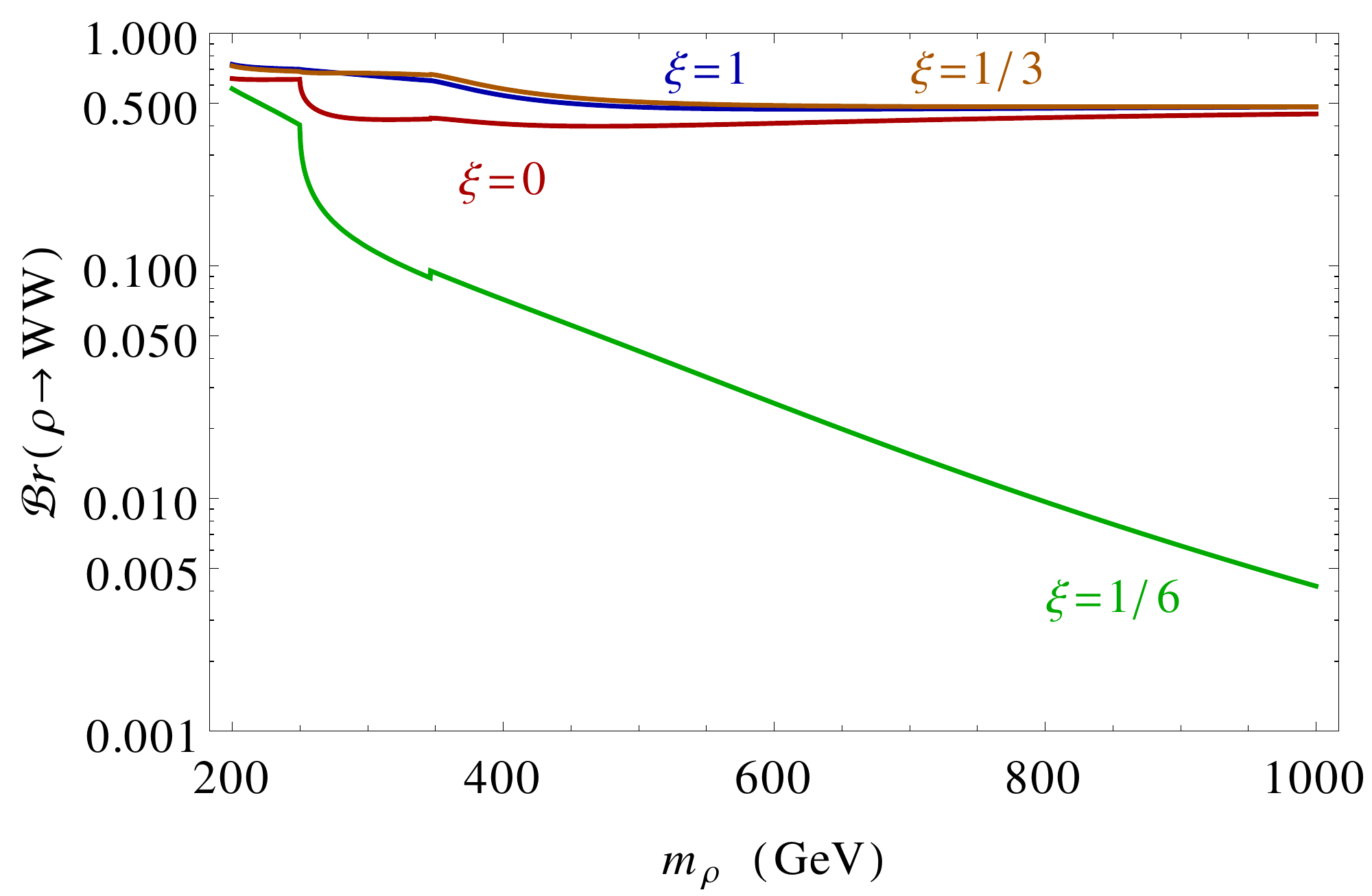}} \hspace{0.3cm}   
\subfigure[]{\includegraphics[width=0.55\linewidth]{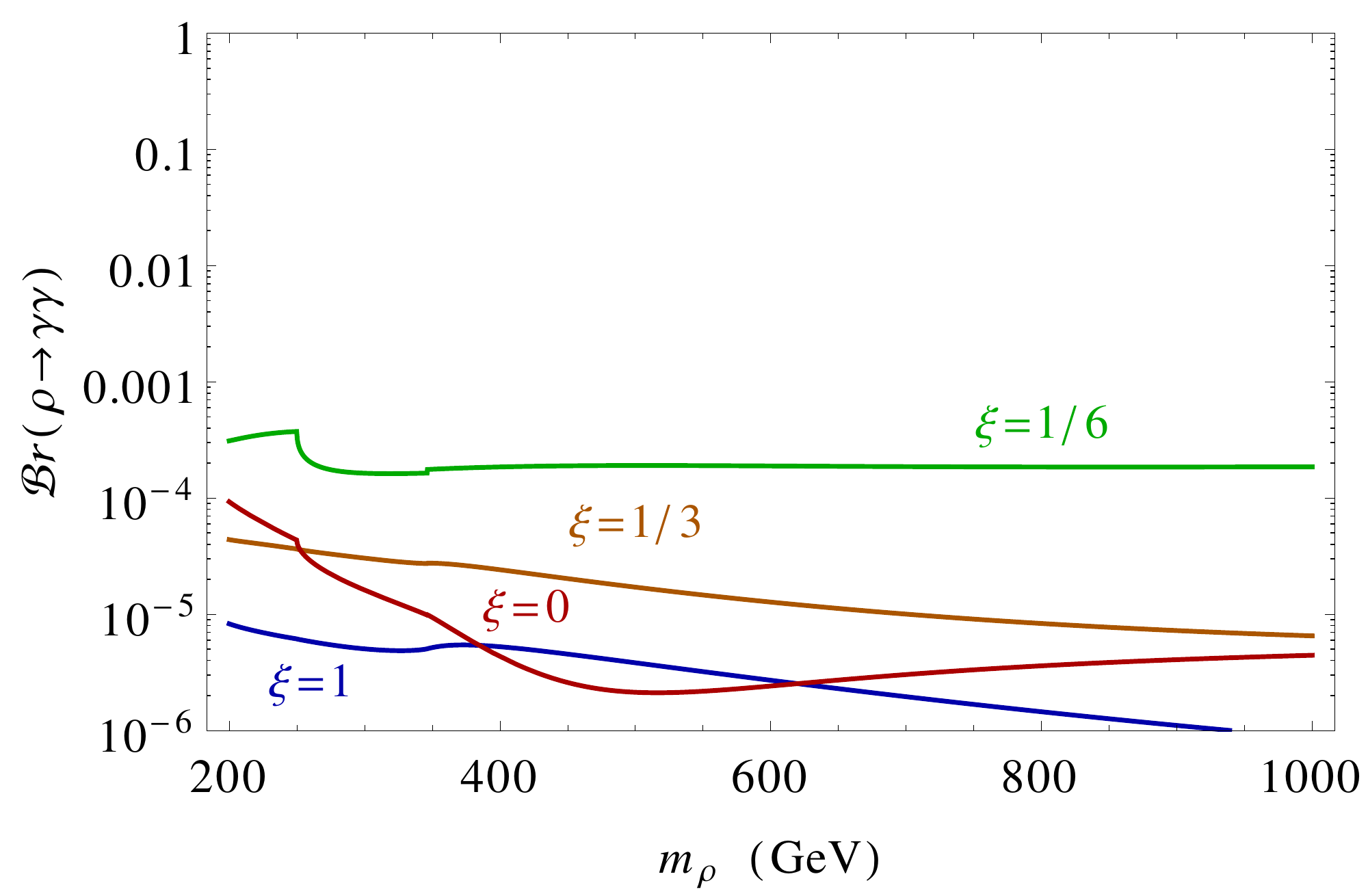}}}
\caption{The decay branching ratios of the dilaton (a) to gluons, (b)-(c) massive gauge bosons 
and (d) photons pairs for different $\xi$ parameters.}\label{diffxi}
\end{center}
\end{figure}

\begin{figure}[thb]
\begin{center}
\hspace*{-1.3cm}
\mbox{\subfigure[]{
\includegraphics[width=0.55\linewidth]{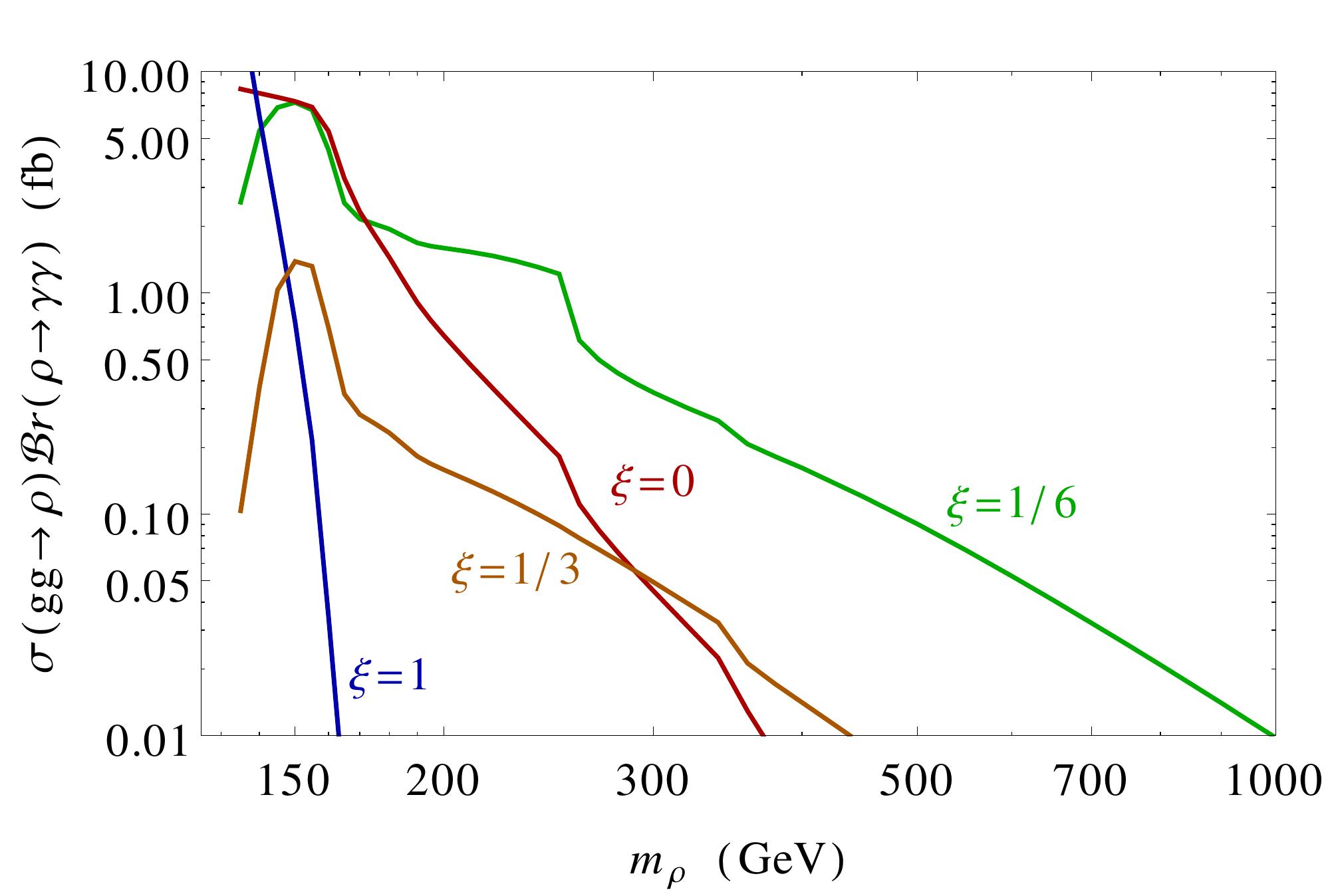}}
\hspace*{0.3cm}
\subfigure[]{
\includegraphics[width=0.55\linewidth]{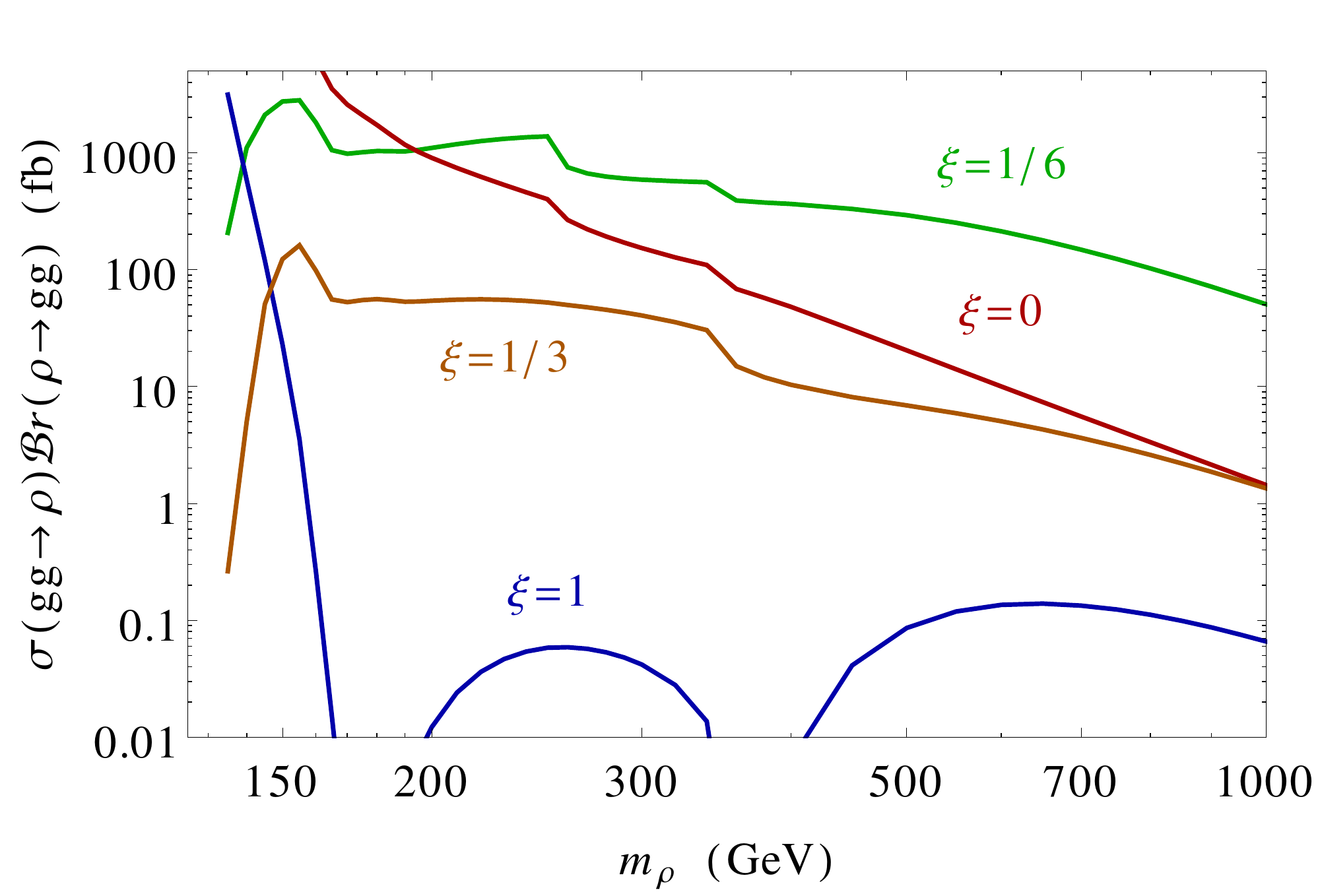}}}
\caption{Di-gluon and di-photon signal of a dilaton for a varying $\xi$.}\label{xsgldi}
\end{center}
\end{figure}

In our analysis the dilaton has been treated as a fundamental state, with interactions which are dictated from 
Eq. (\ref{tmunu}). The perturbative analysis that follows from this interaction does not take into account possible effects of compositeness, which would involve the wave function of this state both in its production and decay. In this respect, this treatment is quite similar to the study of the $\pi\to \gamma\gamma$ decay using only the divergence of the interpolating axial-vector current rather then the pion itself, with its hadronic wave function now replaced by the divergence of the dilatation current $J_D$. 
Those effects could modify the predictions that emerge from our analysis. \\
Another possible modification of our results will be certainly linked to a nonzero value of the $\xi$ parameter. The search for a valuable signal of a nonminimal dilaton at the LHC requires a completely independent calibration of the kinematical cuts that we have discussed. While we hope to address this point in a future work, we can however obtain a glimpse of the dependence of the signal (production/decays) as a function of $\xi$. \\
This behaviour is clearly illustrated in Fig.~\ref{diffxi} where the decay into massless and massive states of a conformal dilaton are  dependent on the  improvement coefficient  $\xi$. Fig.~\ref{diffxi}(a), (d)  show the decay branching fraction to gluon and photon pair respectively.  We see that for $\xi=1/6$ they are enhanced compared to other values of $\xi$. Similarly, the massive gauge bosons modes are suppressed for $\xi=1/6$ as can be seen from  Fig.~\ref{diffxi}(b),(c).  In Fig.~\ref{xsgldi}
we present the production cross-sections for di-gluons and di-photon final states. Notice that for 
$\xi=1/6$ these two modes have much larger rates than for other $\xi$ cases.  Unlike the minimal case of $\xi=0$, the $\xi=1/6$ can be studied via di-jet or di-photon final states. 

It is expected that a dilaton which arises from the breaking of a conformal symmetry should be described by a conformal coupling $\xi=1/6$, at least in the high energy limit. The signature of such a state, if composite, is in the anomaly pole of correlators involving the dilatation current and two vector currents, as pointed out in \cite{CDS}. The dilatation current inherits the same pole from the $TVV$ correlator \cite{Giannotti:2008cv,ACD2,ACD3} while
the explicit/non perturbative breaking of the conformal symmetry would then be responsible for the generation of its mass. \\
In a more general framework, the possibility of having similar states in superconformal theories has been extensively discussed in \cite{CCDS} from a perturbative side. It has been shown, for instance, that classical superconformal theories are characterised by a complete alignment in their conformal anomaly multiplets. An axion/dilaton/dilatino composite multiplet would then be the natural manifestation of this alignment found in the superconformal anomaly action.
%%%%%%%%%%%%%%%%%%%%%%%%%
%%%%%%%%%%%%%%%%%%%%%%

%%%%%%%%%%%%%%%%%%%%%%%%%%%%%%%%%

%%%%%%%%%%%%%%%%%%%%%%%%%%%%%%%%%%%%

\chapter{The Higgs sectors with an $SU(2)$ Triplet in a Superconformal Theory}
\section{Synopsis}
In this and in the following chapter we are going to investigate a superconformal model with an extended Higgs sector, from a strict phenomenological perspective. The model is characterised by a superpotential which includes a triplet superfield of $SU(2)$ and a singlet. We will be investigating this theory in a phase in which supersymmetry is broken, and, from this perspective, the effects of the superconformal symmetry are masked by the presence of soft-breaking mass terms. We anticipate that as a result of this analysis, 
three massless states of the physical spectrum, $H_4, A_1$ and $\chi_{10}$ should be identified as the Goldstone modes of a superconformal symmetry. The first is a scalar state, the second a pseudoscalar, and the third a neutralino. The main features of the scalar sector and the corresponding bounds on the parameter space of this model emerging from the recent experimental data are discussed.
\section{Introduction}
With the recent discovery of the Higgs boson at the Large Hadron Collider, the mechanism responsible for the breaking of the electroweak symmetry has finally been uncovered and it has been shown to involve at least one scalar field along the lines of the Standard Model (SM) description. This discovery has removed, at least in part, previous doubts about the real existence of a scalar with Higgs-like properties in our Universe. 
Both the CMS \cite{CMS, CMS2} and the ATLAS \cite{ATLAS} experimental collaborations have confirmed the discovery of a Higgs boson, by an analysis of the $\gamma \gamma, ZZ^*,$  and $WW^*$ decay channels of the Higgs particle - as predicted by the Standard Model (SM) - at a confidence level of more than $5\sigma$, except for the $WW^*$ decay rate, which has been recorded with a $4.7\sigma$ accuracy by CMS \cite{CMS2}. The fermionic decay modes, instead, have still to reach the $5\sigma$ accuracy, and show some disagreement in the results elaborated by the two experimental collaborations. Clearly, the disagreement of the experimental results with the predictions from the SM opens the possibility of further investigation of the Higgs sector. \\
  For such reasons, it is widely believed that the SM is not a complete theory, being not able, for instance, to account for the 
 neutrino masses, but also for being affected, in the scalar sector, by the gauge hierarchy problem \cite{hierarchy}.
 The widespread interest in the study of a possible supersymmetric extension of this model has always being motivated with the goal of finding a natural and elegant solution to this problem. In fact, supersymmetry protects the Higgs mass from the undesired quadratic divergences introduced by the radiative corrections in the scalar sector of the SM, and it does so  by the inclusion of superpartners. \\
In the minimal supersymmetric extension of the SM (MSSM) the conditions of analyticity of the superpotential and of absence of the gauge anomalies require a minimal extensions of the scalar sector with two Higgs superfields, in the forms of $SU(2)$ doublets carrying opposite hypercharges $(Y)$. Supersymmetric extensions are, in general, characterized by a large set of additional parameters which render their phenomenological study quite involved. For this reason, in the near past, the interest has turned towards models, such as the constrained minimal supersymmetric extension (cMSSM/mSUGRA), with only 5 new parameters, generated at a large supergravity scale, quite close to the Planck scale \cite{cMSSM}. 

Unlike the SM case, in the MSSM the tree-level  mass of the lightest Higgs ($h_1$)  $m_{h_1}$ is not a free parameter but it is constrained to lay below the mass of the Z gauge boson, $m_Z$  ($m_{h_1}\leq m_Z$).   This constraint has been in tension with the results of the experimental searches at LEP-2  which have failed to detect any CP-even Higgs below $m_Z$ and which had established a lower bound of $114.5$ GeV for the SM Higgs boson \cite{LEPb}. With the recent discovery of a CP-even Higgs boson around 125 GeV \cite{CMS, CMS2, ATLAS} the resolution of this conflict is, therefore, mandatory.

To avoid the conflict between the MSSM prediction for the Higgs and 
the LHC results, one needs to consider the effect of the radiative corrections which could lift the bound on the Higgs mass in this model. It has been shown - and it is now well known - that in the case of the MSSM the significant radiative corrections come from the stop-top corrections, specially at low $\tan{\beta}$, due to large Yukawa couplings and to the presence of colour charges. This has triggered analysis envisioning scenarios with a heavy stop, which require a very high supersymmetric (SUSY) scale for the most constrained supersymmetric models like mSUGRA/cMSSM, AMSB, etc \cite{cMSSMb}. In the case of the phenomenological MSSM (pMSSM) there are two possibilities: a very large third generation SUSY mass scale and/or a large splitting between the stop mass eigenstates \cite{pMSSMb}. The second case leads to large soft trilinear  couplings $\gsim 2$ TeV \cite{pMSSMb}, which brings back the fine-tuning problem in a different way. A possible way to address the fine tuning problem is to consider an extended Higgs sector.  In this respect, there are some choices which could resolve it, based on the inclusion of one singlet \cite{nmssm} and of one or more triplet superfields of appropriate hypercharges \cite{tssmHiggsb}. In particular, the addition of a $Y=0$ hypercharge superfield gives large tree-level as well as one-loop corrections to the Higgs masses, and relaxes the fine tuning problem of the MSSM by requiring a lower SUSY mass scale \cite{tssmyzero, DiChiara:2008rg}. There are some special features of these extensions which are particularly interesting and carry specific signatures.  For instance, the addition of a ($Y=0,\pm 2$ hypercharge) - ($SU(2)$ triplet) Higgs sector induces $H^\pm-W^\mp-Z$ couplings mediated by the non-zero vacuum expectation value (vev) of the Higgs triplet, due to the breaking of the custodial symmetry \cite{tssmch1prime,tssmch1}. Other original features of the $Y=\pm 2$ hypercharge triplets are the presence of doubly charged Higgs in the spectrum \cite{tssmch2}. There are also other significant constraints which are typical of these scenarios, and which may help in the experimental analysis. In the supersymmetric Higgs triplet extension, the vev of the triplet $v_T$ is highly constraints by the $\rho$ parameter \cite{rho}, which leads to $v_T\lesssim 5$ GeV in the case of $Y=0$ triplets. In the same case, this value of 
$v_T$ can account for the value of the mixing parameter $\mu_D$ of the 2-Higgs doublets (or $\mu$-term), which remains small in the various possible scenarios. Another dynamical way to generate a $\mu_D$ term is by adding a SM gauge singlet superfield to the spectrum \cite{nmssm}, as in the NMSSM. Thus a triplet-singlet extended supersymmetric SM  built on the superpotential of the MSSM, can address both the fine tuning issue and resolve, at the same time, the problem of the $\mu$-term of the two Higgs doublets \cite{tnssm, FileviezPerez:2012gg, Agashe:2011ia}. We will see that the addition of a discrete symmetry in this model removes the mass terms from the superpotential and its continuum limit generates a Nambu-Goldstone pseudoscalar particle in the spectrum, characterising some of its most significant features.

In the MSSM we have two Higgs doublets giving masses to up and down type quarks respectively.  After EWSB we have two  CP-even light neutral Higgs bosons among which one can be the discovered Higgs around 125 GeV, a CP-odd neutral Higgs boson and a charged Higgs boson pair. Observation of a charged Higgs boson will be a obvious proof of the existence of another Higgs doublet which is necessary in the context of supersymmetry. Searches for the extended Higgs sector by looking for charged Higgs boson at the LHC are not new. In fact, both the CMS and ATLAS collaborations have investigated scenarios with charged Higgs bosons, even under the assumption of these being lighter than the top quark ($m_{H^\pm}\leq m_t$). In this case, the channel in question has been the $pp\to t\bar{t}$ production channel, with one of the top decaying into $b H^\pm$. In the opposite case of a charged Higgs heavier than the top ($m_{H^\pm}\geq m_t$), the most studied channels have been the $bg \to tH^\pm$ and $pp \to  tb H^\pm$, with the charged Higgs decaying into $\tau \nu_\tau$ \cite{ChCMS, ChATLAS}. We recall that both doublet type charged and neutral Higgs bosons couple to fermions with Yukawa interactions which are proportional to the mixing angle of the up and down type $SU(2)$ doublets. The extension of the MSSM with a  SM gauge singlet, i.e. the NMSSM \cite{Ellwanger}, has a scalar which does not couple to fermions or gauge bosons thus changes the search phenomenology. Similar extensions are possible with only $SU(2)$ triplet superfields with $Y=0 \pm 2 $ hypercharges  \cite{pbas1, pbas2, DiChiara, pbas3, EspinosaQuiros}. In the case of $Y=0$, the neutral part of the triplet scalar does not couple to $Z$ boson and does not contribute to $Z$ mass, whereas non-zero hypercharge triplets  contribute both in $W^\pm$ and $Z$ mass. The supersymmetric extensions of the Higgs sectors with $Z_3$ symmetry have the common feature of a light pseudoscalar in the spectrum, known as R-axion in the literature. Such feature is common to NMSSM with $Z_3$ symmetry \cite{Ellwanger} and also to extensions with singlet and triplet(s) with appropriate hypercharges \cite{TNMSSM1, TNMSSM2, tnssm, tnssma}. In this article we consider an extension of the MSSM with $SU(2)$ triplet superfield of $Y=0$ hypercharge and SM gauge singlet superfield, named as TNMSSM \cite{TNMSSM1, TNMSSM2}, with $Z_3$ symmetry. The main motivation to work with $Y=0$ triplet is that it is the simplest triplet extension in supersymmetric context, where the triplet only contribute in $W^\pm$ mass. For a model with non-zero hypercharges we need at least two triplets and also we get constrained from both $W^\pm$ and $Z$ masses \cite{tnssma}.  The light pseudoscalar in this model is mostly singlet and hence does not have any coupling to fermions or gauge bosons.  For this reason such light pseudoscalar is still allowed by the earlier LEP \cite{LEPb} data and current LHC data\cite{CMS, CMS2, ATLAS}. Similarly the triplet type Higgs bosons also do not couple to fermions  \cite{pbas1, pbas2, DiChiara, pbas3} which makes  a light triplet-like charged Higgs still allowed by the charged Higgs searches \cite{ChCMS, ChATLAS} and such Higgs bosons have to looked for in different production as well as decay modes. General features of this model have been presented in \cite{TNMSSM1}, while a more detailed investigation of the hidden pseudoscalar has been discussed by us in \cite{TNMSSM2}. Existence of the light pseudoscalar makes the phenomenology of the Higgs sector very rich for both the neutral and the charged sectors along with other signatures. In the TNMSSM we have three physically charged Higgs bosons $h^\pm_{1,2,3}$, two of which are triplet type in the gauge basis. The neutral part of the Higgs sector has four  CP-even ($h_{1,2,3,4}$) and three CP-odd sectors ($a_{1,2,3}$) states. In the gauge basis two of CP-even states are doublet-like one of which should be the discovered Higgs around $125$ GeV, one triplet type and one singlet type.  For the CP-odd states, there are one doublet type, one triplet type and one singlet type. Often it is the singlet-like pseudoscalar which becomes very light, which makes the phenomenology very interesting. The mass spectrum often splits into several regions with distinctively doublet/triplet blocks. The goal of our analysis will be to address the main features of this complete spectrum, characterising its main signatures in the complex environment of a hadron collider.

\section{The Model}\label{model}
We consider a scale invariant superpotential $W_{TNMSSM}$ with an extended Higgs
sector containing a $Y=0$ SU(2) triplet $\hat{T}$ and a SM gauge singlet ${\hat S}$ (see \cite{tnssm, FileviezPerez:2012gg}) on top of the superpotential of the MSSM. We recall that the inclusion of the singlet superfield on the superpotential of the MSSM realizes the NMSSM superpotential. We prefer to separate the complete superpotential of the model into a MSSM part, 
\begin{equation}
W_{MSSM}= y_t \hat U \hat H_u\!\cdot\! \hat Q - y_b \hat D \hat H_d\!\cdot\! \hat Q - y_\tau \hat E \hat H_d\!\cdot\! \hat L\ ,
\label{spm}
 \end{equation}
 where ''$\cdot$'' denotes a contraction with the Levi-Civita symbol $\epsilon^{ij}$, with $\epsilon^{12}=+1$, and combine the singlet superfield $(\hat{S})$ and the triplet contributions into a second superpotential 
\begin{equation}
W_{TS}=\lambda_T  \hat H_d \cdot \hat T  \hat H_u\, + \, \lambda_S S  \hat H_d \cdot  \hat H_u\,+ \frac{\kappa}{3}S^3\,+\,\lambda_{TS} S  \textrm{Tr}[T^2]
\label{spt}
 \end{equation}
 with 
 \begin{equation}
 W_{TNMSSM}=W_{MSSM} + W_{TS}.
 \end{equation}
  The triplet and doublet superfields are given by 

\begin{equation}\label{spf}
 \hat T = \begin{pmatrix}
       \sqrt{\frac{1}{2}}\hat T^0 & \hat T_2^+ \cr
      \hat T_1^- & -\sqrt{\frac{1}{2}}\hat T^0
       \end{pmatrix},\qquad \hat{H}_u= \begin{pmatrix}
      \hat H_u^+  \cr
       \hat H^0_u
       \end{pmatrix},\qquad \hat{H}_d= \begin{pmatrix}
      \hat H_d^0  \cr
       \hat H^-_d
       \end{pmatrix}.
 \end{equation}
 Here $\hat T^0$ is a complex neutral superfield, while  $\hat T_1^-$ and $\hat T_2^+$ are the charged Higgs superfields. Note that $(\hat{T}_1^-)^*\neq \hat{T}_2^+$. 
 Only the MSSM Higgs doublets couple to the fermion multiplet via Yukawa coupling as in Eq.~(\ref{spm}), while the singlet and the triplet superfields generate the supersymmetric $\mu_D$ term after their neutral parts  acquire  vevs, as shown in Eq.~(\ref{spt}).
 
In any scale invariant supersymmetric theory with a cubic superpotential, the complete Lagrangian with the soft SUSY breaking terms has an accidental  $Z_3$ symmetry, the invariance after the multiplication of all the components of the chiral superfield by the phase $e^{2\pi i/3}$.
Such terms are given by
 \bea\nn
V_{soft}& =&m^2_{H_u}|H_u|^2\, +\, m^2_{H_d}|H_d|^2\, +\, m^2_{S}|S|^2\, +\, m^2_{T}|T|^2\,+\, m^2_{Q}|Q|^2 + m^2_{U}|U|^2\,+\,m^2_{D}|D|^2 \\ \nn
&&+(A_S S H_d.H_u\, +\, A_{\kappa} S^3\, +\, A_T H_d.T.H_u \, +\, A_{TS} S Tr(T^2)\\ 
 &&\,+\, A_U U H_U . Q\, +\, \, A_D D H_D . Q + h.c),
\label{softp}
 \eea
while the D-terms are given by 
 \begin{equation}
 V_D=\frac{1}{2}\sum_k g^2_k ({ \phi^\dagger_i t^a_{ij} \phi_j} )^2 .
 \label{dterm}
 \end{equation}
 
 In this article we assume that all the coefficients involved in the Higgs sector are real in order to preserve CP invariance. The breaking of the $SU(2)_L\times U(1)_Y$ electroweak symmetry is obtained by giving real vevs to the neutral components
 of the Higgs fields
 \be
 <H^0_u>=\frac{v_u}{\sqrt{2}}, \, \quad \, <H^0_d>=\frac{v_d}{\sqrt{2}}, \quad ,<S>=\frac{v_S}{\sqrt{2}} \, \quad\, <T^0>=\frac{v_T}{\sqrt{2}},
 \ee
 which give mass to the $W^\pm$ and $Z$ bosons
 \be
 m^2_W=\frac{1}{4}g^2_L(v^2 + 4v^2_T), \, \quad\ m^2_Z=\frac{1}{4}(g^2_L \, +\, g^2_Y)v^2, \, \quad v^2=(v^2_u\, +\, v^2_d) .
 \ee
 and also generate the $ \mu_D=\frac{\lambda_S}{\sqrt 2} v_S+ \frac{\lambda_T}{2} v_T$ term.
 
 The non-zero triplet contribution to the $W^\pm$ mass leads to a deviation of the tree-level expression of the $\rho$ parameter
 \be
 \rho= 1+ 4\frac{v^2_T}{v^2} .
 \ee
 Thus the triplet vev is strongly  constrained by the global fit on the measurement of the $\rho$ parameter \cite{rho}
 \be
 \rho =1.0004^{+0.0003}_{-0.0004} ,
 \ee 
 which restricts its value to $v_T \leq 5	$ GeV.  In our numerical analysis we have chosen  $v_T =3	$ GeV.

 %%%%%%%%%%%%%%%%Tree level Higgs masses %%%%%%%%%%%%%%%%%%%%%%
\section{Tree-level Higgs masses}\label{treel}

To determine the tree-level mass spectrum, we first consider the tree-level minimisation conditions, 
\be\label{mnc}
\partial_{\Phi_i}V|_{vev}=0; \quad V=V_D\, +\,V_F\,+\,V_{soft}, \quad <\Phi_{i,r}>=\frac{v_{i}}{\sqrt 2},\quad \Phi_i=H^0_{u},H^0_{d}, S, T^0,
\ee
where we have defined the vacuum parameterizations of the fields in the Higgs sector as 
\be
H^0_u=\frac{1}{\sqrt{2}}(H^0_{u,r} + i H^0_{u,i}), \quad H^0_d=\frac{1}{\sqrt{2}}(H^0_{d,r} + i H^0_{d,i}), \quad S=\frac{1}{\sqrt{2}}(S_r + i S_i), \quad T^0=\frac{1}{\sqrt{2}}(T^0_{r} + i T^0_{i}). 
\ee
from which the soft-breaking masses are derived in the form

\begin{align}\label{mnc2}
m^2_{H_u}=& \frac{v_d}{2\,v_u} \left(\sqrt{2} A_S v_S-v_T \left(A_T+\sqrt{2} v_S \lambda _T \lambda_{TS}\right)+\lambda _S \left(\kappa  v_S^2+v_T^2 \lambda_{TS}\right)\right)\nn\\
&-\frac{1}{2}\left(\lambda _S^2 \left(v_d^2-v_S^2\right)+ \frac{1}{2}\lambda _T^2\left(v_d^2+v_T^2\right)+\sqrt{2} \lambda _S v_S \lambda _T v_T\right)\nn\\
&+\frac{1}{8}(v_d^2 - v_u^2)
   \left(g_L^2+g_Y^2\right),
\end{align}
\begin{align}
m^2_{H_d}=& \frac{v_u}{2\,v_d} \left(\sqrt{2} A_S v_S-v_T \left(A_T+\sqrt{2} v_S\lambda _T \lambda _{TS}\right)+\lambda _S \left(\kappa  v_S^2+v_T^2 \lambda _{TS}\right)\right)\nn\\
   &-\frac{1}{2} \left(\lambda _S^2 \left(v_u^2+v_S^2\right)+\frac{1}{2}\lambda _T^2 \left(v_u^2+v_T^2\right)-
   \sqrt{2} \lambda _S v_S \lambda _T v_T\right)\nn\\
   &+\frac{1}{8}(v_u^2 - v_d^2)
   \left(g_L^2+g_Y^2\right),
\end{align}
\begin{align}
m^2_S=& \frac{1}{2 \sqrt{2} v_S}\left(v_T \left(\lambda _T \left(\lambda _S
   \left(v_d^2+v_u^2\right)-2 v_d v_u \lambda _{TS}\right)-2 A_{TS}
   v_T\right)+2 A_S v_d v_u\right)\nn\\
   &-\frac{A_{\kappa} v_S}{\sqrt{2}}+\kappa 
   v_d v_u \lambda _S-\frac{1}{2} \lambda _S^2 \left(v_d^2+v_u^2\right)-\kappa ^2
   v_S^2-\kappa  v_T^2 \lambda _{TS}-2 v_T^2 \lambda _{TS}^2,
\end{align}
\begin{align}
m^2_T=& \frac{1}{4 v_T}\left(\sqrt{2} v_S \lambda _T
   \left(\lambda _S \left(v_d^2+v_u^2\right)-2 v_d v_u \lambda _{TS}\right)-2
   A_T v_d v_u\right)-\sqrt{2} A_{TS} v_S\nn\\
   &+\lambda _{TS} \left(v_d
   v_u \lambda _S-v_S^2 \left(\kappa +2 \lambda _{TS}\right)\right)-\frac{1}{4} \lambda _T^2
   \left(v_d^2+v_u^2\right)-v_T^2 \lambda _{TS}^2.
\end{align}

\begin{figure}[t]
\begin{center}
\includegraphics[width=0.6\linewidth]{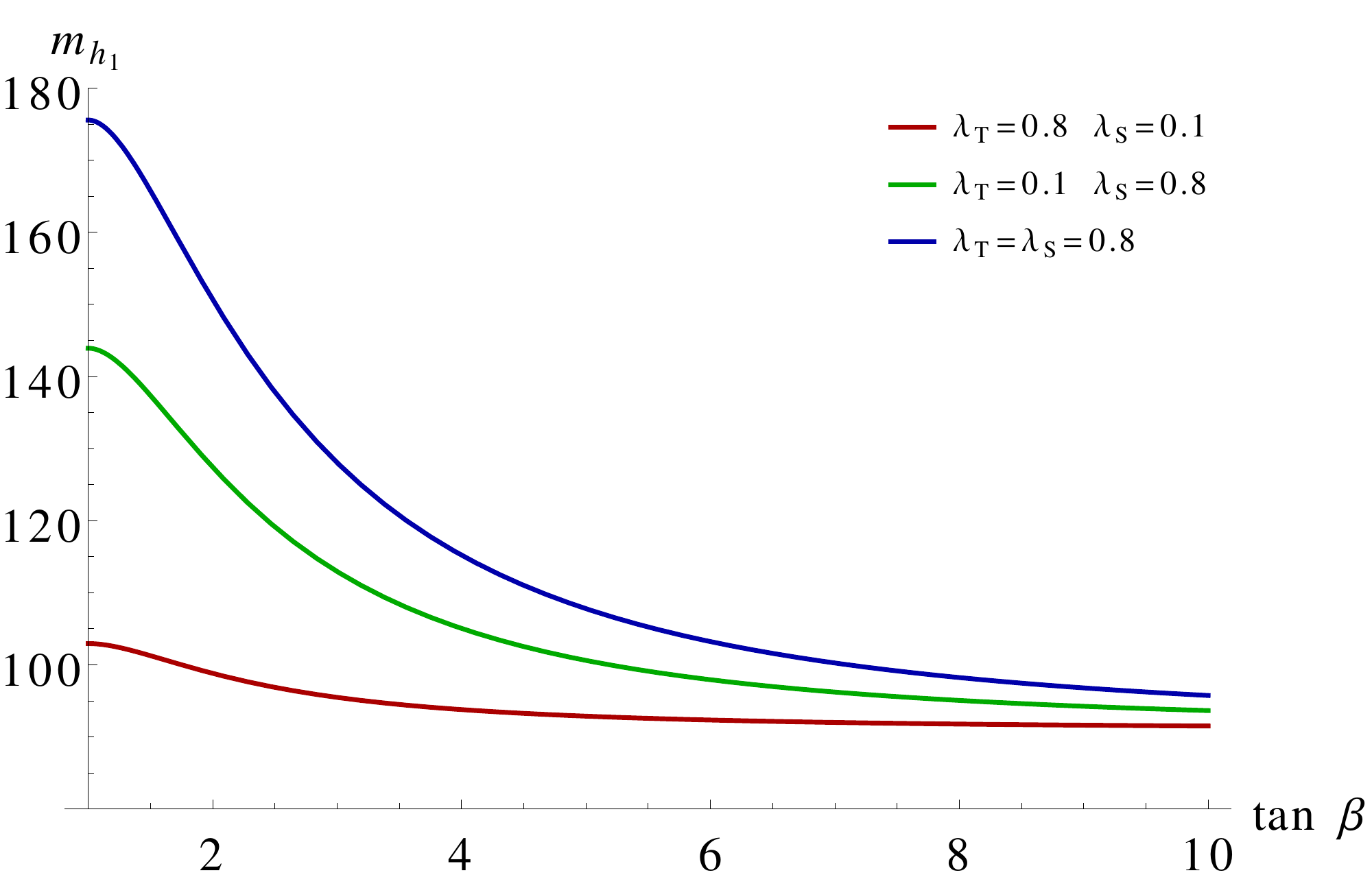}
\caption{Tree-level lightest CP-even Higgs mass maximum values with respect to $\tan{\beta}$ for 
 (i) $\lambda_T=0.8,\, \lambda_S=0.1$ (in red), (ii)$\lambda_T=0.1,\, \lambda_S=0.8$ (in green ) and (iii) $\lambda_T=0.8,\, \lambda_S=0.8$ (in blue).}\label{mht}
\end{center}
\end{figure}

It can be shown that the second derivative of the potential with respect to the fields satisfy the tree-level stability constraints. The neutral CP-even mass matrix in this case is $4$-by-$4$, since the mixing terms involve the two $SU(2)$ Higgs doublets, the scalar singlet $S$ and the neutral component of the Higgs triplet.  After electroweak symmetry breaking, the neutral Goldstone gives mass to the $Z$ boson and the charged Goldstone bosons give mass to the $W^\pm$ boson. Being the Lagrangean CP-symmetric, we are left with four CP-even, three CP-odd  and three charged Higgs bosons as shown below
 \bea\label{hspc}
  \rm{CP-even} &&\quad \quad  \rm{CP-odd} \quad\quad   \rm{charged}\nn \\
 h_1, h_2, h_3, h_4 &&\quad \quad a_1, a_2, a_3\quad \quad h^\pm_1, h^\pm_2, h^\pm_3. 
 \eea
The neutral Higgs bosons are combination of doublets, triplet and singlet, whereas the charged Higgses are a combination of doublets and triplet only. We will denote with $m_{h_i}$ the corresponding mass eigenvalues, assuming that one of them will coincide with the 125 GeV Higgs $(h_{125})$ boson detected at the LHC. The scenarios that we consider do not assume that this is the lightest eigenvalue which is allowed in the spectrum of the theory. Both scenarios with lighter and heavier undetected Higgs states will be considered. In particular, we will refer to those in which one or more Higgses with a mass lower than 125 GeV is present, to {\em hidden Higgs} scenarios.  \\
%%%%%%%%%%%%%%%%%%%%%%%%%%%%%%
At tree-level the maximum value of the lightest neutral Higgs has additional contributions from the triplet and the singlet sectors respectively. The numerical value of the upper bound on the lightest CP-even Higgs can be extracted from the relation
\be\label{hbnd}
m^2_{h_1}\leq m^2_Z(\cos^2{2\beta} \, +\, \frac{\lambda^2_T}{g^2_L\,+\,g^2_Y }\sin^2{2\beta}\, +\, \frac{2\lambda^2_S}{g^2_L\,+\,g^2_Y }\sin^2{2\beta}), \qquad \tan\beta=\frac{v_u}{v_d},
\ee
which is affected on its right-hand-side by two additional contributions from the triplet and singlet. These can raise the allowed tree-level Higgs mass. Both 
contributions are proportional to $\sin{2\beta}$, and thus they can be large for a low value of $\tan{\beta}$, as shown
in Figure~\ref{mht}. The plots indicate that for higher values of $\lambda_{T,S}$ a lightest 
tree-level Higgs boson mass of $\sim 125$ GeV can be  easily achieved. For general parameters,
the required quantum corrections needed in order to raise the mass bound are thus much smaller compared to the MSSM. In the case of the MSSM, as we have already mentioned, at tree-level  $m_h\leq m_Z$, and we need a correction $\gsim  35$ GeV to match the experimental value of the discovered Higgs boson mass, which leads to a fine-tuning of the SUSY parameters. In fact, this requires that the allowed parameter space of the MSSM is characterized either by large SUSY masses or by large splittings among the mass eigenvalues.
%%%%%%%Ligtest  Higgs mass vs tanbeta %%%%%%%
%%%%%%%%%%%%%%%%%%%%%%%%%%%%%%
In fact, this requires that the allowed parameter space of the MSSM is characterized either by large SUSY masses or by large splittings among the mass eigenvalues.
We have first investigated  the tree-level mass spectrum for the Higgs bosons and analysed the prospect of a $\sim 125$ GeV Higgs boson along with the hidden Higgs scenarios.  We
have looked for tree-level mass eigenvalues where at least one of them corresponds to the Higgs
discovered at the LHC. For this purpose we have performed an initial scan of the parameter space  
\bea\label{parat}
|\lambda_{T, S, TS}| \leq 1, \quad |\kappa|\leq 3, \quad |v_s|\leq 1 \, \rm{TeV}, \quad 1\leq \tan{\beta}\leq 10, 
\eea
and searched for a CP-even Higgs boson around $100-150$ GeV, assuming that at least one of the 4 eigenvalues $m_{h_i}$ will fall within  the interval $123$ GeV $\leq m_{h_i}\leq 127$ GeV at one-loop.

%%%%%%%%%%%%%%%%HH

\begin{figure}[t]
\begin{center}
\mbox{\hskip -15 pt\subfigure[]{\includegraphics[width=0.55\linewidth]{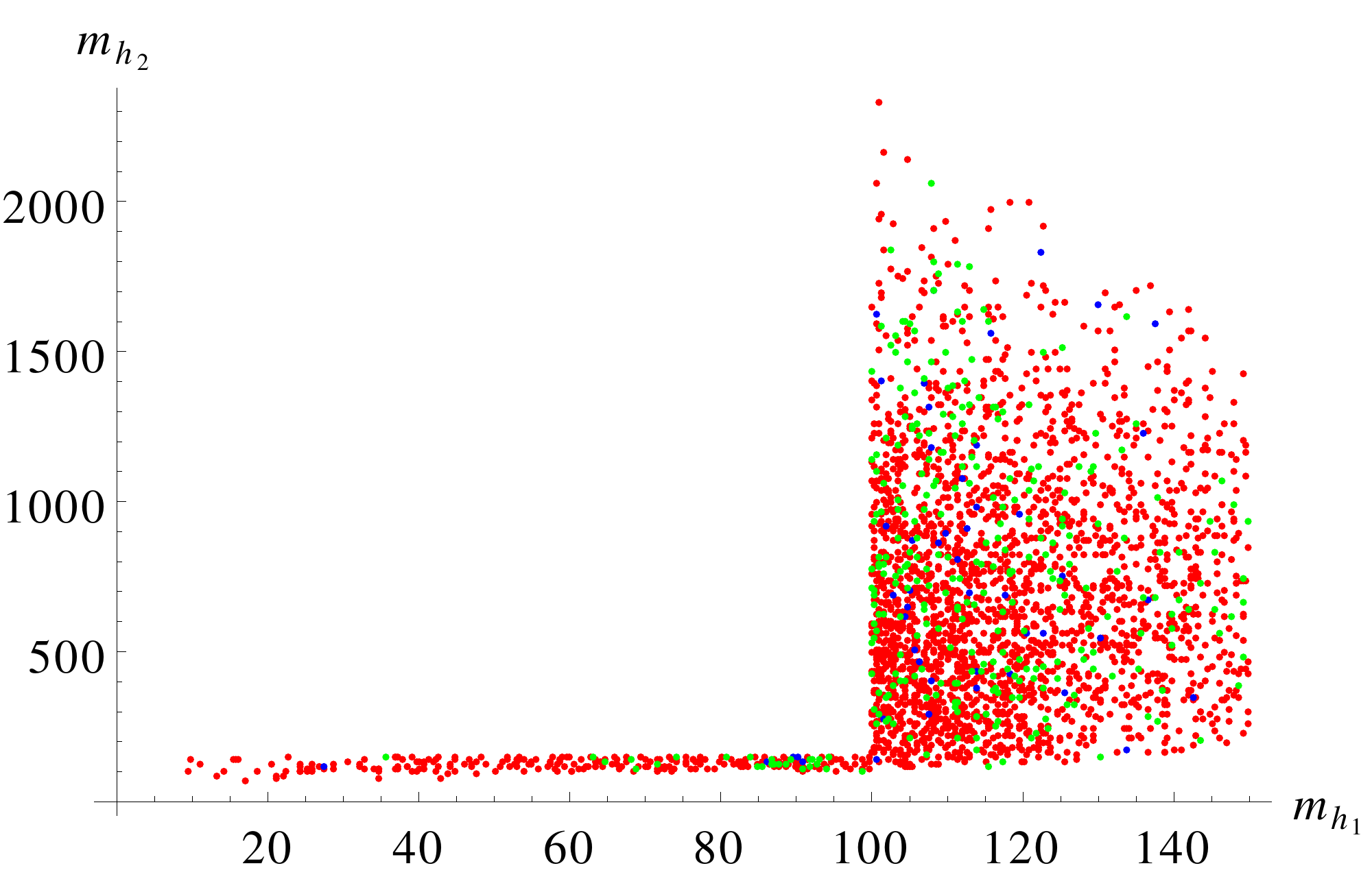}}
%\hskip 25 pt
\subfigure[]{\includegraphics[width=0.55\linewidth]{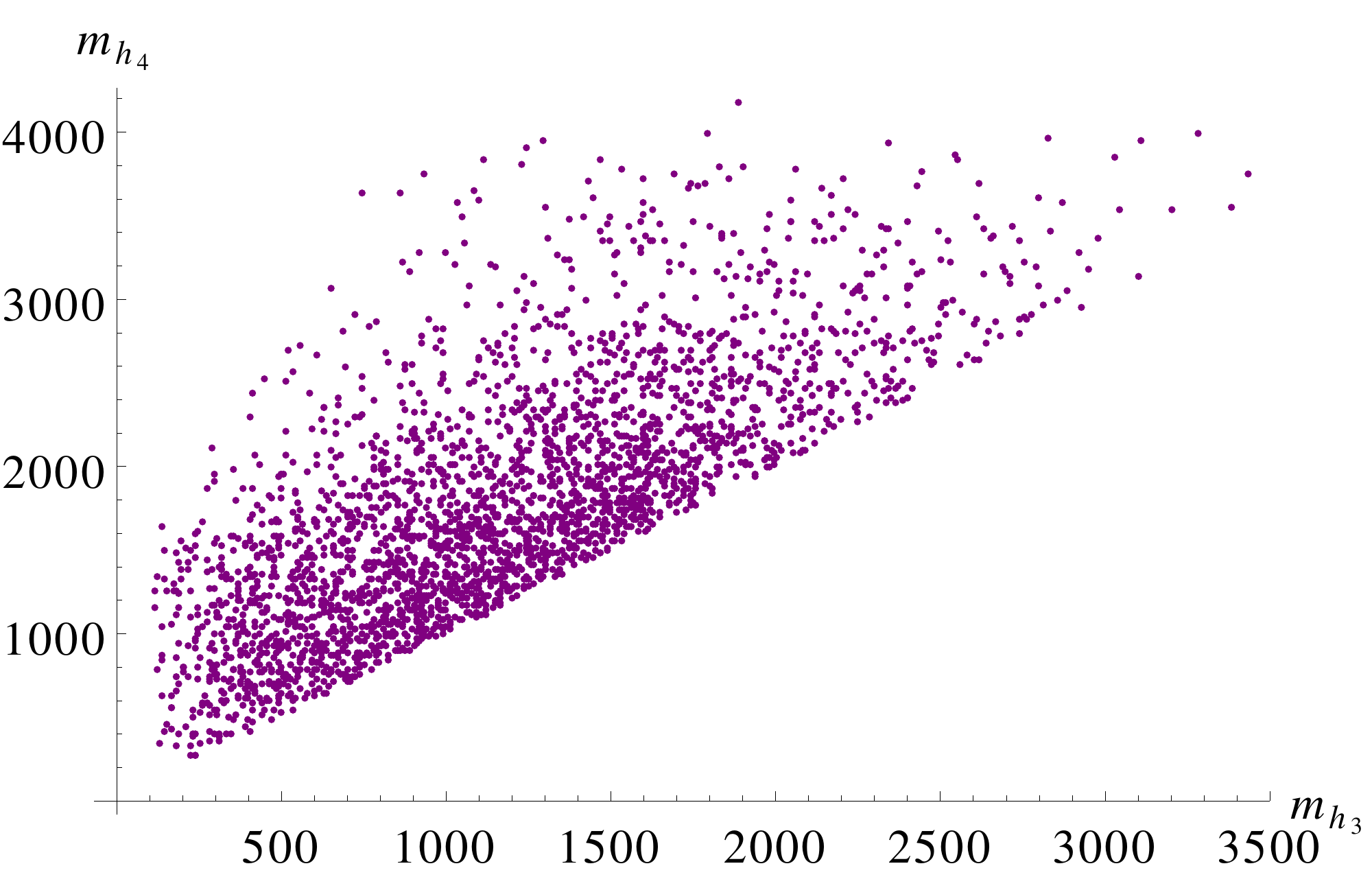}}}
\caption{Tree-level CP-even Higgs mass correlations (a) $m_{h_1}$ vs $m_{h_2}$ and (b) $m_{h_3}$ vs $m_{h_4}$, where we have a candidate $\sim 125$ GeV Higgs boson. The colors refers to the character of the $h_1$ mass eigenstate, describing the weights of the doublet, singlet and triplet  contributions in their linear combinations. Red points are $>90\%$ doublets-like, the green points are either $\geq 90\%$ triplet-like or singlet-like and blue points are mixtures of
doublet and triplet/singlet components. The linear combinations corresponding to green points are chosen to satisfy the constraints from LEP onto Z and lepton final states. }\label{hihj}
\end{center}
\end{figure}

 Figure~\ref{hihj}(a) presents the mass correlations between $m_{h_1}$ and $m_{h_2}$, where we have a CP-even neutral Higgs boson in the $100\leq m_{h_i}\leq 150$ GeV range. The candidate Higgs boson around $125$  GeV will be determined at one loop level by including positive and negative radiative corrections in the next section. The mass correlation plot
at tree-level shows that there are solutions with very light $h_1$, $m_{h_1}\leq 100$ GeV, which should be confronted with LEP data \cite{LEPb}.  At LEP were conducted searches for the Higgs boson via the $e^+ e^- \to Z h$ and 
$e^+ e^- \to  h_1 h_2$ channels (in models with multiple Higgs bosons) and their fermionic decay modes ($h\to b\bar{b},\tau \bar{\tau}$ and $Z\to \ell \ell$). The higher centre of mass energy at LEP II  (210 GeV) allowed to set a lower bound of 114.5 on the SM-like Higgs boson and of 93 GeV for the MSSM-like Higgs boson in the maximal mixing scenario \cite{LEPb}. Interestingly, neither the triplet (in our case) nor the singlet type Higgs boson couple to $Z$ or to leptons (see  Eq.~(\ref{spt})), and as such they are not excluded by LEP data.

We mark such points with $\geq 90\%$ triplet/singlet components, which can evade the LEP bounds, in green. In Figure~\ref{hihj}(a) one can immediately realize that the model allows for some very light Higgs bosons ($m_{h_1}\leq 100$ GeV). We expect that the possibility of such a hidden Higgs would be explored at the LHC with 14 TeV centre of mass energy, whereas the points where $h_1$ is mostly a doublet ($\geq 90\%$) could be ruled out by the LEP data. The points with the mixed scenario for $h_1$ (with doublet, triplet and singlet) are marked in blue. We remark that a triplet of non-zero hypercharge will not easily satisfy the constraints from LEP, due to its coupling to the $Z$ boson. 

For the points with $m_{h_1/a_1}\leq 100$ GeV which are mostly doublet (red ones) it is very hard to satisfy the LEP bounds \cite{LEPb}. This is because, being doublet like, such $h_1$ would have been produced at LEP and decayed to the fermionic pairs, which have been searched extensively at LEP. On the other hand the singlet and triplet like points (green points) are very difficult to produce at LEP due to the non-coupling to $Z$ boson, which was one of the dominant production channel. This is true for both $e^+e^- \to Zh_1$ and  $e^+e^- \to h_1a_1$. Such triplet and singlet like points will reduce the decay widths in charged lepton pair modes due to non-coupling with fermions. These make the green points more suitable candidate for the hidden Higgs bosons, both for the CP-even and CP-odd. However such parameter space would be highly constrained from the data of the discovered Higgs boson around 125 GeV at the LHC. So far the discovery of the Higgs boson at the LHC has reached  $5\sigma$ or more in the channels  $h_{125}\to \gamma\gamma, WW^*, ZZ^*$. Effectively this could be satisfied by the candidate Higgs around 125 GeV which is mostly doublet like and its decay branching fractions should be within the uncertainties give by  CMS and ATLAS experiments at the LHC. Such requirements rule out vast number of parameter points, including some the triplet and/or signet like hidden Higgs boson(s). In section~\ref{Hdata} we consider such constrains coming from the Higgs data at LHC and the existing data from LEP. 

\begin{figure}[t]
\begin{center}
\mbox{\hskip -15 pt\subfigure[]{\includegraphics[width=0.55\linewidth]{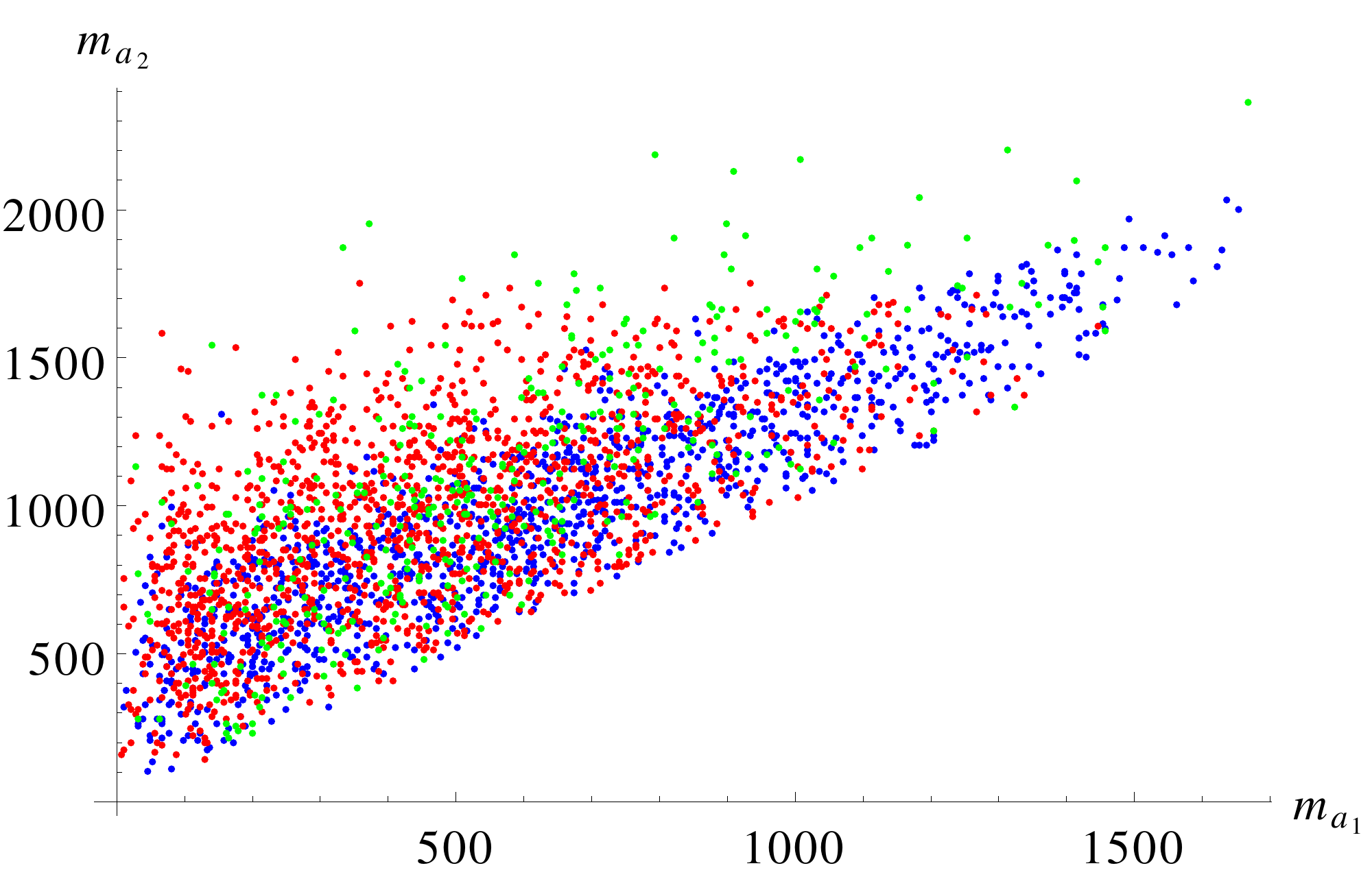}}
%\hskip 25 pt
\subfigure[]{\includegraphics[width=0.55\linewidth]{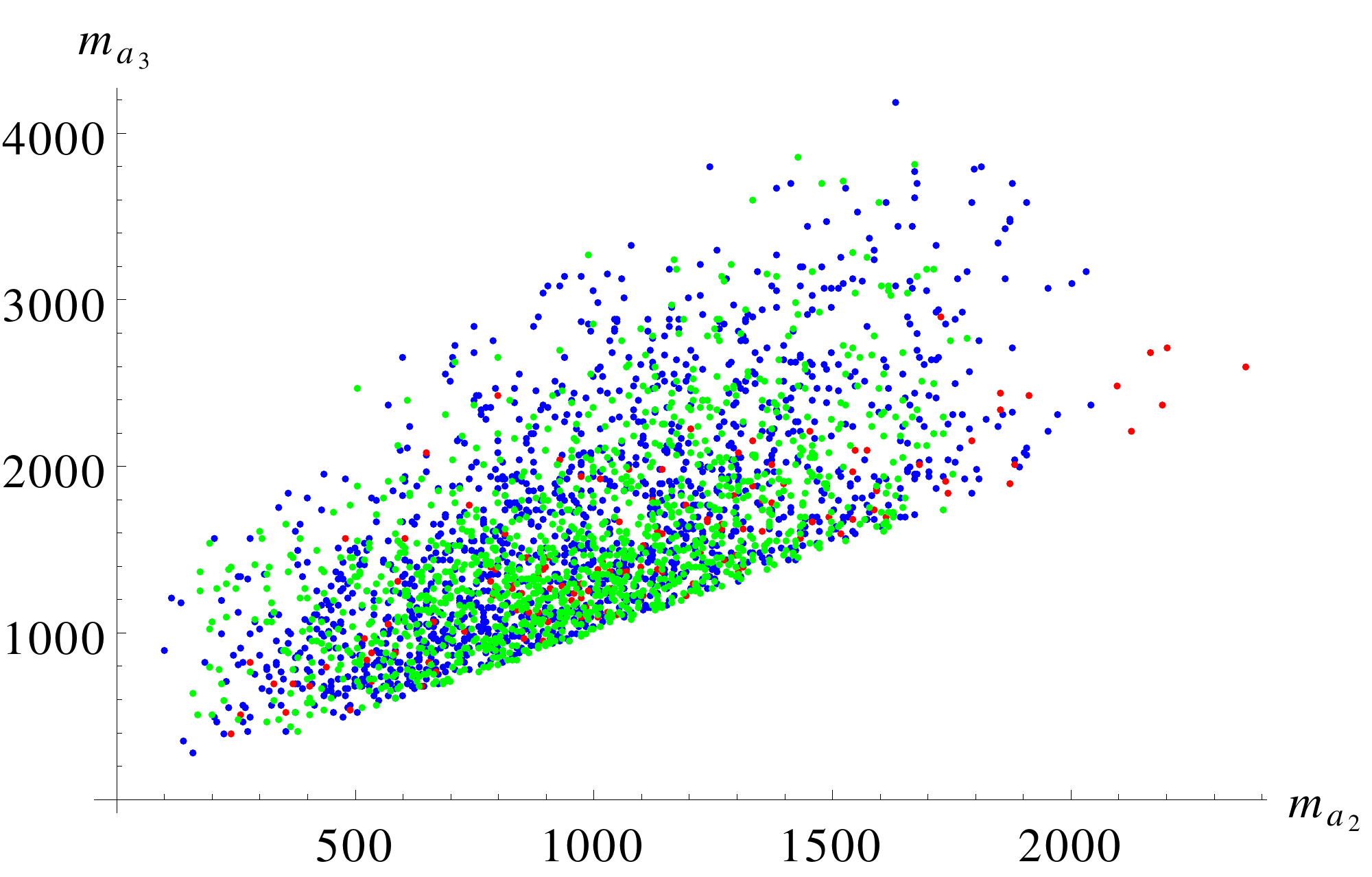}}}

\caption{Tree-level CP-odd Higgs mass correlations (a) $m_{a_1}$ vs $m_{a_2}$ and (b) $m_{a_2}$ vs $m_{a_3}$, where we have a candidate $\sim 125$ GeV Higgs boson. The red points are $>90\%$ doublets-like and the green points are $\geq 90\%$ triplet-like. The blue points are mixtures of
doublet and triplet components for $a_1$ in (a) and for  $a_2$ in (b) respectively.}\label{aiaj}
\end{center}
\end{figure}

\begin{figure}[t]
\begin{center}
\mbox{\hskip -15 pt\subfigure[]{\includegraphics[width=0.55\linewidth]{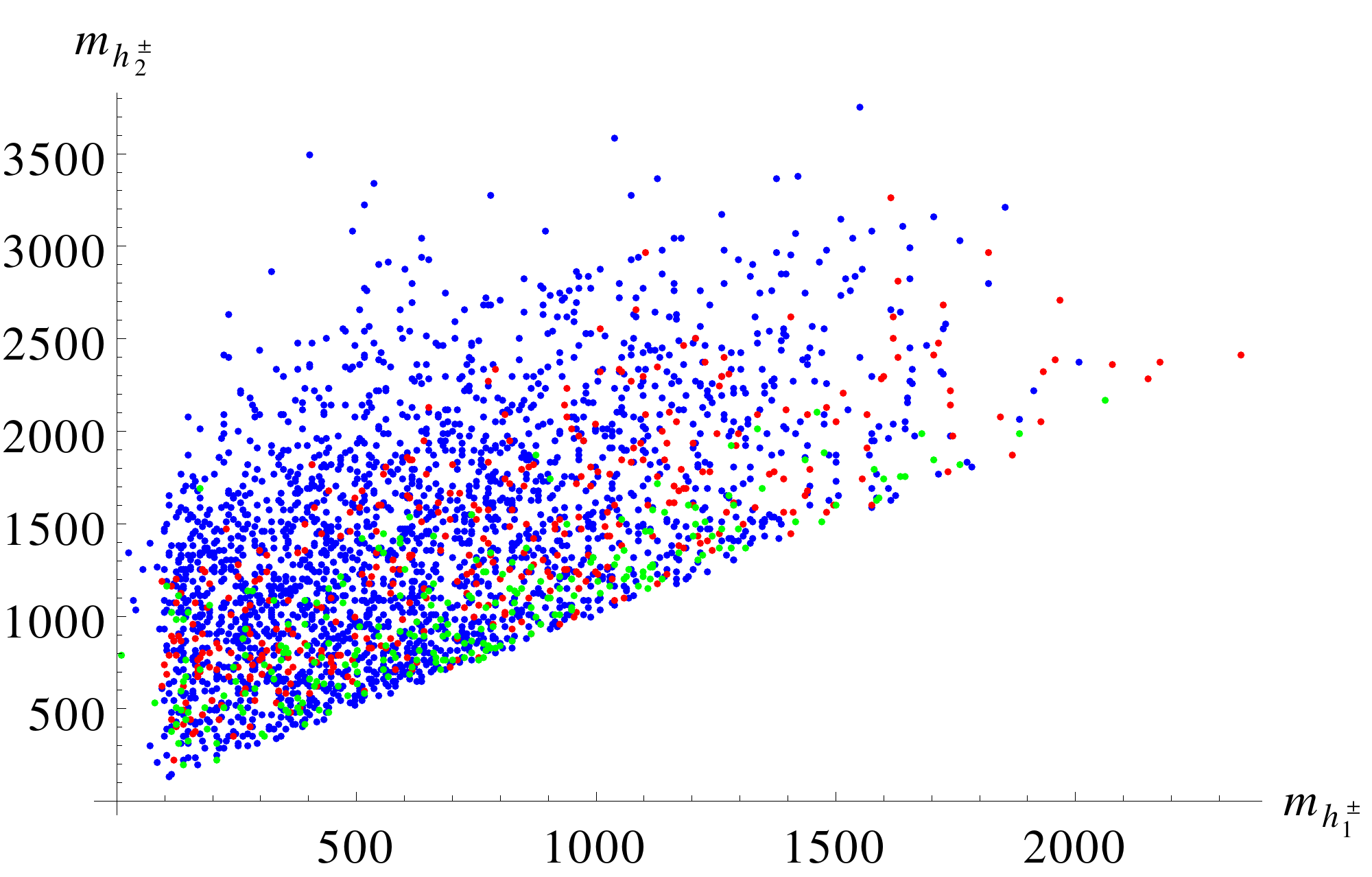}}
\hskip 25 pt
\subfigure[]{\includegraphics[width=0.55\linewidth]{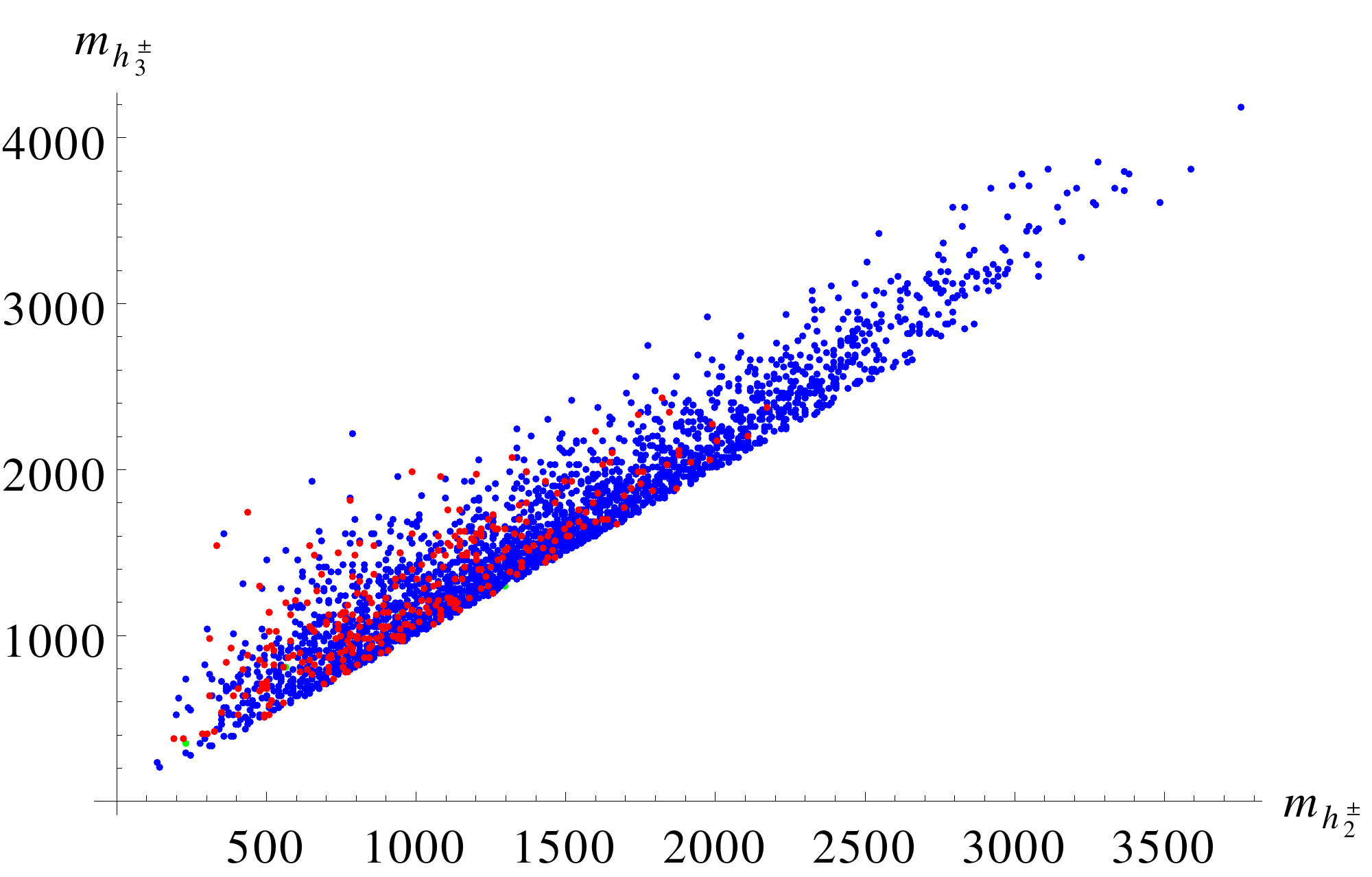}}}

\caption{Tree-level charged Higgs mass correlations (a) $m_{h^\pm_1}$ vs $m_{h^\pm_2}$ and (b) $m_{h^\pm_2}$ vs $m_{h^\pm_3}$, where we have a candidate $\sim 125$ GeV Higgs boson. The red points are $>90\%$ doublets-like and the green points are $\geq 90\%$ triplet- or singlet-like. The blue points are mixture of doublet and triplets/singlets, for $h^\pm_1$ in (a) and for  $h^\pm_2$ in (b) respectively.}\label{chichj}
\end{center}
\end{figure}

Figure~\ref{hihj}(b) shows the mass correlation between ${h_3}$ and $h_4$ for the the same region (\ref{parat}) of the parameter  space.  We see that although there are points characterized by a mass $m_{h_3}$ lighter than 500 GeV, states with $m_{h_4}\leq 500$ GeV are less probable. 
%%%%%%%%%%% m_h1 vs m_h2 and m_h3 vs m_h4 %%%%%%%%
%%%%%%%%%%%%%%%%%%%%%%%%%%%
Figure~\ref{aiaj} shows the mass correlations of the CP-odd neutral Higgs bosons. Specifically, Figure~\ref{aiaj}(a) presents the analysis of the mass correlation between $a_1$ and $a_2$. The plot shows that there exists the possibility of having a pseudo-scalar $a_1$ lighter than 100 GeV, accompanied by a CP-even $\sim 125$ GeV Higgs boson.  Note that a very light pseudoscalar Higgs in the MSSM gets strong bounds from LEP \cite{LEPb}. In this case, for a high $\tan{\beta}$, the pair production process $e^+e^- \to h A$, where $A$ is the pseudoscalar of the MSSM, is the most useful one, providing limits in the vicinity of 93 GeV for $m_A$ \cite{LEPb}.  In the TNMSSM instead, if the light 
pseudoscalar Higgs bosons are either of triplet or singlet type then they do not couple to the $Z$, which makes it easier for these states to satisfy the LEP bounds. For this purpose, points which are mostly-triplet or -singlet ($90\%$) have been marked in green; points which are mostly-doublet ($90\%$) in red, whereas the mixed points have been marked in blue as before. Certainly, mass eigenvalues labelled in green would be much more easily allowed by the LEP data, but they would also be able to evade the recent bounds from the LHC $H\tau \tau $ decay mode for a pseudoscalar Higgs \cite{Htautau}. This occurs because neither the triplet nor the singlet Higgs boson couple to fermions (See Eq.~(\ref{spt})).
Figure~\ref{aiaj}(b) presents the correlation between $a_2$ and $a_3$ where the same colour code applies for the structure of $a_2$. As one can easily realize from the figure, there are plenty of green coloured points which represent  triplet/singlet type $a_2$ states, which can easily evade the recent bounds on pseudoscalar states derived at the LHC  \cite{Htautau}. 
%%%%%%%%%%% m_a1 vs m_a2 and m_a3 vs m_a4 %%%%%%%%
%%%%%%%%%%%%%%%%%%%%%%%%%%%
Figure~\ref{chichj} shows the correlation of the three charged Higgs bosons for the region in parameter space where we can have a $\sim 125$ GeV Higgs candidate.  Figure~\ref{chichj}(a) shows that there are allowed points for a charged Higgs of light mass  ($m_{h^\pm_1} \lsim 200$ GeV) correlated with a heavier charged Higgs $h^\pm_2$.  Only Higgses of doublet and triplet type  can contribute to the charged Higgs sector. We have checked the structure of the lightest charged Higgs $h^\pm_1$ in Figure~\ref{chichj}(a), where the red points correspond to $\geq 90\%$ doublet, while the green points correspond to  $\geq 90\%$ triplet and the blue points to doublet-triplet mixed states. Charged Higgs bosons which are mostly triplet-like in their content (the green points) do not couple to the fermions (see Eq.~(\ref{spt})), and thus can easily evade the bounds on the light charged Higgs derived at the LHC from the $H^\pm  \to \tau \nu$ decay channel \cite{chHb}. This kind of triplet charged Higgs boson would also be hard to produce from the conventional decay of the top quark and the new production modes as well as the decay modes will open up due to the new vertex $h^\pm_i - Z-W^\mp$ \cite{tssmch1}. Thus vector boson fusion (VBF) with the production of a single charged Higgs is a possibility due to a non-zero  $h^\pm_i - Z-W^\mp$ vertex \cite{tssmch1}. Apart from the $h^\pm_i \to ZW^\pm$ channels, the $h^\pm_i \to  a_1(h_1)W^\pm$ channels are also allowed, for very light neutral Higgs bosons ($a_1/h_1$). Figure~\ref{chichj}(b) presents the correlation between $m_{h^\pm_2}$ and $m_{h^\pm_3}$. We have used for $h^\pm_2$ the same colour conventions as in the previous plots. We see that there are only few triplet type $h^\pm_2$ (green points), most of the allowed mass points being doublet-triplet mixed states (blue points).

%%%%%%%%%%% m_h^+1 vs m_h^+2 and m_h^+3 vs m_h^+4 %%%%%%%%
%%%%%%%%%%%%%%%%%%%%%%%%%%%
\section{Strong and weak sectors}\label{storngw}

The TNMSSM scenario has an additional triplet which is colour singlet and electroweak charged and a singlet superfields (see Eq.~(\ref{spt})) not charged under $SU(3)_c\times SU(2)_L\times U(1)_Y$. Therefore, the strong sector of the model is the same of the MSSM, but supersymmetric F-terms affect the fermion mass matrices, and contribute to the off-diagonal terms.  It generates additional terms in the stop mass matrix from the triplet and singlet vevs, which will be shown below. These terms are proportional to $\lambda_T v_T$ and $\lambda_S v_S$ respectively, and allow to generate an effective $\mu_D$-term in the model.  The triplet contribution is of course restricted, due to the bounds coming from the $\rho$ parameter \cite{rho}. Thus, a large effective $\mu_D$ term can be spontaneously generated by the vev of the singlet, $v_S$. 

\begin{figure}[hbt]
\begin{center}
{\includegraphics[width=0.6\linewidth]{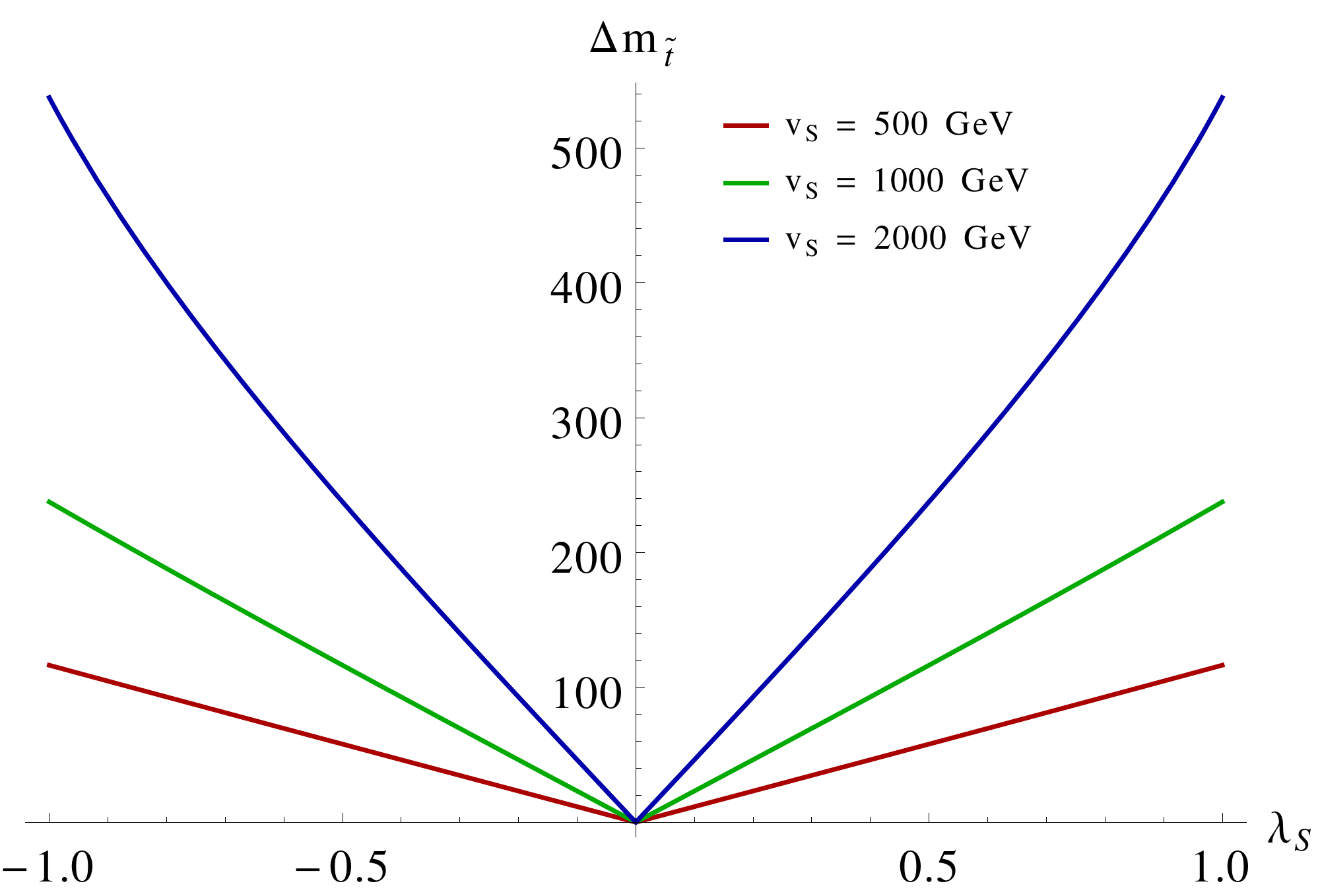}}
\caption{The mass splitting between the stop mass eigen states ($\tilde{t}_{2,1}$) vs $\lambda_S$
for $A_t=0$ with $v_S=500, 1000, 2000$ GeV respectively.}\label{stm}
\end{center}
\end{figure}

Figure~\ref{stm} shows the mass splitting between the $\tilde{t}_2$ and $\tilde{t}_1$ stops versus $\lambda_S$, for several $v_S$ choices and with $A_t=0$. Large mass splittings can be generated without a large parameter $A_t$, by a suitably large $v_S$, which is a common choice if the singlet is gauged respect to an extra $U(1)'$ \cite{uprime}, due the mass bounds for the additional gauge boson $Z'$ \cite{zprime}. The mass matrices for the stop and the sbottom are given by 

\bea\label{stop}
M_{\tilde{t}}=\left(
\begin{array}{cc}
m^2_t+m^2_{Q_3}+\frac{1}{24} \left(g_Y^2-3g_L^2\right) \left(v_u^2-v_d^2\right)\qquad &  \frac{1}{\sqrt{2}}
   A_t v_u+\frac{Y_t v_d}{2} \left( \frac{v_T \lambda _T}{\sqrt 2}- v_S \lambda _S\right) \\
   \\
\frac{1}{\sqrt{2}}
   A_t v_u+\frac{Y_t v_d}{2} \left( \frac{v_T \lambda _T}{\sqrt 2}- v_S \lambda _S\right) & m_t^2+m^2_{\bar{u}_3}+\frac{1}{6} \left(v_d^2-v_u^2\right) g_Y^2
\end{array}
\right)
\eea

%%%%%%%%%%% m_{stop2} - m_{stop1} vs lambda_S%%%%%%%%
%%%%%%%%%%%%%%%%%%%%%%%%%%%
\bea\label{sbt}
M_{\tilde{b}}=\left(
\begin{array}{cc}
m_b^2+m^2_{Q_3}+\frac{1}{24} \left(g_Y^2+3g_L^2\right) \left(v_u^2-v_d^2\right)\qquad & \frac{1}{\sqrt{2}}
   A_b v_d+\frac{Y_b v_u}{2} \left( \frac{v_T \lambda _T}{\sqrt 2}- v_S \lambda _S\right) \\
   \\
 \frac{1}{\sqrt{2}}
   A_b v_d+\frac{Y_b v_u}{2} \left( \frac{v_T \lambda _T}{\sqrt 2}- v_S \lambda _S\right) & m_b^2+m^2_{\bar{d}_3}+\frac{1}{12} \left(v_u^2-v_d^2\right) g_Y^2 \\
\end{array}
\right)
\eea

In the electroweak sector the neutralino ($\tilde{\chi}^0_{i=1,..6}$ ) and chargino ($\tilde{\chi}^\pm_{i=1,2,3}$ ) sector are enhanced due to
the extra Higgs fields in the superpotential given in (\ref{spt}). The neutralino sector is now composed of
$\tilde{B},\, \tilde{W}_3, \,\tilde{H}_u, \, \tilde{H}_d, \, \tilde{T}_0, \, \tilde{S}$. The corresponding mass matrix is thus now 6-by-6 and given by  

\bea\label{ntln}
M_{\tilde{\chi}^0}&=\left(
\begin{array}{cccccc}
 M_1 & 0 & -\frac{1}{2} g_Y v_d & \frac{1}{2}g_Y v_u & 0 & 0 \\
 0 & M_2 & \frac{1}{2}g_L v_d & -\frac{1}{2} g_L v_u & 0 & 0 \\
 -\frac{1}{2} g_Y v_d & \frac{1}{2}g_L v_d & 0 & \frac{1}{2}v_T \lambda _T-\frac{1}{\sqrt{2}}v_S \lambda_S & \frac{1}{2}v_u \lambda _T & -\frac{1}{\sqrt{2}}v_u \lambda _S \\
 \frac{1}{2}g_Y v_u & -\frac{1}{2} g_L v_u & \frac{1}{2}v_T \lambda _T-\frac{1}{\sqrt{2}}v_S \lambda_S & 0 & \frac{1}{2}v_d \lambda _T & -\frac{1}{\sqrt{2}}v_d \lambda _S \\
 0 & 0 & \frac{1}{2}v_u \lambda _T & \frac{1}{2}v_d \lambda _T & \sqrt{2} v_S \lambda _{TS} & \sqrt{2}
   v_T \lambda _{TS} \\
 0 & 0 & -\frac{1}{\sqrt{2}}v_u \lambda _S & -\frac{1}{\sqrt{2}}v_d \lambda _S & \sqrt{2} v_T
   \lambda _{TS} & \sqrt{2} \kappa  v_S. \\
\end{array}
\right)\nn\\
\eea
The triplino ($\tilde{T}_0$) and the 
singlino ($\tilde{S}$) masses and mixings are spontaneously generated by the corresponding vevs.
The triplino and singlino are potential dark matter candidates and have an interesting phenomenology as they do not couple directly to the fermion superfields.  The doublet-triplet(singlet) mixing is very crucial
in determining the rare decay rates as well as the dark matter relic densities.

Unlike the neutralino sector, the singlet superfield does not contribute to the chargino mass matrix, and hence the MSSM chargino mass matrix is extended by the triplets only. The chargino mass matrix in the basis of $\tilde{W}^+, \tilde{H}^+_{u}, \tilde{T}^+_{2} (\tilde{W}^-, \tilde{H}^-_{d}, \tilde{T}^-_{1} )$ takes the form

 \bea\label{chn}
M_{\tilde{\chi}^\pm}=\left(
\begin{array}{ccc}
 M_2 & \frac{1}{\sqrt{2}}g_L v_u & -g_L v_T \\
 \frac{1}{\sqrt{2}}g_L v_d & \frac{1}{\sqrt{2}}v_S \lambda _S+\frac{1}{2}v_T \lambda _T & \frac{1}{\sqrt{2}}v_u
   \lambda _T \\
 g_L v_T & -\frac{1}{\sqrt{2}}v_d \lambda _T & \sqrt{2} v_S \lambda _{TS} \\
\end{array}
\right).
\eea
The chargino decays also have an interesting phenomenology due to the presence of a doublet-triplet mixing. 
%%%%%%%%%%%%%%%%%%%%%

\section{Higgs masses at one-loop}\label{onel}

 To study the effect of the radiative correction to the Higgs masses, we calculate the one-loop
Higgs mass for the neutral Higgs bosons via the Coleman-Weinberg effective 
potential given in Eq.~(\ref{cwe}) 
\begin{align}\label{cwe}
V_{\rm CW}=\frac{1}{64\pi^2}{\rm STr}\left[ \mathcal{M}^4
\left(\ln\frac{\mathcal{M}^2}{\mu_r^2}-\frac{3}{2}\right)\right],
\end{align}
where $\mathcal{M}^2$ are the field-dependent mass matrices, $\mu_r$ is the renormalization scale, and the supertrace includes a factor of $(-1)^{2J}(2J+1)$ for each particle of spin J in the loop. We have omitted additional charge and colour factors which should be appropriately included. The corresponding one-loop contribution to the neutral Higgs mass matrix  is given by Eq.~(\ref{1Lmh})
\begin{align}
(\Delta\mathcal{M}^2_h)_{ij}
&=\left.\frac{\partial^2 V_{\rm{CW}}(\Phi)}{\partial \Phi_i\partial \Phi_j}\right|_{\rm{vev}}
-\frac{\delta_{ij}}{\langle \Phi_i\rangle}\left.\frac{\partial V_{\rm{CW}}(\Phi)}{\partial \Phi_i}\right|_{\rm{vev}}
\nn\\
&=\sum\limits_{k}\frac{1}{32\pi^2}
\frac{\partial m^2_k}{\partial \Phi_i}
\frac{\partial m^2_k}{\partial \Phi_j}
\left.\ln\frac{m_k^2}{\mu_r^2}\right|_{\rm{vev}}
+\sum\limits_{k}\frac{1}{32\pi^2}
m^2_k\frac{\partial^2 m^2_k}{\partial \Phi_i\partial \Phi_j}
\left.\left(\ln\frac{m_k^2}{\mu_r^2}-1\right)\right|_{\rm{vev}}
\nonumber\\
&\quad-\sum\limits_{k}\frac{1}{32\pi^2}m^2_k
\frac{\delta_{ij}}{\langle \Phi_i\rangle}
\frac{\partial m^2_k}{\partial \Phi_i}
\left.\left(\ln\frac{m_k^2}{\mu_r^2}-1\right)\right|_{\rm{vev}}\ ,\quad \Phi_{i,j}=H^0_{u,r},H^0_{d,r},S_r,T^0_r\ .
\label{1Lmh}
\end{align}

Here, $m^2_k$ is the set of eigenvalues of the field-dependent mass matrices given in the equation above, and we remind that the real components of the neutral Higgs fields are defined as 
\bea
&H^0_u=\frac{1}{\sqrt{2}}(H^0_{u,r} + i H^0_{u,i}), \quad H^0_d=\frac{1}{\sqrt{2}}(H^0_{d,r} + i H^0_{d,i}),\nn\\
&S=\frac{1}{\sqrt{2}}(S_r + i S_i), \quad T^0=\frac{1}{\sqrt{2}}(T^0_{r} + i T^0_{i}). 
\eea.

 For simplicity we drop the supertrace expressions in Eq. (\ref{1Lmh}), but for each particle the supertrace coefficient should be taken into account.

Having characterized the entire sector of the TNMSSM, we gear up for the numerical evaluation of the one-loop neutral Higgs masses in the model. We have already seen in Eq.~\ref{hbnd} that for low $\tan{\beta}$ the contribution of the radiative corrections required in order to reach the $\sim 125$ GeV Higgs mass, overcoming the tree-level bound in (\ref{hspc}), is reduced. This is due to the additional Higgs and higgsinos running in the loops. In our analysis we have chosen the following subregion of the parameter space 
\bea\label{scan}
&|\lambda_{T, S, TS}| \leq 1, \quad |\kappa|\leq 3, \quad |v_s|\leq 1 \, \rm{TeV}, \quad 1\leq \tan{\beta}\leq 10,\nn\\
&|A_{T, S, TS, U, D}|\leq 500,\qquad|A_\kappa|\leq1500, \qquad  m^2_{Q_3, \bar{u}_3, \bar{d}_3}\leq1000,\\
&65\leq|M_{1, 2}|\leq1000,\nn
\eea
that we have used in the computation of the Higgs boson mass. In this scan, we have included the radiative corrections to the mass eigenvalues at one-loop order of the neutral sector and retained only those sets of eigenvalues which contain 
one 125 GeV CP-even Higgs. We have selected the range  $65\leq|M_{1, 2}|\leq1000$ in order to avoid the constraints on the Higgs invisible decay and use $\mu_r=500$ GeV for the numerical calculation.

%%%%%%%%% Radiative correction vs parameters %%%%%%%%%
\begin{figure}[]
\begin{center}

\mbox{\hskip -15 pt\subfigure[]{\includegraphics[width=0.55\linewidth]{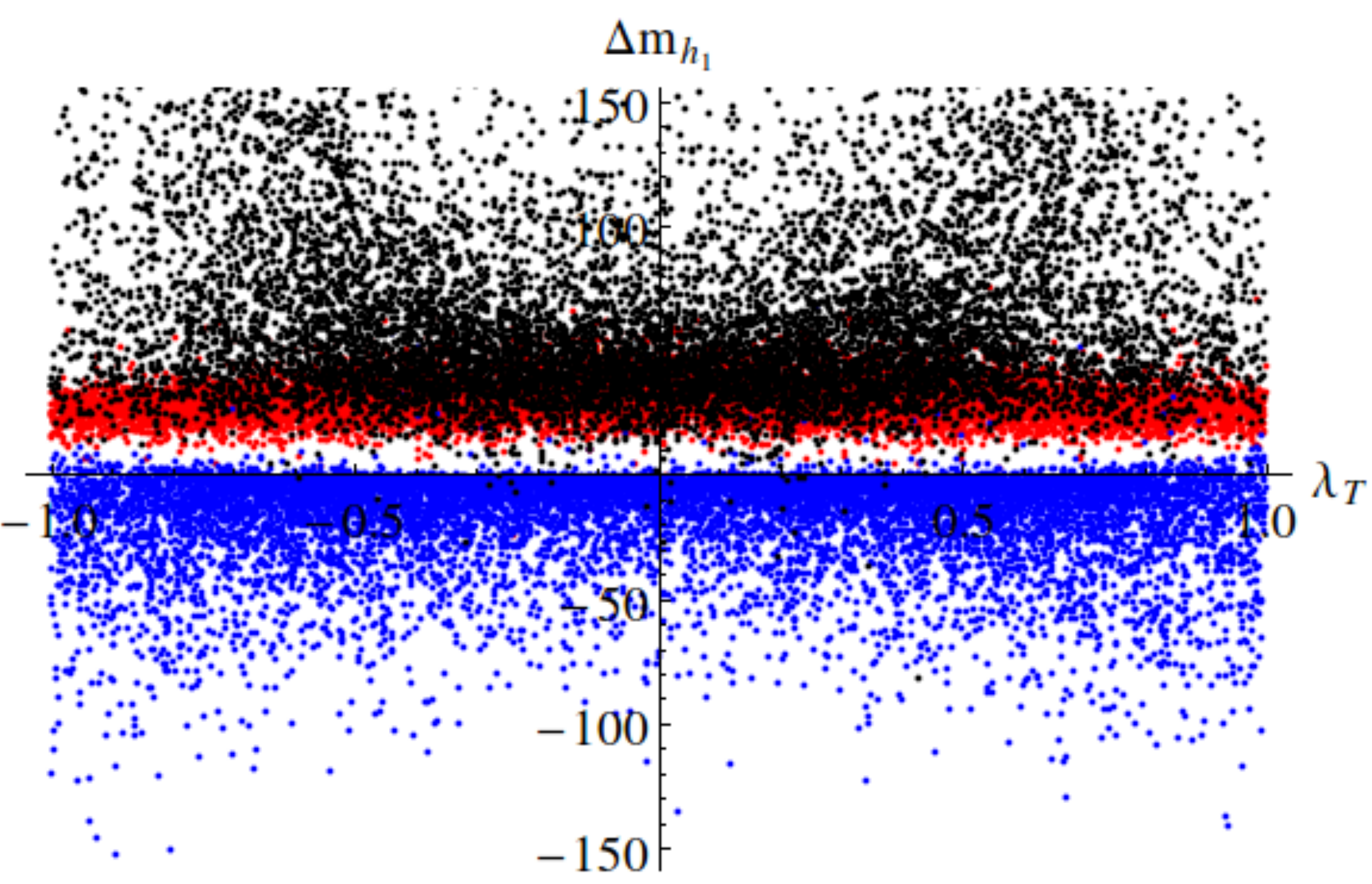}}
%\hskip 25 pt
\subfigure[]{\includegraphics[width=0.55\linewidth]{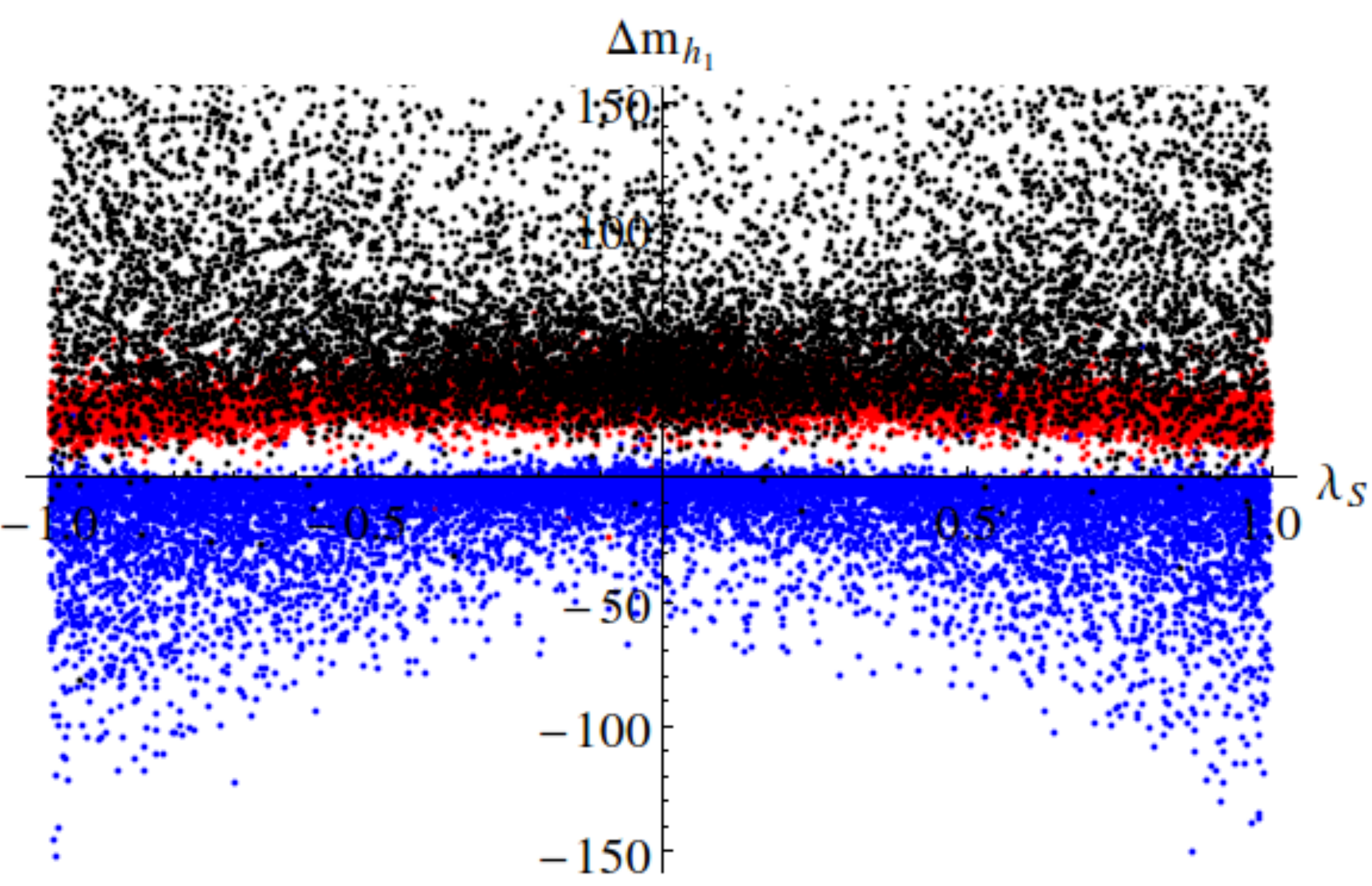}}}
%\hskip 25 pt

\mbox{\subfigure[]{\includegraphics[width=0.7\linewidth]{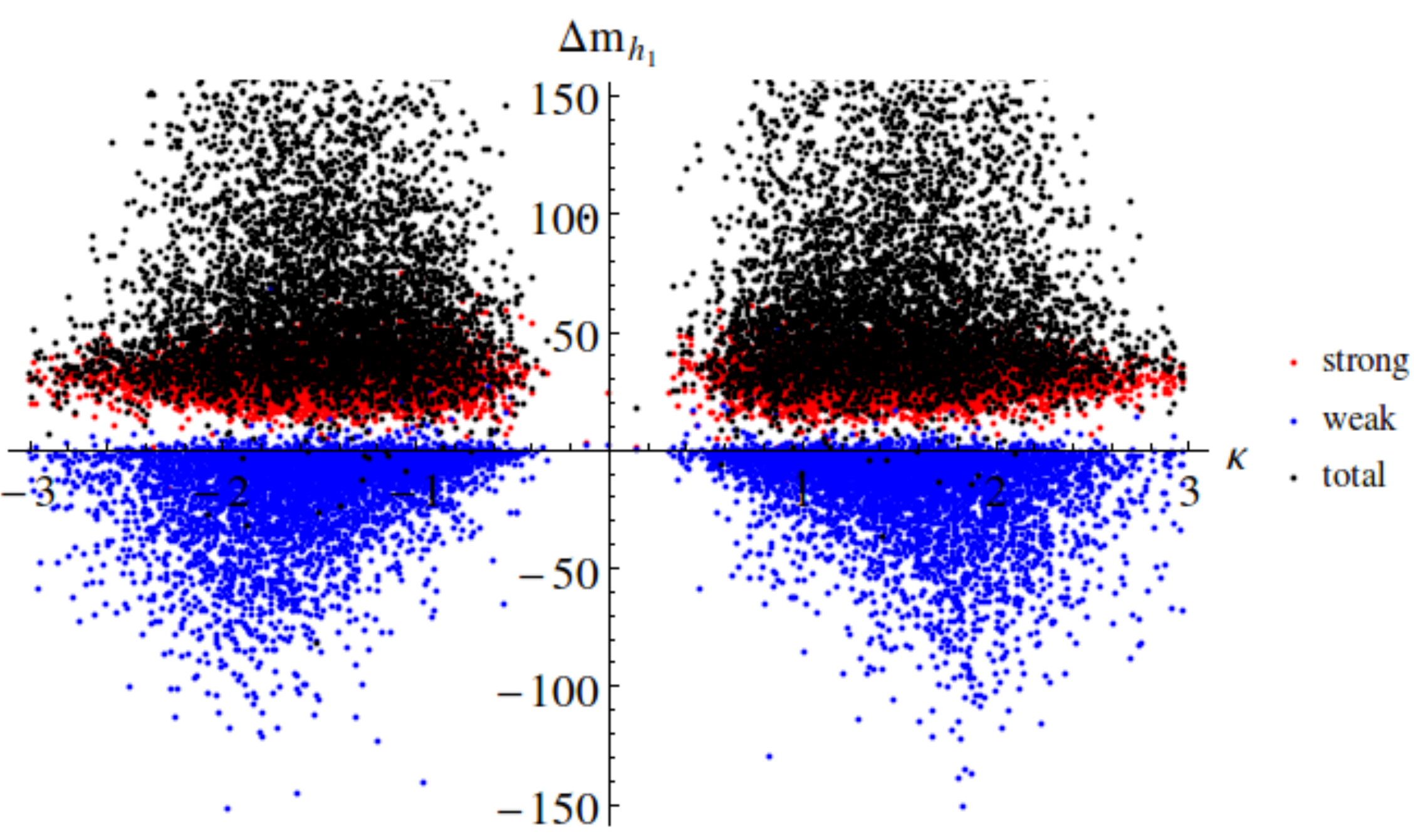}}
}

\caption{The radiative corrections at one-loop 
for $m_{h_1}$ vs (a) $\lambda_T$, (b) $\lambda_S$ and (c) $\kappa$. The red points are
only with strong (top-stop, bottom-sbottom) corrections, blue points are with weak corrections without the Higgs bosons (higgsinos, gauge boson and gauginos), (c) black points are the total (strong +weak + Higgs bosons) corrections. }\label{dmh1vsp}
\end{center}
\end{figure}
%%%%%%%%%%%%%%%%%%%%%%%%%%%%%%%%%%%%%%

Figure~\ref{dmh1vsp} shows the radiative corrections to $m_{h_1}$ as $\Delta m_{h_{1}}=m^{1-\rm{loop}}_{h_1} - m^{\rm{tree}}_{h_1}$, plotted against (a) $\lambda_T$, (b) $\lambda_S$ and (c) $\kappa$ respectively. The red points show the corrections to $m_{h_1}$ from the strong sector, due to the contributions generated by top-stop and bottom-sbottom running in the loops.
The blue points include the corrections from the weak sector with gauge bosons, gaugino and higgsino, and the black points take into account the total corrections which include strong, weak and the contributions from the Higgs sector. As one can deduce from the plots, the corrections (top-stop, bottom-sbottom) coming from the strong interactions are independent of the triplet and singlet Higgs couplings, as expected, with a maximum split of 50 GeV respect to the tree-level mass eigenvalue.\\
In the triplet-singlet extension we have four CP-even, three CP-odd neutral Higgs bosons and three charged Higgs bosons as shown in Eq.~(\ref{hspc}). These enhance both the Higgs and higgsino contributions to the radiative correction. The weak corrections (blue points) are dominated by the large number of higgsinos which contribute negatively to the mass and tend to increase for large values of the Higgs couplings ($\lambda_{T,S}$ and $\kappa$).\\
Finally, the black points show the sum of all the sectors, which are positive in sign, due to the large number of scalars contributing in the loop, with an extra factor of two for the charged Higgs bosons. This factor of two originates from the CW expression of the potential, and accounts for their multiplicity $(\pm)$. Such scalar contributions increase with the values of the corresponding couplings $\lambda_T, \lambda_S, \kappa$. 
From Figure~\ref{dmh1vsp} one can immeditaley notice that the electroweak radiative corrections could be sufficient in order to fulfill the requirement of a $\sim 125$ GeV Higgs mass, without any contribution from the strong sector.

%%%%%%%%%%%%mh1 at one loop vs mstp1 and tan beta %%%%%%
\begin{figure}[t]
\begin{center}
\mbox{\hskip -15 pt
\subfigure[]{\includegraphics[width=0.55\linewidth]{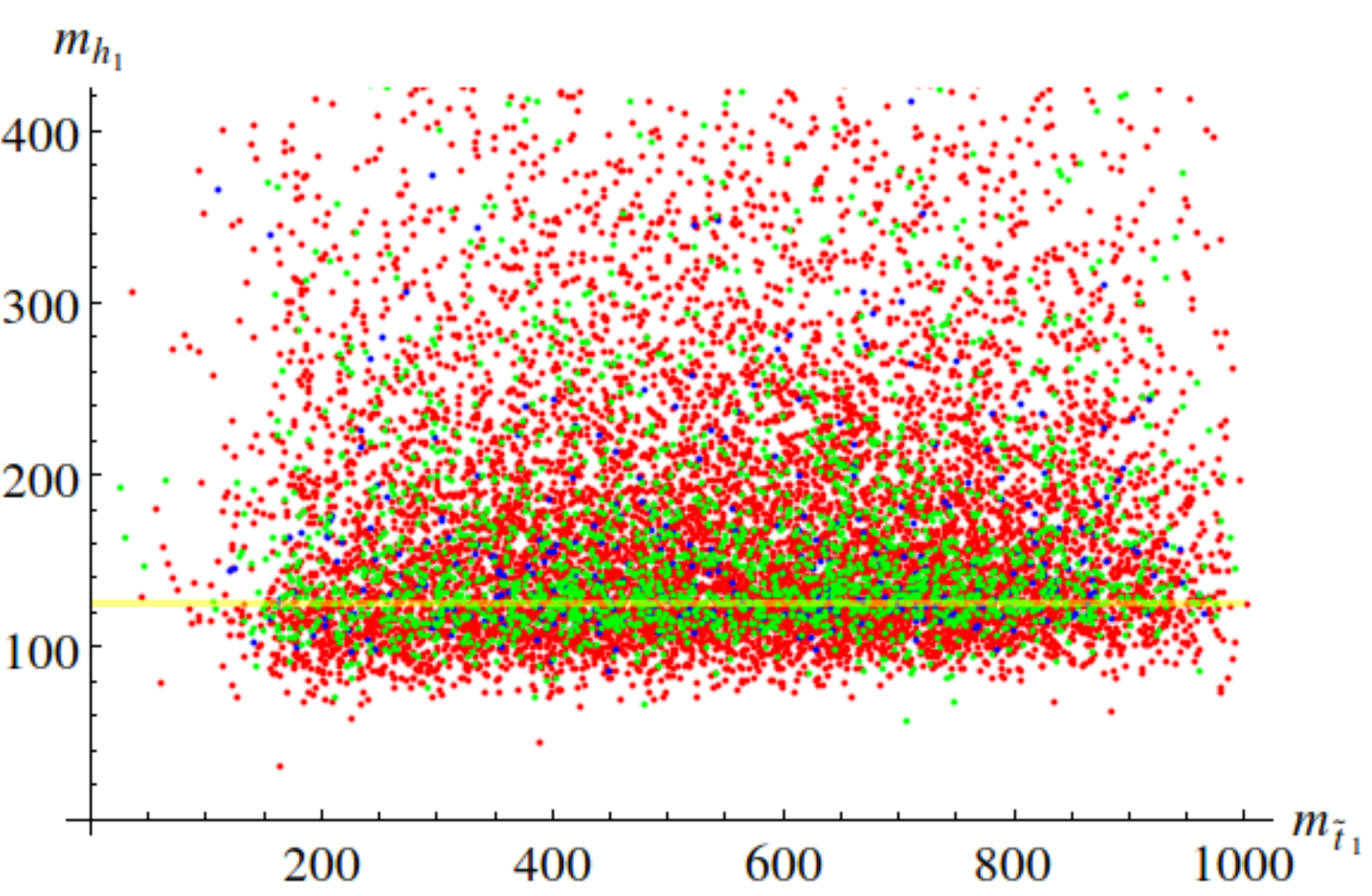}}
%\hskip 25 pt
\subfigure[]{\includegraphics[width=0.55\linewidth]{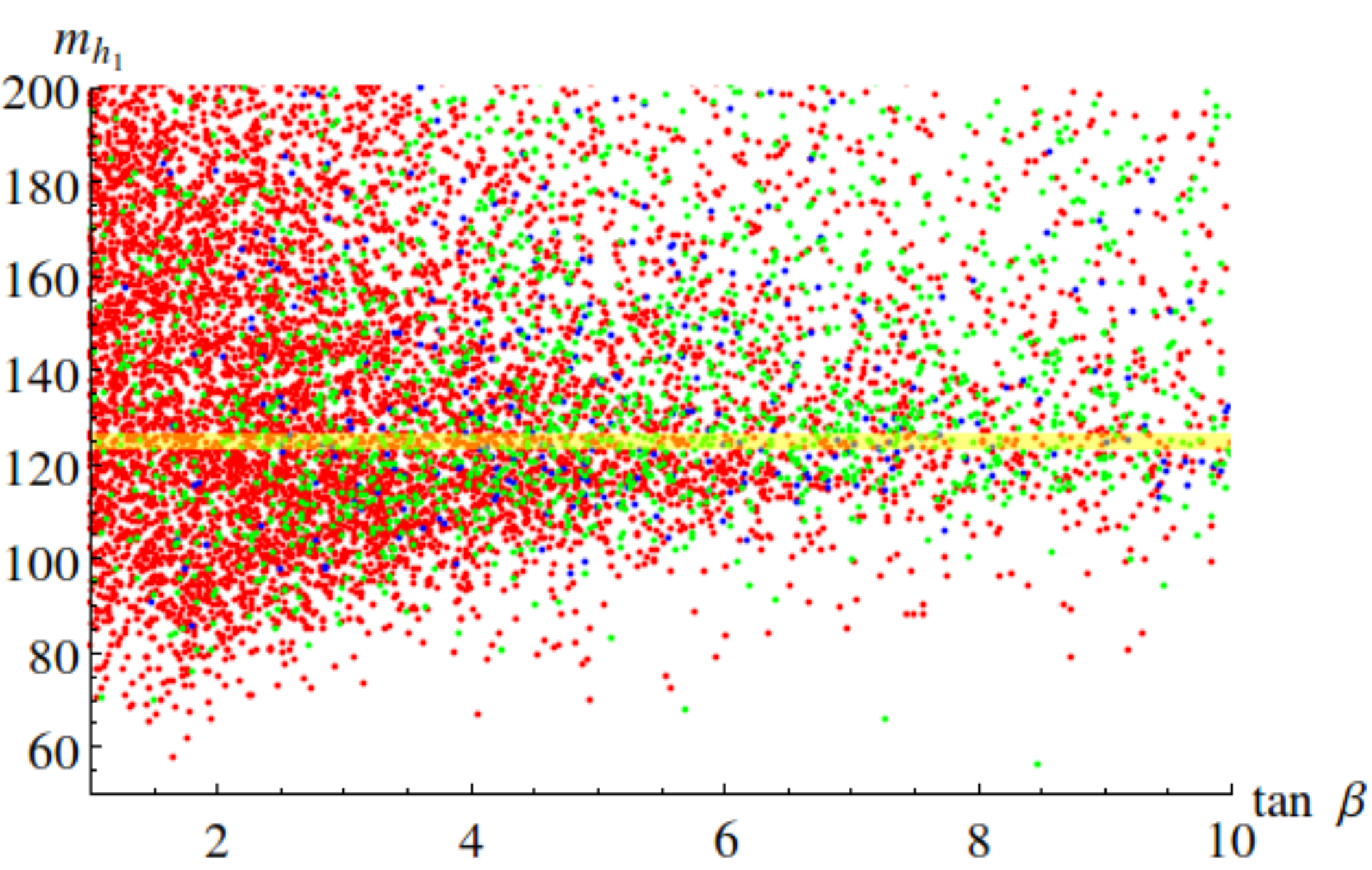}}}

\caption{The variation of  the one-loop lightest CP-even Higgs mass $m_{h_1}$ with (a) the lightest stop mass $m_{\tilde{t}_1}$, and (b) with $\tan{\beta}$, respectively. The yellow band shows the candidate Higgs mass
$123\leq m_{h_1} \leq 127$ GeV. }\label{mh1lvsp}
\end{center}
\end{figure}
%%%%%%%%%%%%%%%%%%%%%%%%%%%%%%%%%%%%%%%%%%%%%%%%%%%

To illustrate this point, in Figure~\ref{mh1lvsp}(a) we have plotted the lightest CP-even neutral Higgs mass at one-loop versus the lighter stop mass ($m_{\tilde{t}_1}$). We have used the same color coding conventions of the tree-level analysis. The red points are mostly doublets ($\geq 90\%$), the green points are mostly triplet/singlet($\geq 90\%$) and blue points are mixed ones, as explained in section~\ref{treel}.  The yellow band shows the Higgs mass range $123\leq m_{h_1} \leq 127$ GeV. 
We notice that a $\sim 125$ GeV CP-even neutral Higgs could be achieved by requiring a stop of very 
low mass, as low as 100 GeV. This is due to the presence of additional tree-level and radiative corrections from the Higgs sectors.  Thus, in the case of extended SUSY scenarios like the TNMSSM, 
the discovery of a $\sim 125$ GeV Higgs boson does not put a stringent lower bound on the required SUSY mass scale, and one needs to rely on direct SUSY searches for that.

In Figure~\ref{mh1lvsp}(b) we present the dependency of the one-loop corrected Higgs mass of the lightest CP-even neutral Higgs on $\tan{\beta}$. The distribution of points is clearly concentrated at low values of $\tan{\beta} \lsim 4$. This is due to the additional contributions on the tree-level Higgs masses, which are maximal in the same region of $\tan{\beta}$ (see Eq.~(\ref{hbnd})). It is then clear that an extended Higgs sector reduces the amount of fine-tuning  \cite{tssmyzero} needed in order to reproduce the mass of the discovered Higgs boson, compared to constrained supersymmetric scenarios. The latter, in general, require much larger supersymmetric mass scales beyond the few TeV \cite{cMSSMb} region. Compared to the pMSSM, this also represents an improvement, as it does not require large mixings in the stop masses in order to have the lighter stop mass below a TeV \cite{pMSSMb}.
%%%%%%%%%%%%%%%%%%%%% Appendix %%%%%%%%%%%%
%%%%%%%%%%%%mh1 -mh2  and ma1-mh2 at one loop vs mstp1 and tan beta %%%%%%
\begin{figure}
\begin{center}
\mbox{\hskip -15 pt
\subfigure[]{\includegraphics[width=0.55\linewidth]{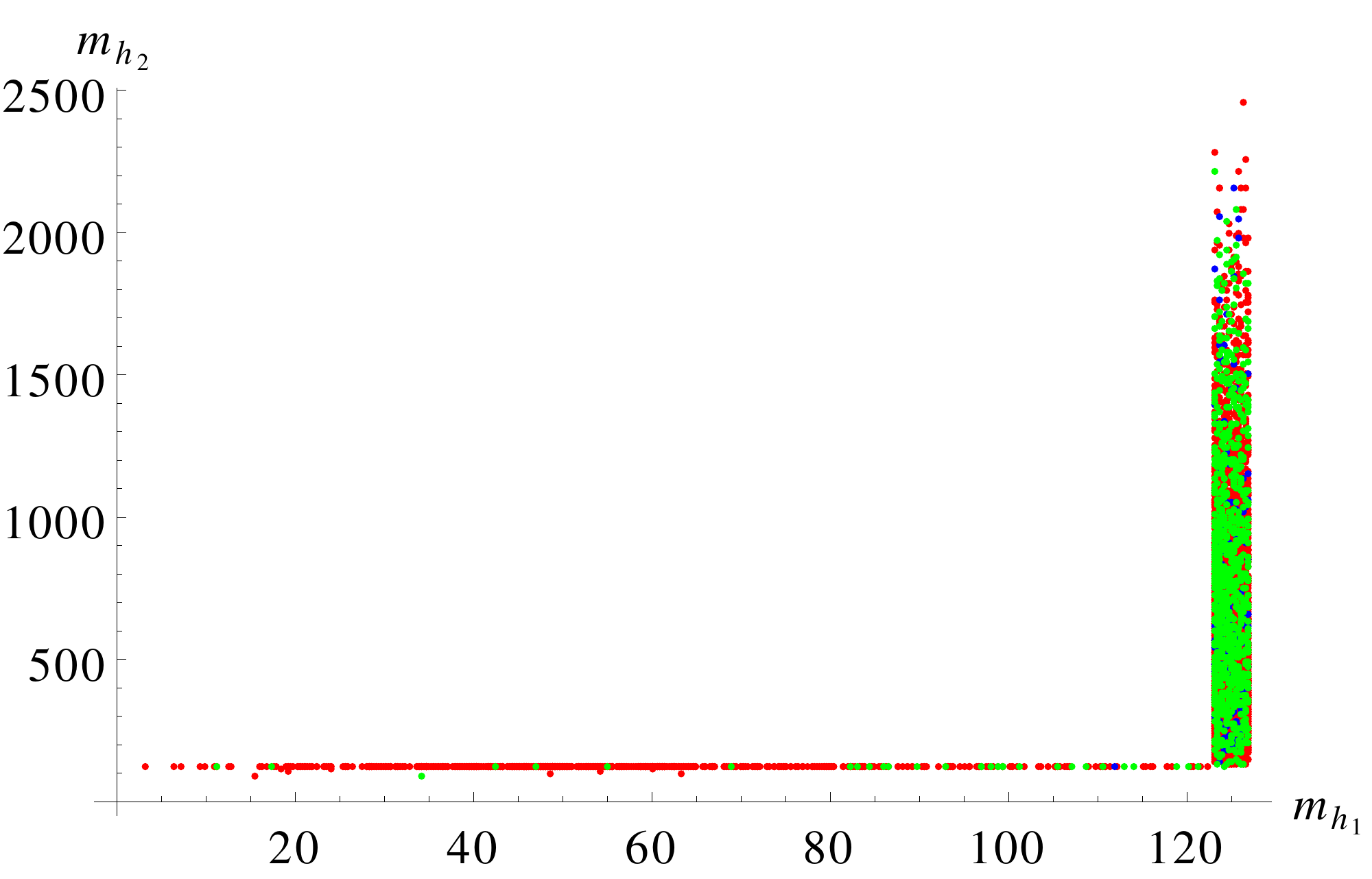}}
%\hskip 25 pt
\subfigure[]{\includegraphics[width=0.55\linewidth]{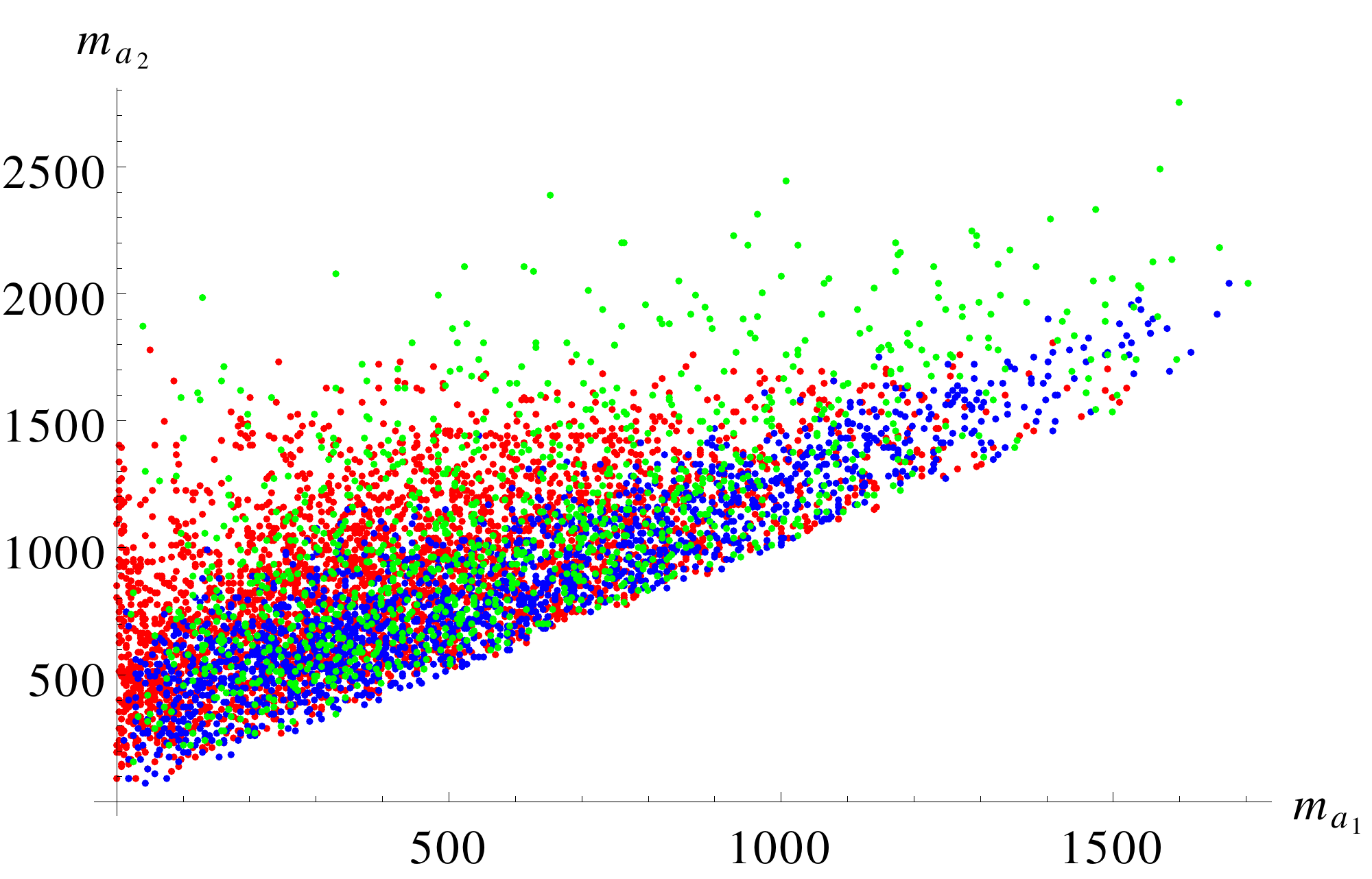}}}
\caption{The mass correlations at one-loop (a) $m_{h_1}-m_{h_2}$ and (b) $m_{a_1}-m_{a_2}$
where we have a CP-even candidate Higgs in the mass range $123\leq m_{h_i} \leq 127$ GeV.  Red, blue and green points are defined as in Figure 2. }\label{mhimhj}
\end{center}
\end{figure}
%%%%%%%%%%%%%%%%%%%%%%%%%%%%%%%%%%%%%%%%%%%%%%%%%%%
\subsection{Hidden Higgs bosons}
Next we investigate the case in which we have one or more hidden Higgs bosons, lighter in mass than 125 GeV,  scalars and/or pseudoscalars.  In Figure~\ref{mhimhj} we present the mass correlations at one-loop for (a) $m_{h_1}-m_{h_2}$ and (b) $m_{a_1}-m_{a_2}$, where we have a CP-even candidate Higgs boson in the mass range $123\leq m_{h_i} \leq 127$ GeV. The red points are mostly doublets ($\geq 90\%$), green points are mostly triplets/singlets ($\geq 90\%$) and blue points are mixed ones, as already explained. The green points have a high chance of evading the LEP bounds \cite{LEPb}, showing that the possibility 
of having a hidden scalar sector is realistic, even after taking into account the radiative corrections to the mass spectrum. A closer inspection of  Figure~\ref{mhimhj}(a) 
reveals that there are points where both $m_{h_1}$ and $m_{h_2}$ are less than $100$ GeV, showing that there is the possibility of having two CP-even hidden Higgs bosons. In that case $h_3$ is the candidate Higgs of $\sim 125$ GeV. Similarly, Figure~\ref{mhimhj}(b) shows the possibility of having two hidden pseduoscalars. The arguments mentioned in section~\ref{treel} will apply to the Higgs masses at one-loop as well. These imply that for $m_{h_1/a_1}\leq 123$ GeV, the green points  could evade the bounds from LEP and LHC, the red points would be ruled out and the blue points need to be carefully confronted with the data. In section~\ref{Hdata} we analyse such scenarios in detail. The lightest pseudoscalar present in the spectrum, as we are going to discuss below, can play a significant role in cosmology. In fact, it is crucial 
in enhancing the dark matter annihilation cross-section, which is needed in order to get the correct dark matter relic in the universe \cite{Arina:2014yna}.

\section{$\beta$-fuctions and the running of the couplings}\label{beta}
We have implemented the model in SARAH (version 4.5.5) \cite{sarah} in order to generate the vertices and the model files for CalcHep \cite{calchep}, and generated the $\beta$ functions for the dimensionless couplings and the other soft parameters. The $\beta$ functions for $\lambda_{T, S, TS}$, $\kappa, g_Y, g_L, g_c, y_{t, b}$ are given in the appendix \ref{RGs}.

%%%%%%%%%%%beta functions %%%%%%%%%%%%%%%
\begin{figure}[t]
\begin{center}
\mbox{\subfigure[]{\hskip -15 pt
\includegraphics[width=0.55\linewidth]{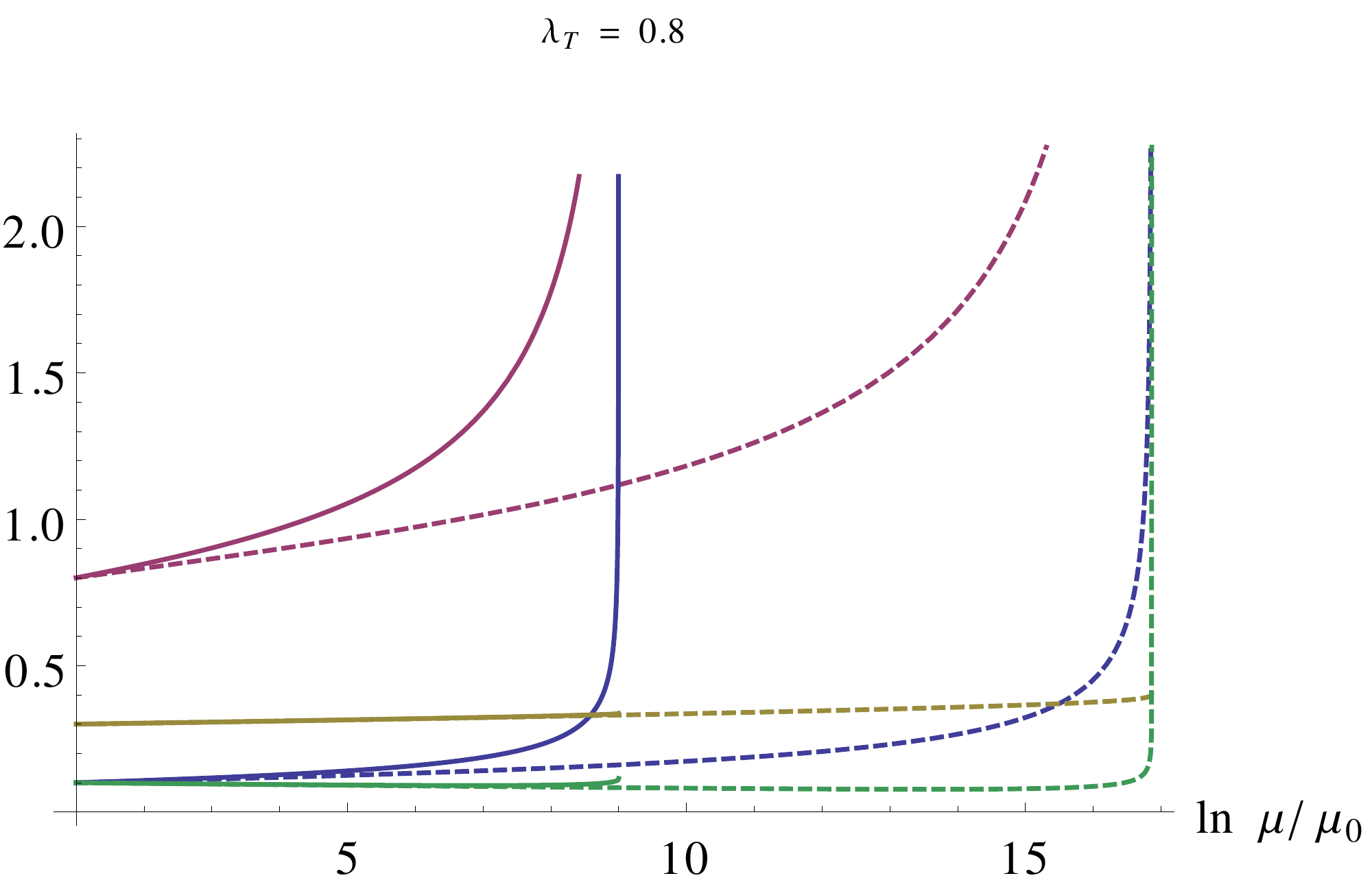}}
%\hskip 25 pt
\subfigure[]{\includegraphics[width=0.55\linewidth]{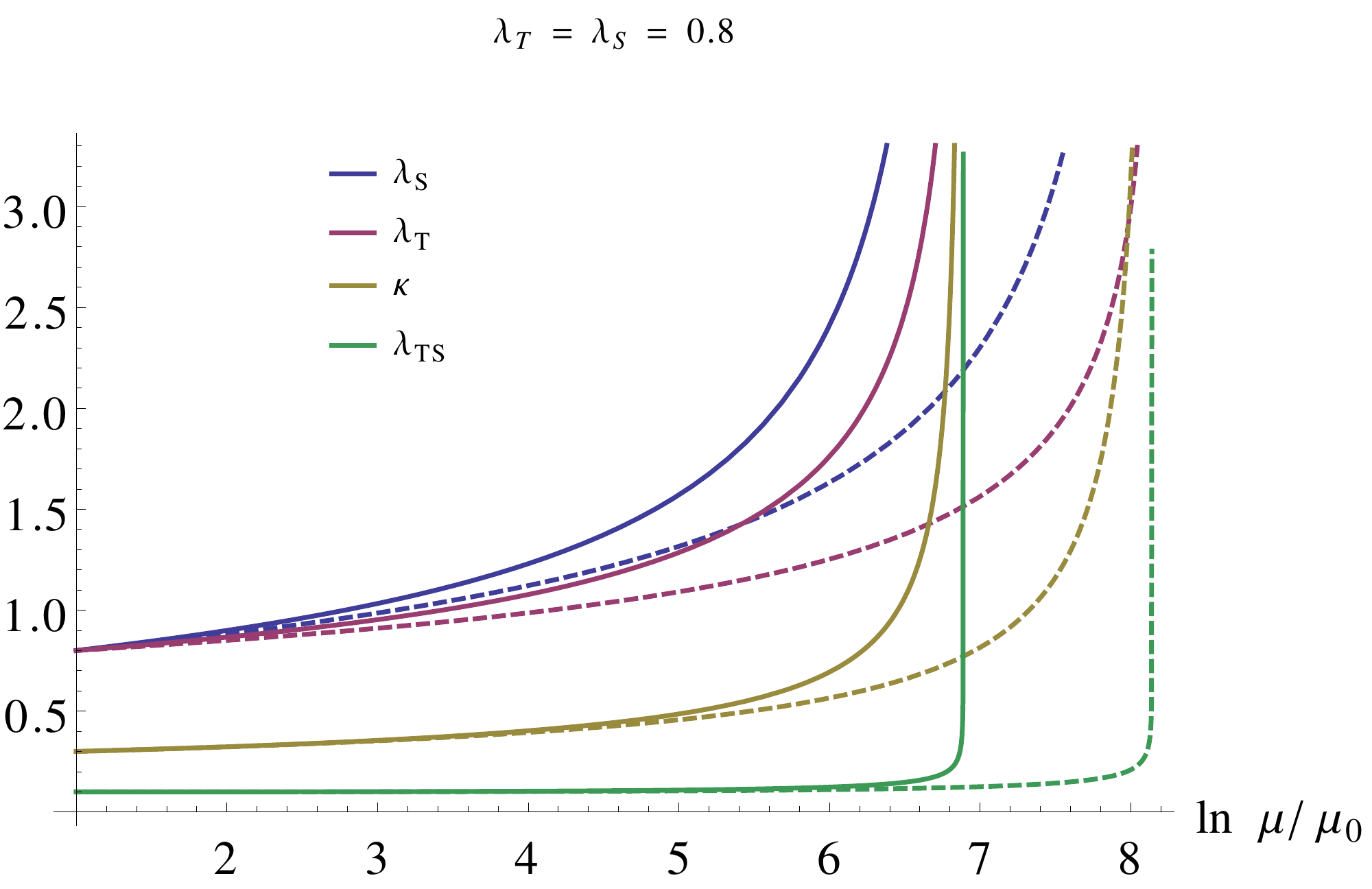}}}
\mbox{\subfigure[]{\hskip -15 pt
\includegraphics[width=0.55\linewidth]{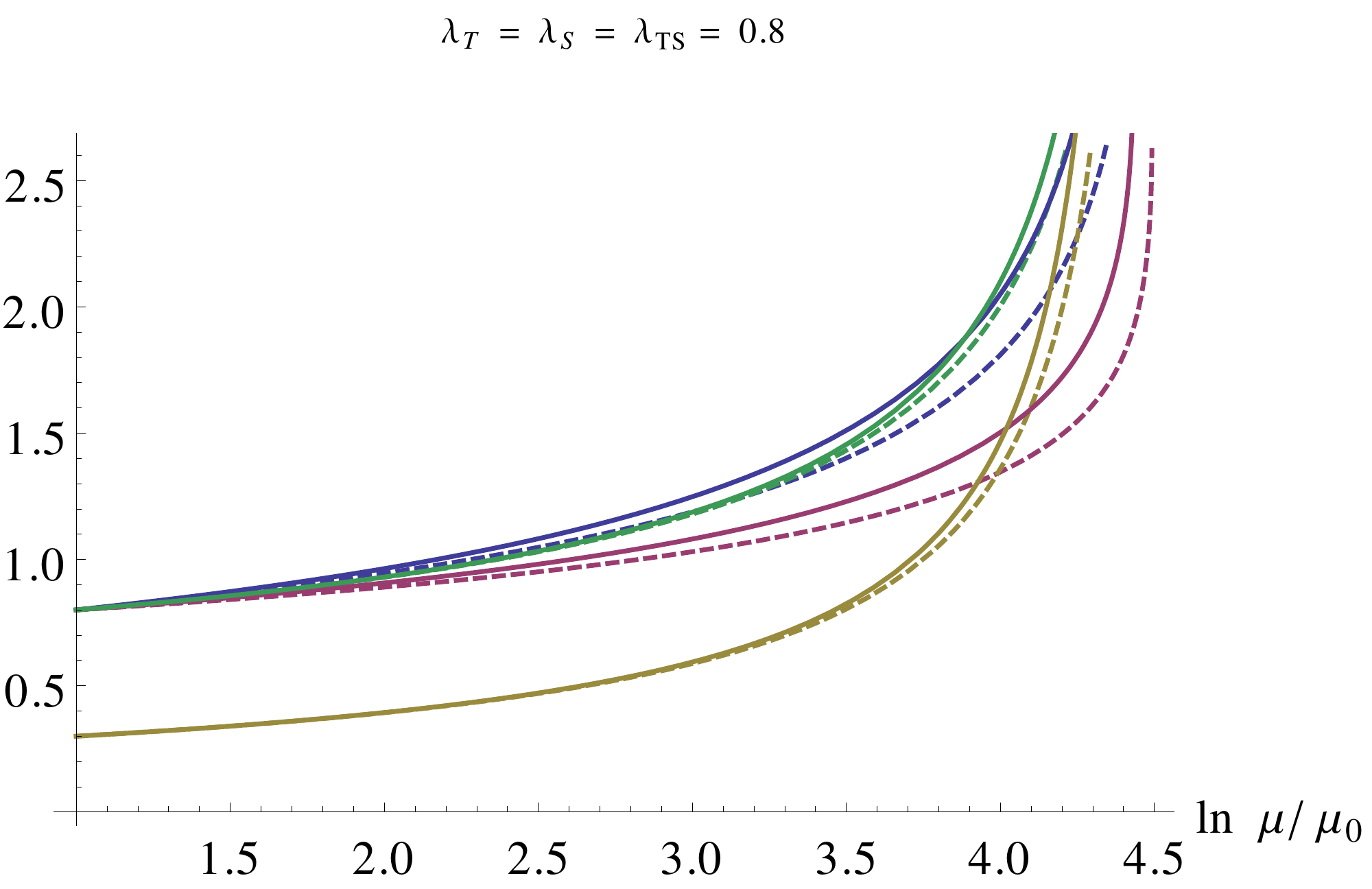}}
%\hskip 25 pt
\subfigure[]{\includegraphics[width=0.55\linewidth]{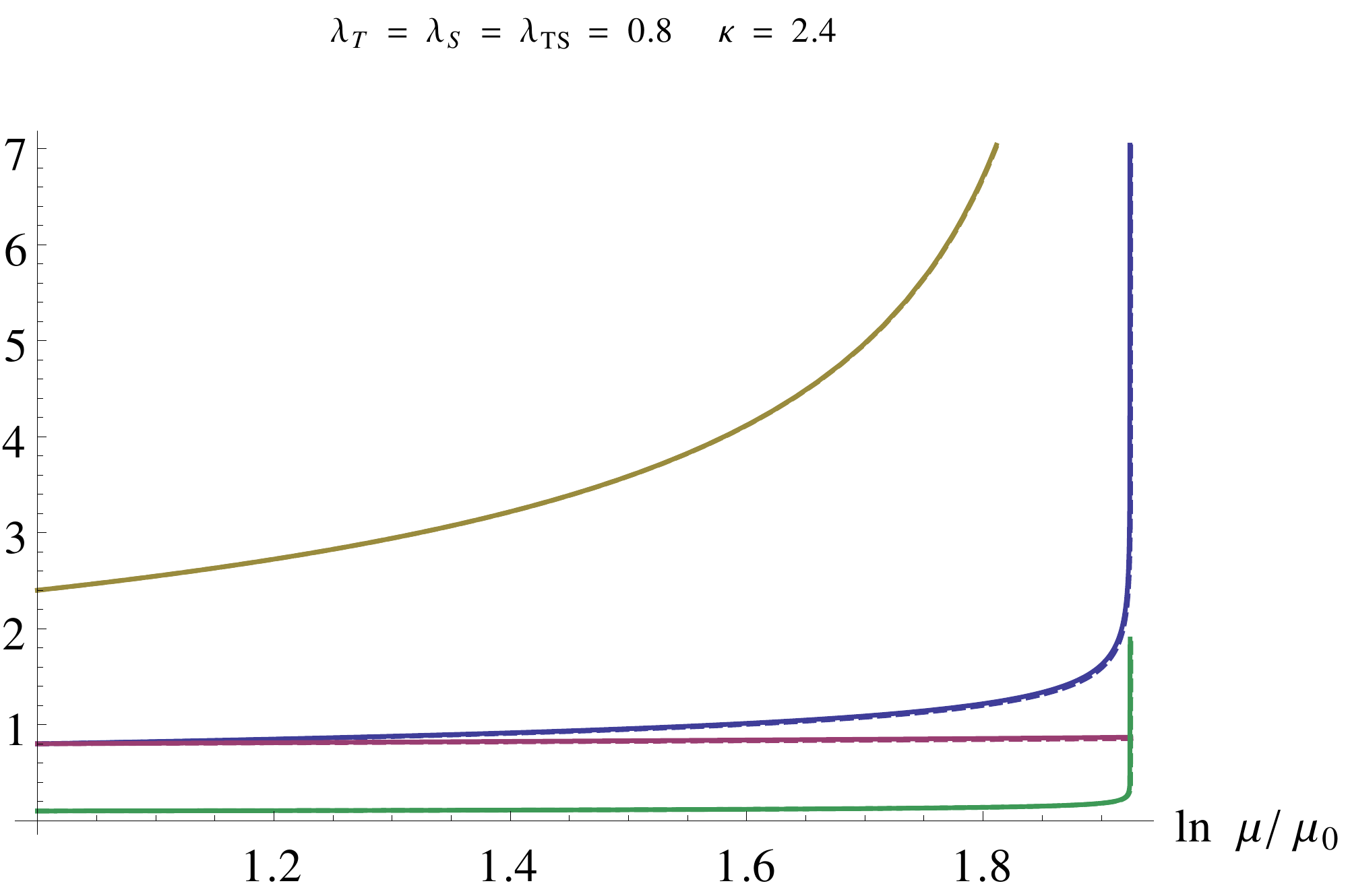}}}
\caption{The running of the dimensionless Higgs couplings $\lambda_{T, S, TS}$
and $\kappa $ with the log of the ratio the scales ($\ln{\mu/\mu_0}$) for $\tan{\beta}=1.5$ (solid lines) and $\tan{\beta}=10$ (dashed lines), where $\mu_0=M_Z$.  We have checked the corresponding variations for (a)$\lambda_{T}=0.8$,  (b)$\lambda_{T, S}=0.8$,  (c)$\lambda_{T, S, TS}=0.8$ and  (d)$\lambda_{T, S,}=0.8, \kappa=2.4$  chosen at scale $\mu_0$  respectively.}\label{rgv}
\end{center}
\end{figure}
%%%%%%%%%%%%%%%%%%%%%%

To analyse the perturbativity of the couplings we have selected four different scenarios and identified the cut-off scale ($\Lambda$) in  the renormalization group evolution, where one of the coupling hits the Landau pole and becomes non-perturbative ($\lambda_i(\Lambda)=4\pi$). Figure~\ref{rgv}(a) presents a mostly-triplet scenario at the electroweak scale as we choose $\lambda_T=0.8$, $\lambda_{S,TS}=0.1, \kappa=0.3$ at the scale $\mu_0=M_Z$, for $\tan{\beta}=1.5$ (solid lines) and $\tan{\beta}=10$ (dashed lines). For lower values of $\tan{\beta}$ ($\tan{\beta}=1.5$) the triplet coupling $\lambda_T$ becomes non-perturbative already at scale of $\Lambda \sim 10^{9-10}$ GeV, similarly to the behaviour shown in the triplet-extended MSSM \cite{FileviezPerez:2012gg, nardini}. For larger values of $\tan{\beta}$ ($\tan{\beta}=10$) all the couplings remain perturbative up to the (Grand Unification) GUT scale ($\Lambda \sim 10^{16}$ GeV).

Figure~\ref{rgv}(b) presents the case where $\lambda_{T, S}=0.8$ at $\mu_0=M_Z$.  We see that although the  $\tan{\beta}$ dependency becomes less pronounced, the theory becomes non-perturbative at a relatively lower scale $\Lambda \sim 10^{8}$ GeV.

From Figure~\ref{rgv}(c) it is evident that on top of $\lambda_T$ and $\lambda_S$ if we also choose $\lambda_{TS}=0.8$ at $\mu_0=M_Z$, the $\tan{\beta}$ dependency almost disappears. In this case the 
theory becomes more constrained with a cut-off scale $\Lambda \sim 10^{6}$ GeV. 

Finally, Figure~\ref{rgv}(d) illustrates the effect of a larger $\kappa$ value, the singlet self-coupling, with $\kappa=2.4$ at $\mu_0=M_Z$. The perturbative behaviour of the theory comes under question at a scale as low as $10^4$ GeV.  Such a large value of $\kappa$ at the electroweak scale thus restricts the upper scale of the theory to lay below 10 TeV, unless one extends the theory with an extra sector\footnote{For the scan in Eq.~\ref{scan} we select $|\kappa|\leq 3 $. The theoretical perturbativity of the parameter points have to be checked explicitly.}.  Choosing relatively lower values of $\lambda_{TS}$ and $\kappa$ would allow the theory to stay perturbative until $10^{8-10}$ GeV even with
$\lambda_{T, S}$ as large as $0.8$. The choice of larger values of $\lambda_{T, S}$ increases the tree-level contributions to the Higgs mass (see Eq.~(\ref{hbnd})) as well as the radiative corrections, via the additional Higgs bosons exchanged in the loops. Both of these contributions reduce the amount of supersymmetric fine-tuning, assuming a Higgs boson of $\sim 125$ GeV in the spectrum,  by a large amount, respect both to a normal and to a constrained MSSM scenario. Obviously, the addition of the triplet spoils the gauge coupling unification under the renormalization group evolution. This features is already evident in the triplet-extended MSSM \cite{FileviezPerez:2012gg, nardini}. 

\section{Fine-tuning}\label{finet}
The minimisation conditions given in Eq.~(\ref{mnc2}) relate the $Z$ boson mass to
the soft breaking parameters in the form  
\bea
M_Z^2&=&\mu^2_{\rm{soft}}-\mu_{\text{eff}}^2\\
\mu_{\text{eff}}&=&v_S \lambda_S-\frac{1}{\sqrt 2}v_T\lambda_T, \quad \mu^2_{\rm{soft}}=2\frac{m_{H_d}^2-\tan^2\beta\, m_{H_u}^2}{\tan^2\beta -1}.
\eea
It is also convenient to introduce the additional parameter
\bea\label{ft}
\mathcal{F}&=&\left|\ln\frac{\mu^2_{\rm{soft}}-\mu_{\text{eff}}^2}{\mu^2_{\text{soft}}}\right|,
\eea
characterizing the ratio between $M^2_Z$ and $\mu^2_{\text{soft}}$, which can be considered a measure of the fine-tuning. Unlike the MSSM, here the $\mu_{\text{eff}}$ parameter is generated spontaneously by the singlet and triplet vevs. Notice that while the triplet contribution is bounded by the $\rho$ parameter \cite{rho}, the singlet vev is unbounded and it may drive $\mu_{\text{eff}}$ to a large value. Similarly, the 
soft parameters $m_{H_u, H_d}$, which are determined by the minimisation condition (\ref{mnc2}), can be very large, and thus can make $\mu^2_{\text{soft}}$ also large. Finally, to reproduce the $Z$ boson mass we need large cancellations between these terms, which leads to the well know fine-tuning problem of the MSSM and of other supersymmetric scenarios. 

%%%%%%%%%%% Fine-tuning at tree-level and at one-loop%%%%%%%%%%%%%%%
\begin{figure}[bht]
\begin{center}
\mbox{\subfigure[]{\hskip -15 pt
\includegraphics[width=0.55\linewidth]{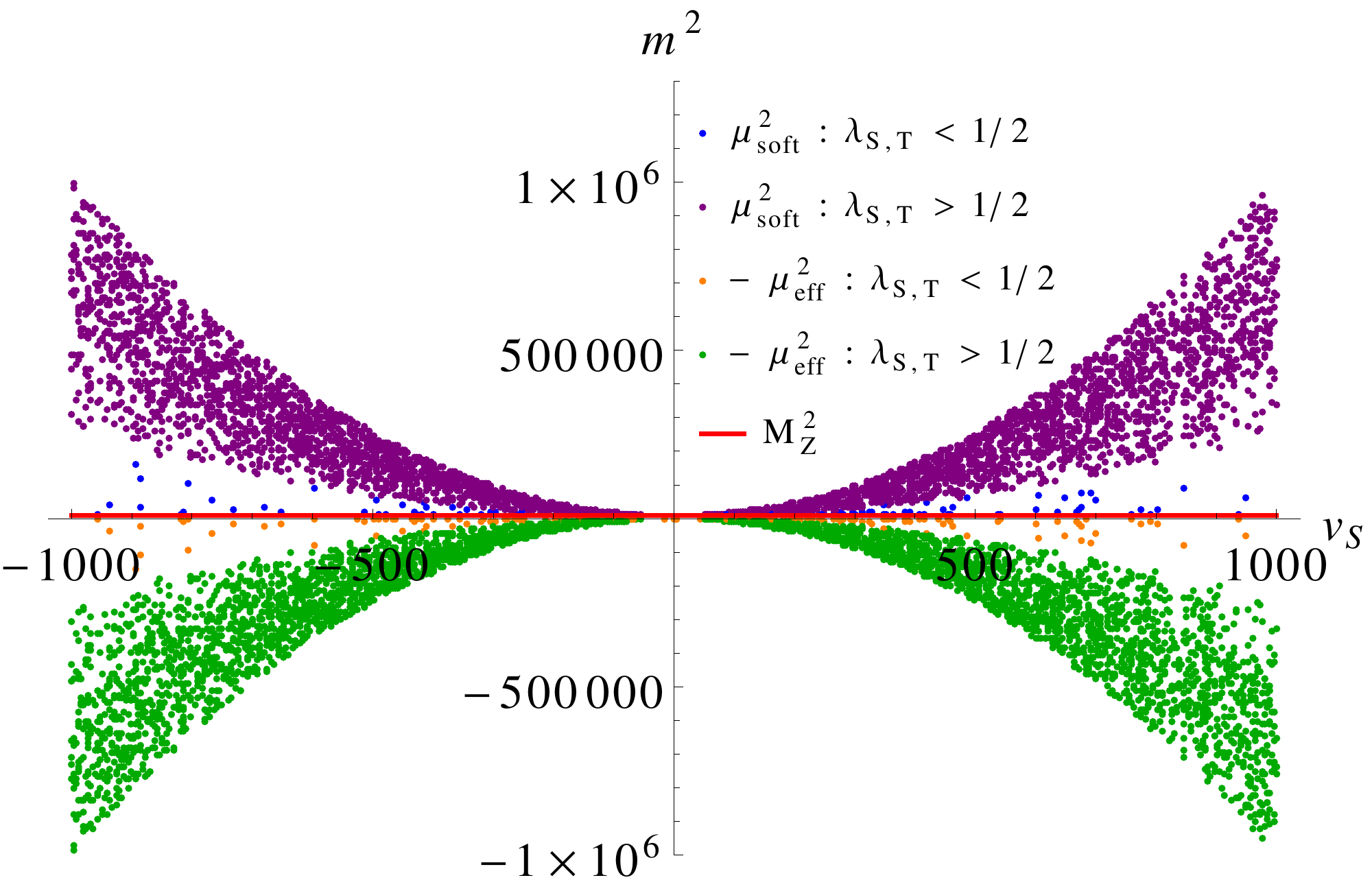}}
%\hskip 15 pt
\subfigure[]{\includegraphics[width=0.55\linewidth]{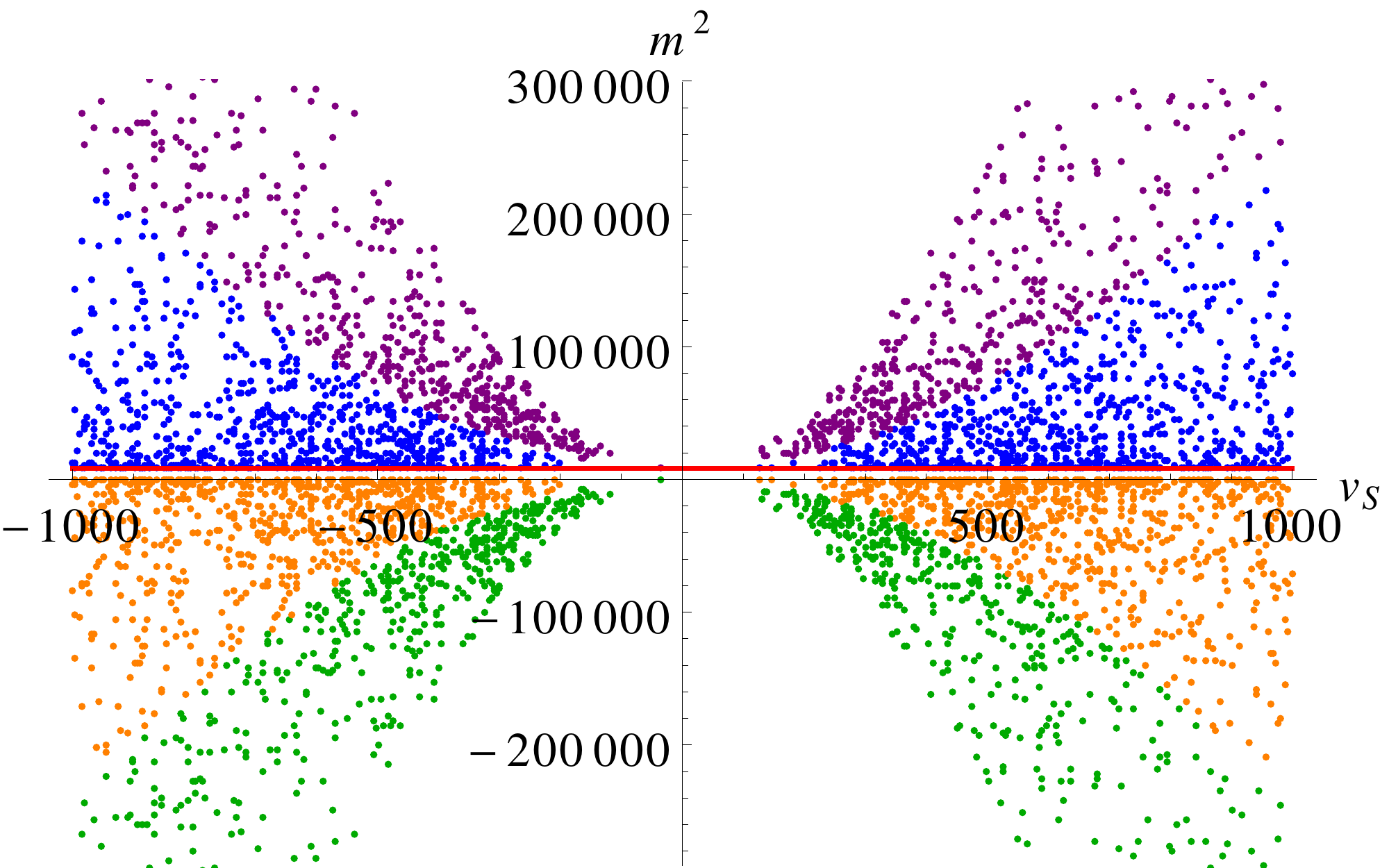}}}
\caption{The (a)tree-level and (b) one-loop level fine-tuning measures $\mu_{\text{soft}}$ and  $-\mu^2_{\text{eff}}$ versus the singlet vev $v_S$ for a candidate Higgs of mass between $120\leq m_{h_1} \leq 130$ GeV respectively. The violet points represent $\mu^2_{\rm{soft}}$ values $\lambda_{S,T}\geq 0.5$ and blue points represent $\mu^2_{\rm{soft}}$ values $\lambda_{S,T}<0.5$. The green points represent $\mu^2_{\rm{eff}}$ values $\lambda_{S,T}\geq 0.5$ and the orange points represent $\mu^2_{\rm{eff}}$ values $\lambda_{S,T}< 0.5$. The red line shows the $Z$ boson mass $M_Z$.}\label{fnt}
\end{center}
\end{figure}
%%%%%%%%%%%%%%%%%%%%%%
We show in Figure~\ref{fnt}(a) plots of $\mu^2_{\text{soft}}$ and  $-\mu^2_{\text{eff}}$ versus the singlet vev $v_S$ for tree-level 
candidate Higgs masses in the interval $120\leq m_{h_1} \leq 130$ GeV. Figure~\ref{fnt}(b) presents the same plots, but with $m_{h_1}$, the candidate Higgs mass, calculated at one-loop. The violet points represent $\mu^2_{\rm{soft}}$ values for which $\lambda_{S,T}\geq 0.5$, and the points in blue refer to values of $\mu^2_{\rm{soft}}$ with $\lambda_{S,T}<0.5$. The green points mark values of $\mu^2_{\rm{eff}}$ with $\lambda_{S,T}\geq 0.5$, and the orange points refer to $\mu^2_{\rm{eff}}$ values with $\lambda_{S,T}< 0.5$. We see that for low $\lambda_{T,S}$ both $\mu^2_{\text{soft}}$ and  $-\mu^2_{\text{eff}}$  (blue and orange points) are small, so that the required cancellation needed in order to reproduce the $Z$ boson mass is also small. This leads to less fine-tuning, measured by $\mathcal{F}< 1$. Unfortunately, such points are small in numbers in the tree-level case,
since they require the extra contributions from the triplet and the singlet in order to reproduce the $\sim 125$ GeV Higgs mass. For  $\lambda_{T,S}\geq 0.5$ both $\mu^2_{\text{soft}}$ and  $-\mu^2_{\text{eff}}$  (the violet and green points) are both very large, leading to large cancellations and thus to a fine-tuning parameter $\mathcal{F}\sim 5$ for $\mu^2_{\text{soft}},-\mu^2_{\text{eff}}  \sim 10^6$. \\
Comparing Figure~\ref{fnt}(a) and Figure~\ref{fnt}(b) we see that the tree-level Higgs mass needs more fine-tuning as $\mu^2_{\text{soft}, \text{eff}} \sim 10^6$  for large $\lambda_{T, S}$. The situation improves significantly at one-loop due to the contributions from the radiative corrections. This is due to the fact that there are more solutions with low values of $\lambda_T$, $\lambda_{T, S}<0.5$, compared to tree-level and, on top of this,  (for high and low $\lambda_{T,S}$) the required fine-tuning is reduced ($\mathcal{F}\lesssim 2$). This fine-tuning measure is a theoretical estimation but it is constrained from the lightest chargino mass bound from LEP ($m_{\tilde{\chi}^\pm_1}> 104$ GeV), which results in $\mu_{\text{eff}}>104$ GeV and $\mathcal{F}\gsim 0.2$.

We have performed a run of $m^2_{H_u, H_d}$ using the corresponding $\beta$-functions for large $\lambda_{T,S}$, from electroweak scale ($M_Z$) up to a high-energy scale $\sim 10^{9,10}$ GeV, where the couplings become non-perturbative. It can be shown that $m^2_{H_u, H_d}, \mu^2_{\text{soft}}$  do not blow up unless the couplings $\lambda_{T,S}$ hit the Landau pole. The requirement of perturbativity of the evolution gives stronger bounds on the range of validity of the theory and the fine-tuning parameter is a good indicator at the electroweak scale.

In the case of MSSM, the large effective quartic coupling 
comes from the storng SUSY sectors which also increase $m^2_{H_u}$ and other parameters. However the situation changes in the case of extended Higgs sectors,
which gives additional tree-level as well as quantum corrections to the Higgs masses.
These reduce the demand for larger $m^2_{H_u}$. In our case there is a singlet and a triplet which contribute largely at the tree-level for low $\tan\beta$ and also contribute at the quantum level. In the case of tree-level Higgs mass, the extra tree-level contributions demand very large $\lambda_{T,S} \sim 0.8$, which in turn make
$\mu_{\text{eff}}$ very large and so the fine-tuning. However in the case of Higgs mass at one-loop, the extra contributions from the extended Higgs sectors are shared by both tree-level and quantum corrections, which reduces the requirement of large $\lambda_{T,S}$. This reduces  $\mu_{\text{eff}}$  and so the fine-tuning $\mathcal{F}$.

\section{A light pseudoscalar in the spectrum}\label{axion}
In the limit when the $A_i$ parameters in Eq.~(\ref{softp}) go to zero, the  discrete $Z_3$ symmetry of the Lagrangian is promoted to  a continuos $U(1)$ symmetry given by Eq.~(\ref{csmy}). 
%%%%%%%%%%%Axion and light pseudo-scalar boson%%%%%%%%%%%%%%%
\begin{figure}
\begin{center}
\mbox{\subfigure[]{ \hskip -15 pt
\includegraphics[width=0.55\linewidth]{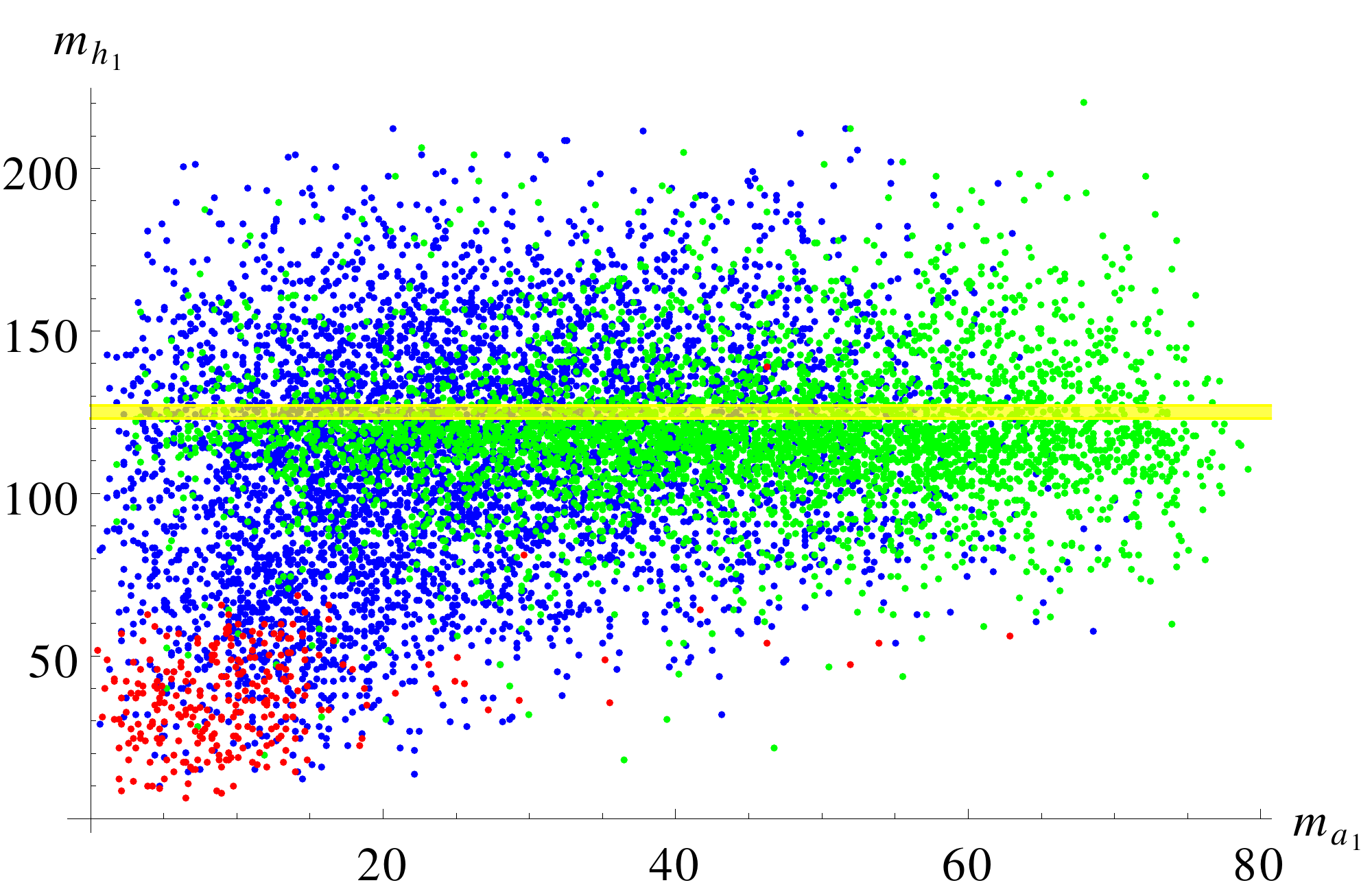}}
%\hskip 25 pt
\subfigure[]{\includegraphics[width=0.55\linewidth]{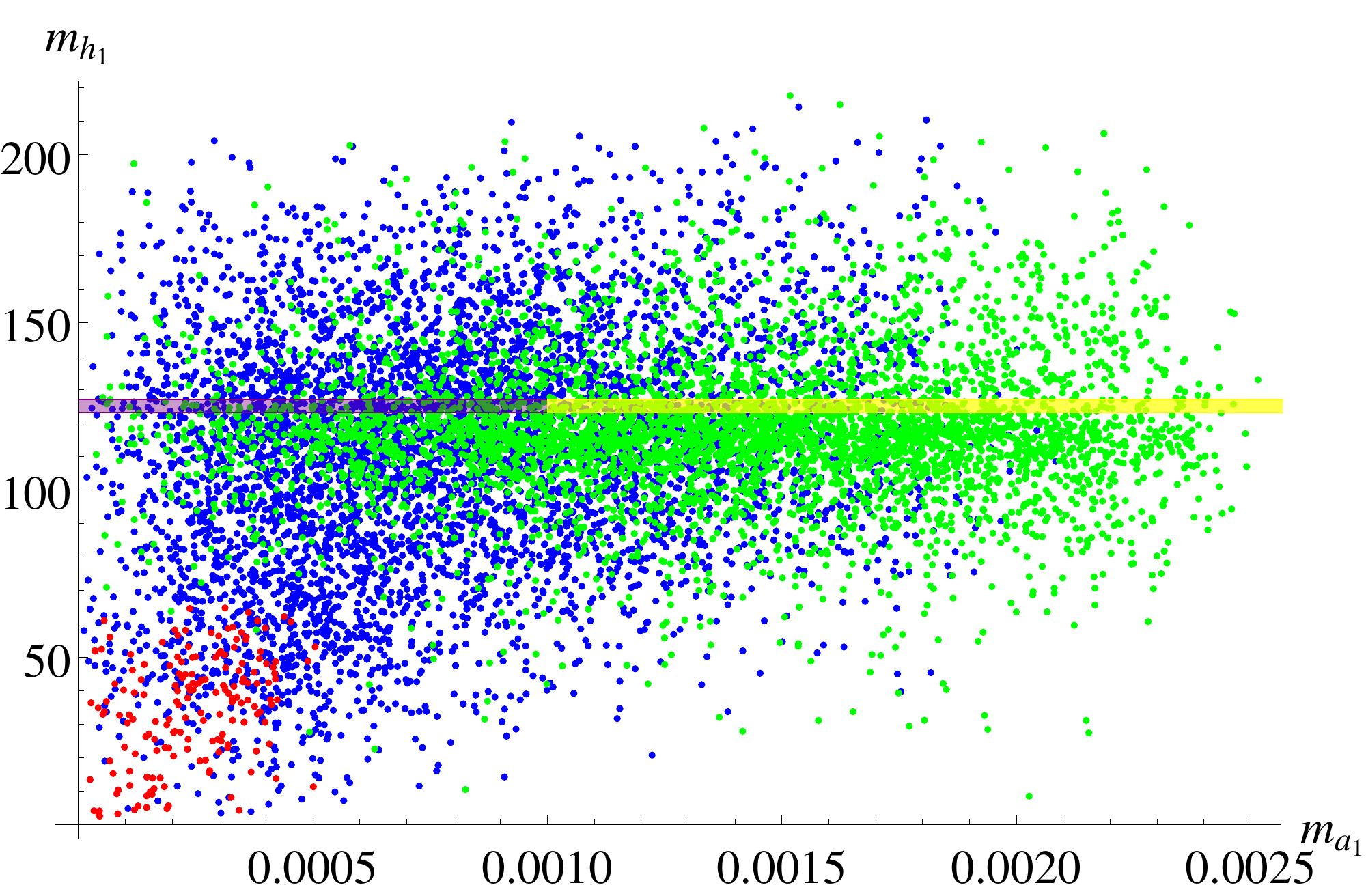}}}
\caption{The lightest CP-even Higgs boson mass $m_{h_1}$ vs the lightest pseudo-scalar mass $m_{a_1}$ at one-loop (top-stop and bottom-bottom corrections). The violet-yellow band presents the candidate Higgs mass $123\leq m_{h_1} \leq 127$ GeV.  The violet band specify the points with $m_{a_1}\leq 1$ MeV, where the $a_1 \to e^+e^-$ decay is kinematically forbidden. Red, blue and green points are defined as in Figure 2.}\label{mAA1}
\end{center}
\end{figure}
%%%%%%%%%%%%%%%%%%%%%%
This symmetry is spontaneously broken by the vevs of the doublets, triplet and the singlet fields and should contain a physical massless pseudoscalar, $a_1$,  the Nambu-Goldstone boson of the symmetry. The soft breaking parameters will then lift the mass of $a_1$, turning it into a pseudo-Goldstone mode whose mass will depend on the $A_i$. The symmetry takes the form 
\bea\label{csmy}
(\hat{H}_u,\hat{H}_d, \hat{T},\hat{ S}) \to e^{i\phi}(\hat{H}_u,\hat{H}_d, \hat{T},\hat{ S}) .
\eea 
If this symmetry is softly broken by very small parameters $A_i$ of $\mathcal{O} (1)$ GeV, we get a very light pseudoscalar
\cite{nmssm, Agashe:2011ia} which could be investigated at cosmological level. Notice that the vector-like nature of the symmetry decouples this pseudoscalar from any anomalous behaviour.
We are going to briefly investigate some features of the $a_1$ state in the context of the recent Higgs discovery and we will consider two different realizations.
In the first case we consider a scenario where such continuous symmetry is broken very softly. In this case we choose the $A_i$ parameters to be $\mathcal{O}(1)$ GeV. We expect the pseudo-Goldstone boson to be very light, with a mass $\mathcal{O}(1)$ GeV. In Figure~\ref{mAA1} we show the mass correlation between 
the lightest CP-even neutral Higgs boson $h_1$ and the lightest massive CP-odd neutral Higgs boson $a_1$. The red points are of doublet type, the green points represent massive states of triplet/singlet type and the blue points represent the mixed contributions to the $a_1$ pseudoscalar. The violet-yellow band presents the region of parameter space where $h_1$ is the  candidate Higgs, with a mass $123\leq m_{h_1} \leq 127$ GeV.  It is rather clear from Figure~\ref{mAA1}(a) that there plenty of points in the parameter space where there could be a hidden pseudoscalar Higgs boson along with or without a CP-even hidden scalar. Such a light pseudoscalar boson gets strong experimental bounds from LEP searches \cite{LEPb} and from the bottomonium decay rates \cite{bottomium}.  Such light pseudoscalar in the mass range of 5.5 -14 GeV, when it couples to fermions, gets strong bound  from the recent CMS data at the LHC \cite{cmsab}. For triplet/singlet green points these bounds can be evaded quite easily since these states do not couple to gauge bosons (the $Z$ boson in the case of a triplet) and to fermions. Of course, for real mass eigenstates the mixing between the doublet-triplet/singlet would be very crucial in the characterization of their allowed parameter space. 

For a mass of the $a_1$ of $\mathcal{O}(100)$ MeV, the decay to $\pi \gamma \gamma$, 
$\pi \pi \pi$ could be an interesting channel to investigate in order to search for this state \cite{infnr}. The simpler 2-particle decay channel $a_1\to \pi\pi$
is not allowed due to the CP conservation of the model. 
Due to the singlet/triplet mixing nature of this state, it decays into fermion pairs $e^+e^-, \mu \bar{\mu}, \tau\bar{\tau}$, if kinematically allowed. Notice that there is no discrete symmetry to protect this state from decaying, preventing it from being a dark matter candidate \cite{Mambrini:2011ri}. 
Now, if we choose the $A_i$ parameters to be of $\mathcal{O}(1)$ MeV then we get a very light pseudoscalar
boson with mass of the same order, as shown in Figure~\ref{mAA1}(b). Such a bosons cannot decay to $\mu \bar{\mu}, \tau\bar{\tau}$ kinematically. Following the same reasoning, if its mass is $<1$ MeV, then even the $a_1 \to e^+e^-$ channel is not allowed and only the photon channel remains open to its decay.
In this case the $a_1$ resembles an axion-like particle, and can be a dark matter candidate only if its lifetime is larger than the age of the universe \cite{infnr, Bernabei:2005ca}. The pseudoscalar, in this case, couples to photons at one-loop, due to its doublet component which causes the state to have a direct interaction with the fermions.\\
We recall that the effective lifetime of a light pseduoscalar decaying into two photons is given by Eq.~(\ref{agg}) \cite{Bernabei:2005ca}

\bea\label{agg}
\tau_a =\frac{64 \pi}{g^2_{a\gamma\gamma}m^3_{a}}
\eea
where $g_{a\gamma\gamma}$ is the effective pseduoscalar /fermion coupling which is proportional 
to the doublet-triplet/singlet mixing. Notice that the $a_1$ shares some of the behaviour of axion-like particles, which carry a mass that is unrelated to their typical decay constant, and as such are not described by a standard Peccei-Quinn construction. They find 
a consistent description in the context of extensions of the SM with extra anomalous abelian symmetries \cite{ga0} \cite{ga1} and carry a direct anomalous (contact) interaction to photons. Such interaction is absent in the case of a $a_1$ state. \\
Along with the lightest neutralino of the TNMSSM, this particle can be a dark matter candidate. In the supersymmetic context a similar scenario, with two dark matter candidates 
has been discussed in \cite{ga2}. The role of this pseudoscalar state, in the context of the recent results by FERMI about the 1-3 GeV excess gamma-ray signal from the galactic center \cite{Arina:2014yna} is under investigation for this model \cite{pb3}.  

\section{$\sim 125$ GeV Higgs and LHC data}\label{Hdata}
In this section we consider the one-loop Higgs mass spectrum, including only the correction coming from quarks and squarks, in light of recent results from the LHC \cite{CMS,CMS2, ATLAS} and the existing data from LEP \cite{LEPb}. In particular, we consider the uncertainties in the decay modes of the Higgs to $WW^*$, $ZZ^*$ and $\gamma\gamma$ in a conservative way \cite{CMS, ATLAS}. We explore the scenario where one of the CP-even neutral scalars is the candidate $\sim 125$ GeV Higgs boson within the mass range $123\leq m_{h_i}\leq 127$ GeV and investigate the possibilities of having one or more light  scalars, CP-even and/or CP-odd, allowed by the LEP data and consistent with the recent Higgs decay branching fractions at the LHC. 
 
We just mention that in the TNMSSM the triplet and the singlet type Higgs bosons do not couple to the $Z$ boson but the triplet couples to the $W^\pm$ bosons, which result in a modified $h_i\, W^\pm\,W^\mp$ vertices given by 
\bea
h_i\,W^\pm\,W^\mp = \frac{i}{2}\,g_L^2\left(v_u\mathcal{R}^S_{i1} +v_d\mathcal{R}^S_{i2} +4\,v_T\mathcal{R}^S_{i4}\right),
\eea
where the rotation matrix $R^S_{ij}$ is defined as $h_i= \mathcal{R}^S_{ij} H_j$. The vertices $h_i\,Z\,Z$ are given by
\bea
h_i\,Z\,Z = \frac{i}{2}\left(g_L\cos\theta_W+g_Y\sin\theta_W\right)^2\left(v_u\mathcal{R}^S_{i1} +v_d\mathcal{R}^S_{i2}\right),
\eea
where $\theta_W$ is the Weinberg angle. The Yukawa part of the superpotential is just the MSSM one. Hence the couplings of the CP-even sector to the up/down-type quarks and to the charged leptons are
\bea
h_i\,u \,\bar u = -\frac{i}{\sqrt2}y_u \mathcal{R}^S_{i1},\\
h_i\,d \,\bar d = -\frac{i}{\sqrt2}y_d \mathcal{R}^S_{i2},\\
h_i\,\ell \,\bar\ell = -\frac{i}{\sqrt2}y_\ell \mathcal{R}^S_{i2},
\eea
respectively. 

\begin{figure}
\begin{center}
\mbox{\subfigure[]{\hskip -15 pt
\includegraphics[width=0.55\linewidth]{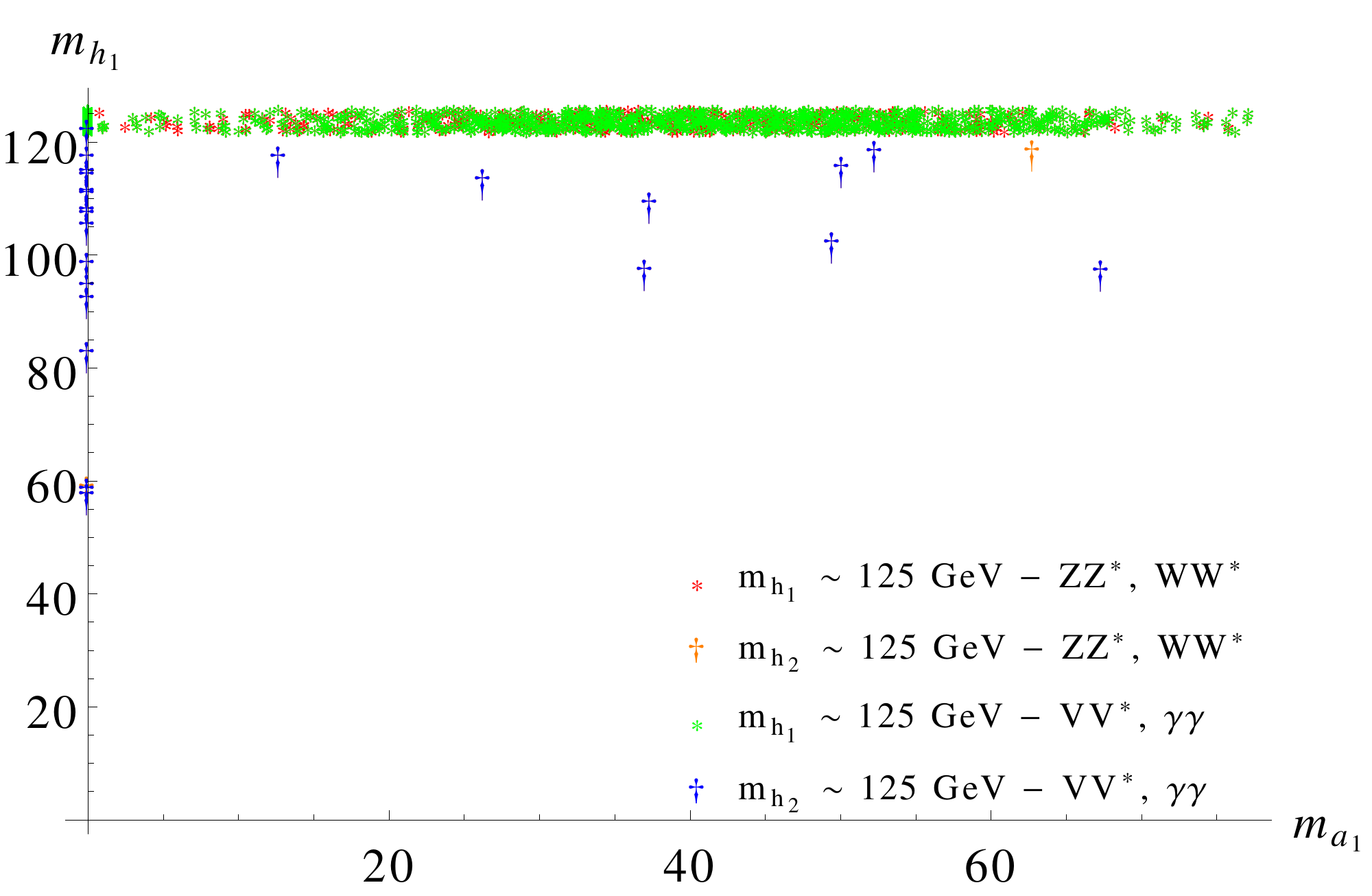}}
%\hskip 25 pt
\subfigure[]{\includegraphics[width=0.55\linewidth]{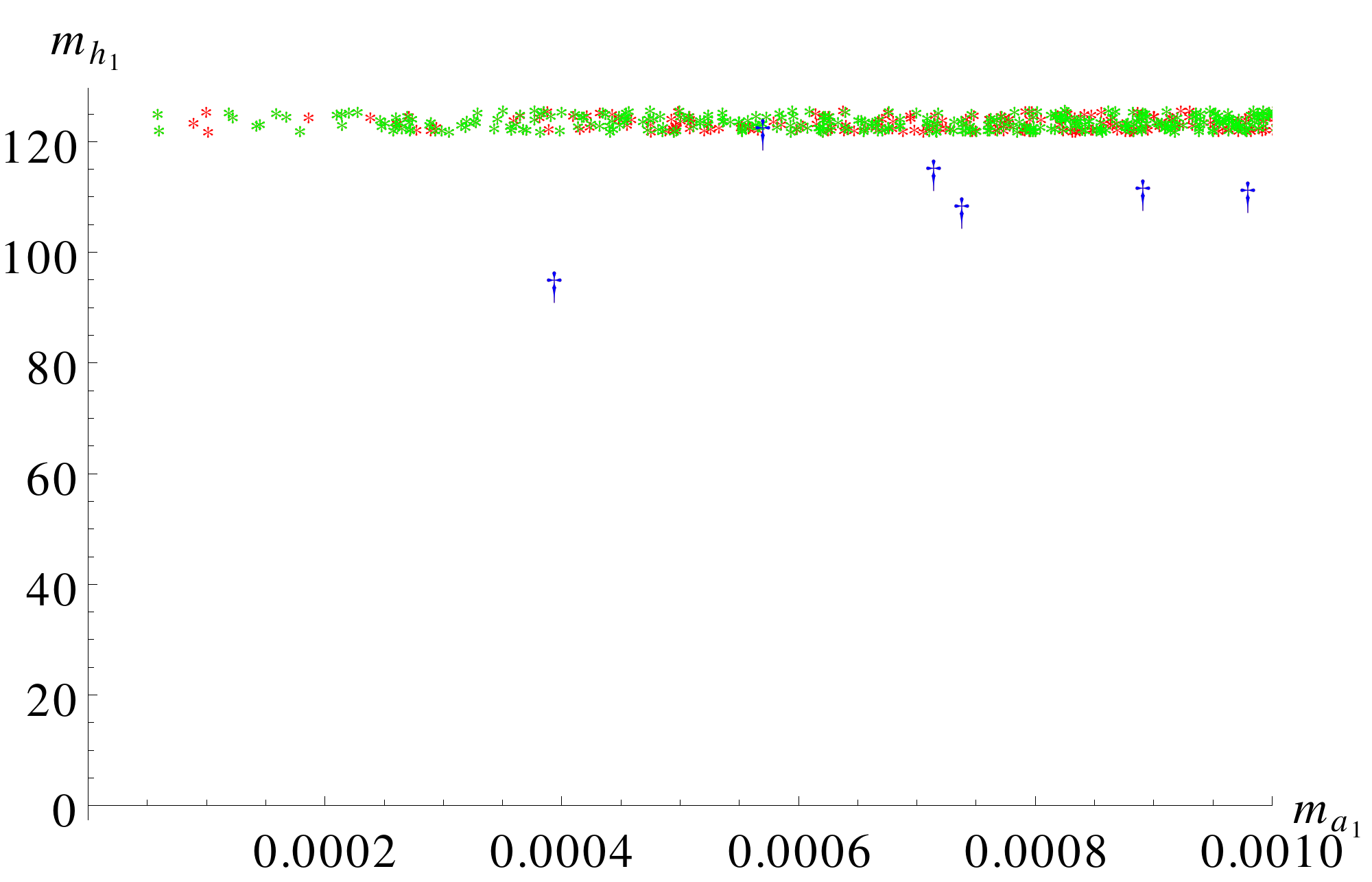}}}
\caption{The lightest CP-even Higgs boson mass $m_{h_1}$ vs the lightest pseudo-scalar mass $m_{a_1}$ at one-loop (top-stop and bottom-bottom corrections) consistent with the Higgs data from CMS, ATLAS and LEP. The red points corresponds to the case where $m_{h_1}\sim m_{125}$, the orange points correspond to mass values of $m_{h_1}$ and $m_{a_1}$  where  $m_{h_2}\sim m_{125}$ and all of them satisfy  the $ZZ^*$, $WW^*$ bounds at $1\sigma$ and $\gamma\gamma$
bound at $2\sigma$ level from both CMS and ATLAS. The red (orange) points which satisfy the  $\gamma\gamma$ result at $1\sigma$ are marked green (blue). Very light pseudoscalar masses $m_{a_1}\leq 1$ MeV are shown in panel (b), which is a zoom of the small mass region of (a).}\label{higgsdata}
\end{center}
\end{figure}

On the other hand, in the Higgs bosons decay into di-photons, 
there are more virtual particles which contribute in the loop compared to the SM. This is due to the
enlarged Higgs and strong sectors which have a non-zero coupling with 
the photon. In particular there are three charginos ($\chi_{1,2,3}^\pm$), three charged Higgs bosons ($h_{1,2,3}^\pm$), the stops ($\tilde{t}_{1,2}$) and the sbottoms ($\tilde{b}_{1,2}$). Compared to the MSSM and the NMSSM we have two additional charged Higgs bosons and one additional chargino which contribute to the decay. The decay rate in the di-photon channel is given by \begin{align}\label{gammagamma}
\Gamma(h\rightarrow\gamma\gamma)&=\frac{\alpha\,m_h^3}{1024\,\pi^3} \Big|\frac{g_{hWW}}{m_W^2}\,A_1(\tau_W)+\sum_{ \chi_i^\pm,\,t,\, b}2\frac{g_{hf\bar{f}}}{m_f}N^c_f\, Q_f^2\, A_{1/2}(\tau_f)\\
&+\sum_{h_i^\pm,\,\tilde{t}_i,\,\tilde{b}_i}\frac{g_{hSS}}{m_S^2}N_S^c\,Q_S^2\,A_0(\tau_S)\Big|^2,\nn
\end{align}
where $N_{f, S}^c$ are the color number of fermion and scalars, $Q_{f, S}$ are the electric charges, in unit of $|e|$, of the fermions and scalars, and $\tau_i=\frac{m_h^2}{4\,m_i^2}$. $A_0, \,A_{1/2}$ and $A_1$ are the spin-0, spin-1/2 and spin-1 loop functions
\bea
&&A_0(x)=-\frac{1}{x^2}\left(x-f(x)\right),\\
&&A_{1/2}(x)=\frac{2}{x^2}\left(x+(x-1)f(x)\right),\\
&&A_1(x)=-\frac{1}{x^2}\left(2\,x^2+3\,x+3(2\,x-1)f(x)\right),
\eea
with the analytic continuations 
\bea
f(x)=\left\{
\begin{array}{lr}
\arcsin^2(\sqrt{x})& x\leq1\\
-\frac{1}{4}\left(\ln\frac{1+\sqrt{1-1/x}}{1-\sqrt{1-1/x}}-i\pi\right)^2& x>1
\end{array}\right.
\eea
In the limit of heavy particles in the loop, we have $A_0\rightarrow 1/3$, $A_{1/2}\rightarrow 4/3$ and $A_1\rightarrow-7$.\\
Using the expression above, we study the discovered Higgs boson ($h_{125}$)  decay rate to di-photon in this model. We also check the consistency of light scalar(s) and/or light pseudoscalar(s)
with the current data at the LHC and the older LEP data.
Such analysis is presented in Figure~\ref{higgsdata}. Figure~\ref{higgsdata}(a) shows such hidden Higgs scenarios with one $a_1$ and/or one $h_1$ below 123 GeV, which find significant realizations. 
%%%%%%%%%Higgs solutions from LHC%%%%%%%%%%%%
%%%%%%%%%%%%%%%%%%%%%%
 We first consider the results  coming from both CMS and ATLAS in the decay of the Higgs to $WW^*$ and $ZZ^*$ modes at $1\sigma$ \cite{CMS, CMS2, ATLAS} and also consider the cross-section bounds from LEP \cite{LEPb}. The allowed mass values are shown as red points for which the lightest CP-even Higgs boson ($h_1$) is the detected Higgs at $\sim 125$ GeV. Cleary we see that there are many light pseudo-scalars ($\leq 100$ GeV) which are allowed. The orange points present the scenario where $m_{h_2}\sim m_{125}$ and which leaves both  $h_1$ and $a_1$ hidden ($< 125$ GeV).

We have performed additional tests of such points and compared them with the results from the decay of the Higgs boson to di-photons at the LHC, both from CMS \cite{cmsgamma} and ATLAS \cite{atlasgamma}. The red points (with one hidden Higgs boson)  which satisfy $h_{125}\to \gamma \gamma$ at $1\sigma$ level, are marked as green points. The orange points (with two hidden Higgs bosons) when allowed at  $1\sigma$ level, have been marked as blue points. Notice that all the points in Figure~\ref{higgsdata} are allowed at  $1\sigma$ by the $WW^*$, $ZZ^*$ channels and at $2\sigma$ by the $\gamma \gamma$ channel. These requirements automatically brings the fermionic decay modes closer to the SM expectation. Of course the uncertainties of these decay widths give us a room for $h_{125}\to a_1a_1/h_1h_1$. 

Notice also the presence of a very light pseudoscalar mass values near $a_1 \sim 0$. Figure~\ref{higgsdata}(b) is a zoom of this region, where such solutions are shown for $m_{a_1} \leq 1$ MeV. The points in this case correspond to possible $a_1$ states which do not decay into any charged fermion pair ($m_{a_1} \leq 2m_{e}$) and have an interesting phenomenology, as briefly pointed out in section~\ref{axion}. The fact that such mass values only allow a decay of this particle to two photons via doublet mixing mediated by a fermion loop, makes the $a_1$ a possible dark matter candidate, being long lived. Two hidden Higgs bosons render the phenomenology very interesting, allowing both the $h_{125} \to a_1 a_1$ and the $h_{125} \to h_1 h_1$ decay channels \cite{hdlh, ehdc}. In Figure~\ref{bps} we show some of the points in this model as benchmark points (BMP's), which are allowed 
both by LHC \cite{CMS, ATLAS} and LEP \cite{LEPb} data.
%%%%%%%%%BPs %%%%%%%%%%%%
\begin{figure}
\begin{center}
\mbox{\hskip -15 pt\subfigure[]{\includegraphics[width=0.55\linewidth]{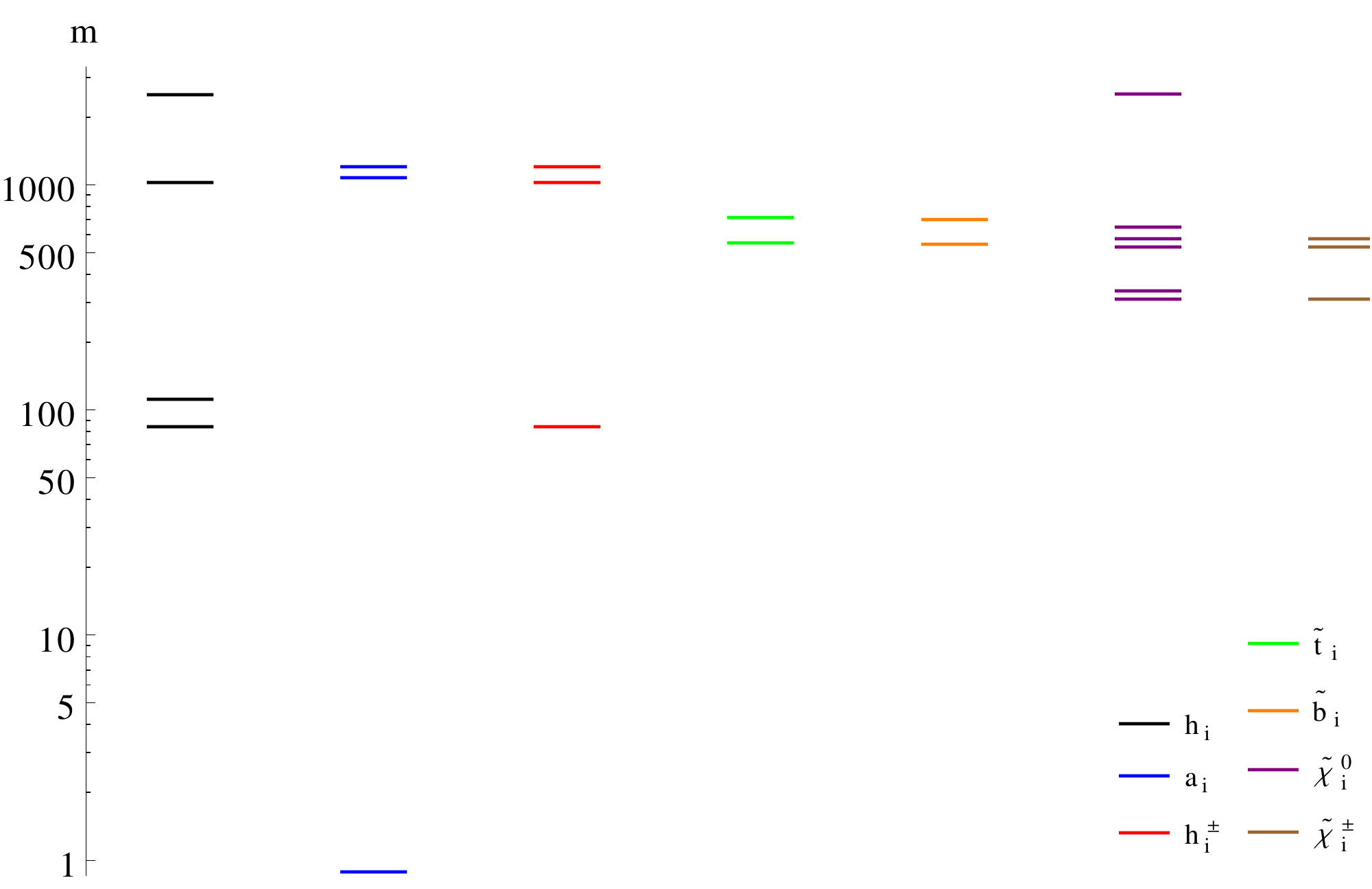}}
%\hskip 25 pt
\subfigure[]{\includegraphics[width=0.55\linewidth]{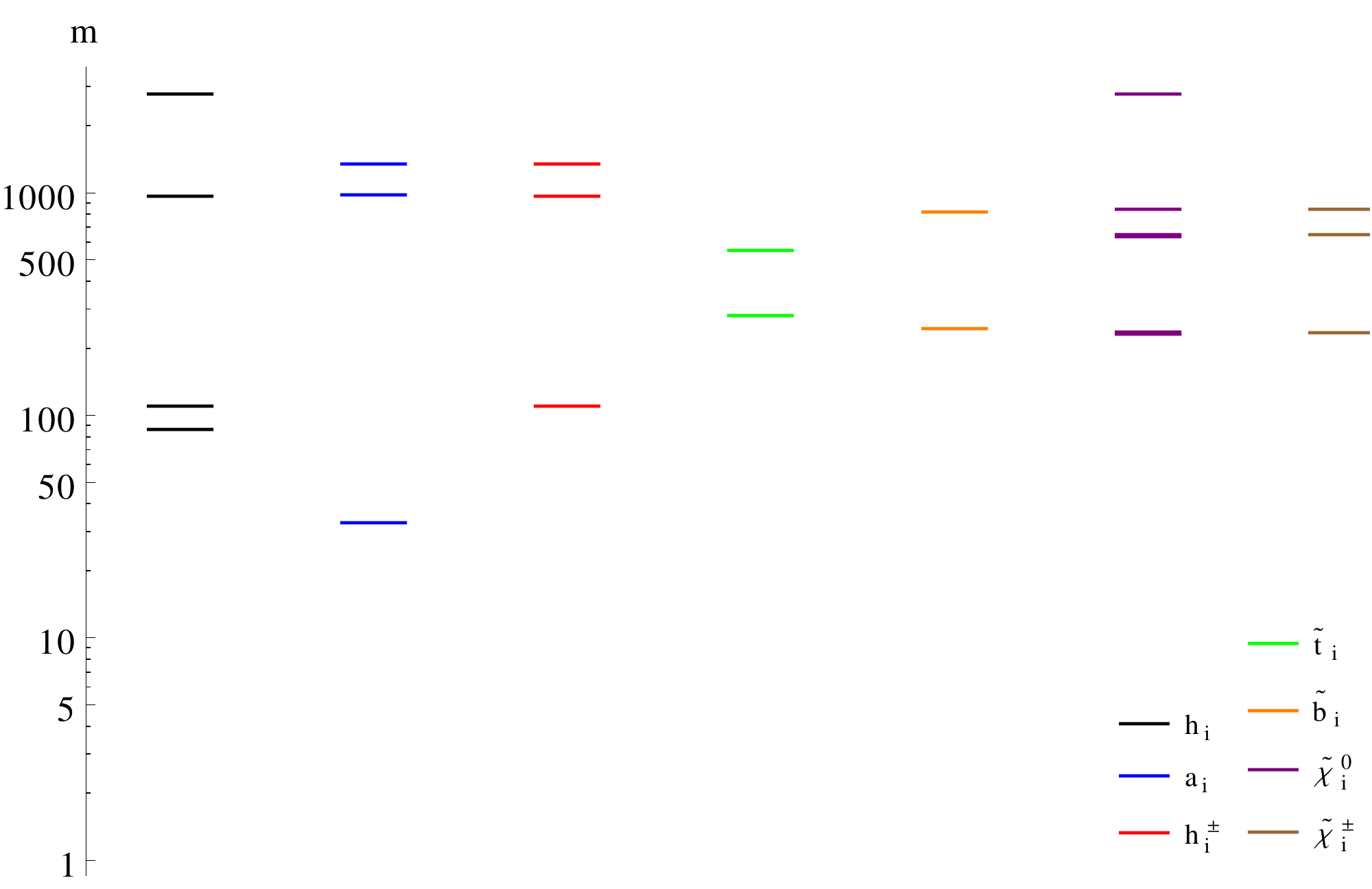}}}
%\hskip 25 pt
\mbox{\subfigure[]{\includegraphics[width=0.55\linewidth]{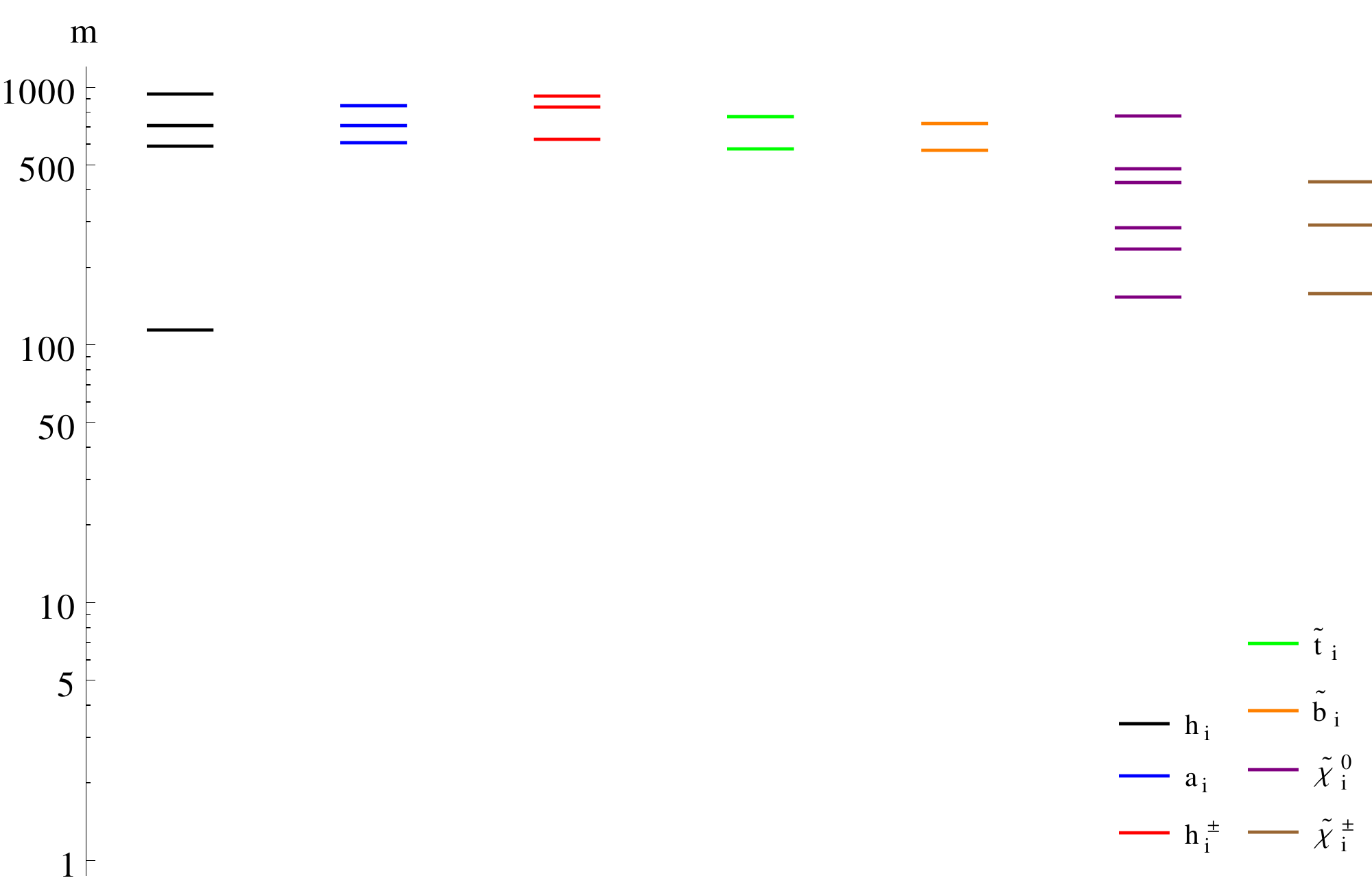}}
}
\caption{We show the benchmark points of the model which are allowed both by LHC \cite{CMS, ATLAS} and LEP\cite{LEPb} data. The neutral Higgs spectrum has been calculated at one-loop and the rest of the spectrum at tree-level.}\label{bps}
\end{center}
\end{figure}
%%%%%%%%%%%%%%%%%%%%%%
The neutral Higgs spectrum has been calculated at one-loop order and the remaining states at tree-level. Figure~\ref{bps}(a) shows a point (BMP1) where we have a hidden pseudoscalar ($a_1$) with mass $\mathcal{O}(10^{-1})$ MeV and another triplet/singlet-like hidden CP-even scalar ($h_1$) with a mass around $\sim 93$ GeV. In this case the candidate Higgs boson is $h_2$, taken around $125$ GeV. This point also have a triplet type very light charged Higgs boson at a mass around 90 GeV, which is not excluded by the recent charged Higgs bounds from the LHC \cite{chHb}. Figure~\ref{bps}(b) shows a benchmark point (BMP2) where we have  a pseudoscalar  around 37 GeV, and the lightest scalar and charged Higgs bosons around 100 GeV. Figure~\ref{bps}(c) shows a trivial (SM-like) solution where we have a doublet-type CP-even Higgs around $\sim 125$ GeV, with the other states decoupled. In the next study we are going to analyse such points through a detailed collider simulation \cite{pb3}.

\section{Phenomenology of the TNMSSM}\label{pheno}
%%%%%%%%%Production channels %%%%%%%%%%%%
\begin{figure}
\begin{center}
\mbox{\subfigure[]{
\includegraphics[width=0.3\linewidth]{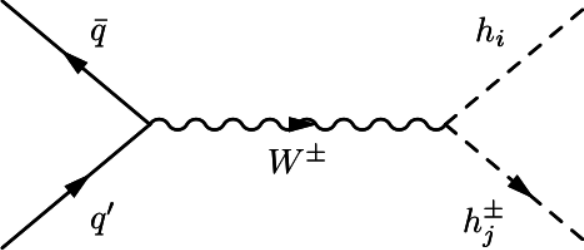}}
\hskip 25 pt
\subfigure[]{\includegraphics[width=0.3\linewidth]{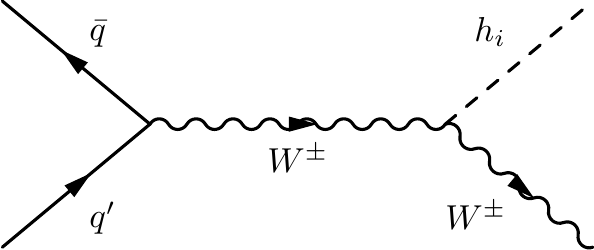}}}
\mbox{\subfigure[]{
\includegraphics[width=0.3\linewidth]{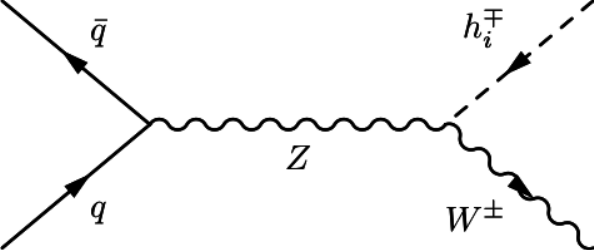}}
\hskip 25 pt
\subfigure[]{\includegraphics[width=0.25\linewidth]{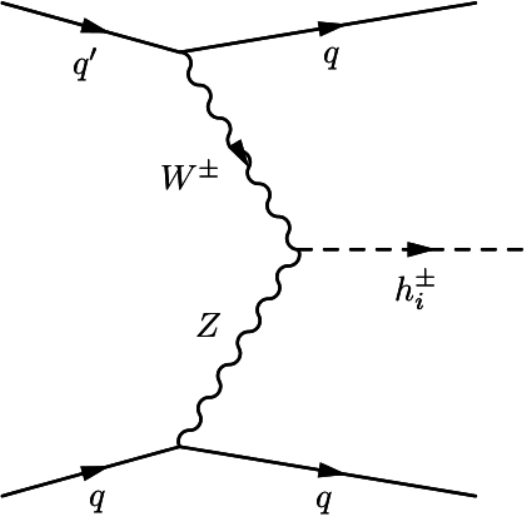}}}
\caption{The new and modified production channels for the Higgs bosons at the LHC.}\label{higgsprd}
\end{center}
\end{figure}
%%%%%%%%%%%%%%%%%%%%%%
The TNMSSM extends the Higgs sector as well as the electroweak chargino-neutralino sectors by additional
higgsino contributions. Both the triplet and singlet fields do not couple to the fermions but affect the
phenomenology to a large extent. In the context of the recent Higgs discovery, searches for additional Higgs bosons, both neutral and charged, are timely. In particular, if an extended Higgs sector will be discovered at the LHC, it will be crucial to determine the gauge representation which such states belong to, by investigating its allowed decays modes. 

We have seen from Eq.~(\ref{spt}), that a $Y=0$ hypercharged triplet couples to the $W^\pm$ bosons and contributes to their mass. On the other hand, the singlet does not directly couple to any of the gauge bosons. In the case of Higgs mass eigenstates which carry a doublet-triplet-singlet mixing, we need to look either for their direct production processes at the LHC or take into consideration the possibility of their cascade production from other Higgses or supersymmetric particles.

\subsection{Productions}
We have detailed a model with a rich Higgs sector with additional Higgs bosons of triplet and singlet type. We recall that the relevant  production processes of a Higgs boson which is a SU(2) doublet at the LHC \cite{anatomy1, anatomy2} are the gluon-gluon fusion (GGF) and vector boson fusion channels, followed by the channels of associated production of gauge bosons and fermions. In our case, the production channels for the new Higgs bosons are different, due to their different couplings to the gauge bosons and fermions. We list below the possible additional production channels for the neutral and charged Higgs bosons at the LHC.

\begin{itemize}
\item \underline{ Neutral Higgs boson production in association with charged Higgs boson:} The triplet only couples to $W^\pm$ boson.
Thus a neutral Higgs (doublet or triplet) can be produced in association with a charged Higgs boson (doublet or triplet) via a $W^\pm$ exchange. As shown in Figure~\ref{higgsprd}(a) a light in mass and charged Higgs boson in the TNMSSM can be easily explored 
by this production channel $\bar{q} q' \to h_i h^\pm_j$.

\item  \underline{ Neutral Higgs boson production in association with $W^\pm$: } A triplet or a doublet type neutral Higgs boson
can be produced via $\bar{q}q' \to W^\pm h_i$ as shown in Figure~\ref{higgsprd}(b). A triplet admixture modifies the $h_i-h^\pm_j-W^\mp$ couplings by an additional term proportional to the vev of the triplet.

\item  \underline{Charged Higgs boson production in association with $W^\pm$: } Triplet of $Y=0, \pm2$ hypercharge 
has a non-zero tree-level coupling to $Z-W^\pm-h^\mp_i$. This leads to additional contributions to 
$q\bar{q} \to W^\pm h^\mp_i$ as shown in Figure~\ref{higgsprd}(c).

\item  \underline{Production of charged Higgs boson in vector boson fusion:} The non-zero $Z-W^\pm-h^\mp_i$ coupling leads to vector boson fusion ($Z, W$ fusion) which produces a charged Higgs boson as shown in Figure~\ref{higgsprd}(d). This mode is absent in 2-Higgs doublet models (2HDM), in the MSSM and in the NMSSM. This is a unique feature of the $Y=0, \pm2$ hypercharge, triplet-extended scenarios. 

\item \underline{Singlet Higgs production:} The singlet in this model is not charged under any of the gauge groups, and hence the direct  production of such a singlet at the LHC is impossible.
Gauging this additional singlet with the inclusion of an extra additional $U(1)'$ gauge group would open new production channels via the additional gauge boson ($Z'$). Most of the extra $Z'$ models get a bound on the $Z'$ mass, $m_{Z'} \gsim 2.79$ TeV \cite{LHCZ'} which makes such channels  less promising at the LHC. In our case such a singlet type Higgs boson would only be produced via mixing with the Higgs bosons of doublet and triplet type.

\end{itemize}
\subsection{Decays}
The smoking gun signatures for the model would be the decays of the doublet, triplet and singlet states that are produced. Different F-term contributions can generate these types of mixing and corresponding decay vertices. 

%%%%%%%%%Decay channels %%%%%%%%%%%%
\begin{figure}
\begin{center}
\mbox{\subfigure[]{
\includegraphics[width=0.23\linewidth]{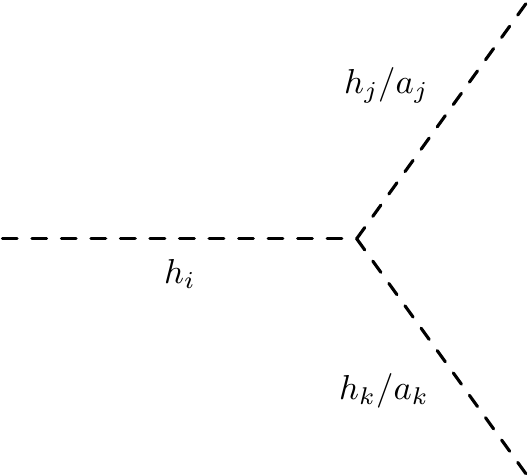}}
\hskip 25 pt
\subfigure[]{\includegraphics[width=0.23\linewidth]{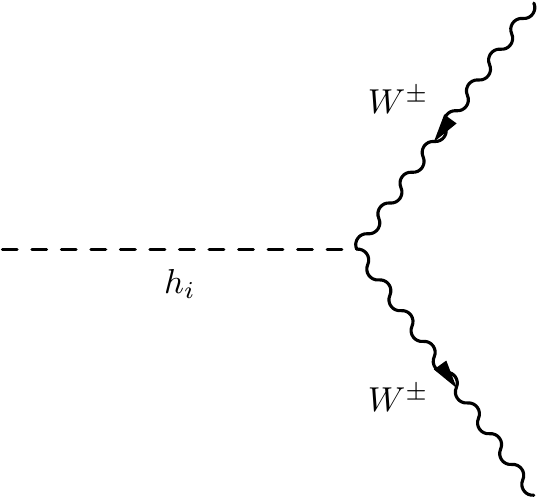}}}
\mbox{\subfigure[]{
\includegraphics[width=0.23\linewidth]{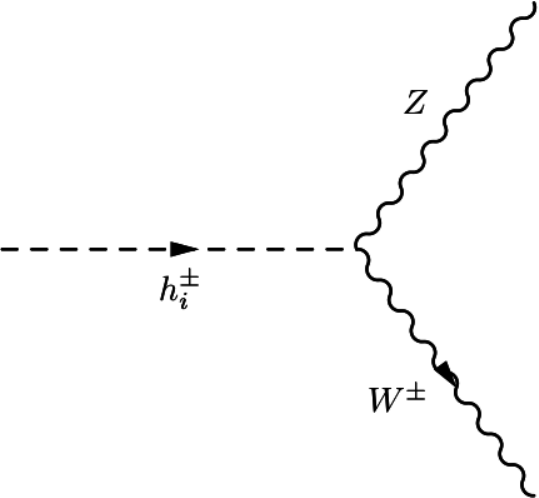}}
\hskip 25 pt
\subfigure[]{\includegraphics[width=0.23\linewidth]{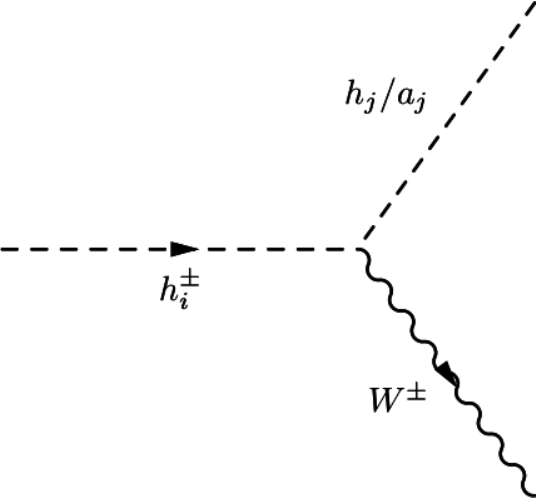}}
\hskip 25 pt
\subfigure[]{\includegraphics[width=0.23\linewidth]{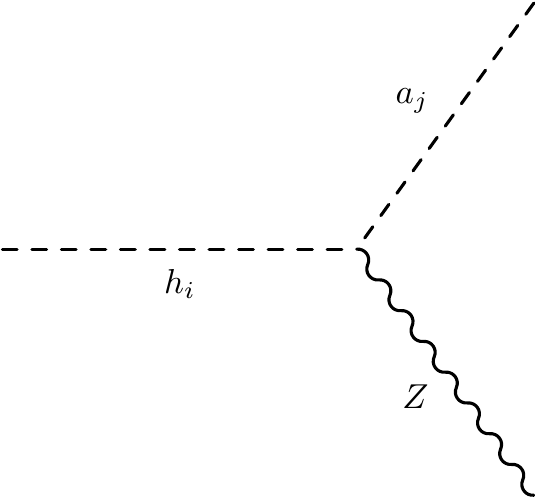}}
}
\caption{The new and modified decay channels of the Higgs bosons at the LHC.}\label{higgdcy}
\end{center}
\end{figure}
%%%%%%%%%%%%%%%%%%%%%%
\begin{itemize}

\item \underline{Higgs decays to Higgs pairs:}  The candidate Higgs around the 125 GeV mass in this case can decay into two hidden Higgs bosons if this channel is kinematically allowed as can be seen in Figure~\ref{higgdcy}(a). Such hidden Higgs boson(s) could be both scalar and pseudoscalar in nature. The discovered Higgs is however $99\%$ CP-even \cite{CMS2}, which forbids any CP-violating decay of the nature $h_{125} \to a_i h_j$.  However, the CP-conserving decays  like  $h_{125} \to a_i a_j$ and/or $h_{125} \to h_i h_j$ are allowed. Such decays should be carefully investigated on the basis of the current Higgs data at the LHC. If the two light Higgs bosons are mostly singlet or triplet, then it is easy to evade the bounds from LEP \cite{LEPb}. As we have already pointed out, a singlet Higgs does not couple to any of SM gauge bosons and even the triplet type does not couple to the $Z$ boson. Such a light Higgs boson could decay into $\tau$ pairs only through the mixing with the doublet type Higgs bosons, since neither the singlet nor the triplet couple to fermions (see Eq.~(\ref{spt})). The mixing angle is also constrained by data on bottomonium decay, for a very light neutral Higgs boson ($\lsim 8$ GeV) \cite{bottomium}. \\
The decay of a Higgs boson into other Higgs bosons depends on the cubic coupling, which is proportional to the vevs of  Higgs fields, and thus it is very sensitive to the values of $v_i$. It is therefore requires an analysis of the allowed decay widths of the Higgs boson into a light Higgs pair using LHC data \cite{hdlh}.

\item \underline{Higgs decays to $W^\pm W^\mp$:} The triplet couples to $W^\pm$ via its non-zero SU(2) charge which is at variance respect to the analogous coupling of the doublet. This will modify the decay width of $h_i \to W W$ (Figure~\ref{higgdcy}(b)).  The recent data show that  there is some disagreement and uncertainties  between the CMS \cite{CMS, CMS2} and  ATLAS \cite{ATLAS} results in the  $h_{125} \to WW^*$ 
channels. The measurement of this decay channel thus becomes even more crucial under the assumption of a triplet mixture. 
\item \underline{Charged Higgs decays to $Z\,W^\pm$:}   We know that the triplet type charged Higgs has a non-zero tree-level coupling to $Z\,W$, for a non-zero triplet vev, as shown in Figure~\ref{higgdcy}(c). This opens up the possible decay modes $h^\pm_i \to Z W^\pm$, which are absent in the 2HDM and in the MSSM at tree-level.

\item \underline{Charged Higgs decays to $h_j(a_j)W^\pm$:} A doublet or triplet type Higgs boson can decay to a lighter neutral Higgs and a $W^\pm$ (Figure~\ref{higgdcy}(d)). A possibility of a very light triplet-singlet-like neutral Higgs makes this decay mode more interesting compared to the case of the CP-violating MSSM \cite{cpv}. 

\item \underline{Higgs decays to $a_j Z$:} In the MSSM the odd and heavy Higgs bosons are almost degenerate, so $h_i \to a_j Z$ is not kinematically allowed. The introduction of a triplet and of a singlet 
adds two more massive CP-odd Higgs bosons, and the degeneracy is lifted. In this case we have a relatively 
lighter CP-odd Higgs state $a_i$ which makes $h_i \to a_j Z$ possible, as shown in Figure~\ref{higgdcy}(e). This scenarios is also possible in the context of the CP-violating MSSM, where we have a very light pseudoscalar Higgs boson due to the large mixing between the Higgs CP eigenstates \cite{pb2} and in the NMSSM, for  having an additional scalar \cite{nmssm}.

\item \underline{Higgs decays to fermion pairs:}  In a scenario where a  triplet or/and singlet type Higgs boson decays to gauge bosons and other Higgses are kinematically forbidden, the only permitted decays are into light fermion pairs, viz, $bb, \tau\tau$ and $\mu\mu$. Even such decays are only 
possible by a mixing with doublet type Higgs bosons. When such mixing angles are very small
this can results into some displaced charged leptonic signatures.

\end{itemize}
\subsection{Possible signatures}\label{sign}
The unusual production and decay channels lead to some really interesting phenomenology which could be tested in the next run of the LHC and at future colliders.  From the testability point of view, one could use the data form the discovered Higgs boson $\sim 125$ GeV in order to get bounds from the Higgs decaying to Higgs boson pair \cite{hdlh}, or the existing bounds from LEP \cite{LEPb} for two Higgs bosons productions. We have already taken into account these bounds by ensuring that the hidden Higgs boson is mostly of singlet or of triplet type. Given the uncertainty in the Higgs decay branching fractions in different modes and the absence of direct bounds on the non-standard decays of Higgs boson to Higgs boson pair ($h_{125}\to a_i a_j/h_i h_j$), this remains phenomenologically an interesting scenario. Below we list different possible signatures that could be tested in the LHC with 13/14 TeV. 

\begin{itemize}

\item 
The singlet and doublet F-terms  generate the doublet-triplet-triplet vertex which is proportional to $\lambda_S \lambda_{TS}$ and $\lambda^2_T$. This would provide a signature of a doublet type Higgs decaying into two triplet type Higgs bosons, which, in turn, do not decay into fermions. Similarly the F-terms of $H_u$ and $H_d$ generate vertices involving triplet-singlet-doublet  which are proportional to 
$\lambda_T \lambda_S$. The F-term of triplet type also contributes to this mixing, which is 
proportional to $\lambda_T \lambda_{TS}$. Thus the relative sign between the two contributions become important.

In the case of a $\sim 125$ GeV Higgs boson, this can decay into two triplet-like hidden scalars or pseudoscalars, which in turn decay into off-shell $W^\pm$s only. This type of decays can be looked for by searching for very soft jets or leptons coming from the off-shell $W^\pm$s.  The signatures could be the $4\ell +\ptmiss$ or $4j+2l +\ptmiss$ channels, where the jets and the leptons are very soft. On the other hand, both the triplet and the singlet hidden Higgses can decay to fermion pairs ($b\bar{b}$, $c\bar{c}, e^+e^-, \mu\bar{\mu}, \tau \bar{\tau}$) via the mixing with doublets. The recent bounds on these non-standard decays has been calculated for the LHC \cite{ehdc}. Such decays give $4\ell, \, 2b+ 2\ell$ final states, where the leptons are very soft. For the triplet type hidden Higgs bosons it would be interesting to analyze the competition between the four-body and the two-body decays (which depend on the triplet-doublet mixing). Demanding for the presence of softer  leptons and jets in the final states, allows to reduce the SM backgrounds at the LHC. If the mixing is very small, this could lead to displaced charged leptonic final states, similar to those of a Higgs boson decay in a $R$-parity violating supersymmetric scenario \cite{pbrp}.  Due to the coupling both with the up and the down type doublets, this coupling could be tested both at a low and a high $\tan{\beta}$.

\item The singlet does not contribute to charged Higgs mass eigenstates, so the charged Higgs bosons  could be either of doublet or triplet nature. In the case of a heavy doublet type, the heavier charged Higgs can decay to a triplet type a neutral Higgs (CP even or odd) and a triplet type charged Higgs ($H^\pm_{u,d} \to T^0 T^\pm_{1,2}$). The coupling is proportional to ($g^2_L -\lambda^2_T$).  The lighter triplet type charged Higgs then mostly decays into on-shell or off-shell $Z\, W^\pm$. This is a generic signature for $Y=0, \pm2$ hypercharge triplets with non-zero triplet vev, which breaks the custodial symmetry of the Higgs potential.  The relatively lighter triplet (either CP-odd or even) neutral Higgs can decay via an on/off-shell $W^\pm$
 boson pair, which leads to leptonic final states. The final states  with multi-lepton($>3\ell$), multi-jet($>4$) and missing energy, could be the signature for this model. Depending on the off-shell decays, few leptons or a jet could be softer in energy.  

\item  In other cases a triplet type heavier charged Higgs can decay into a doublet type neutral Higgs and a triplet type charged Higgs.  These couplings are proportional  to ($\frac{g^2_L}{2} -\lambda^2_T$ ) and can give rise to $3\ell +2b +\ptmiss$ and  $3\ell +2\tau +\ptmiss$ final states. Here the $b$ and $\tau$ pairs expected from the neutral doublet type Higgs boson decay. 

\item Unlike to the neutral Higgs bosons, the up and down type charged Higgs bosons doublet only mix with the triplets. The couplings are again proportional to a combination of $\lambda_S \lambda_{T}$
and $\lambda_T \lambda_{TS}$.  In this case the doublet (triplet)  charged Higgs state will decay into a triplet (doublet) charged Higgs and a singlet neutral Higgs boson. As the singlet is not coupled to any SM particles, it can only decay through mixing with doublets and triplets.  Decays of such singlets to leptons (in the case of mixing with doublets) and off-shell or on-shell $W^\pm$-pair will be determined by the mixing only. In a fine-tuned region where such mixing is very low this decay channel can lead to a displaced vertex of charged leptons, whose measurement can give information about such a mixing. 
\end{itemize}

\section{Higgs spectrum and the experimental constraints}\label{scans}

As already pointed out, in the Higgs sector there are four CP-even neutral ($h_1,h_2,h_3,h_4$), three CP-odd neutral ($a_1,a_2, a_3$) and three charged Higgs bosons ($h_1^\pm,h_2^\pm,h_3^\pm$). In general the interaction eigenstates are obtained via a mixing of the two Higgs doublets, the triplet and the singlet scalar. However, the singlet does not contribute to the charged Higgs bosons, which are mixed states generated only by the $SU(2)$ doublets and triplets.
The rotation from gauge eigenstates to the interaction eigenstates are
\bea\label{chmix}
h_i= \mathcal{R}^S_{ij} H_j\nn\\
a_i= \mathcal{R}^P_{ij} A_j\\
h^\pm_i= \mathcal{R}^C_{ij} H^\pm_j\nn
\eea
where the eigenstates on the left-hand side are interaction eigenstates whereas the eigenstates on th right-hand side are gauge eigensates. Explicitly we have $h_i=(h_1,h_2,h_3,h_4)$, $H_i=(H^0_{u,r},H^0_{d,r},S_r,T^0_r)$, $a_i=(a_0,a_1,a_2,a_3)$, $A_i=(H^0_{u,i},H^0_{d,i},S_i,T^0_i)$, $h_i^\pm=(h_0^\pm,h_1^\pm,h_2^\pm,h_3^\pm)$ and $H_i^+=(H_u^+,T_2^+,H_d^{-*},T_1^{-*})$. { Using these definitions we can write the doublet and triplet fraction for the scalar and pseudoscalar Higgs bosons as
\bea
h_i|_{D}=(\mathcal{R}^S_{i,1})^2+(\mathcal{R}^S_{i,2})^2, \,\, a_i|_{D}=(\mathcal{R}^P_{i,1})^2+(\mathcal{R}^P_{i,2})^2
\eea
\bea
h_i|_{S}=(\mathcal{R}^S_{i3})^2, \,\, a_i|_{S}=(\mathcal{R}^P_{i3})^2
\eea
\bea
h_i|_T=(\mathcal{R}^S_{i4})^2, \,\, a_i|_T=(\mathcal{R}^P_{i4})^2
\eea
and the triplet and doublet fraction of the charged Higgs bosons as
\bea
h_i^\pm|_D=(\mathcal{R}^C_{i1})^2+(\mathcal{R}^C_{i3})^2, \,\, h_i^\pm|_T=(\mathcal{R}^C_{i2})^2+(\mathcal{R}^C_{i4})^2 .
\eea
We call a scalar(pseudoscalar) Higgs boson doublet-like if $h_i|_D(a_i|_D)\geq\,90\%$, singlet-like if $h_i|_S(a_i|_S)\geq\,90\%$ and triplet-like if $h_i|_T(a_i|_T)\geq\,90\%$. Similarly a charged Higgs boson will be doublet-like if $h_i^\pm|_D\geq\,90\%$ or triplet-like if $h_i^\pm|_D\geq\,90\%$.}

If the discovered Higgs is the lightest CP-even boson, $h_1\equiv h_{125}$, then $h_1$ must be doublet-like and the lightest CP-odd and charged Higgses must be triplet/singlet-like, in order to evade the experimental constraint from LEP \cite{LEPb} for the pseudoscalar and charged Higgses.
{ LEP searched for the Higgs boson via the $e^+ e^- \to Z h$ and $e^+e^- \to h_1h_2$ channels (in models with multiple Higgs bosons) and their fermionic decay modes ($h \to \bar{b}b,\bar\tau \tau$ and $Z \to \ell\ell$). The higher centre of mass energy at LEP II (210 GeV) allowed to set a lower bound of 114.5 on the SM-like Higgs boson and of 93 GeV for the MSSM-like Higgs boson in the maximal mixing scenario \cite{LEPb}. Interestingly, neither the triplet nor the singlet type Higgs boson couple to Z or to leptons (see
Eq.~\ref{spt}), and we checked explicitly the demand of $\geq 90\%$ singlet and/or triplet is sufficient 
for the light pseudoscalar to be allowed by LEP data. We also checked explicitly the LHC allowed parameter space for the light pseudoscalar and the details can be found out in \cite{TNMSSM2}. Later we also discuss how  the criteria of $\geq 90\%$ singlet/triplet is enough to fulfill the constraints coming from the B-observables.}
Similar constraints on the structure of the Higgses must be imposed if $h_2\equiv h_{125}$. To scan the parameter space we have used a code written by us, in which we have randomly selected $1.35\times10^6$ points that realize the EWSB mechanism at tree-level. In particular, we have performed the scan using the following criteria for the couplings and the soft parameters
\bea\label{scan}
&|\lambda_{T, S, TS}| \leq 1, \, |\kappa|\leq 3, \, |v_s|\leq 1 \, \rm{TeV}, \, 1\leq \tan{\beta}\leq 10,\nn\\
&|A_{T, S, TS, U, D}|\leq 1\, \rm{GeV},\,|A_\kappa|\leq 3\, \rm{GeV},\\
&65\leq|M_{1, 2}|\leq10^3\,\rm{GeV},\,  3\times10^2\leq m_{Q_3, \bar{u}_3, \bar{d}_3}\leq10^3\,\rm{GeV}.\nn
\eea
We have selected those points which have one of the four Higgs bosons with a one-loop mass of $\sim125$ GeV { with one-loop minimization conditions} and, out of the $1.35\times10^6$ points, over $10^5$ of them pass this constraint. On this set of Higgs candidates we have imposed the constraints on the structure of the lightest CP-even, CP-odd and charged Higgses. The number of points with $h_1\equiv h_{125}$ doublet-like and $a_1$ singlet-like is about 70 \% but we have just one point with $h_1\equiv h_{125}$ which is doublet-like and $a_1$ triplet-like. If we add the requirement on the lightest charged Higgs to be triplet-like, we find that the number of points with $h_1\equiv h_{125}$ doublet-like, $a_1$ singlet-like and $h_1^\pm$ triplet-like is 26 \%. The case of $h_2\equiv h_{125}$ doublet-like allows more possibilities, because in this case we have also to check the structure of $h_1$. However we find 75 points only when $h_1$ is triplet-like, $h_2\equiv h_{125}$ is doublet-like and $a_1$ is singlet-like. This selection is insensitive to the charged Higgs selection, i.e. we still have 75 points with $h_1$ triplet-like, $h_2\equiv h_{125}$ doublet-like, $a_1$ singlet-like and $h_1^\pm$ triplet-like.\\
The LHC constraints have been imposed on those points with $h_1\equiv h_{125}$, because they provide a better statistics. For these points we demand that
\bea\label{LHCdata}
&\mu_{WW^*}=0.83\pm0.21\,\,\mu_{ZZ^*}=1.00\pm0.29\\
&\mu_{\gamma\gamma}=1.12\pm0.24\nn
\eea
at 1$\sigma$ of confidence level \cite{CMS2}. The LHC selection give us 12223 points out of the 26776 points that have  $h_1\equiv h_{125}$ doublet-like, $a_1$ singlet-like and $h_1^\pm$ triplet-like.

 %%%%%%%%%%%%%%% h2 h1+ %%%%%%%%%%%
\begin{figure}[thb]
\begin{center}
\mbox{\hskip -10pt\subfigure[]{\includegraphics[width=0.45\linewidth]{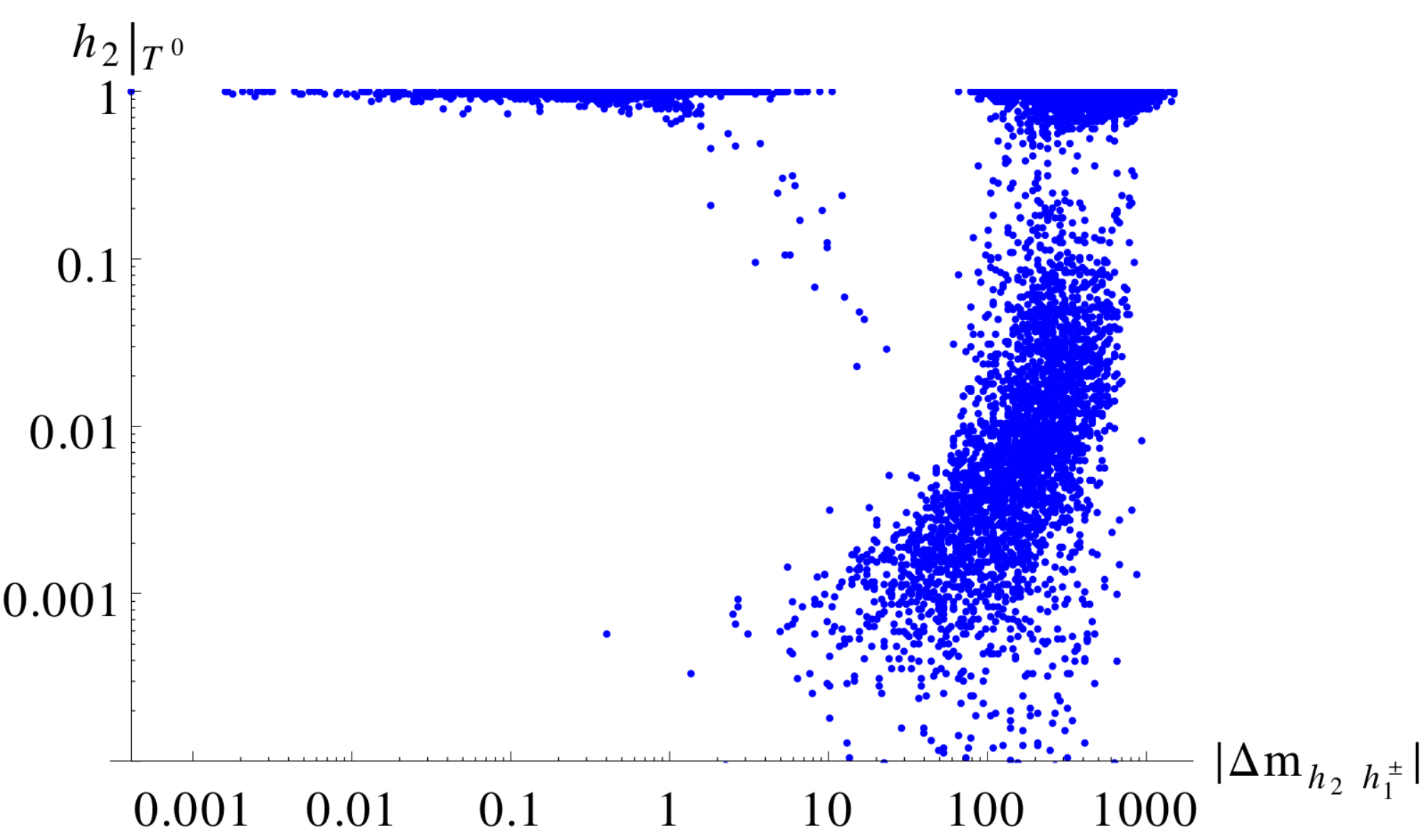}}
\subfigure[]{\includegraphics[width=0.45\linewidth]{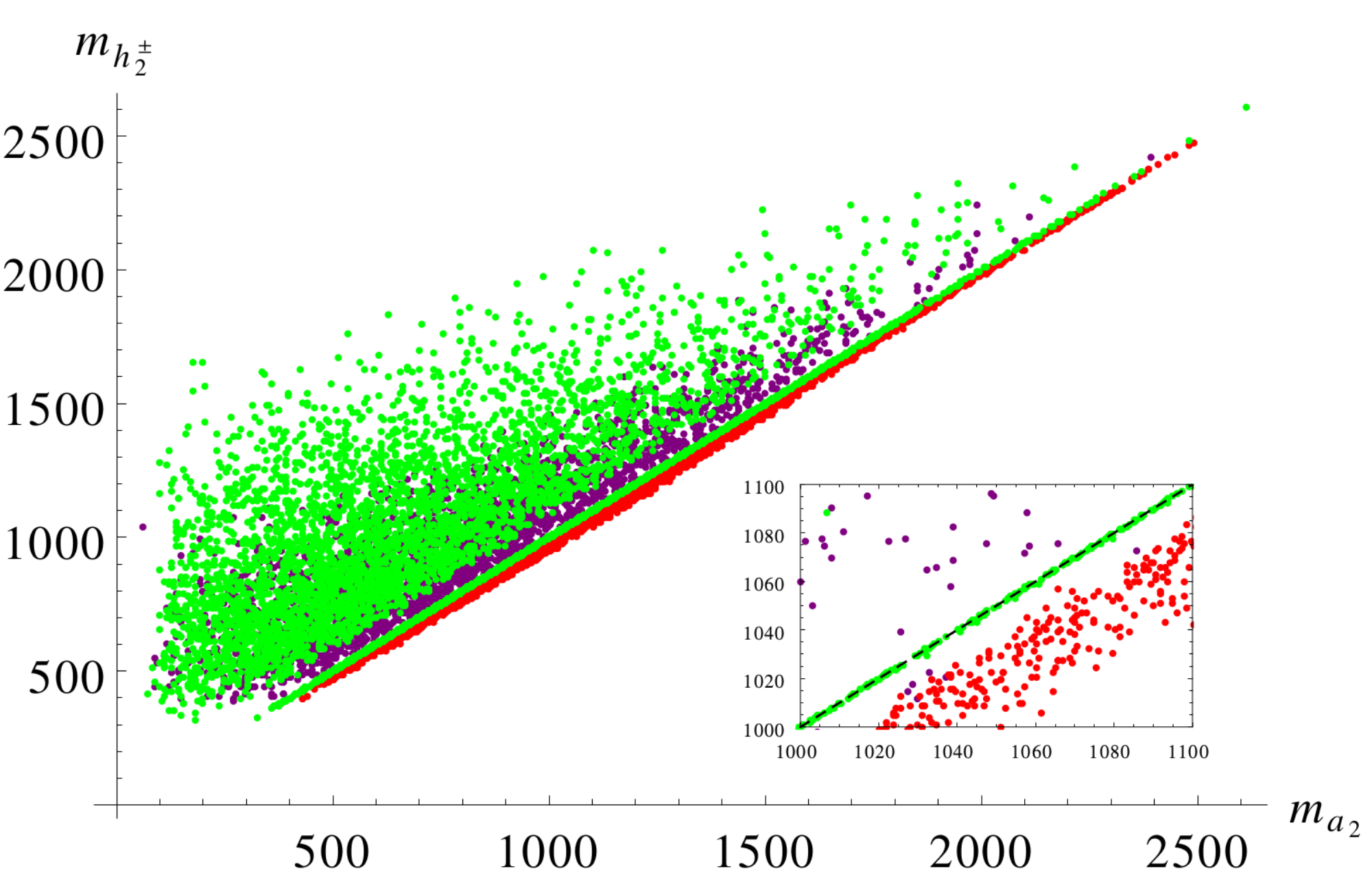}}}
\mbox{\hskip -10pt\subfigure[]{\includegraphics[width=0.45\linewidth]{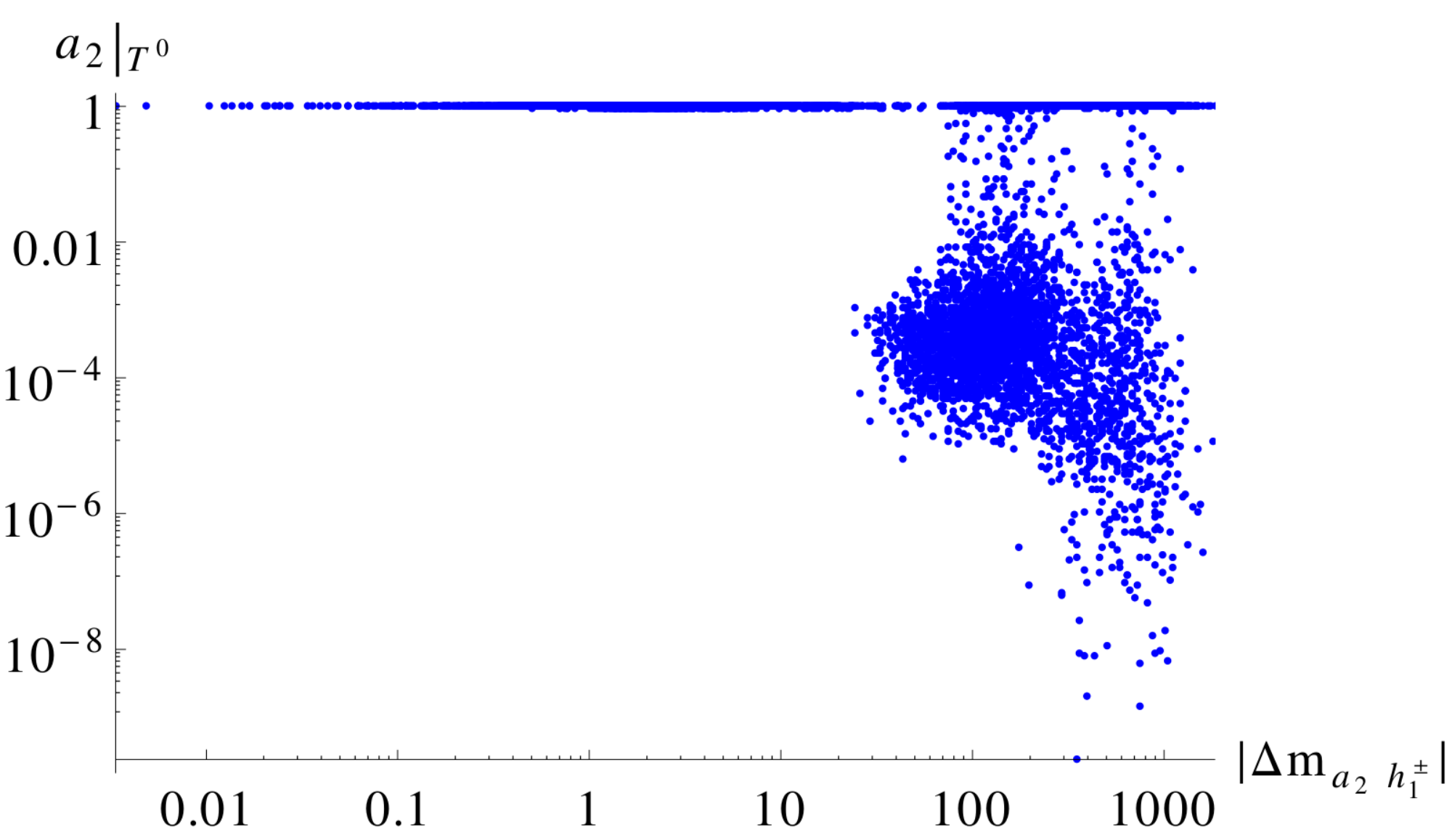}}
\subfigure[]{\includegraphics[width=0.45\linewidth]{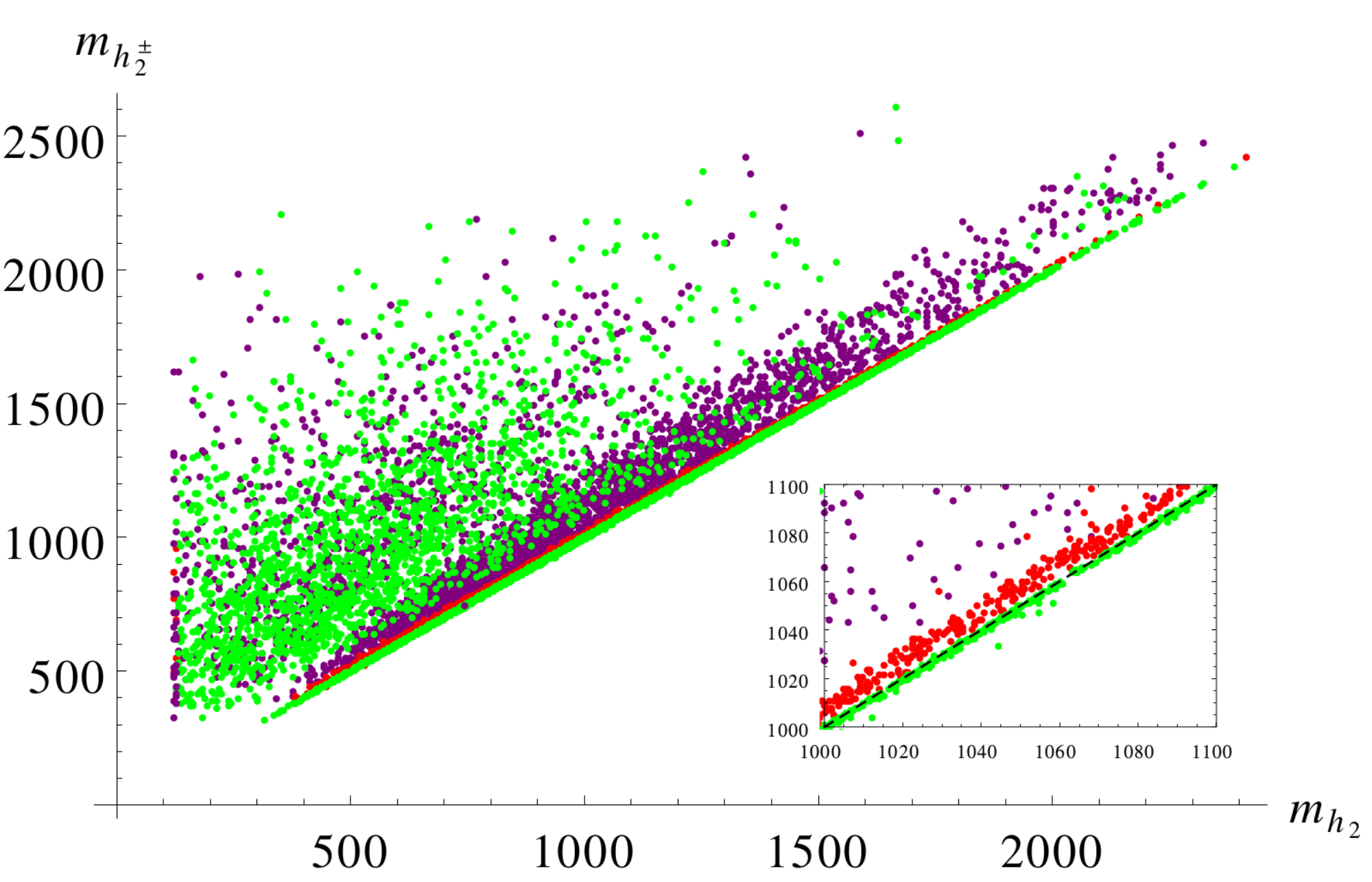}}
}
\caption{We show the fraction of triplets of $h_2$ (a) and $a_2$ (c) as a function of the mass difference $|\Delta m_{h_2/a_2\, h^\pm_1}|$ between $h_2/a_2$ and $h^\pm_1$ respectively. We plot the mass correlation between $a_2$ and $h_2^\pm$ (b) and between $h_2$ and $h_2^\pm$ (d). These exhaust the possible hierarchies for the triplet eigenstates. We mark in red the points with both $a_2$ and $h_2^\pm$ doublet-type, in purple the points with $a_2$ triplet-type and $h_2^\pm$ doublet-type or viceversa, and in green the points with both $a_2$ and $h_2^\pm$ triplet-like.}\label{ch1h2a2}
\end{center}
\end{figure}
%%%%%%%%%%%%%%%%%%%%%%%%%%%%%%
{ Apart from the LEP \cite{LEPb} and LHC  \cite{CMS2} constraints, we also ensure the validity of the constrains coming from the $B$-observables. For this particular reason we claim the light pseudoscalar $a_1$ to be $\geq 90\%$ singlet-type and the light charged Higgs $h^\pm_1$ to be $90\%$ triplet-type. 
A very light scalar or pseudoscalar, with a mass around $1-10$ GeV, gets strong bounds from bottomonium decay to $a_1\gamma$ \cite{bottomonium1}. The decay rate for $\Upsilon \to a_1 \gamma$ can be approximated as follows 
\be
\mathcal{Br}(\Upsilon \to a_1 \gamma)=\mathcal{Br}(\Upsilon \to a_1 \gamma)_{SM}\times g^2_{a_1 b\bar{b}},
\ee
where $g_{a_1 b\bar{b}}$ is the reduced down-type Yukawa coupling with respect to SM \cite{bottomonium}. We checked explicitly that the requirement of more than $90\%$ singlet type $a_1$ and low $\tan{\beta}$ ensure that we are in the region of validity. 

Another important constraint for a light pseudoscalar comes from $\mathcal{Br}(B_s \to \mu \mu)$ which can be summerised as follows \cite{bottomonium}
\be
\mathcal{Br}(B_s \to \mu \mu)\simeq \frac{2\tau_{B_s}M^5_{B_s}f^2_{B_s}}{64\pi}|C|^2( \mathcal{R}^P_{12})^4,
\ee
with 
\bea
C=\frac{G_F\alpha}{\sqrt2\pi}V_{tb}V^*_{ts}\frac{\tan^3\beta}{4\sin^2\theta_w}\frac{m_\mu m_t |\mu_r|}{m_W^2(m^2_{a_1}-m^2_{B_s})}\frac{\sin2\theta_{\tilde t}}{2}\Delta f_3\nn\\
\eea
where $\Delta f_3=f_3(x_2)-f_3(x_1)$, $x_i=m^2_{\tilde t_i}/|\mu_r|^2$, $f_3(x)=x\ln x/(1-x)$, $\theta_{\tilde t}$ is the stop mixing angle and $\mathcal{R}^P_{12}$ is the rotation angle, defined in Eq.~\ref{chmix}, which gives the coupling with the down type Higgs ($H_d$) with leptons and down type quarks.  The demand of mostly singlet $a_1$ ($\geq 90\%$) on the data set ensures that we are well below the current upper limit \cite{lhcb}. 

Other constraint that affects the models with extra Higgs boson, specially the charged Higgs bosons, comes from the rare decay of $B\to X_s \gamma$. The charged Higgs bosons which are doublet in nature couple to quarks via Yukawa couplings and contribute to the rare decay of $B\to X_s \gamma$. Similar contributions also come from the charginos which couple to the quarks, namely doublet-type Higgsinos and Wino. However when we have charged Higgs or charginos which are triplet in nature they do not couple to the fermions and thus do not contribute in such decays \cite{pbas1,pbas2}. If the light charged Higgs bosons are triplet in nature the dominant Wilson coefficients $F_{7,8}$ are suppressed by the charged Higgs rotation angles $\mathcal{R}^C_{11,13}$ as defined in Eq.~\ref{chmix}. The demand of the light charged Higgs boson mostly triplet $\geq 90\%$ enable us to avoid the constraint from $\mathcal{Br}(B\to X_s \gamma)$ \cite{pbas1,pbas2}.

}

%%%%%%%%%%%%%%%%%%%%%%%%%%%%
\begin{figure}[thb]
\begin{center}
\mbox{\subfigure[]{
\includegraphics[width=0.6\linewidth]{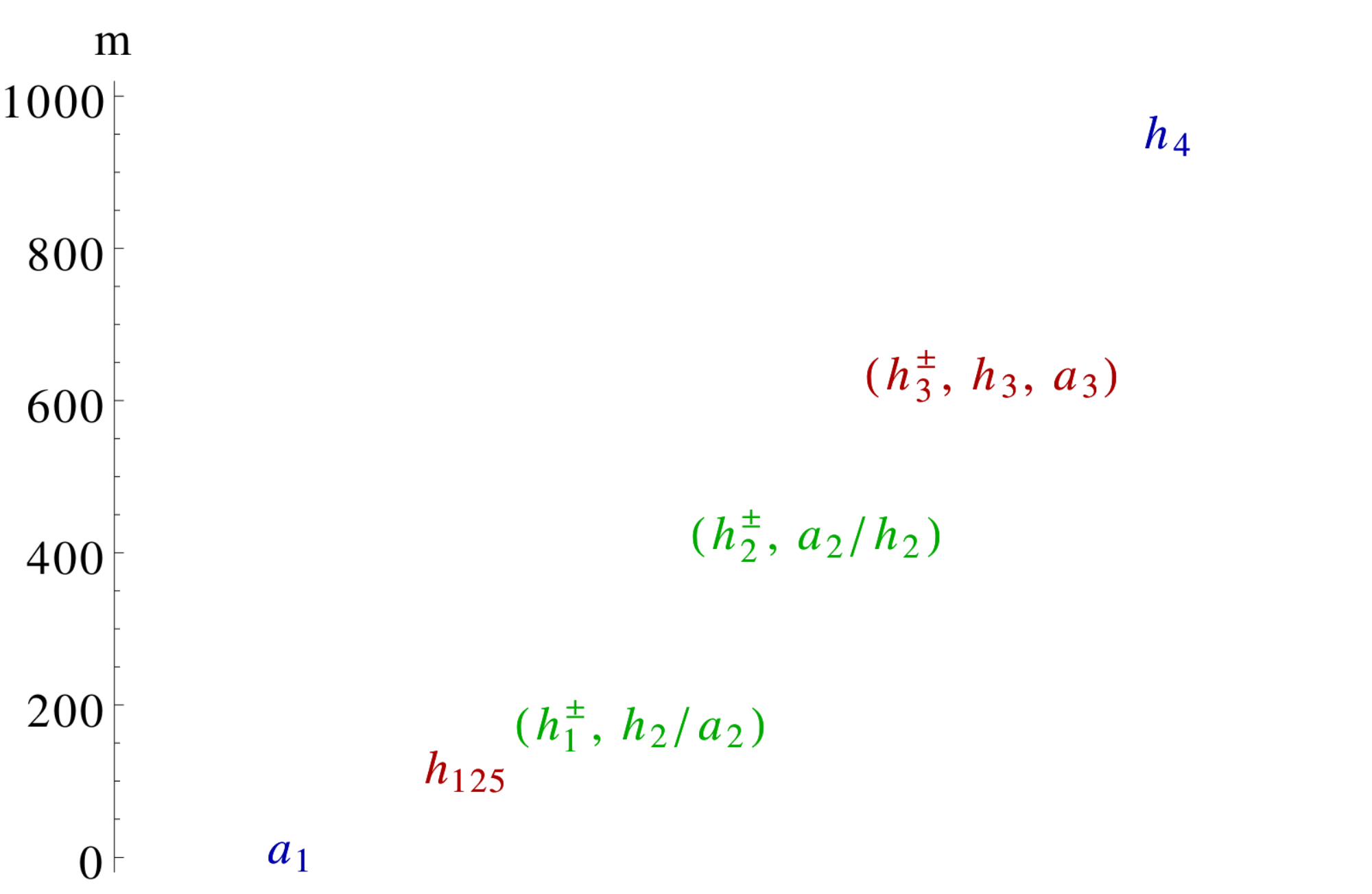}}}
\caption{A typical mass hierarchy of the scalar sector, with the singlet in blue, the doublets in red 
and the triplet Higgs bosons in green colour. The eigenstates of the triplet sector with $a_2/h_2$ or  
$h_2/a_2$ are alternative: if $h_1^\pm$ pairs with the neutral $h_2$, then $h_2^\pm$ is mass degenerate with the pseudoscalar $a_2$ (and viceversa).
}\label{cartoon}
\end{center}
\end{figure}
%%%%%%%%%%%%%%%%%%%%%%
%\end{document}
In Figure \ref{ch1h2a2}(a) we plot the triplet fraction of $h_2$ in function of the mass splitting between $h_2$ and $h_1^\pm$. The lightest charged Higgs is selected to be triplet-like ($\geq 90\%$). It is evident that in the case of mass degeneracy between  $h_2$ and $h_1^\pm$ the triplet-like structure of $h_1^\pm$ is imposed also on $h_2$. In Figure \ref{ch1h2a2}(b) we plot the mass correlation between $a_2$ and $h_2^\pm$. We use the following color code: we mark in red the points with both $a_2$ and $h_2^\pm$ doublet-type, in purple the points with $a_2$ triplet-type and $h_2^\pm$ doublet-type or viceversa, and in green the points with both $a_2$ and $h_2^\pm$ triplet-like. In the zoomed plot the dashed line indicates a configuration of mass degeneracy. It is evident that the mass degeneracy between $a_2$ and $h_2^\pm$ implies that both of them are triplet-like. As we have depicted in Figure \ref{cartoon}, there could be an exchange between $a_2$ and $h_2$ in the triplet pairs, shown in green. For this reason we illustrate also the other possible hierarchy path in Figure \ref{ch1h2a2}(c) and \ref{ch1h2a2}(d). As one may notice, the two sets of plots are qualitatively similar, although there is a quantitative difference between the red points of Figures \ref{ch1h2a2}(b) and \ref{ch1h2a2}(d). The points in the latter are closer than the former to the line of mass degeneracy.
 %%%%%%%%%%%%%%% h4 a1 %%%%%%%%%%%
\begin{figure}[thb]
\begin{center}
\mbox{\hskip -10pt\subfigure[]{\includegraphics[width=0.55\linewidth]{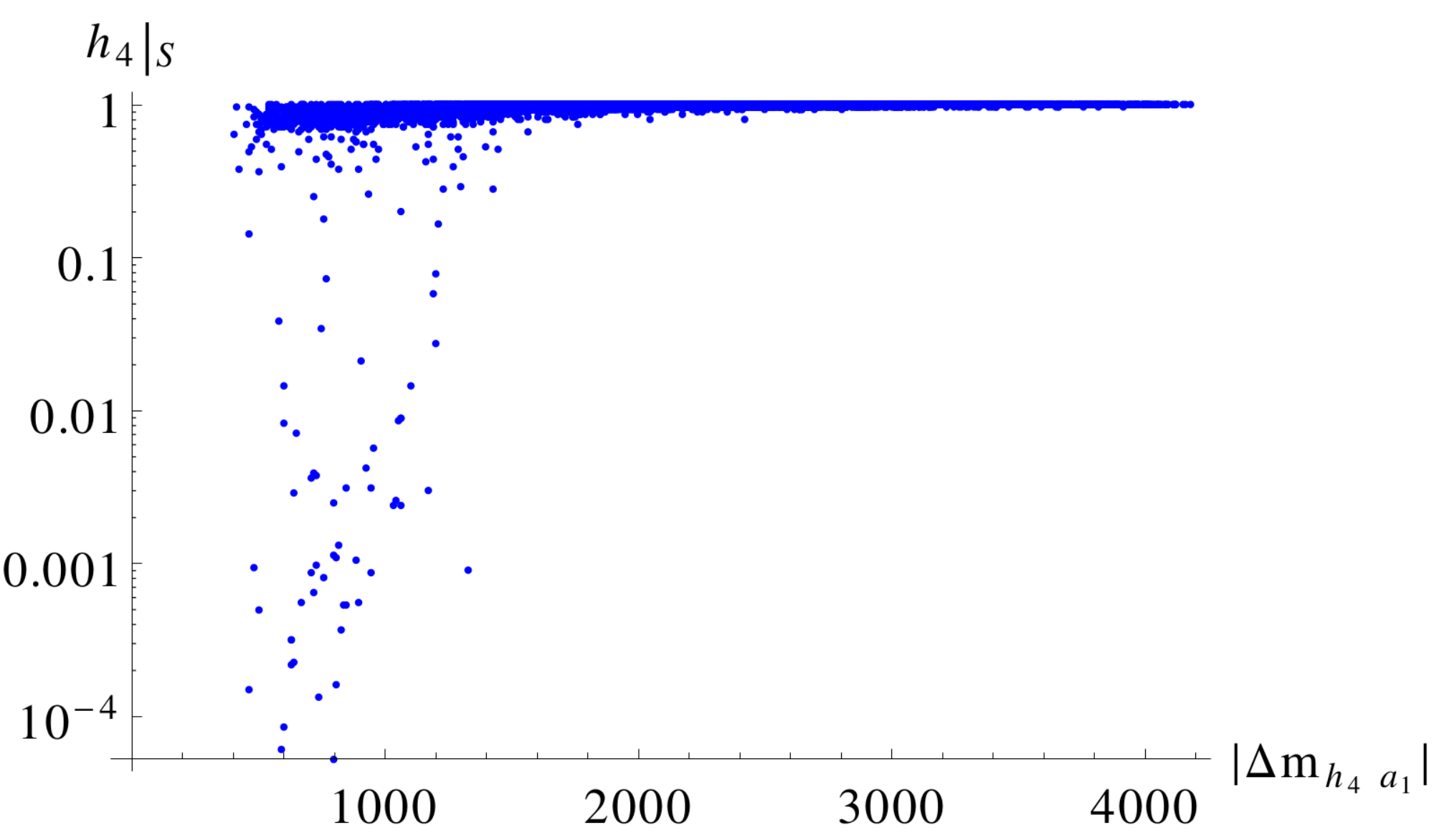}}
\subfigure[]{\includegraphics[width=0.55\linewidth]{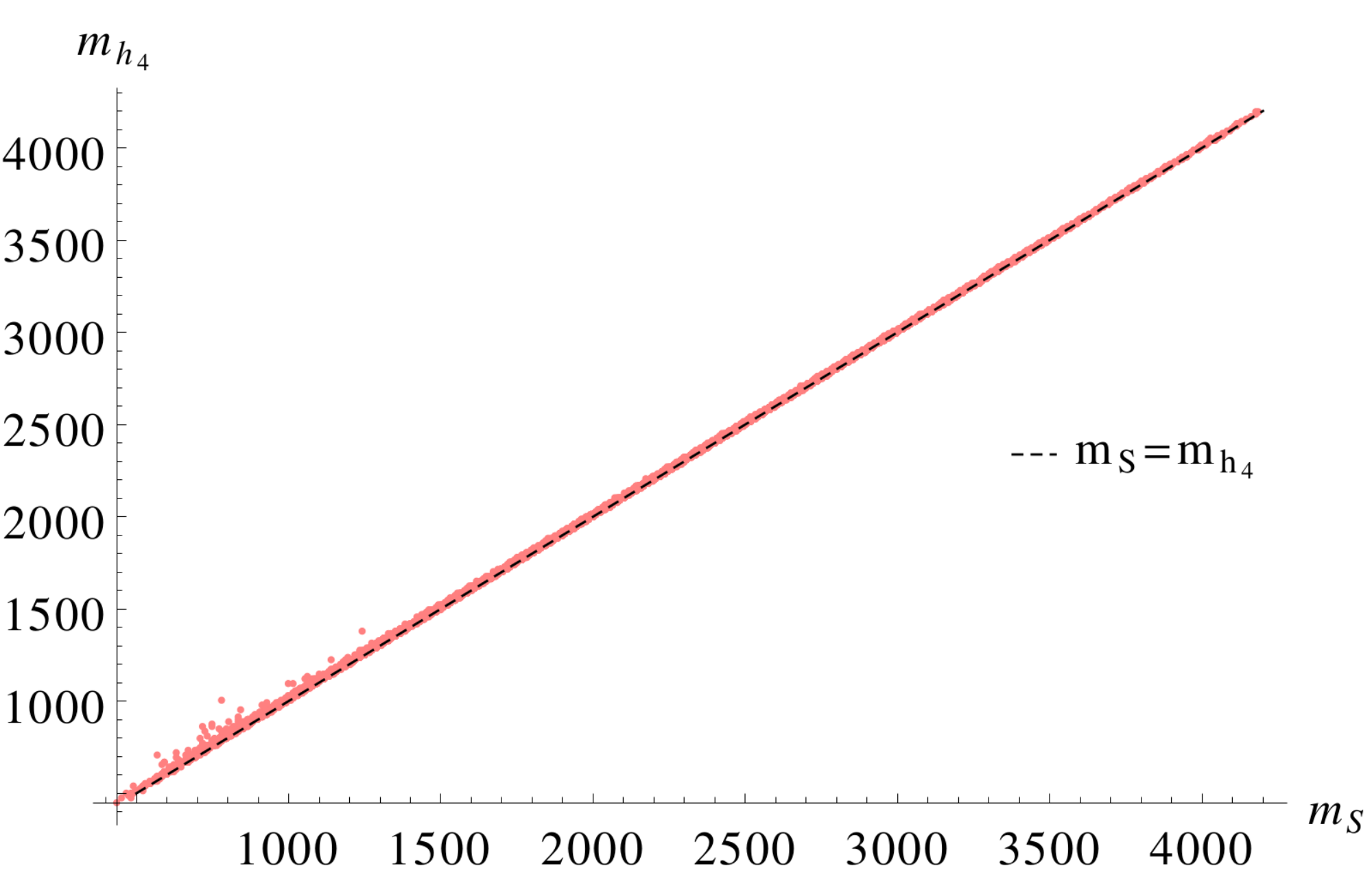}}
}
\caption{We show the singlet fraction of $h_4$ as a function of mass difference $|\Delta m_{h_4\, a_1}|$ 
between the two states $h_4$ and $a_1$ (a), and the mass correlation between $h_4$ and $m_S$ (b).}\label{h4a1}
\end{center}
\end{figure}
%%%%%%%%%%%%%%%%%%%%%%%%%%%%%%
Figure \ref{h4a1}(a)  shows that the more $h_4$ is decoupled,
compared to $a_1$, the more tends to be in a singlet-like eigenstate. We remind that $a_1$ is a pseudo NG mode and hence it is naturally light. From Figure \ref{h4a1}(b) it is evident that $h_4$ takes the soft mass $m_S$ coming from the singlet.
%%%%%%%%%%%%%%%%%%%%%%%%%%%%%%%%%%%%%%%%
\begin{figure}[htb]
\begin{center}
\mbox{\hskip -10pt\subfigure[]{\includegraphics[width=0.55\linewidth]{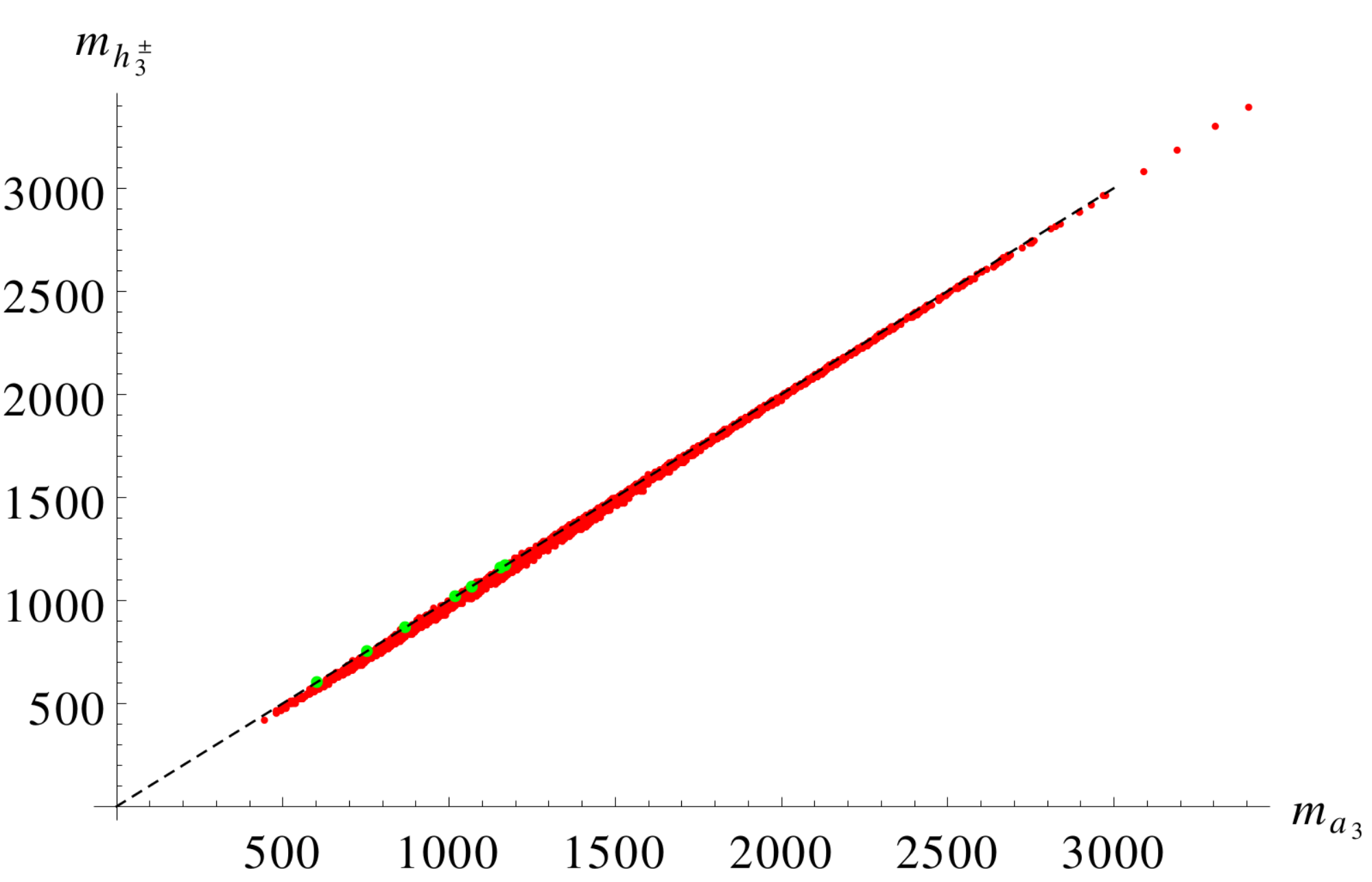}}
\subfigure[]{\includegraphics[width=0.55\linewidth]{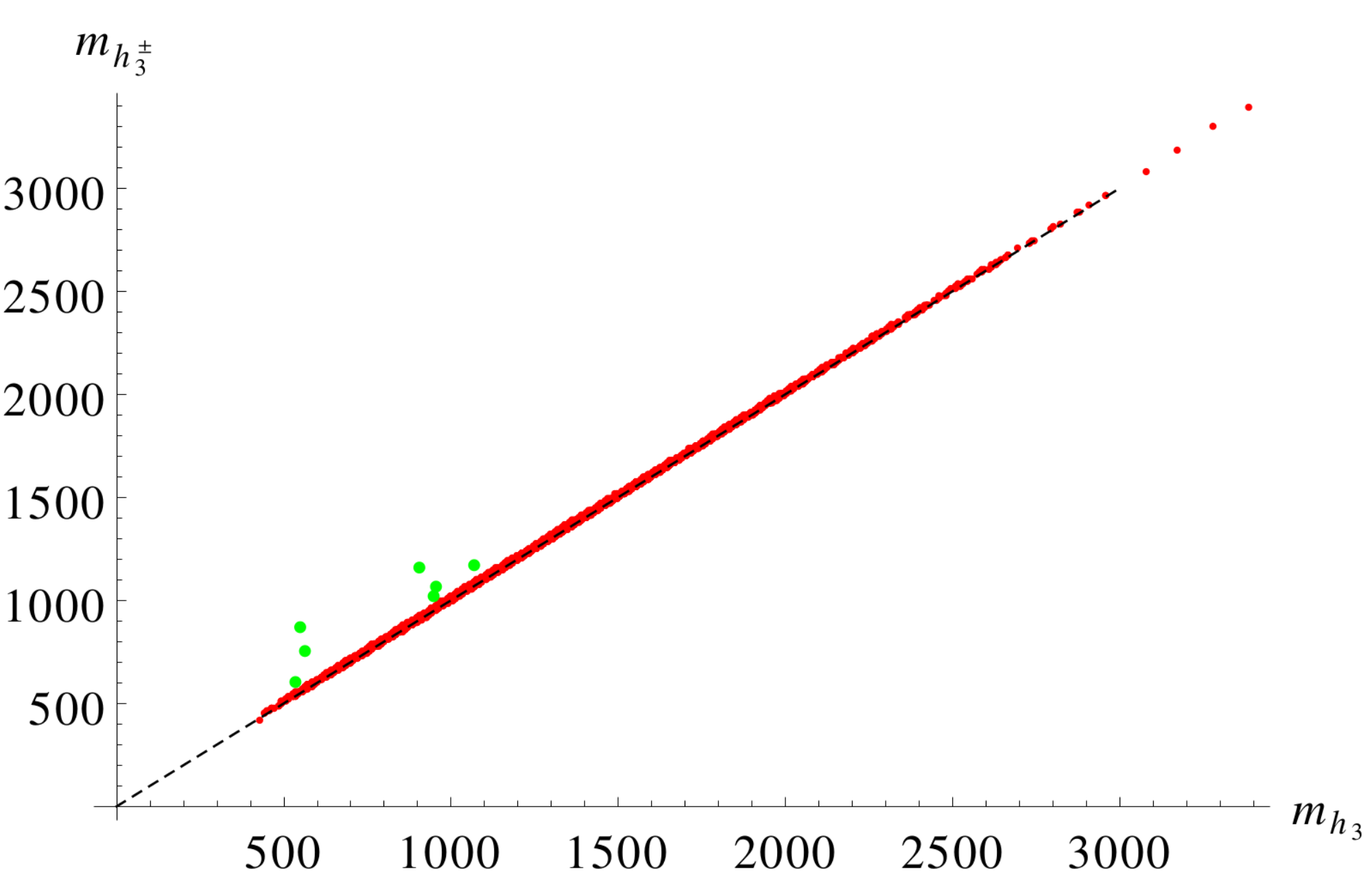}}}
\caption{Scattered plots of the mass correlation between $a_3$ and $h_3^\pm$ (a) and between $h_3$ and $h_3^\pm$ (b). The color code is defined as follows: we mark in red the points where $h_3, a_3, h^\pm_3$ are mostly doublets ($\geq90\%$) and in green the points where they are mostly triplet.}\label{mcrl2}
\end{center}
\end{figure}
%%%%%%%%%%%%%%%%%%%%%%%%%%%%%%%
Figure~\ref{mcrl2}(a) shows the mass correlations between $h^\pm_3$ and $a_3$, while Figure~\ref{mcrl2}(b)  shows the same correlation but between $h^\pm_3$, $h_3$  where 
all of them are of doublet-type nature and are marked in red. It is easily seen that all the
three doublet-like Higgs bosons $h^\pm_3$, $h_3$ and $a_3$ remain degenerate. 
There are only 7 points which behave like triplets and are shown in green. 
Thus it is evident from the above analysis that eigenstates dominated by the same representation (i.e mostly singlet or mostly triplet) tend to be hierarchically clustered. In this case of a $Z_3$ symmetric Lagrangian, the light pseudoscalar is actually a pseudo NG mode of a continuous $U(1)$ symmetry of the Higgs potential, also known as R-axion \cite{Ellwanger}, and remains very light across the entire allowed parameter space.

Though the interaction eigenstates are a mixture of the gauge eigenstates, there seems to be a pattern for the various representations of the Higgs sector.
A given representation tries to keep their masses in the same block, i.e., the masses of scalar, pseudoscalar and charged components of the triplets will form a different mass block than the doublet Higgs sectors. A typical mass hierarchy is shown in Figure~\ref{cartoon}, where  a light pseudoscalar which is a pseudo NG boson lays hidden below $100$ GeV and the scalar state $h_4$ takes a heavy mass $\sim m_S$, and is therefore decoupled from the low energy spectrum. There is a CP-even Higgs boson of doublet type around $125$ GeV and  doublet-like heavy Higgs bosons of larger mass ($h^\pm_3, h_3, a_3$), shown in red. Apart from doublet and singlet interaction eigenstates, we have two triplets $T_1$ and $T_2$ which then forms two different sets, ($h^\pm_1, h_2/a_2$)  and ($h^\pm_2, a_2/h_2$) in the mass hierarchy, shown in green colours. Of course this is not the most general situation but it comes from the phenomenological constraints that should be applied to the scanned points in the parameter space. We remind again that these constraints include a scalar Higgs boson with a mass around 125 GeV which satisfy the LHC constraint of Eq.~\ref{LHCdata} and no light doublet-like pseudoscalar or charged Higgs boson. We take care of the latter requesting that the lightest pseudoscalar  as mostly singlet and lightest charged Higgs boson is mostly triplet.

\section{Charged Higgs bosons and its structure}\label{chcdcy}
In this section we will describe the feature of the charged Higgs sector, emphasizing the role of the rotation angles in the limit $|\lambda_T|\simeq0$. The charged Higgs bosons are a mixture of two doublet and two triplet fields, as can be seen from Eq.~\ref{chH},
\be\label{chH}
h^\pm_i= \mathcal{R}^C_{i1}H_u^+ +  \mathcal{R}^C_{i2}T_2^+ + \mathcal{R}^C_{i3}H_d^{-*} +  \mathcal{R}^C_{i4}T_1^{-*}
\ee
with $\mathcal{R}^C_{i1, i3}$  and $\mathcal{R}^C_{i2, i4}$ determining the doublet and triplet part respectively.  In general $\mathcal{R}^C_{ij}$ is a function of all the vevs, $\lambda_{T, TS, S}$ and the $A_i$ parameters and we can write schematically
\bea\label{rc}
\mathcal{R}^C_{ij} = f^C_{ij}\left(v_u, v_d, v_T, v_S, \lambda_T, \lambda_{TS}, \lambda_S, A_i\right).
\eea
The charged Higgs mass matrix which is given in appendix (Eq.~\ref{chMM}), shows the similar dependency on the parameters. However, the charged Goldstone mode, expressed in terms of the gauge eigenstates, is a function only of the vevs and the gauge couplings, as we expect from the Goldstone theorem. 
\bea\label{gstn}
h_0^\pm=\pm N_T \left(\sin\beta H_u^+\, -\cos\beta H_d^{-*}  \,\mp\sqrt2\,\frac{v_T}{v}(T_2^+ +T_1^{-*})\right)\, ,\quad N_T=\frac{1}{\sqrt{1+4\frac{v_T^2}{v^2}}}
\eea
Eq.~\ref{gstn} presents the explicit expression of the charged Goldstone mode and we can see that it is independent of any other kind of couplings or parameters. Among the three kind of vevs entering in the charged Goldstone mode, the triplet  vev is very small ($v_T\lesssim 5$ GeV) due to its contribution in the $W^\pm$ boson mass, as already discussed. The triplet vev, being restricted by the $\rho$ parameter \cite{rho}, makes the charged Goldstone always doublet-type. However among the massive states in the gauge basis, two of them are triplet-like and one is doublet-like. We shall see later that this small triplet contribution to the Goldstone boson protects one of the three physical charged Higgs bosons from becoming absolute triplet-like.

%%%%%%%%%\lambda_T ~0 feature %%%%%%%%%%%%
\begin{figure}[thb]
\begin{center}
\includegraphics[width=0.6\linewidth]{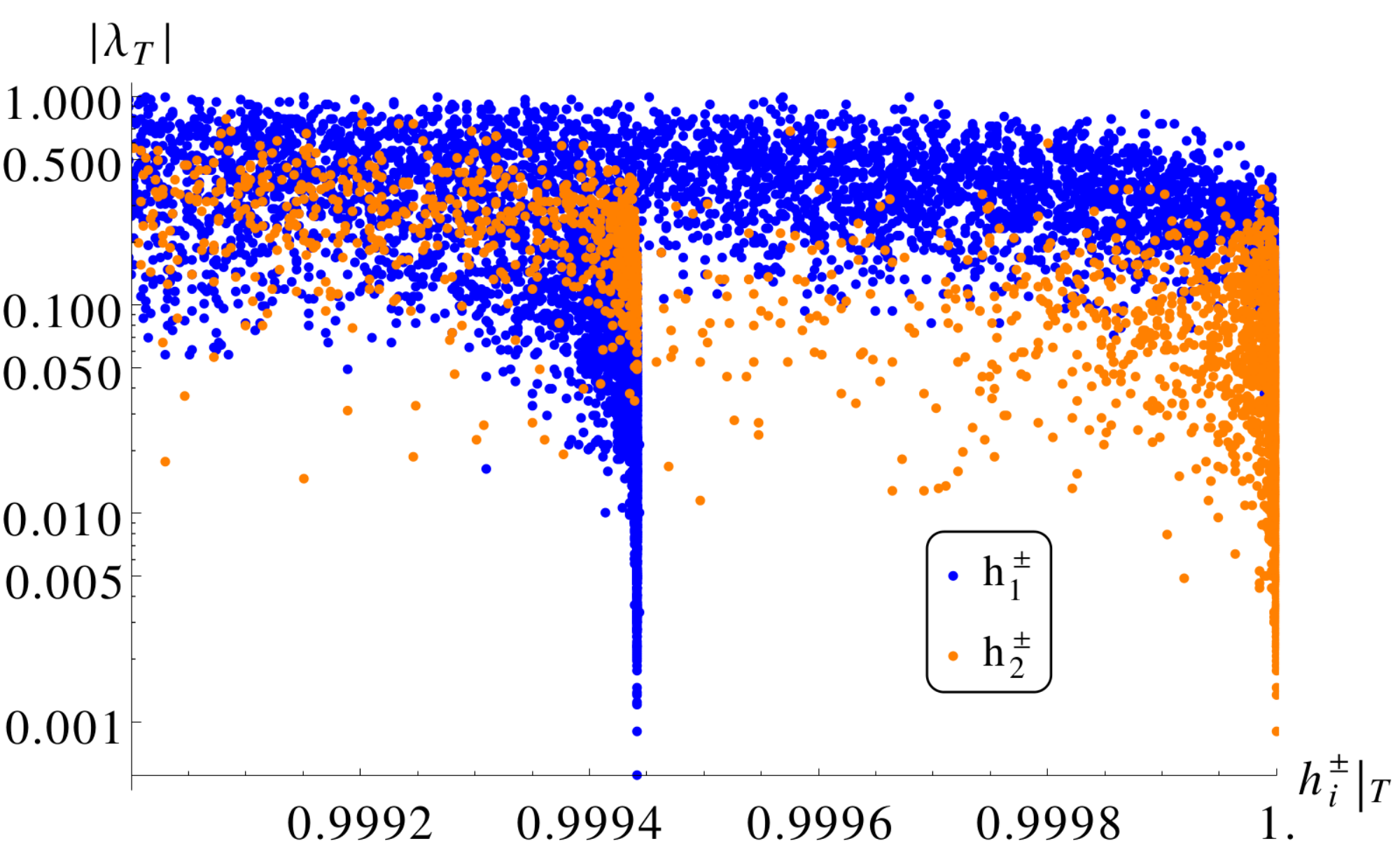}
\includegraphics[width=0.6\linewidth]{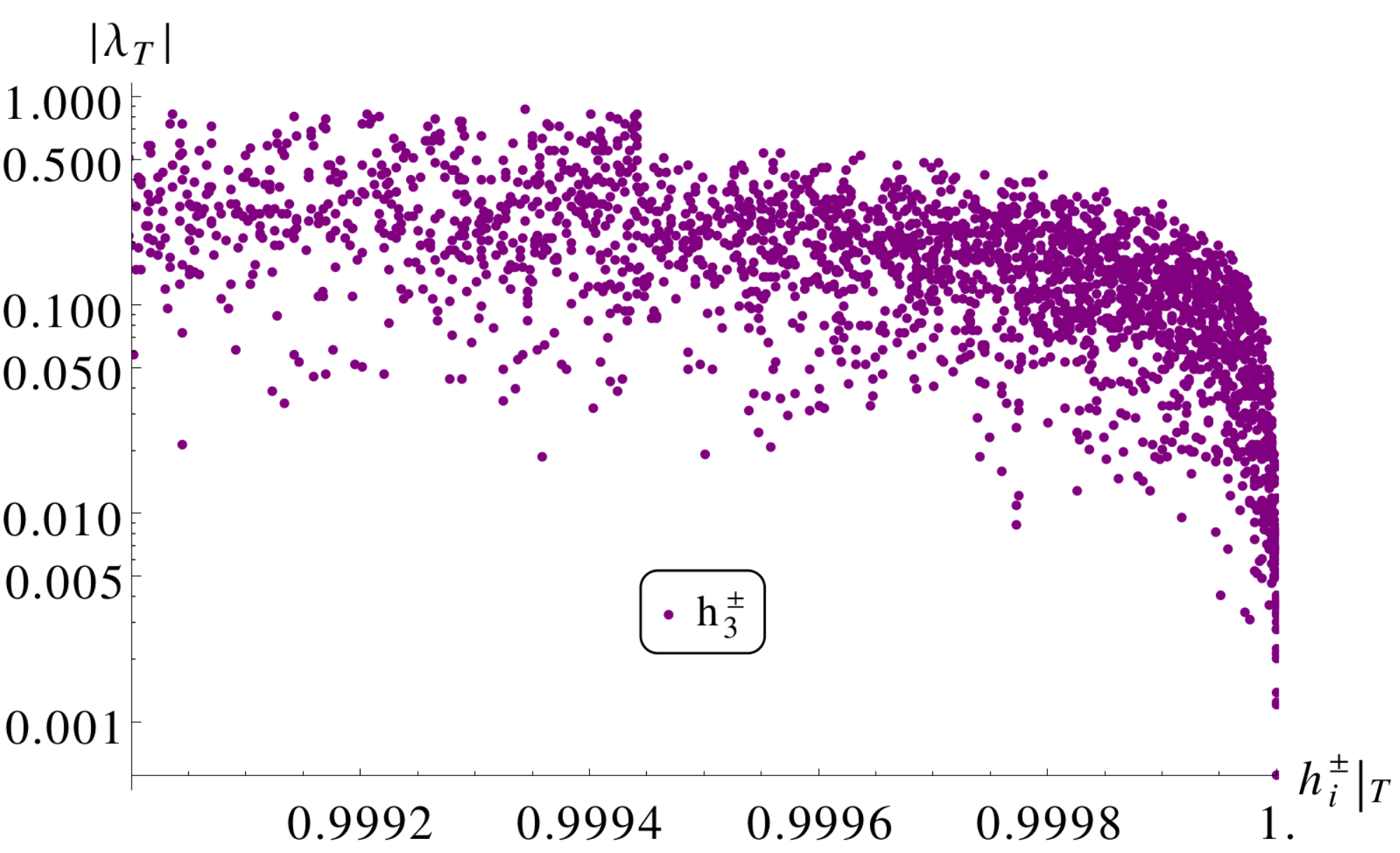}
\caption{Triplet component of the massive charged Higgs bosons versus $\lambda_T$.}\label{chpslmbda}
\end{center}
\end{figure}
%%%%%%%%%%%%%%%%%%%%%%
In Figure~\ref{chpslmbda} we show the structure of the 
charged Higgs bosons as a function of  $|\lambda_T|$, where we demand 
the lightest charged Higgs massive state to be mostly triplet. One can realize that that for a non-zero $\lambda_T$, their tendency is to mix. However, as we move towards the $|\lambda_T|\simeq 0$ region, one of the charged Higgs boson gives away the $\sim (\frac{v_T}{v})^2$ triplet part to the charged Goldstone and fails to become 100\% triplet (see the blue points in Figure~\ref{chpslmbda}). 
{ In the models where $A_T$ parameter is proportional to $\lambda_T$, the mixing induced by the soft parameter $A_T$ automatically goes to zero in this limit. However the mixing of doublet and triplet  in the charged  Goldstone comes from the corresponding vevs and it is independent of $\lambda_T$ or $A_T$ as can be seen from Eq. 21. Now all the other massive charged Higgs bosons are orthogonal to the Goldstone boson, which makes the similar mixing in the massive states as well. This mixing goes to zero only when the triplet does not play any role 
in EWSB, i.e. $v_T=0$. However for non-zero $\lambda_T$ and $A_T$ the additional mixings come for the massive eigenstates.}

Anyone of the three massive charged Higgs boson can show this feature but we see it only for $h_1^\pm$ because in the selection criteria we have demanded that $h_1^\pm$ must be triplet-like. Thus for non-zero triplet vev even with $|\lambda_T|=0$, complete decoupling of doublet and triplet representations is not possible. Therefore by 'decoupling limit' we mean $|\lambda_T|\simeq 0$ here onwards.  In this decoupling limit either the $h^\pm_2$ or the $h^\pm_3$ become completely of triplet-type.  A similar conclusion was shown for the triplet extension of the supersymmetric standard model \cite{EspinosaQuiros}.
%%%%%%%%% SS and OS feature of R2 and R4 %%%%%%%%%%%%
\begin{figure}[thb]
\begin{center}
\includegraphics[width=0.78\linewidth]{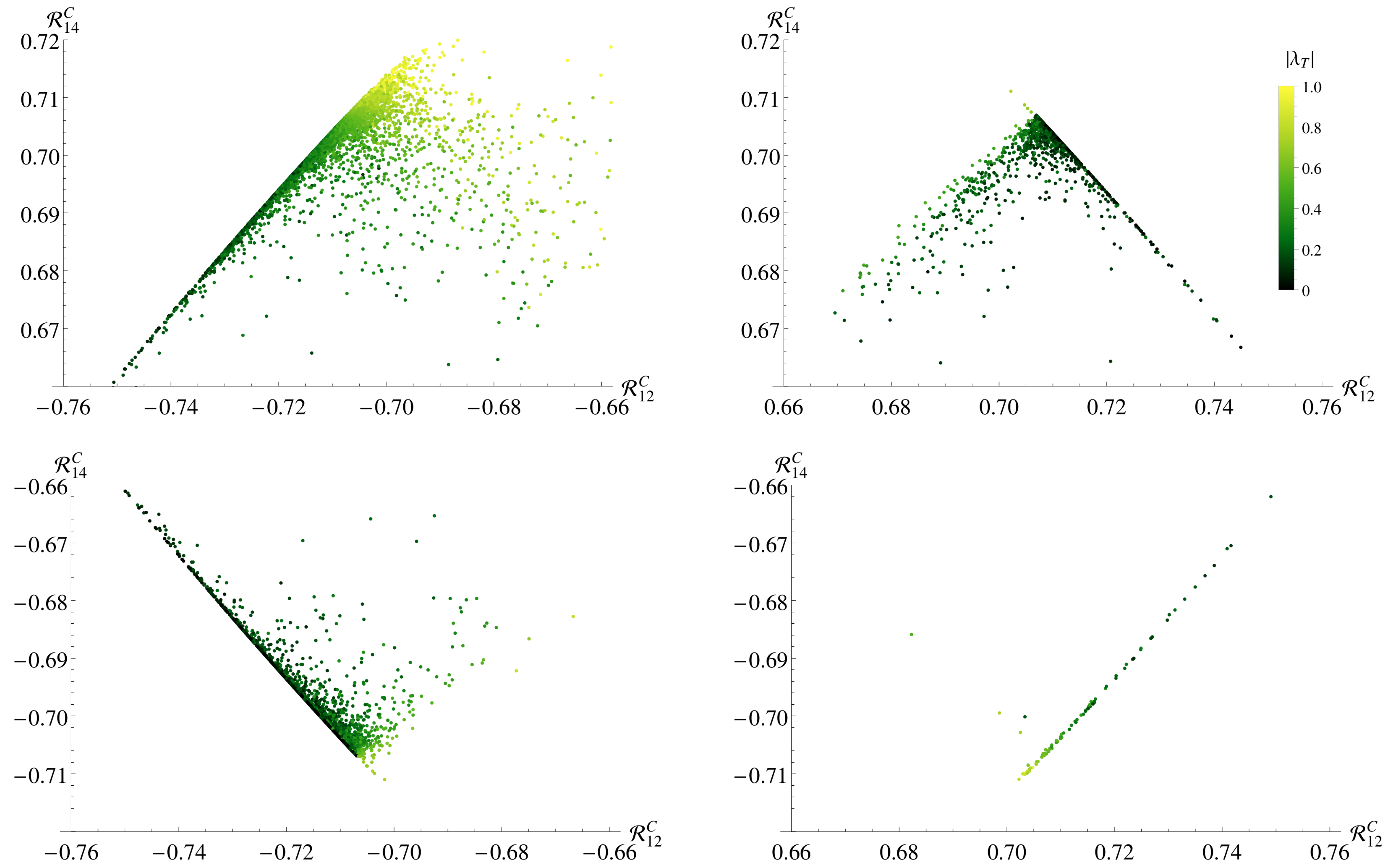}
\caption{Correlations of the rotation angles of the lightest charged Higgs boson $h^\pm_1$ as a function of $\lambda_T$.}\label{ssoslmbda}
\end{center}
\end{figure}
%%%%%%%%%%%%%%%%%%%%%%

%%%%
\begin{table}
\begin{center}
\renewcommand{\arraystretch}{1.4}
\begin{tabular}{||c||c|c||}
\hline%\hline
&$10^{-2}<|\lambda_T|<1$&$|\lambda_T|<10^-2$\\
\hline
\rm{sign} $\mathcal{R}^C_{12}$ $\mathcal{R}^C_{14}$&+ or -&+\\
\hline
%\hline
\end{tabular}
\caption{The sign of the product $\mathcal{R}^C_{12}$ $\mathcal{R}^C_{14}$. The sign of the two rotation angles of the lightest charged Higgs boson plays a crucial role in the interactions of a triplet-like charged Higgs boson. In the limit $|\lambda_T|\sim0$ these two rotation angles have the same sign. This feature has important consequences for the interaction, and hence the cross-section, of the lightest charged Higgs boson in various channels.}\label{r2r4s}
\end{center}
\end{table}

%%%%%%%%%%%%%%%%%%%%%%%%%%%%%%%%%%

The decoupling limit of $|\lambda_T|\sim 0$ not only affects the structure of the charged Higgs bosons, where two of them become triplet-like and one of them doublet-like, but also affects the respective coupling via the corresponding rotation angles. In Figure~\ref{ssoslmbda} we show the rotation matrix elements for the light charged Higgs boson $h^\pm_1$ with respect to $|\lambda_T|$. We can see that when $\lambda_T$ becomes very small the mixing angles in the triplet component of the light charged Higgs boson $h^\pm_1$, $\mathcal{R}^C_{12}$ and $\mathcal{R}^C_{14}$, as defined in Eq.~\ref{chH}, take same signs, unlike the general case. We will see later that the presence of same signs in $\mathcal{R}^C_{12}$ and  $\mathcal{R}^C_{14}$  in the decoupling limit, causes an enhancement of some production channels and decrement for some other ones.

\section{Decays of the charged Higgs bosons}\label{chdcys}

As briefly mentioned above, the phenomenology of the Higgs decay sector of the TNMSSM, as discussed in \cite{TNMSSM1}, is affected by the presence of a light pseudoscalar which induces new decay modes. In this section we consider its impact in the decay of a light charged Higgs boson $h^\pm_1$. Along with the existence of the light pseudoscalar, which opens up the $h^\pm_1 \to a_1 W^\pm$ decay mode, the triplet-like charged Higgs adds new decay modes, not possible otherwise. In particular, a $Y=0$ triplet-like charged Higgs boson gets a new decay mode into $ZW^\pm$ which is a signature of custodial symmetry breaking.  Apart from that, the usual doublet-like decay modes into $\tau\nu$ and $tb$ are present via the mixings with the doublets. 

\subsection{$h_i^\pm  \to W^\pm h_j/a_i$}
The trilinear couplings with charged Higgses, scalar (pseudoscalar) Higgses and $W^\pm$ are given by 
\begin{align}\label{hachW}
g_{h_i^\pm W^\mp h_j}&=\frac{i}{2}g_L\Big(\mathcal R_{j2}^S\mathcal R_{i3}^C-\mathcal R_{j1}^S\mathcal R_{i1}^C+\sqrt2\mathcal R_{j4}^S\left(\mathcal R_{i2}^C+\mathcal R_{i4}^C\right)\Big), \\
g_{h_i^\pm W^\mp a_j}&=\frac{g_L}{2}\Big(\mathcal R_{j1}^P\mathcal R_{i1}^C+\mathcal R_{j2}^P\mathcal R_{i3}^C+\sqrt2\mathcal R_{j4}^P\left(\mathcal R_{i2}^C-\mathcal R_{i4}^C\right)\Big).
\end{align}
Both the triplet and doublet has $SU(2)$ charges so they couple to $W^\pm$ boson. Their coupling in association with neutral Higgs bosons have to be  doublet(triplet) type for doublet(triplet) type charged Higgs bosons. For the phenomenological studies we have considered a doublet-like Higgs boson around $125$ GeV, a light triplet-like charged Higgs boson $\lesssim 200$ GeV and a very light singlet type pseudoscalar $\sim 20$ GeV. Hence the mixing angles become really important. In the next few section we will see how the various rotation angles involved with the charged Higgs bosons and their relative signs determine the strength of the couplings and thus of the decay widths. Eq.~\ref{hachW} shows that for $h_i^\pm  \to W^\pm h_j$ decay the rotation angles $\mathcal R^C_{i2}$ and $\mathcal R^C_{i4}$ come as additive where as for $h_i^\pm  \to W^\pm a_j$ they come as subtractive.

\begin{figure}[thb]
\begin{center}
\includegraphics[width=0.7\linewidth]{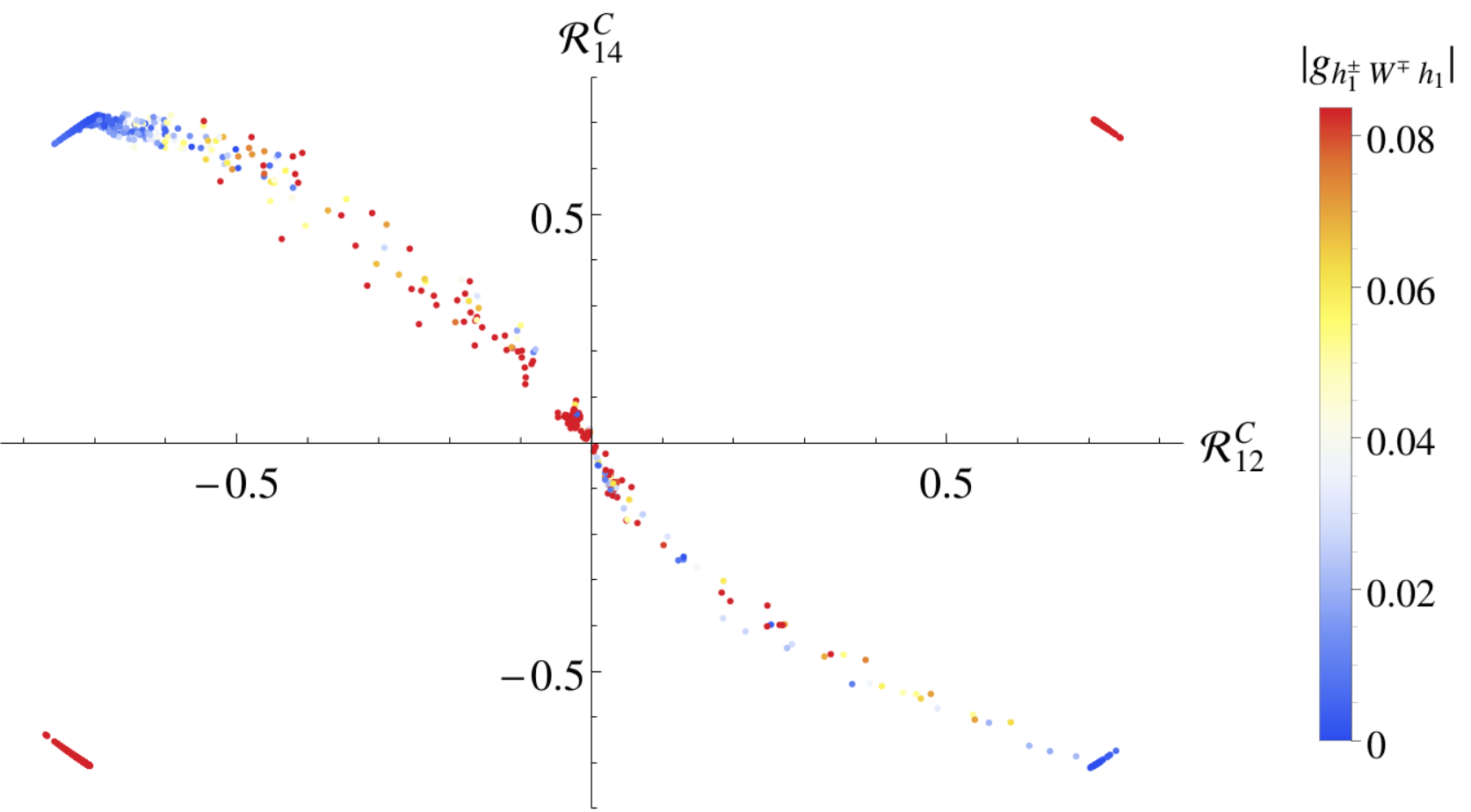}
\caption{Correlation of $g_{h^\pm_1 W^\mp h_1}$ with  $\mathcal{R}^C_{12}$ and $\mathcal{R}^C_{14}$. { The coupling is enhanced when $\mathcal{R}^C_{12}$ and $\mathcal{R}^C_{14}$ are small, i.e. for a doublet-like charged Higgs $h_1^\pm$. The enhancement in the I and III quadrants are related to the same sign of $\mathcal{R}^C_{12}$ and $\mathcal{R}^C_{14}$, cfr. Eq.~\ref{hachW}.}}\label{ghmp1h1W}
\end{center}
\end{figure}

The decay width of a massive charged Higgs boson in a $W$ boson and a scalar (or pseudoscalar) boson is given by
\bea \label{chwah}
\Gamma_{h_i^\pm\rightarrow W^\pm h_j/a_j}&=&\frac{G_F}{8\sqrt2\pi}m^2_{W^\pm}|g_{h_i^\pm W^\mp h_j/a_j}|^2 \,\sqrt{\lambda(1,x_W,x_{h_j/a_j})}\,\lambda(1,y_{h_i^\pm},y_{h_j/a_j})
\eea

where $x_{W,h_j}=\frac{m^2_{W,h_j}}{m^2_{h_i^\pm}}$ and $y_{h_i^\pm,h_j}=\frac{m^2_{h_i^\pm,h_j}}{m^2_{W^\pm}}$ and similarly for $a_j$.
%%%%%%%%% ghmp1A1W with R2 and R4 %%%%%%%%%%%%
\begin{figure}[thb]
\begin{center}
\includegraphics[width=0.7\linewidth]{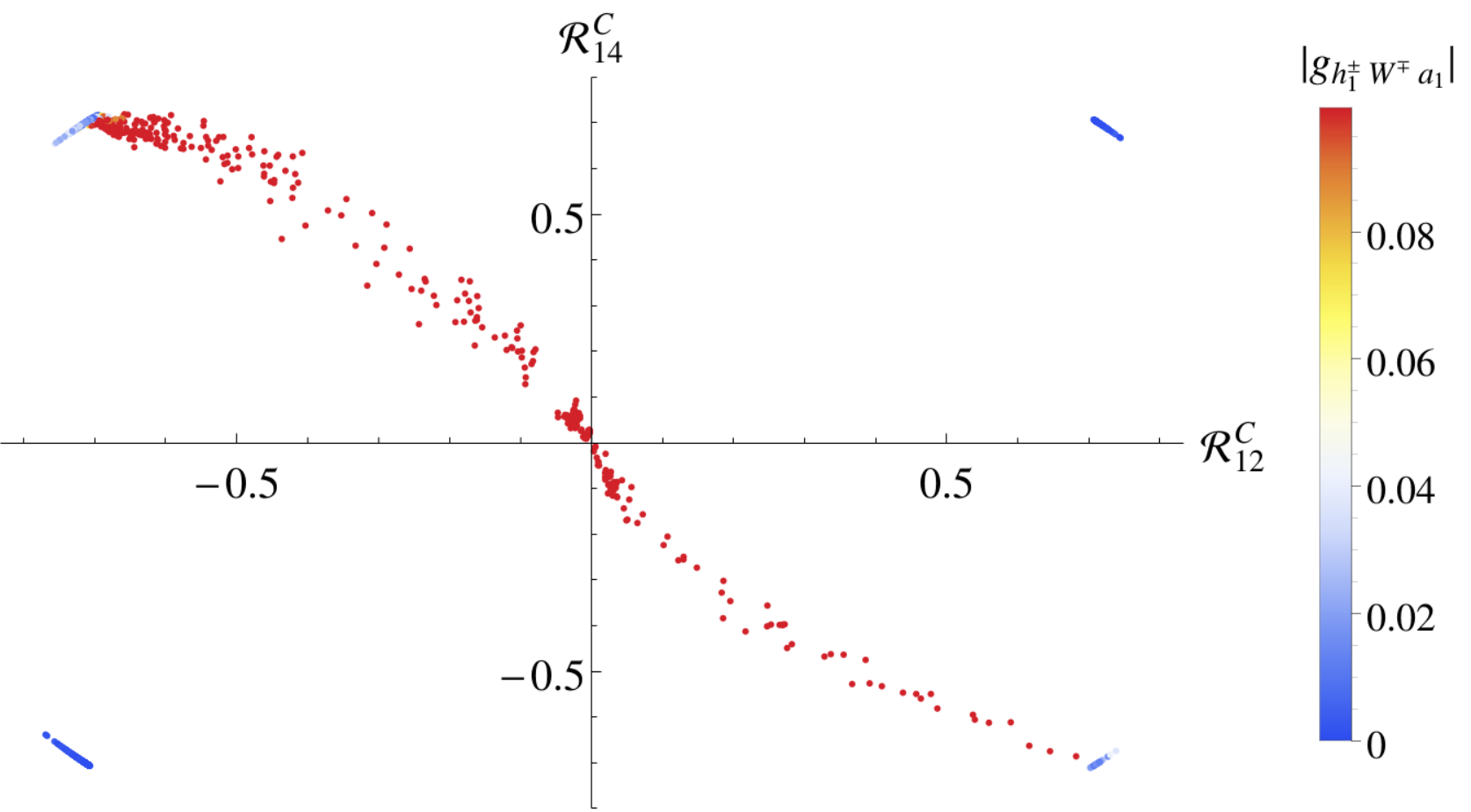}
\caption{Correlation of $g_{h^\pm_1 W^\mp a_1}$ with  $\mathcal{R}^C_{12}$ and $\mathcal{R}^C_{14}$. { For the blue points in II and IV quadrants the low values of the coupling are due to the selection of a singlet-like $a_1$, which means that $\mathcal{R}^P_{13}\sim1$, whereas for the blue points in the I and  III quadrants the low value of $|g_{h^\pm_1 W^\mp a_1}|$ comes from the cancellation between $\mathcal{R}^C_{12}$ and $\mathcal{R}^C_{14}$.}}\label{ghmp1a1W}
\end{center}
\end{figure}
%%%%%%%%%%%%%%%%%%%%%%
Figure~\ref{ghmp1a1W} shows the dependency of the $g_{h^\pm_1 W^\mp a_1}$ coupling with the triplet components of the lightest charged Higgs eigenstate, i.e., $\mathcal{R}^C_{12}$ and $\mathcal{R}^C_{14}$.  We have seen from Figure~\ref{ssoslmbda} and Table~\ref{r2r4s} the behaviour of $\mathcal{R}^C_{12}$ $\mathcal{R}^C_{14}$ as a function of $\lambda_T$, i.e. that for $\lambda_T \sim 0$ they take same sign. We can see that in the decoupling limit, i.e. for $\lambda_T\sim 0$, the coupling decreases because $\mathcal{R}^C_{12}$ and $\mathcal{R}^C_{14}$ take same sign and they tend to cancel, cfr. Eq.~\ref{hachW}. 
{ A low value of this coupling can come even when the pseudoscalar Higgs boson ($a_j$) is singlet-like, which means that $\mathcal{R}^P_{j3}\sim1$.}
The situation is just opposite in the case of $g_{h^\pm_1 W^\mp h_1}$, as one can see from Figure~\ref{ghmp1h1W}. Here in the decoupling limit the coupling  $g_{h^\pm_1 W^\mp h_1}$ is enhanced. { In Figure~\ref{ghmp1h1W} we can also see some blue points with low $\mathcal{R}^C_{12}$, $\mathcal{R}^C_{14}$. In this case the charged Higgs boson is not triplet-like and the suppression in the coupling is due to the accidental cancellation of $\Big(\mathcal R_{12}^S\mathcal R_{13}^C-\mathcal R_{11}^S\mathcal R_{11}^C\Big)$, cfr. Eq.~\ref{hachW}. This cancellation is of course not related to the limit $\lambda_T\sim 0$.} We see later how it affects the corresponding production processes. 
%%%%%%%%% ghmp1h1W with R2 and R4 %%%%%%%%%%%%
%%%%%%%%%%%%%%%%%%%%%%

\subsection{$h_i^\pm  \to W^\pm Z$}
The charged sector of a theory with scalar triplet(s) is very interesting due to the tree-level interactions $h_i^\pm-W^\mp-Z$ for $Y=0, \pm 2$ hypercharge triplets which break the custodial symmetry \cite{pbas3,EspinosaQuiros,tnssm, tnssma}. In the TNMSSM this coupling is given by 
\bea\label{zwch}
g_{h_i^\pm W^\mp Z}&=&-\frac{i}{2}\left(g_L\, g_Y\left(v_u\sin\beta\,\mathcal R^C_{i1}-v_d\cos\beta\,\mathcal R^C_{i3}\right)+\sqrt2\,g_L^2v_T\left(\mathcal R^C_{i2}+\mathcal R^C_{i4}\right)\right),
\eea
where the rotation angles are defined in Eq.~\ref{chmix}. The on-shell decay width is given by
\bea\label{chzw}
\Gamma_{h_i^\pm\rightarrow W^\pm Z}&=&\frac{G_F\,\cos^2\theta_W}{8\sqrt2\pi}m^3_{h_i^\pm}|g_{h_i^\pm W^\mp Z}|^2\,\sqrt{\lambda(1,x_W,x_Z)}\left(8\,x_W\,x_Z+(1-x_W-x_Z)^2\right)
\eea
where $\lambda(x,y,z)=(x-y-z)^2-4\,y\,z$ and $x_{Z,W}=\frac{m^2_{Z,W}}{m^2_{h_i^\pm}}$ \cite{Asakawa}.

%%%%%%%%% ghmp1WZ with R2 and R4 %%%%%%%%%%%%
\begin{figure}[thb]
\begin{center}
\includegraphics[width=0.7\linewidth]{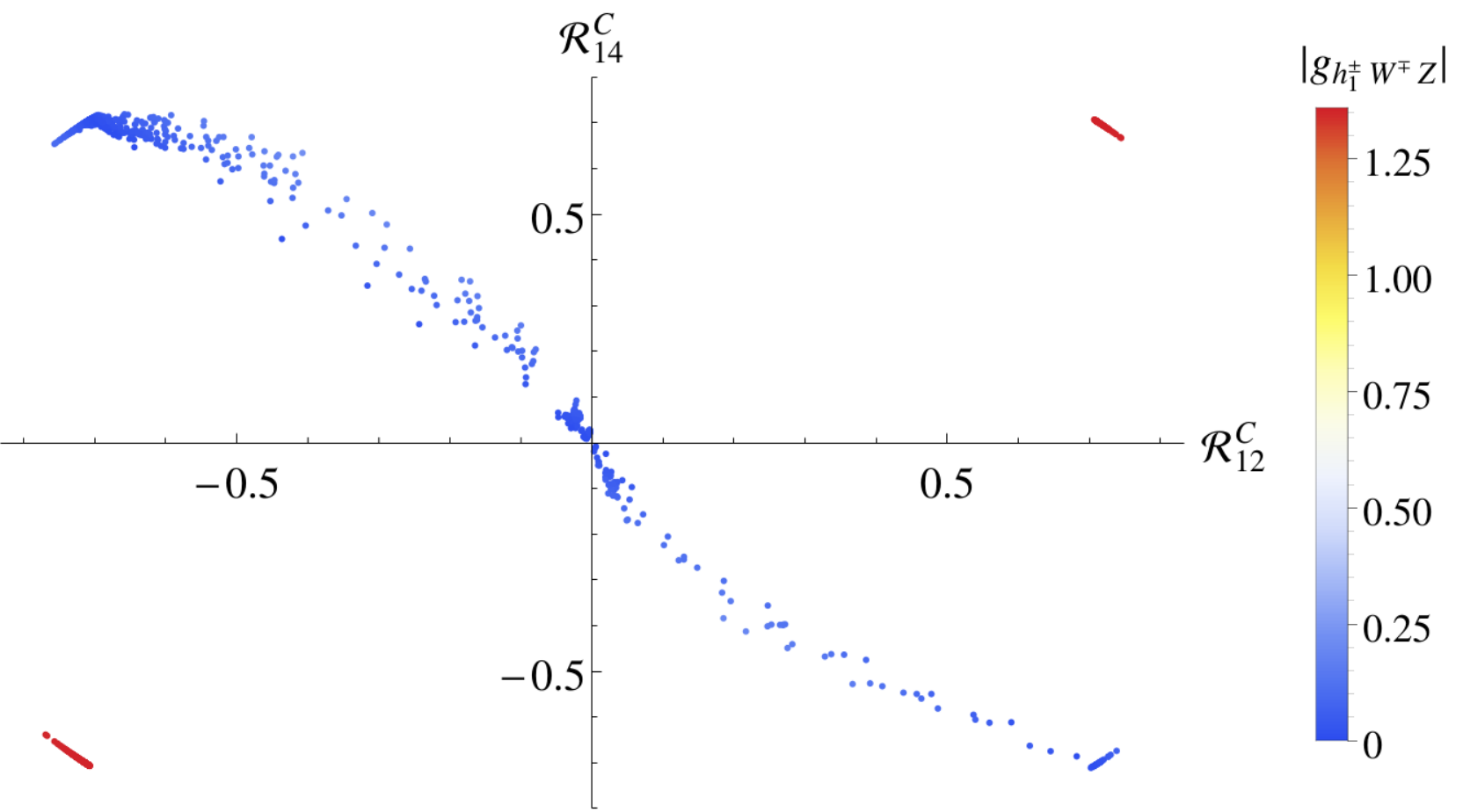}
\caption{Correlation of $g_{h^\pm_1 W^\mp Z}$ with  $\mathcal{R}^C_{12}$ and  $\mathcal{R}^C_{14}$.}\label{ghmp1WZ}
\end{center}
\end{figure}
%%%%%%%%%%%%%%%%%%%%%%

Figure~\ref{ghmp1WZ} shows the dependency of $g_{h_i^\pm W^\mp Z}$ with respect to $\mathcal{R}^C_{12}$ and $\mathcal{R}^C_{14}$. We see that for 
$\lambda_T \sim 0$ $\mathcal{R}^C_{12}$ and $\mathcal{R}^C_{14}$ take the same sign, and hence the $h_i^\pm-W^\mp-Z$  coupling is enhanced.

\subsection{$h_i^\pm \to t b$}
\begin{figure}[thb]
\begin{center}
\includegraphics[width=0.7\linewidth]{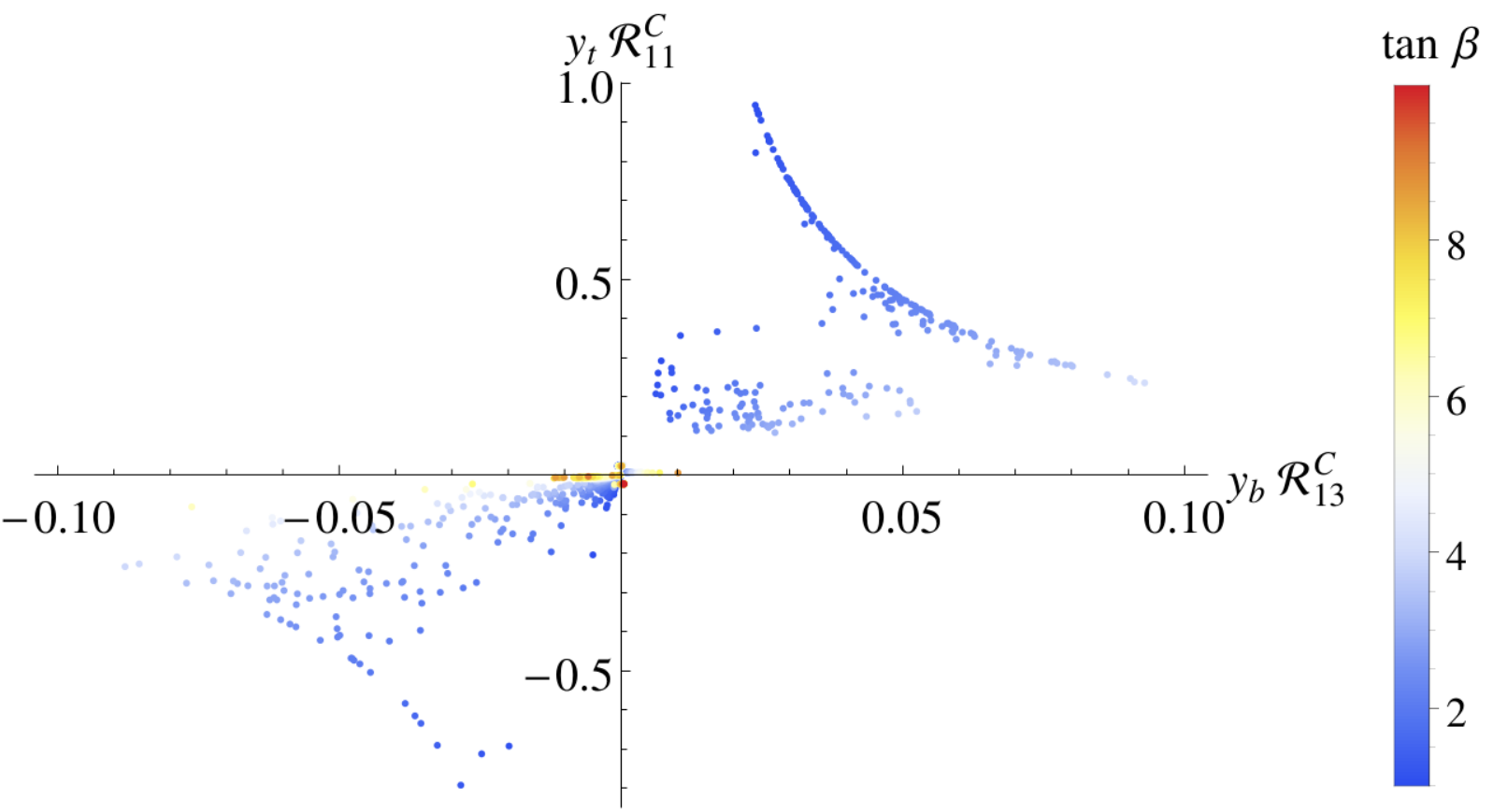}
\caption{Correlation of  $y_t\mathcal{R}^C_{11}$ and $y_b\mathcal{R}^C_{13}$ as a function of $\tan{\beta}$.}\label{ghmp1tb}
\end{center}
\end{figure}

Beside the non-zero $h^\pm_i-W^\mp-Z$ coupling at the tree-level due to custodial symmetry breaking, the charged Higgs bosons can also decay into fermions through the Yukawa interaction given below
\bea
g_{h_i^+ \bar u d}=i\left(y_u\,\mathcal R^C_{i1}\,\mathtt{P_L}+y_d\,\mathcal R^C_{i3}\,\mathtt{P_R}\right)
\eea
governed by doublet part of the charged Higgses. The decay width at leading order is
\begin{align}\label{chtb}
\Gamma_{h_i^\pm\rightarrow u\,d}&=\frac{3}{4}\frac{G_F}{\sqrt2\pi}m_{h_i^\pm}\sqrt{\lambda(1,x_u,x_d)}\Bigg[(1-x_u-x_d)\,\left(\frac{m^2_u}{\sin^2\beta}(\mathcal R^C_{i1})^2+\frac{m_d^2}{\cos^2\beta}(\mathcal R^C_{i3})^2\right)\nn\\
&\hspace{4.5cm}-4\frac{m_u^2m_d^2}{m^2_{h_i^\pm}}\frac{\mathcal R^C_{i1}\mathcal R^C_{i3}}{\sin\beta\cos\beta}\Bigg]
\end{align}
where $x_{u,d}=\frac{m^2_{u,d}}{m^2_{h_i^\pm}}$. The QCD correction to the leading order formula are the same as in the MSSM and are given in \cite{anatomy2}. The decay of the charged Higgs bosons into quarks is then suppressed in the case of triplet-like eigenstates, as one can easily realize from the expression above. In Figure~\ref{ghmp1tb} we show the correlation of the effective Yukawa coupling $(y_u\,\mathcal R^C_{i1}$ and $y_d\,\mathcal R^C_{i3})$ of top and bottom quark respectively as a function of $\tan\beta$. The dominant contribution comes from the top for small $\tan\beta$, as we expected.

%%%%%%%%% ghmp1tb with R1 and R3 %%%%%%%%%%%%
%%%%%%%%%%%%%%%%%%%%%%

\section{Decay branching ratios of the charged Higgs bosons}\label{ch1dcy}
Prepared with the possibilities of new decay modes we finally analyse such scenarios with the data satisfying various theoretical and experimental constraints.  The points here have a CP-even neutral Higgs boson around 125 GeV which satisfies the LHC constraint given in Eq.~\ref{LHCdata}.
To study the decay modes and calculate the branching fractions we have implemented our model in \texttt{SARAH$\_$4.4.6} \cite{sarah} and we have generated the model files for \texttt{CalcHEP$\_$3.6.25} \cite{calchep}.
%%%%%%%%%Decay channels %%%%%%%%%%%%
\begin{figure}[thb]
\begin{center}
\mbox{\subfigure[]{
\includegraphics[width=0.23\linewidth]{plots/hhW.pdf}}
\hspace*{.5cm}
\subfigure[]{
\includegraphics[width=0.23\linewidth]{plots/hZW.pdf}}
\hspace*{.5cm}
\subfigure[]{
\includegraphics[width=0.23\linewidth]{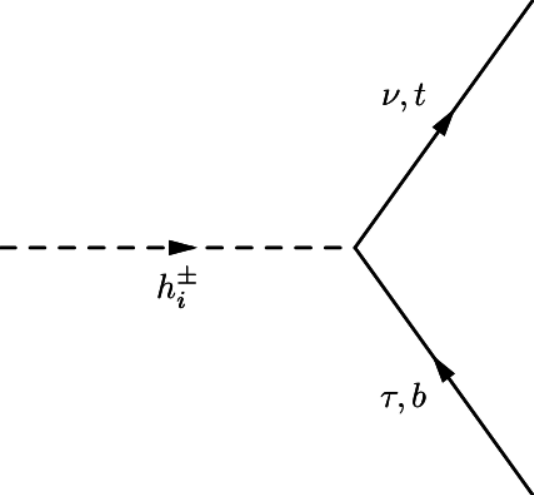}}}
\caption{The new and modified decay channels of the Higgs bosons at the LHC.}\label{higgdcy}
\end{center}
\end{figure}
%%%%%%%%%%%%%%%%%%%%%%
\begin{figure}[thb]
\begin{center}
\mbox{\subfigure[]{\includegraphics[width=.5\linewidth]{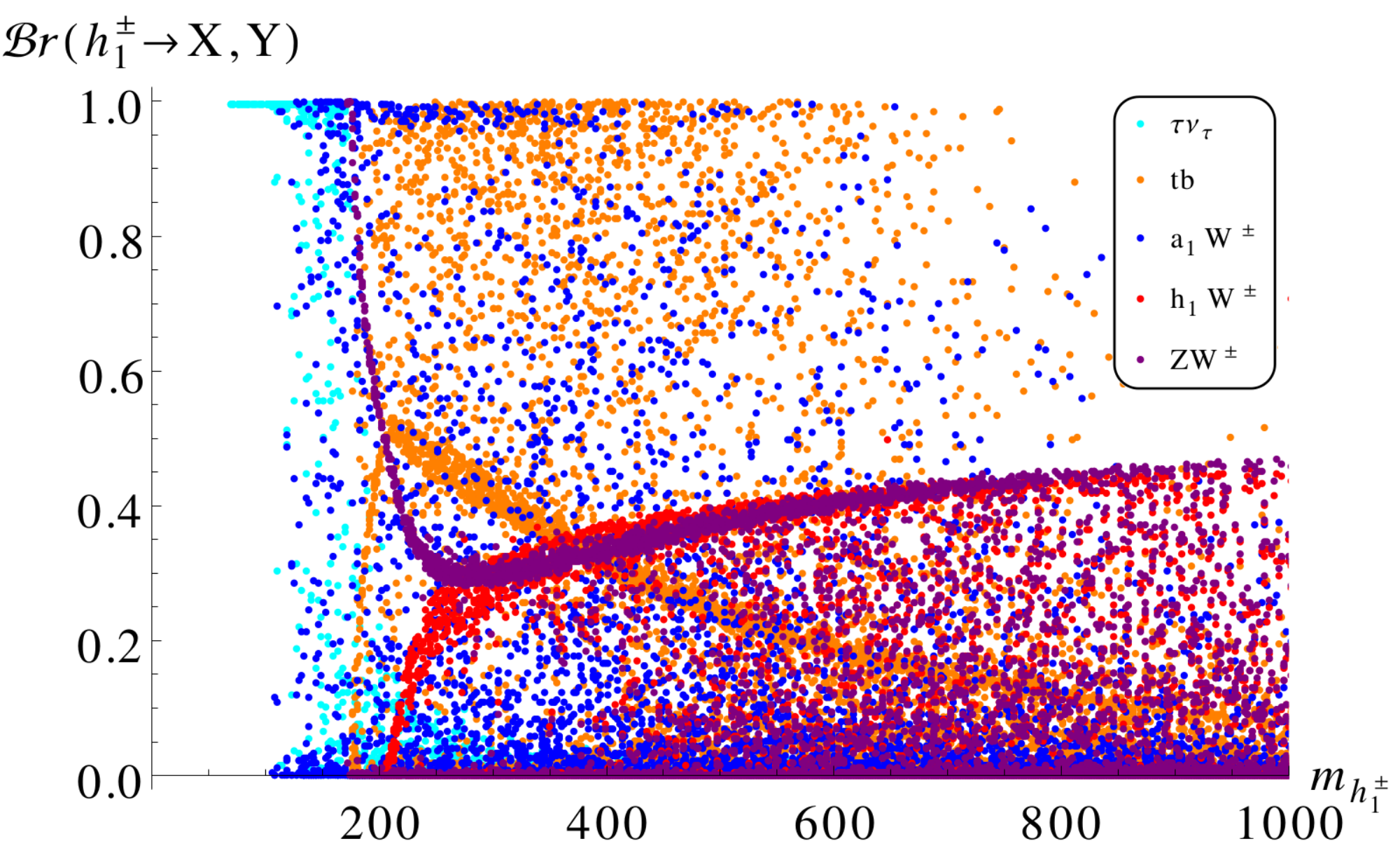}}
\subfigure[]{\includegraphics[width=.5\linewidth]{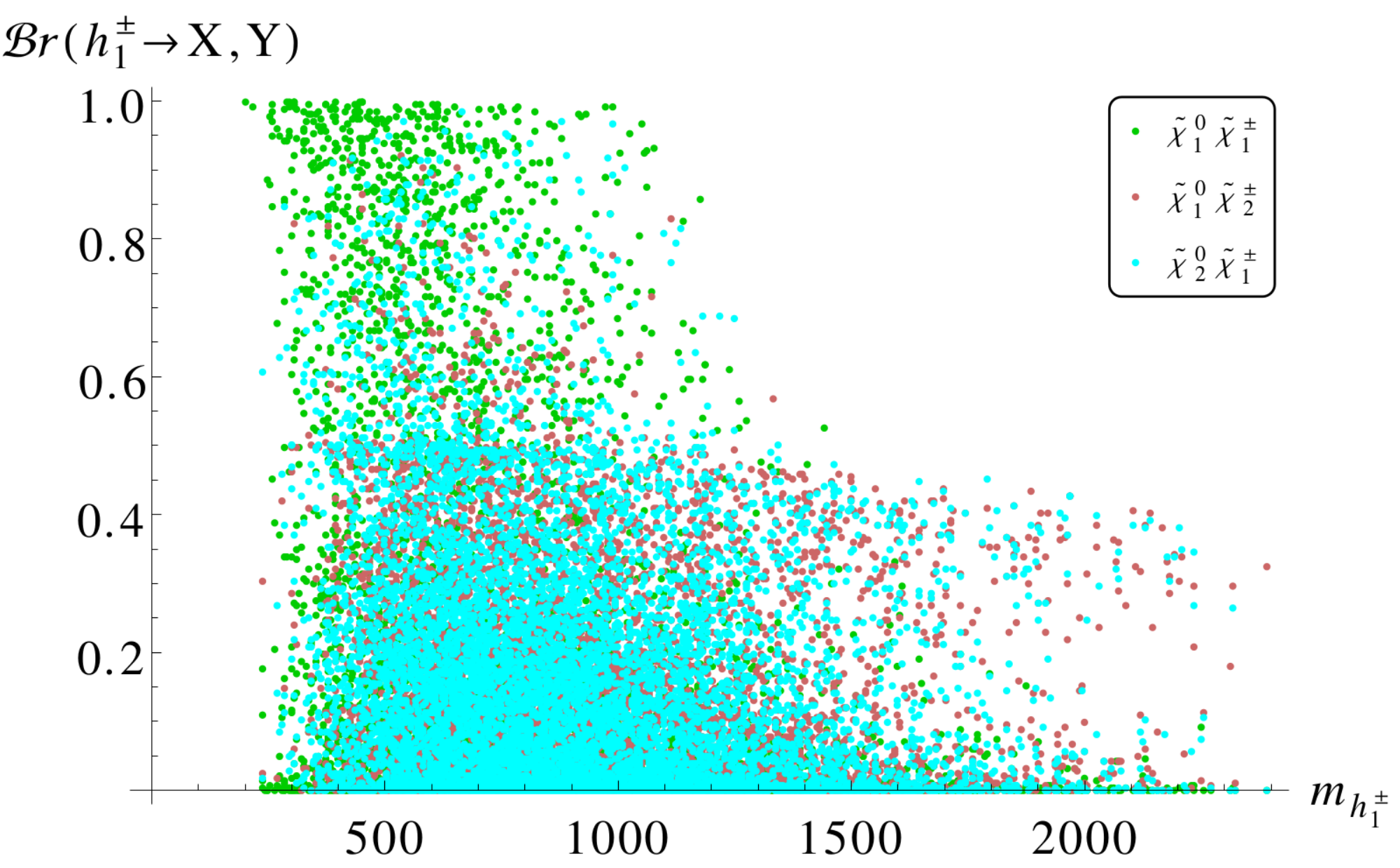}}}
\caption{The branching ratios for the decay of the lightest charged Higgs boson $h^\pm_1$ into non-supersymmetric (a) and supersymmetric modes (b).}\label{ch1br}
\end{center}
\end{figure}
%%%%%%%%%%%%%%%%%%%%%%

\begin{figure}[bht]
\begin{center}
\includegraphics[width=.5\linewidth]{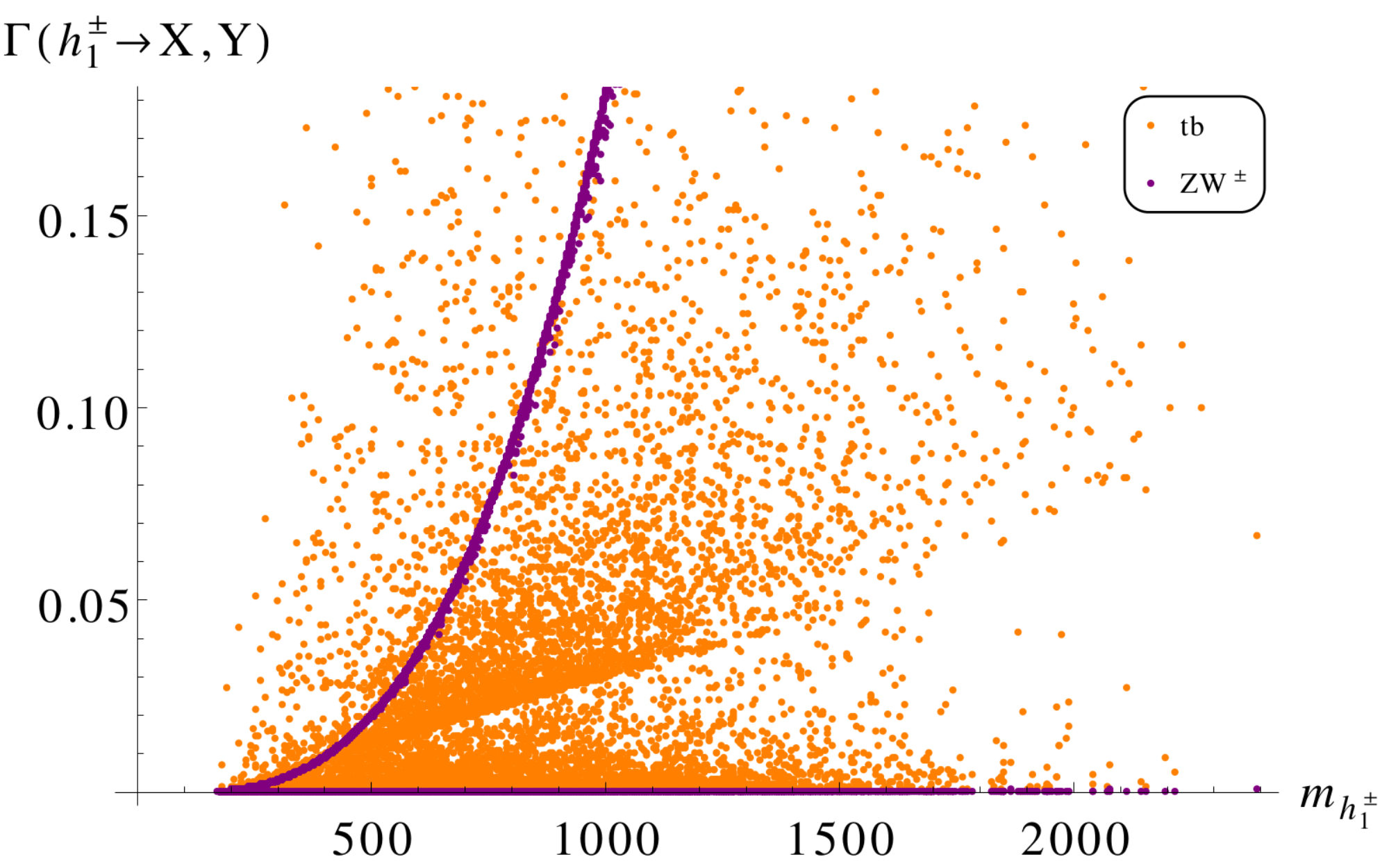}
\caption{The decay widths of the lightest charged Higgs boson $h^\pm_1$ to $tb$ and $ZW^\pm$.}\label{ch1dc}
\end{center}
\end{figure}

Figure~\ref{ch1br}(a) presents the decay branching ratios of the light charged Higgs boson $h^\pm_1$
into non-supersymmetric modes.  This includes the $a_1W^\pm$, $h_1W^\pm$, $ZW^\pm$, $tb$ and $\tau\nu$ channels.
The points in the Figure~\ref{ch1br} include a discovered Higgs boson at $\sim 125$ GeV  and a triplet-like light charged Higgs boson $h_1^\pm$. When $a_1$ is singlet-type, the $a_1W^\pm$ decay mode is suppressed in spite of being kinematically open.  One can notice that, being the $ h^\pm_1$ triplet-like, the decay mode $ZW^\pm$ can be very large, even close to $100\%$. When the $tb$ mode is kinematically open, the $ZW^\pm$ gets an apparent suppression, but it increases again for a charged Higgs bosons of larger mass ($m_{h^\pm_1}\sim 400$ GeV). This takes place because the $h^\pm_i \to ZW^\pm$ decay width is proportional to $m_{h^\pm_i}^3$, unlike the $tb$ one, which is proportional to $m_{h^\pm_i}$ (see Eq.~\ref{chzw} and Eq.~\ref{chtb}). The variation of these two decay widths, as a function of $m_{h^\pm_1}$, are shown in Figure~\ref{ch1dc}.\\
Figure~\ref{ch1br}(b) shows the decays of the lightest charged Higgs boson into the supersymmetric modes, i.e. into charginos $\tilde{\chi}^\pm_i$ and neutralinos $\tilde{\chi}^0_j$, when these modes are are kinematically allowed. We observe that for a charged Higgs boson of a relatively higher mass $m_{h^\pm_i} \gsim 300$ GeV, these modes open up and can have very large branching ratios. 
%%%%%%%%%%%%%%%%%%%%%%%%%%%%
%%%%%%%%%%%%%%%%%%%%%%%%

Apart from the lightest charged Higgs boson, there are two additional charged Higgs bosons, $h^\pm_2$ and $h^\pm_3$. As we have pointed out many times, we have selected data points for which the light charged Higgs boson is triplet-type. Certainly, in the decoupling limit, i.e. when $|\lambda_T|\simeq 0$, either one of $h^\pm_{2,3}$ is triplet-like and the other one is doublet-like. The points that we have generated, which satisfy also the precondition of allowing a $h_{125}$ in the spectrum, have a $h^\pm_2$ as a triplet- and a $h^\pm_3$ as a doublet-like Higgs boson, cfr. Figure~\ref{chpslmbda}.
%%%%%%%%%%%%%%%%%%%%%%
\begin{figure}[thb]
\begin{center}
\mbox{\subfigure[]{
\includegraphics[width=0.5\linewidth]{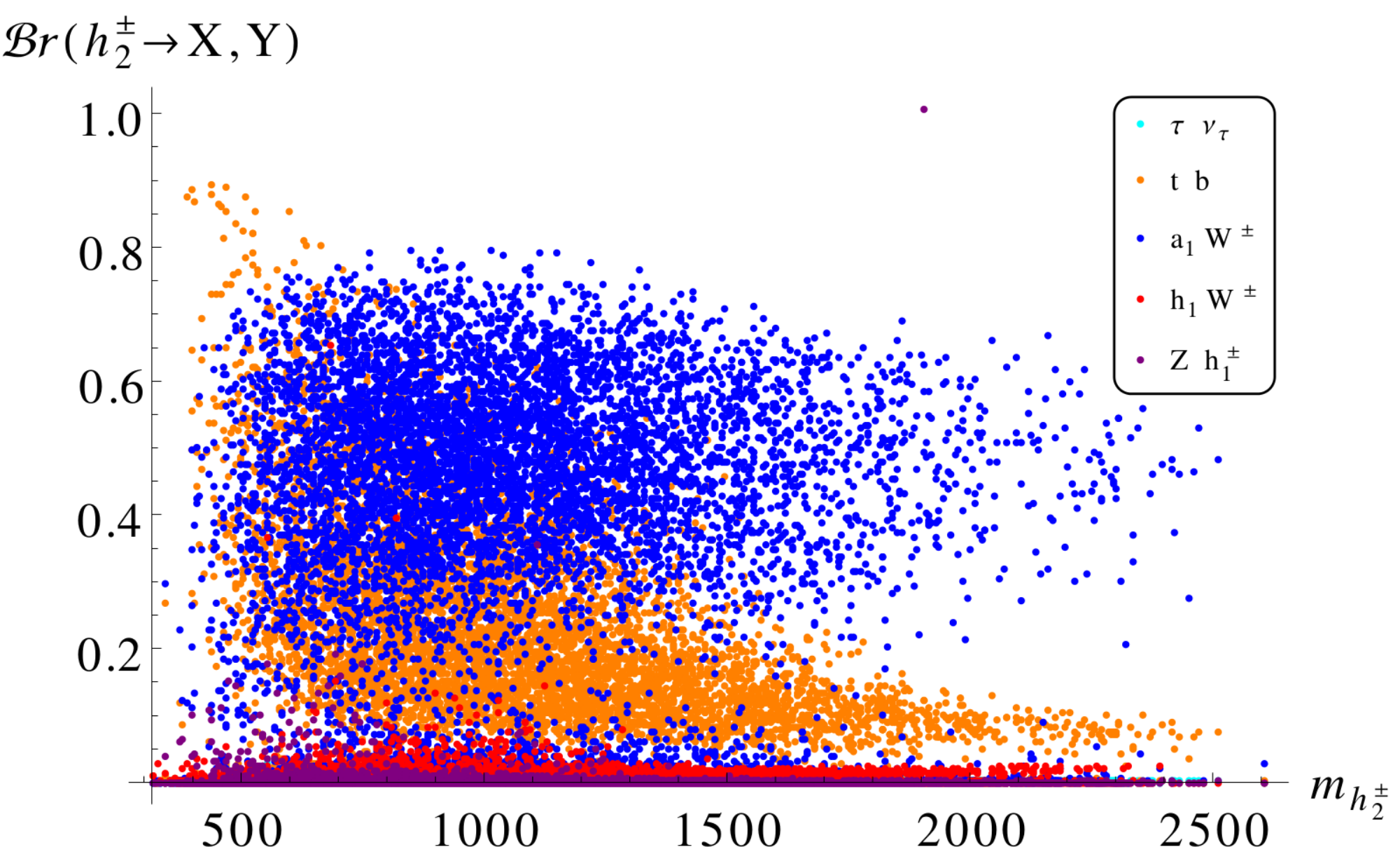}}
\subfigure[]{\includegraphics[width=0.5\linewidth]{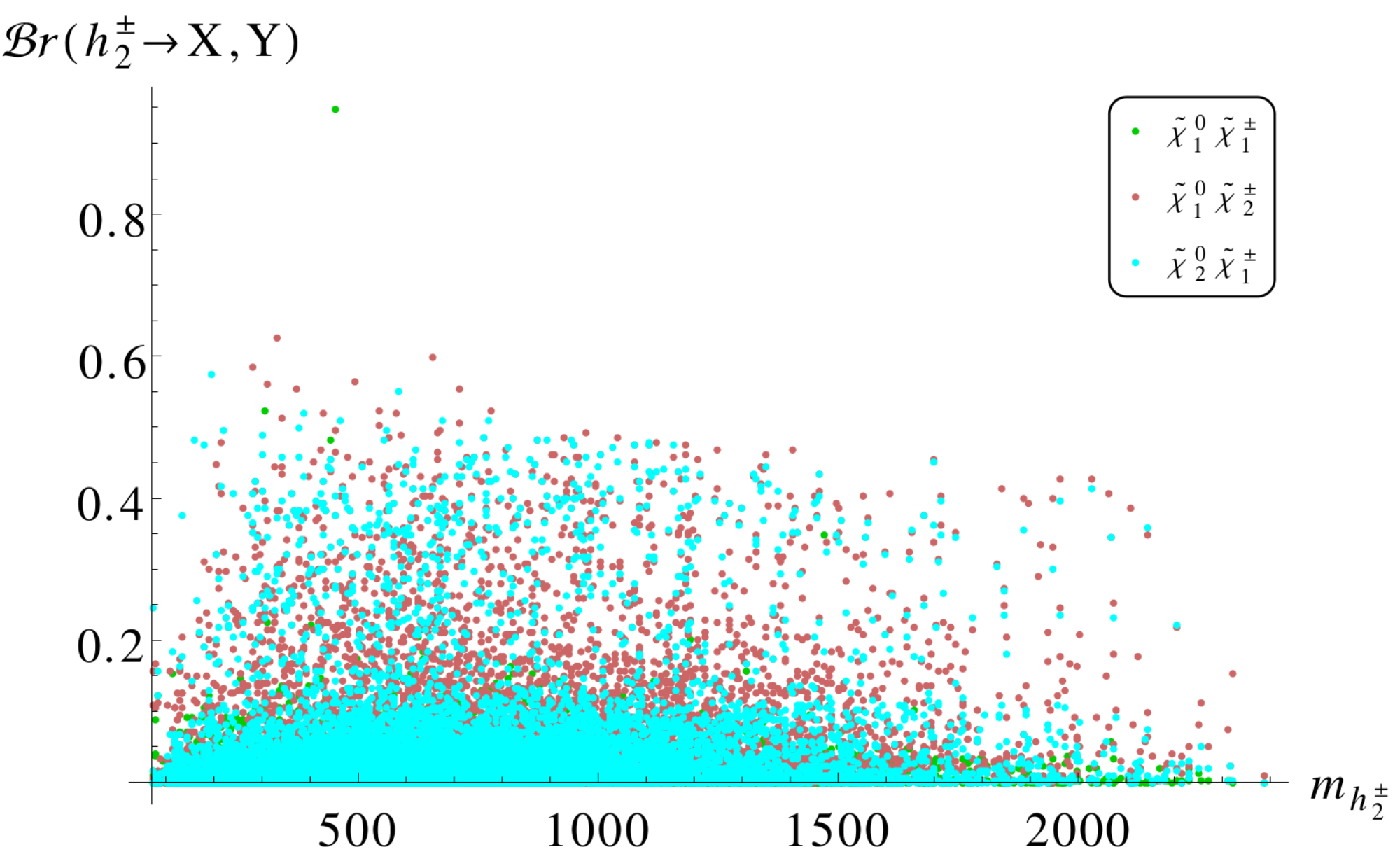}}}
\mbox{\subfigure[]{\includegraphics[width=0.5\linewidth]{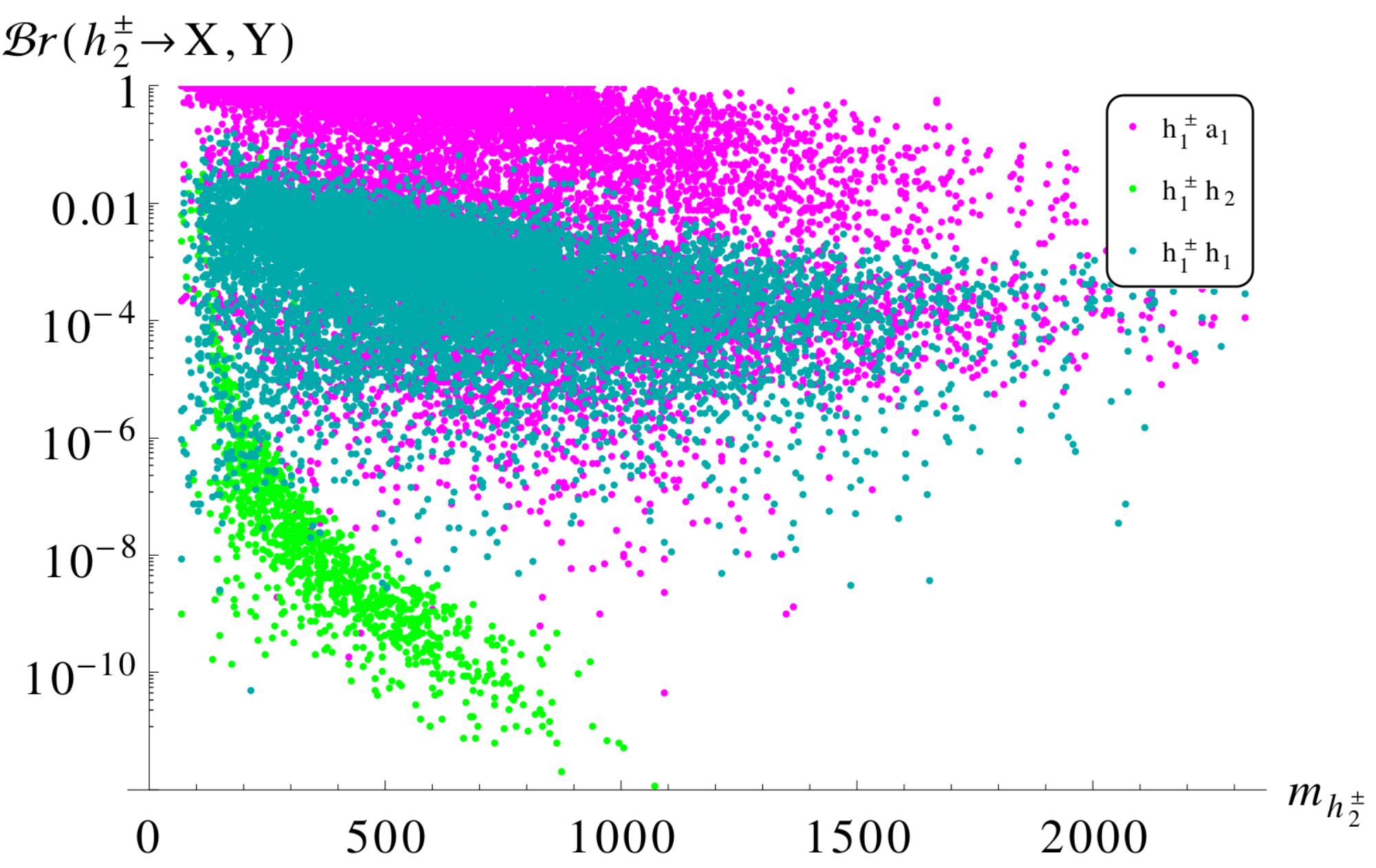}}}
\caption{The branching ratios of the decay of the charged Higgs boson $h^\pm_2$ into non-supersymmetric (a), supersymmetric modes (b) and into Higgs bosons (c).}\label{ch2br}
\end{center}
\end{figure}
%%%%%%%%%%%%%%%%%%%%%%
In Figure~\ref{ch2br} we present the decay branching ratios of the second charged Higgs boson $h^\pm_2$. Figure~\ref{ch2br}(a) shows the ratios in $\tau\nu$, $tb$, $a_1W^\pm$, 
$h_1 W^\pm$ and $Z h^\pm_1$. As one can observe, $tb$ and $a_1 W^\pm$ are the dominant modes reaching up to $\sim 90\%$ and $\sim80\%$ respectively.  Figure~\ref{ch2br}(b) shows the branching ratios into supersymmetric modes with neutralinos and charginos, which are kinematically allowed. For some benchmark points these modes can have decay ratios as large as $\sim 60\%$. Figure~\ref{ch2br}(c)  shows the ratios for $h^\pm_2$ decaying into two scalars, i.e. to $h_1^\pm h_{1,2}$ and $h^\pm_1 a_1$, with the $h^\pm_1 a_1$ final state being the dominant among all. 
%%%%%%%%%%%%%%%%%%%%%%
\begin{figure}[thb]
\begin{center}
\mbox{\subfigure[]{
\includegraphics[width=0.44\linewidth]{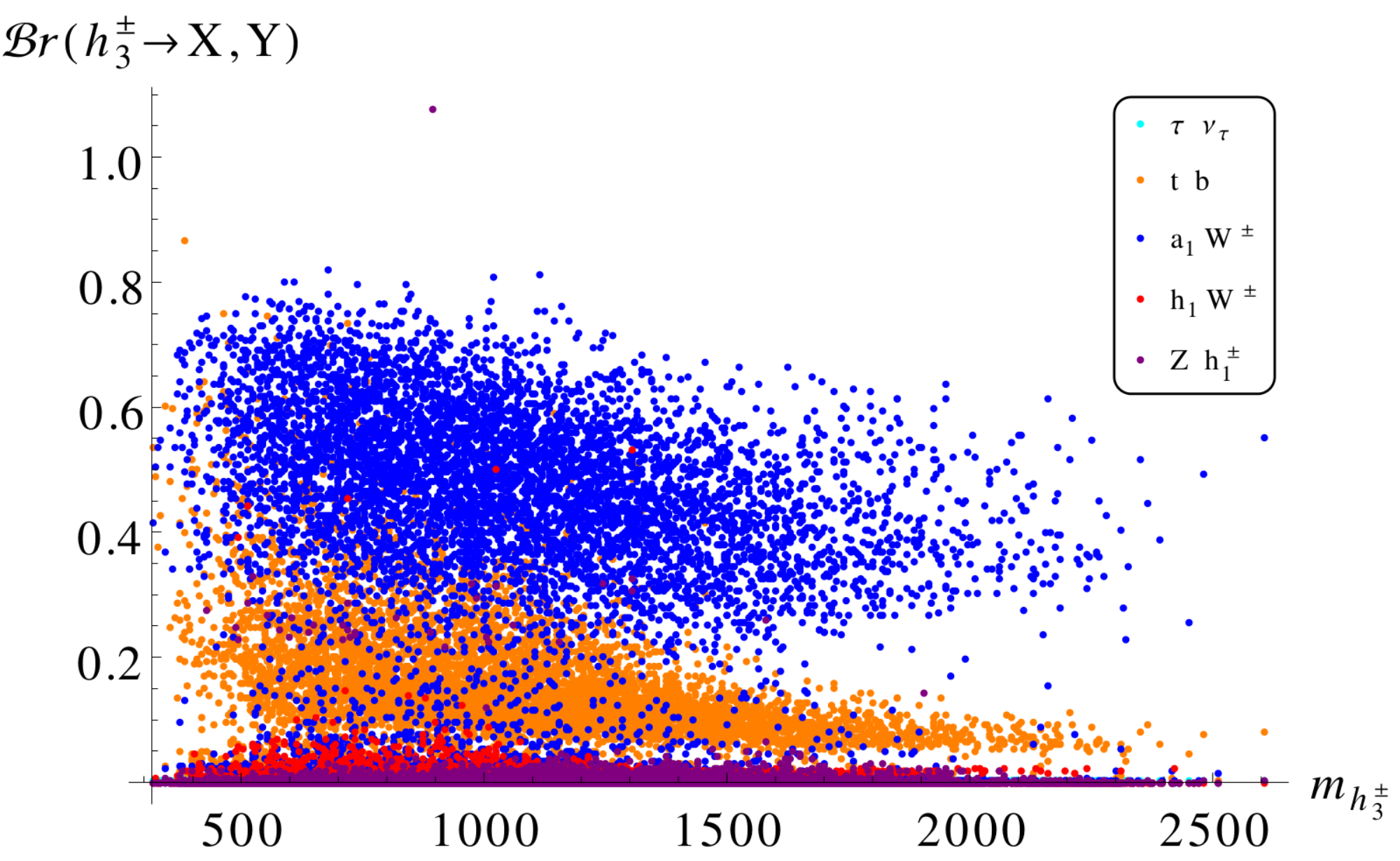}}}
\mbox{\subfigure[]{\includegraphics[width=0.44\linewidth]{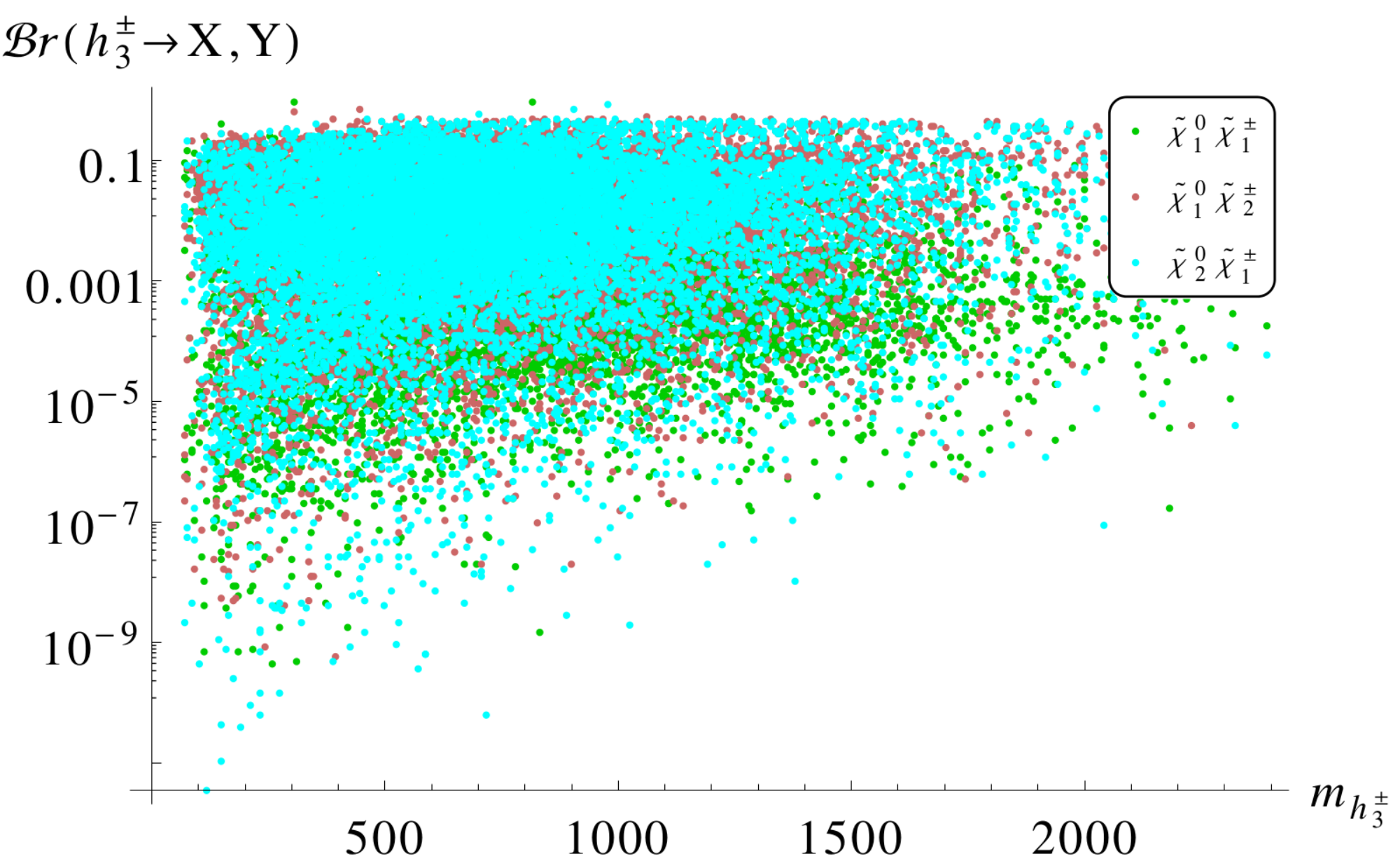}}}
\mbox{\subfigure[]{\includegraphics[width=0.44\linewidth]{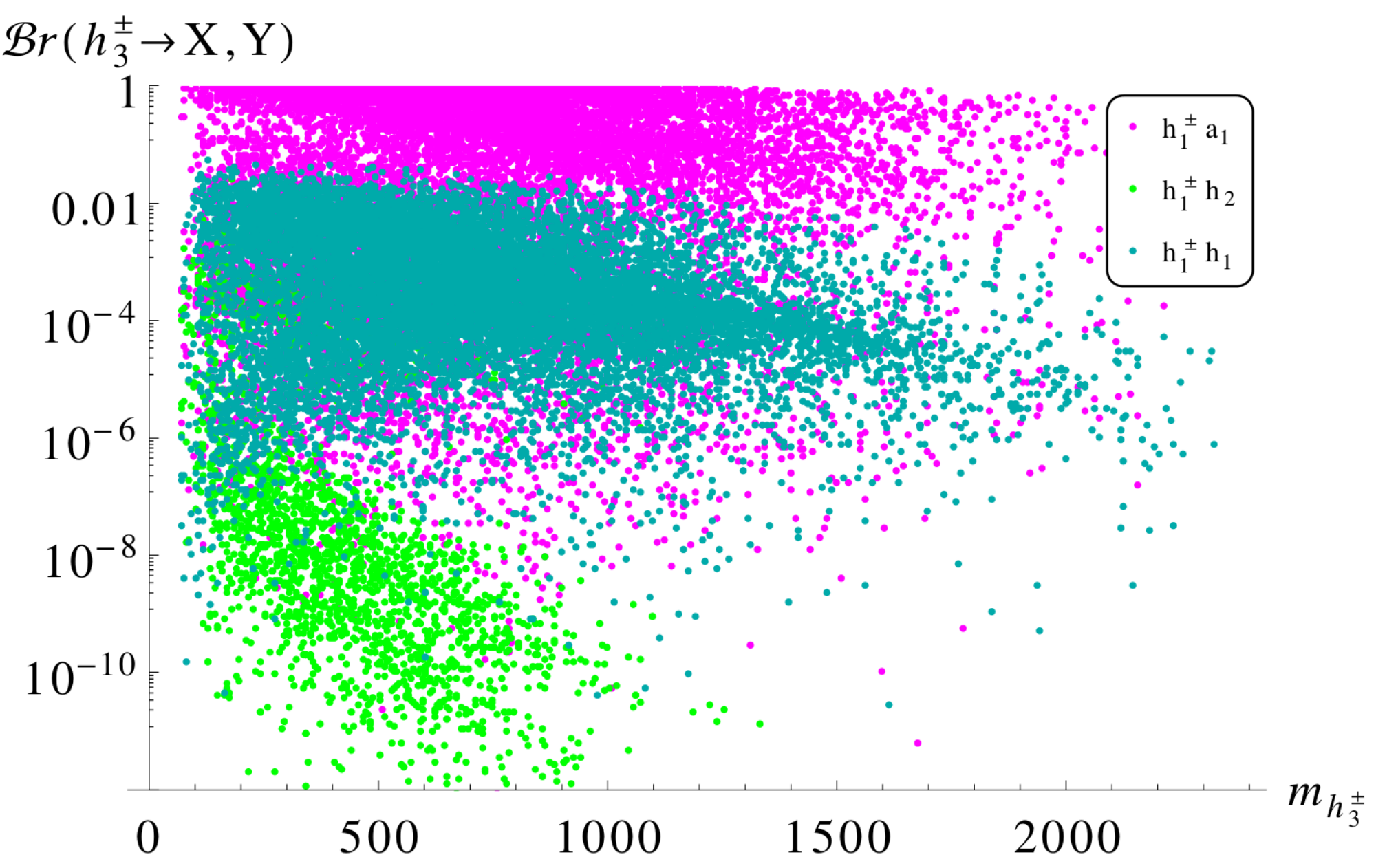}}}
\mbox{\subfigure[]{\includegraphics[width=0.44\linewidth]{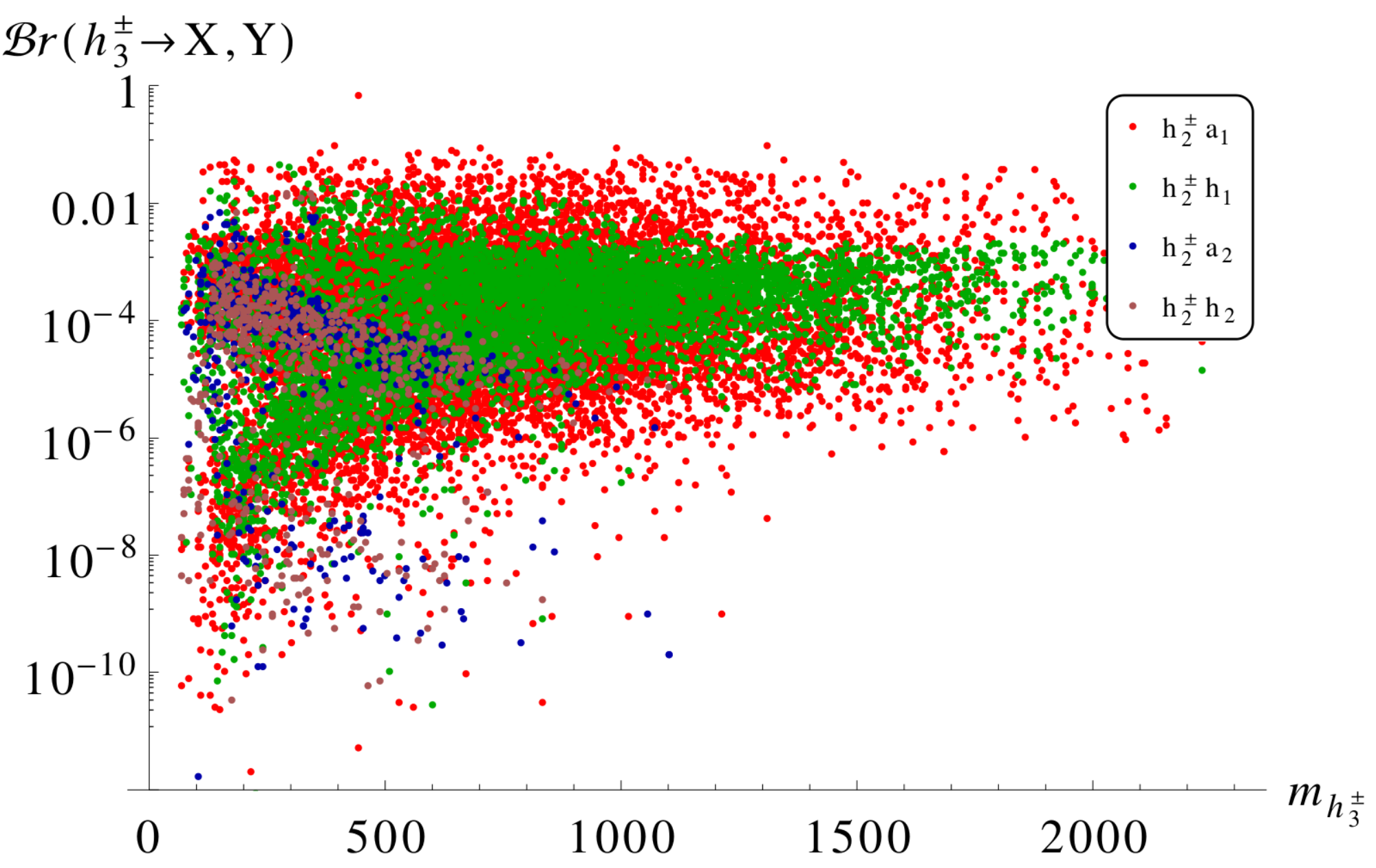}}}
\caption{The branching ratios of the decay of the charged Higgs boson $h^\pm_3$ into non-supersymmetric (a), supersymmetric modes (b), lightest charged Higgs boson $h^\pm_1$ in association with the neutral Higgs bosons (c) and second light charged Higgs boson $h^\pm_2$ in association with the neutral Higgs bosons (d).}\label{ch3br}
\end{center}
\end{figure}
%%%%%%%%%%%%%%%%%%%%%%

Figure~\ref{ch3br} presents the third charged Higgs boson $h^\pm_3$ decays. From Figure~\ref{ch3br}(a) we can see that for a large parameter space the decay branching fraction to $a_1W^\pm$ is the most relevant mode which can be probed at the LHC. Even though $tb$ mode is kinematically open but not the most dominant one. Figure~\ref{ch3br}(b) shows that $\tilde{\chi}^0_2 \tilde{\chi}^\pm_1$ mode is kinematically open and also one of the most important. Figure~\ref{ch3br}(c) shows the decay branching ratios for the decay modes into the lightest charged Higgs boson in association with the neutral Higgs bosons. It is evident that the $h^\pm_1 a_1$ mode is the most important and one can probe more than one charged Higgs boson and also the light pseudoscalar. In Figure~\ref{ch3br}(d) the branching ratios are shown where the heaviest charged Higgs boson $h^\pm_3$ decays to second lightest charged Higgs boson $h^\pm_2$ in  association with the neutral Higgs bosons. Again the light pseudoscalar mode can have large branching ratios.
%%%%%%%%%%Production %%%%%%%%%%%%%%%
\section{Production channels of a light charged Higgs boson}\label{ch1prod}
The triplet nature of the charged Higgs bosons adds few new production processes at the LHC
along with the doublet-like charged Higgs production process. For a doublet-like charged 
Higgs boson the production processes are dominated by the top quark decay for the light charged Higgs boson ($m_{h^\pm_i} < m_t$) or $b g \to t h^\pm_i$ for  ($m_{h^\pm_i} > m_t$) which are governed by the 
corresponding Yukawa coupling and $\tan{\beta}$ viz, in 2HDM, MSSM and NMSSM. In TNMSSM however the charged Higgs bosons can be triplet-like,  and hence do not couple to fermions. Fermionic channels, including top and bottom and in general all the fermions, are then suppressed. The presence of the $h_i^\pm-W^\mp-Z$ vertex generates new production channels and also modifies the known processes for the production of a charged Higgs  boson $h^\pm_i$. In these sections we address the dominant and characteristically different production mechanisms for the light charged Higgs  bosons $h^\pm_1$ at the LHC. For this purpose we select in the parameter space the benchmark points with a discovered Higgs boson around $125$ GeV and with the lightest charged Higgs boson $h^\pm_1$ that is triplet-like ($\geq 90\%$). The cross-sections are calculated at the LHC with a center of mass energy of 14 TeV for such events. We have performed our analysis at leading order, using $\mathtt{CalcHEP\_3.6.25}$ \cite{calchep}, using the CTEQ6L \cite{6teq6l} set of parton distributions and  a renormalization/factorization scale $Q=\sqrt{\hat{s}}$   where $\hat{s}$ denotes the total center of mass energy squared at parton level.

\subsection{Associated $W^\pm$ }
\begin{figure}[thb]
\begin{center}
\mbox{\subfigure[]{
\includegraphics[width=0.35\linewidth]{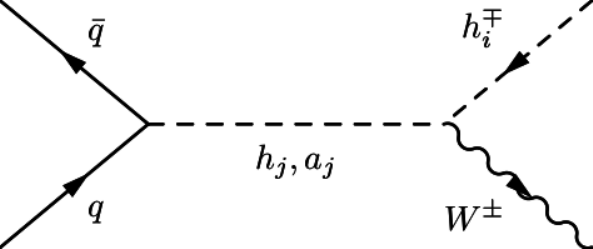}}
\hskip 15pt
\subfigure[]{\includegraphics[width=0.35\linewidth]{plots/ZWhci.pdf}}}
\mbox{\subfigure[]{\includegraphics[width=0.3\linewidth]{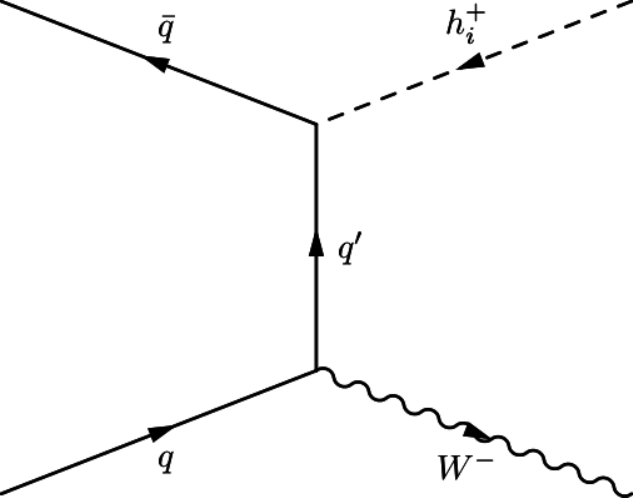}}
}
\caption{The Feynman diagrams for the charged Higgs production in association with $W^\pm$  boson at the LHC.}\label{prodchW}
\end{center}
\end{figure}

The dominant channels are shown in Figure~\ref{prodchW}, which are mediated by
the neutral Higgs bosons, the $Z$ boson and the quarks. Figure~\ref{prodchW}(b) which describe the $Z$ mediation requires the non-zero $h_1^\pm-W^\mp-Z$ vertex which is absent in theories without the $Y=0,\pm2$ triplet-extended Higgs sector. For a  doublet-like charged Higgs, the only contributions comes from the neutral Higgs-mediated diagrams in the s-channel and $t$-quark mediated diagram in the t-channel  (see Figure~\ref{prodchW}(a), (c)). For low $\tan{\beta}$ case the t-channel contribution in $b\bar{b}$ fusion is really large due to large Yukawa coupling. We will see that this admixture of doublet still affects the production cross-section for low $\tan{\beta}$. 
%%%%%%%%%%%%%%%%%%%%%%%%%
%%%%%%%%%%%%%%%%%%%%%%
%%%%%%%%%%%%%%%%%%%%%%%%%%%%
\begin{figure}[thb]
\begin{center}
\includegraphics[width=0.7\linewidth]{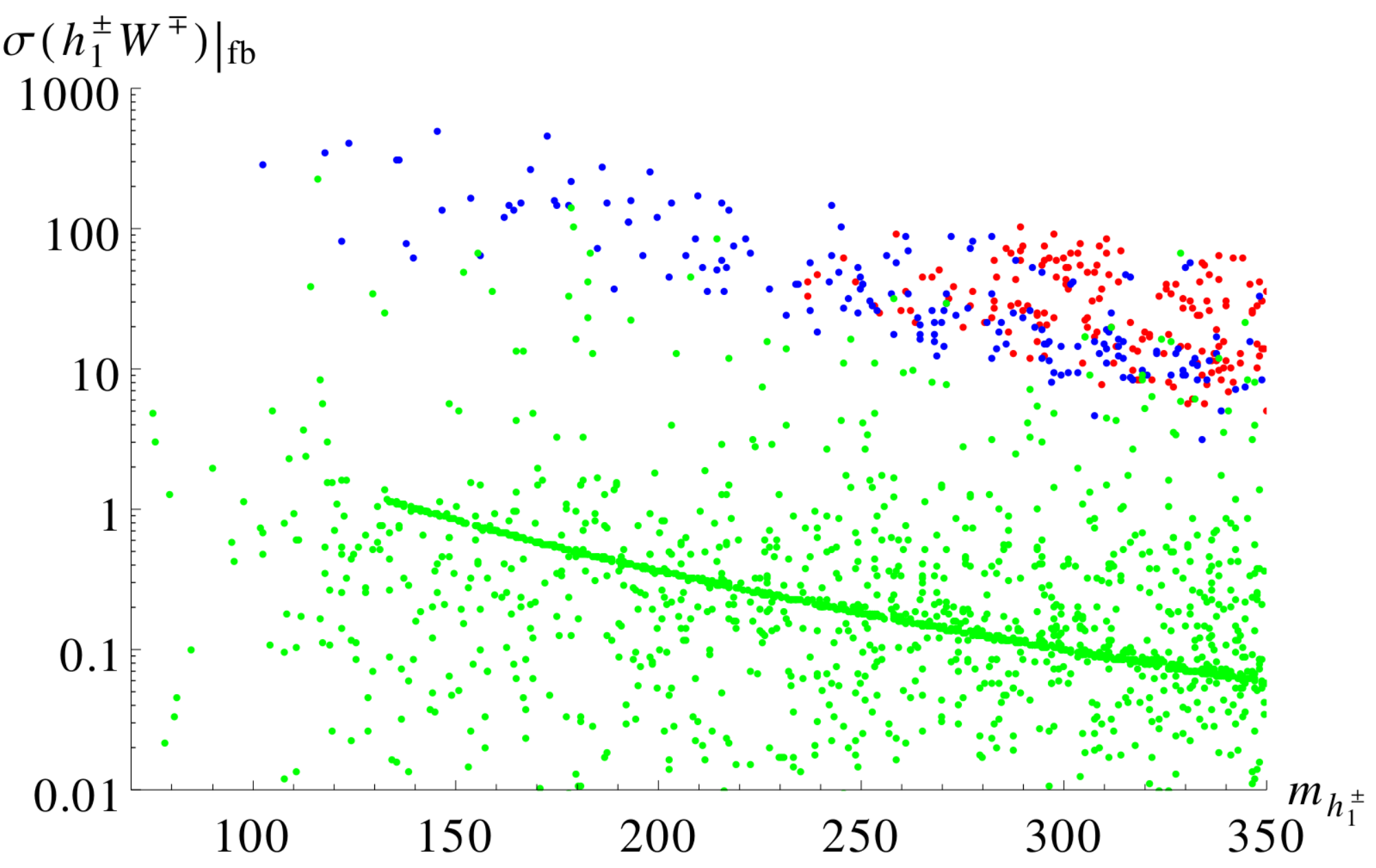}
\caption{The production cross-section of $h^\pm_1W^\mp$ at the LHC versus the lightest charged Higgs boson mass $m_{h^\pm_1}$. The red coloured ones are $\geq 90\%$ doublet-like, green ones are  $\geq 90\%$ triplet-like  and blue ones are mixed type light charged Higgs bosons.}\label{ch1Wcs}
\end{center}
\end{figure}
%%%%%%%%%%%%%%%%%%%%%%

The contribution of $h_1$ is subdominant because $h_1$ and $h_1^\pm$ are selected to be mostly doublet and triplet respectively, in order to satisfies the LHC data. The coupling of a totally triplet charged Higgs boson with a totally doublet neutral Higgs boson and a $W$ boson is not allowed by gauge invariance. For the lightest triplet-like charged Higgs boson,  one of the degenerate neutral Higgs boson, either $h_2$ or $a_2$, is also triplet-like, and fails to contribute as mediator in $b\bar{b}$ fusion mode (Figure~\ref{prodchW}(a)). The other relevant neutral Higgs boson which is not degenerate with  the lightest charged Higgs boson $h^\pm_1$ contributes to $b\bar{b}$ fusion production process via its doublet mixings. Thus doublet-triplet mixing part plays an important role even when we are trying to produce a light charged Higgs boson which is triplet-like.  This feature also has been observed in Triplet Extended Supersymmetric Standard Model (TESSM) \cite{pbas3}. Even the off-shell doublet type neutral Higgs mediation ($h_{125}$) in s-channel via gluon-gluon fusion fails to give sufficient contribution to $h^\pm_1 W^\mp$ final state. We checked such process at the LHC for the center of mass energy of 14 TeV and a triplet-like charged Higgs of mass $\sim 300$ GeV and $h^\pm_1 W^\mp$ cross-section is below $\mathcal{O}(10^{-3})$ fb. 
 
In Figure~\ref{ch1Wcs} we present the associated production cross-section for a light charged Higgs boson $h^\pm_1$ together with the light charged Higgs boson mass $m_{h^\pm_1}$. 
The red coloured ones are $\geq 90\%$ doublet-like, green ones are  $\geq 90\%$ triplet-like  and blue ones are mixed type light charged Higgs bosons. It can be seen that as the doublet 
the fraction grows, the production cross-section also grows. At $\lambda_T\simeq0$ the lightest
charged Higgs cannot be completely triplet-like, due to the doublet fraction $\frac{v_T}{v}$.
In this limit the cross section follows the line given by the green points in Figure~\ref{ch1Wcs}. As we have seen in the previous section, for $\lambda_T\neq 0$ the coupling $g_{h_1^\pm W^\mp Z}$ is very small even if the lightest charged Higgs is completely triplet-like. This means that the $Z$ propagator (cfr. Figure~\ref{prodchW}(b)) does not give contribution. However, since for $\lambda_T\neq 0$ the triplet fraction of $h_1^\pm$ is not fixed, the cross-section can be enhanced or decreased compared to the $|\lambda_T|\simeq0$ one.

\subsection{Associated $Z$ }
%%%%%%%%%%%%%%%%%%%%%%%%%
\begin{figure}[thb]
\begin{center}
\mbox{\subfigure[]{
\includegraphics[width=0.35\linewidth]{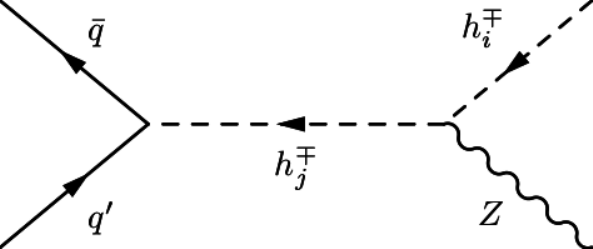}}\hskip 15pt
\subfigure[]{\includegraphics[width=0.35\linewidth]{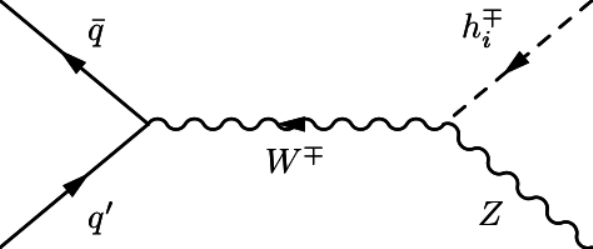}}}
\caption{The Feynman diagrams for the charged Higgs production in association with $Z$  boson at the LHC.}\label{prodch1Z}
\end{center}
\end{figure}
%%%%%%%%%%%%%%%%%%%%%%
Unlike the previous case, the charged Higgs production in association with $Z$ does not have sizeable contributions from the doublet part of the Higgs boson spectrum. For instance, the doublet nature of the charged Higgs allows its exchange in the s-channel, as shown in  Figure~\ref{prodch1Z}(a), via an annihilation process ($q \bar{q}')$ which requires quarks of different flavours. The contributions from the valence $u/\bar{d}, \bar{u}/d$ distributions, in a $pp$ collision are strongly suppressed by the much lower Yukawa couplings. On the other hand contributions from heavier generations such as $c/\bar{b},\bar{c} /b$ are suppressed by CKM mixing angles and the involvement of sea quarks in the initial state.

\begin{figure}[thb]
\begin{center}
\includegraphics[width=0.7\linewidth]{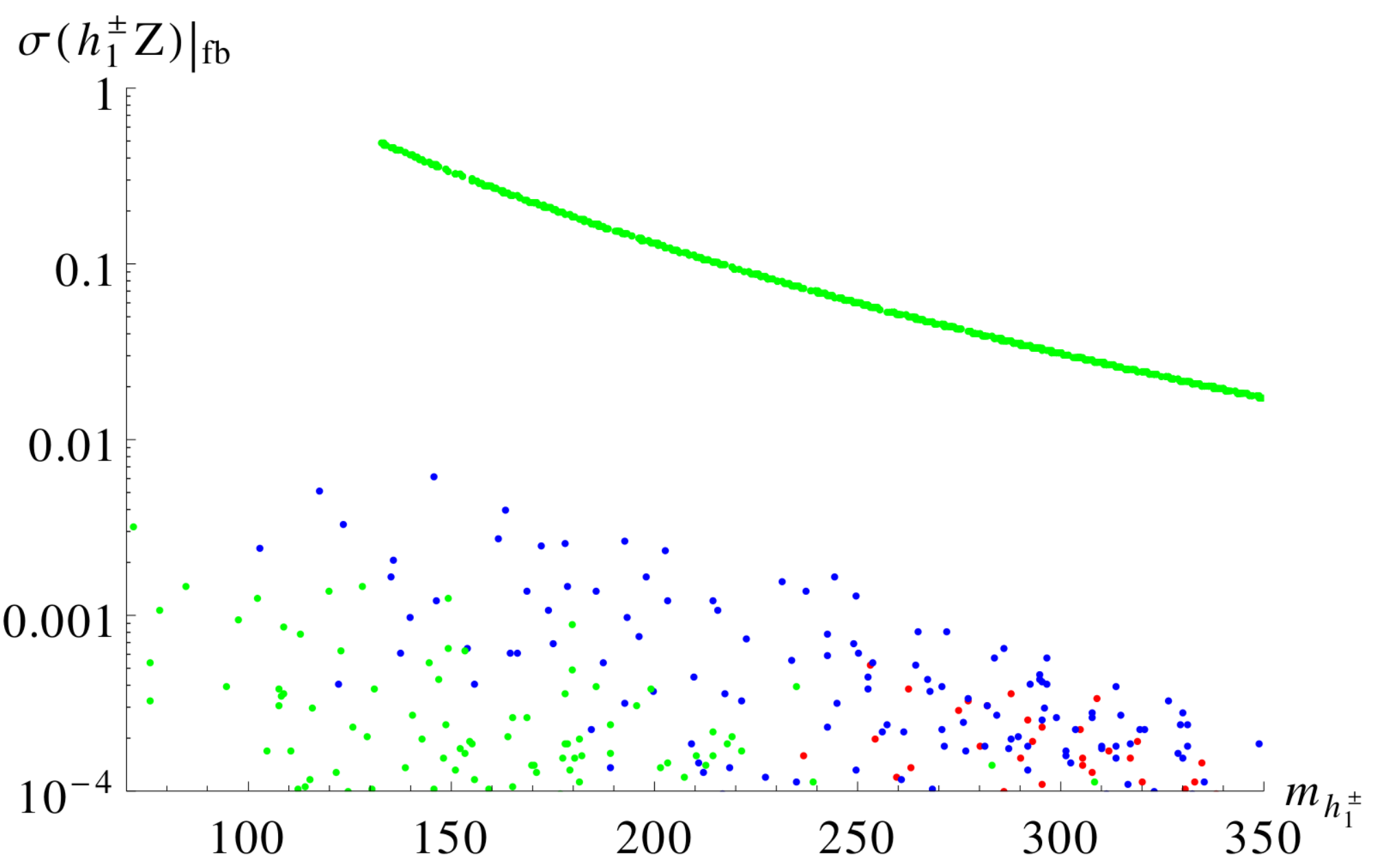}
\caption{The production cross-section of light charged Higgs boson $h^\pm_1$ in association with
$Z$ boson versus the light charged Higgs boson mass $m_{h^\pm_1}$.}\label{ch1Zcs}
\end{center}
\end{figure}

Nevertheless, in the case of the TNMSSM, a non-zero $h_1^\pm-W^\mp-Z$ vertex gives an extra contribution to this production process, which is absent in the case of doublet-like charged Higgs bosons. In fact, for $\lambda_T \simeq 0$, which corresponds to what we have called decoupling limit,  the $T^+_1$ and $T^-_2$ interaction eigenstates contribute additively to the $h_1^\pm-W^\mp-Z$, as can be seen from Eq.~\ref{zwch} and also can be realised from Figure~\ref{ssoslmbda} and Figure~\ref{ghmp1WZ}. However we can see from   Figure~\ref{ch1Zcs} that the $h^\pm_1Z$ production cross-section is smaller than the respective production in association with a $W^\pm$. This is  due to the fact that there are no other efficient contributions beside the channel with the $W^\pm$ in the propagator, as discussed earlier.

%%%%%%%%%%%%%%%%%%%%%%%%%%%%
%%%%%%%%%%%%%%%%%%%%%%

\subsection{Associated $h_1$}
\begin{figure}[thb]
\begin{center}
\mbox{\subfigure[]{
\includegraphics[width=0.35\linewidth]{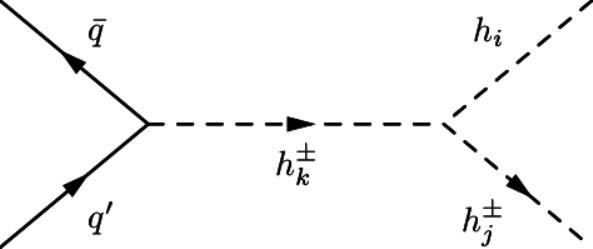}}\hskip 15pt
\subfigure[]{\includegraphics[width=0.35\linewidth]{plots/Whihcj.pdf}}}
\caption{The Feynman diagrams for the charged Higgs production in association with $h_i$  boson at the LHC.}\label{prodchhi}
\end{center}
\end{figure}

We have considered, than, the production of the charged Higgs boson production in association with a scalar Higgs boson, $h_i$. It is clear from Figure~\ref{prodchhi} that there are two contributions to this channel, one via the doublet-type
charged Higgs boson and another mediated by the $W^\pm$ boson.  However the charged Higgs mediated diagrams are suppressed, for the same reasons discussed earlier in the associated $Z$ production. Both the triplet and doublet Higgs bosons couple to $SU(2)$ gauge boson $W^\pm$. However a careful look on the vertex, given in Eq.~\ref{hachW}, shows that their mixing angles can have relative signs. In general their coupling in association with neutral Higgs bosons have to be doublet(triplet) type for doublet(triplet) type charged Higgs bosons. 
 
\begin{figure}[thb]
\begin{center}
\includegraphics[width=0.7\linewidth]{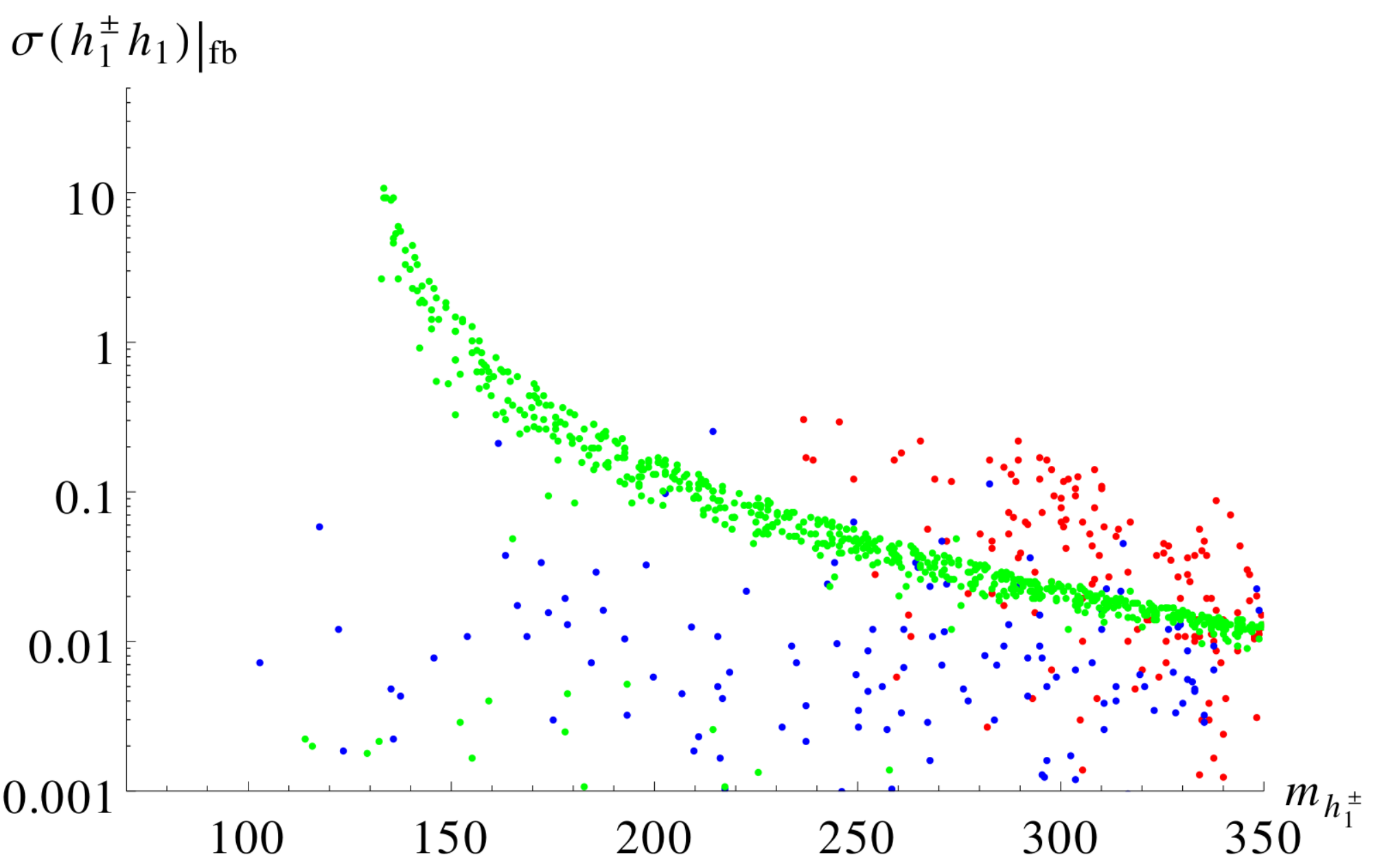}
\caption{The production cross-section of a light charged Higgs boson $h^\pm_1$ in association with
the $h_1$ boson versus the light charged Higgs boson mass $m_{h^\pm_1}$.}\label{ch1h1cs}
\end{center}
\end{figure}

This behaviour can be seen from Figure~\ref{ch1h1cs}, where we plot the production cross-section versus the mass of the lightest charged Higgs boson, $m_{h^\pm_1}$. The colour code for the charged Higgs boson remains as before. It is quite evident that, for a triplet-like charged Higgs boson, the cross-sections in association with $h_1$, which is mostly doublet, are very small, except for the $\lambda_T\simeq0$ points.  We can see the enhanced cross-section for the mostly doublet charged Higgs boson in association with doublet-like $h_1$ (red points). The situation is different for $\lambda_T\simeq0$, where 
 it is easy to produce a mostly triplet charged Higgs boson in this channel due to the enhancement of the $h^\pm_1-W^\mp-h_1$ coupling, given in Eq.~\ref{hachW}. This is due to the fact that  for  $\lambda_T\simeq0$ the rotation angles $\mathcal{R}^C_{12}$ and $\mathcal{R}^C_{14}$ of the triplet sector, which appear in the coupling given in Eq.~\ref{hachW}, take same sign (in the decoupling limit see Figure~\ref{ssoslmbda}).

%%%%%%%%%%%%%%%%%%%%%%%%%
%%%%%%%%%%%%%%%%%%%%%%
%%%%%%%%%%%%%%%%%%%%%%%%%%%%
%%%%%%%%%%%%%%%%%%%%%%

\subsection{Associated $a_1$}
%%%%%%%%%%%%%%%%%%%%%%%%%
\begin{figure}[thb]
\begin{center}
\mbox{\subfigure[]{
\includegraphics[width=0.35\linewidth]{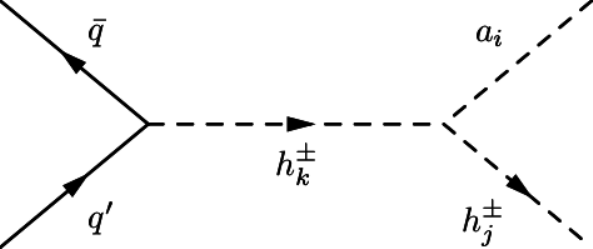}}\hskip 15pt
\subfigure[]{\includegraphics[width=0.35\linewidth]{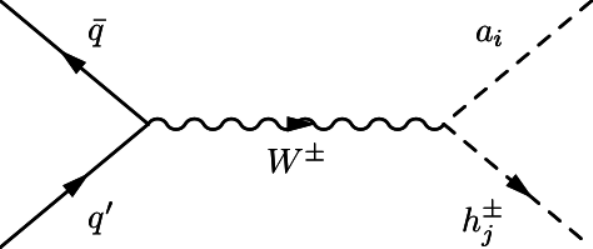}}}
\caption{The Feynman diagrams for the charged Higgs production in association with $a_i$  boson at the LHC.}\label{prodchai}
\end{center}
\end{figure}
\begin{figure}[thb]
\begin{center}
\includegraphics[width=0.7\linewidth]{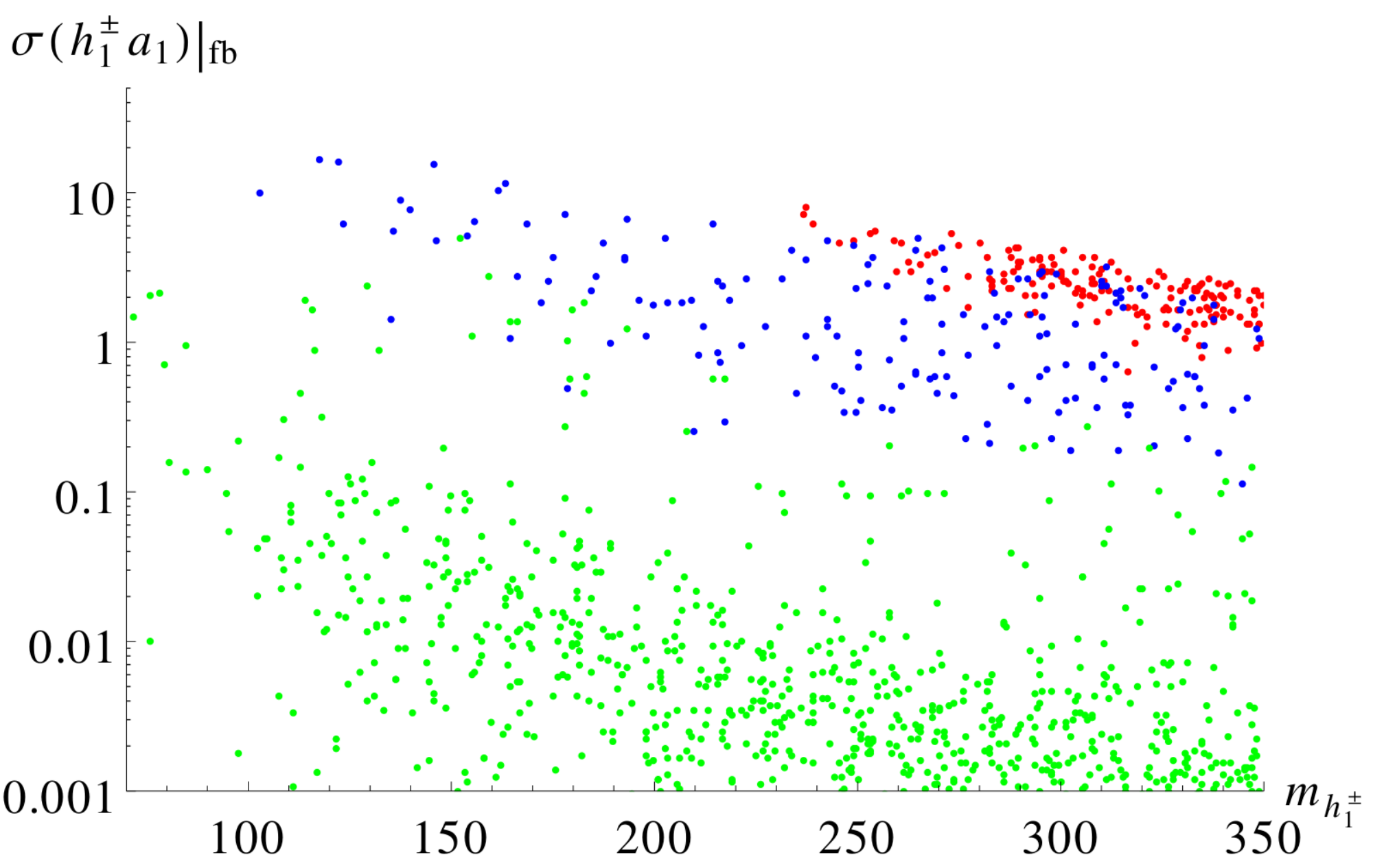}
\caption{The production cross-section of light charged Higgs boson $h^\pm_1$ in association with
the $a_1$ boson versus the light charged Higgs boson mass $m_{h^\pm_1}$.}\label{ch1a1cs}
\end{center}
\end{figure}

%%%%%%%%%%%%%%%%%%%%%%
Similarly, we can also produce the charged Higgs boson in association with a pseudoscalar Higgs boson, as shown in Figure~\ref{prodchai}. Here we also include the two contributions coming from $h^\pm_i$ and $W^\pm$ respectively even though, as before, the contribution from the charged Higgs propagator is negligible. Figure~\ref{ch1a1cs} presents the variation of the cross-section with the mass of the lightest charged Higgs boson. The cross-section stays very low for the triplet-like points (green ones) and reaches a maximum around 10 fb for doublet- and mixed-like points (red and blue ones). For $\lambda_T\simeq0$ points, the triplets ($T^+_1, T^{-*}_2$) rotation angles $\mathcal{R}^C_{i2, i4}$ appear with a relative sign in the coupling $h^\pm_i-W^\mp-a_j$, as can be seen in Eq.~\ref{hachW}. The $h^\pm_1 a_1$ cross-section thus gets a suppression in the decoupling limit, i.e. for $|\lambda_T|\simeq 0$, unlike the $h_ih^\pm_1$ case, as discussed in the previous section.

%%%%%%%%%%%%%%%%%%%%%%%%%%%%
%%%%%%%%%%%%%%%%%%%%%%

\subsection{Charged Higgs pair production }
\begin{figure}[thb]
\begin{center}
\mbox{\subfigure[]{
\includegraphics[width=0.35\linewidth]{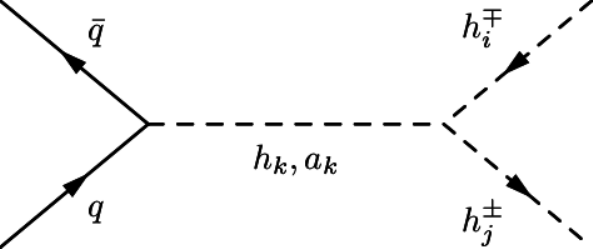}}\hskip 15pt
\subfigure[]{\includegraphics[width=0.35\linewidth]{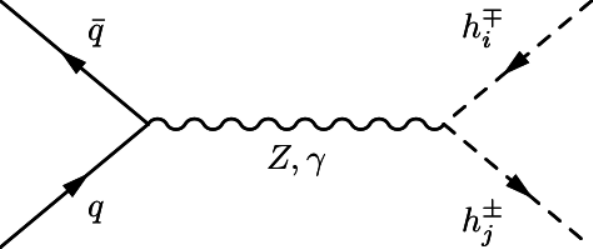}}}
\caption{Feynman diagrams for the production of a charged Higgs boson pair  $h^\mp_i h^\pm_j$ at the LHC, mediated by the Higgs bosons, the $Z$ and the $\gamma$ bosons.}\label{prodchij}
\end{center}
\end{figure}

\begin{figure}[thb]
\begin{center}
\subfigure[]{\hspace{-.5cm}
\includegraphics[width=0.7\linewidth]{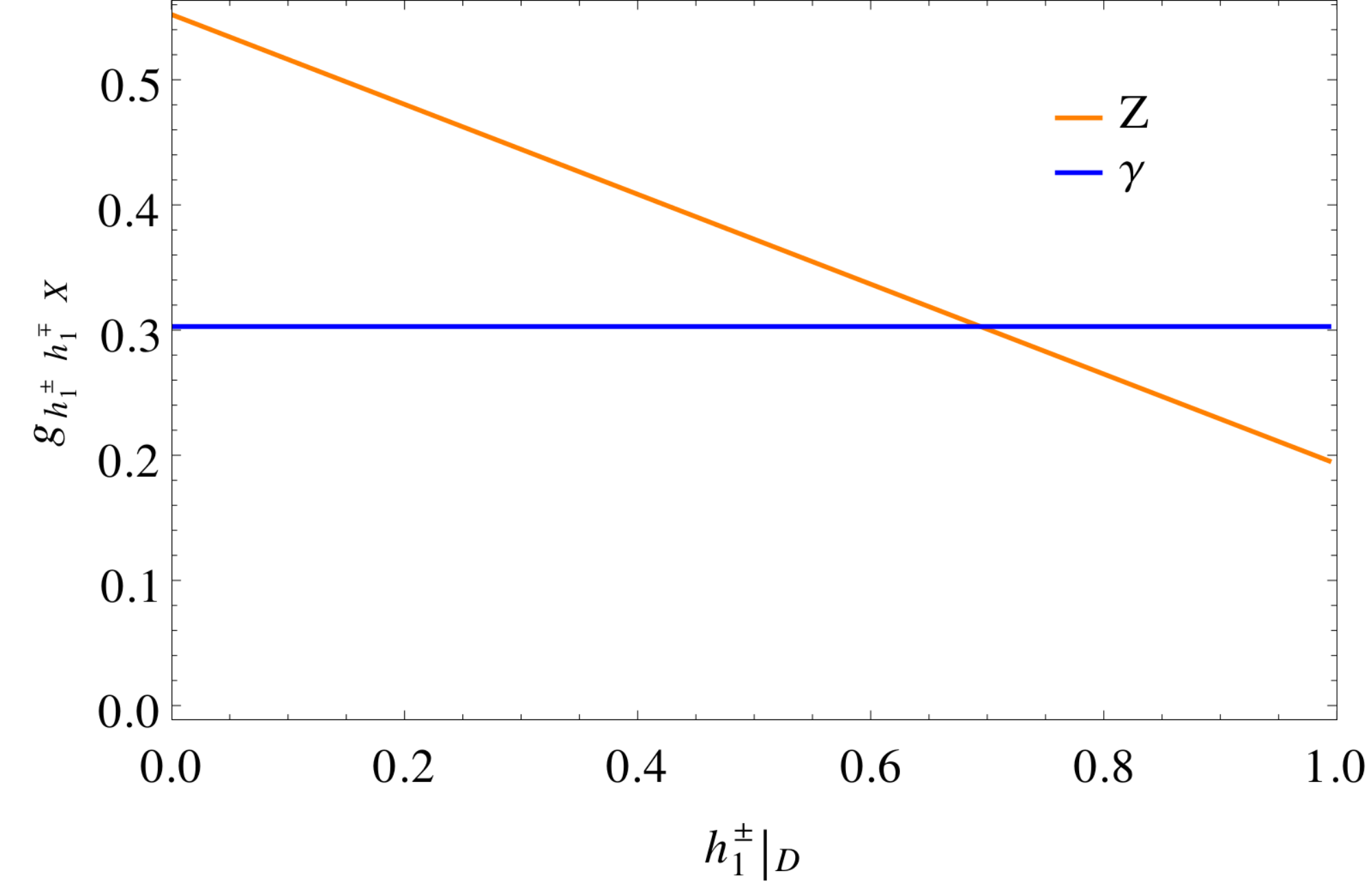}}
\caption{Value of the coupling $g_{h_1^\pm h_1^\mp X}$ as a function of the doublet fraction of the lightest charged Higgs boson. In the case of the photon this coupling is just the value of the electric charge.}\label{ZphoHpmHpm}
\end{center}
\end{figure}

\begin{figure}[thb]
\begin{center}
\includegraphics[width=0.8\linewidth]{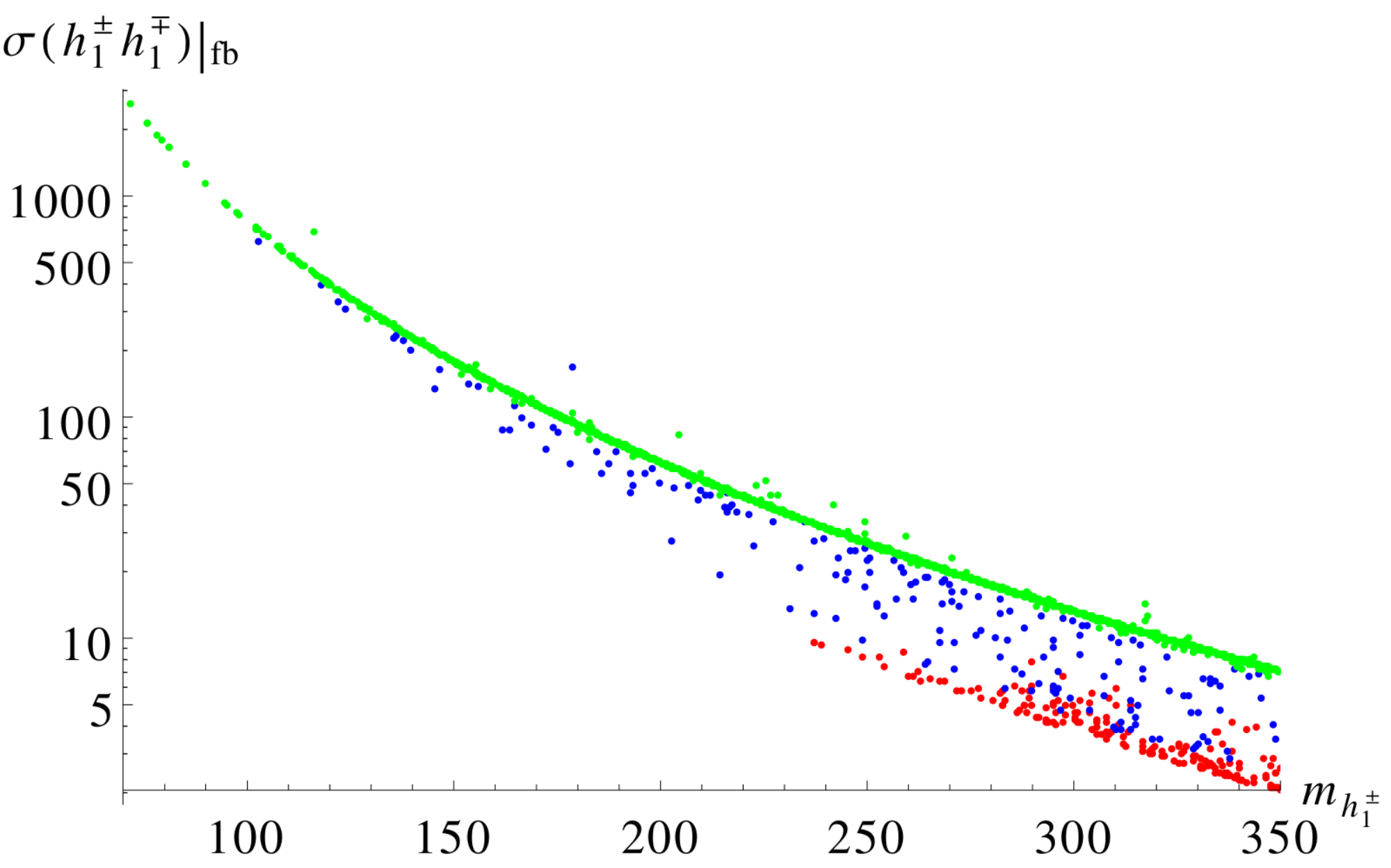}
\caption{The production cross-section of light charged Higgs boson pair  $h^\pm_1h^\mp_1$ versus the light charged Higgs boson mass $m_{h^\pm_1}$.}\label{ch1ch1cs}
\end{center}
\end{figure}

Here we move to the description of the charged Higgs pair production for the lightest charged Higgs boson $h^\pm_1$. The Feynman diagrams for this process are given in Figure~\ref{prodchij}, with the neutral Higgses and $Z,\gamma$ bosons contributing to the process. However, if the lightest charged Higgs boson $h^\pm_1$ is triplet-like, the diagrams of Figure~\ref{prodchij}(a) give less contribution to the cross section. In fact $a_1$ is selected to be singlet-like, so it does not couple to the fermoins, and the diagram with $h_{125}$ in the propagator is subdominant. The reason is that the coupling $g_{h_1^\pm h_1^\mp h_1}$ of a totally doublet scalar Higgs boson with two totally triplet charged Higgs bosons is prevented by gauge invariance. The triplet charged Higgs pair production is more suppressed  than the single triplet-like charged Higgs production via a doublet-like neutral Higgs boson. In that case pair production cross-section via off-shell doublet type neutral Higgs mediation ($h_{125}$) in s-channel via gluon-gluon fusion is below $\mathcal{O}(10^{-6})$ fb. Hence for triplet-like $h_1^\pm$ the diagrams of Figure~\ref{prodchij}(b) are the most relevant ones. The coupling of a pair of $h_1^\pm$ to the $Z$ and the $\gamma$ bosons is shown in Figure~\ref{ZphoHpmHpm} as a function of the doublet fraction. The coupling $g_{h_1^\pm h_1^\mp \gamma}$ is independent of the structure of $h_1^\pm$ as it should be because of the $U(1)_{\rm{em}}$ symmetry. In fact the value of this coupling is just the value of the electric charge. Conversely, the coupling of the $Z$ boson to a pair of charged Higgs depends on the structure of the charged Higgs. When the charged Higgs is totally doublet its coupling approaches the MSSM value $\frac{g_L}{2}\frac{\cos\,2\theta_w}{\cos\theta_w}$. If the charged Higgs is totally triplet the value of the coupling is $g_L\cos\theta_w$, the same of the $W^\pm-W^\mp-Z$ interaction.
%%%%%%%%%%%%%%%%%%%%%%%%%
%%%%%%%%%%%%%%%%%%%%%%
%%%%%%%%%%%%%%%%%%%%%%%%%
%%%%%%%%%%%%%%%%%%%%%%
In Figure~\ref{ch1ch1cs} we show the variation of the cross-sections with respect to the 
 lightest charged Higgs boson mass $m_{h^\pm_1}$. The colour code of the points are as the previous ones. We can see that for triplet-like points with mass around $\sim 100$ GeV the cross-section reach around pb. This large cross-section makes this production a viable channel to be probed at the LHC for the light triplet type charged Higgs boson. We discuss the corresponding phenomenology in section~\ref{pheno}.

%%%%%%%%%%%%%%%%%%%%%%%%%%%%
%%%%%%%%%%%%%%%%%%%%%%

\subsection{Vector boson fusion}
Neutral Higgs boson production via vector boson fusion is second most dominant 
production mode in SM. Even in 2HDM or MSSM this production mode of the neutral Higgs boson is one of the leading ones. However no such channel exist for charged Higgs boson as $h^\pm_i-W^\mp-Z$ vertex is zero at the tree-level, as long as custodial symmetry 
is preserved. The introduction of a $Y=0$ triplet breaks the custodial symmetry at tree-level, giving a non-zero $h^\pm_i-W^\mp-Z$ vertex, as shown in Eq.~\ref{zwch}. This vertex gives rise to the striking production channel of the vector boson fusion into a single charged Higgs boson, which is absent in the MSSM and in the 2-Higgs-doublet model (2HDM) at tree-level. This is a signature of the triplets with $Y=0, \pm 2$ which break custodial symmetry at the tree-level.
%%%%%%%%%%%%%%%%%%%%%%
%%%%%%%%%%%%%%%%%%%%%%%%%
\begin{figure}[t]
\begin{center}
\mbox{\subfigure[]{
\includegraphics[width=0.35\linewidth]{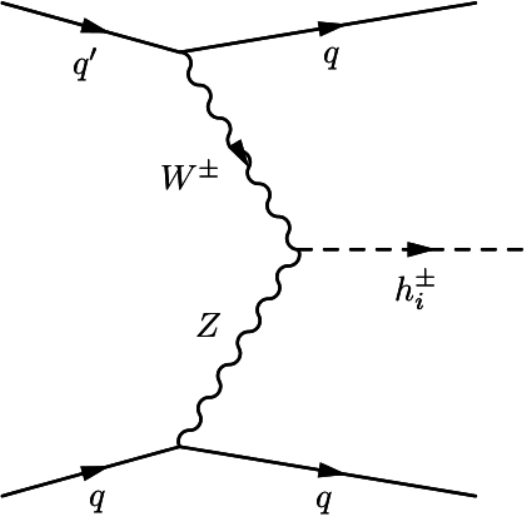}}}
\caption{The Feynman diagram for the charged Higgs production via vector boson fusion at the LHC.}\label{prodvvfch}
\end{center}
\end{figure}
%%%%%%%%%%%%%%%%%%%%%%

%%%%%%%%%%%%%%%%%%%%%%%%%%%%
\begin{figure}[thb]
\begin{center}
\includegraphics[width=0.7\linewidth]{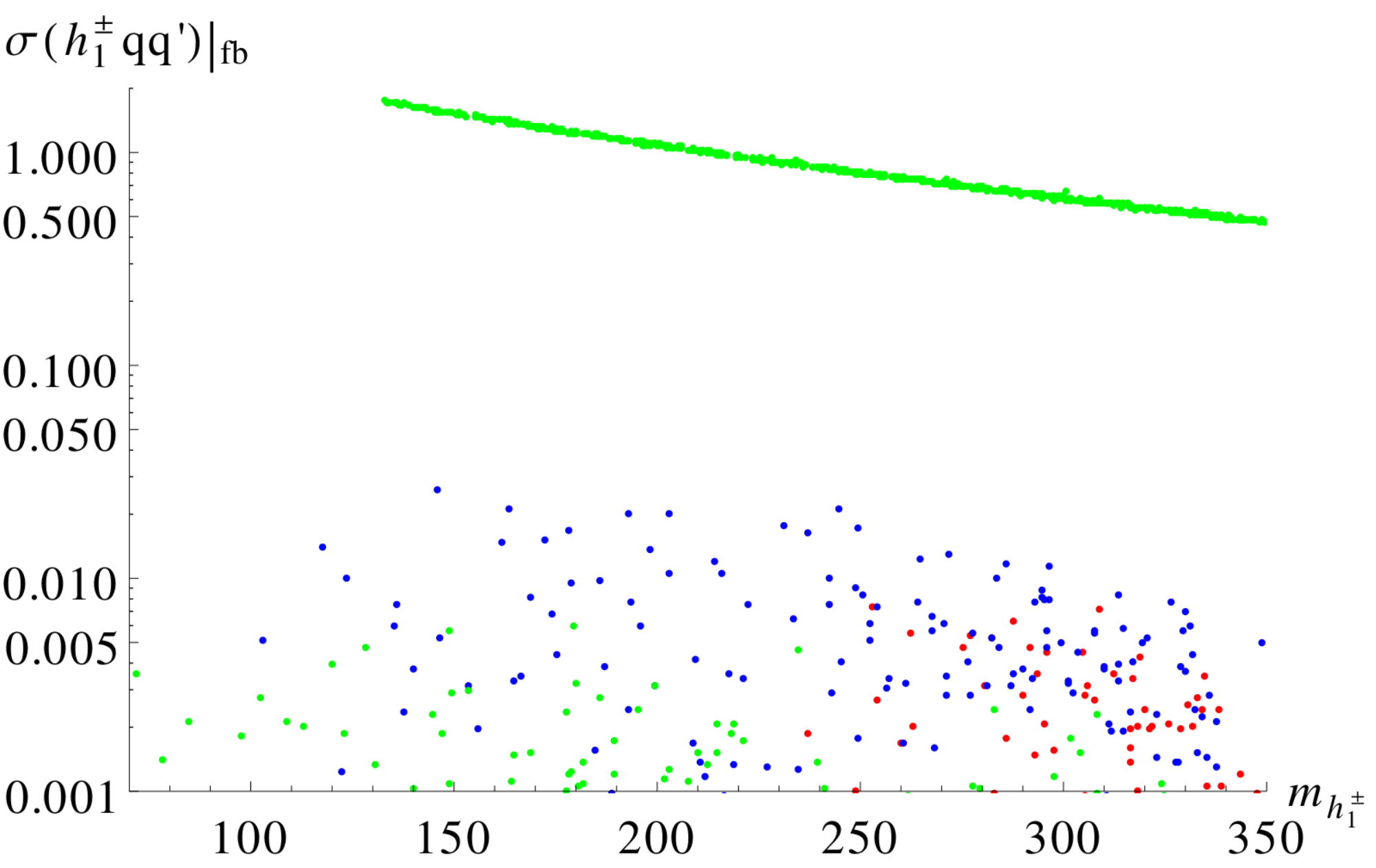}
\caption{The production cross-section of a light charged Higgs boson via  vector boson fusion versus the light charged Higgs boson mass $m_{h^\pm_1}$.}\label{VBFcs}
\end{center}
\end{figure}
%%%%%%%%%%%%%%%%%%%%%%
Figure~\ref{VBFcs} shows the cross-section variation with respect to the lightest charged Higgs boson mass $m_{h^\pm_1}$. As expected, doublet-like points (in red) have very small cross-sections, and for the mixed points (in blue) we see a little enhancement. Green points describe the cross-sections for the triplet-like points. We see that a triplet-like charged Higgs boson does not necessarily guarantee large values for the cross-section. As one can notice from Eq.~\ref{zwch}, the coupling $g_{h_1^\pm W^\mp Z}$ is a function of $\mathcal{R}^{C}_{12}$ and $\mathcal{R}^{C}_{14}$ and their relative sign plays an important role. From Figure~\ref{ghmp1WZ} we see that only in the decoupling limit, where where $\lambda_T=0$, both $\mathcal{R}^{C}_{12}$ and $\mathcal{R}^{C}_{14}$ take the same sign, thereby enhancing the $h_1^\pm- W^\mp -Z$ coupling and thus the cross-section.  It can been seen that only for lighter masses $\sim 150-200$ GeV the cross-sections is around few fb. Such triplet-like charged Higgs bosons can be probed at the LHC as a single charged Higgs production channel without  the top quark. This channel thus can be used to distinguish from other known single charged Higgs production mode in association with the top quark, which characterises  a doublet-like charged Higgs bosons.

\subsection{Associated top quark}
%%%%%%%%%%%%%%%%%%%%%%%%%
\begin{figure}[thb]
\begin{center}
\mbox{\subfigure[]{
\includegraphics[width=0.35\linewidth]{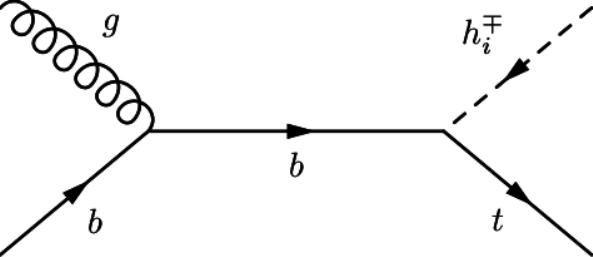}}\hskip 20pt
\subfigure[]{\includegraphics[width=0.3\linewidth]{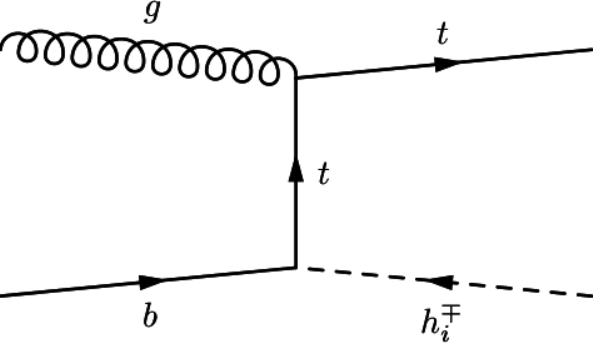}}}
\caption{The Feynman diagrams for the charged Higgs production in association with a top quark at the LHC.}\label{prodtch}
\end{center}
\end{figure}
%%%%%%%%%%%%%%%%%%%%%%
In the TNMSSM the triplet sector does not couple to fermions, which causes a natural suppression of the production of a triplet-like charged Higgs in association with a top quark. The only way for this channel to be allowed is via the mixing with doublets. Figure~\ref{prodtch} shows the Feynman diagrams of such production processes, which are dominant and take place via a $b$ quark and gluon fusion. They are highly dependent on the value of $\tan{\beta}$ \cite{djuadi, moretti}.
%%%%%%%%%%%%%%%%%%%%%%%%%%%%
\begin{figure}[thb]
\begin{center}
\includegraphics[width=0.7\linewidth]{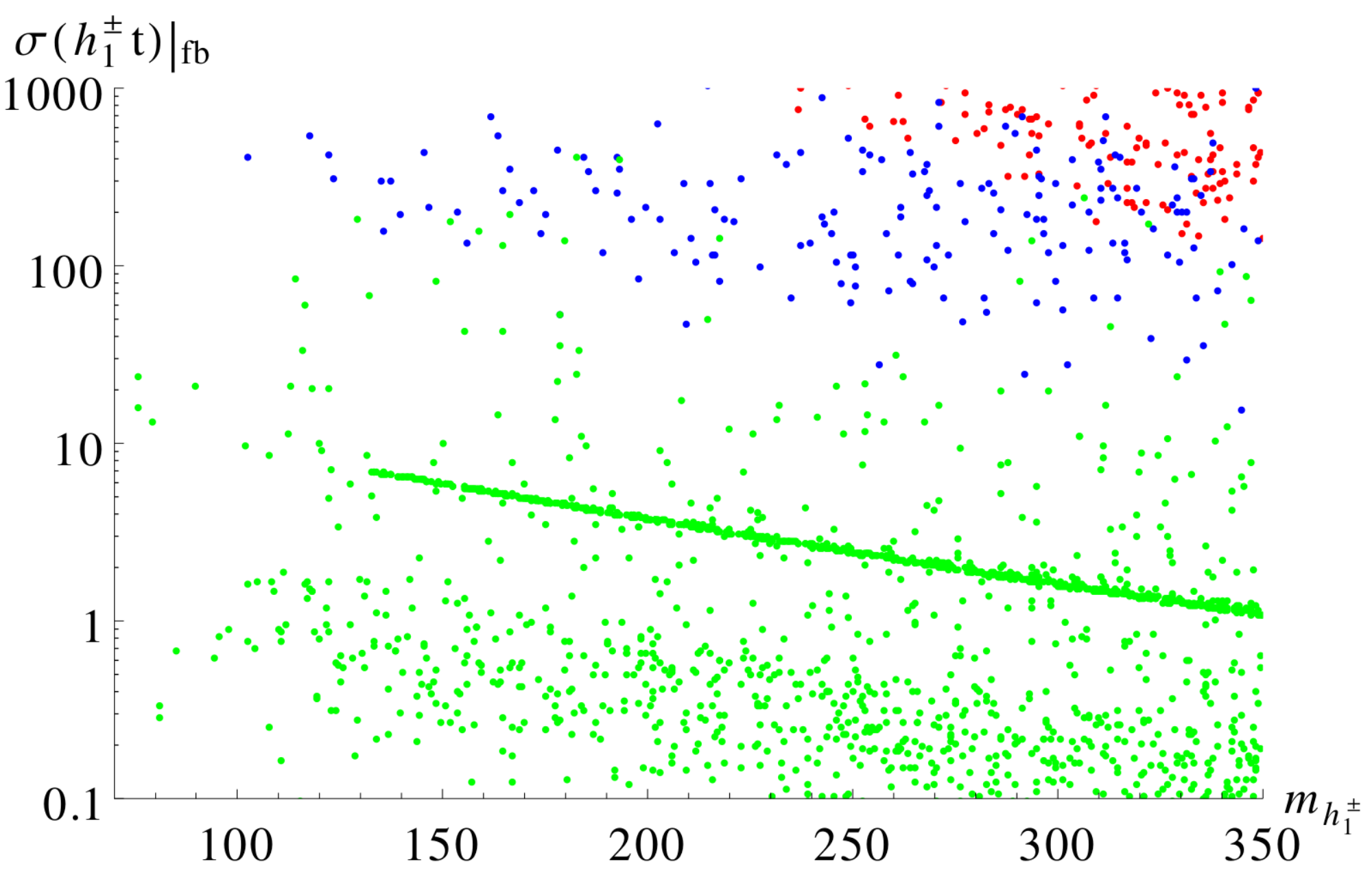}
\caption{The production cross-section of light charged Higgs boson in association with top quark versus the light charged Higgs boson mass $m_{h^\pm_1}$.}\label{tchcs}
\end{center}
\end{figure}
%%%%%%%%%%%%%%%%%%%%%
Figure~\ref{tchcs}(b) shows the production cross-section as a function of the lightest charged Higgs boson mass, where the green points correspond to linear combinations which are mostly triplet ($\gsim 90\%$), while red points correspond to those which are mostly of doublet ($\gsim 90\%$) and the blue points are of mixed type. Triplet-like points have a naturally suppressed cross-section whereas the doublet-like points have a large cross-sections, that can be $\sim$ pb. The mixed points lay in between, with cross-sections $\mathcal{O}(100)$ fb. One can also notice the certain enhanced line in the green points which correspond to  $|\lambda_T| \simeq 0$. As already explained in the previous sections, in this limit some portion ($\sim(\frac{v_T}{v})^2$) of the lightest charged Higgs boson $h^\pm_1$ remains doublet type, as shown in Figure~\ref{chpslmbda}, and is responsible for the enhancement of the cross-section.

Thus not finding a charged Higgs boson in this channel does not mean that it is completely ruled out, simply it can come from higher representation of $SU(2)$.

\section{Charged Higgs boson phenomenology}\label{pheno}
As already pointed out before, the TNMSSM with a $Z_3$ symmetry allows a very light singlet-like pseudoscalar in its spectrum, which turns into a pseudo-NG mode in the limit of small soft parameters $A_i$ \cite{TNMSSM1}. The existence of such a light and still hidden scalar prompts the decay of a light charged Higgs boson $h^\pm_1 \to a_1 W^\pm$. Of course the gauge invariant structure of the vertex further restricts such decay mode, which is only allowed by the mass mixing of the singlet with the doublets or the triplet.  In the extended supersymmetric scenarios with only triplet, one cannot naturally obtain such light triplet-like pseudoscalar, because imposing $Z_3$ symmetry would be impossible due to existence of $\mu$ term, which is necessary to satisfy the lightest chargino mass bound \cite{pbas3}. The existence of a light pseudoscalar mode has been observed and studied in the context of the NMSSM \cite{han, colepa, guchait, pbsnkh}. Unlike NMSSM, in TNMSSM with a $Z_3$ symmetry the decay $h^\pm_1 \to ZW^\pm$ is possible for a triplet-type light charged Higgs boson. Below we discuss the phenomenology of such charged Higgs bosons at the LHC.
 
 {For this phenomenological analysis we have selected three benchmark point, named BP1, BP2 and BP3 given in Table~\ref{BP}.
%%%%
\begin{table}
\begin{center}
\renewcommand{\arraystretch}{1.4}
\begin{tabular}{||c||c|c|c|c|c||}
\hline
\hline
&$m_{h_1^\pm}$&$m_{a_1}$&$\mathcal{Br}(a_1W^\pm)$&$\mathcal{Br}(Z\,W^\pm)$&$\mathcal{Br}(\tau\nu_\tau)$\\
\hline
BP1&179.69&41.22&$9.7\times10^{-1}$&$2.1\times10^{-2}$&$1.3\times10^{-4}$\\
\hline
BP2&112.75&29.77&$9.9\times10^{-1}$&$6.3\times10^{-5}$&$5.5\times10^{-3}$\\
\hline
BP3&172.55&48.94&$6.3\times10^{-5}$&$9.8\times10^{-1}$&$2.4\times10^{-3}$\\
\hline
\hline
%\hline
\end{tabular}
\caption{The mass of $h_1^\pm$, the mass of $a_1$ and the relevant branching ratios for the three benchmark points used in the phenomenological analysis.}\label{BP}
\end{center}
\end{table}
All of them are characterised by a triplet-like charged Higgs boson $h_1^\pm$, which make the charged Higgs branching fractions into fermions, e.g. $\mathcal{Br}(h_1^\pm\to\tau\nu_\tau)$ or $\mathcal{Br}(h_1^\pm\to t\,b)$, strongly suppressed. We choose this scenario of triplet-like charged Higgs boson to look for new physics signals that is not there in two Higgs doublet model (2HDM), MSSM and NMSSM. The benchmark points maximize following decay modes;
 \begin{itemize}
   \item BP1:  \\
   $\sigma_{pp\to h_1^\pm h_1^\mp} \times \mathcal{Br}(h_1^\pm \to a_1W^\pm)\mathcal{Br}(h_1^\mp \to  Z\,W^\mp)$ ,
   \item BP2:\\
    $\sigma_{pp\to h_1^\pm h_1^\mp} \times \mathcal{Br}(h_1^\pm \to a_1W^\pm)\mathcal{Br}(h_1^\mp \to   a_1W^\pm)$ 
    
    \item BP3:\\
    $\sigma_{pp\to h_1^\pm h_1^\mp} \times \mathcal{Br}(h_1^\pm \to Z\,W^\mp)\mathcal{Br}(h_1^\mp \to   Z\,W^\mp)$.
\end{itemize}
 We will discuss the final sate searches along with dominant SM backgrounds  below starting for BP1 to BP3. A detailed collider study is in preparation \cite{pbch}. 
 
 If the lightest charged Higgs boson is pair produced, it can have the following decay topologies
\bea\label{fs1}
pp &\to& h^\pm_1h^\mp_1\nn \\
 &\to & a_1 W^\pm Z W^\mp \nn \\
   &\to & 2\tau (2b)+ 2j+ 3\ell \nn +\etmiss \\
   & \to  & 2\tau (2b)+ 4\ell +\etmiss .
\eea
Eq.~\ref{fs1} shows that when one of the charged Higgs bosons decays to $a_1 W^\pm$, which is a signature of the existence of singlet-type pseudoscalar, and the other one decays to $ZW^\pm$, which is the triplet signature. Thus we end up with $a_1 +2 W^\pm +Z$ intermediate state. Depending on the decays of the gauge bosons; hadronic or leptonic, and that of the light pseudoscalar (into $b$ or $\tau$ pairs), we can have final states with multi-lepton plus two $b$- or $\tau$-jets. The tri-lepton and four-lepton backgrounds are generally rather low in SM. In this case they are further tagged with $b$ or $\tau$-jet pair, which make these channels further clean. As mentioned earlier the detailed signal, backgrounds study is in progress as a separate study in \cite{pbch}. However in Table~\ref{finalstates}
we look for $ \geq3\ell+2\tau+\etmiss$ and $\geq3\ell+2b+\etmiss$ final states event numbers at an integrated luminosity of 1000 fb$^{-1}$ for both BP1 and dominant SM backgrounds. The demand $\geq 3\ell$ over $4\ell$ was chosen to enhance the signal numbers. The  kinematical cuts on the momentum and various isolation cuts  and tagging efficiencies for $b$-jets \cite{btag} and $\tau$-jets \cite{tautag} reduce the final state numbers. The $b$-tagging efficiency has been chosen to be $0.5$ and $\tau$-jet tagging efficiency varies a lot with the momentum of the $\tau$-jet ($30-70\%$) are taken into account while giving the final state numbers. 
  
For $\geq3\ell+2\tau+\etmiss$ and $\geq3\ell+2b+\etmiss$ final states the dominant backgrounds mainly come from triple gauge boson productions $ZZZ$ and $ZWZ$ respectively.  We can see that that  $\geq3\ell+2b+\etmiss$ reaches around $3\sigma$ of signal significance at an integrated luminosity of 1000 fb$^{-1}$. However a point with larger branching to both $aW^\pm$ and $ZW^\pm$ decay modes can be probed with much earlier data.

 In the case of a TESSM \cite{pbas1, pbas3} we have have only the triplet signature of charged Higgs decaying into $ZW^\pm$, which carries a different signature respect to the doublet-like charged Higgs boson. On the other hand, in the NMSSM we
only have $a_1 W^\pm$ decay \cite{han, colepa, guchait, pbsnkh}, which is characterised by a different signature respect to the MSSM \cite{ChCMS,ChATLAS}. In comparison, Eq.~\ref{fs1} provides a golden plated mode in the search of an extended Higgs sector, as predicted by the TNMSSM. Finding out both $a_1 W^\pm$ and $Z W^\pm$ decay modes at the LHC can prove the existence of both a singlet and  a triplet of the model. However, as we can see in Figure~\ref{muCH}, it is very difficult to find out points where both the $\mathcal{Br}( h^\pm_1 \to ZW^\pm)$ and $\mathcal{Br}( h^\pm_1 \to a_1W^\pm)$ are enhanced at the same time. Nevertheless as the final states carry the signatures of both singlet and triplet type Higgs bosons, it is worth exploring for a high luminosity at the LHC or even for higher energy (more than 14 TeV) at the LHC in future.

%%%%%%%%%%%%%%%%%%%%%%
\begin{figure}[thb]
\begin{center}
\mbox{\subfigure[]{
\includegraphics[width=0.5\linewidth]{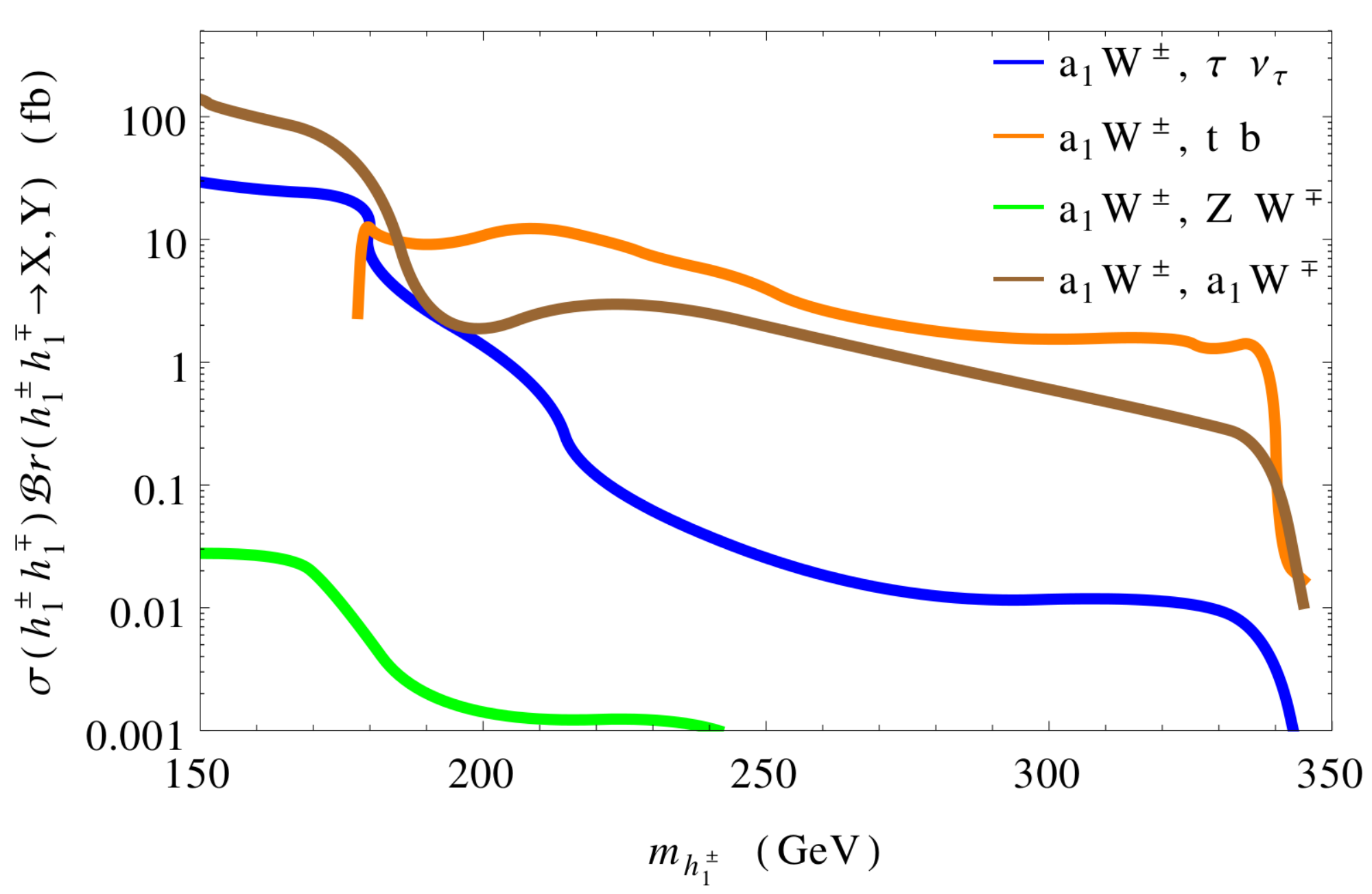}}
\subfigure[]{\includegraphics[width=0.5\linewidth]{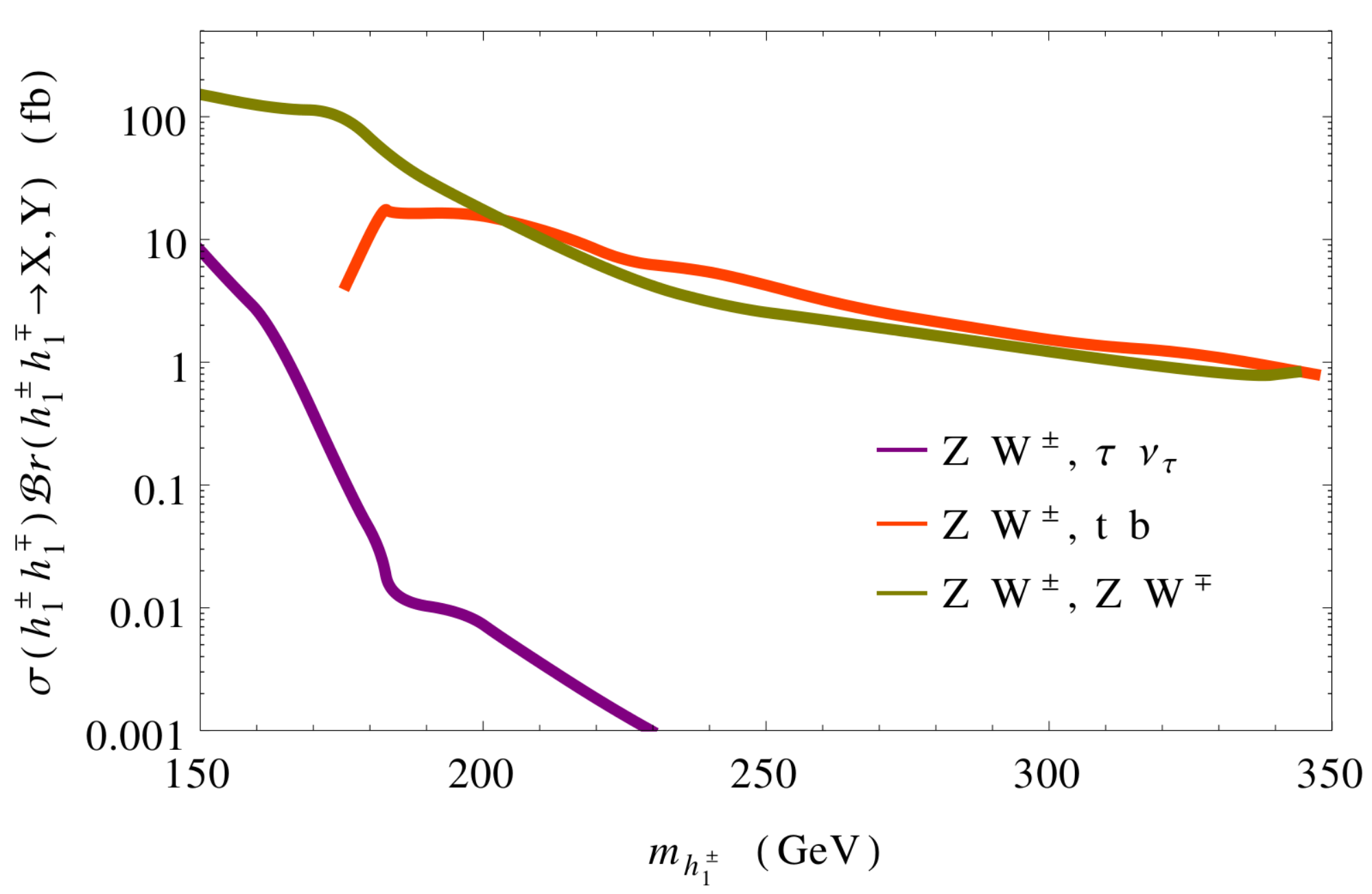}}}

\caption{The signal strength for the pair production of the lightest charged Higgs boson in the intermediate channels of Eq. \ref{fs1}, \ref{fs2}, \ref{fs2b}, \ref{fs2t} (a) and \ref{fs2a}, \ref{fs2c}, \ref{fs2ta} (b) as a function of the mass of the lightest charged Higgs boson.}\label{muCH}
\end{center}
\end{figure}
%%%%%%%%%%%%%%%%%%%%%%

%%%%%%%%%%%%%%%%% h^\pm_1 pair production signal%%%%%%%%%%%%%%%%%%%%
\begin{table}
\begin{center}
\renewcommand{\arraystretch}{1.5}
\begin{tabular}{||c|c|c||c|c||}
\hline\hline
\multicolumn{3}{||c||}{\multirow{2}{*}{Decay Channels}}&\multicolumn{2}{c||}{$\#$ of Events}\\
\cline{4-5}
\multicolumn{3}{||c||}{}&Signal&Backgrounds\\
\hline
\hline
\multirow{2}{*}{\rotatebox{90}{BP1}}&\multirow{2}{*}{$\;a_1W^\pm\,ZW^\mp\;$}
&$\geq3\ell+2\tau+\etmiss$&1&6\\
\cline{3-5}
&&$\geq3\ell+2b+\etmiss$&21&39\\
\hline
\hline
\multirow{2}{*}{\rotatebox{90}{BP2}}&$\;a_1W^\pm\,\tau\nu_\tau\;$
&$3\tau+1\ell+\etmiss$&13&$<1$\\
\cline{3-5}
%&$\;2b+1\tau+1\ell+\etmiss\;$&312&\\
\cline{3-5}

%&$\geq1\ell+4b+\etmiss$&94.51&\\
%\cline{2-4}
&$\;a_1W^\pm\,a_1W^\mp\;$&$\;2b+2\tau+2\ell+\etmiss\;$&164&38\\
\hline
\hline
\multirow{3}{*}{\rotatebox{90}{BP3}}&$\;ZW^\pm\,\tau\nu_\tau\;$
&$1\tau+3\ell+\etmiss$&9&19\\
\cline{3-5}
\cline{3-5}
&\multirow{2}{*}{$\;Z\,W^\pm\,Z\,W^\mp\;$}
&$\geq5\ell+\etmiss$&228&23\\
\cline{3-5}
&&$\;\geq1\ell+2b+2\tau+\etmiss\;$&29&246\\

\hline\hline
\end{tabular}
\caption{The final state numbers for the benchmark points and backgrounds  at an integrated luminosity of 1000 fb$^{-1}$.}\label{finalstates}
\end{center}
\end{table}

}

The light charged Higgs boson can also decay to $\tau\nu$ for $m_{h^\pm_1} < m_t$ and to $tb$ for $m_{h^\pm_1} > m_t$, via its doublet fraction. The charged Higgs pair production then has the signatures given in Eq.~\ref{fs2} and Eq.~\ref{fs2a}, with one of the charged Higgs boson decaying to $\tau\nu$ and the other one to $a_1 W^\pm$ or $Z W^\pm$, respectively. Eq.~\ref{fs2} and Eq.~\ref{fs2a}  probe the existence of singlet, doublet and triplet representations at the same time. The final states with one or more tau-jets along with charged lepton reduce the SM backgrounds but nevertheless $t\bar{t}Z$ and $tZW^\pm$ contribute. 

\bea\label{fs2}
pp &\to & h^\pm_1h^\mp_1 \nn \\
& \to & a_1 W^\pm  \tau\nu \nn \\
 & \to & 3\tau /(2b +1\tau) +1 \ell +\etmiss,  
   \eea

\bea\label{fs2a}
pp &\to & h^\pm_1h^\mp_1 \nn \\
 &\to & Z W^\pm  \tau\nu \nn \\
  & \to & 1(3)\tau + 3(1)\ell  +\etmiss. 
   \eea

Thus these final states would play a very crucial role in determining whether the mechanism of EWSB incorporates a finer structure respect to our current description, with a single Higgs doublet.
{ In Table~\ref{finalstates} we present the number of events in the $3\tau+1\ell+\etmiss$ final state for the channel $a_1W^\pm\,\tau\nu_\tau$ and in the $1\tau+3\ell+\etmiss$ for the channel $ZW^\pm\,\tau\nu_\tau$ at an integrated luminosity of 1000 fb$^{-1}$. As already stated, we chose a triplet-like charged Higgs boson $h_1^\pm$ and hence the branching in $\tau\nu_\tau$ is suppressed, being a signature decay mode for a doublet-type charged Higgs boson. In both the case the dominant backgrounds are the triple gauge bosons $ZZZ$ and $ZWZ$. We can see that that  $3\ell+1\tau+\etmiss$ reaches more than $3\sigma$ of signal significance at an integrated luminosity of 1000 fb$^{-1}$.

}
There are, of course, two other possibilities for the decays of a pair of charged Higgs bosons, that is when both the charged Higgs bosons decays to $a_1W^\pm$ or $ZW^\pm$.
\bea\label{fs2b}
pp &\to& h^\pm_1h^\mp_1\nn \\
 &\to & a_1 W^\pm a_1 W^\mp \nn \\
   &\to & 2\tau +2b+ 2j+ 1\ell \nn +\etmiss \\
   & \to  & 4\tau (4b)+ 2\ell +\etmiss\nn\\
   & \to &  2b+2\tau+2\ell+\etmiss.
\eea
\bea\label{fs2c}
pp &\to& h^\pm_1h^\mp_1\nn \\
 &\to & Z W^\pm Z W^\mp \nn \\
   &\to & 2j+ 4\ell \nn +\etmiss \\
   & \to  & 6\ell +\etmiss \nn\\
   & \to &2b+2\tau+2\ell+\etmiss.
\eea
These channels can prove the existence of singlet and triplet representation separately.
{  For the decay channel $h^\pm_1h^\mp_1\to  a_1 W^\pm a_1 W^\mp$ we have considered the $2b+2\tau+2\ell+\etmiss$ final state for the signal and background analysis. This is because the final states with $\geq1\ell$ have $\bar t t$ as dominant background and hence are strongly suppressed. For $2b+2\tau+2\ell+\etmiss$ the dominant backgrounds are $ZZZ$ and $\bar t tZ$ and we can see from Table~\ref{finalstates} that the signal significance is more than $10\sigma$ for an integrated luminosity of 1000 fb$^{-1}$. A $5\sigma$ of signal significance can be achieved with an integrated luminosity of $\approx$ 200 fb$^{-1}$ at the LHC with 14 TeV center of mass energy.

In the case of $h^\pm_1h^\mp_1\to Z\, W^\pm Z\, W^\mp$ we look into the $\geq5\ell+\etmiss$ and $\geq1\ell+2b+2\tau+\etmiss$ final states where the demand $\geq 1\ell$ over $2\ell$ was chosen to enhance the signal numbers. The $\geq5\ell+\etmiss$ has the triple gauge bosons $ZZZ$ and $ZWZ$ as dominant backgrounds. This is one of cleanest final state and we can see from Table~\ref{finalstates} that it has more than $14\sigma$ of signal significance at an integrated luminosity of 1000 fb$^{-1}$. The integrated luminosity for $5\sigma$ of signal significance is 120 fb$^{-1}$. The dominant backgrounds for the  $\geq1\ell+2b+2\tau+\etmiss$ final state are the triple gauge bosons $ZZZ$ and $ZWZ$ as well as $\bar t tZ$. The $\bar t tZ$ background is the most dominant one in this case and suppress the signal significance, as one can immediately realize looking at Table~\ref{finalstates}.
}

For a charged Higgs bosons heavier than the top quark the channel $h^\pm_1 \to t b$ is kinematically allowed. 
If one of the charged Higgs decays to $tb$ and the other one decays to $ a_1 W^\pm$ we have the final states given by Eq.~\ref{fs2t}. When the other charged Higgs boson decays to $ZW^\pm$, the production of $h^\pm_1h^\mp_1$ results in the final states of  Eq.~\ref{fs2ta} 
\bea\label{fs2t}
pp&\to & h^\pm_1h^\mp_1 \nn \\
 &\to & a_1 W^\pm  t b\nn \\
  & \to & 2\tau + 2b + 2 W \nn \\
     & \to & 2\tau +2b + 2\ell +\etmiss, 
   \eea
   
   \bea\label{fs2ta}
pp&\to & h^\pm_1h^\mp_1 \nn \\
 &\to & Z W^\pm  t b\nn \\
   &\to & 2\tau + 2b + 2 W \nn \\
     & \to & 2\tau +2b + 2\ell +\etmiss\, \nn\\
     & \rm{or}& \,2b+ 4\ell +\etmiss.  
   \eea
The signal related to the intermediate states of the pair production and the decays of the lightest charged Higgs boson in the channels of Eq. \ref{fs1}, \ref{fs2}, \ref{fs2a}, \ref{fs2t} and \ref{fs2ta} is reported in Figure~\ref{muCH}. We can clearly see that for light charged Higgs boson  ($m_{h^\pm_1} \gsim 200$ GeV)  the decay modes in a light pseudoscalar can be probed rather easily at the LHC 
but probing  $a_1W^\pm$ and $ZW^\mp$, i.e., the existence of a light pseudoscalar and the triplet decay modes  together needs higher luminosity.

Another signature of this model could be the existence of the heavier charged Higgs bosons $h^\pm_{2,3}$
which could be produced at the LHC. For our selection points $h^\pm_2$ is triplet-like
and $h^\pm_3$ is doublet-like. Following our discussion in section~\ref{ch1dcy},  
such heavy charged Higgs can decay dominantly to $a_1 h^\pm_1$ or $h_1 h^\pm_1$, as shown in
Eq.~\ref{fs3} and Eq.~\ref{fs4}. The lighter charged Higgs can then decay into final states with  $a_1 W^\pm$
or $Z W^\pm$ giving $2\tau (2b)+  3\ell +\etmiss$ and $4\tau (4b)+  1\ell +\etmiss$ final states

\bea\label{fs3}
pp\to h^\pm_{2,3} +X&  \to & a_1/h_1 h^\mp_1 \nn \\
&  \to & 2\tau (2b)+ Z W^\pm \nn \\
  & \to & 2\tau (2b)+  3\ell + \etmiss,
\eea

\bea\label{fs4}
pp\to h^\pm_{2,3} +X & \to & a_1/h_1 h^\mp_1 \nn \\
 & \to & 2\tau (2b)+ a_1 + W^\pm \nn \\
  & \to & 4\tau (4b)+  1\ell + \etmiss.
\eea

Searching for the above signatures is certainly necessary not only in order to discover a charged Higgs boson but also to determine whether scalars in higher representations of $SU(2)$ are involved in the mechanism of EWSB.

\chapter{Simulating a Light Pseudoscalar in the TNMSSM}
\section{Synopsis} 
This chapter is devoted to the phenomenological analysis of a light pseudoscalar state present in the TNMSS, whose mass is entirely generated by the soft-breaking terms of the theory.  As we have already discussed in the introduction to Chapter 1, modes of this type are associated to the presence of flat directions in the potential of a superconformal theory. The goal of the chapter is to provide an in depth analysis of the possible signatures of this state and the constraints emerging from a comparison oof the prediction of the model against the current LHC data.   
\section{Introduction}
The success of the Standard Model (SM) in explaining the gauge structure of the fundamental interactions has reached its height with the discovery of a scalar particle with {most of} the properties of the SM Higgs boson - as a 125 GeV mass resonance - at the LHC. With this discovery, the mechanism of spontaneous symmetry breaking of the gauge symmetry, which in a gauge theory such as the SM is mediated by a Higgs doublet, has been confirmed, but the possible existence of an extended Higgs sector, at the moment, cannot be excluded. 

The identification by the CMS \cite{CMS, CMS2} and ATLAS \cite{ATLAS} experiments of a new boson exchange, has interested so far only the $WW^*$, $ZZ^*$ and $\gamma\gamma$
channels - using data at 7 and at 8 TeV - at more than $5\sigma$ confidence level for the $Z$ and $\gamma$ cases, and slightly below in the $W$ channel. However, the fermionic decay modes of the new boson, together with other exotic decay modes, are yet to be discovered. Clearly, they are essential in order 
to establish the mechanism of electroweak symmetry breaking (EWSB), which is crucial in the SM dynamics, with better precision.  
The new data collection at the LHC at 13 TeV center of mass energy - which will be upgraded to 14 TeV in the future - will probably provide new clues about some possible extensions of the SM, raising large expectations both at theoretical and at experimental level. \\
The SM is not a completely satisfactory theory, even with its tremendous success, since it does not provide an answer to long-standing issues, most prominently the gauge-hierarchy problem. This is instead achieved by the introduction of supersymmetry, which, among its benefits, allows gauge coupling unification and, in its R-parity conserving version,   
also provides a neutral particle as a dark matter candidate. The absence of any supersymmetric signal at the LHC and the recent observation of  a Higgs boson $(h_{125})$ of 125 GeV in mass, requires either a high SUSY mass scale or larger mixings between the scalar tops \cite{pMSSMb}. The situation is severer for more constrained 
SUSY scenarios like mSUGRA  \cite{cMSSMb}, which merge supersymmetric versions of the SM with minimal 
supergravity below the Planck scale. 

In the current situation, extensions of the Higgs sector with the inclusion of one or more electroweak doublets and/or of triplets of different hypercharges - in combination with {SM gauge} singlets - are still theoretical 
possibilities in both supersymmetric and non-supersymmetric extensions of the SM. We have recently shown that a supersymmetric extension of SM with a  $Y=0$ triplet and a singlet Higgs superfields \cite{TNSSMo}, called the TNMSSM, is still a viable scenario, which is compatible with the recent LHC results and the previous constraints from LEP, while respecting several others direct and indirect experimental limits. Building on our previous analysis, here we are going to show 
that the same model allows a light pseudoscalar in the spectrum, which could have been missed both by older searches at LEP \cite{LEPb} and by the recent ones at the LHC \cite{CMS, CMS2, ATLAS}.\\
Concerning the possible existence of an extended Higgs sector, the observation of a Higgs boson decaying into two light scalar or pseudoscalar states would be one of its direct manifestations. 
This detection would also allows us to gather significant information about the cubic couplings of the Higgs and, overall, about its potential. However, so far neither the CMS nor the ATLAS collaborations have presented direct bounds 
on the decays of the Higgs $h_{125}$ into two scalars.  If such scalars are very light ($m_\Phi\lsim 100$ GeV), then they cannot be part of the spectrum of an ordinary CP-conserving minimal supersymmetric extension of the SM (MSSM). In fact, in that case they are predicted to be accompanied by a heavy pseudoscalar or by a charged Higgs boson. The only possibilities which are left open require CP-violating scenarios where one can have a light scalar  with a mostly CP-odd component \cite{CPVMSSM}. Such scenarios, however, are in tension with the recent observations of the decay mode $h\to \tau \tau$ \cite{CPVMSSMb}.\\
 The natural possibilities for such hidden Higgs bosons are those scenarios characterized by an extended Higgs sector. In the next-to-minimal supersymmetric standard model (NMSSM) with a $Z_3$ symmetry, such a light
 pseudoscalar is part of the spectrum in the form of a pseudo Nambu-Goldstone mode \cite{NMSSMps}. This situation gets even more interesting with the addition of triplets of appropriate hypercharge assignments \cite{TNSSMo,tnssm}, as in the TNMSSM. In the case of  a $Y=0$ Higgs triplet- and singlet-extended scenarios, the triplet does not couple to the $Z$ boson and the singlet to any gauge boson, and both of them do not couple to fermions.\\
At  LEP the Higgs boson was searched in the mass range less than $114.5$ GeV via the production of $e^+e^-\to Zh$ and  $e^+e^-\to h_ia_j$
 (in scenarios with two Higgs doublets), involving scalar $(h_i)$ and pseudoscalar $(a_j)$ with  fermionic final states. The $Y=0$ TNMSSM thus becomes a natural candidate for the
 such hidden Higgs possibility and therefore can evade the LEP bounds \cite{LEPb}.
 However, the situation gets slightly more complicated for Higgs triplets of non-zero hypercharge because they do couple to the $Z$ boson.

In this chapter we will focus our attention on decays of the Higgs boson into light scalars and pseudoscalars ($h_{125} \to h_ih_j/a_ia_j$). Such light scalar or pseudoscalars, when characterized by a mostly triplet or singlet component, do not 
 couple directly to fermions but decay to fermion pairs ($b$ or $\tau$) via their mixing with Higgs bosons of doublet type under $SU(2)$. Thus their final states 
 are often filled up with $b$-quarks, and leptons $\tau$ and $\mu$'s. The corresponding leptons and jets are expected to be rather soft, depending on the masses of the hidden scalars. If the doublet-triplet/singlet mixings in the Higgs sector are very small, they can give rise to the typical leptonic signature of charged displaced vertices. The goal of our analysis is to provide a direct characterization of the final states in the decay of a Higgs-like particle which can be helpful in the search for such hidden scalars at the LHC.

\section{Higgs decays into two gluons}\label{ggh}

\begin{figure}[t]
\begin{center}
\mbox{\hskip -20 pt\subfigure[]{\includegraphics[width=0.55\linewidth]{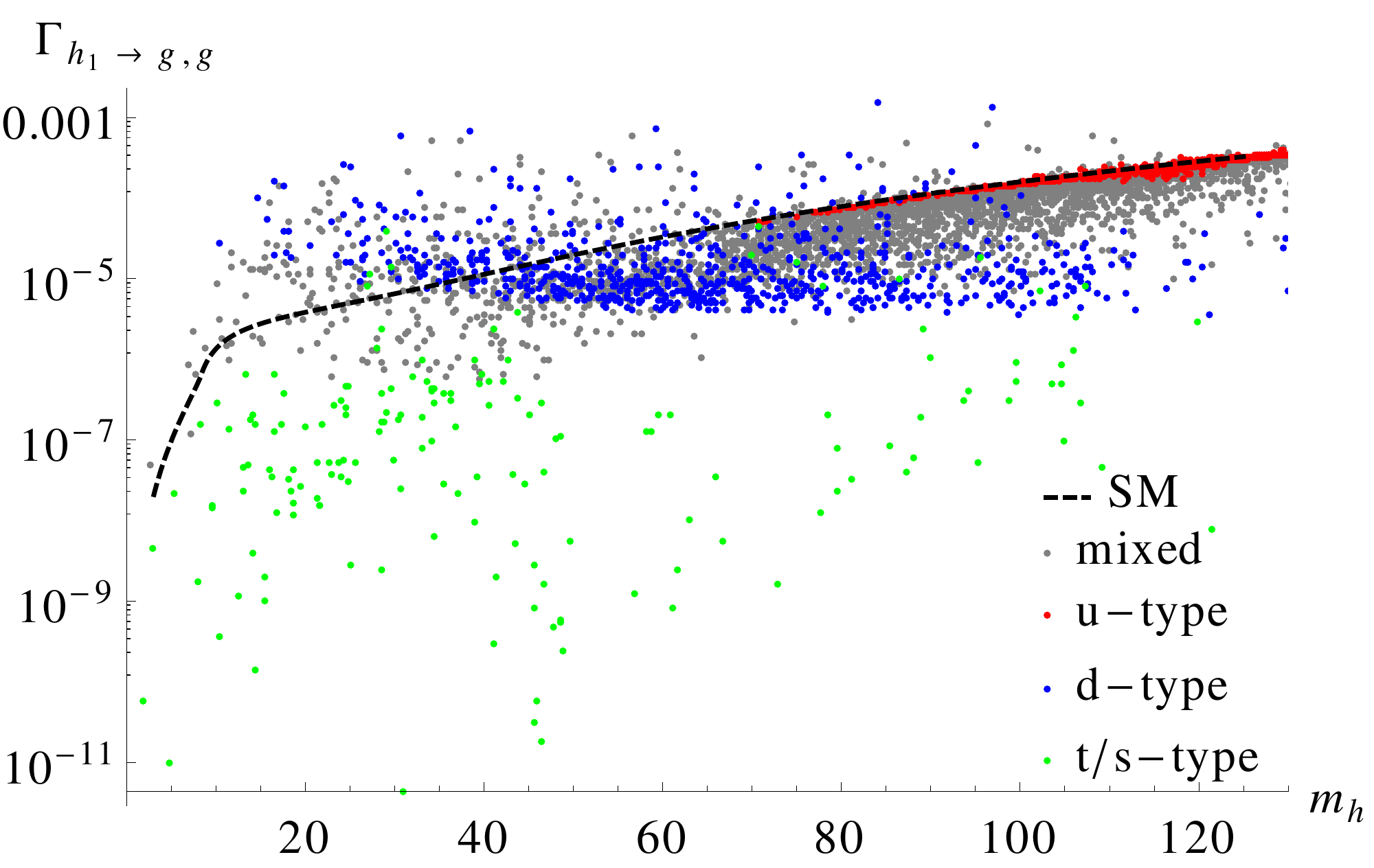}}
\subfigure[]{\includegraphics[width=0.55\linewidth]{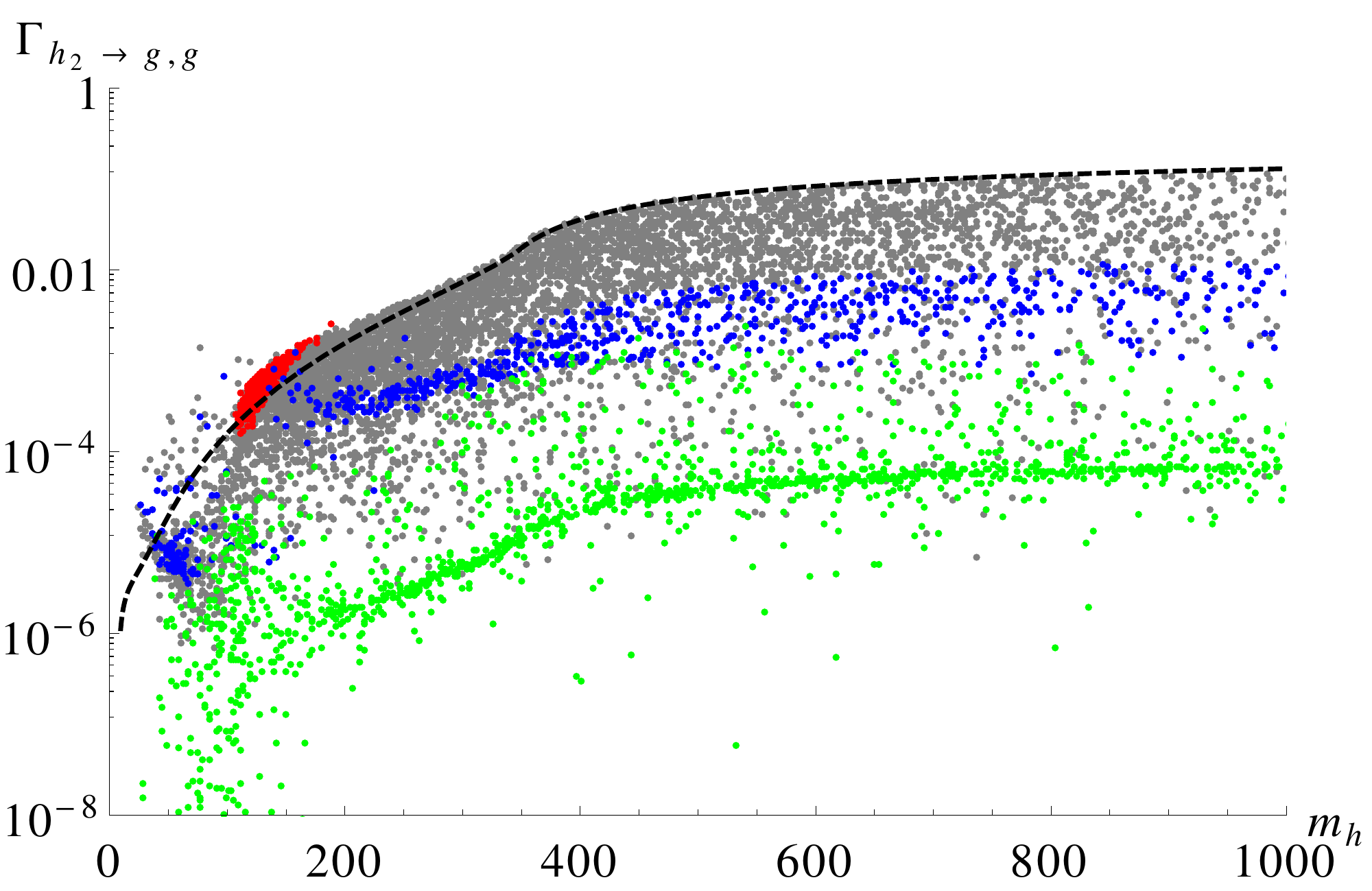}}}
\mbox{\hskip -20 pt\subfigure[]{\includegraphics[width=0.55\linewidth]{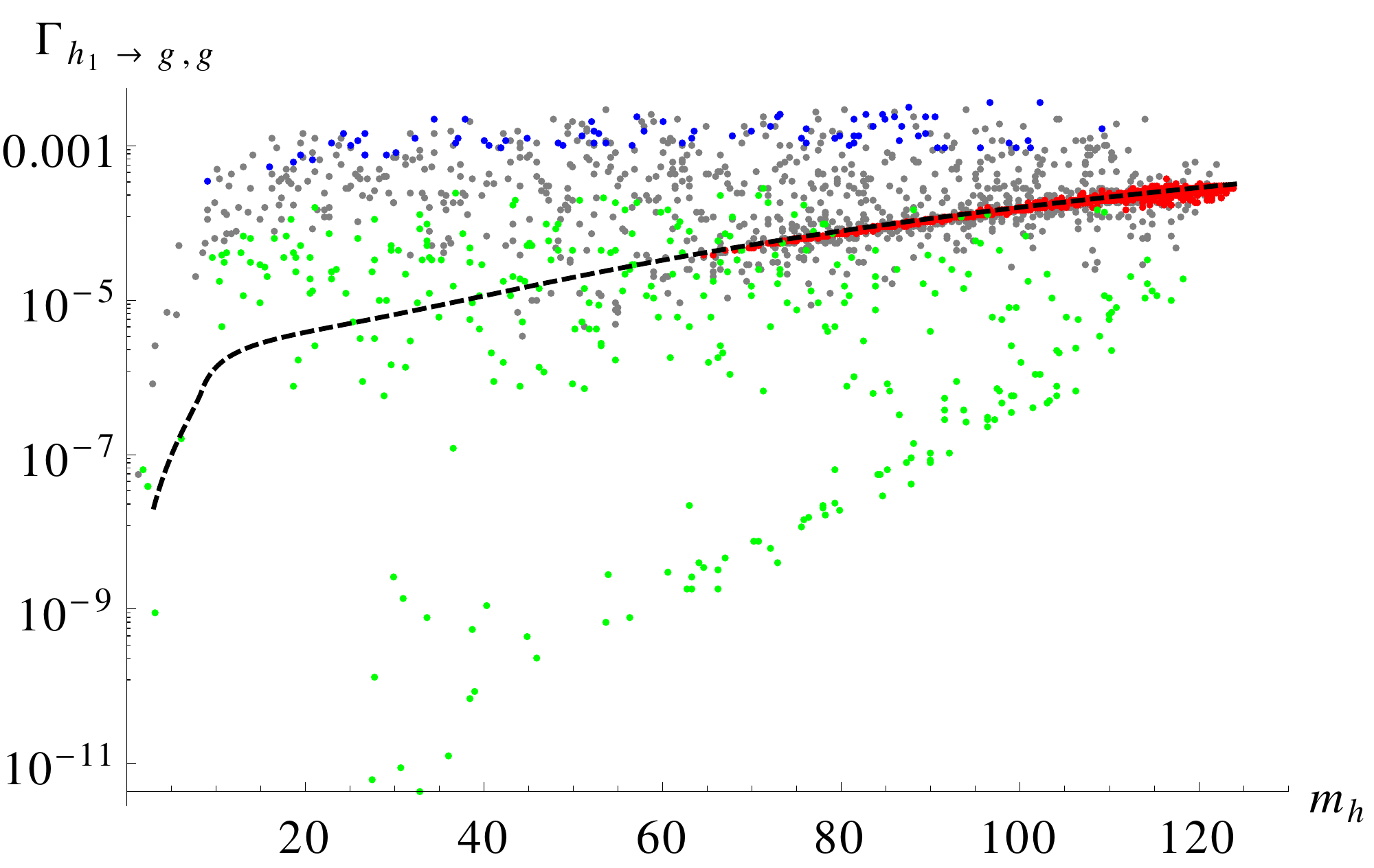}}
\subfigure[]{\includegraphics[width=0.55\linewidth]{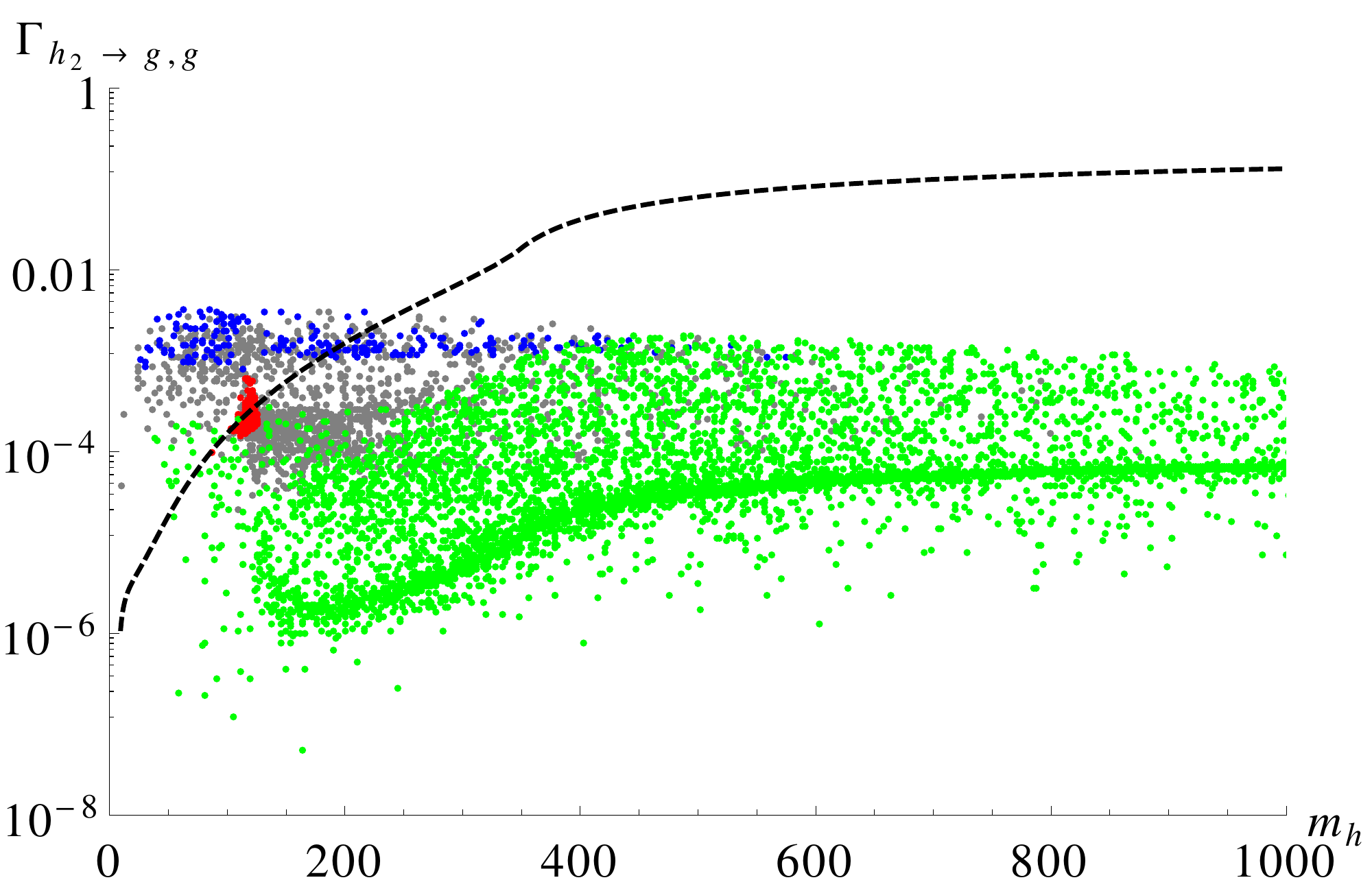}}}
\caption{We show a comparison between the SM and the TNMSSM predictions for the decay width of $h_1\rightarrow g,g$ (a), $h_2\rightarrow g,g$ (b) for $1<\tan\beta<15$ and $h_1\rightarrow g,g$ (c), $h_2\rightarrow g,g$ (d) for $20<\tan\beta<40$. We use the color code to distinguish among the up-type (>90\%) (red), down-type (blue), triplet/singlet-type (green) and mixed type Higgses (gray).}\label{hgg}
\end{center}
\end{figure}

 In the SM the most efficient production process of the Higgs boson is by gluon-gluon $(g)$ fusion (Figure (\ref{glutri})). The amplitude is mediated by a quark loop, which involves all the quarks of the SM, although the third generation, and in particular the top quark, gives the dominant contribution.
\begin{figure}[t]
\begin{center}
\includegraphics[width=0.3\linewidth]{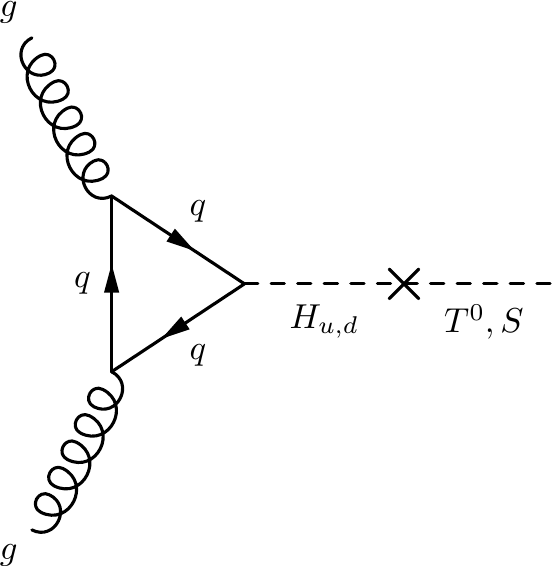}
\caption{A Feynman diagram depicting the coupling of gluons to the triplet/singlet, via their mixing with the doublets.}\label{glutri}
\end{center}
\end{figure}
In supersymmetric theories the situation is slightly different, because there are the up-type and down-type Higgs doublets 
$\hat H_u$ and $\hat H_d$  that couple to the up-type and down-type quarks/squarks respectively. Beside the sparticles contribution, the main difference between the SM and supersymmetric theories comes in the coupling of the Higgs bosons to fermions. These are given by
\bea
g_{h_i u\bar u} = -\frac{i}{\sqrt2}y_u \mathcal{R}^S_{i1},\\
g_{h_i d\bar d} = -\frac{i}{\sqrt2}y_d \mathcal{R}^S_{i2},\\
g_{h_i\ell\bar\ell} = -\frac{i}{\sqrt2}y_\ell \mathcal{R}^S_{i2},
\eea
where $R^S_{ij}$ is the rotation matrix of the CP-even sector. This means that the top/bottom contribution can be suppressed/enhanced, depending on the structure of $h_i$. The production cross section for $g,g\rightarrow h_i$ is related to the decay width of $h_i\rightarrow g,g$. At leading order, this decay width is given by
\begin{align}\label{gluongluon}
\Gamma(h_i\rightarrow g,g)&=\frac{G_F\,\alpha_s\,m_h^3}{36\sqrt2\,\pi^3} \left|\frac{3}{4}\sum_{q=t,\, b} \frac{g_{h_i q\bar q}}{(\sqrt2G_F)^{1/2}m_q}\, A_{1/2}(\tau^i_q)+\sum_{\tilde{q}=\tilde t, \, \tilde b}\frac{g_{h_i\tilde q\tilde q}}{m^2_{\tilde q}}A_0(\tau^i_{\tilde q})\right|^2,
\end{align}
where $A_0$ and $\,A_{1/2}$ are the spin-0 and spin-1/2 loop functions
\bea
&&A_0(x)=-\frac{1}{x^2}\left(x-f(x)\right),\\
&&A_{1/2}(x)=\frac{2}{x^2}\left(x+(x-1)f(x)\right),
\eea
with the analytic continuations 
\bea
f(x)=\left\{
\begin{array}{lr}
\arcsin^2(\sqrt{x})& x\leq1\\
-\frac{1}{4}\left(\ln\frac{1+\sqrt{1-1/x}}{1-\sqrt{1-1/x}}-i\pi\right)^2& x>1
\end{array}\right.
\eea
and $\tau^i_j=\frac{m_{h_i}^2}{4\,m_j^2}$. We show in Figure~\ref{hgg} the decay width of $h_{1,2}\rightarrow g,g$. In general, this decay width can be very different from the SM one in the case of supersymmetric theories with an extended Higgs sector, like the TNMSSM. In fact, in the latter case we have only the doublet Higgs that couples to the fermions, as shown in Eq. (\ref{spm}). This implies that if the Higgs is mostly triplet- or singlet-like, the fermion couplings are suppressed by $\mathcal{R}^S_{i1,2}$, in the limit of low $\tan\beta$. In Figure~\ref{hgg} the dashed line is the SM decay width and the color code is defined as follow:  we mark in red the up-type Higgs (>90\%), in blue the down-type, in green the triplet/singlet-type and in gray the mixed type. A look at Figure~\ref{hgg}(a) and (b) shows that for low $\tan\beta$ the decay width of a triplet/singlet-type Higgs is heavily suppressed. This occurs because the triplet and singlet Higgses couple to fermions only through the mixing with their analogue $SU(2)$ doublets. It is also rather evident that the shape of the decay widths for Higgses of up-type and of mixed-type are similar to those of the SM Higgs, for a large range of the mass of the extra Higgses. In Figure~\ref{hgg}(a) it is shown that for a light Higgs which takes the role of $h_{125}$, the SM decay width can be provided by the down-type Higgs of the TNMSSM, even in the case of low $\tan\beta$. Figure~\ref{hgg}(c) and (d) instead show that for a high value of $\tan\beta$ the decay width is dominated by the down-type Higgs, hence by the bottom quark. However it is still possible to have a SM-like decay width mediated by the top quark. In Figure~\ref{hgg}(d) it is quite evident that the bottom quark contribution has the same shape as in the MSSM \cite{anatomy2}. In this case the TNMSSM decay width of the Higgs is very different from the SM one for $m_h\gsim200$ GeV.
 
\section{Higgs decays into pseudoscalars}\label{psdcy}

The most important consequence of the $Z_3$ symmetry of the potential is that the mass of the pseudoscalar is in the GeV range, $m_{a_1}\sim\mathcal O(10)$ GeV, if we choose $A_{S, T, TS, \kappa, U, D}\sim\mathcal O(1)$ GeV. In this situation the decay $h_{125}\rightarrow a_1,a_1$ can be kinematically allowed. We study the decay of $h_{125}\rightarrow a_1,a_1$ via the decay width, given by

\bea\label{haaWidth}
\Gamma_{h_i\rightarrow a_j,a_j}=\frac{G_F}{16\sqrt2\pi}\frac{M_Z^4}{M_{h_i}}\left(1-\frac{4\,M_{a_j}^2}{M_{h_i}^2}\right)\left|\frac{g_{h_ia_ja_j}}{i M_Z^2/v}\right|^2,
\eea
where the $g_{h_ia_ja_j}$ coupling is given in the appendix. In Figure~\ref{Whaa}(a) and (b) we plot this decay width as a function of $\lambda_S$ and $\lambda_T$ respectively. Figure~\ref{Whaa}(a) shows that for $\left|\lambda_S\right|\gsim0.3$ we have scenarios in which the Higgs of doublet-type decays into pseudoscalars of singlet-type, but Figure~\ref{Whaa}(b) shows no particular structure in the dependence of $\Gamma_{h_1\rightarrow a_1,a_1}$ on $\lambda_T$.

Being interested in the fermionic final states of the decay of the SM-like Higgs into the light pseudoscalar $a_1$, $h_{125}\rightarrow a_1,a_1$, we gather the relevant coupling of the same pseudoscalars to fermions, which are given by  
\bea
g_{a_i u\bar u} = -\frac{\gamma_5}{\sqrt2}y_u \mathcal{R}^P_{i1},\\
g_{a_i d\bar d} = -\frac{\gamma_5}{\sqrt2}y_d \mathcal{R}^P_{i2},\\
g_{a_i\ell\bar\ell} = -\frac{\gamma_5}{\sqrt2}y_\ell \mathcal{R}^P_{i2}.
\eea 
Because the triplet, as well as the singlet, do not couple to the fermions, each $a_i$ will decay into fermions only trough a mixing with the doublet Higgses. This means that if $a_1$ is mostly of triplet or singlet component, its fermionic decay will be suppressed by the rotation elements $\mathcal{R}^P_{i1,2}$. An interesting consequence of this property is that this highly suppressed decay can generate a displaced vertex for the fermionic final states. 
\begin{figure}[t]
\begin{center}
\mbox{\hskip -20 pt\subfigure[]{\includegraphics[width=0.55\linewidth]{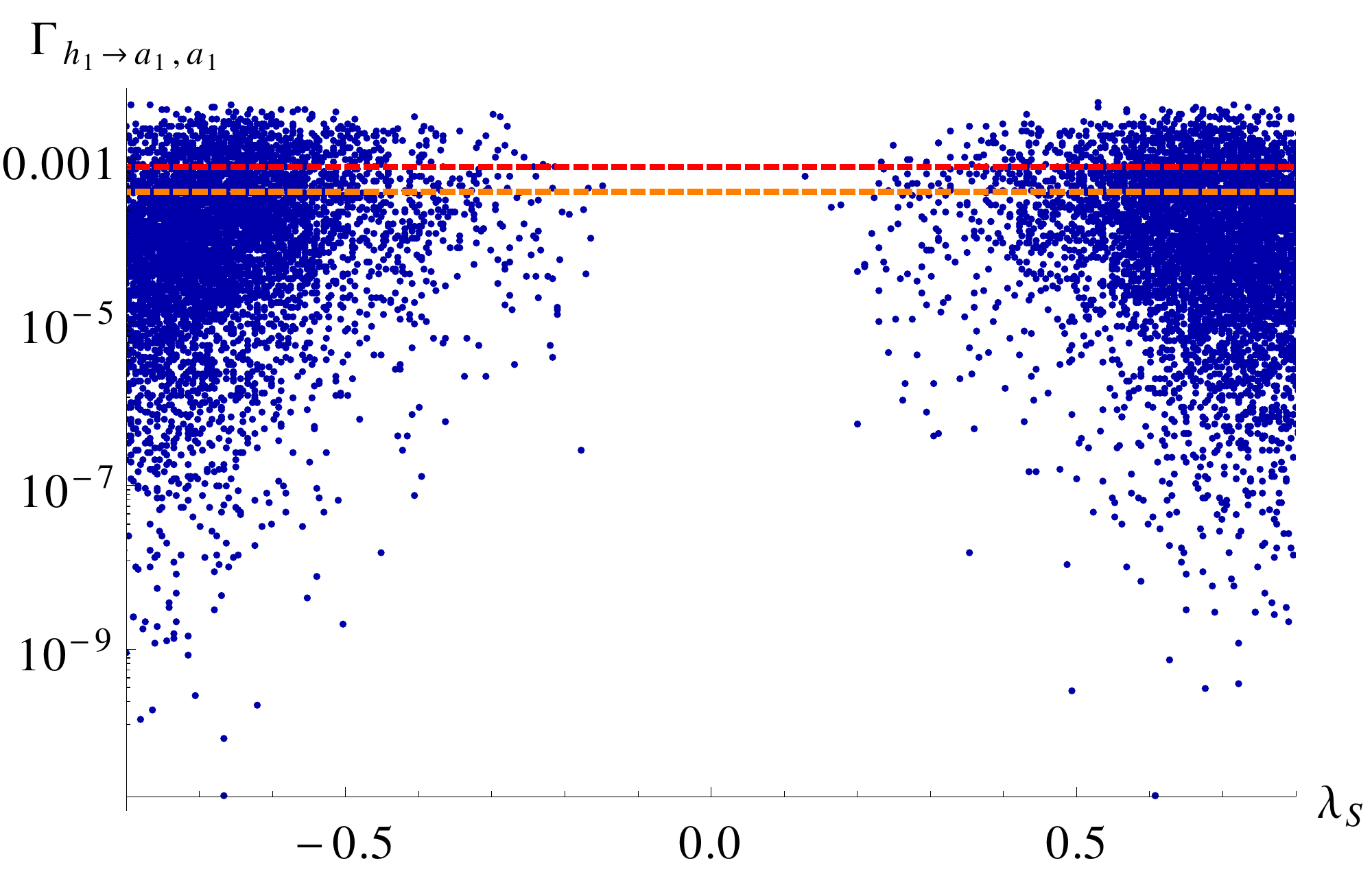}}
\subfigure[]{\includegraphics[width=0.55\linewidth]{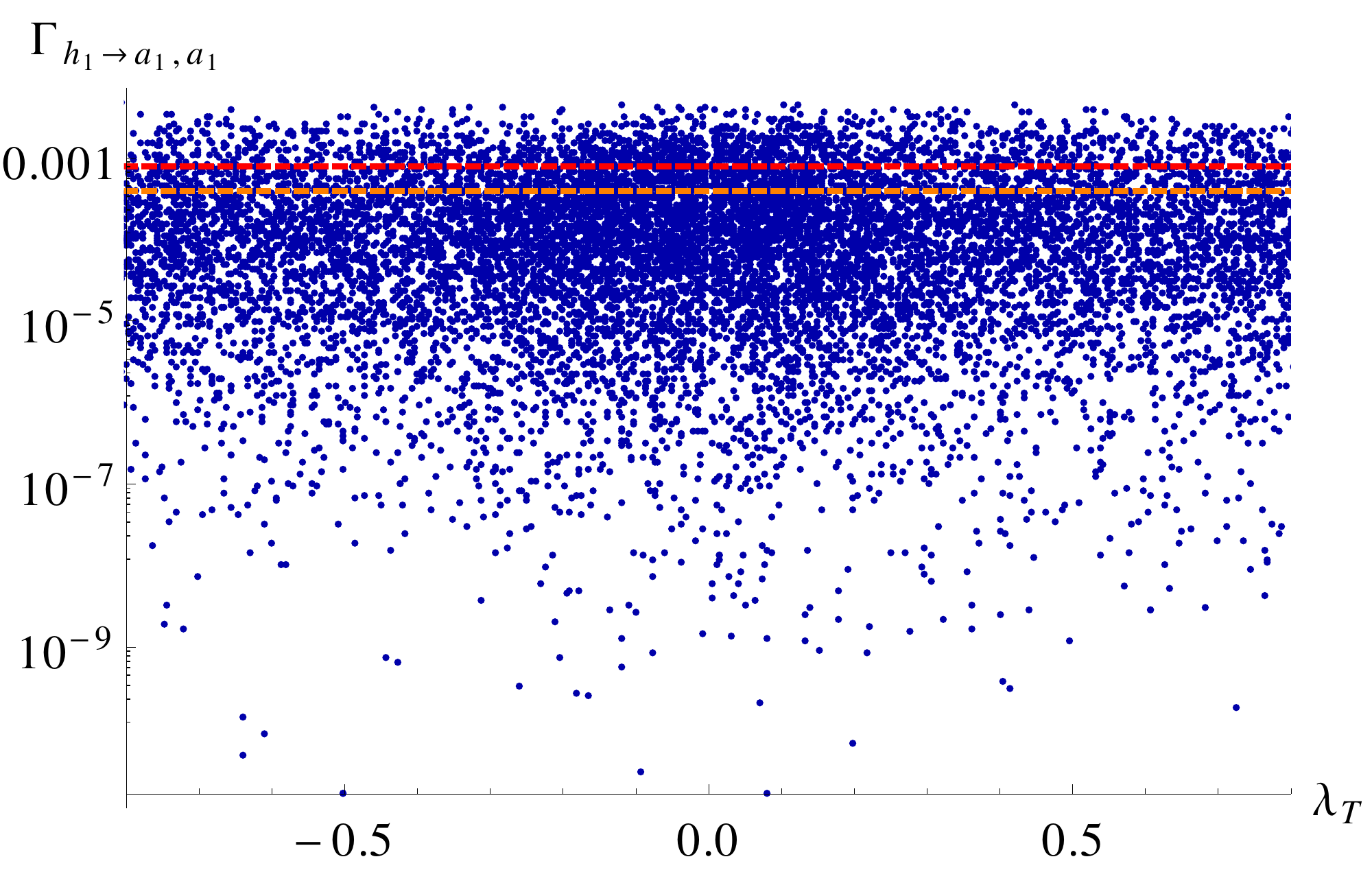}}}
\caption{We plot the decay width of the $h_{125}$ to two pseudoscalars (a) with respect to $\lambda_S$ and (b) with respect to $\lambda_{T}$. The red and orange coloured bands show the  region where $\mathcal{B}(h_{125} \to a_1 a_1)=20\%\, , 10\%$ respectively. }\label{Whaa}
\end{center}
\end{figure}

\section{Phenomenology and benchmark points}\label{secbps}

In Table~\ref{bps} we show the mass spectrum along with the other parameters which are necessary for the identification of three benchmark points. Together with the recent Higgs data we have also considered the recent bounds on the stop and sbottom masses \cite{thridgensusy} and the mass bounds on the lightest chargino from LEP \cite{chargino}. We have also taken into account the recent bounds on the charged Higgs boson mass from both CMS \cite{ChCMS} and ATLAS \cite{ChATLAS}. These have been derived in their searches for light in mass, charged Higgs bosons from the decay of a top quark, and in decays to $\tau \bar{\nu}$. The benchmark points 1 and 2 (BP1 and BP2) are characterized by one hidden Higgs boson, corresponding to a pseudoscalar particle of singlet-type with a mass of $\sim 20$ and $57$ GeV respectively. However BP3 has two hidden Higgs bosons, one of them a pseudoscalar of 
singlet-type around $\sim 37$ GeV and a second (scalar) one of triplet-type, around $\sim 118$ GeV in mass.
In the cases of BP1 \& BP2, $h_1$ is the discovered Higgs boson $h_{125}$, whereas for BP3 it is $h_2$.

%%%%%%%%%%%%%%%%%%%%%%%h_125 decays to a_1 a_1 %%%%%%%%%%%%%%
\begin{figure}[hbt]
\begin{center}
\mbox{\subfigure[]{\includegraphics[width=0.3\linewidth]{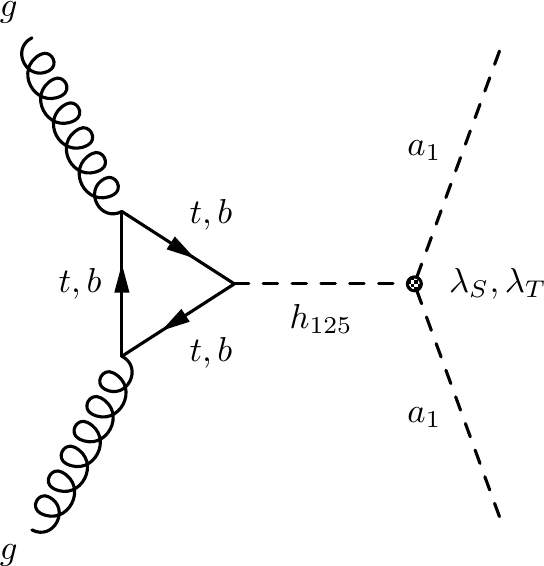}}\hskip 30 pt
\subfigure[]{\includegraphics[width=0.3\linewidth]{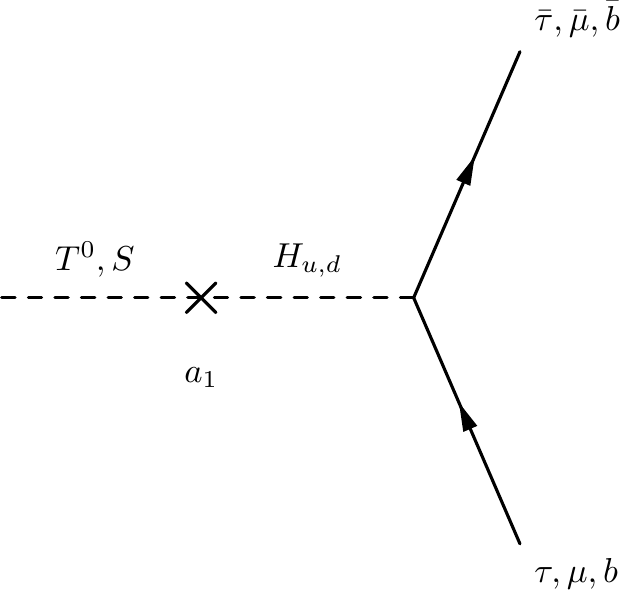}}}
\mbox{\subfigure[]{\includegraphics[width=0.35\linewidth]{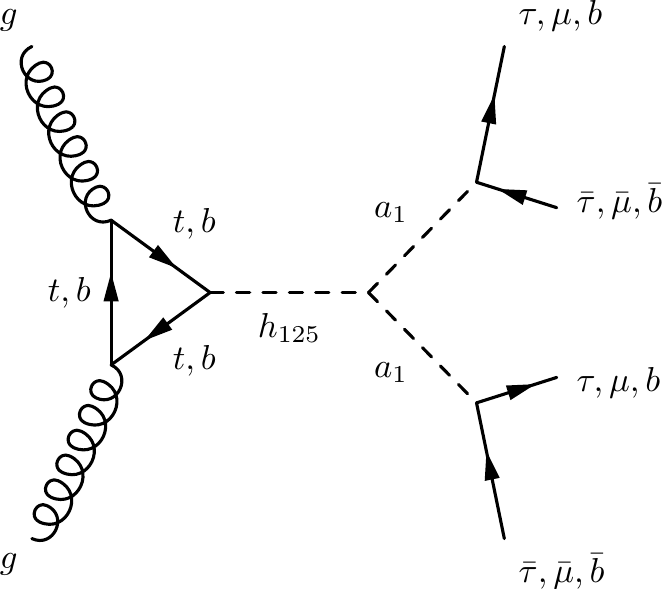}}}
\caption{Pseudoscalar (triplet/singlet) pair production from Higgs boson produced via 
gluon-gluon fusion and their decays, via their mixing with the doublets.}\label{glutri}
\end{center}
\end{figure}
 %%%%%%%%%%%%%%%%% Benchmark points%%%%%%%%%%%%%%%%%%%%
\begin{table}
\begin{center}
\renewcommand{\arraystretch}{1.2}
\begin{tabular}{||c||c|c|c||}
\hline\hline
Benchmark&BP1&BP2&BP3 \\
Points & &&\\ \hline\hline
$m_{h_1}$ & {\color{red}$\sim 125$} & {\color{red}$\sim 125$} & {\color{blue}$117.73 $} \\
\hline
$m_{h_2}$ & 183.58  &$162.59 $ & {\color{red}$\sim 125$} \\
\hline
$m_{h_3}$& 614.14 &$982.59$  & $791.37$ \\
\hline
$m_{h_4}$ & 965.75 &$1560.7$ & $1051.6$  \\
\hline
\hline
$m_{a_1}$ & \color{blue}20.50& \color{blue}$57.02$&  \color{blue}$36.79$ \\
\hline
$m_{a_2}$ &435.83  &$644.50$ & $620.81$  \\
\hline
$m_{a_3}$& 659.20 &$1018.1$  & $831.51$  \\
\hline
\hline
$m_{h^\pm_1}$ & \color{blue}182.84 &\color{blue}$162.25$ &\color{blue}$117.47$ \\
\hline
$m_{h^\pm_2}$ & 436.04 & $644.55$ & $620.86$  \\
\hline
$m_{h^\pm_3}$& 626.23 & $989.77$ & $805.58$   \\
\hline
\hline
\end{tabular}
\caption{Benchmark points for a collider study consistent with the $\sim 125$ GeV Higgs mass, where the $h_{i=1,2,3,4}$, $a_{i=1,2,3}$ are at one-loop and $h^{\pm}_{i=1,2,3}$ masses are calculated at tree level. We color in red the states which are mostly doublets ($>90\%$) and in blue those which are mostly triplet/singlet ($>90\%$). The points are consistent with the $2\sigma$ limits of $h_{125}\to WW^*, ZZ^*, \gamma\gamma$ \cite{CMS, ATLAS}.}\label{bps}
\end{center}
\end{table}
%%%%%%%%%%%%  
We now turn our attention to the decay of the discovered Higgs boson $h_{125}$ into a light pseudoscalar pair $a_1a_1$. Table~\ref{hdcy2} shows the branching ratios for the decay of $h_{125}$, in the case of the three benchmark points that we have selected. The table shows that for BP1 such branching ratio $(\mathcal{B})$ is the lowest $\mathcal{B}(h_{125}\to a_1a_1)\sim 10\%$, while for BP3 it is the highest $\mathcal{B}(h_{125}\to a_1a_1)\sim 18\%$. The discovered decay modes are consistent with the $2\sigma$ 
limits of $h_{125}\to WW^*, ZZ^*, \gamma\gamma$ \cite{CMS, ATLAS}. Such light pseudoscalars - though mostly singlet or
triplet - decay to the fermionic pairs which are kinematically allowed, via the mixing with the $H_u$ and $H_d$ doublets. This is because both singlet and triplet Higgses do not couples to fermions (see Eq.~\ref{spt}). 

For the benchmark point BP3 there is another hidden scalar which is CP-even, with a mass around $\sim 118$ GeV. $h_{125}$ cannot decay into this state $h_1$, as it is kinematically forbidden. If this $h_1$ is produced by other means it can have two-body decays to fermion pairs, as in the case of the $a_1$, via the mixing with the doublets. It will also have three-body decays ($WW^*$, $ZZ^*$) via its $SU(2)$ triplet charge and the mixing with the doublets.

%%%%%%%%%%%%%%%%% h_125 decay branching fraction (with tree level mass)%%%%%%%%%%%%%%%%%%%%
\begin{table}
\begin{center}
\renewcommand{\arraystretch}{1.4}
\begin{tabular}{||c||c|c|c|c|c|c|c||}
\hline\hline
Benchmark&\multicolumn{7}{|c||}{Branching ratios}\\
\hline
Points & $a_1 a_1$& $h_1 h_1$ & $a_1$Z &\; $W^+ W^-$ \;  & \;$b\bar{b}$ \;&\;$\tau \bar{\tau}$&$\mu\bar\mu$ \\
\hline\hline
BP1 &0.106  & - &$4.02\times10^{-7}$  & 0.138 &0.695  & 0.042&$1.50\times10^{-4}$\\
\hline
BP2 & 0.162 & - & $1.43\times10^{-8}$ & 0.136 & $0.645$ &$0.039$&$1.39\times10^{-4}$ \\
\hline
BP3 & $0.178$ & - & $1.93\times10^{-6}$ & 0.137 & $0.628$ & $0.038$&$1.35\times10^{-4}$ \\
\hline\hline
\end{tabular}

\caption{Decay branching ratios of $h_{125}$ for the three benchmark points, where the $h_{125}$ mass is calculated at tree level.
The kinematically forbidden decays are marked with dashes. The points are consistent with the $2\sigma$ limits of $h_{125}\to WW^*, ZZ^*, \gamma\gamma$ \cite{CMS, ATLAS}.}\label{hdcy2}
\end{center}
\end{table}
%%%%%%%%%%%%

%%%%%%%%%%%%%%%%% a_1 decay branching fraction (with tree level mass)%%%%%%%%%%%%%%%%%%%%
\begin{table}
\begin{center}
\renewcommand{\arraystretch}{1.4}
\begin{tabular}{||c||c|c|c||}
\hline\hline
Benchmark&\multicolumn{3}{|c||}{Branching ratios(\%)}\\
\hline
Points& \;$b\bar{b}$ \;&\;$\tau \bar{\tau}$&$\mu\bar\mu$  \\
\hline\hline
BP1 &   $0.939$ & $0.061$&$2.20\times10^{-4}$ \\
\hline
BP2 & $0.943$ & $0.057$&$2.04\times10^{-4}$\\
\hline
BP3 & $0.942$ & $0.058$& $2.07\times10^{-4}$\\
\hline\hline
\end{tabular}

\caption{Decay branching ratios of $a_1$ for the three benchmark points $BP_i$. The kinematically forbidden decays are marked with dashes.}\label{a1dcy2}
\end{center}
\end{table}
%%%%%%%%%%%%

For these benchmark points we have computed the production cross-sections of a $h_{125}$ Higgs boson assuming that  it is mediated by the gluon-gluon fusion channel
at the LHC. Table~\ref{Hcrosssec} presents the cross-sections which include the associated K-factors from the Higgs-Cross-Section
Working Group \cite{HCWG}. In the next section we are going to simulate the production of such light pseudoscalars
produced from the decay of such $h_{125}$. The choice of this particular production process is motivated by its large cross-section and by the rather clean final states ensued, that favour the extraction of the pseudoscalar $a_1$ pair.

%%%%%%%%%%%%%%%%% h_125 Gev Higgs cross-section at 13 TeV and 14 TeV %%%%%%%%%%%%%%%%

%%%%
\begin{table}
\begin{center}
\renewcommand{\arraystretch}{1.4}
\begin{tabular}{||c||c|c|c||}
\hline\hline
ECM&\multicolumn{3}{|c||}{$\sigma(gg\to h_{125}$) in pb}\\
in TeV&\multicolumn{3}{|c||}{for benchmark points}\\
\hline
&BP1 & BP2 &BP3 \\
\hline
13&41.00&41.00&41.00\\
\hline
14&46.18&46.18&46.18\\
\hline
\hline

\end{tabular}
\caption{Cross-section of $gg\to h_{125}$ at the LHC for center of mass energy of 13 and 14 TeV for the three benchmark points.}\label{Hcrosssec}
\end{center}
\end{table}

%%%%%%%%%%%%%%%%%%%%%%%%%%%%%%%%%%

\section{Signature and collider simulation}\label{sigsim}

The discovered Higgs boson $h_{125}$ can decay into two light pseudoscalars, which further decay into $\tau$ or $b$ pairs.  The $b$'s and $\tau$'s channel are therefore the relevant ones to look into, in the search for such hidden decay.  For this purpose we have implemented the model in SARAH \cite{sarah} and we have generated the model files for CalcHEP \cite{calchep}. These have been used to generate the decay file SLHA, containing the decay branching ratios and the corresponding mass spectra. The generated events have then been simulated with {\tt PYTHIA} \cite{pythia} via the the SLHA interface \cite{slha}. The simulation at hadronic level has been performed using the {\tt Fastjet-3.0.3} \cite{fastjet} with the {\tt CAMBRIDGE AACHEN} algorithm. We have selected a jet size $R=0.5$ for the jet formation, with the following criteria:
\begin{itemize}
  \item the calorimeter coverage is $\rm |\eta| < 4.5$

  \item the minimum transverse momentum of the jet $ p_{T,min}^{jet} = 10$ GeV and jets are ordered in $p_{T}$
  \item leptons ($\rm \ell=e,~\mu$) are selected with
        $p_T \ge 10$ GeV and $\rm |\eta| \le 2.5$
  \item no jet should be accompanied by a hard lepton in the event
   \item $\Delta R_{lj}\geq 0.4$ and $\Delta R_{ll}\geq 0.2$
  \item Since an efficient identification of the leptons is crucial for our study, we additionally require  
a hadronic activity within a cone of $\Delta R = 0.3$ between two isolated leptons to be $\leq 0.15\, p^{\ell}_T$ GeV, with 
$p^{\ell}_T$ the transverse momentum of the lepton, in the specified cone.

\end{itemize}

We keep the cuts in $p_T$ of the leptons and  the jets relatively low ($p_T \ge 10$ GeV), as they will be generated from the lighter pseudoscalar decays. $h_{125}$, once produced via  gluon-gluon fusion, will decay into two very light pseudoscalars ($m_{a_1}\sim 20$ GeV for BP1). 
The light pseudoscalars then will decay further into $b$ or $\tau$ pairs (see Table~\ref{a1dcy2}). The parton level signatures would be $4b$, $4\tau$ and $2b+2\tau$. 
In reality, this description is expected to change due to hadronization and to the contributions from the initial- and final-state 
radiation emission in the presence of $b$ quarks and of $\tau$ leptons. The number of jets can indeed
 increase or decrease due to these effects. The efficiency of the jet of the b-quark ($b_{\rm{jet}}$) is determined through the determination of the secondary vertex  and it is therefore momentum dependent. For this purpose
 we have taken - for the $b_{\rm{jet}}$'s from $t\bar{t}$ - the single-jet tagging efficiency equal to $0.5$, while for the remaining components of the final state we have followed closely the treatment of \cite{btag}. 
   Here, in the case of the $\tau_{\rm{jet}}$ we have considered the hadronic decay of the $\tau$ to be characterized by at least one charged track with $\Delta R \leq 0.1$ of the
  candidate $\tau_{\rm{jet}}$ \cite{tautag}.
%%%%%%%%%%%%%%%%%%%%%%%%%%%
\begin{figure}[hbt]
\begin{center}

\includegraphics[width=0.33\linewidth, angle=-90]{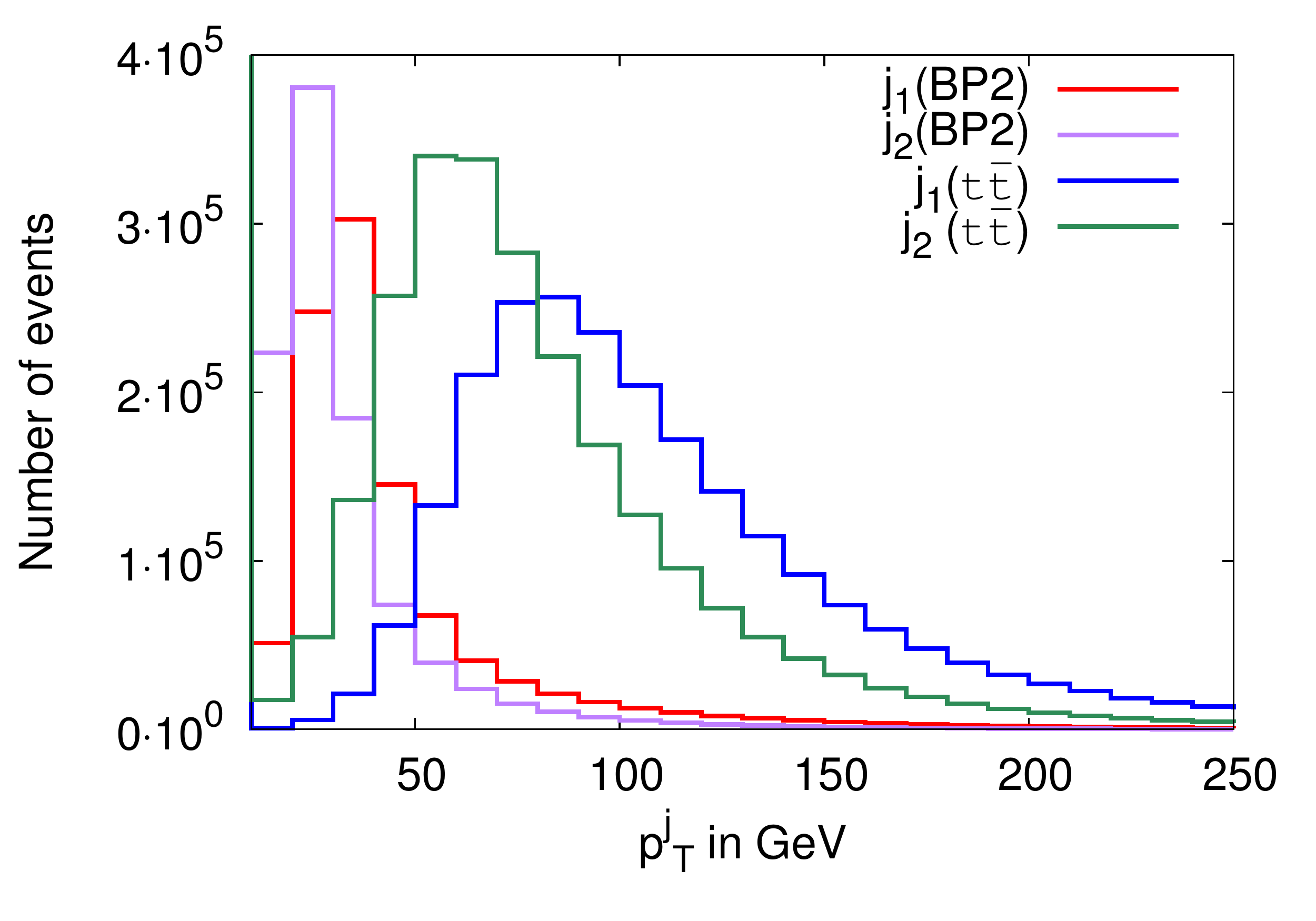}
\includegraphics[width=0.33\linewidth, angle=-90]{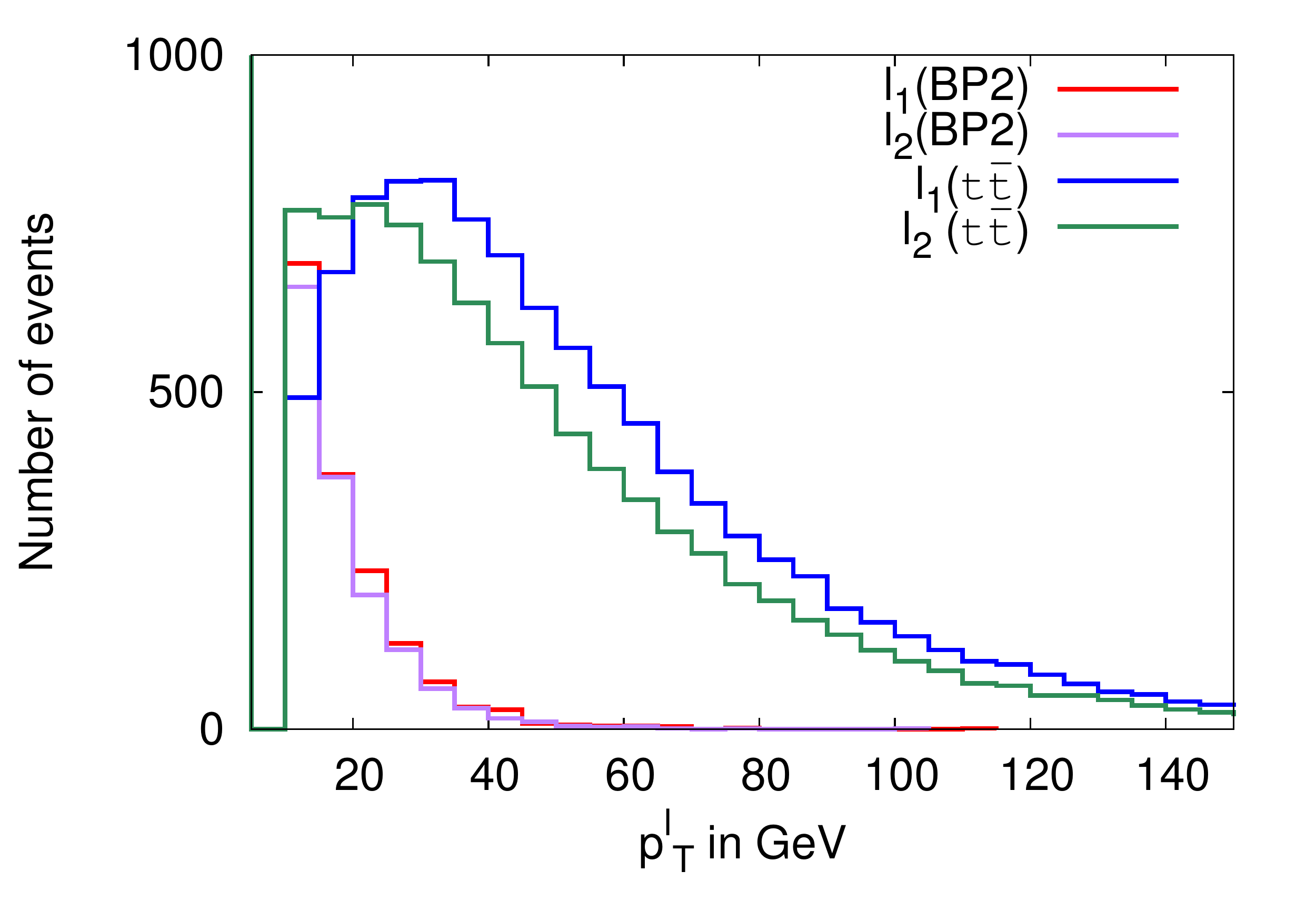}
\caption{ $p^{b_j}_T$ distribution (left) and  $p^{\ell}_T$ distribution (right) for $t\bar{t}$ and for the signal in BP2.}\label{ptjl}

\end{center}
\end{figure}
%%%%%%%%%%%%%%%%%%%%%%%%%%%%%%%%%%%%%%%%%%%%%%%%%%%%%%%%%%%%%%

Figure~\ref{ptjl} (left) shows the $b_{\rm{jet}}$  $p_T$ coming from the pseudoscalar decays in the case of BP2 with the dominant background $t\bar{t}$. 
Clearly one may observe the that $b_{\rm{jet}}$'s coming from the signal (BP2) are rather soft, mostly with $p_T \lesssim 50$ GeV. Figure~\ref{ptjl} (right) shows
the transverse momentum $p_T$ of the lepton coming from the signal (BP2) and the dominant backgrounds $t\bar{t}$ and $ZZ$. This clearly shows that the signal leptons 
are very soft ($p_T \lesssim 40$ GeV) compared to the corresponding backgrounds.

%%%%%%%%%%%%%%%%%%%%%%%%%%%
\begin{figure}[hbt]
\begin{center}

\includegraphics[width=0.33\linewidth, angle=-90]{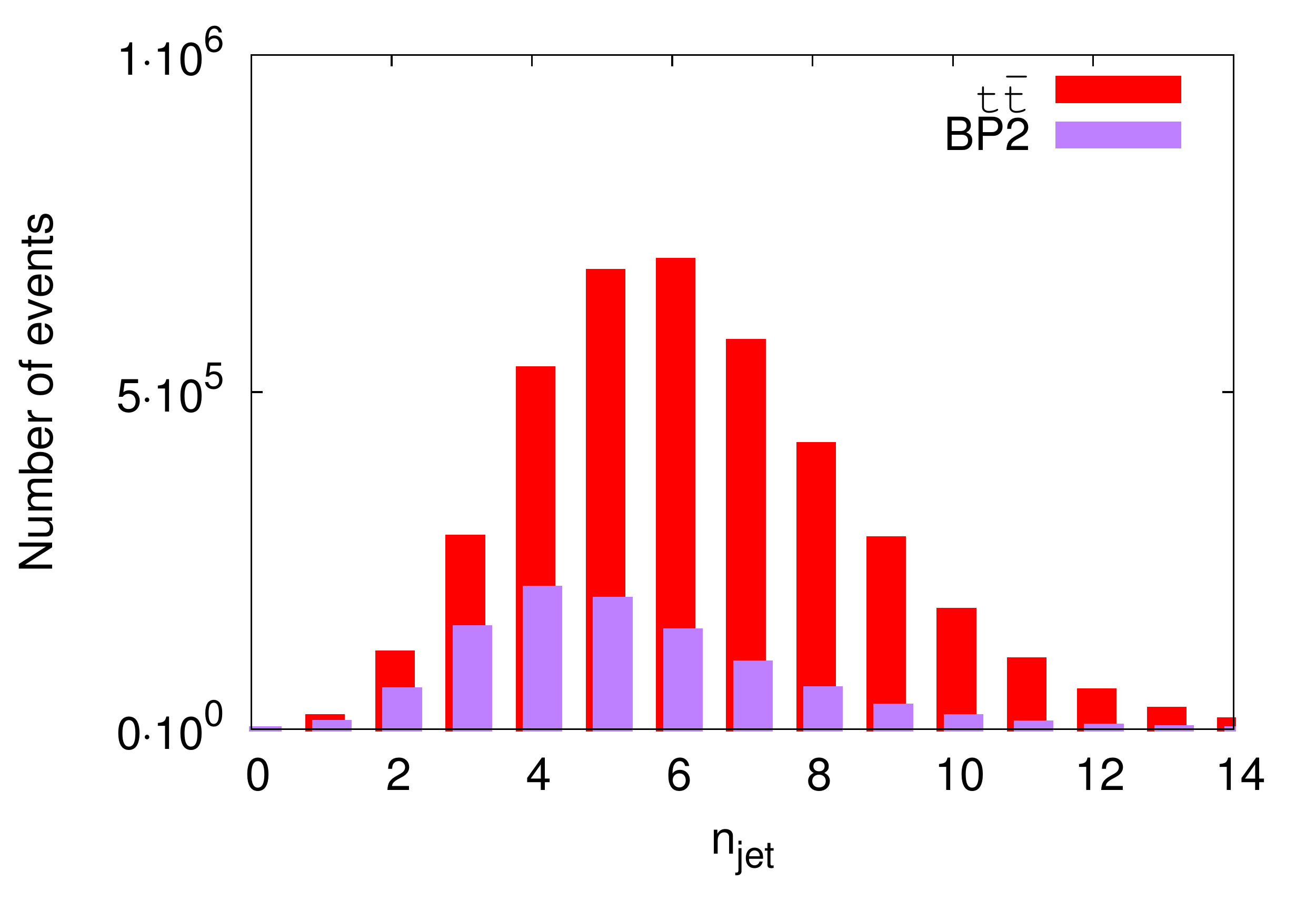}
\includegraphics[width=0.33\linewidth, angle=-90]{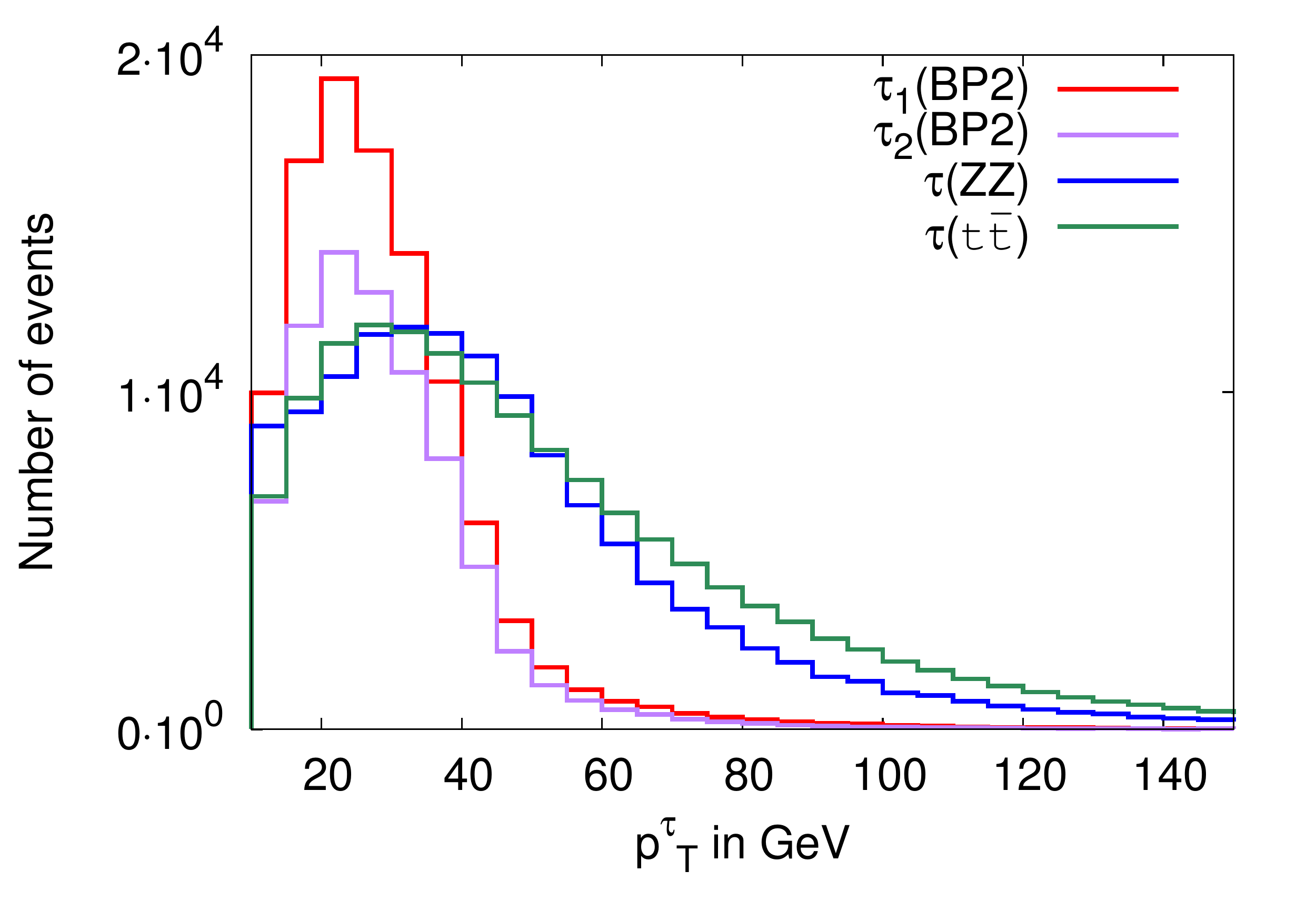}
\caption{(Left) jet-multiplicity ($n_{\textrm{jet}}$) distributions  and (Right) $p^{\tau}_T$ distributions for signal events coming from the  pseudoscalars $a_1$ decays for BP2 and the dominant SM backgrounds $t\bar{t}$, $ZZ$.}\label{njtaupt}

\end{center}
\end{figure}
%%%%%%%%%%%%%%%%%%%%%%%%%%%%%%%%%%%%%%%%%%%%%%%%%%%%%%%%%%%%%%
Next we have investigated the number of jets in the final states after hadronization. Figure~\ref{njtaupt} (left) 
shows the number of jets for the signal (BP2) and for the dominant background $t\bar{t}$. Due to the lower cuts in $p_T$, the number of final state jets has increased, in this case, both for the signal and for the background. The difference is still prominent between the two, where the signal peaks around 4 jets  and $t\bar{t}$ around 6. Thus a requirement of a relatively lower number of jets in the final state will remove the dominant $t\bar{t}$ contribution quite effectively.  

Figure~\ref{njtaupt} (right) shows the transverse momentum ($p^{\tau}_T$) distribution of the $\tau$ at parton level for the signal in BP2 and the dominant $\tau\tau$ backgrounds 
coming from $ZZ$ and $t\bar{t}$. Clearly, the condition of $p^{\tau}_T \lesssim 50$ GeV will reduce effectively the background contributions to the final state.  

\subsection{$2b+2\tau$} 
In the case of the TNMSSM, the discovered Higgs boson can also decay into a pair of  lighter mass eigenstates $a_1a_1$ and/or  $h_1h_1$. 
The possibility of producing such light states specially as singlet-like pseudoscalars has been discussed in \cite{TNSSMo}, and it is shown in Table~\ref{bps}. 
 Table~\ref{hdcy2} presents the branching ratios for the decay of $h_{125}$ for the three benchmark points that we have selected. Notice 
 that the ratios into the pseudoscalar pair $\mathcal{B}(h_{125}\to a_1 a_1)$ is about 10-20\%. The $a_1$ pair
 then decays into $b$ and $\tau$ pairs with rates shown in Table~\ref{a1dcy2}. We have selected a final state with $2b+2\tau$, where one of the $a_1$ decays into a $\tau$ pair  
and the other one decays into a $b$ pair. This also enhances the combinatorial factor and thus the number of events in the final state. The dominant SM backgrounds in this case comes from $t\bar{t}$, $ZZ$ and $b\bar{b}Z$.

 Figure~\ref{njtaupt} (right) shows that the requirement of a lower number of jets ($n_j$) $\leq 5$  will suppress the $t\bar{t}$ backgrounds. A similar effect is generated by requiring a lower $p_T$ on the $\tau_{\rm{jet}}$'s and $b_{\rm{jet}}$'s  ($p_T \lesssim 50$ GeV).  The corresponding $\tau$ decays give rise to very soft neutrinos, and therefore, by demanding a low missing $p_T$ $\leq 30$ GeV, we can reduce the backgrounds even further. The $b$ and $\tau$ tagging come with their own efficiencies \cite{btag} and \cite{tautag}, but this also helps in suppressing the other multi-jet backgrounds present from the SM.

In Table~\ref{2b2tau13} and Table~\ref{2b2tau14} we present the number of events for the three benchmark points coming both from the signal  and the SM backgrounds at the LHC, for a center of mass energy of 13 TeV and 14 TeV respectively.  The tables also show how their values change with each additional cut. We ask for a final state with $n_j\leq 5$, in which we demand the presence of at least two $b_{\rm{jet}}$'s and two $\tau_{\rm{jet}}$'s. In our notations, this request is indicated in the form: $n_j\leq 5\,[2b_{\rm{jet}}\,+ \,2\tau_{\rm{jet}}]$. We will be using the ampersand \& (a logical {\em and}) to combine additional constraints on the event, either in the form of particle/jet multiplicites or kinematical restrictions, and  define the signal as 
$$\rm{sig1}: n_j\leq 5\,[2b_{\rm{jet}}\, + \,2\tau_{\rm{jet}}]\,\&\,\ptmiss \leq 30 \, \rm{GeV.}$$
In the expression above, we have also required that the missing transverse momentum is smaller than 30 GeV 
$(\&\,\ptmiss \leq 30 \, \rm{GeV})$.
 In addition we apply some other cuts on the signal in order to reduce the backgrounds. For instance, in 
Table \ref{2b2tau13} we introduce a long sequence of such cuts (first column). In the case of BP1, for instance, the significance, after these selections, is $4.00\, \sigma$. The two additional conditions $p_1$ and $p_2$ are then applied as alternative clauses, and are enclosed into separate rows. \\
The first sequential cuts include  the $b_{\rm{jet}}$ pair invariant mass veto around $m_Z$,  the conditon that $|m_{bb}-m_Z|>10$ GeV and, around $m_{125}$, the condition$|m_{bb}-m_{h_{125}}|>10$ GeV.  $m_Z$ is the mass of the $Z$ gauge boson and $m_{h_{125}}$ is the Higgs mass (125 GeV).
Similarly, we also put veto on the invariant mass of the $\tau_{\rm{jet}}$ pair as: $|m_{\tau\tau}-m_Z|>10$ GeV and $|m_{\tau\tau}-m_{h_{125}}|>10$ GeV. Finally, since we are searching for hidden Higgs bosons, we demand that $m_{\tau\tau}<125$ GeV and $m_{bb}<125$ GeV respectively, where $m_{bb}$ and $m_{\tau\tau}$ are the invariant masses of the $b$ and $\tau$ pairs.

\begin{table}[htb]
\begin{center}
\hspace*{-1.0cm}
\renewcommand{\arraystretch}{1}
\begin{tabular}{|c||c|c|c||c|c|c|c|c||}
\hline\hline
Final states&\multicolumn{3}{|c||}{Benchmark}&\multicolumn{5}{|c||}{Backgrounds }
\\
\hline
&BP1 & BP2&BP3 & $t\bar{t}$& $ZZ$ & $Z h$  &$b\bar{b}h$& $b\bar{b}Z$\\
\hline
\hline
$n_j\leq 5\,[2b_{\rm{jet}}+ 2\tau_{\rm{jet}}$]&\multirow{2}{*}{220.10}&\multirow{2}{*}{591.46}&\multirow{2}{*}{310.19}&\multirow{2}{*}{1824.08}&\multirow{2}{*}{199.50}&\multirow{2}{*}{39.56}&\multirow{2}{*}{11.87}&\multirow{2}{*}{4903.05}\\
$\&\,\ptmiss \leq 30$ GeV&&&&&&&&\\
&&&&&&&&\\
$\&\, p_T^{bj_{1,2}}\leq 50$GeV&\multirow{2}{*}{211.30}&\multirow{2}{*}{568.14}&\multirow{2}{*}{289.02}&\multirow{2}{*}{410.83}&\multirow{2}{*}{73.04}&\multirow{2}{*}{7.87}&\multirow{2}{*}{3.96}&\multirow{2}{*}{2941.83}\\
$\& \, |m_{bb}-m_Z|>10$ GeV&&&&&&&&\\
&&&&&&&&\\
$ \&\, |m_{bb}-m_{h_{125}}|>10$ GeV&211.30&565.32&289.02&386.18&73.04&7.52&3.96&2614.96\\
&&&&&&&&\\
$ \&\, |m_{\tau\tau}-m_Z|>10$ GeV&211.30&560.37&289.02&312.23&62.13&6.29&3.46&2397.04\\
&&&&&&&&\\
$\&\, |m_{\tau\tau}-m_{h_{125}}|>10$ GeV&211.30&560.37&289.02&287.58&62.13&6.18&2.97&2397.04\\
&&&&&&&&\\
$\&\, m_{\tau\tau}<125$GeV&211.30&560.37&289.02&254.71&62.13&6.18&2.97&2397.04\\
&&&&&&&&\\
$\&\, m_{bb}<125$GeV&211.30&559.66&289.02&230.06&62.13&6.07&2.97&2288.09\\
&&&&&&&&\\
\hline
Significance&4.00&9.98&5.39&\multicolumn{5}{|c||}{}\\
\hline
\hline
\multirow{3}{*}{\&\, $p_1:|m_{bb}-m_{a_1}|\leq 10$GeV}&\multirow{3}{*}{198.82}&\multirow{3}{*}{281.95}&\multirow{3}{*}{216.04}&24.65&0.00&0.22&0.49&326.87\\
&&&&65.73&26.16&1.46&0.49&1307.48\\
&&&&65.73&8.72&1.34&1.00&435.83\\
\hline
Significance&8.47&6.87&8.01&\multicolumn{5}{|c||}{}\\
\hline
\hline
%&&&&&&&&&\\
\multirow{3}{*}{$\&\, p_2:|m_{\tau\tau}-m_{a_1}|\leq 10$GeV}&\multirow{3}{*}{205.29}&\multirow{3}{*}{229.66}&\multirow{3}{*}{203.63}&65.73&3.27&0.33&0.00&0.00\\
&&&&73.95&28.34&1.46&0.49&762.70\\
&&&&41.08&13.08&1.57&1.48&0.00\\
\hline
Significance&12.40&6.94&12.65&\multicolumn{5}{|c||}{}\\
\hline
\hline
%&&&&&&&&&\\
\end{tabular}
\caption{The number of events for a $n_j\leq 5\,[2b_{{\rm{jet}}}+ 2\tau_{\rm{jet}}]\,\&\ptmiss \leq 30$ GeV final state at 100 fb$^{-1}$ of luminosity at the LHC, for a center of mass energy of 13 TeV. We require that the original signal has a number of jets 
$\leq 5$, of which 2 are $b_{\rm{jet}}$'s and 2 are $\tau_{\rm jet}$'s, with a missing $p_T\, (\ptmiss) \leq$ 30 GeV. We have denoted with $p_T^{bj_{1,2}}$ the transverse momentum of the $b_{\rm{jet}}$'s, with the two $b$'s labelled as 1 and 2. The final states are selected by imposing a long list of sequential cuts on the event, indicated with an ampersand (\&). The two additional options $p_1$ and $p_2$ are, however, alternative, and are imposed as additional constraints (a logical {\em or}). For this reason they are enclosed into separate rows.}\label{2b2tau13}
\end{center}
\end{table}
%%%%%%%%%%%%%%%%%%%%%%%%%%%%%%%%%%%%%%%%%%%%%%%%%%%%%%%

From Table~\ref{2b2tau13} and Table~\ref{2b2tau14} we deduce that the most dominant SM backgrounds are those from $t\bar{t}$, $ZZ$, $Zh$, $b\bar{b}h$ and $b\bar{b}Z$ respectively.
Though the 125 GeV bound on the two invariant masses reduces substantially most of the backgrounds, still the $b\bar{b}Z$ rate remains relatively large.  At this stage the signal significances, for the two benchmark points BP2 and BP3, both cross the $5\,\sigma$ value  
at an integrated luminosity 100 fb$^{-1}$, $9.98\, \sigma$ and $5.39 \,\sigma$, for a center of mass energy of 13 TeV.  In the case of BP1 this value is at the level of $4 \,\sigma$.  This is expected, given that in the case of BP2 the branching ratio $\mathcal{B}(h_{125}\to a_1a_1)$ is about $16\%$ (see Table~\ref{hdcy2}) and the 
pseudoscalar is relatively heavy, with a mass around $57$ GeV.  The  $\tau_{\rm jet}$'s and $b_{\rm{jet}}$'s coming from the decays of the $a_1$ are relatively harder (characterized by a larger momentum) compared to the benchmark points BP1 and BP3, so less events are cut out by the threshold on the $p_T$ cuts. Thus for BP2 we can reach a $5\sigma$ level of signal significance at an integrated luminosity of 25 fb$^{-1}$, for a given center of mass energy of 13 TeV. In this case the signal significance stays very similar also at 14 TeV, with little improvement for each of the $BP_i$'s. The signal significances, in this case, are  $4.47 \,\sigma$, $10.18\, \sigma$ and $5.98 \,\sigma$ respectively for BP1, BP2 and BP3.

%%%%%%%%%%%%%%%%%%%%%%%% 2b +2 \tau at 14 TeV%%%%%%%%%%%%%%%%%%%%
\begin{table}
\begin{center}
\hspace*{-1.0cm}
\renewcommand{\arraystretch}{1}
\begin{tabular}{|c||c|c|c||c|c|c|c|c||}
\hline\hline
Final states&\multicolumn{3}{|c||}{Benchmark}&\multicolumn{5}{|c||}{Backgrounds }\\
\hline
&BP1 & BP2&BP3 & $t\bar{t}$& $ZZ$ & $Z h$  &$b\bar{b}h$& $b\bar{b}Z$\\
\hline
\hline
$n_j\leq 5\,[2b_{\rm{jet}}+ 2\tau_{\rm{jet}}$]&\multirow{2}{*}{253.10}&\multirow{2}{*}{641.50}&\multirow{2}{*}{361.69}&\multirow{2}{*}{1530.66}&\multirow{2}{*}{223.72}&\multirow{2}{*}{40.35}&\multirow{2}{*}{19.77}&\multirow{2}{*}{4657.83}\\
$\&\,\ptmiss \leq 30$ GeV&&&&&&&&\\
&&&&&&&&\\
$p_T^{bj_{1,2}}\leq 50$ GeV&\multirow{2}{*}{248.41}&\multirow{2}{*}{605.68}&\multirow{2}{*}{337.04}&\multirow{2}{*}{294.36}&\multirow{2}{*}{85.11}&\multirow{2}{*}{7.80}&\multirow{2}{*}{7.19}&\multirow{2}{*}{3432.09}\\
$\&\, |m_{bb}-m_Z|>10$ GeV&&&&&&&&\\
&&&&&&&&\\
$\&\, |m_{bb}-m_{h_{125}}|>10$ GeV&248.41&604.89&337.04&294.36&85.11&7.43&7.19&3432.09\\
&&&&&&&&\\
$\&\, |m_{\tau\tau}-m_Z|>10$ GeV&248.41&597.73&337.04&255.11&70.52&6.09&5.39&2819.21\\
&&&&&&&&\\
$\&\, |m_{\tau\tau}-m_{h_{125}}|>10$ GeV&248.41&597.73&337.04&255.11&70.52&5.97&2.40&2819.21\\
&&&&&&&&\\
$\&\, m_{\tau\tau}<125$ GeV&248.41&596.93&337.04&255.11&69.30&5.85&2.40&2819.21\\
&&&&&&&&\\
$\&\, m_{bb}<125$ GeV&248.41&596.93&337.04&196.24&69.30&5.85&2.40&2574.07\\
&&&&&&&&\\
\hline
Significance&4.47&10.18&5.98&\multicolumn{5}{|c||}{}\\
\hline
\hline
\multirow{3}{*}{$\&\,p_1:|m_{bb}-m_{a_1}|\leq 10$ GeV}&\multirow{3}{*}{236.43}&\multirow{3}{*}{326.32}&\multirow{3}{*}{279.49}&9.81&2.43&0.37&0.00&490.30\\
&&&&68.68&31.61&1.83&1.20&1348.32\\
&&&&29.43&15.81&1.46&0.00&490.30\\
\hline
Significance&8.70&7.74&9.79&\multicolumn{5}{|c||}{}\\
\hline
\hline
\multirow{3}{*}{\&\,$p_2:|m_{\tau\tau}-m_{a_1}|\leq 10$ GeV}&\multirow{3}{*}{241.64}&\multirow{3}{*}{248.32}&\multirow{3}{*}{279.49}&19.62&6.08&0.49&0.00&0.00\\
&&&&58.87&24.32&1.58&0.00&1103.17\\
&&&&49.06&14.59&1.10&1.80&122.57\\
\hline
Significance&14.78&6.56&12.93&\multicolumn{5}{|c||}{}\\
\hline
\hline
\end{tabular}
\caption{The number of events for a $n_j\leq 5\,[2b_{{\rm{jet}}}+ 2\tau_{\rm{jet}}]\, \& \ptmiss \leq 30$ GeV final state at 100 fb$^{-1}$ of luminosity at the LHC for center of mass energy of 14 TeV. }\label{2b2tau14}.
\end{center}
\end{table}
%%%%%%%%%%%%%%%%%%%%%%%%%%%%%%%%%%
\begin{figure}[bht]
\begin{center}
\includegraphics[width=0.33\linewidth, angle=-90]{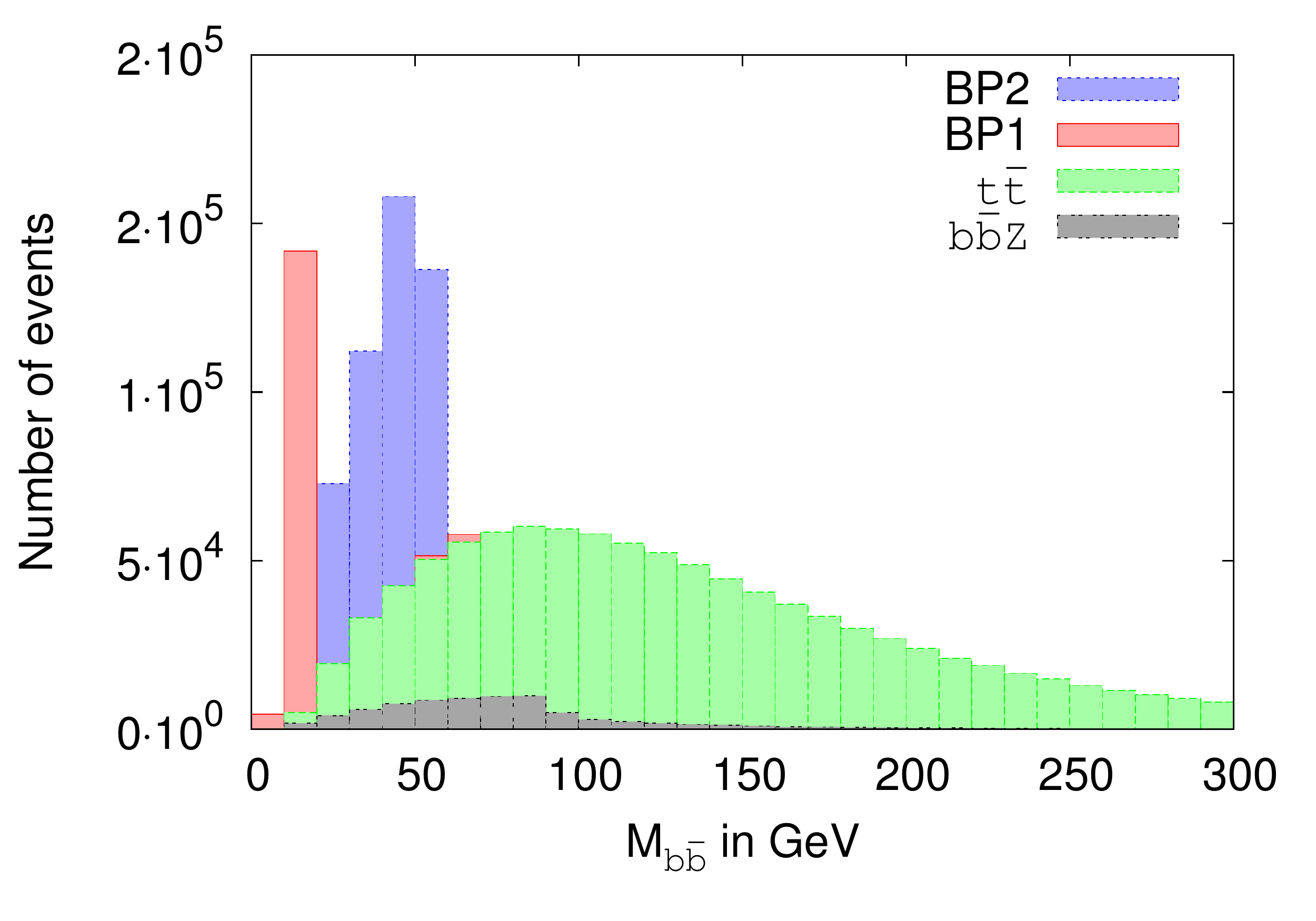}
\includegraphics[width=0.33\linewidth, angle=-90]{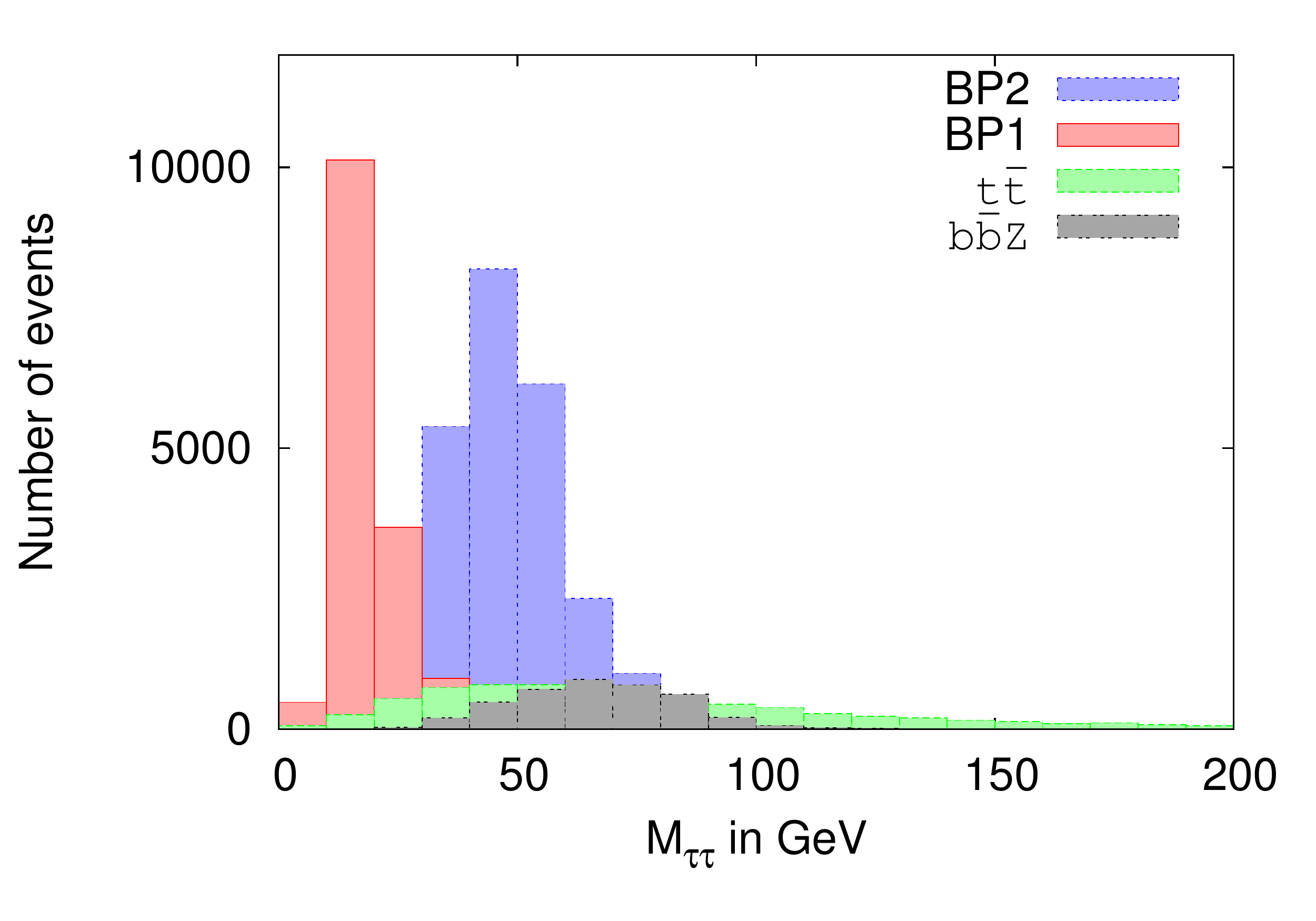}
\caption{ Invariant mass distribution of $b_{\rm{jet}}$'s (left) and $\tau_{\rm{jet}}$'s (right) for $t\bar{t}$ and for the signal in BP2.}\label{invdis}. 
\end{center}
\end{figure}
%%%%%%%%%%%%%%%%%%%%%%%%%%%%%%%%%%%%%%%%%%%%%%%%%%%%%%%%%%%%%%
Next we have analyzed the invariant mass distributions of the $b_{\rm{jet}}$ pair for the same benchmark points. Figure~\ref{invdis} (left) presents the $b_{\rm{jet}}$ pair invariant mass
distributions for the signal in BP1 and BP2, with dominant SM backgrounds coming from $t\bar{t}$ and $b\bar{b}Z$. These results  suggest that, given
the integrated luminosity, it is possible to resolve the resonant peak in the mass distribution of the signal. To further clarify this point, we select events with $|m_{bb}-m_{a_1}|\leq 10$ GeV, that we label as $p_1$.
The resolutions of these peaks depend on the specific benchmark point, but this selection reduces the $b\bar{b}Z$ background drastically, in those cases when $m_{a_1}$  is far separated from the $Z$ gauge boson mass $m_Z$.
The signal significances for all the benchmark points cross the $5\sigma$ level at an integrated luminosity of 100 fb$^{-1}$, and at 13 TeV they are equal to $8.47\, \sigma$,  $6.87\, \sigma$ and $8.01\, \sigma$  for 
BP1, BP2 and BP3 respectively. At a center of mass energy of 14 TeV the significances are $8.70\, \sigma$, $7.74 \,\sigma$ and $9.79 \,\sigma$ in the three cases.

Finally, we simulate the $\tau_{\rm{jet}}$ invariant mass distributions, as they are expected to be cleaner than the $b_{\rm{jet}}$ distributions. Figure~\ref{invdis} (right) shows the invariant mass distributions for both the signals in BP1 and BP3, and the SM backgrounds from $t\bar{t}$ and $b\bar{b}Z$. For this purpose, similarly to the previous case, we select those events with $|m_{\tau\tau}-m_{a_1}|\leq 10$ GeV. For the points which are far away from the $Z$ mass, namely BP1 and BP3, the
signal significance improves significantly, to $12.40 \,\sigma$ and $12.65 \,\sigma$ respectively, whereas for BP2 it is $6.94 \,\sigma$. At a centre of mass energy of 14 TeV these value
 are $14.78\, \sigma$, $6.56 \,\sigma$ and $12.93 \,\sigma$ for BP1, BP2 and BP3 respectively.

\subsection{$3\tau$}
In this subsection we consider the case in which both pseudoscalars decay into $\tau$ pairs.
In this case we expect to see a final state of $4\tau$' s. Of course, due to the lower branching ratio
 in the $a_1\to \tau\bar{\tau}$ mode, the final state numbers are not very promising at low luminosities.
 On top of that, due to a low $\tau$-tagging efficiency for $\tau$'s of low $p_T$, { the final state number is  furtherly reduced.\cite{tautag}.
 Keeping this in mind, we search for final states where we have at least three $\tau$'s. We tag such $\tau$'s
via hadronic $\tau_{\rm{jet}}$'s, as explained earlier.
%%%%%%%%%%%%%%%%% >=3 \tau at 13 TeV  %%%%%%%%%%%%%%%%%%%%
\begin{table}[h]

\begin{center}
\hspace*{-1.0cm}
\renewcommand{\arraystretch}{1}
\begin{tabular}{|c||c|c|c||c|c|c||}
\hline\hline
Final states&\multicolumn{3}{|c||}{Benchmark}&\multicolumn{3}{|c||}{Backgrounds }
\\
\hline
&BP1 & BP2&BP3  &$ZZ$  &$ZW^\pm$&$h Z$\\
\hline
\hline
$n_j\leq 5\,[\geq 3\tau_{\rm{jet}}]$&95.71&199.27&137.21&186.42&437.17&20.68\\
&&&&&&\\
$\&\, |m_{\tau\tau}-m_Z|>10$ GeV&94.79&197.15&135.02&163.53&363.43&17.42\\
&&&&&&\\
$\&\, m_{\tau\tau}\leq125$ GeV&94.79&197.15&135.02&158.07&326.56&16.07\\
&&&&&&\\
 \&\, $p_T^{\tau_{j_1}}\leq 100\, \&\, p_T^{\tau_{j_{2,3}}}\leq 50$ GeV&87.85&184.43&123.34&99.21&210.69&8.31\\
\hline
Significance&4.41&8.22&5.93&\multicolumn{3}{|c||}{}\\
\hline\hline
\multirow{3}{*}{$\&\, p_1:|m_{\tau\tau}-m_{a_1}|\leq 10$ GeV}&\multirow{3}{*}{48.55}&\multirow{3}{*}{54.41}&\multirow{3}{*}{64.96}&4.36&21.07&0.90\\
&&&&44.70&89.54&2.70\\
&&&&26.16&42.14&3.82\\
\hline
Significance&5.61&3.93&5.55&\multicolumn{3}{|c||}{}\\
\hline
\hline
\end{tabular}
\caption{The number of events for a $n_j\leq 5\,[\geq 3\tau_{\rm{jet}}]$ final state at 100 fb$^{-1}$ of luminosity at the LHC with 13 TeV center of mass energy.}\label{3tau13}
\end{center}
\end{table}
%%%%%%%%%%%%%%%%%%%%%%%%%%%%%%%%%%%%%%%%%%%%%%%%%%%%%%%%
The dominant SM backgrounds, in this case, come from the association of $Z$ bosons, i.e. from $ZZ$, $ZW^\pm$, $Zh$
  along with the triple gauge boson productions, namely from $ZZZ$, $ZZW^\pm$, $W^\pm W^\mp W^\pm$, $ZW^\pm W^\mp$ and $WWW$.
  However, the triple gauge boson backgrounds are found to be negligible after imposing the cuts ($\lsim 0.1$) at 100 fb$^{-1}$.
 Table~\ref{3tau13} and Table~\ref{3tau14} show the expected numbers of events for the three benchmark points $BP_i$, together with the dominant backgrounds, at an integrated luminosity of 100 fb$^{-1}$. The final state that we are looking for is characterized by a number of jets $n_j\leq 5$ among which we tag at least  three of them as $\tau_{\rm{jet}}$'s, defined as
 $$\textrm{sig}2: n_j\leq 5\,[\geq 3\tau_{\rm{jet}}].$$
 
 We then add some further kinematical
 cuts to reduce the backgrounds, as before. These cuts include the invariant mass veto on the $\tau_{\rm{jet}}$ pair, $|m_{\tau\tau}-m_Z|>10$ GeV and we also demand that $m_{\tau\tau}\leq125$ GeV, which allows us to search for hidden resonances. Finally, we also demand for softer second and third $\tau_{\rm{jet}}$'s by implementing the cuts $p_T^{\tau_{j_1}}\leq 100\, \&\, p_T^{\tau_{j_{2,3}}}\leq 50$ GeV.

From Table~\ref{3tau13} and Table~\ref{3tau14} one deduces that the $ZW^\pm$ channel remains the most dominant background of all. The signal significance 
at this stage for the three benchmark points are $4.41 \,\sigma$, $8.22\, \sigma$ and $5.93\, \sigma$ for BP1, BP2 and BP3 respectively, at an integrated luminosity of 100 fb$^{-1}$ and a center of mass energy of 13 TeV. At 14 TeV these numbers are  $3.79 \,\sigma$, $8.38 \, \sigma$ and $5.81\, \sigma$.

As in the previous case, also in this case we try to select events around the pseudoscalar mass peak by the constraint $p_1:|m_{\tau\tau}-m_{a_1}|\leq 10$ GeV. The mass resolution depends on the mass value of $a_1$, but BP1 and BP3 now have more than a
$5\sigma$ signal significance. For BP2 $m_{a_1} \sim 57$ GeV, and the multiplicities from the backgrounds involving $ZZ$ and $ZW^\pm$
are more significant than for BP1 and BP3.
 The signal significance at 13 TeV, with an integrated luminosity of 100 fb$^{-1}$ for BP1, BP2 and BP3 are $5.61\, \sigma$, $3.93\, \sigma$ and $5.55\, \sigma$
respectively. These values change for collisions at 14 TeV and equal $5.16\, \sigma$, $4.00 \,\sigma$ and $6.03\, \sigma$ in this second case.
%%%%%%%%%%%%%%%%%%%%%% 14 TeV %%%%%%%%%%%%%%%%%%%%%%%%%%%%%%
\begin{table}

\begin{center}
\hspace*{-1.0cm}
\renewcommand{\arraystretch}{1}
\begin{tabular}{|c||c|c|c||c|c|c||}
\hline\hline
Final states&\multicolumn{3}{|c||}{Benchmark}&\multicolumn{3}{|c||}{Backgrounds }
\\
\hline
&BP1 & BP2&BP3 & $ZZ$  &$ZW^\pm$&$h Z$\\
\hline
\hline
$n_j\leq 5\,[\geq 3\tau_{\rm{jet}}]$&96.34&224.45&146.73&200.62&499.20&18.28\\
&&&&&&\\
$\&\, |m_{\tau\tau}-m_Z|>10$ GeV&94.78&222.85&142.62&178.73&408.70&15.11\\
&&&&&&\\
$\&\, m_{\tau\tau}\leq125$ GeV&94.78&222.06& 141.80&165.36&382.43&13.65\\
&&&&&&\\
$\&\, p_T^{\tau_{j_1}}\leq100\,\& \,p_T^{\tau_{j_{2,3}}}\leq 50$ GeV&82.80&205.34& 133.58&121.59&265.66&7.56\\
\hline
Significance&3.79&8.38&5.81&\multicolumn{3}{|c||}{}\\
\hline\hline
\multirow{3}{*}{$\&\, p_1:|m_{\tau\tau}-m_{a_1}|\leq 10$ GeV}&\multirow{3}{*}{46.35}&\multirow{3}{*}{62.08}&\multirow{3}{*}{79.74}&12.16&20.44&1.71\\
&&&&54.71&122.61&2.44\\
&&&&25.53&67.14&2.56\\
\hline
Significance&5.16&4.00&6.03&\multicolumn{3}{|c||}{}\\
\hline
\hline
\end{tabular}
\caption{The number of events for a $n_j\leq 5\,[\geq 3\tau_{\rm{jet}}]$ final state at 100 fb$^{-1}$ of luminosity at the LHC, for a center of mass energy of 14 TeV.  }\label{3tau14}
\end{center}
\end{table}
%%%%%%%%%%%%%%%%%%%%%%%%%%%%%%%%%%

\subsection{$2b+2\mu$}
The decay rate of the pseudoscalar  to $\mu\bar{\mu}$ is $\mathcal{O}(10^{-4})$, which
makes this channel difficult to observe. If we demand that one of the two pseudoscalars decay into a $b\bar{b}$
pair and the other into a $\mu\bar{\mu}$ pair, the effective cross-section may increase firstly due to
the large branching coming from $a_1\to b\bar{b}$ and, secondly, due to a combinatorial factor of 2, because of the presence of two pseudoscalars. This gives us the option of investigating a final state $2b+2\mu$. 

Table~\ref{2b2mu13} and Table~\ref{2b2mu14} show the corresponding $2\mu$ final states event numbers 
for the benchmark points and the dominant SM backgrounds which include $t\bar t$, $ZZ$, $Zh$, $b\bar b h$ and $b \bar b Z$
at an integrated luminosity of 1000 fb$^{-1}$. We first consider the $2\mu \,\&\, p_T^{\ell_{1,2}}\leq 50$ GeV final state, largely dominated by the SM backgrounds (see Tables~\ref{2b2mu13} and~\ref{2b2mu14}).  Then with impose further requirements on the numbers of jets and their transverse momentum ($p_T$), by defining the signal as 
 $$\textrm{sig}3: \, n_j\leq 3\,[2b_{\rm jet}]\, \& \,n_{\mu}\geq 2 \,[|m_{\mu\mu}-m_Z|> 5 \,\rm{GeV}]\, \& \,p_T^{{\mu,j}_{1,2}}\leq 50\, \rm{GeV}\, \& \,\ptmiss \leq 30\, \textrm{GeV}.$$ 
 
The $\mu$-pair invariant mass veto around the $Z$ mass $(|m_{\mu\mu}-m_Z|> 5 \,\rm{GeV})$, together with the condition of having softer $b_{\rm{jet}}$'s in the final state ($p_T^{j_{1,2}}\leq 50 \,\rm{GeV}$), conspire to reduce the SM backgrounds coming from the $Z$ bosons quite drastically. Finally, since this final state - in an ideal situation - should not have any missing energy, we also demand that
$\ptmiss \leq 30$ GeV. To reduce the backgrounds even further, and to ensure that we select signatures of the light pseudoscalar decay below $125$ GeV, we impose additional constraints on the $\mu$-pair and on the $b_{\rm{jet}}$-pair invariant masses, around the $Z$ mass and the mass of $h_{125}$. These are given by
$|m_{\mu\mu}-m_{h_{125}}|>5\, \textrm{ GeV}$, $|m_{bb}-M_Z|\geq 10\, \textrm{ GeV}$ and |$m_{bb}-m_{h_{125}}|>10 \, \textrm{ GeV}$.\\
At this stage, only in the case of BP2 the signal significance reaches the $3.31 \,\sigma$ value, while for BP1 and BP3 these are $1.03 \,\sigma$,  and $1.83 \,\sigma$ respectively, at 13 TeV. At a center of mass energy of 14 TeV, instead, the values are $1.08 \,\sigma$, $2.64\, \sigma$ and $1.18\, \sigma$ respectively for BP1, BP2 and BP3. 
%%%%%%%%%%%%%%%%% 2 b+2\mu at 13 TeV  %%%%%%%%%%%%%%%%%%%%
\begin{table}[t]
\begin{center}
\hspace*{-2cm}
\renewcommand{\arraystretch}{1}
\begin{tabular}{|c||c|c|c||c|c|c|c|c||}
\hline\hline
Final states&\multicolumn{3}{|c||}{Benchmark}&\multicolumn{5}{|c||}{Backgroounds }\\
\hline
&BP1 & BP2&BP3 & $t\bar t$&$ZZ$& $Zh$ & $b\bar b h$  &$b \bar b Z$\\
\hline
\hline
$2\mu_{\rm{jet}}\, \& \, p_T^{\ell_{1,2}}\leq 50$ GeV& 1877.23&3660.42&3167.55&909080&132161&2669.20&657.71&$6.3\times10^6$\\
&&&&&&&&\\
$\&\,n_j\leq 3\,\&\, b_{\rm jet}\geq 2$&\multirow{3}{*}{69.36}&\multirow{3}{*}{226.13}&\multirow{3}{*}{124.07}&\multirow{3}{*}{4765.60}&\multirow{3}{*}{457.87}&\multirow{3}{*}{15.73}&\multirow{3}{*}{14.83}&\multirow{3}{*}{28.60}\\
$\&\,|m_{\mu\mu}-m_Z|> 5$ GeV&&&&&&&&\\
$\&\,p_T^{j_{1,2}}\leq 50 \rm{GeV}\,\&\, \ptmiss \leq 30$ GeV&&&&&&&&\\
$\&\, |m_{\mu\mu}-m_{h_{125}}|>5$ GeV&\multirow{2}{*}{69.36}&\multirow{2}{*}{226.13}&\multirow{2}{*}{124.07}&\multirow{2}{*}{4190.45}&\multirow{2}{*}{359.76}&\multirow{2}{*}{14.61}&\multirow{2}{*}{14.83}&\multirow{2}{*}{28.60}\\
$\&\,|m_{bb}-M_Z|\geq 10$ GeV&&&&&&&&\\
$\&\, |m_{bb}-m_{h_{125}}|>10$ GeV&69.36&226.13&124.07&4026.11&359.76&13.49&14.83&28.60\\
\hline
Significance&1.03&3.31&1.83&\multicolumn{5}{|c||}{}\\
\hline
\hline
\multirow{3}{*}{$\&\, p_1:|m_{bb}-m_{a_1}|\leq 10$ GeV}&\multirow{3}{*}{64.73}&\multirow{3}{*}{98.93}&\multirow{3}{*}{80.28}&328.66&0.00&0.00&4.94&19.67\\
&&&&1150.32&141.72&5.62&9.89&9.53\\
&&&&492.99&43.61&2.25&0.00&0.00\\
\hline
Significance&3.17&2.63&3.23&\multicolumn{5}{|c||}{}\\
\hline
\hline
\multirow{3}{*}{$\&\, p_2:|m_{\mu\mu}-m_{a_1}|\leq 5$ GeV}&\multirow{3}{*}{41.61}&\multirow{3}{*}{148.40}&\multirow{3}{*}{72.98}&328.66&43.61&1.12&0.00&0.00\\
&&&&575.15&32.70&0.00&0.00&9.53\\
&&&&410.83&21.80&1.12&4.94&0.00\\
\hline
Significance&2.04&5.36&3.22&\multicolumn{5}{|c||}{}\\
\hline
\hline
\end{tabular}
\caption{The number of events for the $n_j\leq 3\,[2b_{\rm{jet}}]\, \&\, \geq 2\mu\, \& \,\ptmiss \leq 30$ GeV final state at 1000 fb$^{-1}$ of luminosity at the LHC, for a center of mass energy of 13 TeV. The constraint
$(\& \geq 2\mu)$ requires the presence of at least 2 muons. The clause ($\&\, b_{\rm jet}\geq 2$) demands at least 2  jets of $b$ quarks, denoted as $b_{\rm jet}$. }\label{2b2mu13}
\end{center}
\end{table}
%%%%%%%%%%%%%%%%%%%%%%%%%%%%%%%%%%
Later we try to enhance the mass peak resolutions on the $bb$ and $\mu\mu$ invariant mass distributions by imposing the two 
constraints (denotes as $p_1,p_2$)
$$p_1:|m_{bb}-m_{a_1}|\leq 10\, \textrm{GeV \,\, and}\,\, p_2:|m_{\mu\mu}-m_{a_1}|\leq 5 \,\textrm{GeV}.$$ At a center of mass energy of 13 TeV, the $m_{bb}$ peaks are characterized by about a $3\, \sigma$ signal significance i.e., $3.17\sigma$, $2.63 \, \sigma$ and $3.23\, \sigma$ respectively for BP1, BP2 and BP3 at an integrated luminosity of of 1000 fb$^{-1}$. At 14 TeV the respective values are $3.17\, \sigma$, $2.63 \, \sigma$ and $3.23 \, \sigma$ respectively for the three benchmarks. \\
The constraint $p_2:|m_{\mu\mu}-m_{a_1}|\leq 5$ GeV, brings BP2 at $5.36 \,\sigma$, BP1 at $2.04\, \sigma$, and BP3 at $3.22 \,\sigma$, for a center of mass energy of 13 TeV. At 14 TeV
the significances are $4.71\, \sigma$, $3.82 \, \sigma$ and $3.00 \, \sigma$ in the three cases, respectively. 
%%%%%%%%%%%%%%%%%% 2 b+2\mu at 14 TeV %%%%%%%%%%%%%%%%
\begin{table}
\begin{center}
\hspace*{-2cm}
\renewcommand{\arraystretch}{1}
\begin{tabular}{|c||c|c|c||c|c|c|c|c||}
\hline\hline
Final states&\multicolumn{3}{|c||}{Benchmark}&\multicolumn{5}{|c||}{Backgrounds }
\\
\hline
&BP1 & BP2&BP3 & $t\bar t$&$ZZ$& $Zh$ & $b\bar b h$  &$b\bar b Z$\\
\hline
\hline
$2\mu_{\rm{jet}}\,\&\, p_T^{\ell_{1,2}}\leq 50$ GeV& \multirow{2}{*}{2281.00}&\multirow{2}{*}{4011.37}&\multirow{2}{*}{3362.13}&\multirow{2}{*}{788683}&\multirow{2}{*}{141428}&\multirow{2}{*}{2926.71}&\multirow{2}{*}{946.42}&\multirow{2}{*}{$7\times10^6$}\\
&&&&&&&&\\
$\&\,n_j\leq 3\,\&\, b_{\rm jet}\geq 2$&\multirow{3}{*}{67.70}&\multirow{3}{*}{167.14}&\multirow{3}{*}{ 73.99}&\multirow{3}{*}{5102.21}&\multirow{3}{*}{583.61}&\multirow{3}{*}{20.72}&\multirow{3}{*}{17.97}&\multirow{3}{*}{10.72}\\
$\&\,|m_{\mu\mu}-m_Z|> 5$ GeV&&&&&&&&\\
$\&\,p_T^{j_{1,2}}\leq 50 \rm{GeV}\,\&\, \ptmiss \leq 30$ GeV&&&&&&&&\\
$|m_{\mu\mu}-m_{h_{125}}|>5$ GeV&\multirow{2}{*}{67.70}&\multirow{2}{*}{167.14}&\multirow{2}{*}{73.99}&\multirow{2}{*}{3630.42}&\multirow{2}{*}{510.66}&\multirow{2}{*}{9.75}&\multirow{2}{*}{11.98}&\multirow{2}{*}{0.00}\\
$\&\,|m_{bb}-M_Z|\geq 10$ GeV&&&&&&&&\\
$\&\,|m_{bb}-m_{h_{125}}|>10$ GeV&67.70&167.14&73.99&3336.06&498.50&9.75&11.98&0.00\\
\hline
Significance&1.08&2.64&1.18&\multicolumn{5}{|c||}{}\\
\hline
\hline
\multirow{3}{*}{$\&\, p_1:|m_{bb}-m_{a_1}|\leq 10$ GeV}&\multirow{3}{*}{67.70}&\multirow{3}{*}{79.60}&\multirow{3}{*}{ 57.54}&196.24&0.00&0.00&0.00&0.00\\
&&&&1373.67&255.33&1.22&0.00&0.00\\
&&&&686.83&24.32&2.44&0.00&0.00\\
\hline
Significance&4.16&1.93&2.08&\multicolumn{5}{|c||}{}\\
\hline
\hline
\multirow{3}{*}{$\&\, p_2:|m_{\mu\mu}-m_{a_1}|\leq 5$ GeV}&\multirow{3}{*}{41.66}&\multirow{3}{*}{103.47}&\multirow{3}{*}{ 45.21}&0.00&36.47&0.00&0.00&0.00\\
&&&&588.72&36.47&0.00&5.99&0.00\\
&&&&98.12&85.11&0.00&0.00&0.00\\
\hline
Significance&4.71&3.82&3.00&\multicolumn{5}{|c||}{}\\
\hline
\hline
\end{tabular}
\caption{The number of events for $n_j\leq 3\,[2b_{\rm{jet}}]\, \&\, \geq 2\mu\, \&\,\ptmiss \leq 30$ GeV final state at 1000 fb$^{-1}$ of luminosity at the LHC for center of a center of mass energy of 14 TeV.}\label{2b2mu14}
\end{center}
\end{table}
%%%%%%%%%%%%%%%%%%%%%%%%%%%%%%%%%%

\subsection{$2\tau+2\mu$}
In this section we discuss a scenario where one of the pseudoscalars decays into a $\tau$ pair and the second one into a $\mu$ pair. Due to the low branching ratios of these two modes, even with a large integrated luminosity, the signal remains small. It is however accompanied by a SM backgrounds for such final states ($2\tau+2\mu$) which is quite suppressed. As in the previous cases, also in this case we tag the $\tau$ via its hadronic decay into a $\tau_{\rm{jet}}$ \cite{tautag}.
The  threshold $p_T$ cuts both for the $\tau_{\rm{jet}}$ and for the muons are kept as low as 10 GeV, since we are considering the decay of a very light pseudoscalar.

%%%%%%%%%%%%%%%%% 2 \tau +2\mu at 13 TeV %%%%%%%%%%%%%%%%%%%%
\begin{table}
\begin{center}
\hspace*{-1cm}
\renewcommand{\arraystretch}{1.2}
\begin{tabular}{|c||c|c|c||c|c||}
\hline\hline
Final states&\multicolumn{3}{|c||}{Benchmark}&\multicolumn{2}{|c||}{Backgrounds }
\\
\hline
&BP1 & BP2&BP3 & $ZZ$& $Zh$ \\
\hline
\hline
$2\mu\, \&\, n_j\leq 3\,[2\tau_{\rm{jet}}$]&\multirow{2}{*}{16.18}&\multirow{2}{*}{14.13}&\multirow{2}{*}{29.19}&\multirow{2}{*}{490.58}&\multirow{2}{*}{28.10}\\
$\&\,p_T^{\ell_{1,2}}\,\&\,p_T^{j_{1,2}}\leq 50$ GeV&&&&&\\
$\&\,|m_{\mu\mu}-m_Z|\geq 5$ GeV&16.18&14.13&29.19&218.03&9.00\\
$\&\,|m_{\tau\tau}-m_Z|>10$ GeV&16.18&14.13&29.19&163.53&9.00\\
$\&\,|m_{\tau\tau}|<125$ GeV&16.18&14.13&29.19&152.62&7.87\\
\hline
Significance&1.22&1.07&2.12&\multicolumn{2}{|c||}{}\\
\hline
\hline
\multirow{3}{*}{$\&\,p_1:|m_{\tau\tau}-m_{a_1}|\leq 10$ GeV}&\multirow{3}{*}{11.56}&\multirow{3}{*}{14.13}&\multirow{3}{*}{21.90}&0.00&0.00\\
&&&&54.51&1.12\\
&&&&32.70&1.12\\
\hline
Significance&3.40&1.70&2.93&\multicolumn{2}{|c||}{}\\
\hline
\hline
\multirow{3}{*}{$\&\, p_2:|m_{\mu\mu}-m_{a_1}|\leq 5$ GeV}&\multirow{3}{*}{6.94}&\multirow{3}{*}{7.07}&\multirow{3}{*}{0.00}&0.00&0.00\\
&&&&0.00&0.00\\
&&&&43.61&2.25\\
\hline
Significance&2.63&2.65&-&\multicolumn{2}{|c||}{}\\
\hline
\hline
\end{tabular}
\caption{The number of events for $n_j\leq 3\,[2\tau_{\rm{jet}}] \,\&\,\geq 2\mu \,\&\,\ptmiss \leq 30$ GeV final state at 1000 fb$^{-1}$ of luminosity at the LHC for a center of mass energy of 13 TeV.}\label{2ta2mu13}
\end{center}
\end{table}
%%%%%%%%%%%%%%%%%%%%%%%%%%%%%%%%%%%%%%%%%%%%%%%%%%
The results of this analysis are reported in Table~\ref{2ta2mu13} and Table~\ref{2ta2mu14}, where we present the number of events for the benchmark points and the dominant SM backgrounds, for a center of mass energy of 13 and 14 TeV and an integrated luminosity of 1000 fb$^{-1}$. We search for a muon pair and at least two $\tau$'s in the final state. Though muons ($\mu$) will be detected as a charged leptons, the $\tau$'s will be 
detected via their hadronic decays as $\tau_{\rm{jets}}$'s \cite{tautag}. Being the two pseduoscalars light, we require  both the $\mu$ and the $\tau$ jets to be rather soft (i.e. ($p_T^{\ell_{1,2}}\&p_T^{j_{1,2}}) \leq 50$ GeV) in the final state. This defines the signal as 
$$\textrm{sig}4: \,n_j\leq 3\,[2\tau_{\rm{jet}}]\, \& \, \geq 2\mu \, \&\, \ptmiss \leq 30\, \rm{GeV}.$$

 Tagging both muons and requiring the cut $p_T \leq 50$ GeV for the transverse momentum $p_T$ of the $\tau_{\rm{jet}}$, will suppress much of the hard SM backgrounds, favouring the search for a low mass resonance, in this case a light 
pseudoscalar. The dominant backgrounds in this case comes from the SM $ZZ$ and $hZ$ channels. The background due to the $a_1 Z$ channel is negligible, due to the mostly-singlet nature of the $a_1$. We have also checked for other triple gauge boson contributions to this final states, but they are all either zero or negligible.  To reduce further the SM backgrounds we apply a veto on the mass peak of the $Z$ boson, by requiring that $|m_{\mu\mu}-m_Z|\geq 5$ GeV and $|m_{\tau\tau}-m_Z|>10$ GeV respectively.  As one may deduce from Table~\ref{2ta2mu13} and Table~\ref{2ta2mu14},  the application of these two cuts, though reduces the SM backgrounds quite drastically, does not affect the signal, which remains unchanged. Finally, we apply the constraint $|m_{\tau\tau}|<125$ GeV to ensure the search for
hidden scalars, i.e., $m_{a_1}< 125$ GeV, which causes an even larger suppression of the background. At this level the signal significances are still below $3\sigma$ at 13 TeV and reach $3.20 \, \sigma$ only in the case of the benchmark point BP3, at 14 TeV. 

Next we apply the constraint $p_1:|m_{\tau\tau}-m_{a_1}|\leq 10$ GeV to favour the search for a possible mass peak of the pseudoscalar and this enhances the signal significance to $3.40\, \sigma$, $1.70\, \sigma$ and $2.93\, \sigma$ respectively for BP1, BP2 and BP3 at 13 TeV. At 14 TeV these numbers are $2.47 \,\sigma$, $2.51\, \sigma$ and $3.27\, \sigma$ respectively. Similar peaks around $\mu$ pair invariant mass distribution, i.e. with $p_2:|m_{\mu\mu}-m_{a_1}|\leq 5$ GeV, give signal significances of $2.63 \,\sigma$ and $2.65\, \sigma$ for BP1 and BP2, at a center of mass energy of 13 TeV. BP3 in this case runs out of statistics. At 14 TeV the signal significances are $2.05 \,\sigma$, $2.82\, \sigma$ and $2.04 \,\sigma$ respectively. The leptonic modes thus need higher luminosities $\gsim 2000$ fb$^{-1}$ in order to reach the discover limit for a light pseudoscalar.
%%%%%%%%%%%%%%%%% 2 \tau +2\mu at 14 TeV %%%%%%%%%%%%%%%%%%%%
\begin{table}
\begin{center}
\hspace*{-1cm}
\renewcommand{\arraystretch}{1.2}
\begin{tabular}{|c||c|c|c||c|c||}
\hline\hline
Final states&\multicolumn{3}{|c||}{Benchmark}&\multicolumn{2}{|c||}{Backgrounds }
\\
\hline
&BP1 & BP2&BP3 & $ZZ$& $Zh$\\
\hline
\hline
$2\mu\,\&\, n_j\leq 3\,[2\tau_{\rm{jet}}$]&\multirow{2}{*}{15.62}&\multirow{2}{*}{31.84}&\multirow{2}{*}{41.10}&\multirow{2}{*}{498.50}&\multirow{2}{*}{20.72}\\
$\&\,p_T^{\ell_{1,2}}\,\&\,p_T^{j_{1,2}}\leq 50$ GeV&&&&&\\
$\&\,|m_{\mu\mu}-m_Z|\geq 5$ GeV&15.62&31.84&41.10&145.90&7.31\\
$\&\, |m_{\tau\tau}-m_Z|>10$ GeV&15.62&31.84&41.10&121.58&3.66\\
$\&\, |m_{\tau\tau}|<125$ GeV&15.62&31.84&41.10&121.58&2.44\\
\hline
Significance&1.32&2.55&3.20&\multicolumn{2}{|c||}{}\\
\hline
\hline
\multirow{3}{*}{$\&\,p_1:|m_{\tau\tau}-m_{a_1}|\leq 10$ GeV}&\multirow{3}{*}{15.62}&\multirow{3}{*}{15.92}&\multirow{3}{*}{28.77}&24.32&0.00\\
&&&&24.32&0.00\\
&&&&48.63&0.00\\
\hline
Significance&2.47&2.51&3.27&\multicolumn{2}{|c||}{}\\
\hline
\hline
\multirow{3}{*}{$\&\,p_2:|m_{\mu\mu}-m_{a_1}|\leq 5$ GeV}&\multirow{3}{*}{5.21}&\multirow{3}{*}{7.96}&\multirow{3}{*}{12.33}&0.00&1.22\\
&&&&0.00&0.00\\
&&&&24.32&0.00\\
\hline
Significance&2.05&2.82&2.04&\multicolumn{2}{|c||}{}\\
\hline
\hline
\end{tabular}
\caption{The number of events for $n_j\leq 3\,[2\tau_{\rm{jet}}]\,\&\,\geq 2\mu\,\&\,\ptmiss \leq 30$ GeV final state at 1000 fb$^{-1}$ of luminosity at the LHC for center of mass energy (ECM) of 14 TeV.}\label{2ta2mu14}
\end{center}
\end{table}
%%%%%%%%%%%%%%%%%%%%%%%%%%%%%%%%%%

\chapter*{Conclusions}\label{concl}

We have considered a scenario with an extended Higgs sector characterized by a $Y=0$ hypercharge $SU(2)$ triplet and a gauge singlet superfields, along with the remaining MSSM superfields. The triplet vev is restricted by the $\rho$ parameter, hence the $\mu_{\text{eff}}$  is generated spontaneously mostly by the singlet vevs. In models with gauged $U(1)'$ symmetry the singlet could be invoked in the mass generation of the extra gauge boson $Z'$ by spontaneous symmetry breaking. This would require a large singlet vev $v_S$, due to the recent bounds on extra $Z'$ coming from the analysis at the LHC \cite{LHCZ'}. 

We have first investigated the masses of the Higgs sector of the model at  tree-level. The lightest tree-level Higgs state, in this case, is not bounded to lay below $M_Z$, due to the additional contributions from the triplet and the singlet, which are proportional to their respective couplings and are enhanced at low $\tan{\beta}$. This allows to reduce the size of the quantum correction needed in order to reach the $\sim 125$ GeV  at one-loop, compared to the MSSM or to others constrained MSSM scenarios. Then we have extended our analysis at one-loop level. The one-loop Higgs with mass around $\sim 125$ GeV puts some indirect bounds on the masses of the particles contributing in the radiative corrections. For this purpose we have included the one-loop contributions using the Coleman-Weinberg potential. We have also presented results for the neutralino, and charginos spectra, together with the stop and sbottom mass matrices. We have calculated full one-loop Higgs masses considering all the weak sectors and the strong sectors. We also showed that the gauge boson-gaugino-higgsino sectors mostly contribute negatively to the mass eigenstates, while the stop-top, sbottom-bottom and Higgs sectors contribute positively. Due to the large number of scalars, seven neutral and three charged Higgs bosons, the Higgs self corrections can be larger than the strong corrections in the large $\lambda_{T,S}$ limit.  This substantially reduces the indirect lower bounds on the stop and sbottom masses. Thus in TNMSSM the discovery of a $\sim 125$ GeV Higgs boson does not put a stringent  lower bound on the stop and sbottom masses, and one has to rely on direct search results  for the lower bounds on the SUSY mass scale.

We have implemented the model in SARAH3.5 \cite{sarah} in order to generate the vertices and other model files for CalcHEP \cite{calchep}. The beta-functions have also been generated at one-loop. We have addressed the issues of perturbativity of the couplings at the higher scale, as we have run the corresponding renormalization group equations from the electroweak scale up.  This has shown that the couplings of the model at the electroweak scale need to be restricted to certain values. For example, even with a value of $\lambda_{T,S}\sim 0.8$ at the electroweak scale, the theory remains perturbative up to $10^{8-10}$ GeV.  Setting all the couplings at a value ($\lambda_{TS} \sim 0.8$, $\kappa\sim 2.4$) the upper scale in the perturbative evolution gets lowered to $10^{4-6}$ GeV. The issue of fine-tuning at the electroweak scale has been discussed in this context.  We have seen that although the tree-level mass spectrum is highly fine-tuned for larger $\lambda_{T,S}$, the amount of fine tuning is reduced after the inclusion of the radiative corrections.

The prospects for hidden Higgs(es), which are scalars and/or pseudoscalars of mass lower than the current Higgs mass, has been discussed quite thoroughly. We have seen that in the rich Higgs spectrum of the model there are several possibilities for having one or more hidden neutral Higgs bosons ($\lesssim125$ GeV) both CP-even and CP-odd. A special scenario emerges when we break the continuous $U(1)$ symmetry softly by the parameters $A_i$. This leads to the appearance of a very light pseudoscalar state of $\mathcal{O}(1)$ GeV to $\mathcal{O}(1)$ MeV in mass, which has its own interesting  phenomenology.  Finally, we have discussed the doublet-triplet-singlet mixing which influences the productions and decays of neutral and charged Higgs bosons at the LHC. The existence of a $h^\pm_i-W^\mp-Z$ tree-level vertex, due to the triplet, impacts both the production as well as the decay channels of the charged Higgs bosons \cite{tssmch1}. In the presence of a light pseudoscalar, the $h_i\to Z a_j$ channel is a possibility due to the very light mass of the pseudoscalar(s). Both the triplet and the singlet states do not couple to the fermions, which leads to some very interesting phenomenology. This property also has an impact on rare decays like $b\to \mu \mu$ and $b\to s \gamma$ \cite{tssmyzero, infnr}. Given the rich phenomenology and the specific predictions of this model, the  current analysis at the LHC and future colliders could be able to test and shed a light on this scenario by looking at its interesting signatures. 

We focus our attention on a typical mass spectrum with a doublet-like CP-even Higgs boson around 125 GeV,  a light triplet-like charged Higgs boson and a light singlet-like pseudoscalar. The existence of light singlet-like pseudoscalar and triplet-like charged Higgs  boson enrich the phenomenology at the LHC and at future colliders. In general we expect to have mixing between doublet and triplet type charged Higgs. We find that in the decoupling limit, $\lambda_T \simeq 0$, one should expect two triplet-like and one doublet-like massive charged Higgs bosons. However since the Goldstone boson is a linear combination which includes a triplet contribution $\sim {v_T}/{v}$ (see Eq.~\ref{gstn}), one of the massive eigenstates triplet cannot be $100\%$ triplet-like.  Recent searches by both CMS \cite{ChCMS} and ATLAS  \cite{ChATLAS} are conducted for a charged Higgs mainly of doublet-type and coupled to fermions. For this reason such a state can be produced in association with the top quark and can decay to $\tau\nu$. Clearly, these searches have to be reinvestigated in order to probe the possibility of triplet representations of $SU(2)$ in the Higgs sector. The breaking of the custodial symmetry via a non-zero triplet vev generates $h_i^\pm-W^\mp-Z$ vertex at the tree-level in TNMSSM. This leads to the vector boson fusion channel for the charged Higgs boson, which is not present in the MSSM or the 2HDM.  On top of that the $Z_3$ symmetric superpotential of TNMSSM has a  light pseudoscalar $a_1$ as a pseudo NG mode of a global $U(1)$ symmetry, known as the "$R$-axion" in the literature. However the later can also be  found in the context of the $Z_3$ symmetric NMSSM. In this case the light charged Higgs boson can decay to $a_1 W^\pm$ \cite{han, colepa, guchait, pbsnkh} just like in the TNMSSM. In the context of a CP-violating MSSM, such modes can arise due to the possibility of a light Higgs boson $h_1$ and of CP-violating interactions. A charged Higgs boson can decay to $h_1 W^\pm$ \cite{CPVMSSM}, just as in our case. Therefore, one of the challenges at the LHC will be to distinguish among such models, once such a mode is discovered. Triplet charged Higgs bosons with $Y=0$, however, have some distinctive features because they do not couple to the fermions, while the fusion channel $ZW^\pm$ is allowed.

The phenomenology of such triplet-like charged Higgs boson has already been studied in the context of TESSM \cite{pbas3}. Such charged Higgs bosons also affect the predictions of $B$-observables \cite{pbas1, pbas2} for missing  the coupling to fermions and to the $Z$ boson.  However in TESSM, even though the charged Higgs boson decays to $ZW^\pm$ \cite{pbas3}, the possibility of a light pseudoscalar is not so natural \cite{pbas1, pbas2, DiChiara, pbas3}. Indeed, one way to distinguish between the TESSM and the TNMSSM is to exploit the prediction of a light pseudoscalar in the second model, beside the light triplet type charged Higgs boson. We expect that such a Higgs in the TNMSSM will be allowed to decay both to $ZW^\pm$ as well as to $a_1 W^\pm$, the former being a feature of the triplet nature of this state, and the latter of the presence of an $R$-axion in the spectrum of the model.

We have investigated the discovery potential of a light pseudoscalar sector which is present in this model.  Our analysis has been performed assuming as a production mechanism the gluon-gluon fusion channel of the 125 GeV Higgs $h_{125}$, and focused on the currents experimental rates on its decay into the $WW^*$, $ZZ^*$
and $\gamma\gamma$ derived at the LHC. Given the current uncertainties in these discovered modes as well as in other (fermionic) modes of the Higgs, we have investigated the possibility that such uncertainties are compatible with the production of two light pseudoscalars, predicted by the TNMSSM, which have so far been undetected.

Benchmarking three points in the parameter space of the model, we have proposed and simulated final states of the form $2b+2\tau$, $3\tau$, $2b+2\mu$ and $2\tau +2\mu$, derived from the decays of such pseudoscalars.
A PYTHIA-FastJet based simulation of the dominant SM backgrounds shows that, depending on the benchmark points, such light pseudoscalars can be probed with 
early LHC data ($\sim 25$ fb$^{-1}$) at 13 and 14 TeV. The $2\tau+2\mu$ decay modes of such states, though much cleaner compared to other channels, need higher luminosity ($\sim 2000$ fb$^{-1}$) in order to be significant. Nevertheless, such muon final states will be crucial for precision mass measurements of the $a_1$. In this case, due to the $Z-a_1-a_1$ coupling, one may consider the production of an $a_1$ pair directly at tree-level, and this can enhance the signal strength by about $10\%$.

The identification of such hidden scalars would be certainly a signal in favour of an extended Higgs sectors, but finding the triplet and singlet $SU(2)$ representations of these extra states would require more detailed searches. Clearly, there are some other distinctive features of this model respect to the NMSSM. The NMSSM does not have any extra charged Higgs bosons compared to the MSSM, while the TNMSSM has an extra triplet-like charged Higgs boson which does not couple to fermions and can decay to $h^\pm \to Z W^\pm$. This possibility changes the direct bounds derived from searches for a charged Higgs at the LHC, as well as the indirect bounds on flavour. These changes are due to the doublet-triplet mixing in the charged Higgs and chargino sectors of the triplet extended model \cite{tripch}. Such sectors can be very useful in order to establish the $SU(2)$ content of the extra scalars, since in this model a very light triplet-like charged Higgs states cannot be ruled out \cite{pbancc}. 
 
 Finally, the superpartners of this triplet- and singlet- like scalars can be dark matter candidates. In particular, a light pseudoscalar sector provides the much needed
 annihilation channel in order to respect the correct dark matter relic density. As we have seen, 
 both direct and indirect constraints can play a significant role in the searches for scalars in higher representations of the $SU(2)$ gauge symmetry, setting a clear distinction respect to the ordinary doublet construction, which is typical of the SM.

%%%%%%%%%%%%%%%%%%%%%%%%%%%%%%%%%%%%%%%%%%%%

%%%%%%%%%%%%%%%%%%%%%%%%%%%%%%%%%%%%%%%%%%%

\part{Applications to Gravitational Lensing of the TVV and TFF correlators}

\chapter{Radiative Effects in Gravitational Lensing }  

\section{Synopsis} 
This chapter develops an application of the trilinear vertices $TVV$ and $TFF$,  where $F$ denotes a fermion, in our case a neutrino, computed at one-loop in the SM, to the case of propagation of photons and neutrinos in a gravitational background. These vertices, as already pointed out in previous chapters, are responsible for the tree level interaction between gravity and the fields of the SM. The study is built around previous analysis of the same topic, with the idea to investigate the role of the conformal anomaly in the process of gravitational lensing. As already pointed out in \cite{Coriano:2014gia}, the effect of the conformal anomaly manifests in some corrections to the classical Einstein's formula for the deflection in General Relativity (GR).
Here we propose a method to incorporate radiative effects in the classical lens equations of neutrinos and photons. The study is performed for a Schwarzschild metric, generated by a point-like source, and expanded in the Newtonian potential at first order. We use a semiclassical approach, where the perturbative corrections to neutrino scattering, evaluated at one-loop in the Standard Model, are compared with the Einstein formula for the deflection using an impact parameter formulation. As  just mentioned, for this purpose we use the renormalized expression of the graviton/fermion/fermion vertex presented in previous studies. We show the agreement between the classical and the semiclassical formulations, for values of the impact parameter $b_h$ of the neutrinos of the order of $b_h\sim 20$, measured in units of the Schwarzschild radius. The analysis is then extended with the inclusion of the post Newtonian corrections in the external gravity field, showing that this extension finds application in the case of the scattering of a neutrino/photon off a primordial black hole. The energy dependence of the deflection, generated by the quantum corrections, is then combined with the standard formulation of the classical lens equations. We illustrate our approach by detailed numerical studies, using as a reference both the thin lens and the nonlinear Virbhadra-Ellis lens.
 
\section{Introduction}
According to classical GR massless particles follow null spacetime geodesics which bend significantly in the presence of very massive sources. The gravitational lensing enforced on their spatial trajectories provides important information on the underlying distributions of matter and, possibly, of dark matter, which act as sources of the gravitational field.\\
Several newly planned weak lensing experiments such as the Dark Energy Survey (DES) \cite{DES}, the Large Synoptic Survey Telescope (LSST)\cite{LSST}, both ground based, or from space with the Wide-Field Infrared Survey Telescope (WFIRST) \cite{WFIRST} and Euclid \cite{EUCLID}, are expected to push forward, in the near future, the boundaries of our knowledge in cosmology.\\
 In the analysis of the deflection by a single compact and spherically symmetric  source, one significant variable, beside the mass of the source, is the impact parameter of the incoming particle beam, measured respect to the center of the source, which determines the size of the deflection. It is very convenient to measure the impact parameter $(b)$, which is typical of a given collision, in units of the Schwarzschild radius $r_s\equiv 2 G M $, denoted as $b_h\equiv b/r_s$. In the Newtonian approximation for the external background, this allows to scale out the entire mass dependence of the lensing event. \\
For an impact parameter of the beam of the order of $10^5-10^6$, the corresponding deflection is rather weak, of the order of 1-2 arcseconds, as in the case of a photon skimming the sun. Stronger lensing effects are predicted as the particle beam nears a black hole, with deflections which may reach 30 arcseconds or more. These are obtained for impact parameters $b_h$ of the order of $2\times 10^4$. Even larger deflections, of 1 to 2 degrees or a significant fraction of them, are generated in scatterings which proceed closer to the event horizon \cite{Coriano:2014gia}. In fact,  as we are going to show, for closer encounters, with the beam located between 20 and 100 $b_h$, such angular deflections are around 
$10^{-2}$ radians in size, as predicted by classical GR. A high energy cosmic ray of 10-100 GeV  will then interact with the field of the source by exchanging momenta far above the MeV region, and will necessarily be sensitive to radiative effects, such as those due to the electroweak corrections.\\
Interactions with such momentum exchanges cannot be handled by an effective Newtonian potential, as derived, for instance, from the (loop corrected) scattering amplitude. We recall that, in general, in the derivation of such a potential, one has to take into account only non-analytic terms in the momentum transfer $q$. These are obtained from a given amplitude and/or gravitational form factor of the incoming particle after an expansion at small momentum. The analytic terms in the expansion correspond to contact interactions which are omitted from the final form of the potential, being them proportional to Dirac delta functions. \\
 As one can easily check by a direct analysis, non-analytic contributions originate from massless exchanges in the loops, 
which approximate the full momentum dependence of the radiative corrections only for momentum transfers far below the MeV region. Therefore, the validity of the method requires that the typical impact parameter of the beam, for a particle with the energy of few GeV's, be of the order of $10^6$ Schwarzschild radii and not less. For such a reason, if we intend to study a lensing event characterized by a close encounter between a cosmic ray and a black hole, we need to resort to an alternative approach, which does not suffer from these limitations. \\
Finally, with the photon sphere located at $b_h\sim 2.5$ for a Schwarzschild metric, one expects that very strong deflections are experienced by a beam for scattering events running close to such a value of the impact parameter. This is also the radial distance from the black hole center at which the scattering angle diverges. A simple expansion of the Einstein formula for the deflection shows that this singularity is logarithmic \cite{Coriano:2014gia}. In such extreme cases the beam circulates around the source one or more times before escaping to infinity, generating a set of relativistic images \cite{Bozza:2001xd}. This is also the region where the simple Newtonian approach, discussed in 
\cite{Coriano:2014gia},  fails to reproduce the classical GR prediction, as expected. \\

\section{Comparing classical and semiclassical effects}
 The analysis of possible extensions of the classical GR prediction for lensing, with the inclusion also of quantum effects 
in the interaction between the particle source and the deflector (lens), has not drawn much attention in the past, except for a couple of very original proposals \cite{Delbourgo:1973xe, Berends:1974gk}. While these effects are expected to be small, even for huge gravitational sources such as massive/supermassive black holes, they could provide, in principle, a way to test the impact of quantum gravity and of other radiative corrections to the propagation of cosmic rays. Close encounters of a beam with a localized source, which could be a large black hole or a neutron star, are expected to be quite common in our universe, although the probability of identifying a lensing event characterized by a close alignment between the source, the lens and an earth based detector, especially for neutrinos, is exceedingly rare \cite{Mena:2006ym}. The situation might be more promising for photons in close encounters with primordial black holes, revealed by resorting to spaceborne detectors. \\ 
Such is the FERMI satellite \cite{FERMI}, with source beams given by Gamma Ray Bursts (GRBs) \cite{Gould1992}, which could detect fringes between primary and secondary paths of the GRBs on its ultra sensitive camera, generated by a gravitational time delay. This approach was termed in 
\cite{Gould1992} "femtolensing", due to the size of the Einstein radius characteristic of these events, which was estimated to be of the order of a femtoarcsecond. As shown in \cite{Gould1992}, a classical GR analysis based on the thin lens equation can be applied quite straightforwardly also to this extreme situation.\\
An important point which needs to be addressed, in this case, concerns the quantum features of these types of lensing events, since the Schwarzschild radius of a primordial black hole, for a gamma ray photon, is comparable to its wavelength. Our analysis draws a path in this direction.

The classical deflections of photons, as pointed out in the past and in a recent work \cite{Coriano:2014gia}, can be compared at classical and quantum
 levels by equating the classical gravitational cross section, written in terms of the impact parameter of the incoming photon beam, to the perturbative cross section. The latter is expanded in ordinary perturbation theory with the inclusion of the corresponding radiative corrections. 
 The result is a differential equation for the impact parameter of the beam, whose solution provides the link between the two descriptions. In particular, the energy dependence, naturally present in the cross section starting at one-loop order, allows to derive a new formula which 
 relates $b_h$ to the energy $E$ of the beam and to the angle of deflection $\alpha$, $b_h(E, \alpha)$. This dependence, which is absent in Einstein's formula, propagates into all the equations for the usual observables of any lensing process: magnifications, cosmic shears, the light curve of microlensing events and Shapiro time delays. Clearly, such a dependence implies, as noted in \cite{Accioly1}, that radiative corrections induce a violation of the classical equivalence principle in General Relativity.  
The violation of the equivalence principle, viewed from a quantum perspective, is not surprising, since this principle is inherently classical and requires the localization of the point particle trajectory on a geodesic. It can be summarized in the statement that an experiment will not be able to determine the nature of the point particle which is subjected to gravity, except for its mass. The notion of a point particle clearly clashes with the quantum description, which is, on the other hand, inherently tight to Heisenberg's indetermination principle. For this reason, one expects that the inclusion of radiative corrections will cause a violation of such principle.
 
 Gravity, in this approach, is treated as an external background and the transition amplitude involves on the quantum side, in the photon case, the $TVV$ vertex, where $T$ denotes the energy momentum tensor (EMT) of the Standard Model and $V$ the electromagnetic current. In the fermion case (f), the corresponding vertex is the $Tff$, with $f$ denoting a neutrino. The comparison between the classical and the semiclassical formula for the deflection derived by this method can then be performed at numerical level, as shown in \cite{Coriano:2014gia} for the photons. The energy dependence of the bending angle, for a given impact parameter of the photon beam, though small, is found to become more pronounced at higher energies, due to the logarithmic growth of the electroweak corrections with the energy.    

The goal of our present work is to propose a procedure which allows to include these effects in the ordinary lens equations, illustrating in some detail how this approach can be implemented in a complete numerical study. We mention that our semiclassical analysis is quite general, and applies both to macroscopic and to microscopic black holes. In the case of macroscopic black holes the procedure has to stop at Newtonian level in the external field. In fact, post-Newtonian corrections, though calculable, render the perturbative expansion in the external (classical) gravitational potential divergent, due to the macroscopic value of the Schwarzschild radius. On the other hand, in the case of primordial black holes, the very same corrections play a significant role in the deflection of a cosmic ray, and bring to a substantial modification of the classical formulas.

We will firstly extend a previous analysis of photon lensing \cite{Coriano:2014gia}, developed along similar lines, to the neutrino case, presenting a numerical study of the complete one-loop corrections derived from the electroweak theory. The formalism uses a retarded graviton propagator with the effects of back reaction of the scattered beam on the source not included, as in a typical scattering problem by a static external potential. In this case, however, because of the presence of a horizon, we search for a lower bound on the size of the impact parameter of the collision where the classical GR prediction and the quantum one overlap. Indeed, above the bound the two descriptions are in complete agreement.
As already mentioned above, both in the fermion as in the photon case \cite{Coriano:2014gia}, this bound can be reasonably taken to lay around 20 ${b_h}$, which is quite close to the horizon of the classical source. For smaller values of $b_h\, (4< b_h <20)$, the two approaches are in disagreement, since the logarithmic singularity in the angle of deflection, once the beam gets close to the photon sphere, starts playing a significant role. This is expected, given the assumption of weak field for the gravitational coupling, which corresponds to the Newtonian approximation in the metric.
 
Than we deal with the implementation of the semiclassical deflection within the formalism of the classical lens equations. We use the energy dependence of the angular deflection to derive new lens equations, which are investigated numerically. We quantify the impact of these effects both in the thin lens approximation, where the trigonometric relations in the lens geometry are expanded to first order, and for a lens with deflection terms of 
 higher order included.
As an example, in this second case, we have chosen the 
Virbhadra-Ellis \cite{Virbhadra:2002ju} lens equation. The observables that we discuss are limited to solutions of these equations and to their magnifications, although time delays, shears and the light curves of a typical microlensing event can be easily included in this framework.  We anticipate that the effects that we quantify are small and cover the milliarcsecond region, remaining quite challenging to detect at experimental level. We hope though,  that the framework that we propose can draw further interest on this topic in the future, both at theoretical and at phenomenological level.

Finally we discuss the post Newtonian formulation of the impact parameter formalism, and apply it to the case of a compact source with a microscopic Schwarzschild radius. This is the only case in which the gravitational corrections to the Newtonian cross section can be consistently included in our approach in a meaningful way. We then summarize our analysis and discuss in the conclusions some possible future directions of possible extensions of our work.

 \section{Gravitational interaction of neutrinos}
\label{Sec.TheorFram}
We start our analysis with a brief discussion of the structure of the gravitational interaction of neutrinos, building on the results of \cite{Coriano:2012cr, Coriano:2013iba}, to which we refer for additional details, and that we are going to specialize to the case of a massless neutrino. An analysis of gravity with the fermion sector is contained in \cite{Degrassi:2008mw}.
We simply recall that the dynamics of the Standard Model in external gravity is described by the Lagrangian  
\beq S = S_G + S_{SM} + S_{I}= -\frac{1}{\kappa^2}\int d^4 x \sqrt{-{g}}\, R+ \int d^4 x
\sqrt{-{g}}\mathcal{L}_{SM} + \chi \int d^4 x \sqrt{-{g}}\, R \, H^\dag H.      \, 
\label{thelagrangian}
\eeq
This includes the Einstein term $\mathcal{S}_G$, the $\mathcal{S}_{SM}$ action and a term $\mathcal{S}_I$ involving the Higgs doublet $H$ \cite{Callan:1970ze}, called the term of improvement. $\mathcal{S}_{SM}$, instead, is obtained by extending the ordinary Lagrangian of the Standard Model to a curved metric background. The term $\chi$ is a parameter which, at this stage, is arbitrary and that at a special value $(\chi\equiv\chi_c=1/6)$ guarantees the renormalizability of the model at leading order in the expansion in $\mathcal{\kappa}$. \\ 
Deviations from the flat metric $\eta_{\mu\nu}=(+,-,-,-)$ will be parametrized in terms of the gravitational coupling $\kappa$, with $\kappa^2= 16 \pi G$ and with $G$ being the gravitational Newton's constant. At this order the metric is given as $g_{\mu\nu}=\eta_{\mu\nu} + \kappa h_{\mu\nu}$, with $h_{\mu\nu}$ describing its fluctuations. We will consider two spherically symmetric and static cases, corresponding to the Schwarzschild and Reissner-Nordstrom metrics. The first, in the weak field limit and in the isotropic form is given by

\beq
ds^2\approx\left(1- \frac{2 G M}{|\vec{x}|}\right)dt^2 -\left(1 + \frac{2 G M}{|\vec{x}|}\right)d\vec{x}\cdot d\vec{x}
\label{SCH3}.
\eeq
In this case the fluctuation tensor takes the form 
\bea
h_{\mu\nu}(x)
&=& \frac{2 G M}{\kappa |\vec{x}|}\bar{S}_{\mu\nu}, \qquad  \bar{S}_{\mu\nu}\equiv \eta_{\mu\nu}-2 \delta^0_{\mu}\delta^0_{\nu}.
\label{hh}
\eea
The inclusion of higher order terms in the weak field expansion will be discussed in the following sections. \\
The coupling of the gravitational fluctuations to the fields of the Standard Model involves the 
 EMT, which is defined as  
\beq
T_{\mu\nu}=\frac{2}{\sqrt{-g}}\frac{\delta \left(S_{SM}+S_{I}\right)}{\delta g^{\mu\nu}} \bigg|_{g=\eta} \,
\label{stmn}
\eeq 
with a tree-level coupling summarized by the action 
\beq
\mathcal{S}_{int}=-\frac{\kappa}{2}\int d^4 x \, T_{\mu\nu} h^{\mu\nu} \,,
\label{inter}
\eeq
where $T_{\mu \nu}$ is symmetric and covariantly conserved. The complete expression of the EMT of the Standard Model, including ghost and gauge-fixing contributions can be found in \cite{Coriano:2011zk}. \\
The Higgs field is parameterized in the form 
\beq
H = \left(\begin{array}{c} -i \phi^{+} \\ \frac{1}{\sqrt{2}}(v + h + i \phi) \end{array}\right)
\eeq
 in terms of $h$, $\phi$ and $\phi^{\pm}$, which denote the physical Higgs and the Goldstone bosons of the $Z$ and $W'$ s respectively. $v$ is the Higgs vacuum expectation value. The terms of the Lagrangian $\mathcal{S}_I$, generate an extra contribution to the EMT which is given by
\bea
\label{Timpr}
T^{\mu\nu}_I = - 2 \chi (\partial^\mu \partial^\nu - \eta^{\mu \nu} \Box) H^\dag H = - 2 \chi (\partial^\mu \partial^\nu - \eta^{\mu \nu} \Box) \left( \frac{h^2}{2} + \frac{\phi^2}{2} + \phi^+ \phi^- + v \, h\right) \, ,
\eea
the term of improvement, which can be multiplied by an arbitrary  constant ($\chi$). As mentioned above, it is mandatory to choose 
the value $\chi=1/6$ for any insertion of the EMT on the correlators of the Standard Model. These are found to be ultraviolet finite only if $T^{\mu\nu}_I$ is included \cite{Callan:1970ze,Coriano:2011zk,Freedman:1974gs}.\\
We will be dealing with the $T f \bar{f}$ vertex, where $T$ denotes the EMT and $f\equiv \nu_f$ a neutrino of flavour $f$, and work in the limit of zero mass of the neutrinos. The vertex, to lowest order, is obtained from the EMT of the neutrino. For instance, the explicit expression of the EMT for the (left-handed, $\nu\equiv \nu_L$) electron neutrino is given by
\begin{equation}
\begin{split}
T^{\nu^e}_{\mu\nu}
=& \frac{i}{4}\bigg\{\bar\nu^e\g_\mu\stackrel{\rightarrow}{\pd}_\nu\nu^e - \bar\nu^e\g_\mu\stackrel{\leftarrow}{\pd}_\nu\nu^e + \frac{2e}{\sin2\th_W}\bar\nu^e \g_\mu\frac{1-\g^5}{2}\nu^e Z_\nu \\
&- 2i\frac{e}{\sqrt{2}\sin\th_W}\bigg(\bar\nu^e \g_\mu \frac{1-\g^5}{2}e\,W^+_\nu
+ \bar e\g_\mu\frac{1-\g^5}{2}\nu^e\,W^-_\nu\bigg) \\
& + (\mu \leftrightarrow \nu) \biggr\} - \eta_{\mu\nu} \mathcal{L}_{\nu^e} \,,
\end{split}
\end{equation}
with 
\begin{equation}
\begin{split} 
\mathcal{L}_{\nu_e}
&= i\bar\nu^e\gamma^\mu\partial_\mu\nu^e + \frac{e}{\sin2\vartheta_{\text{W}}}\bar\nu^e \gamma^\mu\frac{1-\gamma^5}{2}\nu^e Z_\mu \\
&+ \frac{e}{\sqrt{2}\sin\vartheta_{\text{W}}}\bigg(\bar\nu^e \gamma^\mu \frac{1-\gamma^5}{2}e\, W^+_\mu
    + \bar e\gamma^\mu\frac{1-\gamma^5}{2}\nu^e\, W^-_\mu\bigg).\\
\end{split} 
\end{equation}
In momentum space, in the case of a massless fermion, the vertex takes the form
\beq
V^{(0) \mu\nu}=\frac{i}{4}\left( \gamma^\mu(p_1 +p_2)^\nu + \gamma^\nu(p_1 +p_2)^\mu -
2 \eta^{\mu\nu}(\slashed{p}_1+\slashed{p}_2)\right).
\eeq
while in the case of neutrinos we have
\beq
V_\nu^{(0)\mu\nu}=V^{(0)\mu\nu}\,P_L
\eeq
with $P_L=(1-\gamma_5)/2$ being the chiral projector.
We will denote with
\beq
\hat{T}^{(0) \mu\nu}=\bar u(p_2)V^{(0) \mu\nu}u(p_1),
\eeq
the corresponding invariant amplitude, a notation that we will use also at one-loop level in the electroweak expansion.  
We introduce the two linear combinations of momenta  $p = p_1 + p_2$ and $q = p_1 - p_2$ to express our results. %
It has been shown that the general $Tf\bar{f}$ vertex, for any fermion $f$ of the Standard Model, decomposes into six different contributions  \cite{Coriano:2012cr}, but in the case of a massless neutrino only three amplitudes at one-loop level are left, denoted as 
\bea
\label{hatT}
\hat T^{\mu\nu} =\hat T^{\mu\nu}_{Z} + \hat T^{\mu\nu}_{W} + \hat T^{\mu\nu}_{CT}. 
\eea
In the expression above, the subscripts indicate the contributions mediated by virtual $Z$ and $W$ gauge bosons, while $CT$ indicates  the contribution from the counterterm.

We show in Fig. \ref{diagrams} some of the typical topologies appearing in their perturbative expansion.\\
Two of them are characterized by a typical triangle topology, while the others denote terms where the insertion of the EMT and of the fermion field occur on the same point. The computation of these diagrams is rather involved and has been performed in dimensional regularization using the on-shell renormalization scheme. %
\begin{figure}[t]
\centering
\subfigure[]{\includegraphics[scale=0.7]{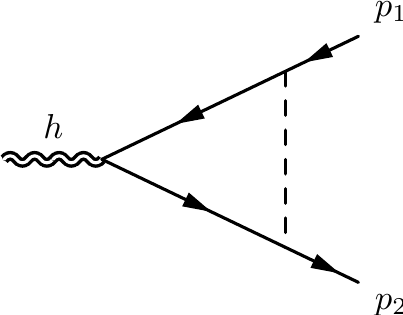}} \hspace{.5cm}
\subfigure[]{\includegraphics[scale=0.7]{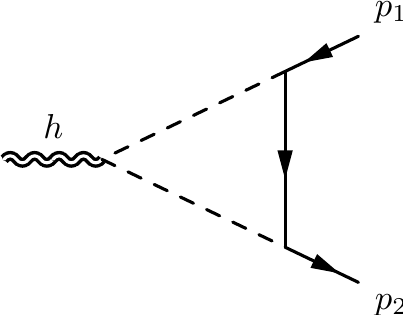}} \hspace{.5cm}
\subfigure[]{\includegraphics[scale=0.7]{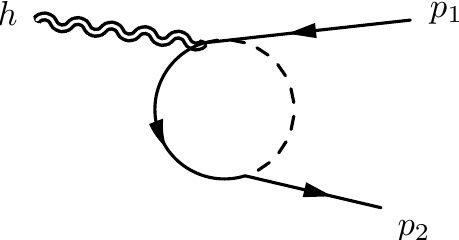}} \hspace{.5cm}
\subfigure[]{\includegraphics[scale=0.7]{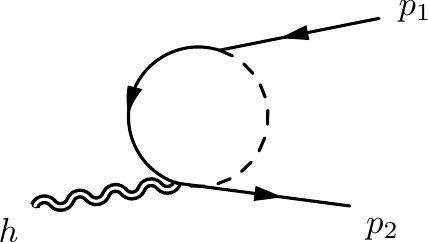}}
\caption{The one-loop Feynman diagrams of the neutrino vertex in a gravitational background. The dashed lines can be $Z$ and $W$.
\label{diagrams}}
\end{figure}
Neutrinos interactions, in the limit of massless neutrinos, involve only few of the structures of the $Tf\bar{f}$ tensor decomposition presented in \cite{Coriano:2012cr}. In this case we are left with only one tensor structure and hence only one form factor for each sector
\bea
\hat T^{\mu\nu}_Z &=& i \, \frac{G_F}{16 \pi^2 \sqrt{2}}   f^{Z}_1(q^2, m_Z) \, \bar u(p_2) \, O^{\mu\nu}_{C  1} \, u(p_1) \,, \nn \\ 
\hat T^{\mu\nu}_W &=& i \, \frac{G_F}{16 \pi^2 \sqrt{2}}  f^{W}_1(q^2, m_f, m_W) \, \bar u(p_2) \, O^{\mu\nu}_{C  1} \, u(p_1) \,, 
\label{expans}
\eea
where we have defined the vertex
\bea
\label{chiralbasis}
O^{\mu\nu}_{C  1} &=& \left( \gamma^\mu \, p^\nu + \gamma^\nu \, p^\mu \right) P_L.
\eea
The counterterms needed for the renormalization of the vertex can be obtained by promoting the counterterm Lagrangian of the Standard Model from a flat spacetime to the curved background, and then extracting the corresponding Feynman rules, as for the bare one. We obtain  
\bea
\label{TCT}
\hat T^{\mu \nu}_{CT}  = - \frac{i}{4}  \Sigma^L(0)\,\bar u(p_2)O^{\mu\nu}_{C 1}u(p_1),
\eea
where we have denoted with $\Sigma^L$ the neutrino self-energy
\bea
\Sigma^L(p^2) = \frac{G_F}{16 \pi^2 \sqrt{2}} \bigg[ \Sigma^L_Z (p^2) +  \Sigma^L_W(p^2) \bigg],
\eea
which is a combination of the self-energy contributions 
\bea
&&\Sigma^L_W (p^2) = - 4\bigg[ \left( m_f^2 + 2 m_W^2 \right) \mathcal B_1 \left( p^2, m_f^2, m_W^2 \right) + m_W^2 \bigg]\\
&& \Sigma^L_Z (p^2) = -2 m_Z^2\bigg[ 2 \, \mathcal B_1 \left( p^2, 0, m_Z^2 \right)  +1 \bigg]  \,,
\eea
with
\bea
\mathcal B_1 \left( p^2, m_0^2, m_1^2 \right) = \frac{m_1^2 -m_0^2}{2 p^2} \bigg[ \mathcal B_0(p^2, m_0^2, m_1^2) -  \mathcal B_0(0, m_0^2, m_1^2) \bigg] -\frac{1}{2} \mathcal B_0(p^2, m_0^2, m_1^2),
\eea
expressed in terms of the scalar form factor $\mathcal{B}_0$. We have denoted with 
$m_Z $ and $m_W$ the masses of the $Z$ and $W$ gauge bosons; with $q^2$ the virtuality of the incoming momentum of the EMT and $m_f$ is the mass of the fermion of flavor $f$ running in the loops. \\
The explicit expressions of the form factors appearing in (\ref{expans}) is given by 
\bea
f^Z_1&=&-2\,m_Z^2-\frac{4\,m_Z^4}{3\,q^2}+\left(2+\frac{7\,m_Z^2}{3\,q^2}\right)\,\mathcal A_0(m_Z^2)-\left(\frac{17\,m_Z^2}{6}+\frac{7\,m_Z^4}{q^2}+\frac{4\,m_Z^6}{q^4}\right)\,\mathcal B_0(q^2, 0, 0)\nn\\
&&+\frac{2}{3\,q^4}\,m_Z^2(2\,m_Z^2+q^2)\,(3m_Z^2+2q^2)\,\mathcal B_0(q^2, m_Z^2, m_Z^2)\nn\\
&&-\frac{4}{q^4}\,m_Z^6\,(m_Z^2+q^2)\,\mathcal C_0(0, m_Z^2, m_Z^2)-\frac{1}{q^4}m_Z^2\,(m_Z^2+q^2)^2(4\,m_Z^2+q^2)\,\mathcal C_0(m_Z^2, 0, 0),
\eea
with $\mathcal C_0$ denoting the scalar 3-point function, and with the form factor $f^W_1$ related to the exchange of the $W$' s given by 
\begin{align}
f^W_1&=\frac{m_f^2}{2}-4\,m_W^2+\frac{4}{3\,q^2}\,(m_f^4+m_f^2\,m_W^2-2\,m_W^4)-\frac{1}{3\,q^2}(m_f^2+2\,m_W^2)\left(\mathcal A_0(m_f^2)-\mathcal A_0(m_W^2)\right)\nn\\
&-\frac{2}{q^2}\Big(m_f^4+m_f^2\,m_W^2-2m_W^2\,(m_W^2+q^2)\Big)\,\mathcal B_0(0, m_f^2, m_W^2)+\frac{1}{6\,q^4}\Big(-24\,m_f^6-10\,m_f^4\,q^2\nn\\
&+m_f^2\,(72\,m_W^4+46\,m_W^2\,q^2+q^4)-2\,m_W^2\,(24\,m_W^4+42\,m_W^2\,q^2+17\,q^4)\Big)\,\mathcal B_0(q^2, m_f^2, m_f^2)\nn\\
&+\frac{1}{3\,q^4}\Big(12\,m_f^6+12\,m_f^4\,q^2+4\,m_W^2\,(2\,m_W^2+q^2)(3\,m_W^2+2\,q^2)\nn\\
&+m_f^2\,(-36\,m_W^4-16\,m_W^2\,q^2+q^4)\Big)\,\mathcal B_0(q^2, m_W^2, m_W^2)+2\Big(m_f^4+\frac{2}{q^4}\,(m_f^2-m_W^2)^3\,(m_f^2+2\,m_W^2)\nn\\
&+\frac{1}{q^2}\left(3\,m_f^6-4\,m_f^4\,m_W^2+5\,m_f^2\,m_W^4+4\,m_W^6\right)\Big)\,\mathcal C_0(m_f^2, m_W^2, m_W^2)\nn\\
&+\frac{1}{q^2}\Big(4\,m_f^8+m_f^6\,(q^2-4\,m_W^2)-2\,m_W^2\,(m_W^2+q^2)^2\,(4\,m_W^2+q^2)\nn\\
&-m_f^4\,(2\,m_W^2+q^2)\,(6\,m_W^2+q^2)\Big)\,\mathcal C_0(m_W^2, m_f^2, m_f^2)\nn\\
&+\frac{m_f^2}{q^2}\,(20\,m_W^6+25\,m_W^4\,q^2+6\,m_W^2\,q^4)\,\mathcal C_0(m_W^2, m_f^2, m_f^2).
\end{align}
Being the computations rather involved, the correctness of the results above has been secured by appropriate Ward identities, whose general structure has been discussed in \cite{Coriano:2011zk}. 
As an example, by requiring the invariance of the generating functional of the theory under a diffeomorphic change of the spacetime metric, one derives the following Ward identity
\bea
\label{WI}
q_{\mu} \, \hat T^{\mu\nu}& =& \bar u(p_2) \bigg\{ p_2^{\nu} \,  \Gamma_{\bar f f}(p_1) -  p_1^{\nu} \, \Gamma_{\bar f f}(p_2)
 + \frac{q_\mu}{2} \left( \Gamma_{\bar f f}(p_2) \, \sigma^{\mu\nu} - \sigma^{\mu\nu} \, \Gamma_{\bar f f}(p_1) \right) \bigg\} u(p_1) \,,
\eea
where $ \Gamma_{\bar f f}(p)$ is the fermion two-point function, diagonal in flavor space \cite{Coriano:2012cr}.
From this equation one obtains 
\bea
0 &=&  f^Z_1 - \frac{1}{4} \Sigma^L_Z(0) \nn \\
0 &=&  f^W_1 - \frac{1}{4} \Sigma^L_W(0),
\eea
which, as one can check, are identically satisfied by the explicit expressions of $f_Z$ and $f_W$ given above. \\
In the case of MeV neutrinos, the expressions of the two form factors simplify considerably, since the typical momentum transfer $q^2=-4 E^2 \sin^2(\theta/2)$ may be small. These expansions, in fact, are useful in the case of scattering and lensing of neutrinos far from the region of the event horizon, of the order of $10^3-10^6$ horizon units.  As we are going to see, an expansion in $q^2$ provides approximate analytical expressions of the $b_h(\alpha)$ relation, connecting the impact parameter to the angle of deflection $\alpha$, 
valid at momentum transfers which are smaller compared to the electroweak scale, i.e.  $q^2/m_W^2 \ll1$. 
We will come back to illustrate this point more closely in the following sections. \\
In these cases the expression of the renormalized $f_Z$ form factor takes the form
\bea
f^{Z\,(ren)}_{low\,q}=-\frac{11}{18} q^2,
\eea
while the $W$ form factor is slightly lengthier
\bea
f^{W\,(ren)}_{low\,q}=& -  \dfrac{q^2}{36\,(m_f^2-m_W^2)^4}\Biggl[ 5\,m_f^8-98\,m_f^6m_W^2+243\,m_f^4m_W^4-194\,m_f^2m_W^6+44\,m_W^8  \nn\\ 
&+6\left(10\,m_f^6m_W^2-15\,m_f^4m_W^4+2\,m_f^2m_W^6 \right) \ln \left( m_f^2 / m_W^2\right) \Biggr]  .
\eea

\section{Cross Sections for photons, massive fermions and scalars}

\begin{figure}[t]
\centering
\includegraphics[width=0.65\textwidth]{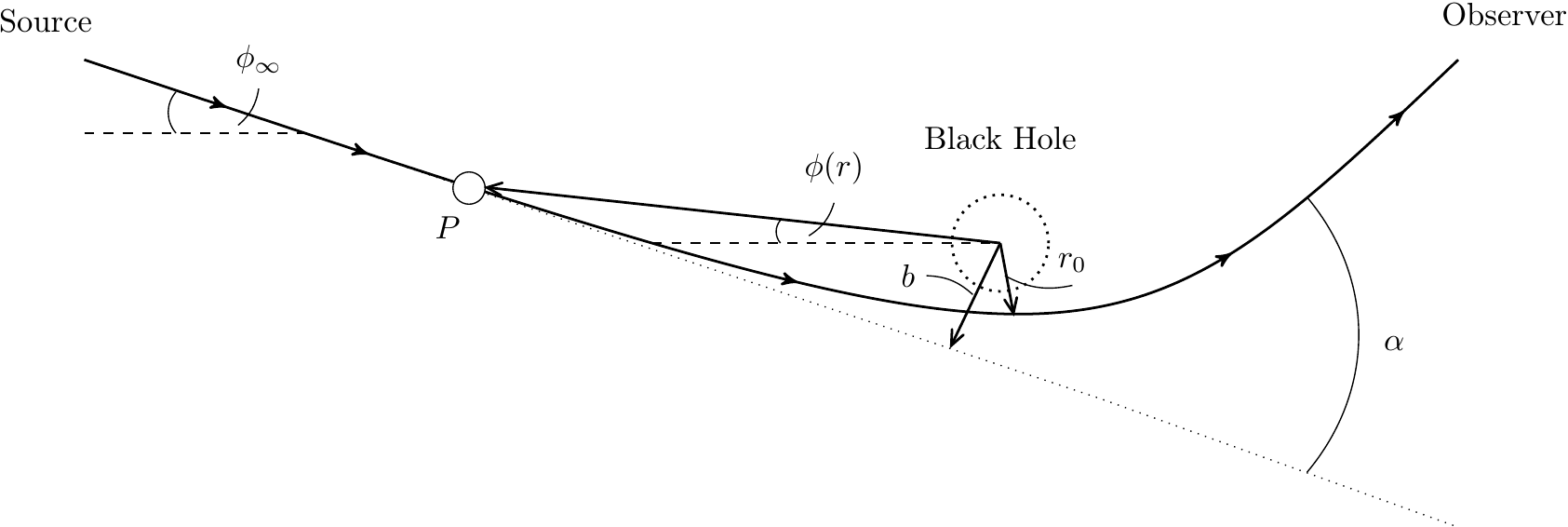}
\caption{The deflection of the trajectory of a massless particle $P$ approaching a black hole. }
\label{picx}
\end{figure}

Before coming to a discussion of the 1-loop effects in the scattering of neutrinos, we briefly summarize the result for the 
leading order cross sections for fermions, photons and scalars using in an external static background \cite{Accioly1} \cite{Coriano:2012cr,Coriano:2013iba}. We just recall that the scattering matrix element is written as 
\beq
i\mathcal{S}_{if}=-\frac{\kappa}{2}\int_ V d^4 x \langle p_2 | h_{\mu\nu}(x) T^{\mu\nu}(x)| p_1 \rangle 
\label{volume},
\eeq
where ${ V}$ is the integration volume where the scattering occurs, which gives
\bea
 \langle p_2 |h_{\mu\nu}(x) T^{\mu\nu}(x)| p_1 \rangle&=& h_{\mu\nu}(x) \bar{\psi}(p_2)V^{\mu\nu}\psi(p_1) e^{i q\cdot x}.
  \eea
 Denoting with $i$ and $f$ the initial and final neutrino, we have introduced plane waves normalized as 
\beq
\psi_{i}(p_{1})={\mathcal{N}_{i} }u(p_{1}), \qquad \mathcal{N}_{i}=\sqrt{\frac{1}{E_{1} V}}, \qquad \bar{u}(p_1)u(p_1)=1,
\eeq
and similarly for $\psi_f$,  while $V$ denotes a finite volume. The $E_1$ ($E_2$) are the energy of the incoming (outgoing) particle respectively.\\
In momentum space the matrix element is given by
 \bea
i \mathcal{S}_{fi}= -\frac{\kappa}{2} h_{\mu\nu}(q) \bar{\psi}(p_2)V^{\mu\nu}\psi(p_1)
=  -\frac{\kappa}{2}  h_{\mu\nu}(q)  \mathcal{N}_i\mathcal{N}_f \hat{T}^{\mu\nu}
\label{sfi}
\eea
in terms of the gravitational fluctuations in momentum space $h_{\mu\nu}(q)$. For a static external field the energies of the incoming/outgoing fermions are conserved ($E_1=E_2\equiv E$).  \\
The Fourier transform of $h_{\mu\nu} $ in momentum space is given by 
\bea
 h_{\mu\nu}(q_0,\vec{q})&=&\int d^4 x e^{i q\cdot x} h_{\mu\nu}(x),
 \eea
 which for a static field can be expressed as 
  \beq
 h_{\mu\nu}(q_0,\vec{q})=2 \pi \delta(q_0) h_{\mu\nu}(\vec{q} ),
 \label{h1}
 \eeq
 in terms of a single form factor $h_0(\vec{q})$
 \beq
 h_{\mu\nu}(\vec{q})\equiv h_0(\vec{q}) \bar{S}_{\mu\nu} \qquad \textrm{with}\qquad  h_0(\vec{q})\equiv \left(\frac{\kappa M}{2 \vec{q}^2}\right).
 \label{h2}
 \eeq
The squared matrix element in each case takes the general form
\begin{equation}
\label{eq:sfi}
\left|iS_{fi}\right|^2=\frac{\kappa^2}{16 V^2 E_1 E_2}\, 2\pi \delta(q_0)\, \,\mathcal{T} \,\frac{1}{2} \, \mathcal{J}^{\mu\nu\rho\sigma}(p_1,p_2)  \,h_{\mu\nu}(\vec{q}\,) \,h_{\rho\sigma}(\vec{q}\,), \,
  \end{equation} 
where $\mathcal{T}$ is the transition time. Specifically, in the case of a massive (Dirac) fermion one obtains 
\begin{equation}
\mathcal{J}^{\mu\nu\rho\sigma}_f(p_1,p_2)= \text{tr} \left[ (\slashed{p}_2+m) V_m^{\mu\nu}(p_1,p_2)( \slashed{p}_1+m) V_m^{\rho\sigma}(p_1,p_2) \right] \,,
\end{equation}
where the $V_m^{\mu\nu}$ vertex is in this case given by
\begin{equation}
\label{eq:hff}
V_m^{\mu\nu}(p_1,p_2)=\frac{i}{4} \Bigl( \gamma^{\mu} (p_1+p_2)^{\nu} + \gamma^{\nu}(p_1+p_2)^{\mu} - 2\eta^{\mu\nu} (\slashed{p}_1+\slashed{p}_2 - 2 m )\Bigr) \,
\end{equation}
which gives a cross section 
\begin{equation}
\label{eq:crosssecFerm}
\left.\frac{d \sigma}{d \Omega}\right|^{(0)}_f= \Biggl( \frac{G M}{\sin^2 (\theta/2)} \Biggr)^{\!2} \left( \cos^2\vartheta/2 + \frac{1}{4} \frac{m^2}{|\vec{p}_1|^2} + \frac{1}{4} \frac{m^4}{|\vec{p}_1|^4} + \frac{3}{4} \frac{m^2}{|\vec{p}_1|^2} \cos^2\vartheta/2      \right)\,.
\end{equation}
In the case of a neutrino, the corresponding cross section is obtained by sending the fermion mass $m$ of the related Dirac cross section to zero, giving  
\beq
\left.\frac{d \sigma}{d \Omega}\right|_\nu^{(0)} =\left(\frac{G M }{\sin^2\frac{\theta}{2}}\right)^2\cos^2\frac{\theta}{2}
\label{leading},
\eeq
which is energy independent. Notice that the inclusion of the chiral projector $P_L$ in the expression of the neutrino amplitude, which carries a factor $1/2$, makes the neutrino and Dirac cross sections coincide. The same $1/2$ factor, in the Dirac case, appears in the average over the two states of helicity, while the axial-vector terms induced by $P_L$ are trivially zero (see \cite{ChangCorianoGordon} for typical studies of polarized processes).\\
In the photon case one obtains
\begin{equation}
  \mathcal{J}^{\alpha\beta\rho\sigma}_{\gamma}(k_1,k_2)= \sum_{\lambda_1,\lambda_2} V^{\alpha\beta\kappa\lambda} (k_1,k_2) e_{\kappa}(k_1,\lambda_1) e_{\lambda}^{*}(k_2,\lambda_2) V^{\rho\sigma\mu\nu}(k_1,k_2)  e_{\mu} (k_2,\lambda_2) e^{*}_{\nu}(k_1,\lambda_1)\,,
  \end{equation}
  where $e_{\mu}$ denotes the polarization vector of the photon, with an interaction vertex which is given by
\begin{equation}
\label{eq:hAA}  
V^{\mu\nu\alpha\beta}(k_1,k_2)=i \bigg\{ \left( k_1 \cdot k_2 \right) C^{\mu\nu\alpha\beta}
+ D^{\mu\nu\alpha\beta}(k_1,k_2) \bigg\}\,,
\end{equation}
where
\begin{gather*}
 C_{\mu\nu\rho\sigma} = \eta_{\mu\rho}\, \eta_{\nu\sigma} +\eta_{\mu\sigma} \, \eta_{\nu\rho} -\eta_{\mu\nu} \, \eta_{\rho\sigma} \,,   \\
 D_{\mu\nu\rho\sigma} (k_1, k_2) = \eta_{\mu\nu} \, k_{1 \, \sigma}\, k_{2 \, \rho} - \biggl[\eta^{\mu\sigma} k_1^{\nu} k_2^{\rho} + \eta_{\mu\rho} \, k_{1 \, \sigma} \, k_{2 \, \nu}
  - \eta_{\rho\sigma} \, k_{1 \, \mu} \, k_{2 \, \nu}  + (\mu\leftrightarrow\nu)\biggr] \,.
\end{gather*} 
The cross section for a photon is then given by 
\begin{equation}
\label{eq:crsechVV}
\left.\frac{d \sigma}{d \Omega}\right|^{(0)}_\gamma= (G M)^2\cot^4 (\theta/2) \,.
\end{equation}
Finally, in the case of a scalar the relative expression is given by
\begin{equation}
  \mathcal{J}^{\alpha\beta\rho\sigma}_{s}(p_1,p_2)=V_s^{\alpha\beta}(p_1,p_2)V_s^{\rho\sigma}(p_1,p_2)\,,
  \end{equation} 
  with 
\begin{equation}
V^{\mu\nu}_s=-i\left\{ p_{1\,\rho} p_{2\,\sigma}  C^{\mu\nu\rho\sigma} - 2\chi \left[ \left( p_1+p_2 \right)^{\mu} \left( p_1+p_2 \right)^{\nu} - \eta^{\mu\nu} (p_1+p_2)^2 \right]  \right\} \,,
\end{equation}
where we have included the minimal and the term of improvement \cite{Coriano:2011zk}. For a conformally coupled scalar $\chi=1/6$.
The cross sections, in this case, are given by
\begin{equation}
\label{eq:crsechSS}
\left.\frac{d \sigma}{d \Omega}\right|^{(0)}_s= \left\{
\begin{array}{l}
(GM)^2\csc^4 (\theta/2)\qquad\chi=0\\
\left(\frac{ G M}{3}\right)^2\cot^4 (\theta/2)\qquad\chi=1/6
\end{array}
\right. \\
\end{equation}
We show in Fig. \ref{fig:TreeLevel} the expressions of these three cross sections at different energies, normalized by $1/(2 GM)^2$ and denoted as $\tilde{\sigma}$. In panel (a) we consider the scattering of a massive fermion, together with the massless limit, which applies in the neutrino case. We have included in (b) and (c) 
two enlargements of (a) which show how the massive and the massless cross sections tend to overlap for energies of the order of 1 GeV. In panel (d)  we show the cross sections for the photon ($s=1$), for the neutrino ($s=1/2$) and for the conformally coupled scalar
($s=0$). 

\begin{figure}[t]
\centering
\subfigure[]{\includegraphics[scale=.52]{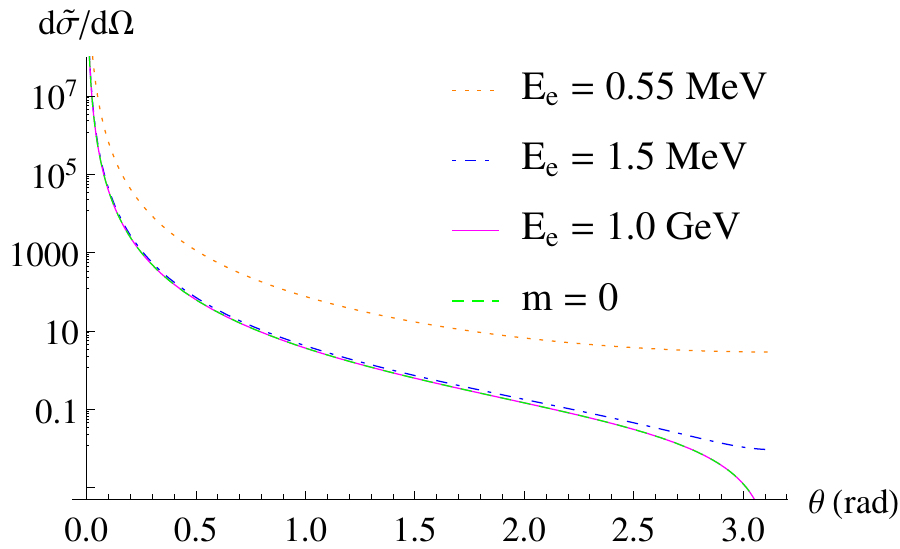}} \hspace{.5cm}
\subfigure[]{\includegraphics[scale=.52]{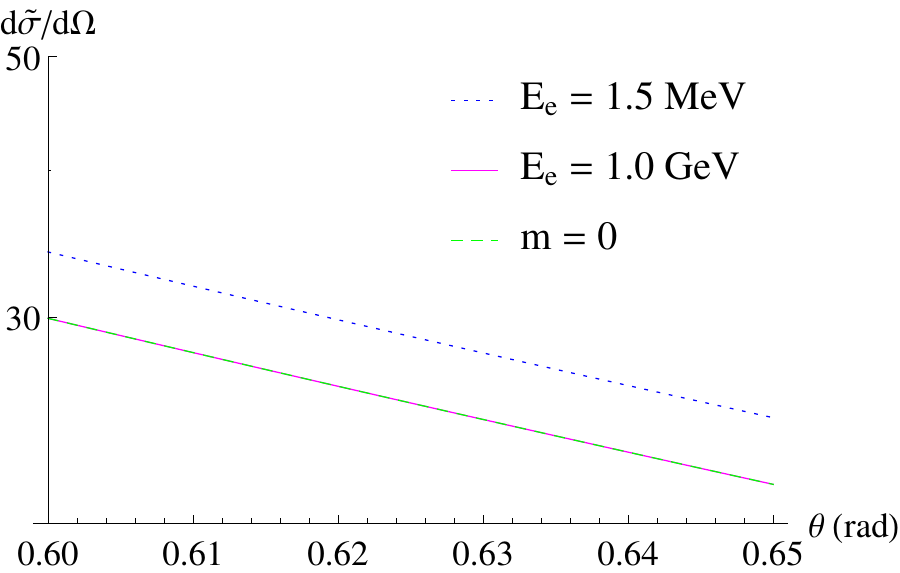}}
\subfigure[]{\includegraphics[scale=.52]{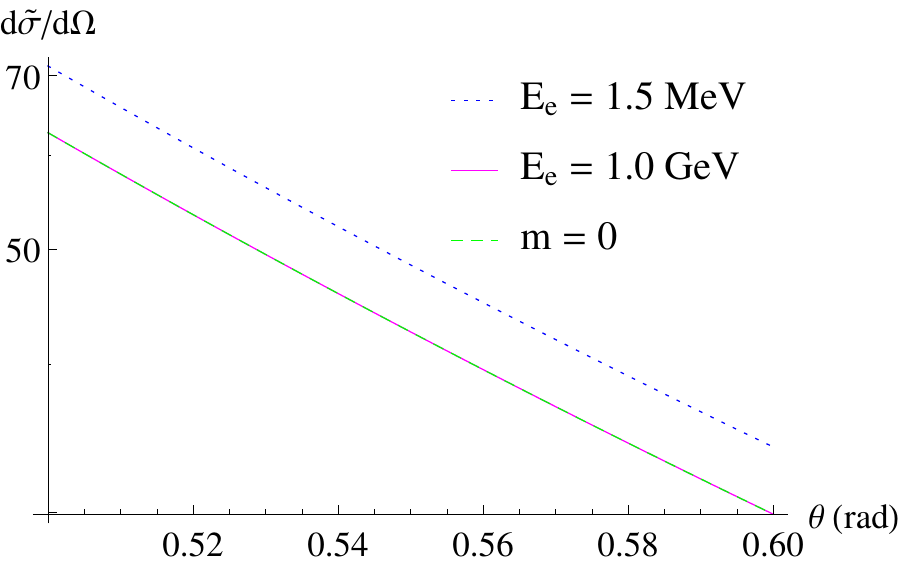}}
\subfigure[]{\includegraphics[scale=.65]{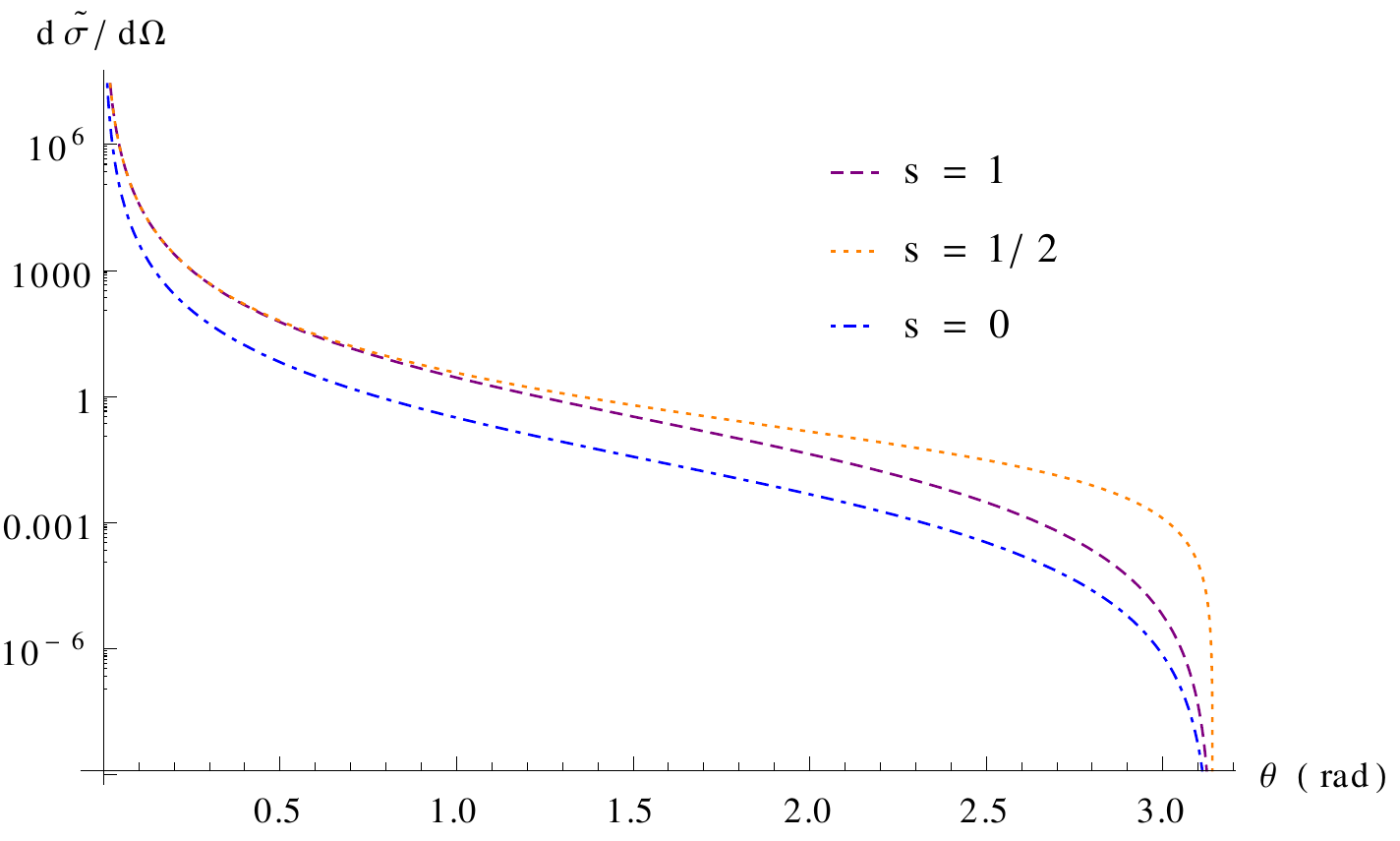}}
\caption{Normalized ($\tilde{\sigma}=\sigma/(2 G M)^2$) cross sections for massive and massless fermions. In the massive case $m$ is the electron mass (a). Two enlargements of (a) are in (b) and (c). Panel (d) 
shows the cross sections for photons ($s=1$), massless neutrinos ($s=1/2$) and conformally coupled scalars ($s=0$).  }
\label{fig:TreeLevel}
\end{figure}

\subsection{The neutrino cross section at 1-loop}
In the neutrino case, at 1-loop level, Eq. (\ref{leading}) is modified in the form 
\beq
\label{sigmaOL}
\frac{d \sigma}{d \Omega}=G^2M^2\frac{\cos^2\theta/2}{\sin^4\theta/2}\left\{1+\frac{4\,G_F}{16\,\pi^2 \sqrt{2}}  \left[ \, f_W^1(E,\theta) + f_Z^1(E,\theta) - \frac{1}{4} \Sigma_Z^L - \frac{1}{4} \Sigma_W^L\right] \right\}.
\eeq
In the massless approximation for the neutrino masses, loop corrections do not induce flavor transition vertices, such as those computed in  \cite{Coriano:2013iba}.\\
\begin{figure}[t]
\centering
\subfigure[]{\includegraphics[scale=.45]{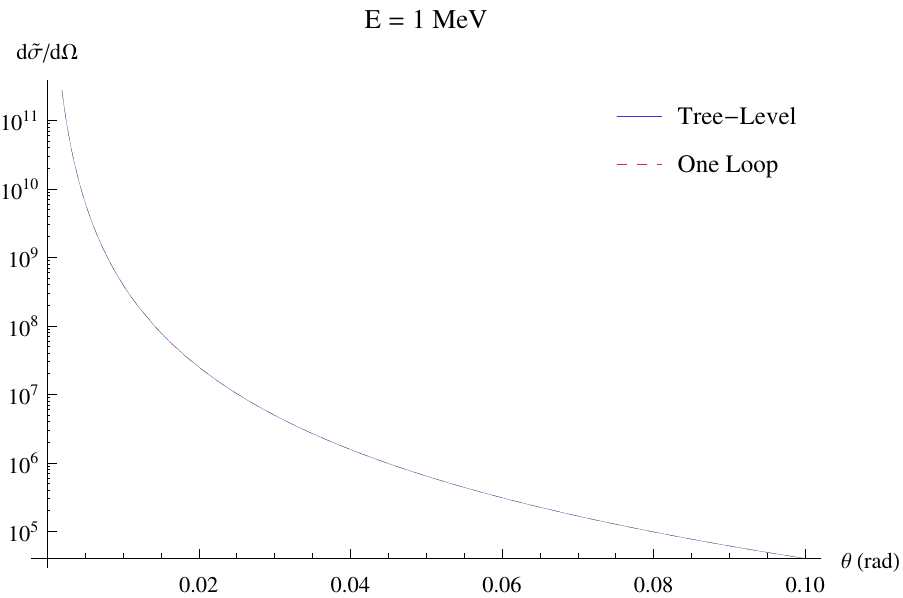}} \hspace{.5cm}
\subfigure[]{\includegraphics[scale=.35]{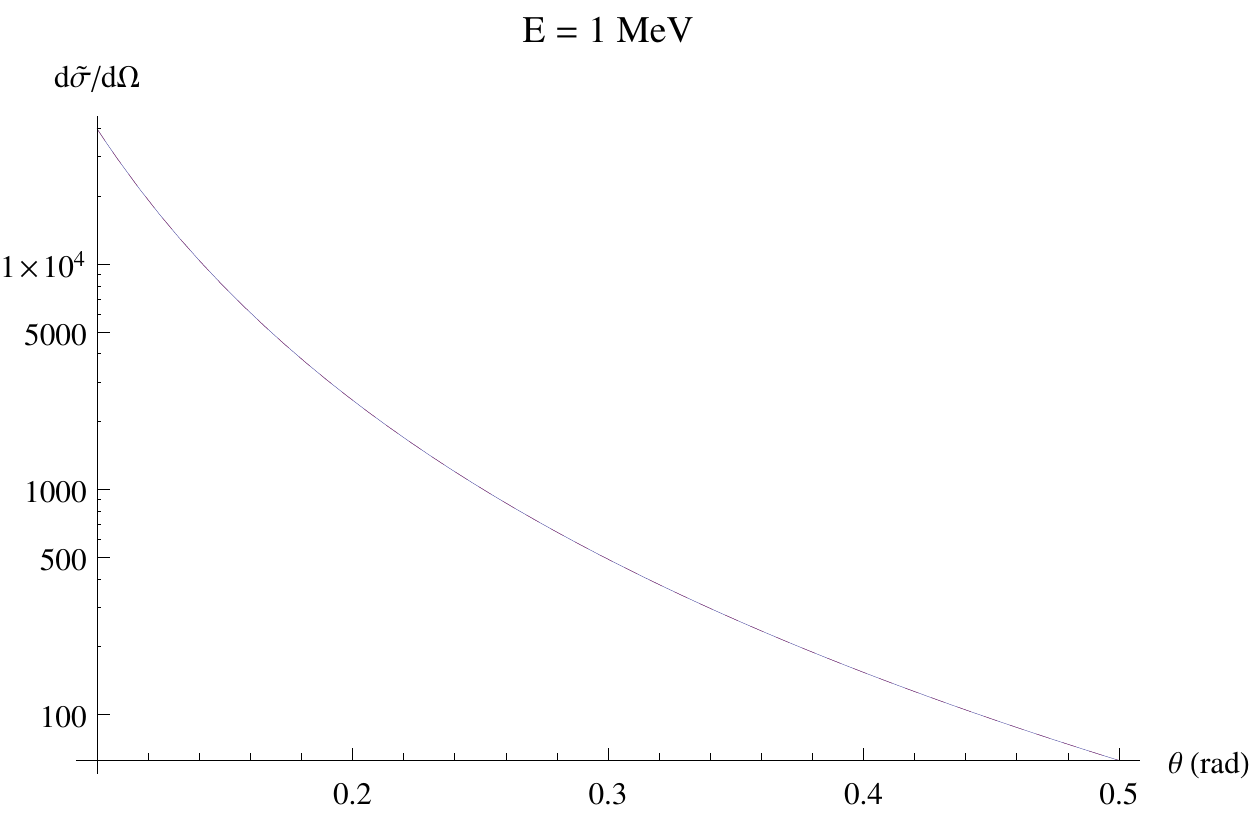}} \hspace{.5cm}
\subfigure[]{\includegraphics[scale=.35]{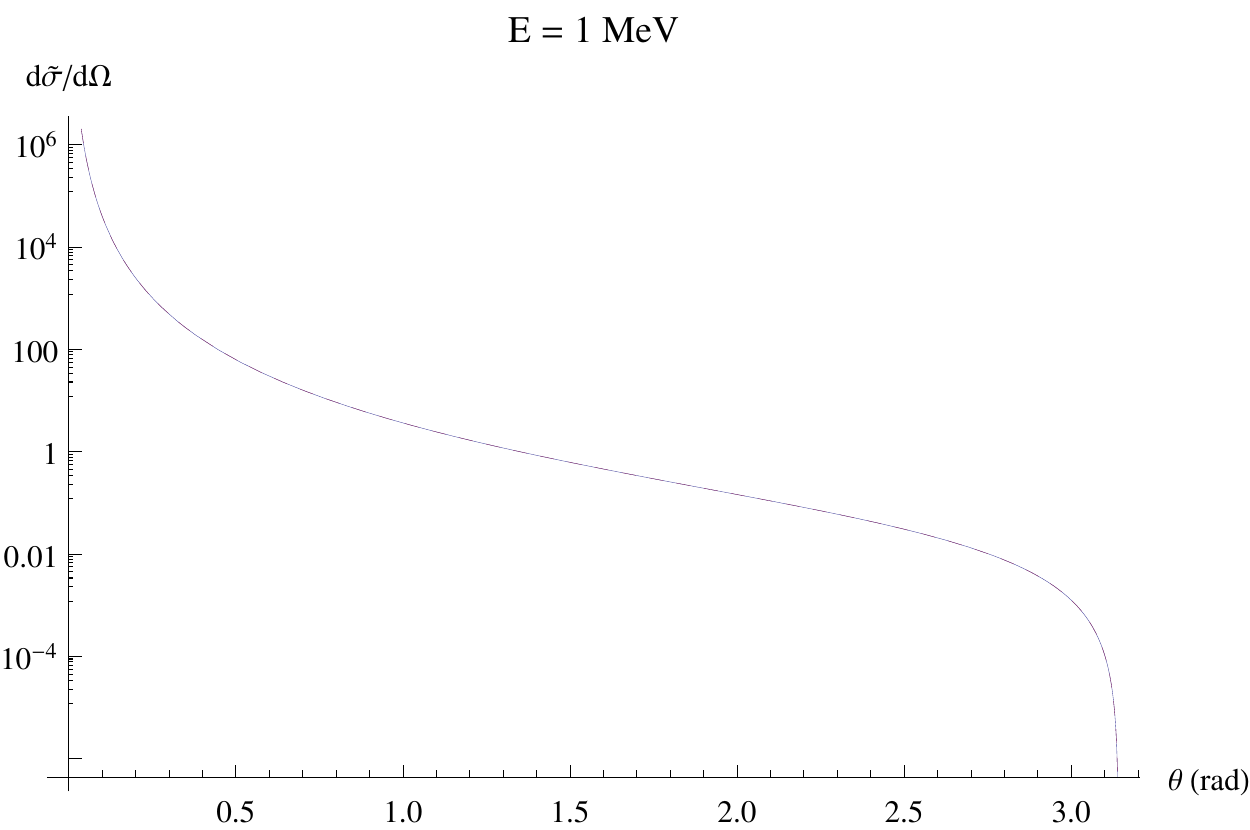}}
\caption{Differential cross section for MeV neutrinos in units of $r_s^2$, with $r_s$ the Schwarzschild radius.\label{OLMeV}}
\label{diff1}
\end{figure}
In the case of neutrinos of an energy $E$ in the MeV range, the expression above simplifies considerably and takes the form
\begin{align}
\frac{d\sigma}{d\Omega}=&\,G^2 M^2 \frac{\cos^2\theta/2}{\sin^4\theta/2} \Biggl\{ 1 + \frac{G_F}{\pi^2 \sqrt{2}} \Biggl[  \frac{11}{18}  +  \frac{1}{36\,(m_f^2-m_W^2)^4} \Biggl(  5\,m_f^8-98\,m_f^6m_W^2+243\,m_f^4m_W^4 \nonumber\\ 
 &  -194\,m_f^2m_W^6  +44\,m_W^8  +\,\, 6\, \Bigl( 10\,m_f^6m_W^2-15\,m_f^4m_W^4+2\,m_f^2m_W^6 \Bigr) \ln \dfrac{m_f^2}{m_W^2} \Biggr)  \Biggr] E^2 \sin^2\frac{\theta}{2}  \Biggr\} .
\end{align}
We show in Fig. \ref{diff1} three plots of the tree level and one-loop cross sections for an energy of the incoming neutrino beam of 1 MeV, for 2 different angular regions (plots $(a)$ and $(b)$), together with a global plot of the entire cross section (plot ($c$)) for the rescaled differential cross section $d\tilde{\sigma}/d\Omega\equiv 1/r_s^2\,\, d\sigma/d\Omega$. Notice that the tree-level and one-loop results are superimposed. We can resolve the differences between the two by zooming-in in some specific angular regions of the two results, varying the energy of the incoming beam. The result of this analysis is shown in Fig. \ref{diff2}, where in plots $(a)$ and $(b)$ we show the rescaled cross section $d\tilde{\sigma}/d\Omega$ as a function of the scattering angle 
$\theta$, for three values of the incoming neutrino beam equal to $1$ GeV,  $1$ TeV and $1$ PeV.  PeV neutrinos events are 
rare, due to the almost structureless cosmic ray spectrum, which falls dramatically with energy. They could be 
produced, though, as secondaries from the decays of primary protons of energy around the GZK
\cite{Greisen:1966jv, Zatsepin:1966jv} cutoff, and as such they are part of our analysis, which we try to keep as general as possible. \\ 
It is clear from these two plots that the tree-level and the one-loop result are superimposed at low energies, with a difference which becomes slightly more remarked at higher energies.
A similar behaviour is noticed in the cross section for scatterings at larger angles. Also in this case the radiative corrections tend to raise as the energy of the incoming beam increases. This  behaviour is expected to affect the size of 
the angle of deflection $\alpha$ as we approach the singular region of a black hole. In fact, $\alpha$ is obtained by integrating the semiclassical equation (\ref{semic}), introduced below, and large deviations are expected as the impact parameter $b_h$ reaches the photon sphere. As we are going to illustrate in the next sections, the $b_h(E, \alpha)$ relation is significantly affected by the behaviour of the cross section at large $\theta$ as $b_h\to 3/2 r_s$. This is the closest radial distance allowed to a particle approaching the black hole from infinite distance without being trapped. Therefore, these differences in $\tilde{\sigma}$ for large $\theta$ are going to render $b_h$ sensitive on the changes in energy of the neutrino beam for such close encounters of the neutrinos with a black hole.

\begin{figure}[t]
\centering
\subfigure[]{\includegraphics[scale=.6]{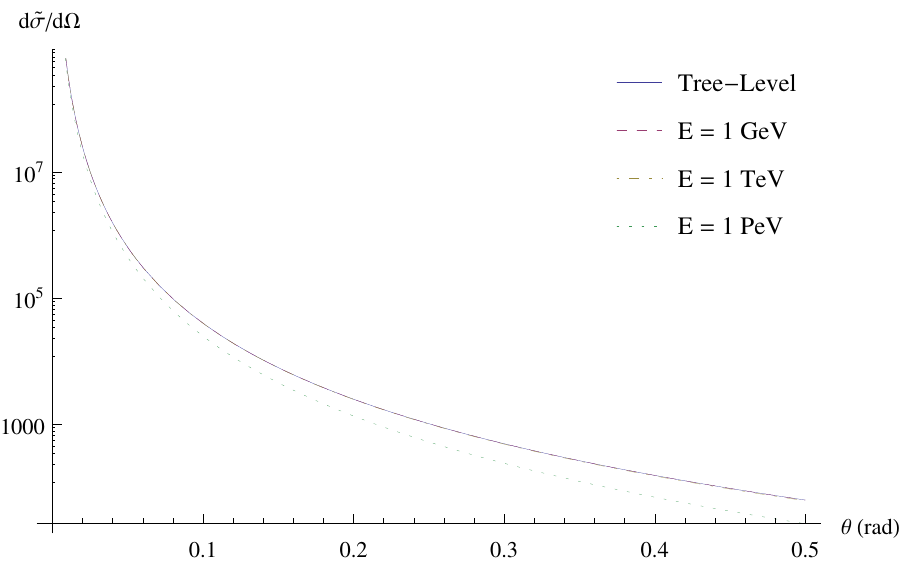}} \hspace{.5cm}
\subfigure[]{\includegraphics[scale=.4]{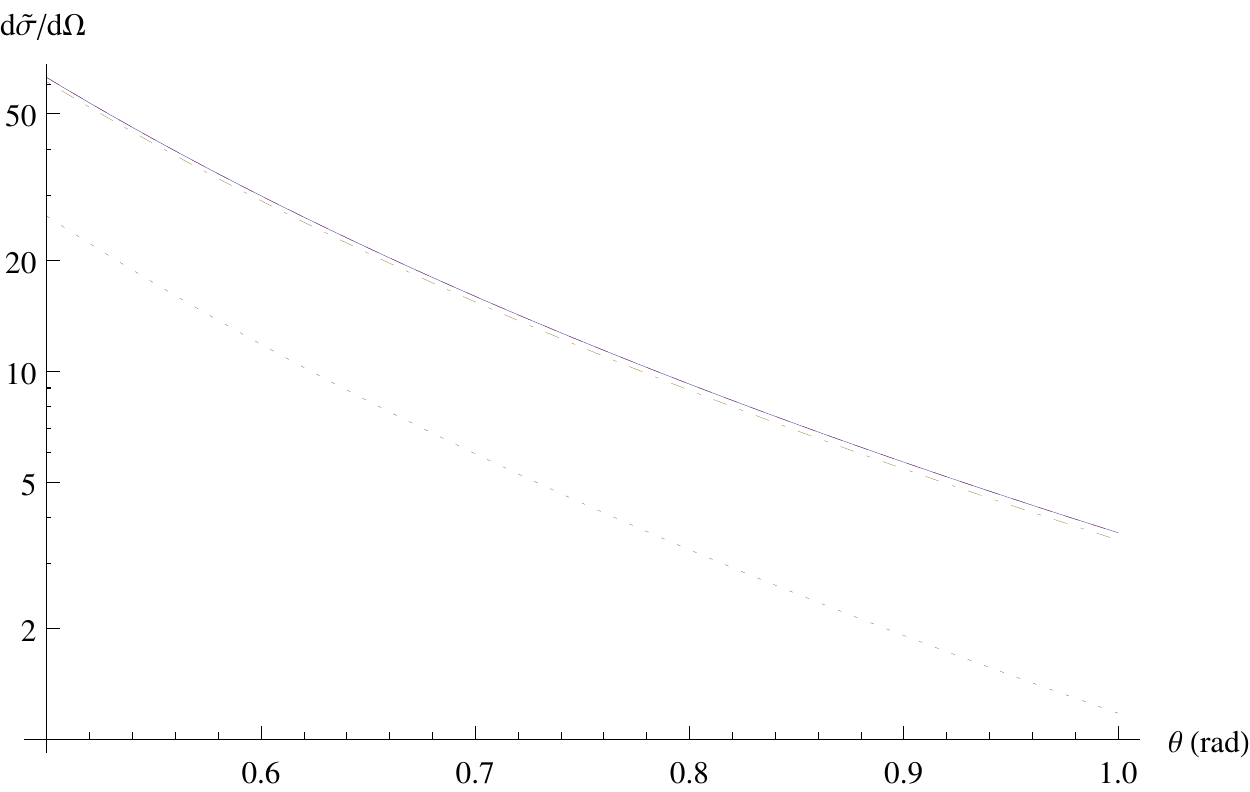}}
\caption{Differential cross section: tree level and one-loop contribution for a wide range of energies.\label{energies}}
\label{diff2}
\end{figure}

\section{Impact parameter formulation of the semiclassical scattering} 
As pointed out in previous studies \cite{Coriano:2014gia,Delbourgo:1973xe,Coriano:2013iba, Berends:1975ah}, the computation of the angle of deflection for a fermion or a photon involves a simple semiclassical analysis, in which one introduces the 
impact parameter representation of the specific classical cross section and equates it to the quantum one. The classical/semiclassical scattering process is illustrated in Fig.~\ref{picx}, with $\alpha$ denoting the angle of deflection.
By assuming that the incoming particle is moving along the $z$ direction, with the source localized at the origin, and denoting with $\theta$ the azimuthal scattering angle present in the quantum cross section, we have the relation
\beq
\frac{b}{\sin\theta}\vline\frac{d b}{d\theta}\vline=\frac{d \sigma}{d\Omega}
\label{semic}
\eeq
between the impact parameter $b$ and $\theta$, as measured from the $z$-direction. This semiclassical equation \cite{Delbourgo:1973xe, Berends:1975ah} allows to relate the quantum and the classical features of the interaction between the particle beam and the gravitational source. The explicit expression of $b(\alpha)$, at least for small deflection angles, which correspond to large values of the impact parameter, can be found either analytically, such as at Born level and, for small momentum transfers also at one-loop, but it has to be obtained numerically otherwise.     
The solution of (\ref{semic}) takes the general form 
\beq
b_h^2({\alpha})=b_h^2(\bar{\theta}) +2\int_{\alpha}^{\bar{\theta}} d\theta' \sin\theta' \frac{d \tilde\sigma}{d\Omega'}, 
\label{intg}
\eeq
with $b_h^2(\bar{\theta})$ denoting the constant of integration. The semiclassical scattering angle $\alpha$ is obtained from (\ref{intg}) as a boundary value of the integral in $\theta$ of the quantum cross section. As discussed in \cite{Coriano:2014gia}, the integration constant derived from (\ref{intg}) has to be set to zero (for $\bar\theta=\pi$) in order for the solution of (\ref{semic}) to match the classical GR result for a very large $b_h$.

In the case of a point-like gravitational source and of neutrino deflection, one obtains from (\ref{intg}) the differential equation
\bea
\frac{d b^2}{d\theta}&=&- 2  \left(\frac{G M }{\sin^2\frac{\theta}{2}}\right)^2\cos^2\frac{\theta}{2}\, \sin\theta.
\label{semic1}
\eea

Notice that the variation of $b$ with the scattering angle $\theta$ is negative, since the impact parameter decreases as $\theta$ grows, as we approach the center of the massive source. A comparison of this expression with the analogous relation in the photon case $(\gamma)$ shows that the two equations differ by a simple prefactor 
\bea
\frac{d b^2}{d\theta} =\frac{1}{\cos^2\frac{\theta}{2}}\frac{d b^2}{d\theta}\Big|_\gamma \qquad  \textrm{with} \qquad
\frac{d b^2}{d\theta}\Big|_\gamma = - 2 \,G^2 M^2\,\cot^4\frac{\theta}{2}\, \sin\theta.
\eea
The solution of (\ref{semic1}) takes the form 
\beq
\label{classb}
b^2(\alpha)=4\,G^2M^2\left(-1+\csc^2\frac{\alpha}{2}+2\ln\left(\sin\frac{\alpha}{2}\right)\right),
\eeq
and in the small $\alpha$ (i.e. large $b$) limit takes the asymptotic form  
\beq
b = G M\left(\frac{4}{\alpha} +\frac{\alpha}{3}(1+\ln\,8-3\ln\alpha)\right) + {\cal O}(\alpha^2)
\label{blocal}
\eeq
which allows us to identify the deflection angle as 
\beq
\alpha\sim 4 \frac{G M}{b}
\label{impact}
\eeq
in agreement with Einstein's prediction for the angular deflection.
This is the result expected from the classical (GR) analysis. The inversion of the asymptotic expansion (\ref{blocal}) generates the asymptotic behaviour
\bea
\alpha=\frac{2}{b_h} -\frac{2}{b_h^3}(\ln b_h +\frac{1}{3})+\frac{3}{b_h^5}(\ln^2 b_h-\frac{1}{5})+ \mathcal{O}(1/b_h^7)
\label{inv1}
\eea
which corresponds to the general functional form 
\beq
\label{genex}
\alpha= \frac{2}{b_{h}} + \sum_{k\geq1} \frac{a_{2k}}{b_{h}^{2k}} + \sum_{k\geq1} \frac{1}{b_h^{2k+1}} \left( a_{2k+1} + d_1 \ln b_h + d_2 \ln^2 b_h + \cdots +d_k\ln^k b_h \right) \,.       
\eeq
The analytic inversion of (\ref{blocal}), given by (\ref{inv1}), is very stable under an increase of the order of the asymptotic expansion over a pretty large interval of $b_h$, from low to very high values. Solutions (\ref{inv1}) and (\ref{genex}) can be obtained by an iterative (fixed point) procedure, 
which generates a sequence of approximations $\alpha_0\to\alpha_1\to\ldots\to\alpha_n$ to $\alpha(b_h$) implemented after a Laurent expansion of (\ref{blocal}) and the use of the initial condition $\alpha_0=2/b_h$. The approach can be implemented also at one-loop and with the inclusion of the post-Newtonian corrections, if necessary.

The logarithmic corrections present in (\ref{genex}) are a genuine result of the quantum approach and, as we are going to discuss below, are not present in the classical formula for the deflection. Radiative and post-Newtonian effects, not included in (\ref{inv1}), give an expression for 
$\alpha(b_h)$ which coincides with the form (\ref{genex}), with specific coefficients $(a_n, d_n)$ which are energy dependent. This is at the origin of the phenomenon of light dispersion (gravitational rainbow) induced by the quantum corrections, which is absent at classical level \cite{Accioly1}. 

Eq. (\ref{genex}) will play a key role in our proposal for the inclusion of the radiative corrections in the classical lens equation. Such equation will relate the angular position of the source in the absence of lensing, $\beta$, to $\alpha(b)$.\\
We give, for completeness, the analogous expressions in the case of the scalar and for a massive fermion. For a massless scalar we have the relation 
\bea
\alpha = \frac{2}{3\,b_h}-\frac{1}{b^3_h}\left(\frac{12\,\ln 3 - 1}{243}+\frac{4}{81}\ln b_h\right)+\mathcal O (1/b_h^5),
\eea
while for a massive fermion the corresponding expression becomes more involved and takes the form
\begin{align}\label{massf}
\alpha &= \frac{8\,E^4}{4\,E^4-2\,E^2m_f^2+m_f^4}\frac{1}{b_h}-\frac{1}{b_h^3}\left[\frac{8\,E^4}{3(2\,E^2-m_f^2)(4\,E^4-2\,E^2m_f^2+m_f^4)^2}\times \right.\nn\\
&\left.\times \left(m_f^6+8\,E^6(1+\ln 8)+E^4m_f^2\ln 64 -6E^4(4\,E^2+m^2)\ln \frac{2}{1-\frac{m_f^2}{2\,E^2}}\right)\right.\nn\\
&\left.+\frac{4\,E^4(4\,E^2+m_f^2)}{8\,E^6-8\,E^4m_f^2+4\,E^2m_f^4-m_f^6}\ln b_h\right]
+\mathcal O(1/b_h^5),
\end{align}
where $E$ and $m_f$ are the energy and the mass of the fermion respectively.  One can easily check that in the limit $E\gg m_f$ Eq.~(\ref{massf}) reproduce the formula for the massless fermion (neutrino). The angular deflection is much less enhanced in the scalar case compared to the remaining cases, showing a systematic difference respect to the classical prediction form Einstein's deflection integral. The angular deflection in the scalar case is significantly affected by the choice of $\chi$ the free coupling factor of a scalar field to the external curvature $R$.

\subsection{Bending at 1-loop}
Moving to the one-loop expression given in (\ref{sigmaOL}), we can derive an analytic solution of the corresponding semiclassical equation (\ref{semic}) for $b=b(E,\alpha)$, in the limit of small momentum transfers. For this reason we perform an expansion of (\ref{sigmaOL}) in $q^2/m_W^2$ up to $\mathcal{O}((q^2/m_W^2)^2)$ and solve (\ref{semic}) in this approximation for $b_h^2(E, \alpha)$, obtaining
\bea
\label{OLb}
b_h^2(E, \alpha)&=&\left[-1+\csc^2\frac{\alpha}{2}+2\ln\left(\sin\frac{\alpha}{2}\right)\right]+C_1(E)\left[1+\cos\alpha+4\ln\left(\sin
\frac{\alpha}{2}\right)\right]+C_2(E)\cos^4 \frac{\alpha}{2}\nn\\
&&+20\,D_2(E)\ln\left(\sin\frac{\alpha}{2}\right)-4\,F_2(E)\cos \alpha-8\,D_2(E)\cos \alpha\ln\left(\sin^2\frac{\alpha}{2}\right)-G_2(E)\cos 2\alpha\nn\\
&&-2\,D_2(E)\cos 2\alpha\ln\left(\sin^2\frac{\alpha}{2}\right)-E_2(E), 
\eea
with the coefficients $C,D$, $F$ and $G$ are functions of the energy and of the masses of the weak gauge bosons.
The impact parameter $b_h(\alpha)$, as shown in the same appendix, has a dependence on the angular deflection $\alpha$ which can be summarized by an expression of the form 
\bea
\label{btheta}
b_h(E, \alpha) &=& \frac{2}{\alpha}+c(E)\,\alpha+d(E)\,\alpha\,\ln(\alpha)+f(E)\,\alpha^3+g(E)\,\alpha^3\ln \alpha+h(E)\,\alpha^3\ln^2 \alpha + \mathcal{O}(\alpha^5)\nn\\
\eea
that we can invert in order to get $\alpha(E, b_h)$. This is given by
\beqa
\alpha(E, b_h)&=&\frac{2}{b_h}-\frac{1}{b_h^3}\Big[\big(2+4\,C_1(E)\big)\log b_h + \mathcal{A}(E)\Big] + \mathcal{O}(1/b_h^5)\nonumber \\
\mathcal{A}(E) &=& -2\,C_1(E)-C_2(E)+E_2(E)+4F_2(E)+G_2(E)+\frac{2}{3} .
\eea

\begin{figure}[t]
\centering
\subfigure[]{\includegraphics[scale=.45]{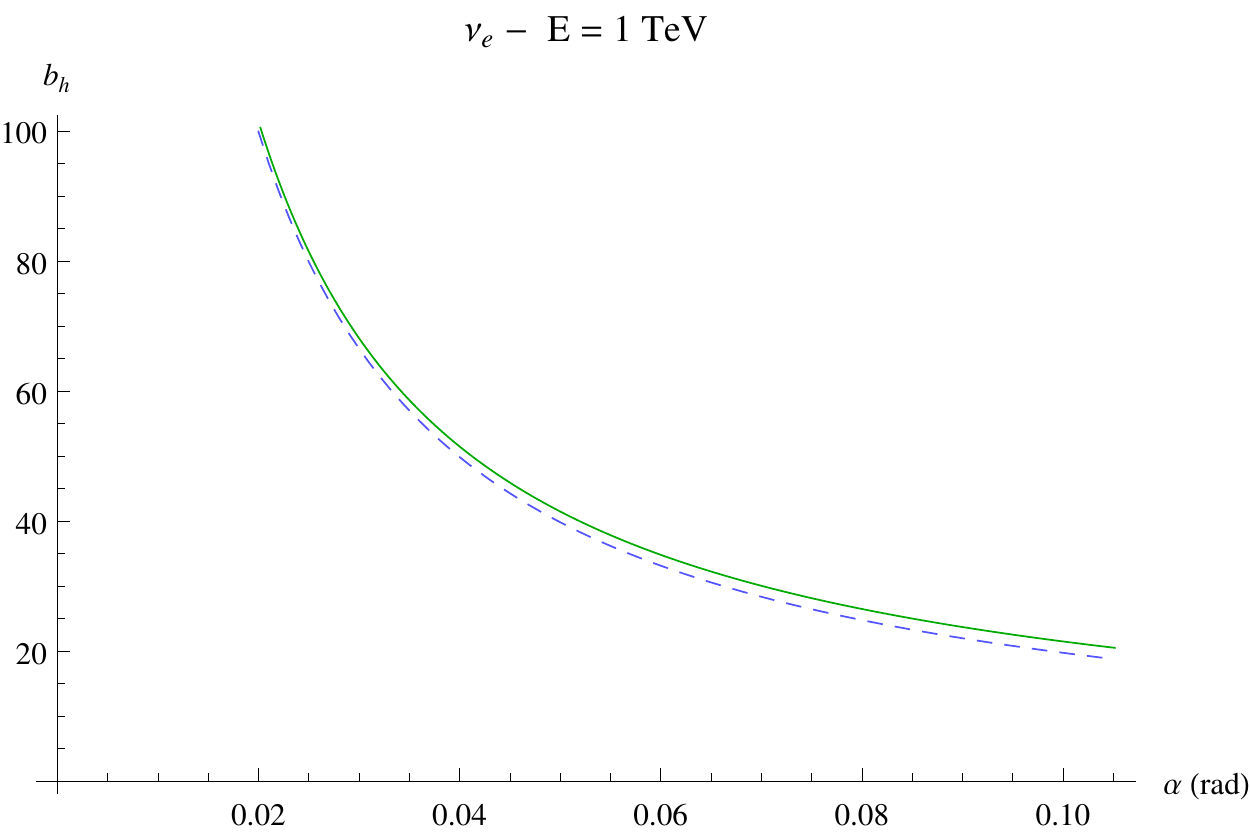}} \hspace{.5cm}
\subfigure[]{\includegraphics[scale=.6]{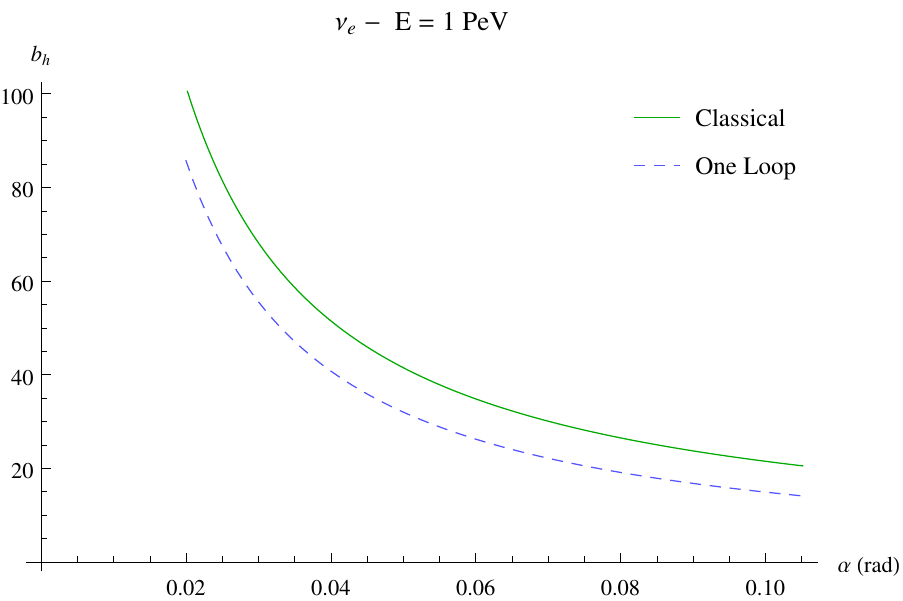}}
\caption{Plots of  the impact parameter $b_h$ versus $\alpha$, the angle of deflection, for $20<b_h<100$ for the classical and quantum solution. \label{bthetamev1}}
\end{figure}

We show in Fig.~\ref{bthetamev1} some plots of the impact parameter $b_h$ as a function of the deflection angle in a range closer to the horizon of a black hole, computed using the Newtonian approximation derived from the metric (\ref{SCH3}). The region involved covers the interval 
between 20 and 100 horizons. The numerical results refer to the GR solution and to the full one-loop prediction respectively.
 The classical expression and the quantum one start differing as we approach the value of $b_h\sim 20$, and are characterized by a certain dependence on the energy of the incoming beam. Shown are the plots corresponding to neutrinos of energies in the TeV and the PeV range respectively. In these regions the lensing is very strong, corresponding to $10^3$ arcseconds and larger. 
  As the neutrino (or the photon) beam gets closer to the photon sphere ($x_0=3/2 r_s$), which is the point of maximum approach, the angular deflection diverges. This is the impact parameter region where one expects the formation of relativistic images. The divergence can be parameterized by an integer $n$, with $\alpha_n=2 \pi n$, and $n$ tending to infinity. The integer is the winding number of the beam path around the photon sphere. In the external neighborhood of the point of closest approach the beam still escapes to infinity, forming an infinite set of images which are parameterized by the same integer $n$ \cite{Bozza:2001xd}. 
\section{$1/b^n$ contributions to the deflection }
It is interesting to compare the classical GR prediction for the deflection with the result of (\ref{genex}), by resorting to a similar expansion for the deflection integral. This has been studied quite carefully in the literature, especially in the limit of strong lensing \cite{Amore:2006pi,Keeton:2005jd}. The $1/b_h^n$ expansion 
has been shown to appear quite naturally in the post-Newtonian approach applied to the Einstein integral 
for light deflection. \\
We recall that Einstein's expression in GR is given by the integral 
\beq
\alpha(r_0)=\int_{r_0}^\infty dr \frac{2}{r^2}\left[ 
\frac{1}{r_0^2}\left(1 -\frac{2 M}{r_0}\right) -
\frac{1}{r^2}\left(1 -\frac{2 M}{r}\right)\right]^{-1/2} -\pi
\label{exactT}
\eeq
and can be re-expressed in the form 

\beq
\alpha=2\int_0^1 \frac{dy}{\sqrt{1 - 2 s - y^2 + 2 s y^3}} -\pi, 
\eeq
with the variable $s\equiv  r_s/ (2 r_0)$ being related to the ratio between the Schwarzschild radius and the distance of closest approach between the particle and the source, $r_0$.  Additional information on 
$\alpha(r_0)$ is obtained via an expansion of the integrand in powers of $s$ and a subsequent integration. This method shows that the result can be cast in the form 
\beq
\alpha(b_h)=\frac{a_1}{b_h} +\frac{a_2}{b_h^2} +\frac{a_3}{b_h^3} + \frac{a_4}{b_h^4} + \frac{a_5}{b_h^5}\ldots
\label{exp}
\eeq
with 
\beq
a_1=2, \qquad a_2=\frac{15}{16}\pi, \qquad a_3=\frac{16}{3}, \qquad a_4=\frac{3465}{1024}\pi, \qquad a_5=\frac{112}{5}.
\eeq
The coefficients $a_i$ differ from those given in \cite{Keeton:2005jd} (up to $a_7$) just by a normalization. They are obtained by re-expressing 
$s=s(r_0)$ in terms of the impact parameter $b_h$ using the relation 
\beq
\label{change}
b_h=x_0 \left( 1 -\frac{1}{x_0}\right)^{-1/2}
\eeq
between the impact parameter and the radial distance of closest approach, having redefined $x_0 \equiv r_0/(2 \, G M)$. This can also be brought into the form 
\bea
\label{form1}
x_0=\frac{2\,b_h}{\sqrt 3}\cos\left[\frac{1}{3}\cos^{-1}\left(-\frac{3^{3/2} }{2\,b_h}\right)\right].
\eea
An expression equivalent to (\ref{form1}) can be found in \cite{Coriano:2014gia}. Eq. (\ref{form1}) can be given in a $1/b_h$ expansion
\beq
x_0=b_h-\frac{3}{8 \, b_h} -\frac{1}{2\,
   b_h^2} -\frac{105}{128\, b_h^3} -\frac{3}{2\, b_h^4} +\mathcal{O}(1/b_h^5),
\label{inversion}
\eeq
which will turn useful below.

We can invert (\ref{exp}) obtaining the relation
\begin{align}
b_h(\alpha)=&\frac{2}{\alpha}+\frac{a_2}{2}+\frac{\alpha}{8}\big(2\,a_3-a_2^2\big)+\frac{\alpha^2}{16}\big(a_2^3-3a_2\,a_3+2\,a_4\big)\nn\\
&+\frac{\alpha^3}{128}\big(8\,a_5-16\,a_2\,a_4-8\,a_3^2+20\,a_2^2a_3-5a_2^4\big)+\mathcal{O}(\alpha^4),
\end{align}
which differs from (\ref{genex}) by the absence of logarithmic terms in the impact parameter $b_h$ and by the energy independence of the coefficients. The inclusion of the extra contributions mentioned above, 
in the classical GR expression, becomes relevant in the case of strong lensing.   
The inclusion of the additional $1/b_h^n $ terms in the expansion of the angular deflection can be extended to the case of a continuous distribution of sources/deflectors. This provides a simple generalization of the standard approach to classical lensing for such distributions.  

\section{Lens equations and $1/b^n$ corrections}
The standard approach to gravitational lensing in GR is based on an equation, derived from a geometrical construction, which relates the angular position of the image ($\theta_I$) to that of the source ($\beta$), with an intermediate angular deflection ($\alpha$) generated on the lens plane. In this section we are going to briefly review this construction, which is based on the asymptotic expression for the angular deflection ($\alpha\sim 2/b_h$), and discuss its extension when one takes into account more general expansions of $\alpha(b_h)$ of the form given by Eq. (\ref{exp}). The extension that we consider covers the case of a thin lens and concerns only the extra $1/b_h^n$ terms derived from classical GR. The discussion is preliminary to the analysis  of the next section, where we will consider the inclusion of the radiative effects, parameterized by 
(\ref{genex}), into the classical lens equation. 

\begin{figure} 
\centering 
\includegraphics[width=0.4\textwidth]{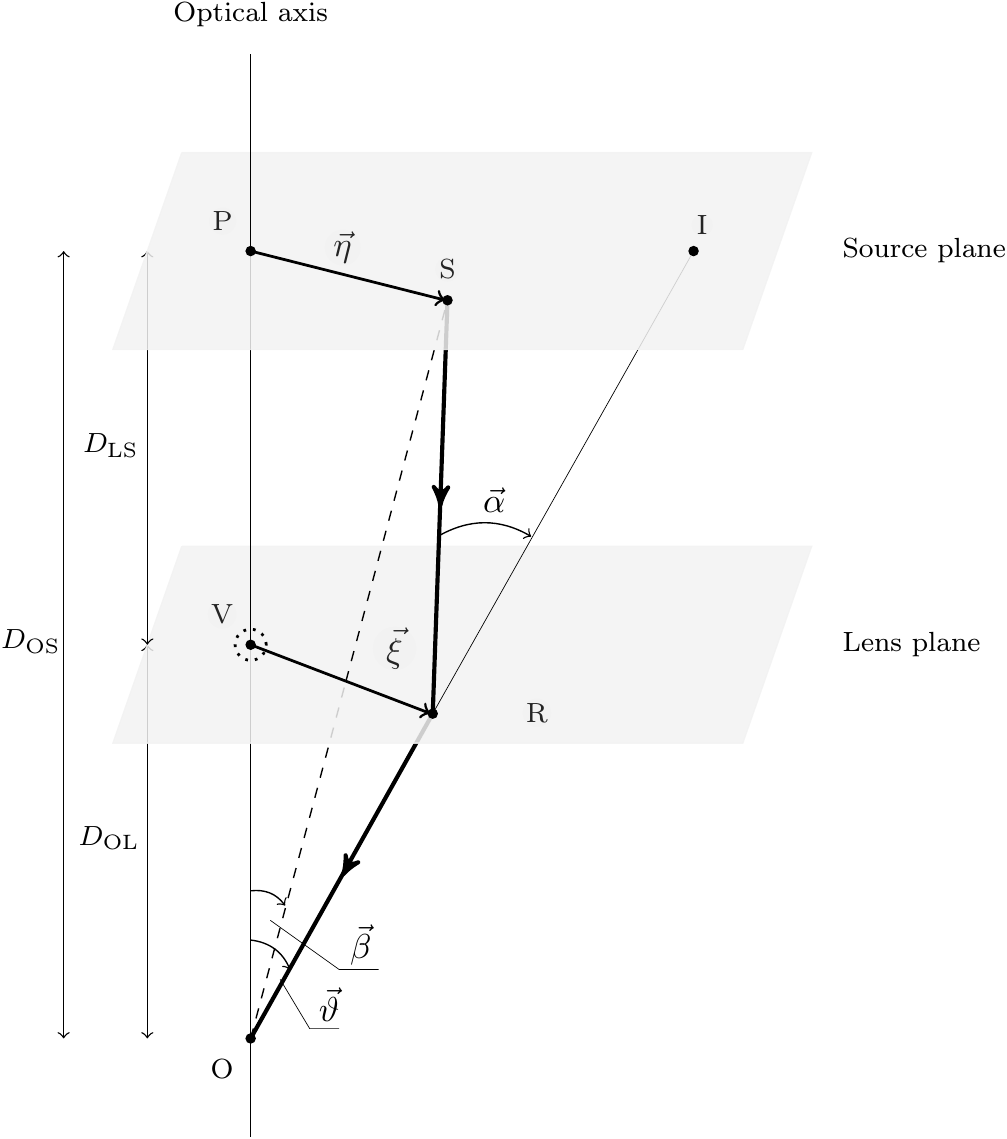}
\caption{Geometric construction of the lens for a continuous distribution of sources. Shown are the plane $S$ of the source distribution and the 
plane of the lens $L$. The line $OI$ identifies the direction at which the observer sees the image after the angular deflection $\alpha$.}
\label{lenspicture}
\end{figure}

\begin{figure}[t]
\centering
\includegraphics[width=0.35\textwidth]{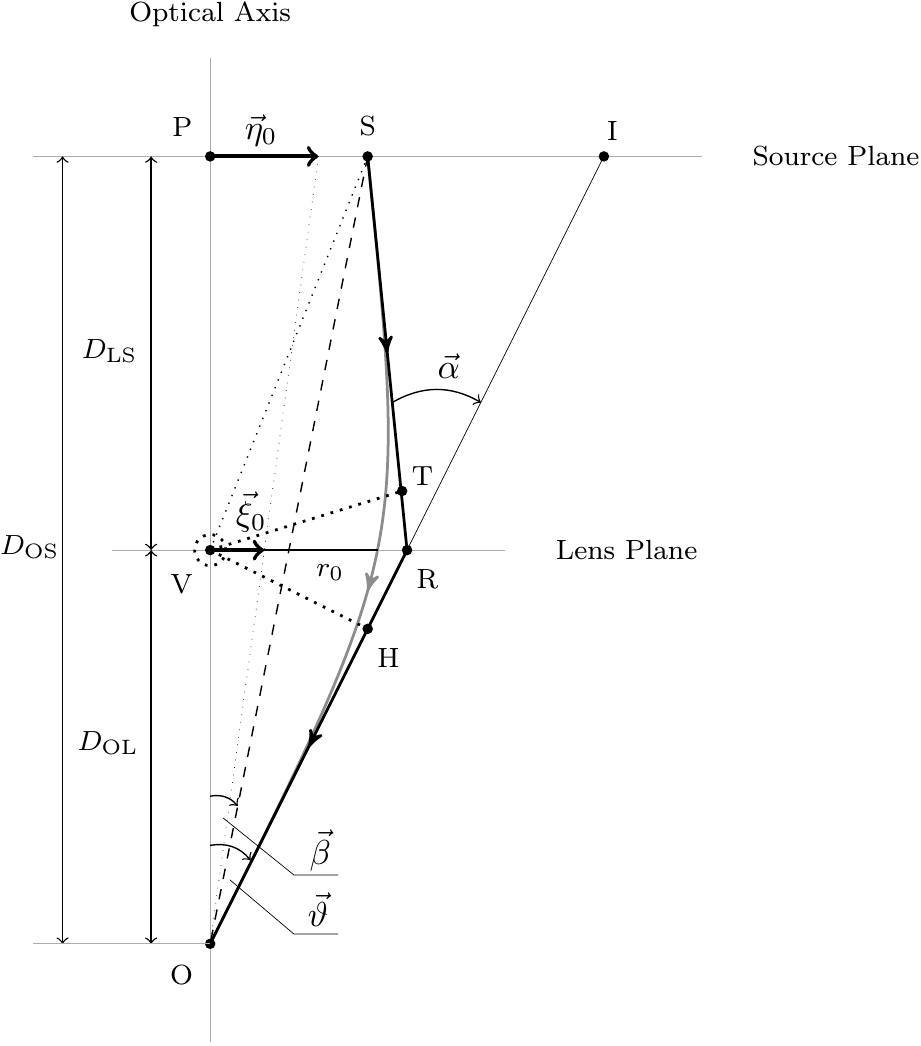}
\caption{The thin lens geometric construction where the source S, the lens V and the observer O lie on the same plane. Notice that figure is not to scale, since $D_{OL}$ and $D_{LS}$ are far larger than the length of $VR$.}
\label{geo}
\end{figure}

\begin{figure}
\centering 
\includegraphics[width=0.4\textwidth]{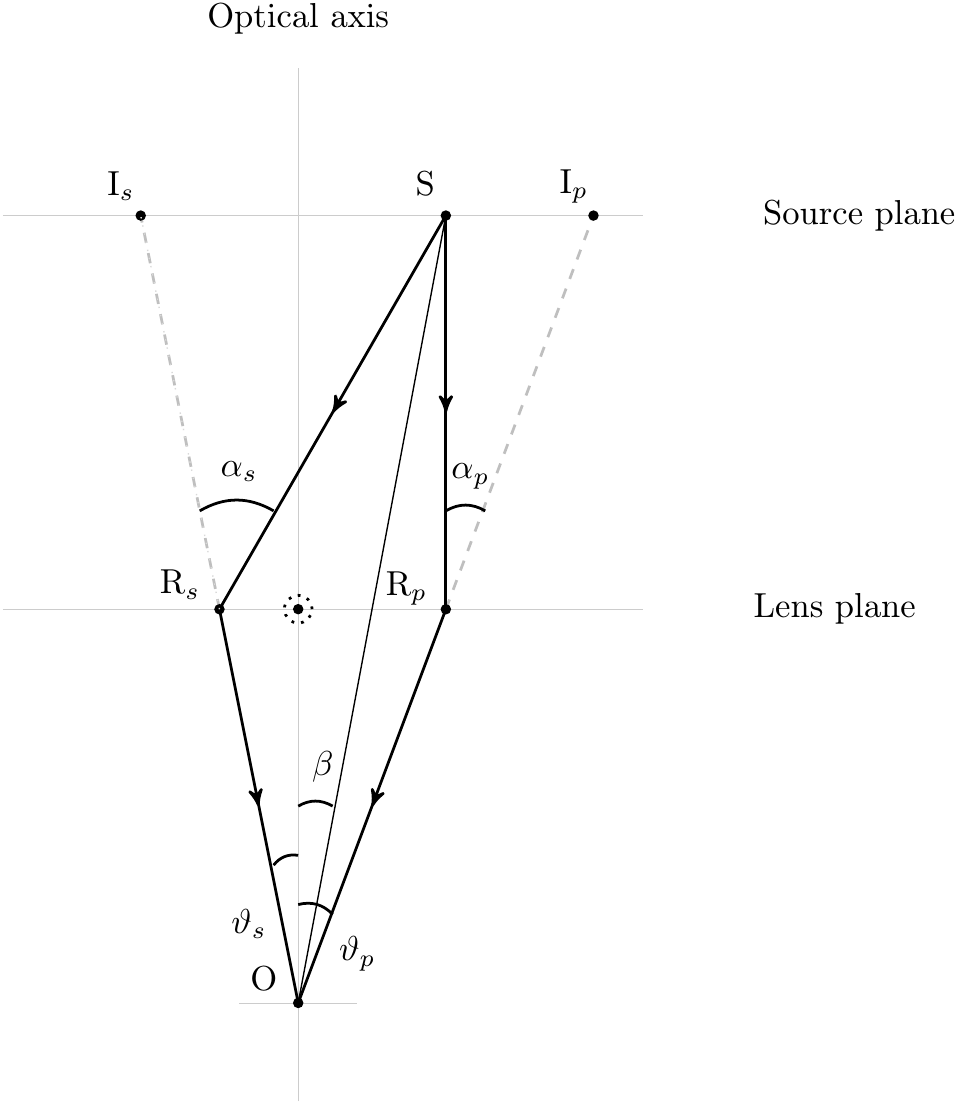} 
\caption{Geometrical construction for the primary I$_p$ and secondary I$_s$ images generated by the two geodesics of the isotropic emission. 
Shown are the source S, the lens, represented by the dotted circle, the observer O and the primary $I_p$ and secondary $I_s$ angular positions involved in the discussion.}
\label{prim}
\end{figure}

\subsection{The lens geometry}
We show in Fig.~\ref{lenspicture} the lens geometry in the case of a continuous distributions of sources and deflectors. A simplified picture of the geometry, with pointlike source and deflector is shown in Fig.~\ref{geo}. 
We indicate with $\vec{\beta}$ the oriented angle between the optical axis $(OP)$ (taken as the $z$ axis) and the unlensed direction of the source $(OS)$. $\vec{\theta}_I$ denotes the angle formed by the visual line of the image $(OI)$ with the optical axis. We also denote with $D_{OL}$  the distance between the observer and the lens plane; with $D_{LS}$ the distance between the lens plane and the source plane and with $D_{OS}$ the distance of the source plane from the observer. $\hat{\alpha}$ is the (oriented) angle of deflection, measured clockwise as all the other angles appearing in the geometrical construction. We also introduce the relations, valid for $D_{LS}, D_{OL}$ much larger than the size of the lens, typical of a linear lens,
\beq
\vec{\eta}\equiv \vec{PS}=\vec{\beta} D_{OS} \qquad \qquad \vec{SI}=\hat{\vec{\alpha}} D_{LS} \qquad \qquad \vec{PI}=\vec{\theta}_I D_{OS}.
\eeq
 The thin lens equation follows from the approximate geometrical relation
 \beq
\label{lin}
\vec{PI}=\vec{PS} +\vec{SI} \qquad \textrm{ i.e.} \qquad \vec{\beta}=\vec{\theta}_I -\hat{\vec{\alpha}}\frac{D_{LS}}{D_{OS}}.
\eeq
Denoting with $\vec{\xi}$ a 2-D vector in the lens plane, it is convenient to introduce two scales $\eta_0$ and $\xi_0$ defined as
\bea
\vec{\eta}=\eta_0\,\vec y\qquad \qquad\vec \xi\equiv\vec{VR}=\xi_0\vec x\qquad \qquad \frac{\eta_0}{\xi_0}=\frac{D_{OS}}{D_{LS}}.
\eea
Using the lens equation in the geometric relation
\bea
\frac{|\vec{PI}|}{|\vec{VR}|}=\frac{D_{OS}}{D_{OL}},
\eea
we find the relation
\bea
\label{thin}
\vec y=\vec x - \hat{\vec \alpha} \frac{D_{LS}\,D_{OL}}{D_{OS}\,\xi_0}\equiv \vec x -\vec \alpha
\qquad \qquad \textrm{with} \qquad\qquad\vec \alpha=\hat{\vec \alpha} \frac{D_{LS}\,D_{OL}}{D_{OS}\,\xi_0},
\eea
which defines the thin lens equation. It is possible to give a simpler expression to the equation above if we go back to (\ref{lin}) and perform simple 
manipulations on the angular dependence. On the lens plane (Fig.~\ref{geo}) the equation takes the scalar form 
\bea
\label{thin1}
\beta=\theta_I- \alpha \frac{D_{LS}}{D_{OS}}, 
\eea
which can be extended to the case of stronger lensing by the inclusion of the contributions of the $1/b^n$ corrections in $\alpha(b)$. Use of the Einstein relation $\alpha=4 G M/b$ and of the relation $b\sim\theta_I D_{OL}$ 
brings (\ref{thin1}) into the typical form
\beq
\beta=\theta_I -\frac{\theta_E^2}{\theta_I} \qquad   \theta_E^2 =\frac{D_{LS}}{D_{OS}}\frac{4 G M}{D_{OL}},
\eeq
which defines the thin lens approximation, with $\theta_E$ being the Einstein radius. For a source $S$ aligned on the optical axis together with the deflector and the observer $O$ (see Fig.~\ref{prim}) - which is defined by the segment connecting the observer, the lens and the plane of the source (with $\beta=0$) - the images will form radially at an opening $\theta_I=\theta_E$ and appear as a circle perpendicular to the lens plane. For a generic $\beta$, instead, the primary and secondary image solutions are given by the well-known expressions 
\beq
\label{img}
\theta_{I {\pm}}=\frac{\beta}{2} \pm \frac{1}{2}\left(\beta^2 + 4 \theta_E^2\right)^{1/2}.
\eeq
It is quite straightforward to extend this derivation with the inclusion of the $1/b^n$ corrections in the $\alpha(b)$ relation and test their effect numerically \cite{Keeton:2005jd}. This is part of a possible improvement of the ordinary (quadratic) thin lens equation which can be investigated more generally in conditions of strong lensing. In that case one can also adopt an equation which includes deflections of higher orders, as we will discuss in the following sections. For the moment we just mention that the inclusion of the higher order $1/b^n$ contributions given by (\ref{exp}) modifies (\ref{thin1}) into the form 
\beq
\label{thin2}
\beta=\theta_I - \frac{\theta_E^2}{\theta_I}-\sum_{n\geq 2} \frac{\theta_E^{(n)}}{\theta_I^n}, 
\eeq
with 
\beq
\theta_E^{(n)}\equiv r_s^n a_n \frac{D_{LS}}{D_{OS} D_{OL}^n}.
\eeq
Another observable that we will investigate numerically is going to be the lens magnification. For this purpose we recall that light beams are subject to deflections both as a whole but also locally, due to their bundle structure. Rays which travel closer to the deflector are subject to a stronger deflection compared to those that travel further away. This generates a difference in the solid angles under which the source is viewed by the observer in the unlensed and in the lensed cases. In the simple case of an axi-symmetric lens the ratio between the two solid angles can be defined in the scalar form 
\bea
\mu=\left|\left(\frac{\partial\beta}{\partial\theta_I}\frac{\sin\beta}{\sin\theta_I}\right)\right|^{-1}.
\eea
In the case of a thin lens (\ref{thin}), the analogous expression is given by 
\bea
\mu^{(0)}_\pm\equiv\left(\frac{\partial\beta}{\partial\theta_I}\frac{\beta}{\theta_I}\right)^{-1}.
\eea
For this lens the analysis simplifies quite drastically. Using the expression of the two images $\theta_{I\pm}$ given in (\ref{img}) one obtains  the simple expression for the primary and secondary images
\beq
\mu_{\pm}=\pm\left(1 -\left(\frac{\theta_E}{\theta_{I\pm}}\right)^4\right)^{-1},
\label{magni}
\eeq
where the Einstein angle is defined as usual
\bea
\theta_E=\sqrt{4\,G M\frac{x}{D_{OL}}}\qquad\qquad \textrm{with}\qquad \qquad x=\frac{D_{LS}}{D_{OS}}.
\eea
It is convenient to measure the angular variables in terms of the Einstein angle $\theta_E$, as $\bar{\beta}\equiv \beta/\theta_E$, 
$\bar{\theta}\equiv \theta_I/\theta_E$, with 
\beq
\bar{\theta}_{I \pm}=\frac{\bar{\beta}}{2} \pm \sqrt{1 + \frac{\bar{\beta}^2}{4}},
\eeq
then the total magnification takes a rather simple form 
\beq
\mu\equiv \mu_+  + \mu_-=\frac{2 + \bar{\beta}^2}{\bar\beta\sqrt{4 + \bar\beta^2}}.
\eeq
This equation is commonly used to calculate the light curve in the microlensing case. We refer to \cite{Mao:2008kp} for a short review on this point.
\subsection{Nonlinear effects in strong deflections}
In conditions of strong lensing, the linear approximations in the trigonometric expressions are not accurate enough and one has to turn to a fully nonlinear description of the geometry, expressed in terms of the angular variables which are involved. 
  We illustrate this point by taking as an example a typical lens equation, which in our case is given by the Virbhadra-Ellis construction (VE) \cite{Virbhadra:1999nm}. \\
 Following Fig.~\ref{geo}, we recall that the VE lens equation is based on the geometrical relation \cite{Virbhadra:1999nm,Bozza:2008ev} 

\begin{figure}[t]
\centering
\subfigure[]{\includegraphics[scale=0.6]{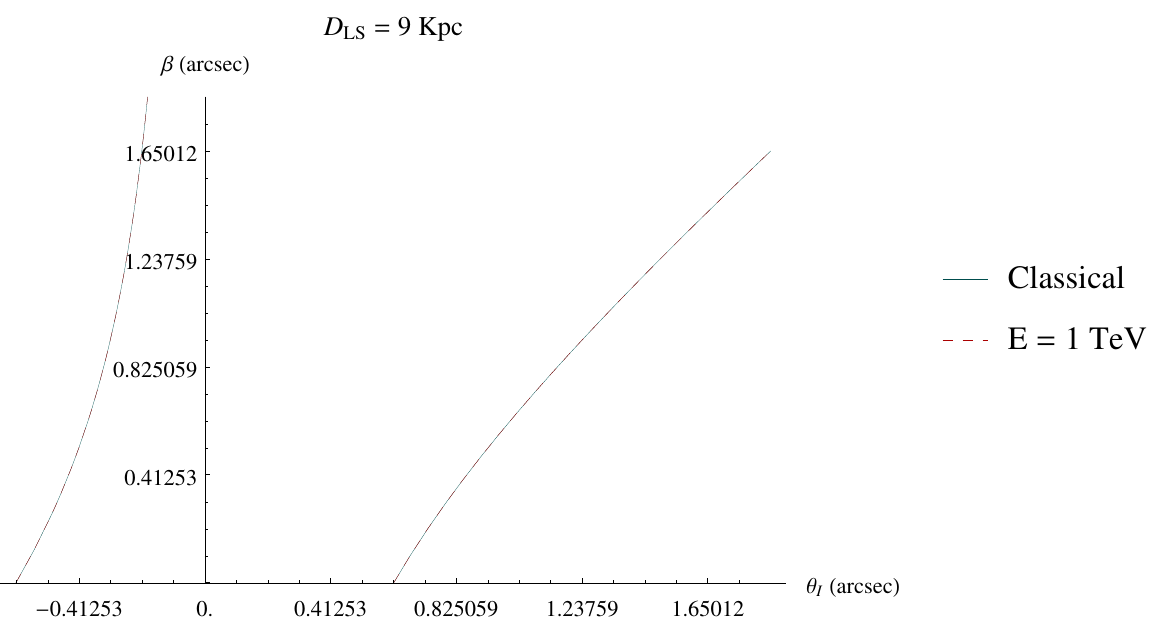}}\hspace{.5cm}
\subfigure[]{\includegraphics[scale=0.6]{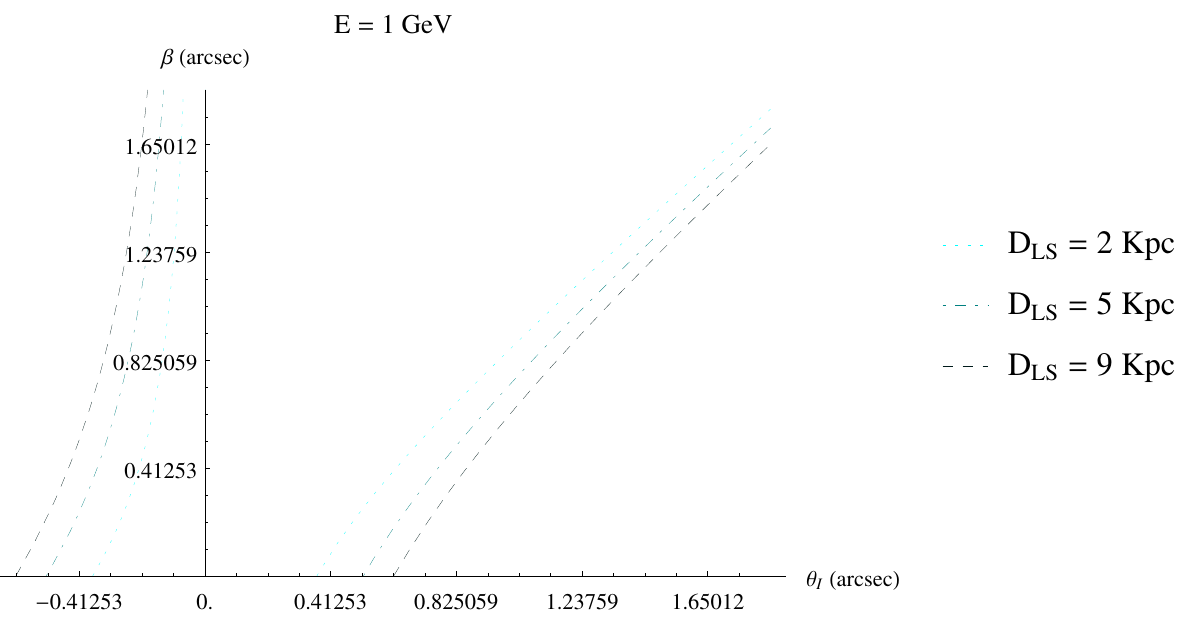}}
\caption{(a): $\beta(\theta_I)$ for the Virbadhra-Ellis lens equation in the neutrino case, for a black hole with M $=10^6\,\text{M}_{\odot}$ and with $D_{OL}$=10 Kpc, $D_{OS}$=19 Kpc. The numerical solution for the classical and the energy-dependent result. (b): $\beta(\theta_I)$ as in (a) but for a 1 GeV neutrino beam.}
\label{VHc1}
\end{figure}

\beq 
\overline{PS}=\overline{PI} -\overline{SI}, 
\eeq
which gives
\bea
\label{lenseq}
D_{OS}\tan\beta=D_{OS}\tan{\theta}_I-D_{LS}(\tan{\theta_I} +\tan(\alpha-{\theta}_I)),
\eea
under the assumption that the point $R$ in Fig. \ref{geo} lies on the vertical plane of the lens. ${\theta}_I$ is the angle at which the image is viewed by the observer and $\beta$ is the unlensed angular position of the source. 
Within this approximation we can use the geometric relation 
\beq
b=D_{OL}\sin{\theta}_I,
\label{bt}
\eeq
which allows to relate the image position ${\theta}_I$ to the angular deflection of the beam $\alpha$. Notice that this approximate relation is justified by the fact that the distances $D_{OL}$ and $D_{OS}$ are very large compared to the radius of closest approach $r_0$. In this limit the two segments $\overline{VH}$ and $\overline{VT}$ are treated as equal.\\
We remind that (\ref{lenseq}) is not the unique lens equation that one can write down, but, differently from Eq. (\ref{thin}), it can be used in the case of strong lensing. It  takes into account the nonlinear contributions to the angular deflection by the introduction of the $\tan(\beta)$ and $\tan(\theta_I)$ terms, which in (\ref{thin}) are not included. We refer to \cite{Bozza:2008ev} for a review of possible lens equations. 

\section{Radiative effects and the geometry of lensing }
Turning to our case study, radiative effects in the lens equations can be introduced by replacing the expression of the angular deflection generated by the source on the source plane, which is a function of the impact parameter 
$b$ $(\alpha=\alpha(b))$ with the new, energy dependent relation $\alpha(b,E)$ whose general form is given by (\ref{genex}). \\
 For simplicity we consider a pointlike source, and a pointlike deflector, as shown in Fig. \ref{geo}.
We recall that for a massless particle the geodesic motion is determined in terms of the energy $E$ and of the angular momentum $L$ at the starting point of the trajectory. The gravitational deflection, however, can be written only as a function of the impact parameter $b$ of the source, with $b=E/L$, which is an important result of the classical approach. For a further clarification of this aspect, which differs from the semiclassical analysis we are interested in, 
we briefly overview the classical case, using the lens geometry as a reference point for our discussion.\\
For a source located on the source plane at an angular opening $\beta$ (in the absence of the deflector), the initial conditions can be expressed in terms of the two components of the initial momentum $ \vec p=(p_r,p_\phi)$ on the plane of the geodesic, or, equivalently, by the pairs $(p_r, E)$ or $(p_\vartheta,E)$, with $E$ the initial energy of the beam. We recall that for a Schwarzschild metric these are defined as 
\beq
p_r = \left(  1- \frac{2 G M }{r} \right)^{\!-1} \dot{r},    
\qquad p_{\vartheta} = -r^2 \dot{\vartheta}, \qquad
p_t =  \left(  1- \frac{2 G M}{r} \right) \dot{t}, 
\qquad p_{\phi} = -  r^2 \sin^2 \vartheta \dot{\phi} \,.
\eeq
We have denoted with $\dot{x}\equiv dx/d s$ the derivative respect to the affine parameter. 
$p_t$ and $p_\phi$ related to the energy and to the angular momentum as 
$p_t=E$ and $p_\phi=-L$, and with the motion taking place on the plane $\vartheta=\pi/2$ $(p_\vartheta=0)$. They are constrained by the mass-shell condition 

\beq
 \left(1-\frac{2 G M }{r} \right) (p^t)^2 - \left( 1-\frac{2 G M}{r} \right)^{\!-1} (p^r)^2 - r^2 (p^\phi)^2=0,
\eeq
with $(p^r=\dot r, p^\phi=\dot \phi,p^t=\dot t)$.

The lens equation, usually written as 
\beq
L(\beta,\theta_I)=0, 
\eeq
can also be written, equivalently, in the form of a constraint between $\beta$ and $b$ using (\ref{bt}). We can use any of the independent variables mentioned above.
For a given initial momentum of the beam, emitted from the plane of the source, the lens equation will then determine the position of the source in such a way that the geodesic motion will reach the observer at its location on the optical axis. In particular, an interesting description emerges if we choose as initial conditions the angular position of the source $(\beta)$ and the value of the impact parameter $b$. These two conditions fix the direction of the trajectory of the beam at its origin on the source plane. In these last variables, the lens equation will then determine one of the two in terms of the other in such a way that outgoing geodesic will reach the observer. \\
The inclusion of an energy dependence in the angle of deflection $\alpha$ renders this picture slightly more complex. For instance, the lens equation will now depend on 3 parameters, which can be chosen to to be $(\beta, \theta_I, E)$ or $(\beta,p_r,p_\phi)$ or any other equivalent combination, with one of the three fixed in terms of the other two by the equation itself. For a monochromatic and spherical source of energy $E$, fixed at a position $\beta$, emitting a beam with a given impact parameter $b$ respect to the deflector, the lens equation may not have a real solution, since the deflector may disperse the beam in such a way that it will never reach the observer. For a fixed spherical source which emits photons or neutrinos of any energy, one can look for solution in the reduced variables $b, E$. Being $b$ related to the primary and secondary images $\theta_{I\pm}$, the beam that reaches the observer will be characterized by a unique energy $E$, assuming that the images are detected at angular positions $\theta_{I\pm} $. \\
The argument above can be repeated by using any triple combination of independent kinematic variables among those mentioned above. \\
Having clarified this point, we now move to a description of the actual implementation of the lens equation is this extended framework. The angular location of the image $\theta_I$ and the impact parameter are related in the geometry of the lens by Eq. (\ref{bt}), and this allows to search for solutions of the lens equation (\ref{lenseq}) in regions characterized by smaller values of the impact parameter ($ 20 <b_h <100$) where the angular deflections are stronger. \\
The key to the derivation of the radiative lens equation are Eqs. (\ref{genex}) and (\ref{bt}). Combining the two relations we obtain 

\begin{align}
\label{genexp}
\alpha(b(\theta_I,E))=&\frac{4 G M}{D_{OL} \sin\theta_I} + \sum_{n\geq 1} \frac{A_{2n}}{\left(D_{OL} \sin\theta_I\right)^{2n}}\nn\\
&+ \sum_{n\geq 1}\left( \frac{2 G M}{D_{OL} \sin\theta_I}\right)^{2n+1}\left(A_{2n+1} + D_1 \ln^n \left(\frac{D_{OL}}{2 G M} \sin\theta_I\right) +\ldots \right),
\end{align}
where the ellipsis refer to the extra logarithmic contributions present in Eq.(\ref{genex}). The expression 
above is known analytically if we manage to solve explicitly the semiclassical equation (\ref{semic}), otherwise it has to be found by a numerical fit. However, it is clear that the ansatz for the fit has, in any case, to coincide with Eqs.~(\ref{genex}) and (\ref{genexp}), due to the typical functional forms of the solutions of Eq. (\ref{semic}). For instance, in the case of a thin lens, the modifications embodied in (\ref{genexp}) can be incorporated into the new equation 

\bea
\label{thin3}
\beta=\theta_I- \alpha(b(\theta_I,E)) \frac{D_{LS}}{D_{OS}}, 
\eea
which is an obvious generalization of (\ref{thin2}), the latter being valid only in the classical GR case.
As we are going to illustrate below, (\ref{thin3}) can be studied numerically for several geometrical configurations, which are obtained by varying the lensing parameters $D_{LS}$ and $D_{OL}$.\\
A similar approach can be followed for the VE or for any other classical lens equation.
The insertion of $\alpha({\theta_I},E)$ given by (\ref{genex}) into (\ref{lenseq}) generates the radiative lens equation 
\beq
\label{lens1}
D_{OS}\tan\beta=D_{OS}\tan{\theta}_I-D_{LS}(\tan{\theta}_I +\tan(\alpha({\theta}_I,E)-{\theta}_I)),
\eeq
which takes into account also the quantum corrections and is now, on the contrary of (\ref{lenseq}), energy dependent. At this point it is clear that all the lens observables, such as magnifications, shears, light curves of microlensing etc. descend rather directly by this general prescription. \\
For instance, we can determine for the Virbadhra-Ellis lens the expression for the magnification using the 
radiative (semiclassical) expression
\bea
\label{mueq}
\mu&=&\frac{\chi_1}{\chi_2} \\
\chi_1&=&D_{OS}\sin\theta_I\,(1+((D_{OL}\tan\theta_I+(D_{OL}-D_{OS})\tan(\alpha(\theta_I,E)-\theta_I))/D_{OS})^2)^{3/2}, \nonumber 
\eeqa
\beqa
\chi_2 &=&(D_{OL}\tan\theta_I+(D_{OL}-D_{OS})\tan(\alpha(\theta_I,E)-\theta_I))\nonumber \\
&&\times (\sec^2\theta_I+(D_{OL}-D_{OS})/D_{OS}(\sec^2\theta_I
+\sec^2(\alpha(\theta_I,E)-\theta_I)\,(\alpha^{\prime}(\theta_I,E)-1)))\nn,
\eea
where $\alpha^{\prime}\equiv \partial \alpha/\partial \theta_I$. As clear from Eqs. (\ref{lens1}) and (\ref{mueq}), both equations are very involved, although they can be investigated very accurately at numerical level. It is also possible to discuss the analytical form of the solutions within the formalism of the $1/b^n$ expansion. In fact, we are entitled to expand all the observables of the fully nonlinear lens in the angular deflection $\alpha$, and work at a certain level of accuracy in the angular parameters. In this work, however, we prefer to proceed with a direct numerical analysis of the full equations, both for the thin  and for the VE lens, leaving the discussion of the explicit solutions to a future work.

\section{Post-Newtonian corrections: the case of primordial black holes}
We have seen in the previous sections that the $b_h(\alpha)$ expression for the deflection 
does not suffer from any apparent divergence (from the gravity or external field side) due to well-defined structure of the Newtonian cross section. The expression given in (\ref{leading}), in fact, is similar to the ordinary Rutherford scattering encountered in electrodynamics.\\
 The dependence of the resulting cross section on the scale $G M/c^2$, the Schwarzschild radius, manifests as an overall dimensionful constant. 
Therefore, the inclusion of the electroweak corrections - and the logarithmic dependence on the energy of the terms in the expansion that follows - do not appear in combination with the macroscopic scale $r_s$. This allows, in principle, an extension of the perturbative computation up to any order in the electroweak coupling constant $\alpha_w$. It is also clear that this result is expected to be valid for any renormalizable field theoretical model, when combined with an external static gravitational field of Coulomb type, as in the case of the Newtonian limit of GR.\\
From now on, we will be using the notation nPN to indicate the (post-Newtonian) order in the potential at which we expand the Schwarzschild metric. For instance, contributions of a certain nPN order involve corrections in the external field proportional to $\Phi^{n+1}$, with 0PN denoting the ordinary (lowest order) Newtonian (i.e. zeroth post-Newtonian) contributions proportional to $\Phi$, as given in Eq. (\ref{h2}).
The inclusion of the higher order corrections in the external potential modifies this simple picture due 1) to the need of introducing a cutoff regulator in the computation of the Fourier transform of the higher powers of the Newtonian potential and 2) to the presence of the Schwarzschild radius $r_s$ in the actual expansion. These features emerge already at the first post-Newtonian order (1PN) for an uncharged black hole and at order 0PN for the Reissner-Nordstrom (RN) metric (charged black hole). \\
Both points 1) and 2) are, in a way, expected, since the microscopic expression for the transition matrix element given by (\ref{volume}), in fact, cannot be extrapolated to the case of a macroscopic source, with the presence of a macroscopic scale such as the black hole horizon. This seems to indicate that the 
success of the Newtonian approximation is essentially due to the rescaling of $r_s$ found in the expression of the cross section, which is a feature of this specific order, and is therefore limited to a $1/r$ potential. It is then natural to ask if there is any other realistic case in which the post-Newtonian corrections can be included in an analysis of this type. Obviously, the answer is affirmative, as far as we require that $r_s$ is microscopic and that the energy of the beam, which is an independent variable of a scattering event, is at most of the order of $1/r_s$. Under these conditions, we are then allowed to extend our analysis through higher orders in $\Phi$, with scatterings in which the dimensionless parameter $r_s q$ with $q$ the impact parameter, is at most of $\mathcal{O}(1)$. 
This specific situation is encountered in the case of primordial black holes, where $r_s$ can be microscopic. 
We are going to illustrate this point in some detail, since it becomes relevant in the case of primordial black holes. 
\subsection{Post Newtonian contributions in classical GR} 
To illustrate this point we extend the expansion of the Schwarzschild metric at order 0PN given in (\ref{SCH3}). A similar expansion will be performed on the RN metric. \\
For this purpose, it is convenient to perform a 
change of coordinates on the Schwarzschild metric 
\bea
\label{ds}
ds^2=\left(1-\frac{2\,G\,M}{r}\right)dt^2-\left(1-\frac{2\,G\,M}{r}\right)^{-1}dr^2-r^2d\Omega
\eea
in such a way that this takes an isotropic form. The radial change of coordinates is given by 
\bea
r=\rho\left(1+\frac{G\,M}{2\rho}\right)^2
\eea
which allows to rewrite (\ref{ds}) as 
\bea
ds^2=A(\rho)dt^2-B(\rho)(d\rho^2+\rho^2\,d\Omega)
\eea
with
\bea
\label{ABsch}
A(\rho)=\frac{(1-G\,M/2\rho)^2}{(1+G\,M/2\rho)^2}\qquad\qquad B(\rho)=(1+G\,M/2\rho)^4 \,.
\eea
Post-Newtonian (weak field) corrections can be obtained by an expansion of $A$ and $B$ taking $M/\rho\ll1$. Up to third order in $\Phi$ this is given by
\bea
&&A(\rho)=1+2\,\Phi+2\,\Phi^2+\frac{3}{2}\,\Phi^3\\
&&B(\rho)=1-2\,\Phi+\frac{3}{2}\,\Phi^2-\frac{1}{2}\,\Phi^3.
\eea
In the RN spacetime for a charged black hole the analysis runs similar. The interest in this metric is due to the fact that the lowest order potential, in this case, involves charge-dependent $1/r^2$ contributions which, for an uncharged black hole, appear at first post-Newtonian order (1PN). The metric, in this case, is given by the expression
\bea
\label{RN}
ds^2=\left(1-\frac{2\,G\,M}{r}+\frac{G\,Q^2}{r^2}\right)dt^2-\left(1-\frac{2\,G\,M}{r}+\frac{G\,Q^2}{r^2}\right)^{-1}dr^2-r^2d\Omega,
\eea
with $Q$ denoting the overall charge of the black hole. It has two concentric horizons which become degenerate in the maximally charged case. The two horizons are the solution of the equation
\bea
\left(1-\frac{2\,G\,M}{r}+\frac{G\,Q^2}{r^2}\right)=0
\eea
with solutions $r=G\,M\pm\sqrt{G^2M^2-G\,Q^2}$.
The RN black hole has a maximum allowed charge $Q=M\sqrt G$, in order to avoid a naked singularity.
In this case, the radial change of variables which brings the metric into a symmetric form is given by
\bea
r=\rho\left(1+\frac{G\,M+\sqrt G\,Q}{\rho}\right)\left(1+\frac{G\,M-\sqrt G\,Q}{\rho}\right),
\eea
so that the RN spacetime in isotropic coordinates is
\begin{align}
\label{RNiso}
ds^2=&\frac{\left(1-\frac{G^2\,M^2-G\,Q^2}{4\rho^2}\right)^2}{\left(1+\frac{G\,M+\sqrt G\,Q}{2\rho}\right)^2\left(1+\frac{G\,M-\sqrt G\,Q}{2\rho}\right)^2}dt^2\nn\\
&-\left(1+\frac{G\,M+\sqrt G\,Q}{2\rho}\right)^2\left(1+\frac{G\,M-\sqrt G\,Q}{2\rho}\right)^2(d\rho^2+\rho^2\,d\Omega).
\end{align}
 We just recall that for a massless particle in this metric background the angle of deflection and the impact parameter are given by the expressions
\bea
&&\alpha(r_0)=2\,\int^\infty_{r_0}\frac{dr}{r\sqrt{\frac{r^2}{r_0^2}\left(1-\frac{2\,GM}{r_0}+\frac{GQ^2}{r_0^2}\right)-\left(1-\frac{2\,GM}{r}+\frac{GQ^2}{r^2}\right)}}-\pi\\
&&b(r_0)=\frac{r_0}{\sqrt{1-\frac{2\,GM}{r_0}+\frac{GQ^2}{r_0^2}}}
\eea
where $r_0$ is the closest distance of approach. It's convenient to normalize $r$, $r_0$ and $Q$ to the Schwarzshild radius $r_s=2\, G M$ and introduce the variables 
\bea
x=\frac{r}{2\, G M}\qquad x_0=\frac{r_0}{2\, G M}\qquad q=\frac{Q}{2\, G M}.
\eea
With this redefinitions the deflection can be expressed in the form \cite{Eiroa:2002mk}
\bea
\alpha(x_0)=G(x_0)\,\mathrm{F}(\phi_0, \lambda)-\pi
\label{elli}
\eea
with
\bea
G(x_0)=\frac{4\,x_0}{\sqrt{1-\frac{1}{x_0}+\frac{q^2}{x_0^2}}}\frac{1}{\sqrt{(r_1-r_3)(r_2-r_4)}}
\eea
and with
\bea
\mathrm{F}(\phi_0,\lambda)=\int_0^{\phi_0}(1-\lambda\,\sin^2\phi)^{-1/2}d\phi
\eea
being an elliptic integral of the first kind with arguments
\bea
&&\phi_0=\arcsin \sqrt{\frac{r_2-r_4}{r_1-r_4}}\\
&&\lambda=\frac{(r_1-r_4)(r_2-r_3)}{(r_1-r_3)(r_2-r_4)}.
\eea
The $r_i$ are the roots of the fourth order polynomial
\bea
P(x)=x^4+\frac{x_0^2}{1-\frac{1}{x_0}+\frac{q^2}{x_0^2}}\,(x-x^2-q^2)
\eea
ordered so that $r_1>r_2>r_3>r_4$. The comparison between Schwarzschild and RN deflection angle is shown in Figure \ref{schRN}. The plots 
describe the behaviour of the angular deflection as a function of the impact parameter $b_h$ for a RN and Schwarzschild metric in the region with 
$10 < b_h < 50$ (top left) and $4 < b_h < 10$ (top right) for the maximally charged case. The differences tend to be very pronounced as we approach the horizon of the Schwarzschild metric.\\
As pointed out in \cite{Amore:2006pi} in the Schwarzschild case, the $1/b$ expansion for the deflection angle does not reproduce the photon sphere singularity of the Schwarzschild metric, which is achieved using the exact GR expression in terms of elliptic function given in (\ref{elli}), but it represents nevertheless an improvement respect to the $0PN$ order.
Expanding the RN metric in $M/\rho\ll1$ up to the third order, the $2PN$ approximation gives
\bea
&&A(\rho)=1-\frac{2\,G\,M}{\rho}+\frac{2\,G^2\,M^2+G\,Q^2}{\rho^2}-\frac{3\,G^3\,M^3+5\,G^2\,M\,Q^2}{2\,\rho^3}\\
&&B(\rho)=1+\frac{2\,G\,M}{\rho}+\frac{3\,G^2\,M^2-G\,Q^2}{2\,\rho^2}+\frac{G^3\,M^3-G^2\,M\,Q^2}{2\,\rho^3}.
\eea
Inserting this expansion into the deflection integral, we can account in a systematic way of the $1/b$ corrections in the angle of deflection $\alpha$  
\bea
\label{alphaRN}
\alpha(b)=4\,\frac{G\,M}{b}+\left(5-\frac{G\,Q^2}{M^2}\right)\frac{3\pi}{4}\,\frac{G^2M^2}{b^2}+\left(\frac{128}{3}-16\frac{G\,Q^2}{M^2}\right)\,\frac{G^3M^3}{b^3}.
\eea
 The deflection (\ref{alphaRN}) in the maximally charged case is given by the expression
\bea
\alpha^{\textit{m.c.}}=4\,\frac{GM}{b}+3\pi\,\frac{(GM)^2}{b^2}+\frac{80}{3}\,\frac{(GM)^3}{b^3}.
\eea
In the next subsection we are going to illustrate how the inclusion of these expansions at nPN order affects the computation of the quantum corrections to the angular deflection. The corrections are embodied in a geometric form factor whose expression is entirely controlled by the $1/b$ expansion.

\begin{figure}[t]
\centering
\subfigure[]{\includegraphics[scale=.5]{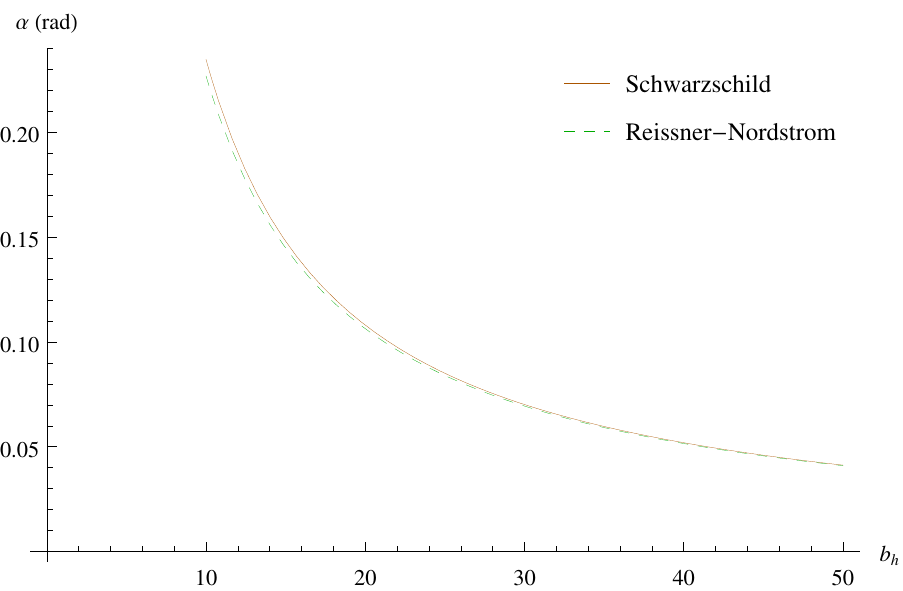}}\hspace{.5cm}
\subfigure[]{\includegraphics[scale=.5]{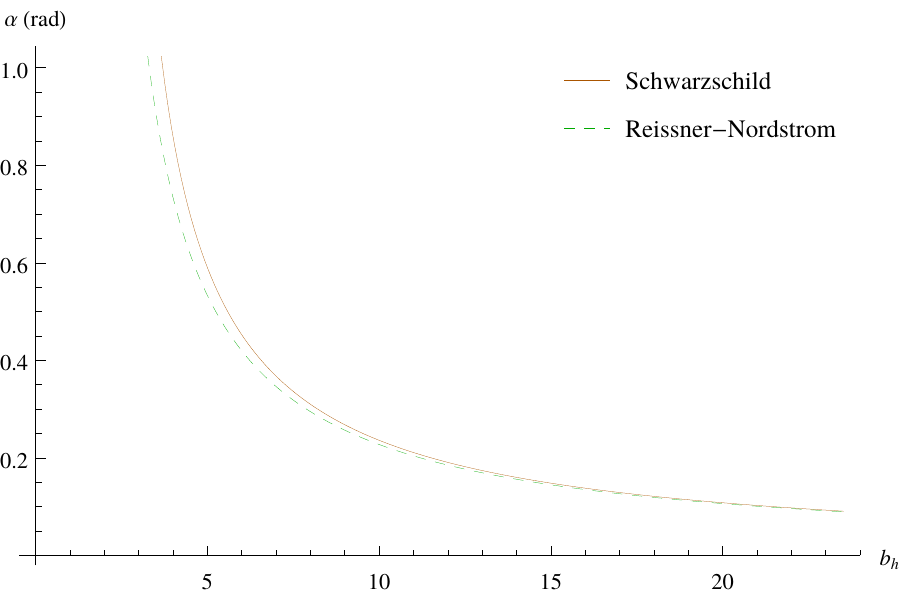}}
\caption{Comparison of the deflection angle for the Schwarzshild case and the maximally charged Reissner-Nordstrom case in the near-horizon (a), in the very-near-horizon (b).\label{schRN}}
\end{figure}
\subsection{Quantum effects at 2nd PN  order} 
The inclusion of the PN corrections to the external background requires a recalculation of the cross section, with the inclusion of the additional 
terms in the fluctuation of the metric in momentum space. 
As usual we consider a static source, so that the metric is written as
\begin{equation}
h_{\mu\nu}(q)=2\pi\delta(q_0) h_{\mu\nu}(\vec{q}\,).
\end{equation}
At leading order in the external field $\Phi$ both the timelike and the spacelike components are equal ( $h_{00}\equiv h_{ii}$), while at higher orders they are expressed in terms of two form factors $h_0$ and $h_1$
 \begin{equation}
h_{\mu\nu}(\vec{q}\,)=h_0(\vec{q}\,) \delta_{0\mu}\delta_{0\nu}+h_1(\vec{q}\,)\bigl( \eta_{\mu\nu}-\delta_{0\mu}\delta_{0\nu} \bigr),
\end{equation}
which at higher order in the weak external field are given by 
\begin{equation}
\label{eq:ffh0h1}
\begin{split}
h_0(\vec{q}\,)&=-\frac{2}{\kappa} \int d^3\vec{x}\, \biggl[  \frac{\Phi}{c^2}  + \biggl( \frac{\Phi}{c^2} \biggl)^{\!\!2} + \frac{3}{4} \biggl( \frac{\Phi}{c^2} \biggl)^{\!\!3\,}   \biggr] \text{e}^{i \vec{q} \cdot \vec{x}}  \\
h_1(\vec{q}\,)&= -\frac{2}{\kappa} \int d^3\vec{x}\,  \biggl[  -\frac{\Phi}{c^2}  +\frac{3}{4} \biggl( \frac{\Phi}{c^2} \biggl)^{\!\!2} - \frac{1}{4} \biggl( \frac{\Phi}{c^2} \biggl)^{\!\!3\,}  \biggr] \text{e}^{i \vec{q} \cdot \vec{x}}, 
\end{split}
\end{equation}
where we have explicitly reinstated the dependence on the speed of light. Below we will conform to our 
previous notations in natural units, with $c=1$.
\begin{itemize} 
\item{\bf \large Neutrinos}
\end{itemize}
The computation, at this stage, follows rather closely the approach of the previous sections, giving for 
 the averaged squared matrix element in the neutrino case 
\begin{equation}
\left|iS_{fi}\right|^2=\frac{\kappa^2}{16 V^2 E_1 E_2} 2\pi \delta(q_0) \,\mathcal{T} \,\frac{1}{2} \, \text{tr}\biggl[ \slashed{p}_2 V^{\mu\nu}(p_1,p_2) \slashed{p}_1 V^{\rho\sigma}(p_1,p_2) \biggr] h_{\mu\nu}(\vec{q}\,) h_{\rho\sigma}(\vec{q}\,),
\end{equation} 
with $\mathcal{T}$ being the time of the transition,
and the differential cross section 
\begin{equation}
d \sigma=\frac{d \mathcal{W}}{\mathcal{T}}=\frac{\left|S_{fi}\right|^2}{j_i \mathcal{T}}dn_f.
\end{equation}
We have denoted with $d n_f$ the density of final states in the transition amplitude, and with $j_i$ the incoming flux density. 
After integration over the final states, and using $|\vec{p}_1|=|\vec{p}_2|$, we obtain the expression
\begin{equation}
\left.\frac{d\sigma}{d\Omega}\right|_{\text{2PN}}^{(0)}=\frac{\kappa^2}{16 \pi^2} E^4 \cos^2\frac{\theta}{2} F_g(q)^2,
\end{equation}
where we have introduced the gravitational form factor of the external source
\beq
\label{fg}
F_g(q)\equiv \Bigl( h_0(\vec{q}\,)-h_1(\vec{q}\,) \Bigr).
\eeq
Notice the complete analogy between the corrections coming from a distributed source charge, for a potential scattering in quantum mechanics, and the gravity case. 
In the evaluation of $F_g$ in momentum space we are forced to introduce a cutoff $\Lambda$, being the Fourier transforms of the cubic contributions in $\Phi$ divergent. The singularity is generated by the integration around the region of $r\sim 0$ in the Fourier transform of the potential. The relevant integrals in this case are given by 
\begin{equation}
\begin{split}
I_n&=\int d^3\vec{x}\,\, \frac{1}{{|\vec {x}|}^n} \text{e}^{i\vec{q}\cdot \vec{x}}\\
\end{split}
\end{equation}
with 
\beq
I_1=\frac{4\pi}{\vec{q}\,^2}, \qquad  I_2 = \frac{2\pi^2}{|\vec{q}|}.  
\eeq
and with $I_3$ requiring a regularization with an ultraviolet cutoff  in space ($\Lambda$) 
\beq
I_3 =  \frac{4\pi}{|\vec{q}\,| }  \int_\Lambda^{\infty} dr\,\, \frac{\sin(|\vec{q}\,| r)}{r^2}.\\
\eeq
The choice of $\Lambda$ is dictated by simple physical considerations. Given the fact that consistency of the expansion requires that $r_s q \lesssim \mathcal{O}(1)$, it is clear the appropriate choice in the regulator is given by the condition that this coincides with the Scwarzschild radius, i.e. $\Lambda\sim r_s$.  Expressed in terms of the cutoff, we obtain for the geometric form factors the expressions 
\begin{equation}
\begin{split}
h_0(\vec{q}\,)&=-\frac{2}{\kappa} \biggl[ - \frac{4 \pi}{|\vec{q}\,|^2}GM  + \frac{2\pi^2}{|\vec{q}\,|}(GM)^2 - \frac{3}{4} \frac{4\pi}{|\vec{q}\,|}\biggl( \frac{\sin(\Lambda |\vec{q}\,|)}{\Lambda} - |\vec{q}\,| \text{Ci}(\Lambda |\vec{q}\,| )\biggl)(GM)^{3}   \biggr]  \\
h_1(\vec{q}\,)&= -\frac{2}{\kappa} \biggl[    \frac{4 \pi}{|\vec{q}\,|^2}GM  +\frac{3}{4} \frac{2\pi^2}{|\vec{q}\,|}(GM)^2 + \frac{1}{4} \frac{4\pi}{|\vec{q}\,|}\biggl( \frac{\sin(\Lambda |\vec{q}\,|)}{\Lambda} - |\vec{q}\,| \text{Ci}(\Lambda |\vec{q}\,|) \biggl)(GM)^{3}  \biggr] , \\
\end{split}
\end{equation}
where we have indicated with $\text{Ci}$ the cosine integral function 
\begin{equation}
\text{Ci}(x)= \int_{\infty}^x dt\, \frac{\cos t}{t}.
\end{equation}
From the previous equations we obtain the cross section
\begin{equation}
\left.\frac{d\sigma}{d\Omega}\right|_{\text{2PN}}^{(0)}=\frac{1}{4 \pi^2}E^4 \cos^2\frac{\theta}{2} \biggl[ \frac{8 \pi}{|\vec q\,|^2}GM  - \frac{\pi^2}{2|\vec{q}\,|}(GM)^2 + \frac{4\pi}{|\vec{q}\,|} \biggl( \frac{\sin(\Lambda |\vec{q}\,|)}{\Lambda} - |\vec{q}\,| \mathrm{Ci}(\Lambda |\vec{q}\,|) \biggl)(GM)^{3}   \biggr]^2,
\end{equation}
which is valid at Born level and includes the weak field corrections up to the third order in $\Phi$. In the expression of the cross sections, we use the subscript nPN, with $n=0,1,2$ to indicate a n-{th} order expansion of the metric in the gravitational potential, while the superscripts ((0), (1) and so on) label the perturbative order in $\alpha_w$. The leading order cross section at order 2PN, for instance, takes the form 

\begin{equation}
\left.\frac{d\sigma}{d\Omega}\right|_{\text{2PN}}^{(0)}=\left.\frac{d\sigma}{d\Omega}\right|_{\text{0PN}}^{(0)}\,\mathcal{PN}_2(E,\theta) ,
\end{equation}
with
\begin{align}
\mathcal{PN}_2(E, \theta)\equiv&\biggl[1-\frac{\pi}{8}(GM)\,E\sin\frac{\theta}{2}+\frac{1}{2}(GM)^2\,E \sin\frac{\theta}{2}\biggl( \frac{1}{\Lambda}\sin\left(2\,\Lambda\,E\sin\frac{\theta}{2}\right)\nn\\
&- 2E\sin\frac{\theta}{2} \mathrm{Ci}\biggl(2\,\Lambda\,E\sin\frac{\theta}{2}\biggr)\biggr)\biggr]^2,
\label{PN}
\end{align}
where we have factorized the tree level result ${d\sigma}/{d\Omega}|_{\text{0PN}}^{(0)}$ given in (\ref{leading}). The post-Newtonian form factor $\mathcal{PN}_2(E, \theta)$ induces an energy dependence of the cross section which is unrelated to the electroweak corrections. 
The analysis, in fact, can be extended at one loop in the electroweak theory. In this case, a lengthy computation gives the 2PN result 
\begin{equation} 
 \frac{d\sigma}{d\Omega}\biggr|_{\text{2PN}}^{\text{(1)}} =\frac{d\sigma}{d\Omega }\biggr|_{\text{0PN}}^{(0)}  \left[ 1 + \frac{4 G_F}{16 \pi^2\sqrt{2}}  \left( f_W^1(E,\theta) + f_Z^1(E,\theta) - \frac{1}{4} \Sigma^L_W - \frac{1}{4} \Sigma^L_Z  \right)  \right]\,\mathcal{PN}_2(E, \theta),
 \label{third}
\end{equation}
where we have inserted the one loop expression given in \eqref{sigmaOL}.\\
We can obtain an explicit solution of the corresponding semiclassical equation at order 1PN. Using the expression of the $\mathcal{PN}$ function at this order
\bea
\mathcal{PN}_1(E, \theta)&\equiv&\biggl[1- \frac{\pi}{8}(GM)\,E\sin\frac{\theta}{2}\biggr]^2
\eea
 on the right hand side of (\ref{PN}) in order to generate the 1PN cross section at Born level, and solving the 
 corresponding semiclassical equation (\ref{semic}) we obtain 
\begin{align}
\label{nbpn}
\left.b^2\right|_{\text{1PN}}^{(0)}&=4\,(GM)^2\Big(-1+\csc^2\frac{\alpha}{2}+2\ln\sin\frac{\alpha}{2}\Big) + E\, (GM)^3\pi\Big(4+(\cos\alpha-3)\csc\frac{\alpha}{2}\Big)\nn\\
&-\frac{1}{32}E^2(GM)^4\pi^2\Big(1+\cos\alpha+4\ln\sin\frac{\alpha}{2}\Big).
\end{align}
At this point, we can invert Eq.~(\ref{nbpn}) for $\alpha(b)$ obtaining
\begin{align}
\left.\alpha\right|_{\text{1PN}}^{(0)}&=\frac{2}{b_h}- \frac{\pi}{2}\frac{1}{b^2_h}E\,(GM)-\frac{1}{b^3_h}\Big(\ln b_h\,\Big(2-\frac{\pi^2}{32}\,E^2(GM)^2\Big)+\frac{2}{3}- E\,(GM)\,\pi\nn\\
&-\frac{3\pi^2}{64}\,E^2(GM)^2\Big) + \mathcal O(b_h^4)
\end{align}
for the tree level post Newtonian one.\\
For the Reissner-Nordstrom geometry the situation is similar. The post-Newtonian form factor is then given by
\begin{align}
\left.\mathcal{PN}(E, \theta)\right|^{RN}=&\left[1- \frac{\pi}{8}(GM)\big(1+3\frac{Q^2}{G\,M^2}\big)\,E\sin\frac{\theta}{2}\right.\nn\\
&\left.+(GM)^2\big(1+\frac{Q^2}{G\,M^2}\big)\,E \sin\frac{\theta}{2}\biggl( \frac{1}{\Lambda}\sin\left(2\,\Lambda\,E\sin\frac{\theta}{2}\right) - 2E\sin\frac{\theta}{2} \mathrm{Ci}\biggl(2\,\Lambda\,E\sin\frac{\theta}{2}\biggr)\biggr)\right]^2,
\label{PNrn}
\end{align}
and the impact parameter in the 1PN approximation is
\begin{align}
\label{nbpnRN}
\left.b^2\right|_{\text{1PN}}^{(0)\,RN}&=4\,(GM)^2\Big(-1+\csc^2\frac{\theta}{2}+2\ln\sin\frac{\theta}{2}\Big)+E\, (GM)^3\big(1+3\frac{Q^2}{G\,M^2}\big)\pi\Big(4+(\cos\theta-3)\csc\frac{\theta}{2}\Big)\nn\\
&-\frac{1}{32}E^2(GM)^4\big(1+3\frac{Q^2}{G\,M^2}\big)^2\pi^2\Big(1+\cos\theta+4\ln\sin\frac{\theta}{2}\Big)
\end{align}
The inversion formula in this case is
\begin{align}
\left.\alpha\right|_{\text{1PN}}^{(0)\,RN}=&\frac{2}{b_h}-\frac{\pi}{2}\frac{1}{b^2_h}E\,(GM)\big(1+3\frac{Q^2}{G\,M^2}\big)-\frac{1}{b^3_h}\Big(\ln b_h\,\Big(2-\frac{\pi^2}{32}\,E^2(GM)^2\big(1+3\frac{Q^2}{G\,M^2}\big)^2\Big)\nn\\
&+\frac{2}{3}-E\,(GM)\big(1+3\frac{Q^2}{G\,M^2}\big)\,\pi-\frac{3\pi^2}{64}\,E^2(GM)^2\big(1+3\frac{Q^2}{G\,M^2}\big)^2\Big)+ \mathcal O(b_h^4).
\end{align}
\begin{itemize} 
\item{\bf \large Photons} 
\end{itemize}
We can extend the analysis presented above for neutrinos to the photon case. Here the cross section takes the form

\begin{equation}
\left.\frac{d\sigma}{d\Omega}\right|_{\gamma,\text{2PN}}^{(0)}=\frac{\kappa^2}{16\pi^2} E^4 \cos^4\frac{\theta}{2}  F_g(q)^2 
\end{equation}
and, as in the neutrino case, we have
\begin{equation}
\left.\frac{d\sigma}{d\Omega}\right|_{\gamma,\text{2PN}}^{(0)}=\left.\frac{d\sigma}{d\Omega}\right|_{\gamma,\text{0PN}}^{(0)}\,\mathcal{PN}_2(E, \theta)
\end{equation}
where we inserted the tree level cross section for the photon
\begin{equation}
\left.\frac{d\sigma}{d\Omega}\right|_{\gamma,\text{0PN}}^{(0)}= (GM)^2 \cot^4 \frac{\theta}{2} \,.
\end{equation}
 In the 0PN Newtonian limit, this cross section has been computed in \cite{Coriano:2014gia}, and takes the form
\bea
\label{phDCS}
\left.\frac{d \sigma}{d \Omega}\right|_{\gamma,\text{0PN}}^{\text{(1)}}=\left.\frac{d\sigma}{d\Omega}\right|_{\gamma,\text{0PN}}^{(0)}\left\{1+2\left[ \sum_{{f_k}} N^c_{f_k} F^3_{f_k}(E, \theta, m_{f_k}, Q_{f_k})+F^3_W(E, \theta)\right]\right\}
\eea
where 
\begin{align}
F^3_{f_k}(E, \theta)=&\frac{1}{36}\frac{\alpha_w}{\pi}\frac{Q^2_{f_k}}{E^2}(35\,E^2-39\,m^2_{f_k}\csc^2\theta/2)\nonumber\\
&-\frac{1}{12}\frac{\alpha_w}{\pi}\frac{Q^2_{f_k}}{E^2}(4\,E^2-5\,m^2_{f_k}\csc^2\theta/2)\sqrt{1+m^2_{f_k}\frac{\csc^2\theta/2}{E^2}}\log\left(\frac{1+\sqrt{1+m^2_{f_k}\frac{\csc^2\theta/2}{E^2}}}{-1+\sqrt{1+m^2_{f_k}\frac{\csc^2\theta/2}{E^2}}}\right)\nonumber\\
&+\frac{1}{16}\frac{\alpha_w}{\pi}\frac{m^2_{f_k}Q^2_{f_k}}{E^4}\csc^4\theta/2\left(E^2\cos\theta-E^2+m^2_{f_k}\right)\log^{2}\left(\frac{1+\sqrt{1+m^2_{f_k}\frac{\csc^2\theta/2}{E^2}}}{-1+\sqrt{1+m^2_{f_k}\frac{\csc^2\theta/2}{E^2}}}\right)
\end{align}
and
\begin{align}
F^3_W(E, \theta)=&-\frac{1}{24}\frac{\alpha_w}{\pi}\frac{1}{E^2}(125\,E^2-39\,m^2_W\csc^2\theta/2)\nonumber\\
&+\frac{1}{8}\frac{\alpha_w}{\pi}\frac{1}{E^2}(14\,E^2-5\,m^2_W\csc^2\theta/2)\sqrt{1+m^2_W\frac{\csc^2\theta/2}{E^2}}\log\left(\frac{1+\sqrt{1+m^2_W\frac{\csc^2\theta/2}{E^2}}}{-1+\sqrt{1+m^2_W\frac{\csc^2\theta/2}{E^2}}}\right)\nonumber\\
&-\frac{1}{32}\frac{\alpha_w}{\pi}\frac{1}{E^4}\left(16\,E^4-16\,E^2\,m^2_W\csc^2\theta/2+3\,m_W^4\csc^4\theta/2\right)\log^2\left(\frac{1+\sqrt{1+m^2_W\frac{\csc^2\theta/2}{E^2}}}{-1+\sqrt{1+m^2_W\frac{\csc^2\theta/2}{E^2}}}\right)
\end{align}
are the relevant electroweak form factors entering in the computation. In the previous equations the sum $f_k$ is over all Standard Model fermions, with $m_{fk}$ and $Q_{f_k}$ their masses and charges. $N^c_{f_k}$ is 1 for leptons and 3 for quarks.
Proceeding similarly to the neutrino case, the one loop cross section in the 2PN approximation takes the form
\begin{equation}
\left.\frac{d \sigma}{d \Omega}\right|_{\gamma,\text{2PN}}^{\text{(1)}}=\left.\frac{d \sigma}{d \Omega}\right|_{\gamma,\text{0PN}}^{\text{(1)}}\,\mathcal{PN}_2(E, \theta),  
\label{twop}
\end{equation}
with $\mathcal{PN}_2$ given by (\ref{PN}), which can be inserted again in (\ref{semic}) and investigated numerically. Solving at order 1PN the analogous of (\ref{twop}), the solution of (\ref{semic}) gives
\begin{align}
\label{pbpn}
\left.b^2\right|_{\gamma,\text{1PN}}^{(0)}&=2\,(GM)^2\Big(-1+2\,\csc^2\frac{\alpha}{2}+\cos\alpha+8\ln\sin\frac{\alpha}{2}\Big)-\frac{2}{3}E\,(GM)^3\pi\Big(1+3\,\csc\frac{\alpha}{2}\Big)\Big(\cos\frac{\alpha}{4}-\sin\frac{\alpha}{4}\Big)^6\nn\\
&-\frac{1}{256}E^2(GM)^4\pi^2\Big(11+12\cos\alpha+\cos2\alpha+32\ln\sin\frac{\alpha}{2}\Big).
\end{align}
In the photon case the inversion formulae at orders 0PN and 1PN are given by
\bea
\left.\alpha\right|_{\gamma, \text{0PN}}^{(0)}=\frac{2}{b_h}-\frac{1}{b_h^3}\Big(4\,\ln b_h-\frac{1}{3}\Big)-\frac{1}{b_h^5}\Big(12\,\ln^2 b_h+10\,\ln b_h+\frac{17}{20}\Big) + \mathcal O(b_h^7)
\eea
and
\begin{align}
\left.\alpha\right|_{\gamma, \text{1PN}}^{(0)}=&\frac{2}{b_h}-\frac{1}{b_h^2}\frac{\pi}{2}E\,(GM)-\frac{1}{b_h^3}\Big(\ln b_h\big(4-\frac{1}{16}\,\pi^2E^2(GM)^2\big)-\frac{1}{64}\,\pi^2E^2(GM)^2\nn\\
&-\frac{4}{3}\,\pi\,E\,(GM)-\frac{1}{3}\Big) + \mathcal O(b_h^4)
\end{align}
respectively.
\subsection{Range of applicability}
The structure of the one-loop 2PN result for neutrinos and photons shows the complete factorization between the quantum corrections and the background-dependent contributions. While the former 
are process dependent, the latter are general. Obviously, this result is not unexpected, and follows rather closely other typical similar cases in potential scattering in quantum mechanics. An example is the case of an electron scattering off a finite charge distribution characterized by a geometrical size $R$, where the finite size corrections are all contained in a geometric form factor. \\
We recall that for a Coloumb interaction of the form $V( r )=e^2/r$, the cross section is given in terms of the pointlike $( p )$ amplitude   
\beq
f(\theta)_{\textrm{p}} =- 2 \frac{m e^2}{\vec{q}^{\,2}}
\label{point}
\eeq
with $\vec{q}= \vec{k} - \vec{k}'$ and $|\vec{q}|=2 |\vec{k}|\sin\theta/2$
being the momentum transfer of the initial (final) momentum of the electron $\vec{k}$ ($\vec{k}'$) and charge $e$. The scattering angle is measured with respect to the z-direction of the incoming electron. The charge of the static source has also been normalized to $e$. The corresponding cross section is given by 
\beq
\frac{d\sigma}{d\Omega}_\textrm{p}=|f(\theta)_{\textrm{p}}|^2=  \frac{(2 m) ^2 e^4}{16 k^4 \sin^4\theta/2},
\eeq
and the modification induced by the size of the charge distribution $(\rho(x))$ is contained in 
\beq
F( \vec{q})=\int d \vec{x}\rho( \vec{x}) e^{i  \vec{q}\cdot  \vec{x}} 
\eeq
with 
\beq
\frac{d\sigma}{d\Omega}=\frac{d\sigma}{d\Omega}_\textrm{p}|F( q)|^2 .
\eeq
For a uniform charge density, for instance, the geometrical form factor $F(\vec{q})$, which is the transform of the charge distribution, introduces a dimensionless variable $ q R$ in the cross section which is absent in the point-like (Coulomb) case, of the form 
\beq
F(q)=3 \frac{\sin(q\, R) -  q\, R \cos(q\, R)}{(q\, R)^3}.
\label{coulomb}
 \eeq
The validity of the expression above is for $q\, R\lesssim 1$, and the presence of the geometrical form factor is responsible for the fluctuations measured in the cross section as a result of the finite extension of the charged region. \\
In the analysis of the nPN corrections in gravity, the situation is clearly analogous, with the size of the horizon taking the role of the classical charge radius $R$. For ordinary (macroscopic) horizons 
(e.g. of a km size) $r_s\sim G M$ invalidates the perturbative expansion due to the 
appearance of the $G M E$ parameter in the expression of the post Newtonian factor 
$\mathcal{P N}(E,\theta)$, which is small only if $E\sim 1/GM$, a choice which is not relevant for our analysis, since it applies to particle beams whose energy is in the very far infrared. \\
\begin{figure}[t]
\centering
\subfigure[]{\includegraphics[scale=.7]{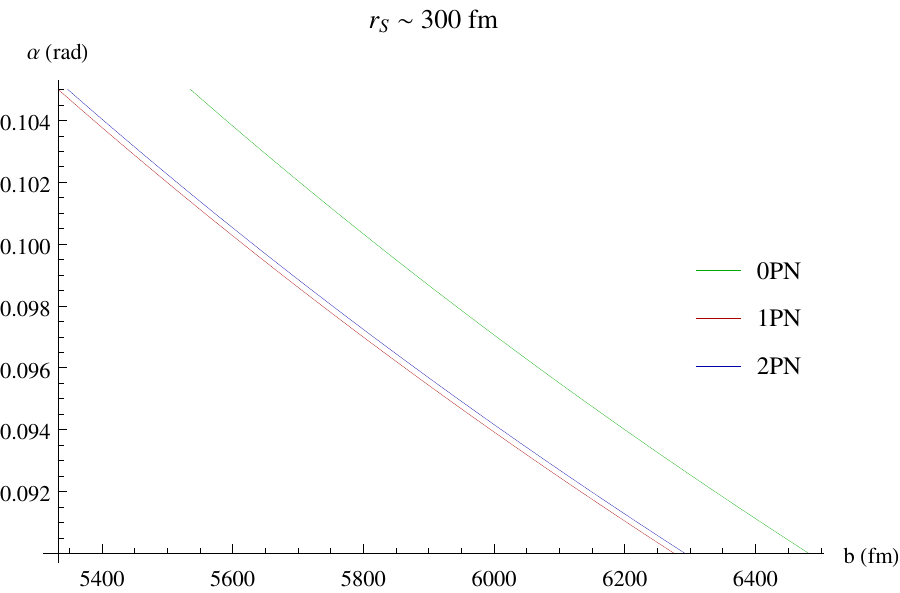}}\hspace{.5cm}
\subfigure[]{\includegraphics[scale=.7]{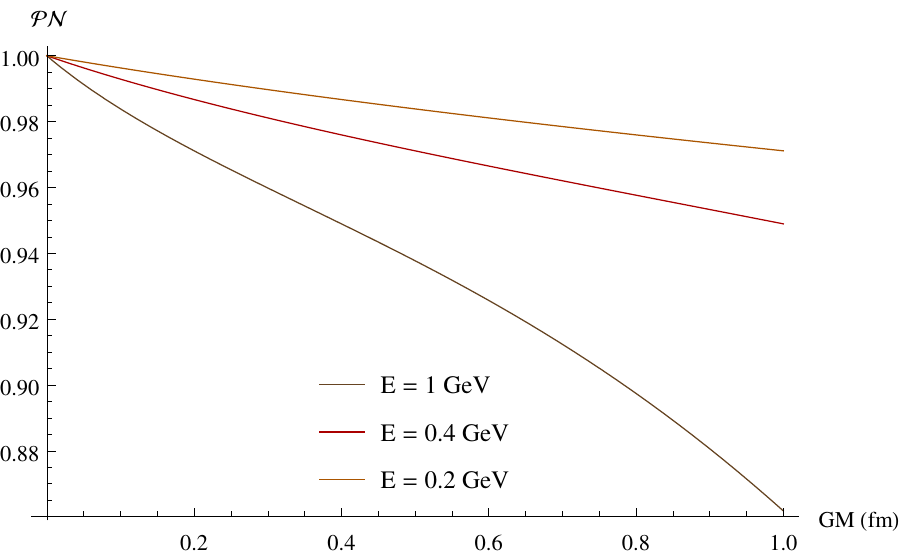}}\hspace{.5cm}
\subfigure[]{\includegraphics[scale=.7]{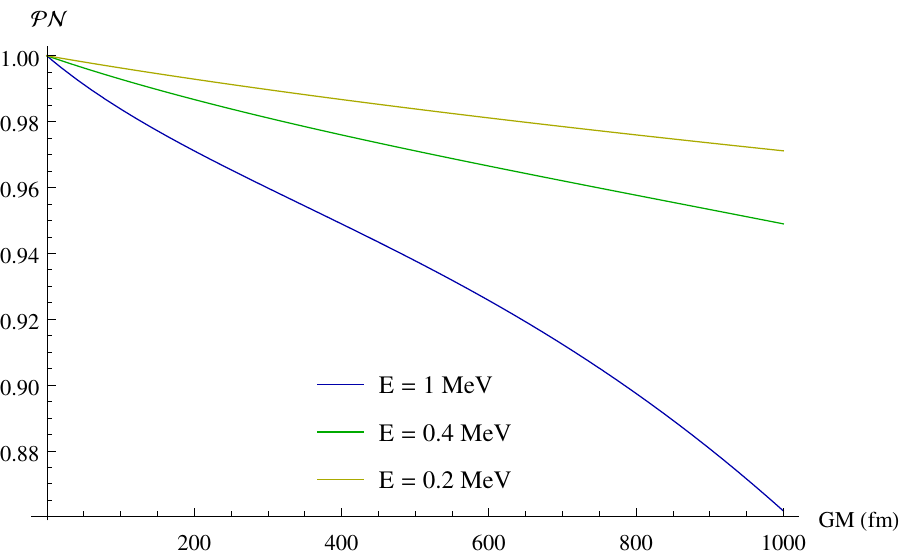}}\hspace{.5cm}
\caption{Comparison of nPN approximations for $\alpha(b)$ in the photon case with $\text{M}_{PBH}=10^{-16}\,\text{M}_\odot$ and for $E G M \approx 1$ (a). In (b) and (c) we show the $\mathcal{PN}$ function for different energies.\label{pnPBH}}
\end{figure}
By imposing that the cutoff $\Lambda$ coincides with the Schwarzschild radius 
($\Lambda\equiv r_s$), one can immediately realize that the post-Newtonian expansion gets organized only in terms of this parameter ($G M E$). In the regions of strong deflections, which are those that concern our analysis, we can reasonably assume that $y\equiv \sin\theta/2\sim \mathcal{O}(1)$, if we use the GR prediction to estimate the bending angle. This allows to discuss the convergence of the PN expansion only in terms of the energy $E$ of the incoming beam and of the size of the horizon. The analogous of the charge oscillations given by (\ref{coulomb}), in the gravitational case, are then uniquely related to the post-Newtonian function 
$\mathcal{PN}$, and hence to the size of the parameter $ \Lambda E\sim 
r_s E$ which defines its expansion in powers of the gravitational potential. Assuming a small value of $x\equiv G M E$, we can indeed rewrite (\ref{PN}) via a small-x expansion, obtaining
 \beq
\mathcal{PN}(x,y)= 1 -\frac{\pi}{8} x y + x^2 y^2\left( 1 -\gamma_E - \log x - \log 4 y\right).
\eeq
This expression can be used to investigate the range of applicability of these corrections in terms of the two factors appearing in $x$, the energy of the incoming beam and the size of the horizon of the gravitational source. The requirement that such a parameter be small defines a unique range of applicability of such corrections in the quantum case. \\
One possible application of the formalism which renders the PN corrections to the gravitational scattering quite sizable is in the context of primordial black holes \cite{Carr:1974nx}, which have found a renewed interest in the current literature \cite{Carr:2014eya,Khlopov:2008qy}. \\ 
We just mention that primordial black holes (PBHs) have been considered a candidate component of dark matter since the 70's, and conjectured to have formed in the early universe by the gravitational collapse of large density fluctuations, with their abundances and sizes tightly constrained by various theoretical arguments. These range from Hawking radiation, which causes their decay to occur at a faster rate compared to a macroscopic black hole (of solar mass); bounds from their expected microlensing events;  their influence on the CMB, just to mention a few \cite{Clesse:2015wea}. For instance, the mechanism of thermal emission by Hawking radiation sets a significant lower bound on their mass ($\sim5\times 10^{14} g$), in order for them to survive up to the present age of the universe. This bound  satisfies also other constraints, such as those coming from the possible interference of their decay with the formation of light elements at the nucleosynthesis time.  
With the launch of the FERMI gamma ray space telescope \cite{FERMI}, the interest in this kind of component has found new widespread interest. The unprecedented sensitivity of its detector in the measurement of interferometric patterns generated by high energy cosmic rays (femtolensing events), such as Gamma Ray Bursts \cite{Gould1992}, has allowed to consider new bounds on their abundances \cite{Barnacka:2012bm}. The hypothesis of having PBHs as a dominant component of the dark matter of the universe provides remarkable constrains on their allowed mass values, except for a 
mass range $10^{18} \textrm{kg} < M_{\textrm{PBH}} <10^{23} \textrm{kg}$, where it has been argued that they can still account for the majority of it. In other mass ranges several analyses indicate that the PBH fraction of dark matter cannot exceed $1\% $ of the total \cite{Clesse:2015wea}. \\
PN corrections turn out to be significant for PBH in this mass range, due to the large variation induced on the  $\mathcal{P N}$ function by the 1PN and 2PN terms. These may play a considerable role in a PBH mediated lensing event. We illustrate this behaviour by showing plots of the post Newtonian behaviour of the relevant expressions for lensing. In Fig. \ref{pnPBH} (a) we plot the angular deflection as a function of the impact parameter for the Newtonian 0PN, and relative post Newtonian corrections. We have considered a primordial black hole with a mass of $10^{-16}$ $M_{\odot}$, which carries a microscopic Schwarzschild radius (300 fm) and chosen $E=1/(GM)=0.6$ MeV for the incoming photon beam. The impact of the corrections on the gravitational cross section are quite large, as one can easily figure out from panel (b), where we plot the factor $\mathcal{P N}$ as a function of the Schwarzschild radius for these compact massive objects, for $b_h\sim$ 1 fm. For a more massive primordial black hole, with $200 < b_h <1000$, the pattern is quite similar, as shown in panel (\textrm{c}). In both cases the post-Newtonian corrections appear to be significant, of the order of 15-20 $\%$ and could be included in a more accurate analysis of lensing for these types of dark matter candidate solutions. 

\section{Conclusions} 
We have presented a discussion of neutrino lensing at 1-loop in the electroweak theory.
In our approach the gravitational field is a static background, and the propagating matter fields are obtained by embedding the Standard Model Lagrangian on a curved spacetime, as discussed in previous works \cite{Coriano:2013iba,Coriano:2011zk}. As in a previous study \cite{Coriano:2014gia}, also in our current case the field theoretical corrections to the gravitational deflection are in close agreement with the predictions of general relativity. The agreement holds both asymptotically, for very large distances from the center of the black hole, of the order of $10^6$ horizon sizes $(b_h)$, but also quite close to the photon sphere $(\sim 20\, b_h)$. In this respect, the similarity of the results for photons and neutrinos indicates the consistency of the semiclassical approach that we have implemented. 
As noticed in \cite{Accioly1}, the inclusion of the quantum effects causes the appearance of an energy dependent dispersion of a particle beam, which implies a violation of the classical equivalence principle. 

Various types of lens equations have been formulated in the past using classical GR, and we have illustrated the modifications induced on their expressions by the 
We have then developed a formalism which allows to include the semiclassical results, due to the radiative effects in the propagation of a photon or a neutrino, in a typical lensing event.
We have considered both the case of a thin lens, which is quadratic in the deflection angle, and the fully nonlinear case, taking as an example the Virbhadra-Ellis lens equation. Radiative and post-Newtonian effects induce 
a dependence of the angle of deflection with the appearance of extra $1/b^n$ suppressed contributions and of extra logarithms of the impact parameter, that we have studied numerically for 
some realistic geometric configurations. In general, radiative effects are significant only for configurations of the source/lens/observer which involve small impact parameters in the deflection $(b_h\sim 20)$, and require angular resolutions in the region of few milliarcseconds. 
Our results are valid for a Schwarzschild metric, considered both in the Newtonian and in the post-Newtonian approximation, but they can be extended to other metrics as well. \\
We have also discussed the consistency of the post-Newtonian approach. We have shown that such corrections can be consistently taken into account in the case of microscopic horizon sizes, such as primordial black holes. These corrections have been shown to factorize and be accounted for by a post-Newtonian function. Our analysis can be extended in several directions, from the case of Kerr-Newman metrics to the study of microlensing and Shapiro delays, and to dynamical gravity.

%%%%%%%%%%%%%%%%%%%%%%%%%%%%%%%%%%%%%%%%%

\chapter*{Conclusions}
The aim of this thesis has been to present some consequences of classical conformal symmetry in some extensions of the SM, both in the supersymmetric and in the non supersymmetric cases. In all the cases we have been stressing on the possible physical implications the scenarios that we have investigated. 
The analysis of Chapter 1 has been centered on general features of supersymmetric theories, where we have analysed some important aspects of anomaly actions, proving that the manifestation of a conformal anomaly is in the presence of massless effective degrees of freedom which interpolate as intermediate states in an anomaly vertex. We have shown that this universal feature is typical both of chiral and of conformal anomalies. For a dilatation symmetry, broken by the trace anomaly, the intermediate state is interpreted as an effective dilaton. Dilatons, axions and dilatinos, interpreted as composite states, are identified in the UV as anomaly poles of fundamental theories, but their manifestation in the IR could be related to possible nonlinear realizations of the same theories. More studies are obviously necessary in order to come up with more conclusive results in regard to the role played by these degrees of freedom in a low energy context. \\
In Chapter 2 we have turned to a phenomenological analysis of a dilaton state in the context of the SM. 
The study, in this case concerns a fundamental state, on which we have quantified the impact of the current constraints coming from Higgs searches at the LHC. The bounds on the conformal scale that we have extracted are not too restrictive, showing that a dilaton can still be searched for at the LHC and is not excluded. \\
Chapter 3 and 4 deal with a specific superconformal theory, the TNMSSM, which extends the NMSSM with one extra triplet and a scalar singlet superfield. We have focused these two chapters on the Higgs sector of the model, characterising its spectrum and the possible implications for the discovery of hidden Higgses, predicted by it, at the LHC. Although we have not much commented upon, we have verified that the model allows a massless supermultiplet, associated to the superconformal symmetry of the model. Chapter 4 has been entirely dedicated to a study of the pseudoscalar state present in the model, and we have discussed several constraints on the allowed parameter space coming from recent ATLAS and CMS data. \\
Finally in Chapter 5,  we have been discussing  an application of the TVV anomaly vertex to the gravitational deflection of photons in a Schwarzschild background. This original analysis, which has been developed as a by-product of previous investigations of conformal anomalies, shows how a fundamental result derived from the study of specific interactions carries far reaching implications in astroparticle physics and semiclassical GR. In particular, we have introduced for the first time the notion of a 
"radiative lens equation" using an original approach which we hope can be further study and extended in the future. 

%%%%%%%%%%%%%%%%%%%%
\appendix

\chapter{Scalar integrals}
\label{AppScalarIntegrals}
One-, two- and three- point functions are denoted respectively as $\mathcal A_0$, $\mathcal B_0$ and $\mathcal C_0$ with
\bea
\mathcal A_0(m^2) &=& \frac{1}{i \pi^2} \int d^n l \frac{1}{l^2 - m^2} \,, \nn \\
\mathcal B_0(p_1^2, m_0^2,m_1^2) &=& \frac{1}{i \pi^2} \int d^n l \frac{1}{(l^2 -m_0^2)((l+p_1)^2 -m_1^2)} \,, \nn \\ 
\mathcal C_0((p+q)^2, p^2, q^2,m_0^2,m_1^2,m_2^2) &=&  \frac{1}{i \pi^2} \int d^n l \frac{1}{(l^2 -m_0^2)((l-p)^2 -m_1^2)((l-p-q)^2 -m_2^2)} \,.
\eea
Moreover, for equal internal masses and for $p^2=q^2=0$ we have used the more compact notation
\bea
\mathcal B_0(p_1^2,m^2) \equiv \mathcal B_0(p_1^2,m^2,m^2)\,, \qquad \mathcal C_0((p+q)^2,m^2) \equiv \mathcal C_0((p+q)^2,0,0,m^2,m^2,m^2) \,.
\eea
In the spacelike region ($k^2 < 0$), using two regulators for the ultraviolet and infrared singularities ($n=4-\epsilon_{UV} = 4 + \epsilon_{IR}$), where $n$ denotes the spacetime dimensions, the relevant 2-point functions appearing in the computation are
\bea
\mathcal B_0(k^2, 0) &=& \frac{2}{\epsilon_{UV}} + 2 - \log\frac{-k^2}{\mu^2} \,, \\
\mathcal B_0(k^2, m^2) &=& \frac{2}{\epsilon_{UV}} + 2 - \log\frac{m^2}{\mu^2} + \sqrt{\tau(k^2,m^2)} \log \frac{\sqrt{\tau(k^2,m^2)} -1}{\sqrt{\tau(k^2,m^2)} +1} \,, 
\eea
with $\tau(k^2,m^2) = 1-4m^2/k^2$,
while for $k^2$ null we obtain
\bea
\mathcal B_0(0, 0) &=& \frac{2}{\epsilon_{UV}} + \frac{2}{\epsilon_{IR}} \,, \\
\mathcal B_0(0, m^2) &=& \frac{2}{\epsilon_{UV}} - \log\frac{m^2}{\mu^2}. \, 
\eea
In the QCD computations we have also used the following finite two-point scalar integrals
\bea
\mathcal D(k^2,m^2) = \mathcal B_0(k^2,m^2) - \mathcal B_0(0,m^2) \,,
\eea
and we have renormalized all the divergent $\mathcal B_0$ functions in the $\overline{MS}$ scheme in which the $1/\eps_{UV}$ divergences have been subtracted. \\
The massless scalar 3-point function, for $k^2<0$,  is given by
\bea
\mathcal C_0(k^2,0) &=& \frac{1}{k^2} \left[ \frac{4}{\epsilon_{IR}} + \frac{2}{\epsilon_{IR}} \log \frac{-k^2}{\mu^2} +\frac{1}{2} \log^2 \frac{-k^2}{\mu^2} - \frac{\pi^2}{12} \right]\,,
\eea
while the massive $\mathcal C_0(k^2,m^2)$ is given in Eq. (\ref{ti}).

\chapter{The $\langle TVV \rangle$ and $\langle AVV \rangle$ vertices in an ordinary gauge theory}
 \label{TVVsection}
The on-shell expansion of the $\langle TVV \rangle$ correlator in a non-abelian gauge theory is expressed in terms of just 3 independent form factors \cite{Armillis:2010qk}
\bea
\Gamma^{\mu\nu\alpha\beta}_{(T)}(p,q) = f_1(k^2) \, \phi_1^{\mu\nu\alpha\beta}(p,q) + f_2(k^2) \, \phi_2^{\mu\nu\alpha\beta}(p,q) + f_3(k^2) \, \phi_3^{\mu\nu\alpha\beta}(p,q) \,,
\eea
where the tensor structures are defined by
\bea
\label{phitensors}
\phi_1^{\mu\nu\alpha\beta}(p,q) &\equiv& t_1^{\mu\nu\alpha\beta}(p,q) = (k^2 \eta^{\mu\nu} - k^\mu k^\nu) u^{\alpha\beta}(p,q)\,, \nn \\
\phi_2^{\mu\nu\alpha\beta}(p,q) &\equiv& t_3^{\mu\nu\alpha\beta}(p,q) + t_5^{\mu\nu\alpha\beta}(p,q) -4 t_7^{\mu\nu\alpha\beta}(p,q) = - 2 u^{\alpha\beta}(p,q) [k^2 \eta^{\mu\nu} + 2 (p^\mu p^\nu + q^\mu q^\nu) \nn \\
&-& 4 (p^\mu q^\nu + q^\mu p^\nu)] \,, \nn \\
\phi_3^{\mu\nu\alpha\beta}(p,q) &\equiv& t_{13}^{\mu\nu\alpha\beta}(p,q) = (p^\mu q^\nu + p^\nu q^\mu) \eta^{\alpha\beta} + p \cdot q (\eta^{\alpha\nu} \eta^{\beta\mu} + \eta^{\alpha\mu} \eta^{\beta\nu}) - \eta^{\mu\nu} u^{\alpha \beta}(p,q) \nn \\
&-&  (\eta^{\beta\nu}p^\mu + \eta^{\beta\mu}p^\nu)q^\alpha - (\eta^{\alpha\nu}q^\mu + \eta^{\alpha\mu}q^\nu)p^\beta,
\eea
with 
\bea
\label{utensor}
u^{\alpha\beta}(p,q) = \eta^{\alpha\beta} p \cdot q - p^\beta q^\alpha \,.
\eea
Here $k=p+q$ is the momentum incoming in the EMT line while $p^\alpha$ and $q^\beta$ are the momenta outgoing from the two vector currents.\\ 
In the on-shell and massless case, for a Dirac fermion $(f)$ in the representation $R_f$ running in the loops, the form factors are given by
\bea
\label{FFfermions}
f_1^{(f)}(k^2) &=& - \frac{g^2 \, T(R_f)}{18 \pi^2 \, k^2} \,, \qquad
f_2^{(f)}(k^2) = - \frac{g^2 \, T(R_f)}{144 \pi^2 \, k^2} \,, \nn \\
f_3^{(f)}(k^2) &=& \frac{g^2 \, T(R_f)}{144 \pi^2} \left\{ 11 + 12 \, \mathcal B_0(k^2,0)\right\}.
\eea
Analogous results hold for a conformally coupled complex scalar $(s)$ in the representation $R_s$
\bea
\label{FFscalars}
f_1^{(s)}(k^2) &=&  - \frac{g^2 \, T(R_s)}{72 \pi^2 \, k^2} \,, \qquad
f_2^{(s)}(k^2) =  \frac{g^2 \, T(R_s)}{288 \pi^2 \, k^2} \,, \nn \\
f_3^{(s)}(k^2) &=& \frac{g^2 \, T(R_s)}{288 \pi^2} \left\{ 7 + 6 \, \mathcal B_0(k^2,0)\right\} \,
\eea
while for a gauge field ($A$) in the adjoint representation one obtains
\bea
\label{FFgauge}
f_1^{(A)}(k^2) &=&  \frac{11 g^2 \, T(A)}{72 \pi^2 \, k^2} \,, \qquad
f_2^{(A)}(k^2) =  \frac{g^2 \, T(A)}{288 \pi^2 \, k^2} \,, \nn \\
f_3^{(A)}(k^2) &=& - \frac{g^2 \, T(A)}{8 \pi^2} \left\{ \frac{65}{36} - \mathcal B_0(0,0) + \frac{11}{6} \mathcal B_0(k^2,0) + k^2 \, \mathcal C_0(k^2,0) \right\}.
\eea 
We recall that for an axial anomaly, the usual Rosenberg parameterization in terms of six form factors $(A_1,\ldots, A_6)$, and the use of the Ward identities and on-shellness conditions on the vector lines, reduce the anomaly amplitude $\Delta^{\lambda\mu\nu}$ in the simple form \cite{Armillis:2009sm}
\beq
\Delta^{\lambda\mu\nu}=A_6(k^2,m^2) k^{\lambda}\epsilon[p,q,\nu,\mu] + (A_4(k^2,m^2) + A_6(k^2,m^2)) ( q^\nu\eps[p,q,\mu,\lambda] - p^{\mu} \epsilon[p,q,\nu,\lambda]) 
\label{anom1}
\eeq
with $k$ denoting the incoming momentum of the axial-vector line (of Lorentz index $\lambda$), and with $p$ and $q$ denoting the outgoing momenta of the $(\mu,\nu)$ vector lines. Note that in this case the transversality condition for the vector currents removes the second combination of form factors, leaving only a nonzero $A_6$, which is given by 
\beq
A_6(k^2,m^2)=\frac{1}{2 \pi^2 k^2}\left(1 +\frac{m^2}{k^2}\log^2\left(\frac{\sqrt{\tau(k^2,m^2)}+1}{\sqrt{\tau(k^2,m^2)}-1}\right)\right) \qquad k^2< 0
\label{a6}
\eeq
with $\tau(k^2,m^2)=1- 4 m^2/k^2$. In the massless limit, the spectral density of this form factor, for $k^2>0$, is proportional to a Dirac $\delta$-function, since the logarithmic term vanishes, and is accompanied by a sum rule. In any case, the spectral density of the $A_4 + A_6$ form factor is not integrable, and the link between the chiral anomaly and the corresponding pole is again unique.

\chapter{Mass deformations and the spectral densities flow } 
We start by introducing the spectral density $\rho(k^2)$, which is the discontinuity of $\mathcal C_0$ along the cut $(k^2>4 m^2)$, as 
\beq
\rho(k^2,m^2)=\frac{1}{2 i} \textrm{Disc}\, \mathcal C_0(k^2, m^2) \,,
\label{spect}
\eeq
with 
\beq
\label{discdef}
 \textrm{Disc} \, \mathcal C_0(k^2, m^2)\equiv \mathcal C_0(k^2 +i \epsilon,m^2) - \mathcal C_0(k^2 -i \epsilon,m^2).
\eeq
\begin{figure}[t]
\centering
\subfigure{\includegraphics[scale=1.3]{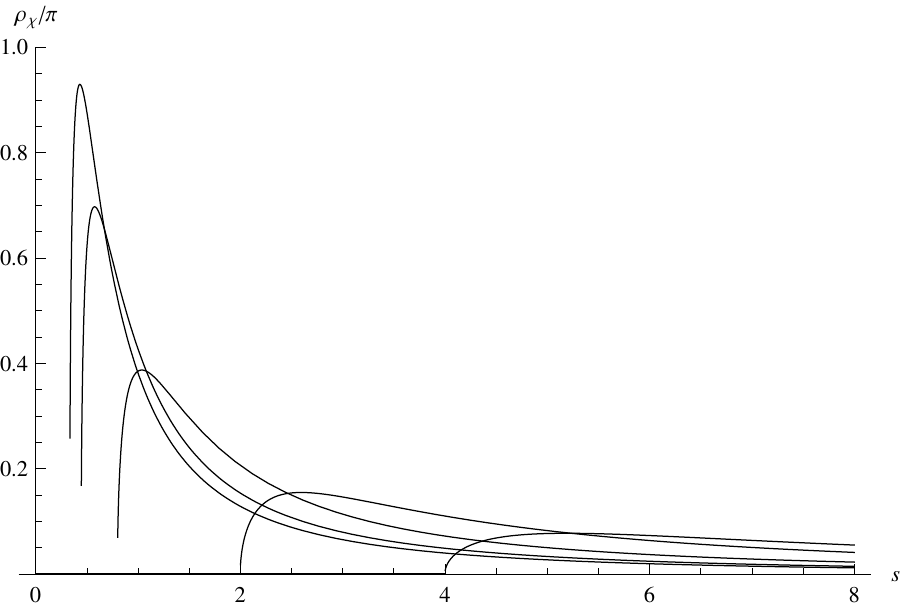}} \hspace{.5cm}
\caption{Representatives of the family of spectral densities $\frac{{{\rho}_\chi}^{(n)}}{\pi}(s)$ plotted versus $ s $ in units of $m^2$. The family "flows"  towards the $s=0$ region becoming a $\delta(s)$ function as $m^2$ goes to zero. }
\label{seq}
\end{figure}
To determine the discontinuity above the two-particle cut we can proceed in two different ways. We can use the unitarity cutting rules and therefore compute the integral
\bea
\label{disco}
\textrm{Disc} \, \mathcal C_0(k^2, m^2)&=& \frac{1}{i \pi^2} 
 \int d^4 l \frac{2 \pi i \delta_+(l^2 - m^2) 2 \pi i \delta_+((l - k)^2- m^2)}{(l-p)^2 - m^2 + i \epsilon} \nn\\
 &=&  \frac{2 \pi}{i k^2}\log\left(\frac{1 + \sqrt{\tau(k^2 ,m^2)}}{1-\sqrt{\tau(k^2 ,m^2)}}\right)\theta(k^2 - 4 m^2) \,,
\label{cut}
 \eea
where $\tau(k^2,m^2) = \sqrt{1 -4m^2/k^2}$.
 The integral has been computed by sitting in the rest frame of the off shell line of momentum $k$.
Alternatively, we can exploit directly the analytic continuation of the explicit expression of the $\mathcal C_0(k^2,m^2)$ integral in the various regions. This is given by
\bea  
\label{ti}
 \mathcal C_0(k^2\pm i\epsilon,m^2) = \left\{ 
\begin{array}{ll} 
\frac{1}{2 k^2}\log^2 \frac{\sqrt{\tau(k^2,m^2)}+1}{\sqrt{\tau(k^2,m^2)}-1} & \mbox{for} \quad k^2 < 0 \,, \\
- \frac{2}{k^2} \arctan^2{\frac{1}{\sqrt{-\tau(k^2,m^2)}}} & \mbox{for} \quad 0 < k^2 < 4 m^2 \,, \\
\frac{1}{2 k^2} \left( \log \frac{1 + \sqrt{\tau(k^2,m^2)}}{1 - \sqrt{\tau(k^2,m^2)} } \mp i \, \pi\right)^2 & \mbox{for} \quad k^2 > 4 m^2 \,.  
\end{array} 
\right.
\eea
From the two branches encountered with the $\pm i \epsilon$ prescriptions, the discontinuity is then present only for $k^2> 4m^2$, as expected from unitarity arguments, and the result for the discontinuity, obtained using the definition in Eq. (\ref{discdef}), 
clearly agrees with Eq. (\ref{disco}), computed instead by the cutting rules. \\
The dispersive representation of $\mathcal C_0(k^2,m^2)$ in this case is written as
\beq
\mathcal C_0(k^2,m^2)=\frac{1}{\pi} \int_{4 m^2}^{\infty} d s \frac{\rho(s, m^2)}{s - k^2 }, 
\eeq
which, for $k^2 < 0$ gives the identity  
\beq
\label{loop}
 \int_{4 m^2}^{\infty} \frac{d s} {(s - k^2) s}\log\left(\frac{1 + \sqrt{\tau(s,m^2)}}{1-\sqrt{\tau(s,m^2)}}\right)=- \frac{1}{2 k^2}\log^2 \frac{\sqrt{\tau(k^2,m^2)}+1}{\sqrt{\tau(k^2,m^2)}-1}
 \eeq
with $\rho(s,m^2)$ given by Eqs. (\ref{spect}) and (\ref{cut}). The identity in Eq. (\ref{loop}) allows to reconstruct the scalar integral 
$\mathcal C_0(k^2,m^2)$ from its dispersive part. 

Having determined the spectral function of the scalar integral $\mathcal C_0(k^2,m^2)$, we can extract the spectral density associated with the anomaly form factors in Eqs. (\ref{RChiralOSMassive}), (\ref{SChiralOSMassive}), (\ref{TChiralOSMassive}), which is given by
\beq
 \chi(k^2, m^2)\equiv \Phi_1(k^2,m^2)/k^2, 
 \eeq
and which can be computed as
\bea
\textrm{Disc}\, \chi(k^2, m^2)= \chi(k^2+i\epsilon,m^2)-\chi(k^2-i\epsilon, m^2) = - \textrm{Disc}\left( \frac{1}{k^2} \right) - 2 m^2 \textrm{Disc}\left(\frac{\mathcal C_0(k^2,m^2)}{k^2}\right).
\label{cancel1}
\eea
Using the principal value prescription 
\beq
\frac{1}{x\pm i\epsilon}=P\left(\frac{1}{x} \right) \mp i\pi \delta(x) \qquad \epsilon >0 \,,
\eeq
we obtain
\bea
&& \textrm{Disc} \left( \frac{1}{k^2} \right) = - 2 i\pi\delta(k^2) \nn\\
&& \textrm{Disc}\left( \frac{\mathcal C_0(k^2,m^2)}{k^2}\right) = P\left(\frac{1}{k^2}\right)\textrm{Disc}\,\mathcal C_0(k^2,m^2) - i\pi \delta(k^2) A(0) \,,
\eea 
where we have defined  
\beq
\label{As}
A(k^2)\equiv C_0(k^2+i\epsilon,m^2)+C_0(k^2-i\epsilon,m^2),
\eeq
and
\bea
A(0) &=& \lim_{k^2\to 0} A(k^2) =  -\frac{1}{m^2}. 
\eea
This gives, together with the discontinuity of $\mathcal C_0(k^2,m^2)$ which we have computed previously in Eq. (\ref{cut}),
\beq
\label{discC0}
\textrm{Disc}\left( \frac{\mathcal C_0(k^2,m^2)}{k^2}\right)=-2 i\frac {\pi}{(k^2)^2} \log \frac{1 + \sqrt{\tau(k^2,m^2)}}{1 - \sqrt{\tau(k^2,m^2)}}\theta(k^2-4 m^2)
+i\frac{\pi}{m^2}\delta(k^2).
\eeq
The discontinuity of the anomalous form factor $\chi(k^2,m^2)$ is then given by
\beq
\textrm{Disc} \, \chi(k^2,m^2)=4 i \pi \frac{m^2}{ (k^2)^2}\log \frac{1 + \sqrt{\tau(k^2,m^2)}}{1 - \sqrt{\tau(k^2,m^2)} }\theta(k^2- 4 m^2).
 \eeq 
The total discontinuity of $\chi(k^2,m^2)$, as seen from the result above, is characterized just by a single cut for $k^2> 4 m^2$, since the $\delta(k^2)$ (massless resonance) contributions cancel between the first and the second term of Eq. (\ref{cancel1}). This result proves the {\em decoupling } of the anomaly pole at $k^2=0$ in the massive case due to the disappearance of the resonant state. \\
The function describing the anomaly form factor, $\chi(k^2,m^2)$, then admits a dispersive representation over a single branch cut
\beq
\chi(k^2,m^2)=\frac{1}{\pi}\int_{4 m^2}^{\infty}\frac{{\rho}_\chi(s,m^2)}{s- k^2 }ds
\eeq
corresponding to the ordinary threshold at $k^2=4 m^2$,
with 
\bea
\label{spectralrho}
{\rho}_\chi(s,m^2) = \frac{1}{2 i} \textrm{Disc} \, \chi(s,m^2)  
=\frac{2 \pi m^2}{s^2}\log\left(\frac{1 + \sqrt{\tau(s,m^2)}}{1-\sqrt{\tau(s,m^2)}}\right)\theta(s- 4 m^2). 
\eea
From the spectral function given above and from the corresponding integral representation one can extract a new nontrivial integral relation 
\beq
\int_{4 m^2}^{\infty} \frac{1}{s^2 (s - k^2)}\log\left(\frac{1 + \sqrt{\tau(s,m^2)}}{1-\sqrt{\tau(s,m^2)}}\right) ds 
=- \frac{1}{2 k^2 m^2} -\frac{1}{2 (k^2)^2} \log^2 \frac{\sqrt{\tau(k^2,m^2)}+1}{\sqrt{\tau(k^2,m^2)}-1},
\eeq
which is the analogue of Eq. (\ref{loop}). 

\begin{figure}[t]
\centering
\subfigure[]{\includegraphics[scale=0.6]{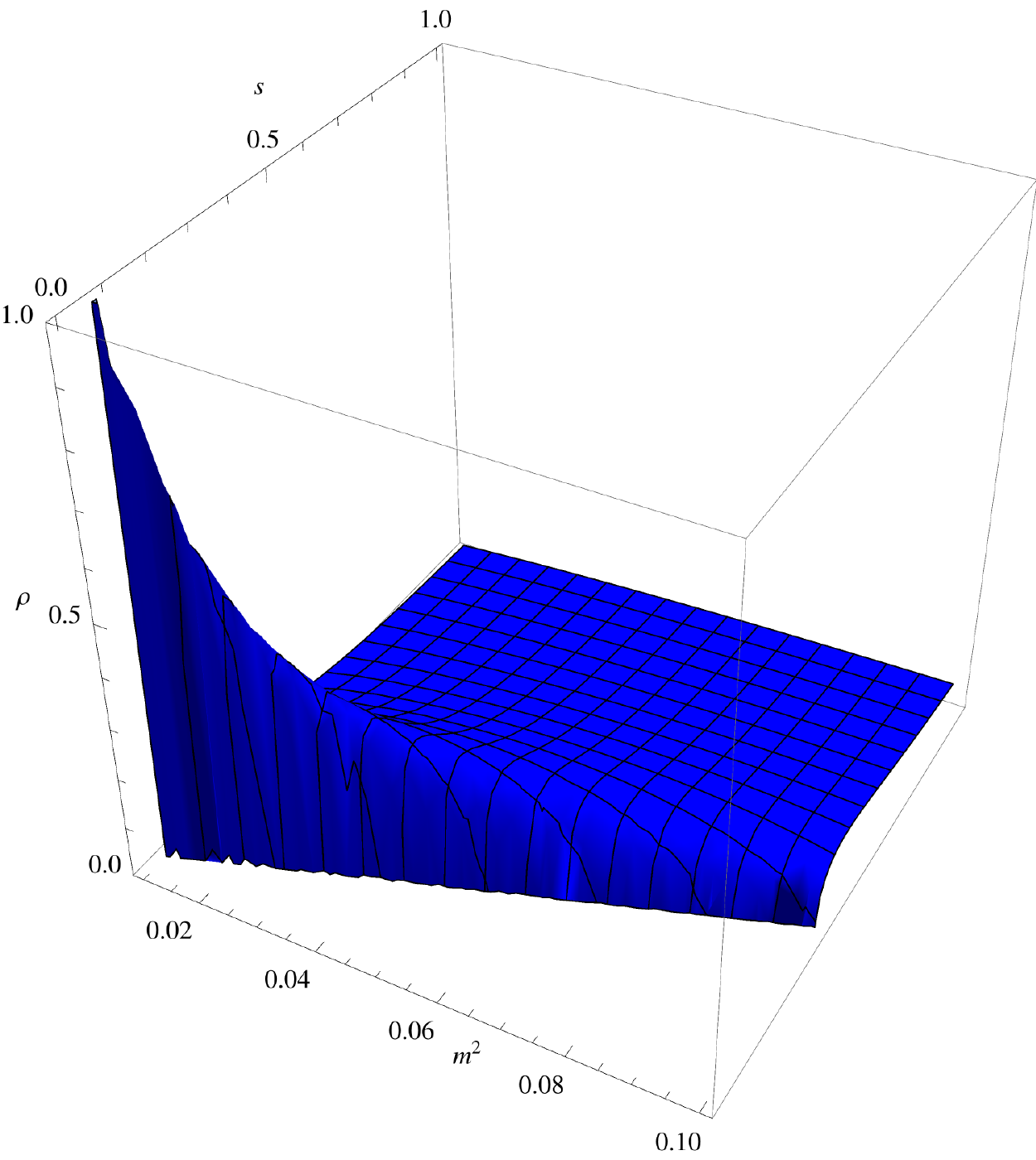}} \hspace{.5cm}
\caption{3-D Plot of the spectral density $\rho_\chi$ in the variables $s$ and $m^2$. }
\end{figure}
As we have anticipated above, a crucial feature of these spectral densities is the existence of a sum rule. In this case it is given by
\beq
\label{superc}
\frac{1}{\pi}\int_{4 m^2}^{\infty}ds{{\rho}_\chi(s,m^2)}=1.
\eeq
At this point, to show the convergence of the family of spectral densities to a resonant behaviour, it is convenient to extract a  
discrete sequence of functions, parameterized by an integer $n$ 
\bea
\rho^{(n)}_\chi(s) &\equiv& \rho_\chi(s,m_n^2)  \qquad \mbox{with} \quad m_n^2 =\frac{4 m^2}{n}.
\eea
One can show that this sequence $\{\rho^{(n)}_\chi\}$ converges to a Dirac delta function 
$\delta(s)$ as $n\to \infty$, which can be written as 
\bea
\lim_{m\to 0} \rho_\chi(s,m^2) = \lim_{m\to 0} \frac{2 \pi m^2}{s^2}\log\left(\frac{1 + \sqrt{\tau(s,m^2)}}{1-\sqrt{\tau(s,m^2)}}\right)\theta(s- 4 m^2) = \pi \delta(s)
\eea
in a distributional sense. 
\section{The analytic structure of $\Phi_2$}
Here we discuss the spectral representation of the second of the form factors appearing in the same $\Gamma_{T}$ and $\Gamma_{S}$ correlators, which is proportional to the renormalized function 
$\Phi_2$ 
\bea
\label{exp11}
\Phi_2(k^2,m^2)&= &  1 - \mathcal B_0(0,m^2) + \mathcal B_0(k^2,m^2) + 2 m^2 \mathcal C_0(k^2,m^2) 
\eea
and, as we have already shown, needs the subtraction of the UV singularities. \\
Clearly, in this case, 
$\Phi_2$ does not admit a dispersive representation, due to its logarithmic divergence at large $k^2$, and, as we are going to show, it is 
characterized just by an ordinary cut for $k^2 > 4 m^2$. We briefly illustrate this point. 

As for $\mathcal C_0(k^2,m^2)$ also in this case we give the three branches of $\mathcal B_0(k^2,m^2)$ in the $k^2<0, 0<k^2<4 m^2$ and $k^2> 4 m^2$ regions
\bea  
 \mathcal B_0(k^2\pm i\epsilon,m^2) = \left\{ 
\begin{array}{ll} 
\frac{2}{\epsilon_{UV}} + 2 - \log\frac{m^2}{\mu^2} + \sqrt{\tau(k^2,m^2)} \log \frac{\sqrt{\tau(k^2,m^2)} -1}{\sqrt{\tau(k^2,m^2)} +1} & \mbox{for}  \quad k^2 < 0 \,, \\
\frac{2}{\epsilon_{UV}} + 2 - \log\frac{m^2}{\mu^2} - 2 \sqrt{- \tau(k^2,m^2)} \arctan{\frac{1}{\sqrt{- \tau(k^2,m^2)}}} & \mbox{for} \quad 0 < k^2 < 4 m^2 \,, \\
\frac{2}{\epsilon_{UV}} + 2 - \log\frac{m^2}{\mu^2} - \sqrt{ \tau(k^2,m^2)} \left( \log \frac{1 + \sqrt{\tau(k^2,m^2)}}{1 - \sqrt{\tau(k^2,m^2)}} \mp i \pi \right )& \mbox{for} \quad k^2 > 4 m^2 \,.  
\end{array} 
\right.
\eea
The discontinuity of the two-point scalar integral $\mathcal B_0(k^2,m^2)$ is then easily computed and it is given by
\bea
 \textrm{Disc}\, \mathcal B_0(k^2,m^2) = \mathcal B_0(k^2 + i\epsilon,m^2) - \mathcal B_0(k^2 - i\epsilon,m^2) = 2 i \pi \sqrt{\tau(k^2,m^2)} \, \theta(k^2 - 4 m^2) \,.
\eea
From the previous equation and from Eq. (\ref{cut}) we extract the discontinuity of $\Phi_2$ which reads as 
 \beq
 \textrm{Disc}\, \Phi_2(k^2,m^2)=   2 i \pi \left( \sqrt{\tau(k^2,m^2)}- \frac{2 m^2}{k^2} \log\frac{1+ \sqrt{\tau(k^2,m^2}) }{1-\sqrt{\tau(k^2,m^2} )}\right)\theta(k^2- 4 m^2).
\eeq
This shows that both $\Phi_1/k^2$ and $\Phi_2$ are characterized by a single 2-particle cut for a nonzero mass deformation.  
It is important to observe that the spectral density of $\Phi_2$ tends to a uniform distribution
\beq
\lim_{m\to 0}\rho_{\Phi_2}(k^2,m^2)= \pi 
\eeq
in the massless limit.
It is obvious, from this analysis, that the spectral density $\rho_{\Phi_2}$ of $\Phi_2$, which characterizes all the non anomalous form factors 
of the correlators that we have investigated, does not satisfy an unsubtracted dispersion relation. There is however a sort of duality between the spectral densities of the two form factors, since while $\rho_{\chi}$ becomes more and more localized 
at $k^2=0$ as $m\to 0$, the opposite is true for the spectral density of the non anomallous form factor $\rho_{\Phi_2}$, as clear from Fig. \ref{phi2spectral}. In this case, as $m$ goes to zero, the flow singles out - in the form factor which is relevant for the anomaly - {\em a single massless state}, while the opposite is true for $\rho_{\Phi_2}$.
\begin{figure}[t]
\centering
\subfigure{\includegraphics[scale=.8]{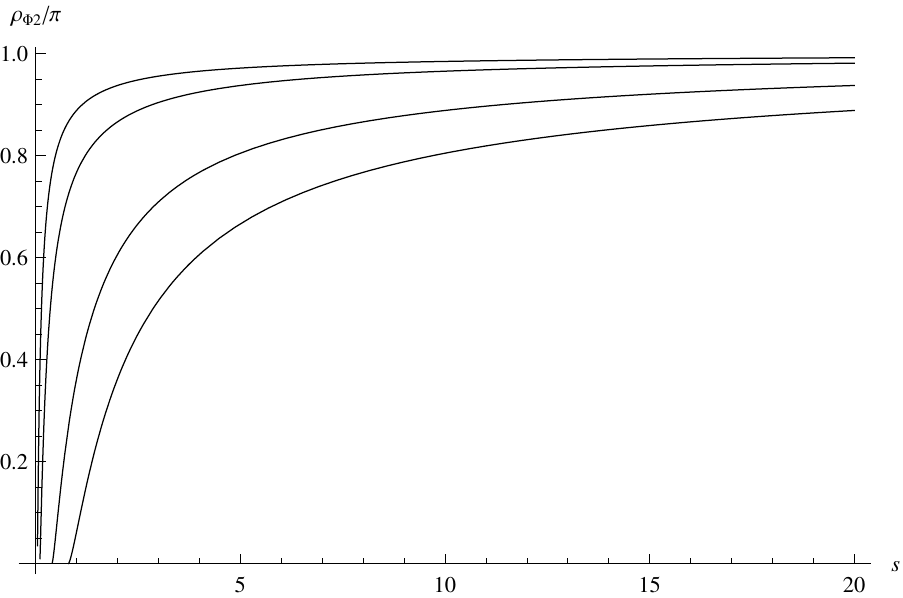}}
\caption{Spectral density flow of $\frac{{{\rho}_{\Phi_2}}}{\pi}(s,m^2)$ versus $s$. As $m^2$ decreases they turn to a unit step function $\theta(s)$. \label{phi2spectral}}
\end{figure}
 \chapter{Constraining the flow: scaling behaviour and sum rules}  
For $k^2$ approaching zero we have
\bea
\label{smallk2limit}
\Phi_1(k^2,m^2) \sim  \frac{1}{12} \frac{k^2}{m^2} + O(k^4/m^4) \,, \qquad \qquad \Phi_2(k^2,m^2) \sim  \frac{1}{12} \frac{k^2}{m^2} + O(k^4/m^4) \,,
\eea
while for a large and negative $k^2$ we find
\bea
\label{largek2limit}
\Phi_1(k^2,m^2) &\sim& - 1 - \frac{m^2}{k^2} \log^2 \frac{-k^2}{m^2} + O(m^4/k^4) \,, \nn \\
\Phi_2(k^2,m^2) &\sim& 3 - \log \frac{-k^2}{m^2} + \frac{m^2}{k^2} \left( 2 + 2  \log \frac{-k^2}{m^2} +  \log^2 \frac{-k^2}{m^2}\right) + O(m^4/k^4) \,.
\eea
The anomaly form factor $\chi=\Phi_1/k^2$ satisfies the relation under rescaling with a constant $\lambda$
\beq
\chi(\lambda k^2,\lambda m^2)=\frac{1}{\lambda}\chi(k^2,m^2)
\eeq
being a homogeneous function. A similar property of homogeneity holds for the spectral density itself 
\beq
\rho_\chi(\lambda s,\lambda m^2)=\frac{1}{\lambda}\rho_\chi(s,m^2),
\label{scale}
\eeq
which under a partial rescaling, involving only the mass parameter $m$, with $m^2 \to m^2/\lambda$ and $\lambda$ large (which is the same as $m\to 0$) has the resonant behaviour
\beq
\label{scaling}
\lim_{\lambda\to\infty}\rho_\chi(s,\frac{m^2}{\lambda})= \pi \delta(s). 
\eeq
At this point, using Eq. (\ref{scale}) a large rescaling of the invariant mass $s$ gives 
\beq
\label{cone}
\rho_\chi(\lambda s, m^2)=\frac{1}{\lambda}\rho_\chi(s, \frac{m^2}{\lambda})\sim \frac {\pi}{\lambda}\delta(s)= \pi \delta(\lambda s),
\eeq
showing that the asymptotic behaviour of $\rho_\chi$ under a rescaling of $s$ with $\lambda$ identifies its support on the $s=0$ region. Notice that Eq. (\ref{cone}) should be interpreted as a light-cone dominance ($s\to 0$) of the asymptotic limit of the correlator as $\lambda$ goes to infinity. \\
One has the constraint
\beq
\label{one1}
k^2 \frac{\partial \chi(k^2,m^2) }{\partial k^2} +m^2 \frac{\partial \chi(k^2,m^2) }{\partial m^2} + \chi(k^2,m^2)=0.
\eeq
Similar conditions are satisfied by the related spectral density $(\rho_\chi)$
\bea
s \frac{\partial \rho_\chi}{\partial s} + m^2 \frac{\partial \rho_\chi }{\partial m^2} +\rho_\chi&=& 0.
\eea
The combination of scaling behaviour and of the sum rule, together with the vanishing of $\rho_\chi(s,m^2)$ at the threshold (i.e. at $s=4 m^2$), induces further constraints on its functional form, for instance 
\beq
\label{eq22}
\frac{1}{\pi} \int_{4 m^2}^\infty \frac{\partial \rho_\chi(s,m^2)}{\partial s} ds =0, \qquad \frac{1}{\pi} \int_{4 m^2}^\infty \frac{\partial \rho_\chi(s,m^2)}{\partial m^2} ds =0,\qquad \frac{1}{\pi}\int_{4 m^2}^\infty s\frac{\partial \rho_\chi(s,m^2)}{\partial s} ds =-f.
\eeq
In the previous equation, and in the following ones, $f$ is a nonzero constant which normalizes the sum rule of the spectral density. For $\rho_\chi$ introduced in the previous section $f=1$. \\
Eq. (\ref{eq22}) can be generalized to give an infinite set of ordinary and superconvergent sum rules 
\bea
\frac{1}{\pi} \int_{4 m^2}^{\infty} ds \,(s - 4 m^2)^n \frac{\partial^n \rho_\chi}{\partial s^n} &=&(-1)^{n} n! f, \qquad n\geq 1\nn\\
\frac{1}{\pi} \int_{4 m^2}^{\infty} ds\,(s - 4 m^2)^n \frac{\partial^{n+1} \rho_\chi}{\partial s^{n+1}} &=& 0.
\eea
Additional constraints come from the scaling relation expanded to second order, 
\beq
s^2  \frac{\partial^2 \rho_\chi}{\partial s^2} + m^4  \frac{\partial^2 \rho_\chi}{\partial ({m^2})^2} + 2 s\, m^2 \frac{\partial^2 \rho_\chi}{\partial s\, \partial m^2}=2 \pi \, f.
\label{duesim}
\eeq
Using the information that the density has only a branch cut for nonzero $m$, integrating over the cut Eq. (\ref{duesim}) we get 
\beq
m^4 \int_{4m^2}^{\infty} ds    \frac{\partial^2 \rho_\chi}{\partial ({m^2})^2} =-2  m^2 \int_{4m^2}^{\infty} ds \, s\frac{\partial^2 \rho_\chi}{\partial s\, \partial m^2}.
\label{pm}
\eeq
At this point, the sign of the dispersive integrals above can be determined by exploiting the derivative form of the sum rule 
\beq
\frac{1}{\pi} \int_{4m^2}^{\infty} ds s \frac{\partial \rho_\chi}{\partial s}=-f,
\eeq
which is satisfied because of the convergence condition of the integral of $\rho_\chi$. Differentiated respect to $m^2$ the sum rule above gives
\beq
 \int_{4m^2}^{\infty} ds \frac{\partial^2 \rho_\chi}{\partial s\, \partial m^2}= 16 m^2 \frac{\partial \rho_\chi}{\partial s}\bigg|_{s=4m^2} \,, 
\eeq
which relates the integral of the mixed derivatives to the spectral density at the threshold.  
Notice that as $m$ goes to zero, 
the density is saturated by the pole behaviour, and it is then clear that it implies the local positivity relation 
\beq
 \frac{\partial^2 \rho_\chi}{\partial s\, \partial m^2} >0  \qquad m\sim 0,
 \eeq
 being the integral dominated just by the region around the threshold  $s\sim 4 m^2 $. Clearly this implies that 
\beq
\label{one2}
 \int_{4m^2}^{\infty} ds    \frac{\partial^2 \rho_\chi}{\partial ({m^2})^2} < 0,
 \eeq
 having used Eq. (\ref{pm}). Also in this case, in the $m\to 0$ limit, the inequality becomes a local condition  $\frac{\partial^2 \rho_\chi}{\partial ({m^2})^2}<0$ which has to be satisfied by the flow. 
 For this purpose we define the contributions of each field to the $\beta$ function of a theory at 1-loop, which for a Dirac fermion and a complex scalar in the representation $R_f$ and $R_s$ respectively, and for a spin 1 in the adjoint are 
 \beq
 \beta(g)=\sum_n \frac{g^3}{16 \pi^2} c^{(n)},
 \eeq
 with
 \beq
 c^{(D)} = \frac{4}{3} T(R_f) \qquad  c^{(A)} =- \frac{11}{3} T(A) \qquad c^{(\phi)} = \frac{1}{3} T(R_s)
 \eeq 
with $T(R_f)$, $T(A)$, $T(R_s)$ being the Dynkin indices of the respective representations. Real scalars and Weyl fermions contribute with an additional factor of $1/2$ respect to complex scalars and Dirac fermions. 
We recall that in a $SU(N)$ $\mathcal{N}=1$ theory, the vector multiplet contributes with
$-11/3 \,T(A)$ and $2/3\, T(A)$ for the gauge field and the gaugino respectively, while the chiral supermultiplet gives $2/3 \, T(R)$ and $1/3 \, T(R)$ for the Weyl fermion and the complex scalar.\\
We use the notation 
\beq
\rho (s,\{m_n^2\}) = \sum_n c^{(n)} \rho_\chi (s,m_n^2)
\eeq
to refer to the total spectral density of a certain theory, with intermediate thresholds at increasing mass values $\{m_n^2\}\equiv(m_1^2,m_2^2, \ldots , m_I^2)$ with
$(m_1< m_2< \ldots < m_I)$, where $I$ counts the total number of degrees of freedom. The corresponding anomaly form factor will be given by 
\beq
\label{th1}
F(Q^2,\{m_n^2\})=\frac{-2}{3 g} \frac{g^3}{16 \pi^2}\sum_n  c^{(n)}\frac{1}{\pi}\int_{4 m_n^2}^\infty ds \frac{\rho_\chi(s,m_n^2)}{s + Q^2}. 
 \eeq
 Notice that if $Q^2 \gg 4 m^2_n$, for a certain mass threshold $n$, then we can set $Q^2=4 m_n^2\lambda$, with $1/\lambda= 4 m_n^2/Q^2 \ll 1$.
Due to scaling, the $n_{th}$ threshold will then contribute to the total form factor with the amount
\beq
F_n(Q^2,m_n^2)=\frac{-2}{3 g} \frac{g^3}{16 \pi^2} c^{(n)} \frac{1}{\pi}\int_{4 m_n^2/\lambda}^\infty ds \frac{\rho_\chi(s,m_n^2/\lambda)}{s + 4 m^2 \lambda}, 
\eeq
which in the $1/\lambda \ll 1$ limit will give
\beq
\label{inter}
F_n(Q^2,m_n^2) \sim \frac{-2}{3 g} \frac{g^3}{16 \pi^2} c^{(n)} \int_{0}^\infty ds\frac{\delta(s)}{\lambda(s + 4 m_n^2)} 
= \frac{-2}{3 g}\beta^{(n)}(g)\frac{1}{Q^2}.
\eeq
Eq. (\ref{inter}) reduces to the anomaly pole contribution times the contribution of the state $(n)$ to the expression of the total $\beta$ function. 
 As $Q^2$ grows larger than any intermediate scale, the total spectral density $\rho$ in the dispersive integral is asymptotically given by the expression
 \bea
 \rho(s, \{m_n^2\}) \sim  \sum_n  c^{(n)}  \delta(s)  = \frac{16 \pi^2}{g^3} \beta(g) \pi \delta(s) \,,
 \eea
where we have used Eq. (\ref{scaling}).
 Notice that  $\rho(s, \{m_n^2\})$ satisfies a total sum rule to which contribute all the intermediate thresholds for $0<s<\infty$ 
 \bea
\frac{1}{\pi} \int_{0}^\infty ds \, \rho(s, \{m_n^2\})&=&\sum_n c^{(n)} \frac{1}{\pi} \int_{4 m_n^2}^\infty ds \, \rho_\chi(s,m_n^2) =  \frac{16 \pi^2}{g^3} \beta(g) 
\eea
 In supersymmetric theories this function is the only one which developes a resonant behaviour at the conformal point and satisfies a sum rule, as we have pointed out. The sum of the densities stripped of the gauge factors, integrated over the thresholds 
  \beq
  \label{th2}
\frac{1}{\pi}\sum_n \int_{4 m_n^2}^\infty ds \, \rho_\chi (s,m_n^2)= I  
\eeq
simply count the number of degrees of freedom $(I)$.\\
Notice that the analysis of this section related to Eqs. (\ref{one1}-\ref{one2}) remains valid also for any form factor which is characterized by a finite (non superconvergent) sum rule. The asymptotic analysis discussed in Eqs. (\ref{th1}-\ref{th2}), can be also easily extended to cases unrelated to the anomaly, with coefficients $c^{(n)}$ replaced by some new coefficients, not related to the $\beta$ function.

\chapter{RG equations}
\label{RGs}
We list the RG equations at one-loop for the dimensionless coupling $\lambda_{T, S, TS}$, $\kappa, g_Y, g_L, g_c, y_{t, b}$. Here $t=\ln{\frac{\mu}{\mu_0}}$, where $\mu$ is the running scale and $\mu_0$ is the initial scale. 
\begin{align}
g_Y'(t)&=\frac{33}{80 \pi ^2}g^3_y(t),\\
g_L'(t)&=\frac{3}{16 \pi ^2}g^3_w(t),\\
g_c'(t)&=-\frac{3}{16 \pi ^2}g^3_c(t)
\end{align}
\begin{align}
y_t'(t)&=\frac{1}{16 \pi ^2}\Bigg(3
   y^3_ t(t)+y^2_b(t) y_t(t)\\
   &+\Bigg(-\frac{13}{15} g^2_Y(t)-3
   g^2_L(t)-\frac{16 g^2_c(t)}{3}+\frac{3 \lambda^2_T(t)}{2}+\lambda^2_S (t)+3y_t^2(t)\Bigg)
   y_t(t)\Bigg),\nn\\
 y_b'(t)&=\frac{1}{16 \pi ^2}\Bigg(3
   y^3_ t(t)+y^2_t(t) y_b(t)\\
   &+\Bigg(-\frac{13}{15} g^2_Y(t)-3
   g^2_L(t)-\frac{16 g^2_c(t)}{3}+\frac{3 \lambda^2_T(t)}{2}+\lambda^2_S (t)+3y_b^2(t)\Bigg)
   y_b(t)\Bigg)\nn,
\end{align}
\begin{align}
\lambda_S '(t)&=\frac{1}{16 \pi^2} \Bigg(4 \lambda^3_S (t)-\frac{3}{5}
   g^2_Y(t) \lambda_S (t)-3 g^2_L(t) \lambda_S (t)\\
   &+3 \lambda^2_T(t) \lambda_S (t)+6 \lambda^2_{TS}(t)
   \lambda_S (t)+2 \kappa^2(t) \lambda_S (t)+3
   \Big(y_t^2 (t)+y^2_b (t)\Big) \lambda_S (t)\Bigg),\nn
\end{align}
\begin{align}
\lambda_T'(t)&=\frac{1}{16 \pi^2} \Bigg(4 \lambda^3_T(t)-\frac{3}{5} g^2_Y(t) \lambda_T(t)-7
   g^2_L(t) \lambda_T(t)\\
   &+4 \lambda^2_{TS}(t) \lambda_T(t)+2 \lambda^2_S (t) \lambda_T(t)+3
   \Big(y_t^2 (t)+y^2_b (t)\Big) \lambda_T (t)\Bigg),\nn\\
\kappa '(t)&=\frac{1}{8 \pi
   ^2}3  \kappa (t) \Bigg(3 \lambda^2_{TS}(t)+\kappa^2(t)+\lambda^2_S (t)\Bigg),\\
\lambda_{TS}'(t)&=\frac{1}{8 \pi ^2}\lambda_{TS}(t)  \Bigg(-4 g^2_L(t)+\lambda^2_T(t)+7
   \lambda^2_{TS}(t)+\kappa^2(t)+\lambda^2_S (t)\Bigg).\nn
\end{align}

\chapter{Mass matrices of the Higgs sector}\label{Higgsm}
The symmetric mass matrices of the Higgs sector are given by
\bea\label{sMM}
\renewcommand{\arraystretch}{1.2}
\mathcal{M}^S=\left(
\begin{array}{cccc}
m^S_{11}&m^S_{12}&m^S_{13}&m^S_{14}\\
 &m^S_{22}&m^S_{23}&m^S_{24}\\
 & &m^S_{33}&m^S_{34}\\
 & & &m^S_{44}
\end{array}
\right),
\eea
where
\footnotesize
\begin{align}
m^S_{11}&=\frac{1}{4 v_u}\Big(2 v_d \big(\sqrt{2} A_S v_S-v_T \left(A_T+\sqrt{2} v_S \lambda _T \lambda_{TS}\right)+\lambda _S \big(\kappa  v_S^2+v_T^2 \lambda _{TS}\big)\big)+v_u^3
   \big(g_L^2+g_Y^2\big)\Big)\nn\\
m^S_{12}&=\frac{1}{2} \Big(-\sqrt{2} A_S v_S+v_T \left(A_T+\sqrt{2} v_S \lambda _T \lambda _{TS}\right)-\lambda_S \left(\kappa  v_S^2+v_T^2 \lambda _{TS}\right)\Big)-\frac{1}{4} v_d v_u
   \left(g_L^2+g_Y^2-2 \left(2 \lambda _S^2+\lambda _T^2\right)\right)\nn\\
m^S_{13}&=-\frac{A_S v_T}{\sqrt{2}}+v_d \left(\frac{\lambda _T v_T \lambda _{TS}}{\sqrt{2}}-\kappa 
   \lambda_S v_S\right)+\frac{1}{2} v_u \lambda _S \left(2 \lambda _S v_S-\sqrt{2} \lambda _T v_T\right)\nn\\
m^S_{14}&=\frac{1}{2} \Big(v_d \left(A_T-2 \lambda _S v_T \lambda _{TS}\right)+\sqrt{2} v_S \lambda_T \left(v_d \lambda _{TS}-v_u \lambda _S\right)+v_u \lambda _T^2 v_T\Big)\nn\\
m^S_{22}&=\frac{1}{4 v_d}\Big(2 v_u \Big(\sqrt{2} A_S v_S-v_T \left(A_T+\sqrt{2} v_S \lambda _T \lambda _{TS}\right)+\lambda _S \left(\kappa  v_S^2+v_T^2 \lambda _{TS}\right)\Big)+v_d^3
   \left(g_L^2+g_Y^2\right)\Big)\nn\\
m^S_{23}&=-\frac{A_S v_u}{\sqrt{2}}+\frac{1}{2} v_d \lambda_S \left(2 \lambda_S v_S-\sqrt{2} \lambda _T   v_T\right)+v_u \left(\frac{\lambda _T v_T \lambda _{TS}}{\sqrt{2}}-\kappa  \lambda _S   v_S\right)\nn\\
m^S_{24}&=\frac{1}{2} \left(v_u \left(A_T-2 \lambda _S v_T \lambda _{TS}\right)+\sqrt{2} v_S \lambda_T \left(v_u \lambda _{TS}-v_d \lambda _S\right)+v_d \lambda _T^2 v_T\right)\nn\\
m^S_{33}&=\frac{1}{4 v_S}\Big(\sqrt{2} v_T \left(\lambda _T \left(\lambda _S \left(v_d^2+v_u^2\right)-2 v_d
   v_u \lambda _{TS}\right)-2 A_{TS} v_T\right)+2 \sqrt{2} A_S v_d v_u+2
   \sqrt{2} A_{\kappa} v_S^2+8 \kappa ^2 v_S^3\Big)\nn\\
m^S_{34}&=\frac{1}{4} \Big(4 \sqrt{2} A_{TS} v_T-\sqrt{2} \lambda _S \lambda _T
   \left(v_d^2+v_u^2\right)+2 \lambda _{TS} \left(\sqrt{2} v_d v_u \lambda _T+4
   v_S v_T \left(\kappa +2 \lambda _{TS}\right)\right)\Big)\nn\\
m^S_{44}&=\frac{1}{4 v_T}\Big(-2 v_d v_u \left(A_T+\sqrt{2} v_S \lambda _T \lambda _{TS}\right)+\sqrt{2}
   v_d^2 \lambda _S v_S \lambda _T+\sqrt{2} v_u^2 \lambda _S v_S \lambda _T+8 v_T^3 \lambda
   _{TS}^2\Big)\nn
\end{align}

\normalsize
\bea\label{pMM}
\renewcommand{\arraystretch}{1.2}
\mathcal{M}^P=\left(
\begin{array}{cccc}
m^P_{11}&m^P_{12}&m^P_{13}&m^P_{14}\\
 &m^P_{22}&m^P_{23}&m^P_{24}\\
 & &m^P_{33}&m^P_{34}\\
 & & &m^P_{44}
\end{array}
\right),
\eea
with
\footnotesize

\begin{align}
m^P_{11}&=\frac{v_d}{2 v_u}\Big( \left(\sqrt{2} A_S v_S-v_T \left(A_T+\sqrt{2} v_S \lambda _T \lambda
   _{TS}\right)+\lambda _S \left(\kappa  v_S^2+v_T^2 \lambda _{TS}\right)\right)\Big)\nn\\
m^P_{12}&=\frac{1}{2} \left(\sqrt{2} A_S v_S-v_T \left(A_T+\sqrt{2} v_S \lambda _T \lambda _{TS}\right)+\lambda
   _S \left(\kappa  v_S^2+v_T^2 \lambda _{TS}\right)\right)\nn\\
m^P_{13}&=\frac{1}{2} v_d \left(\sqrt{2} A_S-2 \kappa  \lambda _S v_S+\sqrt{2} \lambda _T v_T \lambda
   _{TS}\right)\nn\\
m^P_{14}&=-\frac{1}{2} v_d \left(A_T+\lambda _{TS} \left(2 \lambda _S v_T-\sqrt{2} v_S \lambda
   _T\right)\right)\nn\\
m^P_{22}&=\frac{v_u}{2 v_d}\Big( \left(\sqrt{2} A_S v_S-v_T \left(A_T+\sqrt{2} v_S \lambda _T \lambda
   _{TS}\right)+\lambda _S \left(\kappa  v_S^2+v_T^2 \lambda _{TS}\right)\right)\Big)\nn\\
m^P_{23}&=\frac{1}{2} v_u \left(\sqrt{2} A_S-2 \kappa  \lambda _S v_S+\sqrt{2} \lambda _T v_T \lambda
   _{TS}\right)\nn\\
m^P_{24}&=-\frac{1}{2} v_u \left(A_T+\lambda _{TS} \left(2 \lambda _S v_T-\sqrt{2} v_S \lambda
   _T\right)\right)\nn\\
m^P_{33}&=\frac{v_T}{4 v_S} \Big(\Big(\sqrt{2} \lambda _T \left(\lambda _S \left(v_d^2+v_u^2\right)-2 v_d
   v_u \lambda _{TS}\right)-2 v_T \left(\sqrt{2} A_{TS}+4 \kappa  v_S \lambda
   _{TS}\right)\Big)+2 \sqrt{2} A_S v_d v_u-6 \sqrt{2} A_\kappa v_S^2\nn\\
   &+8 \kappa 
   v_d v_u \lambda _S v_S\Big)\nn\\
m^P_{34}&=\frac{1}{4} \left(-4 \sqrt{2} A_{TS} v_T-\sqrt{2} \lambda _T \left(\lambda _S
   \left(v_d^2+v_u^2\right)+2 v_d v_u \lambda _{TS}\right)+8 \kappa  v_S
   v_T \lambda _{TS}\right)\nn\\
m^P_{44}&=\frac{-2 v_d v_u}{4
   v_T}\Big( \left(A_T+\lambda _{TS} \left(\sqrt{2} v_S \lambda _T-4 \lambda _S
   v_T\right)\right)+v_S \left(\sqrt{2} v_d^2 \lambda _S \lambda _T-8 v_T \left(\sqrt{2}
   A_{TS}+\kappa  v_S \lambda _{TS}\right)\right)+\sqrt{2} v_u^2 \lambda _S v_S \lambda _T\Big)\nn
\end{align}

\normalsize
\bea\label{chMM}
\renewcommand{\arraystretch}{1.2}
\mathcal{M}^C=\left(
\begin{array}{cccc}
m^C_{11}&m^C_{12}&m^C_{13}&m^C_{14}\\
 &m^C_{22}&m^C_{23}&m^C_{24}\\
 & &m^C_{33}&m^C_{34}\\
 & & &m^C_{44}
\end{array}
\right),
\eea
where 
\footnotesize
\begin{align}
m^C_{11}&=\frac{1}{4} \Big(2 \Big(\sqrt{2} v_S \left(A_S \cot\beta+\lambda _T v_T \left(2 \lambda_{S}-\cot\beta \lambda
   _{TS}\right)\right)+\cot\beta v_T \left(\lambda_{S} v_T \lambda
   _{TS}-A_{T}\right)+\kappa  \cot\beta \lambda_{S} v_S^2\Big)\nn\\
   &+\cos ^2\beta\, v^2 \left(g_L^2-2
   \lambda_{S}^2+\lambda _T^2\right)\Big)\nn\\
m^C_{12}&=\frac{1}{4} v \Big(\lambda _T \left(2 v_S \left(\sin\beta \lambda_{S}-2 \cos\beta \lambda _{TS}\right)+\sqrt{2} \sin\beta \lambda _T v_T\right)-\sqrt{2} \sin\beta g_L^2 v_T\Big)\nn\\
m^C_{13}&=\frac{1}{4} \Big(2 \Big(v_T \left(A_{T}+\lambda_{TS} \left(\lambda_{S} v_T+\sqrt{2} v_S
   \lambda _T\right)\right)+\sqrt{2} A_S v_S+\kappa  \lambda_{S} v_S^2\Big)+\sin\beta \cos\beta v^2
   \left(g_L^2-2 \lambda_{S}^2+\lambda _T^2\right)\Big)\nn\\
m^C_{14}&=\frac{v}{4}\left(\sin\beta \left(\sqrt{2} v_T \left(g_L^2-\lambda _T^2\right)+2 \lambda_{S} v_S
   \lambda _T\right)-2 \sqrt{2} A_{T} \cos\beta\right)\nn\\
m^C_{22}&=\frac{1}{4 v_T}\Big(v_T \Big(v^2 \left(\cos (2 \beta ) \left(g_L^2-\lambda _T^2\right)+2 \sin (2 \beta ) \lambda_{S}
   \lambda_{TS}\right)-4 v_S \left(\sqrt{2} A_{TS}+\kappa  v_S \lambda_{TS}\right)\Big)-A_{T}
   \sin (2 \beta ) v^2\nn\\
   &+2 v_T^3 \left(g_L^2-2 \lambda_{TS}^2\right)+\sqrt{2} v^2 v_S
   \lambda _T \left(\lambda_{S}-\sin (2 \beta ) \lambda_{TS}\right)\Big)\nn\\
m^C_{23}&=\frac{v}{4}\left(2 \sqrt{2} A_{T} \sin\beta+\cos\beta \left(\sqrt{2} v_T \left(\lambda
   _T^2-g_L^2\right)-2 \lambda_{S} v_S \lambda _T\right)\right)\nn\\
m^C_{24}&=\sqrt{2} A_{TS} v_S-\frac{1}{2} g_L^2 v_T^2+\lambda_{TS} \big(\kappa  v_S^2+v_T^2 \lambda_{TS}- \sin\beta \cos\beta
   v^2 \lambda_{S}\big)\nn\\
m^C_{33}&=\frac{1}{4} \Big(2 \Big(\sqrt{2} v_S \left(A_S \tan\beta+\lambda _t v_T \left(2 \lambda_{S}-\tan\beta \lambda_{TS}\right)\right)+\tan\beta v_T \left(\lambda_{S} v_T \lambda
   _{TS}-A_{T}\right)+\kappa  \tan\beta \lambda_{S} v_S^2\Big)\nn\\
   &+\sin ^2\beta g_L^2 v^2+\sin^2\beta v^2 \left(\lambda _T^2-2 \lambda_{S}^2\right)\Big)\nn\\
m^C_{34}&=\frac{v}{4}\Big(\cos\beta \left(\sqrt{2} v_T \left(g_L^2-\lambda _T^2\right)-2 \lambda_{S} v_S
   \lambda _T\right)+4 \sin\beta v_S \lambda _T \lambda_{TS}\Big)\nn\\
m^C_{44}&=\frac{1}{4 v_T}\Big(v_T \Big(v^2 \left(\cos (2 \beta ) \left(\lambda _T^2-g_L^2\right)+2 \sin (2 \beta ) \lambda_{S}
   \lambda_{TS}\right)-4 v_S \left(\sqrt{2} A_{TS}+\kappa  v_S \lambda_{TS}\right)\Big)-A_{T}
   \sin (2 \beta ) v^2\nn\\
   &+2 v_T^3 \left(g_L^2-2 \lambda_{TS}^2\right)+\sqrt{2} v^2 v_S
   \lambda _T \left(\lambda_{S}-\sin (2 \beta ) \lambda_{TS}\right)\Big)\nn
\end{align}
\normalsize
As already explained, the massive eigenvectors of the charged mass matrix are function of all the parameters of the model, including the parameters that are related to the singlet, e.g. $v_S$, $\lambda_S$, $\kappa$, whereas the Goldstone eigenvector is a function of the doublets and triplet vev only.
This is also true for for the eigenvectors of the pseudoscalar mass matrix. In this case the Goldstone eigenvector is a function of the doublets vev only.

%%%%%%%%%%%%%%%

%%%%%%%%%%%%%%
\end{document}